%% file: main.tex
\documentclass[numbers]{assets/matterlab}

\usepackage{amsmath}
\usepackage{amssymb}
\usepackage{array}
\usepackage{longtable}
\usepackage{tabularx}
\usepackage{pifont}
\usepackage{colortbl}

\definecolor{bhblue}{HTML}{B14F4F}
\definecolor{lihorange}{HTML}{443983}
\definecolor{behgreen}{HTML}{64A800}
\definecolor{h2ored}{HTML}{2A788E}

\providecommand{\doi}[1]{}
\renewcommand{\doi}[1]{\href{https://doi.org/#1}{\nolinkurl{doi:#1}}}
\providecommand{\mydoi}[1]{}
\renewcommand{\mydoi}[1]{\href{https://doi.org/#1}{\nolinkurl{doi:#1}}}
\newcommand{\citelist}[1]{\cite{#1}}
\newcolumntype{L}[1]{>{\raggedright\arraybackslash}p{#1}}
\newcolumntype{C}[1]{>{\centering\arraybackslash}p{#1}}
\newcolumntype{Y}{>{\raggedright\arraybackslash}X}

\newif\ifincludeappendixlogs
\includeappendixlogsfalse
\newcommand{\externallogsupplement}{external log supplement}
\DeclareRobustCommand{\externallogsreference}{the \externallogsupplement~\cite{mantillacalderon2026gsoptlogs}}
\newcommand{\vqelogreference}{\ifincludeappendixlogs Appendix~B1\else \externallogsreference\fi}
\newcommand{\tnlogreference}{\ifincludeappendixlogs Appendix~B2\else \externallogsreference\fi}
\newcommand{\afqmclogreference}{\ifincludeappendixlogs Appendix~B3\else \externallogsreference\fi}
\newif\ifmatterlabfigureplacement
\matterlabfigureplacementtrue

\title{Optimizing ground state preparation protocols\\ with autoresearch}

\author[1,2,3,*]{Luis Mantilla Calder\'on}
\author[1]{J\'er\^ome F. Gonthier}
\author[4]{Ignacio Gustin}
\author[2,4,7,*]{Varinia Bernales}
\author[1,2,3,4,5,6,7,*]{Al\'an Aspuru-Guzik}

\affiliation[1]{NVIDIA Corporation, Santa Clara, CA, USA}
\affiliation[2]{Department of Computer Science, University of Toronto, Toronto, ON, Canada}
\affiliation[3]{Vector Institute for Artificial Intelligence, Toronto, ON, Canada}
\affiliation[4]{Department of Chemistry, University of Toronto, Toronto, ON, Canada}
\affiliation[5]{Department of Materials Science \& Engineering, University of Toronto, Toronto, ON, Canada}
\affiliation[6]{Department of Chemical Engineering \& Applied Chemistry, University of Toronto, Toronto, ON, Canada}
\affiliation[7]{Acceleration Consortium and CIFAR, Toronto, ON, Canada}

\abstract{
\input{includes/manuscript-abstract.tex}
}

\date{\today}
\correspondence{\parbox[t]{0.50\linewidth}{\raggedright \href{mailto:luis@cs.toronto.edu}{luis@cs.toronto.edu}\\ \href{mailto:varinia@bernales.org}{varinia@bernales.org}\\ \href{mailto:aspuru@nvidia.com}{alan@aspuru.com}}}
\metadata[Keywords]{\parbox[t]{0.50\linewidth}{\raggedright Scientific machine learning, quantum computing, tensor networks, quantum Monte Carlo, coding agents}}

\begin{document}

\maketitle

\input{includes/manuscript-body.tex}

\section*{Acknowledgments}
\input{includes/manuscript-acknowledgments.tex}

\section*{Code Availability}
\input{includes/manuscript-code-availability.tex}

{
\small
\bibliographystyle{assets/unsrtnat}
\bibliography{bibliography}
}

\input{includes/manuscript-supplement.tex}

\end{document}

%% file: includes/manuscript-abstract.tex
Artificial intelligent language-model based coding agents have significantly changed the way we interact with computers in our day-to-day, as it is common to use them to create, improve, and run programming scripts only using natural language. Agent code updates can be better guided when such programs can be executed and scored automatically rather than judged by human preference. In quantum computing and classical quantum simulation settings, ground-state preparation has a parallel structure: candidate protocols can be ranked by estimated energies and other proxies indicating proper quantum-state convergence. In this work, we study how autoresearch, a code optimization strategy based on coding agents, can be used to optimize hyperparameter choices of different ground-state preparation and sampling protocols, including the variational quantum eigensolver (VQE), density matrix renormalization group (DMRG), and auxiliary-field quantum Monte Carlo (AFQMC). We validate the viability and capacity of this method on simple spin models and molecular Hamiltonians. Across all three settings, the agent mutates simple baselines into complex protocols with improved energy proxies while operating under constrained space-time computational budgets. We conclude with discussions of other quantum routines that support executable scalar scoring, enabling evolutionary coding agents to automate a substantial portion of the protocol-tuning work that would otherwise be required manually.

%% file: includes/manuscript-body.tex
\section{Introduction}
Recent progress in evolutionary coding agents that employ large language models (LLMs) has clarified a simple principle: once candidate programs can be executed and scored automatically, search and editions over codebases can produce real algorithmic improvement. AlphaTensor~\cite{fawzi2022alphatensor} and AlphaDev~\cite{mankowitz2023alphadev} used reinforcement learning to improve matrix multiplication and low-level sorting routines, while FunSearch~\cite{romeraparedes2024funsearch} showed that large language models can serve as mutation operators inside an evolutionary loop over executable programs. Related work on reinforcement learning with verifiable rewards argues that outcome-level reward signals can themselves incentivize correct reasoning when evaluation is automatically checkable~\cite{wen2026rlvr}. Follow-on systems extended the same logic to prompt evolution, heuristic search, reflective mutation, and codebase-scale editing, including EvoPrompting~\cite{chen2023evoprompting}, Evolution of Heuristics~\cite{liu2024eoh}, GEPA~\cite{agrawal2025gepa}, ReEvo~\cite{ye2024reevo}, AlphaEvolve~\cite{novikov2025alphaevolve, openevolve2025}, TTT-Discover~\cite{yuksekgonul2026tttdiscover}, and Darwin G\"odel Machine~\cite{zhang2025darwin}.

That coding-agent trajectory is now blending into a broader class of scientific co-worker systems that plan, execute, and interpret domain workflows rather than only emitting standalone code. Recent examples span physics, chemistry, and general scientific discovery, including Coscientist~\cite{boiko2023autonomouschem}, Organa~\cite{darvish2025organa}, AI co-scientist~\cite{gottweis2025aicoscientist}, El Agente~\cite{zou2025elagente,gustin2025cuantico,perezsanchez2026quntur,choi2026estructural,kumar2026solido,bai2026grafico}, and many others capable of automating research~\cite{melnikov2018active, menke2019automated, cao2024agents, mitchener2025kosmos, arlt2025towards, krenn2025digital, lu2026automationairesearch, arlt2026meta}. These domain-specific AI agents have been made possible in large part by significant improvements in foundation models, such as GPT Thinking~\cite{openai2026gpt54systemcard}, Claude Opus~\cite{anthropic2026opus46systemcard}, Nemotron Super~\cite{nvidia2026nemotron3superopen}, and similar models framed around long-horizon coding and agentic tool use.

This paper explores how coding agents can optimize quantum and quantum-classical many-body state preparation protocols. The main goal is to propose benchmark problems for which an AI agent can make code changes, run the resulting program, and receive an honest scalar signal from the underlying physics. Andrej Karpathy's autoresearch loop is a useful software template for that setup because it separates candidate generation from evaluation and keeps the optimization trace auditable~\cite{karpathy2026autoresearch}. We adapt this software to the needs of ground-state optimization and use it as an agent skill to improve simple state-preparation protocols.

We instantiate that idea in three benchmark families. First, we let the agent edit the code defining a variational quantum eigensolver (VQE)~\citelist{peruzzo2014vqe,mcclean2016theory,cao2019quantumchem, bharti2022noisy} ansatz and its optimizer for molecular active-space problems. Second, we let it tune one-dimensional density matrix renormalization group (DMRG)~\citelist{ostlund1995mps,schollwock2011dmrg,orus2014tnreview} protocols for spin-chain Hamiltonians. Third, we let it tune auxiliary-field quantum Monte Carlo (AFQMC)~\citelist{blankenbecler1981bss,scalapino1981bss, zhang2003phafqmc} trial-states and propagation choices under a fixed compute budget. Across all three settings, the workflow is the same; we packaged it into an installable agent skill called \texttt{gsopt}. Such skill mutates executable code, obtains energy and energy-related metrics, logs meaningful values, and retains those code changes that improve a physically meaningful score.

\section{Methods}
The autoresearch framework~\cite{karpathy2026autoresearch} follows a simple pattern: generate a candidate configuration or code edit, execute the benchmark, record a scalar score, and retain improvements in a persistent execution history. Specifically, each scored candidate is stored with its outcome, so the optimization trace can be inspected later. A practical advantage of coding-agent systems in this setting is that they can also search online documentation, codebases, and related technical references for implementation ideas, which helps the search adapt to the constraints and opportunities of a specific hardware and software stack. The important design decision is not the specific proposal heuristic but the evaluator. Each task in this paper is constructed so that the score is relevant and scientifically interpretable. That is what makes autonomous search meaningful rather than ambiguous.

Ground-state preparation is particularly well-suited to this loop because the solver typically exposes a scalar proxy for overlap with the ground state. At a coarse level, most scalable ground-state methods are built from a few recurring principles: variational minimization, projector or filtering dynamics in Krylov space or imaginary time, continuation or annealing along a Hamiltonian path, and self-consistent or renormalization-style fixed-point updates~\citelist{lanczos1950iteration,blankenbecler1981bss,scalapino1981bss,kato1950adiabatic,farhi2000adiabatic,kirkpatrick1983annealing,hartree1928scf,fock1930hf,white1992dmrg,white1993dmrg}. These algorithmic families have several similarties: Hartree--Fock is variational but solved self-consistently, DMRG is variational with a renormalization interpretation, and quantum Monte Carlo (QMC) combines projection with stochastic sampling. What matters for autoresearch is that each family yields an executable quantity that improves as the state improves.

For variational methods such as the variational quantum eigensolver (VQE) and tensor-network ground-state solvers, that quantity is the Rayleigh quotient~\cite{ritz1909variationsprobleme},
\begin{equation}
E[\psi] = \frac{\langle \psi | H | \psi \rangle}{\langle \psi | \psi \rangle} \ge E_0,
\end{equation}
where $E_0$ is the exact ground-state energy of the target Hamiltonian $H$. This makes the online objective simple: lower variational energy means a better candidate state under the same evaluation budget.

AFQMC uses the same philosophy but a different numerical mechanism. Starting from a trial state $|\Psi_T\rangle$, the benchmark solver projects toward the ground state in imaginary time,
\begin{equation}
|\Psi_0\rangle \propto \lim_{\beta \to \infty} e^{-\beta(H-E_T)} |\Psi_T\rangle.
\end{equation}
After a Hubbard--Stratonovich transformation~\citelist{blankenbecler1981bss,scalapino1981bss}, AFQMC becomes a stochastic propagation of Slater-determinant walkers whose stability and residual bias depend on the trial state and on the constrained-path~\cite{zhang1995cpmc} or phaseless~\citelist{zhang2003phafqmc,alsaidi2006afqmc} approximation. The online objective is therefore not a formal upper bound. Instead, at a fixed stochastic budget, we prefer candidates with lower post-equilibration energy and smaller fluctuations in the final sampled blocks.

At the level of the autoresearch loop, both variational and projector-based benchmarks fit the same constrained-search picture. If $\mathcal{B}$ denotes a chosen hardware and software stack, the search can be written as the constrained minimization problem
\begin{equation}
\begin{aligned}
\min_{\theta,c}\quad & \mathcal{L}_H(\theta,c;\mathcal{B}) \\
\text{s.t.}\quad & |\psi(\theta,c;\mathcal{B})\rangle \in \mathcal{S}(\mathcal{B}), \\
& \tau(\theta,c;\mathcal{B}) \le T,
\end{aligned}
\end{equation}
where $c$ is a collection of code-level choices such as the ansatz, update rule, compilation settings, or schedule, $\mathcal{S}(\mathcal{B})$ is the set of states realizable on backend $\mathcal{B}$, $T$ is the wall-time budget, and $\mathcal{L}_H$ is an energy proxy for closeness to the ground state. This makes explicit that the goal is not unconstrained energy minimization, but the search for the best score that can actually be realized on a chosen backend within a fixed execution budget. Although Eq.~(3) is written with a wall-time cap because all experiments in this manuscript are scored in that manner, the same constrained-search view can absorb other scarce resources as well, including memory footprint, qubit count, quantum circuit shot budget, non-Clifford and T-gate count, or any other expensive resource of relevance. This broader perspective is the same one used in resource-aware algorithm and compiler optimization, where search is organized around constrained cost rather than a single unconstrained runtime objective~\citelist{fawzi2022alphatensor,mankowitz2023alphadev,amy2014tdepth,nakaji2025qcidaemon}. In the current VQE and tensor-network experiments, the live score is simply the final energy achieved within the allotted budget, while AFQMC uses a fixed-budget post-equilibration score that favours lower energy and smaller fluctuations. This is the quantity actually optimized by the autoresearch loop across all benchmarks.

Related work already shows that the problems we consider in this paper can benefit from automated design. In VQE, these techniques include: adaptive ansatz growth (ADAPT-VQE)~\cite{grimsley2019adaptvqe}, evolutionary eigensolvers (EVQE)~\cite{rattew2019evqe}, circuit-architecture search~\cite{du2022qcas}, Meta-VQE~\cite{cerveralierta2021metavqe}, generative quantum eigensolvers~(GQE)~\cite{nakaji2024gqe}, LLM-optimized VQE~\cite{finger2026automated}, all of which optimize an ansatz based on an energy signal. In tensor networks, the density matrix renormalization group (DMRG)~\citelist{white1992dmrg,white1993dmrg} and its reinterpretation as matrix product states (MPS)~\citelist{ostlund1995mps,schollwock2011dmrg} motivate broader ansätze such as tree tensor networks (TTN)~\cite{shi2006ttn}, multi-scale entanglement renormalization ansatz (MERA)~\cite{vidal2007mera}, and projected entangled pair states (PEPS)~\cite{verstraete2008peps,orus2014tnreview}. Tensor-network methods can also be adapted via automated parameter and structural optimization, including entanglement-based topology selection~\cite{murg2015tree}, automatic structural updates~\cite{hikihara2023automatic}, and differentiable-programming frameworks for end-to-end gradient-based optimization~\cite{liao2019differentiable}. In AFQMC, determinant auxiliary-field formulations~\citelist{blankenbecler1981bss,scalapino1981bss} and later constrained-path and phaseless variants~\citelist{zhang1995cpmc,zhang2003phafqmc,alsaidi2006afqmc} make the trial wavefunction central to bias and efficiency, motivating automated refinement strategies from symmetry-aware trial states~\cite{shi2013symmetry} to self-consistent constraint optimization~\cite{qin2016coupling} and algorithmic construction of multideterminant trial states~\cite{mahajan2022selected, qin2023self, sukurma2025self}.

\ifmatterlabfigureplacement
\newcommand{\mattervqeprogressfigure}{%
\begin{figure}[!htbp]
\centering
\begin{minipage}[c]{0.50\textwidth}
\centering
\vspace{0pt}
\includegraphics[width=\linewidth]{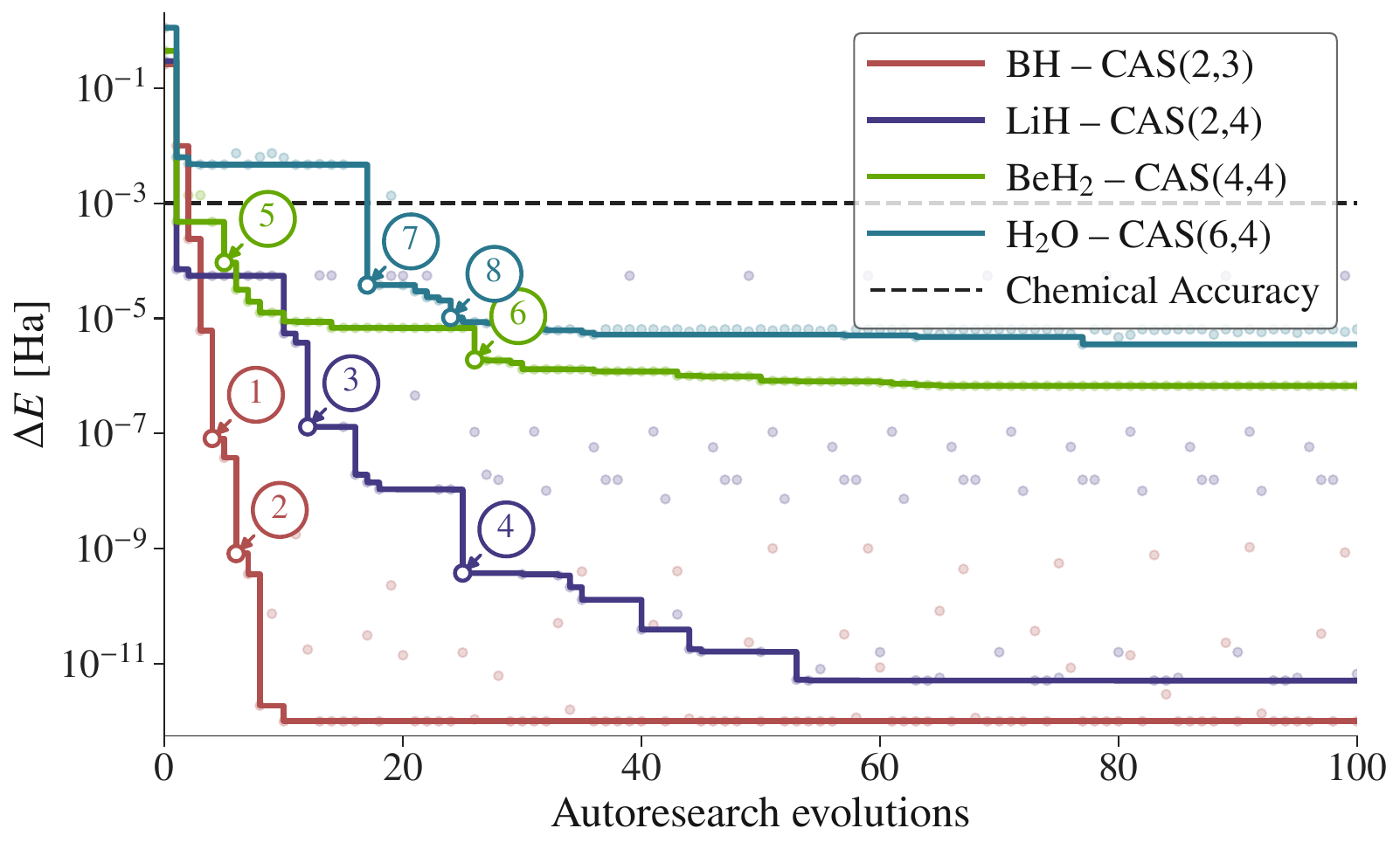}
\end{minipage}\hfill
\begin{minipage}[c]{0.46\textwidth}
\centering
\vspace{0pt}
\small
\setlength{\tabcolsep}{4pt}
\renewcommand{\arraystretch}{1.12}
\begin{tabular}{c l}
\toprule
\textbf{Flag} & \multicolumn{1}{c}{\textbf{Description}} \\
\midrule
\cellcolor{bhblue!10}\parbox[c]{0.05\linewidth}{\centering\textbf{1}} & \parbox[c]{0.84\linewidth}{\raggedright Powell $\rightarrow$ COBYLA; $\texttt{rhobeg}$ $0.5 \rightarrow 0.02$, $\texttt{tol}$ $10^{-4} \rightarrow 10^{-8}$, steps $1024 \rightarrow 4096$.} \\
\midrule
\cellcolor{bhblue!10}\parbox[c]{0.05\linewidth}{\centering\textbf{2}} & \parbox[c]{0.84\linewidth}{\raggedright COBYLA tighten: $\texttt{rhobeg}$ $0.02 \rightarrow 0.005$, $\texttt{tol}$ $10^{-8} \rightarrow 10^{-10}$.} \\
\midrule
\cellcolor{lihorange!10}\parbox[c]{0.05\linewidth}{\centering\textbf{3}} & \parbox[c]{0.84\linewidth}{\raggedright 9-parameter \texttt{spin\_paired\_symmetric} plus COBYLA local refine.} \\
\midrule
\cellcolor{lihorange!10}\parbox[c]{0.05\linewidth}{\centering\textbf{4}} & \parbox[c]{0.84\linewidth}{\raggedright \texttt{step\_size}, \texttt{min\_step\_size} $\div 2$; $\texttt{rhobeg}$ $0.005 \rightarrow 0.0025$.} \\
\midrule
\cellcolor{behgreen!10}\parbox[c]{0.05\linewidth}{\centering\textbf{5}} & \parbox[c]{0.84\linewidth}{\raggedright Seed a 26-amplitude UCCSD warm start.} \\
\midrule
\cellcolor{behgreen!10}\parbox[c]{0.05\linewidth}{\centering\textbf{6}} & \parbox[c]{0.84\linewidth}{\raggedright COBYLA tighten: $\texttt{rhobeg}$ $0.02 \rightarrow 0.002$, $\texttt{tol}$ $10^{-8} \rightarrow 10^{-12}$.} \\
\midrule
\cellcolor{h2ored!10}\parbox[c]{0.05\linewidth}{\centering\textbf{7}} & \parbox[c]{0.84\linewidth}{\raggedright \texttt{uccsd\_sd\_triplet} $\rightarrow$ 9-parameter \texttt{uccsd\_spin\_paired\_symmetric}.} \\
\midrule
\cellcolor{h2ored!10}\parbox[c]{0.05\linewidth}{\centering\textbf{8}} & \parbox[c]{0.84\linewidth}{\raggedright \texttt{init\_scale} $0.005 \rightarrow 0.0025$ before COBYLA refine.} \\
\bottomrule
\end{tabular}
\end{minipage}
\caption{Combined log-scale VQE milestone figure for the four completed 100-iteration archival campaigns in the fixed-budget study. Left: pale points denote every evaluated candidate, solid lines denote the running best $\Delta E$, and the dashed horizontal line marks chemical accuracy. Numbered callouts highlight later, energetically meaningful changes in the search trajectory rather than the trivial first ansatz swap. Right: short technical summaries of the corresponding code edits, with flag tint matching the molecule colour used in the plot. The full VQE mutation logs are listed in \vqelogreference{}.}
\label{fig:vqe_progress}
\end{figure}
}
\else
\begin{figure*}[!t]
\centering
\begin{minipage}[c]{0.50\textwidth}
\centering
\vspace{0pt}
\includegraphics[width=\linewidth]{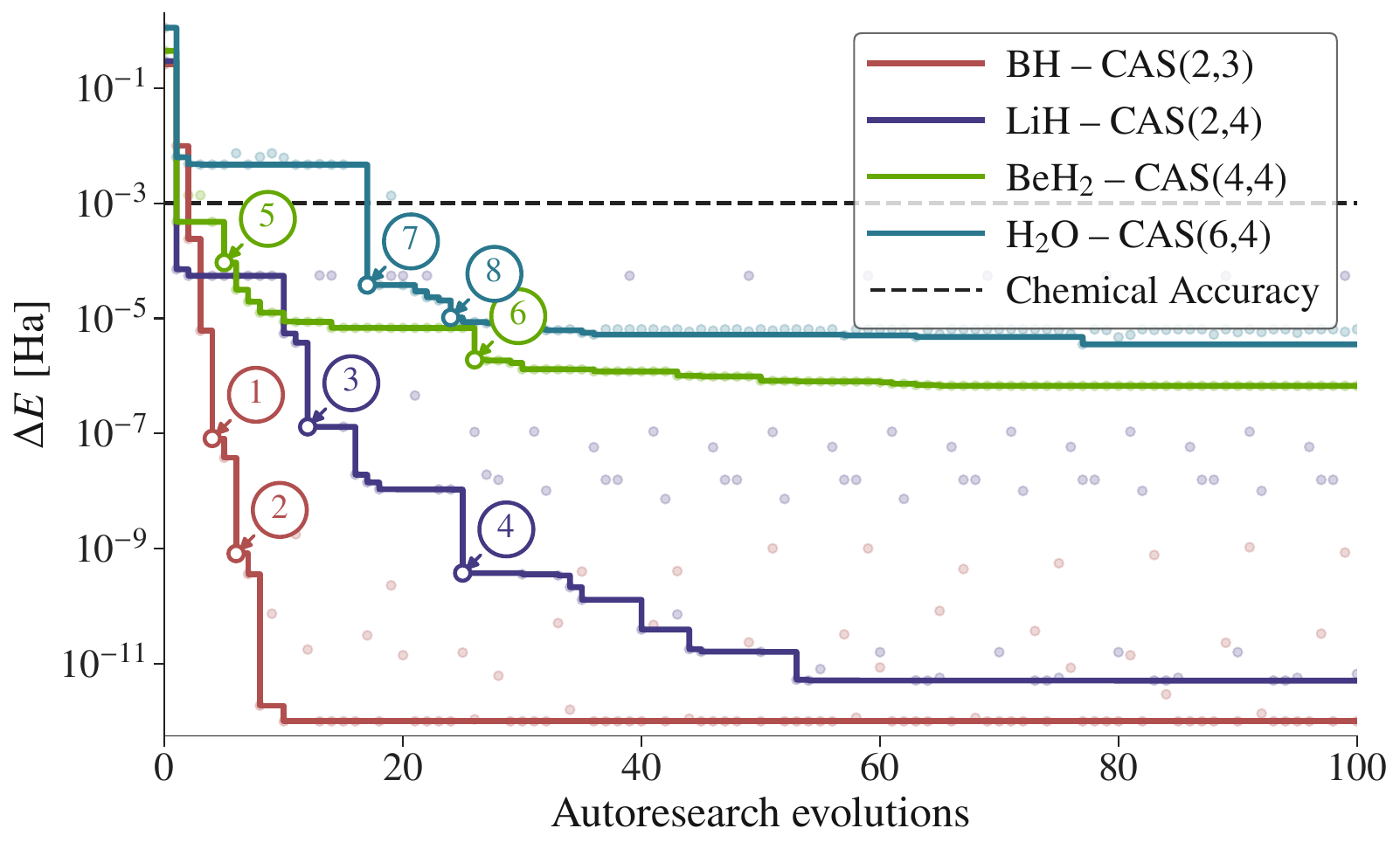}
\end{minipage}\hfill
\begin{minipage}[c]{0.46\textwidth}
\centering
\vspace{0pt}
\small
\setlength{\tabcolsep}{4pt}
\renewcommand{\arraystretch}{1.12}
\begin{tabular}{c l}
\toprule
\textbf{Flag} & \multicolumn{1}{c}{\textbf{Description}} \\
\midrule
\cellcolor{bhblue!10}\parbox[c]{0.05\linewidth}{\centering\textbf{1}} & \parbox[c]{0.84\linewidth}{\raggedright Powell $\rightarrow$ COBYLA; $\texttt{rhobeg}$ $0.5 \rightarrow 0.02$, $\texttt{tol}$ $10^{-4} \rightarrow 10^{-8}$, steps $1024 \rightarrow 4096$.} \\
\midrule
\cellcolor{bhblue!10}\parbox[c]{0.05\linewidth}{\centering\textbf{2}} & \parbox[c]{0.84\linewidth}{\raggedright COBYLA tighten: $\texttt{rhobeg}$ $0.02 \rightarrow 0.005$, $\texttt{tol}$ $10^{-8} \rightarrow 10^{-10}$.} \\
\midrule
\cellcolor{lihorange!10}\parbox[c]{0.05\linewidth}{\centering\textbf{3}} & \parbox[c]{0.84\linewidth}{\raggedright 9-parameter \texttt{spin\_paired\_symmetric} plus COBYLA local refine.} \\
\midrule
\cellcolor{lihorange!10}\parbox[c]{0.05\linewidth}{\centering\textbf{4}} & \parbox[c]{0.84\linewidth}{\raggedright \texttt{step\_size}, \texttt{min\_step\_size} $\div 2$; $\texttt{rhobeg}$ $0.005 \rightarrow 0.0025$.} \\
\midrule
\cellcolor{behgreen!10}\parbox[c]{0.05\linewidth}{\centering\textbf{5}} & \parbox[c]{0.84\linewidth}{\raggedright Seed a 26-amplitude UCCSD warm start.} \\
\midrule
\cellcolor{behgreen!10}\parbox[c]{0.05\linewidth}{\centering\textbf{6}} & \parbox[c]{0.84\linewidth}{\raggedright COBYLA tighten: $\texttt{rhobeg}$ $0.02 \rightarrow 0.002$, $\texttt{tol}$ $10^{-8} \rightarrow 10^{-12}$.} \\
\midrule
\cellcolor{h2ored!10}\parbox[c]{0.05\linewidth}{\centering\textbf{7}} & \parbox[c]{0.84\linewidth}{\raggedright \texttt{uccsd\_sd\_triplet} $\rightarrow$ 9-parameter \texttt{uccsd\_spin\_paired\_symmetric}.} \\
\midrule
\cellcolor{h2ored!10}\parbox[c]{0.05\linewidth}{\centering\textbf{8}} & \parbox[c]{0.84\linewidth}{\raggedright \texttt{init\_scale} $0.005 \rightarrow 0.0025$ before COBYLA refine.} \\
\bottomrule
\end{tabular}
\end{minipage}
\caption{Combined log-scale VQE milestone figure for the four completed 100-iteration archival campaigns in the fixed-budget study. Left: pale points denote every evaluated candidate, solid lines denote the running best $\Delta E$, and the dashed horizontal line marks chemical accuracy. Numbered callouts highlight later, energetically meaningful changes in the search trajectory rather than the trivial first ansatz swap. Right: short technical summaries of the corresponding code edits, with flag tint matching the molecule colour used in the plot. The full VQE mutation logs are listed in \vqelogreference{}.}
\label{fig:vqe_progress}
\end{figure*}
\fi

\section{Autoresearch experiments}
All \texttt{gsopt} autoresearch runs discussed in this section use GPT-5.4 in ``xhigh'' reasoning mode as the proposal engine. The suite is intentionally small: the point is not scale by itself, but exact or near-exact scoring under fixed resources. VQE and DMRG candidates are each scored under a fixed $20\,\mathrm{s}$ budget on the 10 performance cores of a local Apple M4 Pro chip with BLAS/OpenMP threads clamped to ten, AFQMC candidates are scored under a fixed 5-minute budget using two MPI processes, and in every case the untouched starting protocol is archived as iteration $0$ before the mutation loop begins.

\subsection{VQE ansatz and optimizer search}
VQEs cast electronic-structure ground-state search as minimization of the energy of a parameterized circuit~\citelist{aspuruguzik2005simulated,peruzzo2014vqe,mcclean2016theory,cao2019quantumchem,schleich2025quantumcomputing}. In the actual benchmark implementation, the active-space molecular Hamiltonian is mapped to qubits with the Jordan--Wigner transform~\cite{jordan1928equivalenzverbot} and evaluated as a Pauli-string operator,
\begin{equation}
H_{\mathrm{JW}} = \sum_{\ell} c_{\ell} P_{\ell},
\qquad
P_{\ell} \in \{I,X,Y,Z\}^{\otimes n_q},
\end{equation}
where the real coefficients $c_{\ell}$ are determined by the active-space one- and two-electron integrals and $n_q$ is the number of Jordan--Wigner qubits for the chosen molecule.

The benchmark uses four molecules ordered roughly from easier to harder active spaces: BH with CAS(2,3), LiH with CAS(2,4), BeH$_2$ with CAS(4,4), and H$_2$O with CAS(6,4). OpenFermion and PySCF generate the active-space Hamiltonians, while the parameterized ansatz kernels are implemented in CUDA-Q and executed on its CPU-only \texttt{qpp-cpu} target~\citelist{mcclean2020openfermion,sun2018pyscf}. All four molecules are treated as neutral singlets in the STO-3G basis. The active spaces are built from contiguous windows of closed-shell RHF canonical spatial orbitals after freezing the lowest doubly occupied orbitals, and the circuit is initialized from the corresponding Hartree--Fock occupation bitstring. These choices give $6$, $8$, $8$, and $8$ Jordan--Wigner qubits for BH, LiH, BeH$_2$, and H$_2$O, respectively, with each benchmark scored against a frozen active-space FCI reference energy obtained by exact diagonalization of the corresponding Jordan--Wigner qubit Hamiltonian.

\ifmatterlabfigureplacement
\newcommand{\mattertnprogressfigure}{%
\begin{figure}[!htbp]
\centering
\includegraphics[width=0.92\textwidth]{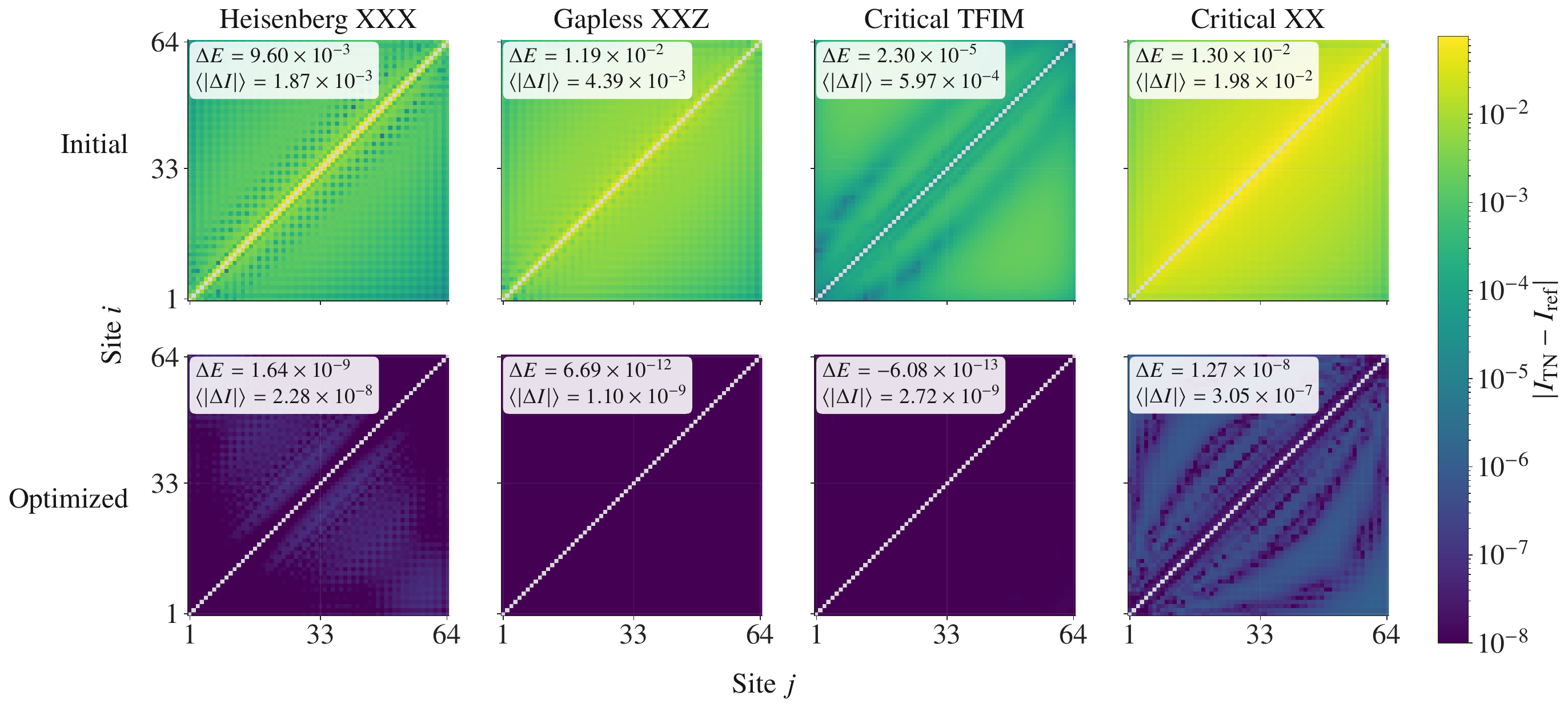}
\caption{Tensor-network mutual-information error matrices for the four one-dimensional critical $L=64$ benchmarks: Heisenberg XXX, gapless XXZ, critical TFIM, and critical XX. Columns denote Hamiltonians; the top row shows the archived baseline protocol and the bottom row shows the current best optimized protocol from the autoresearch run. Each panel plots the absolute entrywise error $|I_{ij}^{\mathrm{TN}} - I_{ij}^{\mathrm{ref}}|$ relative to a frozen high-accuracy \texttt{DMRG2} MPS reference, with the diagonal masked in gray because self-information entries are not compared.}
\label{fig:tn_progress_draft}
\end{figure}
}
\else
\begin{figure*}[!t]
\centering
\includegraphics[width=0.92\textwidth]{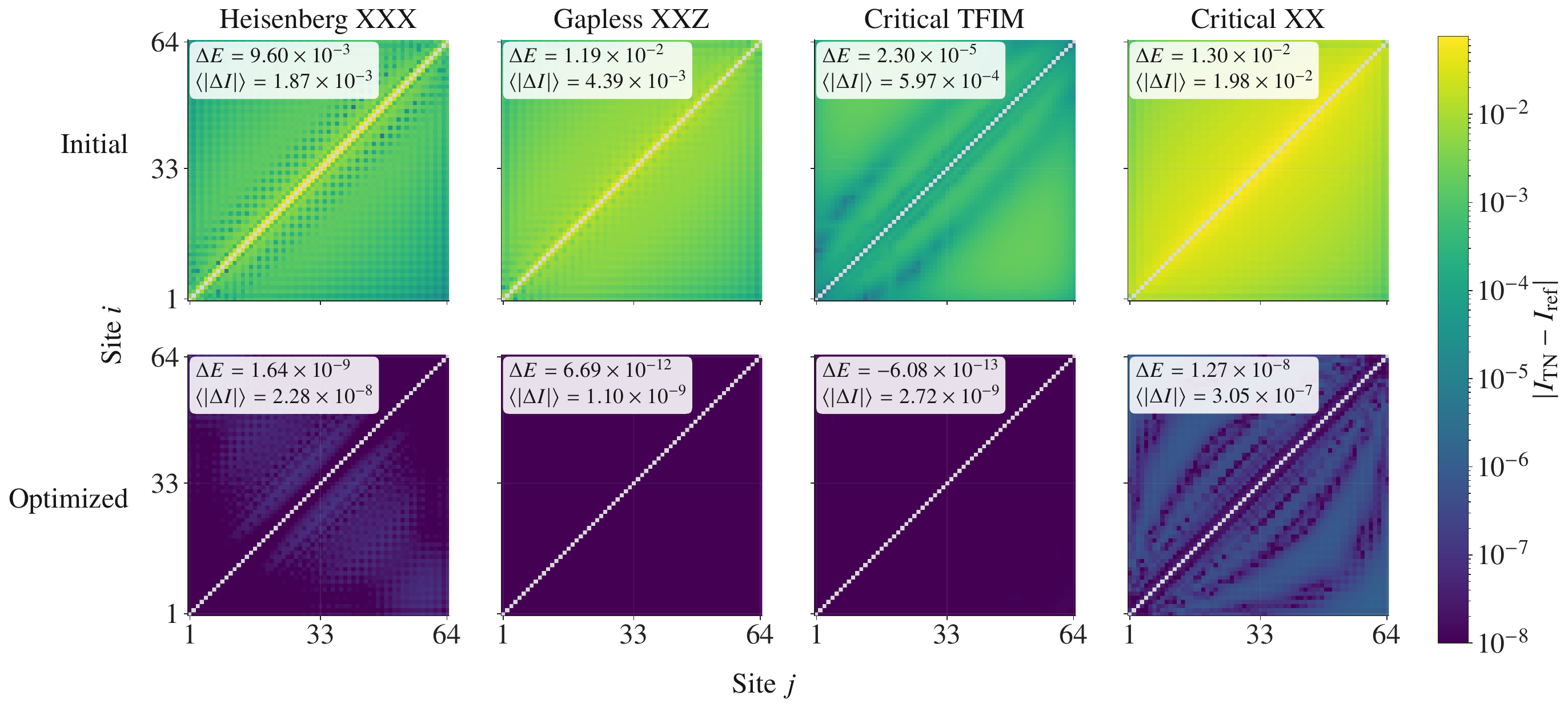}
\caption{Tensor-network mutual-information error matrices for the four one-dimensional critical $L=64$ benchmarks: Heisenberg XXX, gapless XXZ, critical TFIM, and critical XX. Columns denote Hamiltonians; the top row shows the archived baseline protocol and the bottom row shows the current best optimized protocol from the autoresearch run. Each panel plots the absolute entrywise error $|I_{ij}^{\mathrm{TN}} - I_{ij}^{\mathrm{ref}}|$ relative to a frozen high-accuracy \texttt{DMRG2} MPS reference, with the diagonal masked in gray because self-information entries are not compared.}
\label{fig:tn_progress_draft}
\end{figure*}
\fi

One reason to pose this as a fixed-budget search problem is trainability. For deep hardware-efficient ans\"atze that approach unitary 2-design behaviour, barren-plateau theory predicts exponentially small gradients~\citelist{mcclean2018barren,cerezo2021costbp}. A complementary Lie-algebraic view makes the dependence more explicit: for sufficiently deep circuits in the Lie-algebraic setting~\citelist{ragone2024liealgebraic,fontana2024adjointbp},
\begin{equation}
\mathrm{Var}[\partial_{\theta_k} E]
= O\!\left(\frac{\|\rho_{\mathfrak g}\|_F^2\,\|O_{\mathfrak g}\|_F^2}{d_{\mathfrak g}}\right),
\end{equation}
where $\rho_{\mathfrak g}$ and $O_{\mathfrak g}$ denote the Lie-algebra-supported components of the input state and observable, and $d_{\mathfrak g}=\dim(\mathfrak g)$ is the dynamical Lie algebra dimension of the ansatz. When $d_{\mathfrak g}$ itself grows exponentially with system size, this reproduces barren-plateau behaviour. This is the main computational bottleneck in the VQE benchmark: trainability can degrade rapidly even before one reaches large active spaces. The mutable efficiency levers are therefore the standard trainability choices rather than an exhaustive catalog of all possible circuit changes: ansatz family and connectivity, parameter initialization or warm starts, optimizer choice, and optimizer hyperparameters such as step size, trust-region radius, tolerance, and iteration budget. These are exactly the code-level choices the agent mutates under a fixed wall-time budget.

The VQE benchmark is therefore a code-editing search over molecule-specific CUDA-Q method files. In practice, the search space is restricted to a common non-exhaustive subset of these levers: ansatz kernel, initialization or warm-start policy, optimizer choice, and optimizer hyperparameters. All four molecular benchmarks start from the same intentionally weak baseline: a two-layer hardware-efficient ring ansatz with per-qubit $R_y$ and $R_z$ rotations, nearest-neighbour CNOT entanglers, Gaussian initialization scale $0.1$, a Hartree--Fock computational-basis reference, and COBYLA capped at $4096$ steps. Figure~\ref{fig:vqe_progress} shows that the completed 100-iteration campaigns consistently move from this baseline to UCCSD-style warm starts plus tighter local refinement, cutting the final absolute energy error by 5.5--12.9 orders of magnitude and reaching chemical accuracy in all four cases. The numbered callouts identify some of the later code changes that matter most energetically, and Appendix~A1 together with \vqelogreference{} records the archived protocols and full mutation logs.
\ifmatterlabfigureplacement
\mattervqeprogressfigure
\FloatBarrier
\fi

\subsection{Tensor-network ground-state search}

Tensor networks provide a broad family of ans\"atze for lattice ground states, ranging from matrix product states to tree tensor networks, MERA, and PEPS~\citelist{ostlund1995mps,schollwock2011dmrg,shi2006ttn,vidal2007mera,verstraete2008peps,orus2014tnreview}. A compact generic spin-lattice Hamiltonian for this benchmark family is
\begin{equation}
H_{\mathrm{1D}} = \sum_{i=1}^{L-1}\sum_{\alpha=x,y,z} J_{\alpha}\, S_i^\alpha S_{i+1}^\alpha - \sum_{i=1}^{L}\left(h_x S_i^x + h_z S_i^z\right),
\end{equation}
which covers the four active open-boundary spin-$1/2$ chains used here. In this notation the benchmark instances are Heisenberg XXX with $(J_x,J_y,J_z,h_x,h_z)=(1,1,1,0,0)$, gapless XXZ with $(1,1,0.5,0,0)$, critical TFIM with $(0,0,1,0.5,0)$, and critical XX with $(1,1,0,0,0)$. The benchmark uses the Quimb tensor-network package~\cite{gray2018quimb} to tune practical one-dimensional DMRG protocols on these four critical chains at length $L=64$. In this package, \texttt{DMRG1} and \texttt{DMRG2} denote the standard one-site and two-site DMRG local-update variants, respectively~\citelist{white1992dmrg,white1993dmrg,schollwock2011dmrg}. We use \texttt{DMRG2} with each model to obtain a frozen offline reference energy and a frozen pairwise mutual-information matrix extracted from a separate high-accuracy MPS run with a larger bond-dimension schedule, tighter cutoffs and eigensolver tolerances, and a much longer one-shot wall budget. The reference mutual information is then evaluated from the one-site and two-site reduced density matrices of that reference MPS.

The tensor-network bottleneck is different from that of VQE: the central issue is not a flat optimization landscape but entanglement-driven bond growth. Giufre Vidal's original MPS analyses show that slightly entangled one-dimensional states admit representations with ${\cal O}(nd\chi^2)$ parameters, while local updates still inherit a steep bond-dimension cost and hence
\begin{equation}
C_{\mathrm{TN}} = O\!\left(n(d\chi)^3 \frac{T}{\delta}\right)
= O\!\left(nd^3 2^{3S}\frac{T}{\delta}\right), \qquad \chi \ge 2^S,
\end{equation}
where $\chi$ is the bond dimension and $S$ is the bipartite entanglement entropy~\citelist{vidal2003slightly,vidal2004efficient}. This is the main computational challenge in the tensor-network benchmark: the search is fundamentally about how to spend a small bond-dimension budget well. The mutable levers are therefore practical DMRG protocol choices rather than an exhaustive catalog of all tensor-network designs: one-site versus two-site updates, initialization, bond-dimension schedule, truncation cutoff, eigensolver tolerance, and sweep schedule.

\ifmatterlabfigureplacement
\newcommand{\matterafqmcprogressfigure}{%
\begin{figure}[!htbp]
\centering
\includegraphics[width=0.98\textwidth]{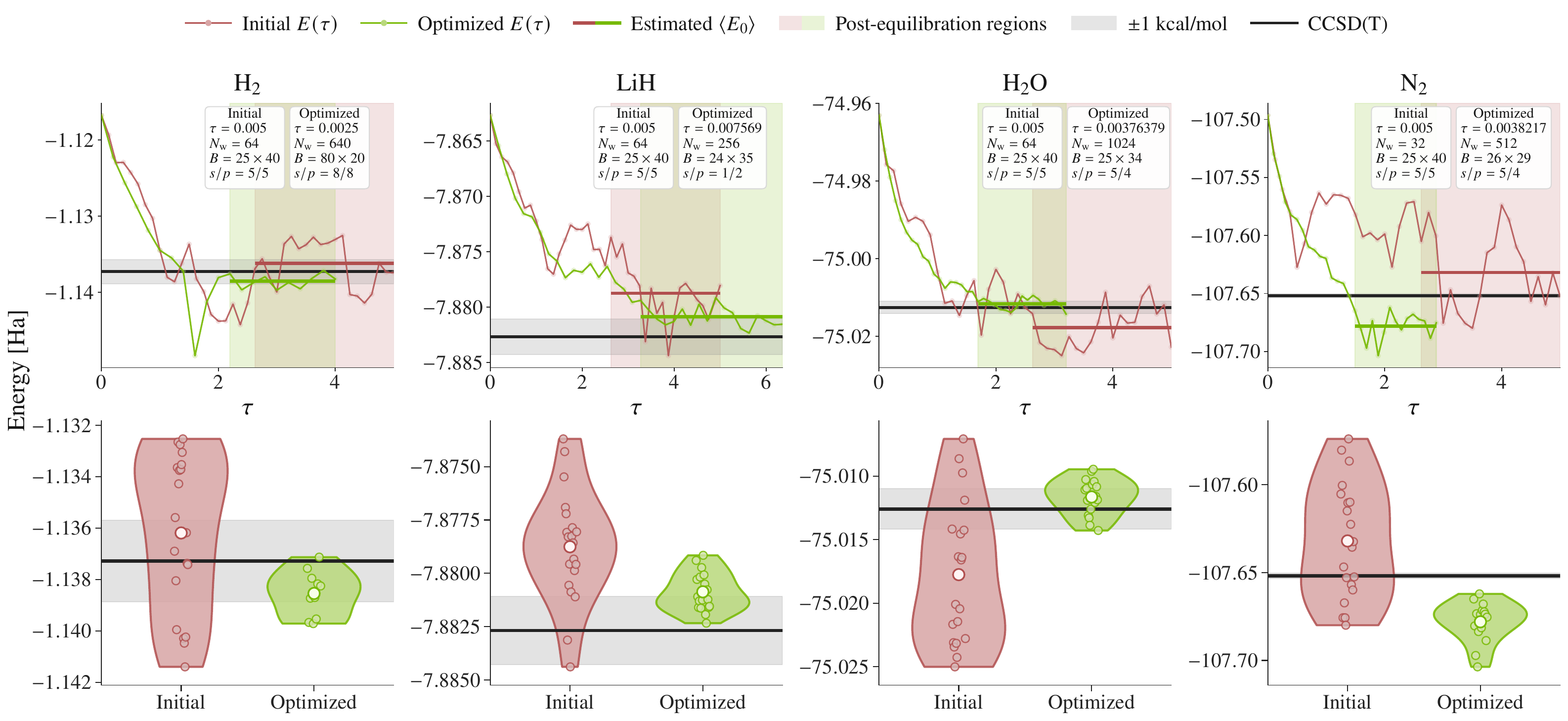}
\caption{AFQMC comparison across H$_2$, LiH, H$_2$O, and N$_2$ for the molecular benchmark suite under the fixed post-equilibration live scorer. The top row shows the sampled mixed-estimator traces $E(\tau)$ for the initial and best archived optimized configurations, with the post-equilibration sampling regions shaded and the post-equilibration energy estimates $\langle E_0\rangle$ drawn as bold horizontal segments. The bottom row shows the corresponding distributions of stored post-equilibration block energies. The shared horizontal guides mark the frozen CCSD(T) reference and the $\pm 1$ kcal/mol chemical-accuracy window. The parameter boxes summarize the initial and optimized AFQMC controls for each molecule: $\tau$ is the imaginary-time step, $N_{\mathrm{w}}$ is the total walker count, $B=N_{\mathrm{steps}}\times N_{\mathrm{blocks}}$ is the block geometry, and $s/p$ denotes the stabilization and population-control cadences.}
\label{fig:afqmc_trace_progress_draft}
\end{figure}
}
\else
\begin{figure*}[t!]
\centering
\includegraphics[width=0.98\textwidth]{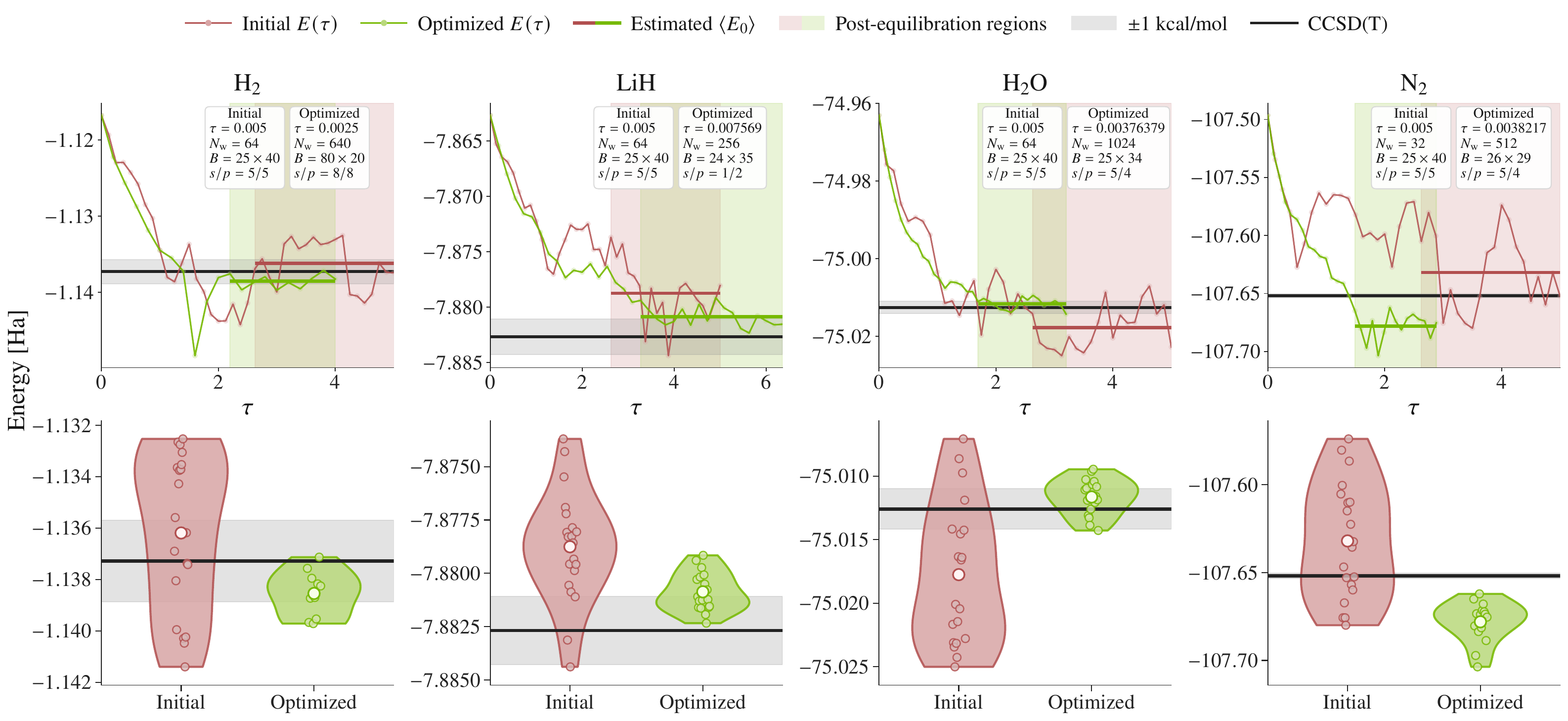}
\caption{AFQMC comparison across H$_2$, LiH, H$_2$O, and N$_2$ for the molecular benchmark suite under the fixed post-equilibration live scorer. The top row shows the sampled mixed-estimator traces $E(\tau)$ for the initial and best archived optimized configurations, with the post-equilibration sampling regions shaded and the post-equilibration energy estimates $\langle E_0\rangle$ drawn as bold horizontal segments. The bottom row shows the corresponding distributions of stored post-equilibration block energies. The shared horizontal guides mark the frozen CCSD(T) reference and the $\pm 1$ kcal/mol chemical-accuracy window. The parameter boxes summarize the initial and optimized AFQMC controls for each molecule: $\tau$ is the imaginary-time step, $N_{\mathrm{w}}$ is the total walker count, $B=N_{\mathrm{steps}}\times N_{\mathrm{blocks}}$ is the block geometry, and $s/p$ denotes the stabilization and population-control cadences.}
\label{fig:afqmc_trace_progress_draft}
\end{figure*}
\fi
Figure~\ref{fig:tn_progress_draft} summarizes the result structurally rather than only energetically. Each panel plots the absolute entrywise mutual-information error $|I_{ij}^{\mathrm{TN}} - I_{ij}^{\mathrm{ref}}|$ for the archived baseline and the best optimized protocol relative to the frozen reference matrix, so darker off-diagonal patches mark site pairs whose correlations are still least accurate. The completed 100-iteration campaigns consistently move from low-bond random-start baselines to higher-bond \texttt{DMRG2} or staged \texttt{DMRG1}$\rightarrow$\texttt{DMRG2} schedules, lowering the final energy and visibly improving the correlation pattern across all four critical chains. Appendix~A2 summarizes the archived protocols, and \tnlogreference{} gives the full mutation logs.
\ifmatterlabfigureplacement
\mattertnprogressfigure
\FloatBarrier
\fi

\subsection{Molecular AFQMC trial-state and propagation search}

Auxiliary-field quantum Monte Carlo~\citelist{blankenbecler1981bss,scalapino1981bss,foulkes2001qmc,zhang1995cpmc,zhang2003phafqmc,alsaidi2006afqmc} is a projector method that evolves a population of Slater-determinant walkers in imaginary time while using a trial wavefunction to control the sign or phase problem. For the molecular electronic targets in the current benchmark family, the relevant Hamiltonian is the standard second-quantized electronic-structure form
\begin{equation}
H_{\mathrm{elec}} = E_{\mathrm{nuc}} + \sum_{pq} h_{pq}\, a_p^\dagger a_q + \frac{1}{2}\sum_{pqrs} h_{pqrs}\, a_p^\dagger a_q^\dagger a_s a_r.
\end{equation}
The present AFQMC suite uses four small molecules in the STO-3G basis: H$_2$, LiH, H$_2$O, and N$_2$, all treated as neutral singlets and compared offline against frozen CCSD(T) references in the same basis. All AFQMC experiments in this work are executed with the open-source \texttt{ipie} codebase~\citelist{malone2023ipie,jiang2024ipie}, whose modular trial-state, walker, and propagation interfaces expose the practical AFQMC controls to the outer mutation loop. In simple terms, the method repeatedly propagates and reweights a walker population so that the constrained imaginary-time dynamics moves the ensemble toward the ground state.

The central AFQMC bottleneck is statistical stability: without constraints, the underlying phase or sign problem causes the stochastic signal to deteriorate exponentially with system size~\citelist{troyer2005sign,zhang1995cpmc,zhang2003phafqmc}. To summarize the stochastic work in one scored run, we define
\begin{equation}
C_{\mathrm{stoch}} = N_{\mathrm{walkers}}\,B = N_{\mathrm{walkers}}\,N_{\mathrm{steps}}\,N_{\mathrm{blocks}},
\end{equation}
where $B$ is the block geometry written in the figure as $N_{\mathrm{steps}}\times N_{\mathrm{blocks}}$. The bottleneck can then be summarized schematically as
\begin{equation}
\frac{\sigma_E}{\langle E \rangle} \propto \frac{e^{\alpha N}}{\sqrt{C_{\mathrm{stoch}}}},
\end{equation}
with $\alpha>0$ capturing the severity of the underlying phase/sign instability~\citelist{loh1990sign,troyer2005sign}. The benchmark therefore exposes the standard practical levers that trade bias, variance, and runtime: the trial family, the orbital basis or single-particle representation used when generating AFQMC inputs, the importance-sampling or weight-update policy, the Cholesky cutoff, the imaginary-time step, the walker count, the number of steps per block, stabilization and population-control cadence, and molecule-specific trial settings. Candidates are then ranked under the fixed-budget live objective
\begin{equation}
\mathcal{L}_{\mathrm{AFQMC}} = \langle E \rangle_{\mathrm{PE}} + \lambda\,\sigma_{E,\mathrm{PE}},
\end{equation}
where $\langle E \rangle_{\mathrm{PE}}$ is the mean of the post-equilibration region, defined here as the final 50\% of sampled AFQMC blocks, $\sigma_{E,\mathrm{PE}}$ is the sample standard deviation over that same region, and in this work we set $\lambda = 5$. This regularizes toward candidates whose final sampled tail is both lower in energy and less noisy at the same effective stochastic cost. Figure~\ref{fig:afqmc_trace_progress_draft} shows the resulting trace-and-distribution comparison for the archived baseline and best optimized protocols. Across the four completed campaigns, the search consistently retunes walker population, timestep, and block geometry to produce lower live scores and lower post-equilibration energies under the same fixed budget. Appendix~A3 summarizes the archived AFQMC protocols, and \afqmclogreference{} gives the full mutation logs.
\ifmatterlabfigureplacement
\matterafqmcprogressfigure
\FloatBarrier
\fi

\section{Conclusions}
This work shows that ground-state preparation can be framed as an automated search problem for state-of-the-art coding agents. We show this optimization across quantum circuits, tensor networks, and Monte Carlo sampling methods, where the same agent loop can improve the accuracy of the underlying ground state. The experiments aim to improve code across different backends and Hamiltonians, spanning molecule-specific CUDA-Q VQE routines, one-dimensional DMRG schedules, and molecular AFQMC protocols under fixed compute budgets. The same template should extend to other simulators, compilers, and quantum-hardware settings whenever execution yields a reliable scalar objective. A promising direction is to optimize the $T$-gates in fault-tolerant quantum algorithms---if this is successfully executed, we could significantly reduce computational resources to achieve a desired quantum advantage.

This work is a proof-of-concept of the methodology and, therefore, was applied to relatively small molecular and spin-lattice systems. The next step in the journey is to explore the scaling of this approach. Ideally we want to find optimal protocols that are capable of preparing large many-body quantum states. To solve this issue, extrapolating performance from a few small- and medium-sized systems to larger systems can serve as a proxy to be optimized with autoresearch---in a similar spirit to scaling laws~\cite{kaplan2020scaling}. This way, we might be able to extrapolate value to unexplored territory.  In terms of finding better-optimized protocols, a natural extension of this work is to allow the underlying model parameters to vary. Reinforcement learning with verifiable rewards (RLVR), test-time adaptation, and related self-improving-agent ideas suggest that the mutation policy itself could be tuned online against the same autoresearch score~\citelist{wen2026rlvr,yuksekgonul2026tttdiscover,zhang2025darwin}. We believe these ideas will enable ground state autoresearch to find protocols that will achieve beyond state-of-the-art results in this field, given that they have already done so in other fields.

%% file: includes/manuscript-acknowledgments.tex
The authors thank Yuri Alexeev for his helpful feedback on the manuscript and thank Kohei Nakaji, Alex Zook, Taylor Patti, and Zijian Zhang for fruitful discussions. L.M.C. acknowledges funding from the Novo Nordisk Foundation, Grant number NNF22SA0081175, NNF Quantum Computing Programme. I.G. acknowledges funding of this project by the National Sciences and Engineering Research Council of Canada (NSERC) Alliance Grant \#ALLRP587593-23. This work was supported by the AI2050 program of Schmidt Sciences. This work was supported by the Defense Advanced Research Projects Agency (DARPA) under Agreement No. HR0011262E022. A.A.-G. thanks Anders G. Fr{\o}seth, for his generous support. A.A.-G. also acknowledges the generous support of Natural Resources Canada and the Canada 150 Research Chairs program.

%% file: includes/manuscript-code-availability.tex
Code for the benchmarks and experiments discussed in this manuscript is available at \url{https://github.com/bestquark/gsopt}.

%% file: includes/manuscript-supplement.tex
\clearpage
\onecolumn
\section*{Supplemental Material}
\setcounter{table}{0}
\renewcommand{\thetable}{S\arabic{table}}
\renewcommand{\theHtable}{supp.\arabic{table}}

\section*{Appendix A. Autoresearch details}

\subsection*{A1. VQE autoresearch details}
In this variational-circuit benchmark, the ground-state search follows the Rayleigh--Ritz principle: a parameterized circuit prepares $|\psi(\theta)\rangle = U(\theta)|\phi_{\mathrm{HF}}\rangle$, a classical optimizer updates $\theta$, and lower energy implies a better ground-state approximation. The Hamiltonian, basis, and active space are fixed; the agent instead mutates the reusable building blocks of the solver itself, namely the ansatz family and connectivity, parameter-tying scheme, initialization scale and reference-state details, and the optimizer policy with its tolerances, step sizes, step cap, and seed. Table~\ref{tab:supp_vqe_protocols} should therefore be read as edits to those few components rather than as changes to the physical target.

\medskip\noindent\textbf{Experimental setup}
The four VQE targets are neutral singlets in the STO-3G basis built from closed-shell RHF orbitals and mapped to qubits with the Jordan--Wigner transform~\cite{jordan1928equivalenzverbot}. The molecular geometries (in Angstrom) are BH: B$(0,0,0)$, H$(0,0,1.23)$; LiH: Li$(0,0,0)$, H$(0,0,1.57)$; BeH$_2$: H$(0,0,-1.326)$, Be$(0,0,0)$, H$(0,0,1.326)$; and H$_2$O: O$(0,0,0)$, H$(0,0.757,0.587)$, H$(0,-0.757,0.587)$. The corresponding active spaces are CAS(2,3), CAS(2,4), CAS(4,4), and CAS(6,4), respectively, all built from contiguous RHF orbital windows. Fig.~\ref{fig:supp_vqe_orbitals} shows the active orbital windows, Table~\ref{tab:supp_vqe_protocols} stores the baseline and best archived protocol values, and the full 100-iteration mutation logs are collected in \vqelogreference{}.

\begin{figure}[h]
\centering
\includegraphics[width=0.95\textwidth]{figures/vqe_active_orbital_gallery.pdf}
\caption{Active-orbital gallery for the four VQE benchmark molecules used in the main text: BH, LiH, BeH$_2$, and H$_2$O. Each panel shows the active-space molecular orbitals used to construct the corresponding CAS Hamiltonian, with positive and negative orbital lobes colored separately.}
\label{fig:supp_vqe_orbitals}
\end{figure}

\input{generated_vqe_summary_table.tex}
\FloatBarrier

\subsection*{A2. Tensor-network autoresearch details}
Here, the wavefunction is represented as a one-dimensional matrix product state, and the energy is reduced via local variational DMRG sweeps, with intermediate Schmidt spectra truncated to control the bond dimension. In the current fixed-budget benchmark, the mutable building blocks are the DMRG workflow itself (\texttt{DMRG1}, \texttt{DMRG2}, and staged warm-start variants, where applicable), the initial state, the bond-dimension schedule, the truncation cutoff, the eigensolver tolerance, and the sweep schedule. In Quimb, \texttt{DMRG1} and \texttt{DMRG2} denote one-site and two-site DMRG updates, respectively~\citelist{gray2018quimb,white1992dmrg,white1993dmrg}. Table~\ref{tab:supp_tn_protocols} therefore summarizes edits to initialization, bond control, and update policy under a fixed $L=64$ spin-chain Hamiltonian.

\begin{figure}[h]
\centering
\includegraphics[width=0.98\textwidth]{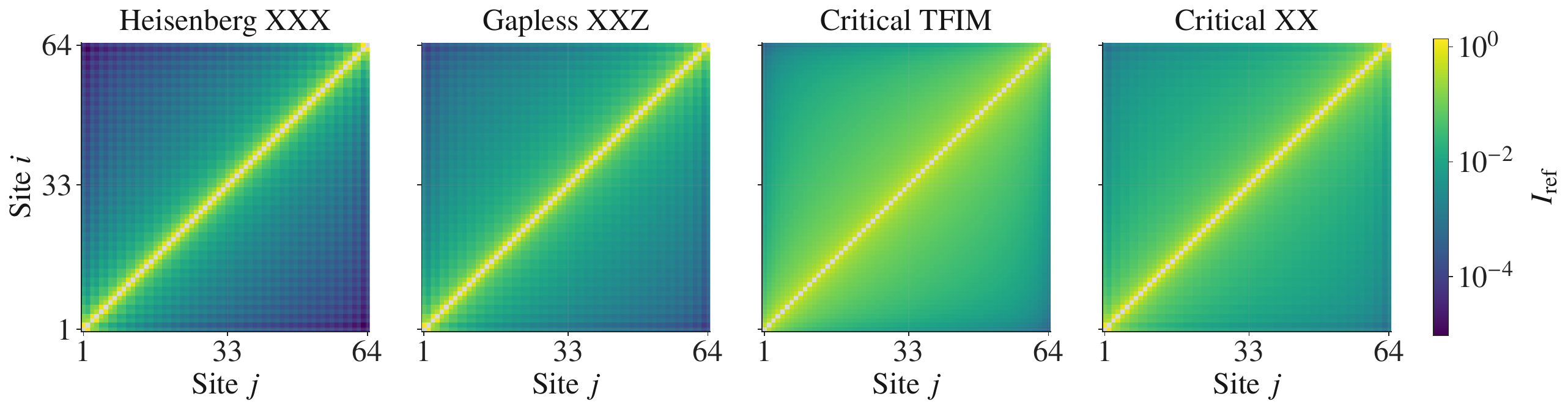}
\caption{Frozen reference mutual-information matrices for the four one-dimensional critical $L=64$ tensor-network benchmarks. Each panel shows $I_{ij}^{\mathrm{ref}}$ for the corresponding high-accuracy \texttt{DMRG2} MPS reference state, with the diagonal masked in gray.}
\label{fig:supp_tn_reference_mi}
\end{figure}

\medskip\noindent\textbf{Experiment setup.}
The active tensor-network suite uses four open-boundary spin-$1/2$ chains at $L=64$: Heisenberg XXX with $(J_x,J_y,J_z,h_x,h_z)=(1,1,1,0,0)$, gapless XXZ with $(1,1,0.5,0,0)$, critical TFIM with $(0,0,1,0.5,0)$, and critical XX with $(1,1,0,0,0)$. Each model is paired with a frozen high-accuracy \texttt{DMRG2} MPS reference state, from which both the reference energy and the mutual-information matrix are extracted. Fig.~\ref{fig:supp_tn_reference_mi} shows the frozen reference mutual-information structure, Table~\ref{tab:supp_tn_protocols} stores the baseline and best archived protocol values, and the full 100-iteration mutation logs are collected in \tnlogreference{}.

\input{generated_tn_summary_table.tex}
\FloatBarrier

\subsection*{A3. AFQMC autoresearch details}
This benchmark uses an imaginary-time projector method built around a population of Slater-determinant walkers. In this work, the \emph{auxiliary-field} quantum Monte Carlo (AFQMC) implementation is provided by \texttt{ipie}~\citelist{malone2023ipie,jiang2024ipie}, which exposes modular trial, walker, and propagation components and supports both CPU execution and CUDA/CuPy-based GPU acceleration on NVIDIA hardware~\cite{jiang2024ipie}. In an orbital basis, one first writes the electronic Hamiltonian as a one-body part plus a factorized two-body part,
\begin{equation}
\hat H = \hat H_1 - \frac{1}{2}\sum_\gamma \hat v_\gamma^2 + E_0,
\end{equation}
where each $\hat v_\gamma$ is a one-body operator obtained from a Cholesky or density-fitted factorization of the two-electron integrals, for example $\hat v_\gamma = \sum_{pq} L^\gamma_{pq}\, a_p^\dagger a_q$, so $\hat v_\gamma^2$ means the operator product $\hat v_\gamma \hat v_\gamma$ whose sum over $\gamma$ reconstructs the factorized two-body term. Constant or residual one-body shifts are absorbed into $\hat H_1$ and $E_0$. The method is called \emph{auxiliary-field} quantum Monte Carlo or AFQMC because the quadratic two-body term is linearized by a Hubbard--Stratonovich transformation~\citelist{stratonovich1958distribution,hubbard1959partition,blankenbecler1981bss,scalapino1981bss},
\begin{equation}
e^{-\Delta\tau \hat H_2}
= e^{\frac{\Delta\tau}{2}\sum_\gamma \hat v_\gamma^2}
= \int d\mathbf{x}\, p(\mathbf{x})\,
e^{\sqrt{\Delta\tau}\sum_\gamma x_\gamma \hat v_\gamma},
\end{equation}
with $p(\mathbf{x})$ a standard normal density over the auxiliary fields $\mathbf{x}$.

Using the second-order Trotter factorization~\cite{trotter1959product}, one short imaginary-time step has the schematic form
\begin{equation}
e^{-\Delta\tau \hat H}
= e^{-\frac{\Delta\tau}{2}\hat H_1}
\left[
\int d\mathbf{x}\, p(\mathbf{x})\,
e^{\sqrt{\Delta\tau}\sum_\gamma x_\gamma \hat v_\gamma}
\right]
e^{-\frac{\Delta\tau}{2}\hat H_1}
 + O(\Delta\tau^3),
\end{equation}
so each walker update is
\begin{equation}
|\phi'\rangle = \hat B(\mathbf{x}) |\phi\rangle,
\qquad
\hat B(\mathbf{x}) =
e^{-\frac{\Delta\tau}{2}\hat H_1}
e^{\sqrt{\Delta\tau}\sum_\gamma x_\gamma \hat v_\gamma}
e^{-\frac{\Delta\tau}{2}\hat H_1}.
\end{equation}
Because $\hat B(\mathbf{x})$ is a one-body propagator for fixed $\mathbf{x}$, a Slater determinant remains a Slater determinant after each step. This is the practical content of the Thouless theorem in the AFQMC setting: exponentials of one-body operators generate the manifold of nonorthogonal Slater determinants from a reference determinant~\cite{thouless1960hf}. In the autoresearch loop (c.f. Table~\ref{tab:supp_afqmc_protocols}), the agent mutates the trial family, orbital representation, Cholesky cutoff, imaginary-time step $\Delta\tau$, walker population, block geometry $B=N_{\mathrm{steps}}\times N_{\mathrm{blocks}}$, stabilization and population-control cadence $s/p=a/b$, and molecule-specific trial settings. Here \texttt{walkers/rank} denotes the local walker population per MPI rank or GPU worker, so $N_{\mathrm{w}} = N_{\mathrm{rank}}\times(\texttt{walkers/rank})$, while \texttt{chol cut} is the truncation threshold used to build the Cholesky vectors defining $\hat v_\gamma$. 

\medskip\noindent\textbf{Experiment setup.}
The active AFQMC suite uses H$_2$, LiH, H$_2$O, and N$_2$ as neutral singlets in the STO-3G basis, with offline CCSD(T) references used only for post hoc comparison. Table~\ref{tab:supp_afqmc_protocols} stores the baseline and best archived protocol values, and the full mutation logs are collected in \afqmclogreference{}.

\input{generated_afqmc_summary_table.tex}
\FloatBarrier

%% file: generated_vqe_summary_table.tex
\begin{table}[!htbp]
\caption{Baseline and best archived VQE protocols for the four molecular campaigns. A/R denotes accepted/rejected counts.}
\label{tab:supp_vqe_protocols}
\centering
\setlength{\tabcolsep}{3pt}
\renewcommand{\arraystretch}{1.12}
\renewcommand{\tabularxcolumn}[1]{m{#1}}
\begin{tabularx}{\textwidth}{>{\centering\arraybackslash}m{0.075\textwidth} >{\centering\arraybackslash}m{0.07\textwidth} >{\centering\arraybackslash}m{0.095\textwidth} Y Y >{\centering\arraybackslash}m{0.105\textwidth}}
\toprule
\textbf{Molecule} & \textbf{A/R} & \textbf{Protocol} & \textbf{Ansatz / Parameterization} & \textbf{Optimizer and Key Settings} & \textbf{Final $|\Delta E|$ [Ha]} \\
\midrule
\multirow[c]{2}{=}{\centering BH} & \multirow[c]{2}{=}{\centering 49/51} & Initial & \texttt{hea\_ryrz\_ring}; $L=2$. & \texttt{cobyla}; steps 4096; init scale 0.1; seed 42. & $2.564\times 10^{-1}$ \\
\arrayrulecolor{black!28}\cmidrule(lr){3-6}\arrayrulecolor{black}
 &  & \shortstack[c]{Best\\(Iter. 40)} & \texttt{uccsd}; warm start $n=8$. & \texttt{cobyla}; steps 4096; init scale 0; seed 1040; rhobeg 1.0e-4. & $3.553\times 10^{-14}$ \\
\midrule
\multirow[c]{2}{=}{\centering LiH} & \multirow[c]{2}{=}{\centering 23/77} & Initial & \texttt{hea\_ryrz\_ring}; $L=2$. & \texttt{cobyla}; steps 4096; init scale 0.1; seed 42. & $2.956\times 10^{-1}$ \\
\arrayrulecolor{black!28}\cmidrule(lr){3-6}\arrayrulecolor{black}
 &  & \shortstack[c]{Best\\(Iter. 64)} & \texttt{uccsd}; param. \texttt{uccsd\_full}; warm start $n=15$. & \texttt{powell}; steps 4096; init scale 0; seed 42; xtol 1.0e-7. & $5.085\times 10^{-12}$ \\
\midrule
\multirow[c]{2}{=}{\centering BeH$_2$} & \multirow[c]{2}{=}{\centering 77/23} & Initial & \texttt{hea\_ryrz\_ring}; $L=2$. & \texttt{cobyla}; steps 4096; init scale 0.1; seed 42. & $4.446\times 10^{-1}$ \\
\arrayrulecolor{black!28}\cmidrule(lr){3-6}\arrayrulecolor{black}
 &  & \shortstack[c]{Best\\(Iter. 100)} & \texttt{uccsd}; warm start $n=26$. & \texttt{cobyla}; steps 4096; init scale 0; seed 5100; rhobeg 1.0e-4. & $6.721\times 10^{-7}$ \\
\midrule
\multirow[c]{2}{=}{\centering H$_2$O} & \multirow[c]{2}{=}{\centering 30/70} & Initial & \texttt{hea\_ryrz\_ring}; $L=2$. & \texttt{cobyla}; steps 4096; init scale 0.1; seed 42. & $1.116\times 10^{0}$ \\
\arrayrulecolor{black!28}\cmidrule(lr){3-6}\arrayrulecolor{black}
 &  & \shortstack[c]{Best\\(Iter. 77)} & \texttt{uccsd}; param. \texttt{uccsd\_full}. & \texttt{cobyla}; steps 4096; init scale 6.3e-4; seed 257. & $3.543\times 10^{-6}$ \\
\bottomrule
\end{tabularx}
\end{table}

%% file: generated_tn_summary_table.tex
\begin{table}[!htbp]
\caption{Baseline and best archived tensor-network protocols for the four $L=64$ campaigns. A/R denotes accepted/rejected counts. Energies are total Hamiltonian energies in the dimensionless model units set by each spin-chain Hamiltonian.}
\label{tab:supp_tn_protocols}
\centering
\setlength{\tabcolsep}{3pt}
\renewcommand{\arraystretch}{1.12}
\renewcommand{\tabularxcolumn}[1]{m{#1}}
\begin{tabularx}{\textwidth}{>{\centering\arraybackslash}m{0.105\textwidth} >{\centering\arraybackslash}m{0.07\textwidth} >{\centering\arraybackslash}m{0.095\textwidth} Y Y >{\centering\arraybackslash}m{0.11\textwidth}}
\toprule
\textbf{Model} & \textbf{A/R} & \textbf{Protocol} & \textbf{Method / Initial State} & \textbf{Bond and Solver Settings} & \shortstack[c]{\textbf{Final Energy}\\\textbf{[model units]}} \\
\midrule
\multirow[c]{2}{=}{\centering Heisenberg XXX} & \multirow[c]{2}{=}{\centering 12/88} & Initial & \texttt{dmrg1}; init \texttt{random}, init bond 2. & bond 4-6-8; cutoff 1.0e-6; tol 0.001; sweeps 6; ncv 4. & -28.165820 \\
\arrayrulecolor{black!28}\cmidrule(lr){3-6}\arrayrulecolor{black}
 &  & \shortstack[c]{Best\\(Iter. 88)} & \texttt{dmrg2}; init \texttt{neel}, init bond 8. & bond 8-12-16-24-32-48-64-80-96; cutoff 1.0e-14; tol 1.0e-7; sweeps 19; ncv 6. & -28.175425 \\
\midrule
\multirow[c]{2}{=}{\centering Gapless XXZ} & \multirow[c]{2}{=}{\centering 54/46} & Initial & \texttt{dmrg1}; init \texttt{random}, init bond 2. & bond 4-6-8; cutoff 1.0e-6; tol 0.001; sweeps 6; ncv 4. & -23.816607 \\
\arrayrulecolor{black!28}\cmidrule(lr){3-6}\arrayrulecolor{black}
 &  & \shortstack[c]{Best\\(Iter. 100)} & \texttt{dmrg1}$\rightarrow$\texttt{dmrg2}; init \texttt{neel}$\rightarrow$\texttt{neel}. & warm: bond 8-12-16-24-32-48; cutoff 1.0e-9; tol 1.0e-5; sweeps 5; refine: bond 148-192; cutoff 4.8e-14; tol 1.0e-6; sweeps 5. & -23.828520 \\
\midrule
\multirow[c]{2}{=}{\centering Critical TFIM} & \multirow[c]{2}{=}{\centering 6/94} & Initial & \texttt{dmrg1}; init \texttt{random}, init bond 2. & bond 4-6-8; cutoff 1.0e-6; tol 0.001; sweeps 6; ncv 4. & -20.281472 \\
\arrayrulecolor{black!28}\cmidrule(lr){3-6}\arrayrulecolor{black}
 &  & \shortstack[c]{Best\\(Iter. 34)} & \texttt{dmrg2}$\rightarrow$\texttt{dmrg1}; init \texttt{plus}$\rightarrow$\texttt{minus}. & warm: bond 8-12-16-24-32-48-64-80-96; cutoff 1.0e-10; tol 1.0e-7; sweeps 18; refine: bond 48-64-80; cutoff 1.0e-10; tol 5.0e-7; sweeps 12. & -20.281495 \\
\midrule
\multirow[c]{2}{=}{\centering Critical XX} & \multirow[c]{2}{=}{\centering 17/83} & Initial & \texttt{dmrg1}; init \texttt{random}, init bond 2. & bond 4-6-8; cutoff 1.0e-6; tol 0.001; sweeps 6; ncv 4. & -20.179132 \\
\arrayrulecolor{black!28}\cmidrule(lr){3-6}\arrayrulecolor{black}
 &  & \shortstack[c]{Best\\(Iter. 94)} & \texttt{dmrg2}; init \texttt{random}, init bond 1. & bond 8-12-16-24-32-48-64-80-96; cutoff 1.0e-10; tol 1.0e-5; sweeps 23; ncv 12. & -20.192157 \\
\bottomrule
\end{tabularx}
\end{table}

%% file: generated_afqmc_summary_table.tex
\begin{table}[!htbp]
\caption{Baseline and best archived AFQMC protocols for the four molecular campaigns. A/R denotes accepted/rejected counts.}
\label{tab:supp_afqmc_protocols}
\centering
\setlength{\tabcolsep}{3pt}
\renewcommand{\arraystretch}{1.12}
\renewcommand{\tabularxcolumn}[1]{m{#1}}
\begin{tabularx}{\textwidth}{>{\centering\arraybackslash}m{0.075\textwidth} >{\centering\arraybackslash}m{0.07\textwidth} >{\centering\arraybackslash}m{0.095\textwidth} Y Y >{\centering\arraybackslash}m{0.11\textwidth}}
\toprule
\textbf{Molecule} & \textbf{A/R} & \textbf{Protocol} & \textbf{Trial / SCF Settings} & \textbf{Walker and Propagation Settings} & \textbf{Live Score} \\
\midrule
\multirow[c]{2}{=}{\centering H$_2$} & \multirow[c]{2}{=}{\centering 9/91} & Initial & \texttt{rhf}; scf tol 1.0e-7; max cycle 64; diis 8. & 32 walkers/rank; $B=25\times 40$; $\Delta\tau=0.005$; chol cut 1.0e-5; $s/p=5/5$. & -1.120937 \\
\arrayrulecolor{black!28}\cmidrule(lr){3-6}\arrayrulecolor{black}
 &  & \shortstack[c]{Best\\(Iter. 80)} & \texttt{rhf}; scf tol 1.0e-7; max cycle 64; diis 8. & 320 walkers/rank; $B=80\times 20$; $\Delta\tau=0.0025$; chol cut 1.0e-8; $s/p=8/8$. & -1.134068 \\
\midrule
\multirow[c]{2}{=}{\centering LiH} & \multirow[c]{2}{=}{\centering 30/70} & Initial & \texttt{rhf}; scf tol 1.0e-7; max cycle 64; diis 8. & 32 walkers/rank; $B=25\times 40$; $\Delta\tau=0.005$; chol cut 1.0e-5; $s/p=5/5$. & -7.865619 \\
\arrayrulecolor{black!28}\cmidrule(lr){3-6}\arrayrulecolor{black}
 &  & \shortstack[c]{Best\\(Iter. 96)} & \texttt{uhf}; scf tol 1.0e-7; max cycle 64; diis 8. & 128 walkers/rank; $B=24\times 35$; $\Delta\tau=0.007569$; chol cut 1.0e-5; $s/p=1/2$. & -7.876347 \\
\midrule
\multirow[c]{2}{=}{\centering H$_2$O} & \multirow[c]{2}{=}{\centering 32/68} & Initial & \texttt{rhf}; scf tol 1.0e-7; max cycle 64; diis 8. & 32 walkers/rank; $B=25\times 40$; $\Delta\tau=0.005$; chol cut 1.0e-5; $s/p=5/5$. & -74.989938 \\
\arrayrulecolor{black!28}\cmidrule(lr){3-6}\arrayrulecolor{black}
 &  & \shortstack[c]{Best\\(Iter. 94)} & \texttt{uhf}; scf tol 1.0e-7; max cycle 64; diis 8. & 512 walkers/rank; $B=25\times 34$; $\Delta\tau=0.003764$; chol cut 1.0e-5; $s/p=5/4$. & -75.004519 \\
\midrule
\multirow[c]{2}{=}{\centering N$_2$} & \multirow[c]{2}{=}{\centering 15/85} & Initial & \texttt{rhf}; scf tol 1.0e-7; max cycle 64; diis 8. & 32 walkers/rank; $B=25\times 40$; $\Delta\tau=0.005$; chol cut 1.0e-5; $s/p=5/5$. & -107.465564 \\
\arrayrulecolor{black!28}\cmidrule(lr){3-6}\arrayrulecolor{black}
 &  & \shortstack[c]{Best\\(Iter. 89)} & \texttt{rhf}; scf tol 1.0e-7; max cycle 64; diis 8. & 512 walkers/rank; $B=26\times 29$; $\Delta\tau=0.003822$; chol cut 1.0e-5; $s/p=5/4$. & -107.620504 \\
\bottomrule
\end{tabularx}
\end{table}

%% file: generated_vqe_log_tables.tex
\refstepcounter{table}
\label{tab:supp_vqe_log_bh}
\noindent\textbf{Table \thetable.} VQE mutation log for BH full 100-iteration archival campaign. Relative gain is measured against the immediately previous scored iteration using the absolute energy error.
\par\medskip
\noindent\rule{\linewidth}{0.5pt}
\noindent{\setlength{\fboxsep}{0pt}\colorbox{black!4}{\parbox{\linewidth}{\parbox[c][2.2em][c]{0.05\linewidth}{\centering \textbf{Iter.}}\hspace{0.006\linewidth}\parbox[c][2.2em][c]{0.04\linewidth}{\centering \textbf{A/R}}\hspace{0.006\linewidth}\parbox[c][2.2em][c]{0.10\linewidth}{\centering \textbf{Relative gain}}\hspace{0.006\linewidth}\parbox[c][2.2em][c]{0.115\linewidth}{\centering \textbf{$|\Delta E|$ [Ha]}}\hspace{0.006\linewidth}\parbox[c][2.2em][c]{0.671\linewidth}{\centering \textbf{Mutation summary}}}}}\par\smallskip
\noindent\rule{\linewidth}{0.5pt}
\noindent{\setlength{\fboxsep}{0pt}\colorbox{black!4}{\parbox{\linewidth}{\parbox[t]{0.05\linewidth}{\centering\strut 1}\hspace{0.006\linewidth}\parbox[t]{0.04\linewidth}{\centering\strut \raisebox{0.08ex}{\ding{51}}}\hspace{0.006\linewidth}\parbox[t]{0.10\linewidth}{\centering\strut ---}\hspace{0.006\linewidth}\parbox[t]{0.115\linewidth}{\centering\strut $9.932\times 10^{-3}$}\hspace{0.006\linewidth}\parbox[t]{0.671\linewidth}{switch to UCCSD ansatz with cached callback energy}}}}\par\smallskip
\noindent\parbox[t]{0.05\linewidth}{\centering\strut 2}\hspace{0.006\linewidth}\parbox[t]{0.04\linewidth}{\centering\strut \raisebox{0.08ex}{\ding{51}}}\hspace{0.006\linewidth}\parbox[t]{0.10\linewidth}{\centering\strut +97.58\%}\hspace{0.006\linewidth}\parbox[t]{0.115\linewidth}{\centering\strut $2.4\times 10^{-4}$}\hspace{0.006\linewidth}\parbox[t]{0.671\linewidth}{warm-start Powell with small UCCSD perturbation}\par\smallskip
\noindent{\setlength{\fboxsep}{0pt}\colorbox{black!4}{\parbox{\linewidth}{\parbox[t]{0.05\linewidth}{\centering\strut 3}\hspace{0.006\linewidth}\parbox[t]{0.04\linewidth}{\centering\strut \raisebox{0.08ex}{\ding{51}}}\hspace{0.006\linewidth}\parbox[t]{0.10\linewidth}{\centering\strut +97.47\%}\hspace{0.006\linewidth}\parbox[t]{0.115\linewidth}{\centering\strut $6.08\times 10^{-6}$}\hspace{0.006\linewidth}\parbox[t]{0.671\linewidth}{tight Powell refinement from best UCCSD warm start}}}}\par\smallskip
\noindent\parbox[t]{0.05\linewidth}{\centering\strut 4}\hspace{0.006\linewidth}\parbox[t]{0.04\linewidth}{\centering\strut \raisebox{0.08ex}{\ding{51}}}\hspace{0.006\linewidth}\parbox[t]{0.10\linewidth}{\centering\strut +98.65\%}\hspace{0.006\linewidth}\parbox[t]{0.115\linewidth}{\centering\strut $8.196\times 10^{-8}$}\hspace{0.006\linewidth}\parbox[t]{0.671\linewidth}{small-radius COBYLA polish from best UCCSD point}\par\smallskip
\noindent{\setlength{\fboxsep}{0pt}\colorbox{black!4}{\parbox{\linewidth}{\parbox[t]{0.05\linewidth}{\centering\strut 5}\hspace{0.006\linewidth}\parbox[t]{0.04\linewidth}{\centering\strut \raisebox{0.08ex}{\ding{51}}}\hspace{0.006\linewidth}\parbox[t]{0.10\linewidth}{\centering\strut +53.79\%}\hspace{0.006\linewidth}\parbox[t]{0.115\linewidth}{\centering\strut $3.787\times 10^{-8}$}\hspace{0.006\linewidth}\parbox[t]{0.671\linewidth}{tight Powell pass from COBYLA-polished UCCSD point}}}}\par\smallskip
\noindent\parbox[t]{0.05\linewidth}{\centering\strut 6}\hspace{0.006\linewidth}\parbox[t]{0.04\linewidth}{\centering\strut \raisebox{0.08ex}{\ding{51}}}\hspace{0.006\linewidth}\parbox[t]{0.10\linewidth}{\centering\strut +97.84\%}\hspace{0.006\linewidth}\parbox[t]{0.115\linewidth}{\centering\strut $8.195\times 10^{-10}$}\hspace{0.006\linewidth}\parbox[t]{0.671\linewidth}{tiny-radius COBYLA micro-step from best UCCSD point}\par\smallskip
\noindent{\setlength{\fboxsep}{0pt}\colorbox{black!4}{\parbox{\linewidth}{\parbox[t]{0.05\linewidth}{\centering\strut 7}\hspace{0.006\linewidth}\parbox[t]{0.04\linewidth}{\centering\strut \raisebox{0.08ex}{\ding{51}}}\hspace{0.006\linewidth}\parbox[t]{0.10\linewidth}{\centering\strut +55.93\%}\hspace{0.006\linewidth}\parbox[t]{0.115\linewidth}{\centering\strut $3.612\times 10^{-10}$}\hspace{0.006\linewidth}\parbox[t]{0.671\linewidth}{adaptive polish: Powell no-perturb warm start}}}}\par\smallskip
\noindent\parbox[t]{0.05\linewidth}{\centering\strut 8}\hspace{0.006\linewidth}\parbox[t]{0.04\linewidth}{\centering\strut \raisebox{0.08ex}{\ding{51}}}\hspace{0.006\linewidth}\parbox[t]{0.10\linewidth}{\centering\strut +99.48\%}\hspace{0.006\linewidth}\parbox[t]{0.115\linewidth}{\centering\strut $1.872\times 10^{-12}$}\hspace{0.006\linewidth}\parbox[t]{0.671\linewidth}{adaptive polish: COBYLA tiny-radius warm start}\par\smallskip
\noindent{\setlength{\fboxsep}{0pt}\colorbox{black!4}{\parbox{\linewidth}{\parbox[t]{0.05\linewidth}{\centering\strut 9}\hspace{0.006\linewidth}\parbox[t]{0.04\linewidth}{\centering\strut \raisebox{0.08ex}{\ding{55}}}\hspace{0.006\linewidth}\parbox[t]{0.10\linewidth}{\centering\strut -3856.55\%}\hspace{0.006\linewidth}\parbox[t]{0.115\linewidth}{\centering\strut $7.408\times 10^{-11}$}\hspace{0.006\linewidth}\parbox[t]{0.671\linewidth}{adaptive polish: Powell tiny perturbation}}}}\par\smallskip
\noindent\parbox[t]{0.05\linewidth}{\centering\strut 10}\hspace{0.006\linewidth}\parbox[t]{0.04\linewidth}{\centering\strut \raisebox{0.08ex}{\ding{51}}}\hspace{0.006\linewidth}\parbox[t]{0.10\linewidth}{\centering\strut +99.33\%}\hspace{0.006\linewidth}\parbox[t]{0.115\linewidth}{\centering\strut $4.974\times 10^{-13}$}\hspace{0.006\linewidth}\parbox[t]{0.671\linewidth}{adaptive polish: COBYLA tiny perturbation}\par\smallskip
\noindent{\setlength{\fboxsep}{0pt}\colorbox{black!4}{\parbox{\linewidth}{\parbox[t]{0.05\linewidth}{\centering\strut 11}\hspace{0.006\linewidth}\parbox[t]{0.04\linewidth}{\centering\strut \raisebox{0.08ex}{\ding{55}}}\hspace{0.006\linewidth}\parbox[t]{0.10\linewidth}{\centering\strut -356637.17\%}\hspace{0.006\linewidth}\parbox[t]{0.115\linewidth}{\centering\strut $1.774\times 10^{-9}$}\hspace{0.006\linewidth}\parbox[t]{0.671\linewidth}{adaptive polish: Powell medium perturbation}}}}\par\smallskip
\noindent\parbox[t]{0.05\linewidth}{\centering\strut 12}\hspace{0.006\linewidth}\parbox[t]{0.04\linewidth}{\centering\strut \raisebox{0.08ex}{\ding{55}}}\hspace{0.006\linewidth}\parbox[t]{0.10\linewidth}{\centering\strut +99.01\%}\hspace{0.006\linewidth}\parbox[t]{0.115\linewidth}{\centering\strut $1.762\times 10^{-11}$}\hspace{0.006\linewidth}\parbox[t]{0.671\linewidth}{adaptive polish: COBYLA medium perturbation}\par\smallskip
\noindent{\setlength{\fboxsep}{0pt}\colorbox{black!4}{\parbox{\linewidth}{\parbox[t]{0.05\linewidth}{\centering\strut 13}\hspace{0.006\linewidth}\parbox[t]{0.04\linewidth}{\centering\strut \raisebox{0.08ex}{\ding{55}}}\hspace{0.006\linewidth}\parbox[t]{0.10\linewidth}{\centering\strut +97.18\%}\hspace{0.006\linewidth}\parbox[t]{0.115\linewidth}{\centering\strut $4.974\times 10^{-13}$}\hspace{0.006\linewidth}\parbox[t]{0.671\linewidth}{adaptive polish: ultra-tight Powell retry}}}}\par\smallskip
\noindent\parbox[t]{0.05\linewidth}{\centering\strut 14}\hspace{0.006\linewidth}\parbox[t]{0.04\linewidth}{\centering\strut \raisebox{0.08ex}{\ding{51}}}\hspace{0.006\linewidth}\parbox[t]{0.10\linewidth}{\centering\strut +18.57\%}\hspace{0.006\linewidth}\parbox[t]{0.115\linewidth}{\centering\strut $4.05\times 10^{-13}$}\hspace{0.006\linewidth}\parbox[t]{0.671\linewidth}{adaptive polish: ultra-tight COBYLA retry}\par\smallskip
\noindent{\setlength{\fboxsep}{0pt}\colorbox{black!4}{\parbox{\linewidth}{\parbox[t]{0.05\linewidth}{\centering\strut 15}\hspace{0.006\linewidth}\parbox[t]{0.04\linewidth}{\centering\strut \raisebox{0.08ex}{\ding{51}}}\hspace{0.006\linewidth}\parbox[t]{0.10\linewidth}{\centering\strut +0\%}\hspace{0.006\linewidth}\parbox[t]{0.115\linewidth}{\centering\strut $4.05\times 10^{-13}$}\hspace{0.006\linewidth}\parbox[t]{0.671\linewidth}{adaptive polish: Powell no-perturb warm start}}}}\par\smallskip
\noindent\parbox[t]{0.05\linewidth}{\centering\strut 16}\hspace{0.006\linewidth}\parbox[t]{0.04\linewidth}{\centering\strut \raisebox{0.08ex}{\ding{51}}}\hspace{0.006\linewidth}\parbox[t]{0.10\linewidth}{\centering\strut +18.42\%}\hspace{0.006\linewidth}\parbox[t]{0.115\linewidth}{\centering\strut $3.304\times 10^{-13}$}\hspace{0.006\linewidth}\parbox[t]{0.671\linewidth}{adaptive polish: COBYLA tiny-radius warm start}\par\smallskip
\noindent{\setlength{\fboxsep}{0pt}\colorbox{black!4}{\parbox{\linewidth}{\parbox[t]{0.05\linewidth}{\centering\strut 17}\hspace{0.006\linewidth}\parbox[t]{0.04\linewidth}{\centering\strut \raisebox{0.08ex}{\ding{55}}}\hspace{0.006\linewidth}\parbox[t]{0.10\linewidth}{\centering\strut -9317.2\%}\hspace{0.006\linewidth}\parbox[t]{0.115\linewidth}{\centering\strut $3.111\times 10^{-11}$}\hspace{0.006\linewidth}\parbox[t]{0.671\linewidth}{adaptive polish: Powell tiny perturbation}}}}\par\smallskip
\noindent\parbox[t]{0.05\linewidth}{\centering\strut 18}\hspace{0.006\linewidth}\parbox[t]{0.04\linewidth}{\centering\strut \raisebox{0.08ex}{\ding{51}}}\hspace{0.006\linewidth}\parbox[t]{0.10\linewidth}{\centering\strut +99.5\%}\hspace{0.006\linewidth}\parbox[t]{0.115\linewidth}{\centering\strut $1.563\times 10^{-13}$}\hspace{0.006\linewidth}\parbox[t]{0.671\linewidth}{adaptive polish: COBYLA tiny perturbation}\par\smallskip
\noindent{\setlength{\fboxsep}{0pt}\colorbox{black!4}{\parbox{\linewidth}{\parbox[t]{0.05\linewidth}{\centering\strut 19}\hspace{0.006\linewidth}\parbox[t]{0.04\linewidth}{\centering\strut \raisebox{0.08ex}{\ding{55}}}\hspace{0.006\linewidth}\parbox[t]{0.10\linewidth}{\centering\strut -146825.01\%}\hspace{0.006\linewidth}\parbox[t]{0.115\linewidth}{\centering\strut $2.297\times 10^{-10}$}\hspace{0.006\linewidth}\parbox[t]{0.671\linewidth}{adaptive polish: Powell medium perturbation}}}}\par\smallskip
\noindent\parbox[t]{0.05\linewidth}{\centering\strut 20}\hspace{0.006\linewidth}\parbox[t]{0.04\linewidth}{\centering\strut \raisebox{0.08ex}{\ding{55}}}\hspace{0.006\linewidth}\parbox[t]{0.10\linewidth}{\centering\strut +93.88\%}\hspace{0.006\linewidth}\parbox[t]{0.115\linewidth}{\centering\strut $1.407\times 10^{-11}$}\hspace{0.006\linewidth}\parbox[t]{0.671\linewidth}{adaptive polish: COBYLA medium perturbation}\par\smallskip
\noindent{\setlength{\fboxsep}{0pt}\colorbox{black!4}{\parbox{\linewidth}{\parbox[t]{0.05\linewidth}{\centering\strut 21}\hspace{0.006\linewidth}\parbox[t]{0.04\linewidth}{\centering\strut \raisebox{0.08ex}{\ding{51}}}\hspace{0.006\linewidth}\parbox[t]{0.10\linewidth}{\centering\strut +99.12\%}\hspace{0.006\linewidth}\parbox[t]{0.115\linewidth}{\centering\strut $1.243\times 10^{-13}$}\hspace{0.006\linewidth}\parbox[t]{0.671\linewidth}{adaptive polish: ultra-tight Powell retry}}}}\par\smallskip
\noindent\parbox[t]{0.05\linewidth}{\centering\strut 22}\hspace{0.006\linewidth}\parbox[t]{0.04\linewidth}{\centering\strut \raisebox{0.08ex}{\ding{51}}}\hspace{0.006\linewidth}\parbox[t]{0.10\linewidth}{\centering\strut +40\%}\hspace{0.006\linewidth}\parbox[t]{0.115\linewidth}{\centering\strut $7.461\times 10^{-14}$}\hspace{0.006\linewidth}\parbox[t]{0.671\linewidth}{adaptive polish: ultra-tight COBYLA retry}\par\smallskip
\noindent{\setlength{\fboxsep}{0pt}\colorbox{black!4}{\parbox{\linewidth}{\parbox[t]{0.05\linewidth}{\centering\strut 23}\hspace{0.006\linewidth}\parbox[t]{0.04\linewidth}{\centering\strut \raisebox{0.08ex}{\ding{51}}}\hspace{0.006\linewidth}\parbox[t]{0.10\linewidth}{\centering\strut +0\%}\hspace{0.006\linewidth}\parbox[t]{0.115\linewidth}{\centering\strut $7.461\times 10^{-14}$}\hspace{0.006\linewidth}\parbox[t]{0.671\linewidth}{adaptive polish: Powell no-perturb warm start}}}}\par\smallskip
\noindent\parbox[t]{0.05\linewidth}{\centering\strut 24}\hspace{0.006\linewidth}\parbox[t]{0.04\linewidth}{\centering\strut \raisebox{0.08ex}{\ding{51}}}\hspace{0.006\linewidth}\parbox[t]{0.10\linewidth}{\centering\strut +0\%}\hspace{0.006\linewidth}\parbox[t]{0.115\linewidth}{\centering\strut $7.461\times 10^{-14}$}\hspace{0.006\linewidth}\parbox[t]{0.671\linewidth}{adaptive polish: COBYLA tiny-radius warm start}\par\smallskip
\noindent{\setlength{\fboxsep}{0pt}\colorbox{black!4}{\parbox{\linewidth}{\parbox[t]{0.05\linewidth}{\centering\strut 25}\hspace{0.006\linewidth}\parbox[t]{0.04\linewidth}{\centering\strut \raisebox{0.08ex}{\ding{55}}}\hspace{0.006\linewidth}\parbox[t]{0.10\linewidth}{\centering\strut -20852.38\%}\hspace{0.006\linewidth}\parbox[t]{0.115\linewidth}{\centering\strut $1.563\times 10^{-11}$}\hspace{0.006\linewidth}\parbox[t]{0.671\linewidth}{adaptive polish: Powell tiny perturbation}}}}\par\smallskip
\noindent\parbox[t]{0.05\linewidth}{\centering\strut 26}\hspace{0.006\linewidth}\parbox[t]{0.04\linewidth}{\centering\strut \raisebox{0.08ex}{\ding{55}}}\hspace{0.006\linewidth}\parbox[t]{0.10\linewidth}{\centering\strut +93.09\%}\hspace{0.006\linewidth}\parbox[t]{0.115\linewidth}{\centering\strut $1.08\times 10^{-12}$}\hspace{0.006\linewidth}\parbox[t]{0.671\linewidth}{adaptive polish: COBYLA tiny perturbation}\par\smallskip
\noindent{\setlength{\fboxsep}{0pt}\colorbox{black!4}{\parbox{\linewidth}{\parbox[t]{0.05\linewidth}{\centering\strut 27}\hspace{0.006\linewidth}\parbox[t]{0.04\linewidth}{\centering\strut \raisebox{0.08ex}{\ding{55}}}\hspace{0.006\linewidth}\parbox[t]{0.10\linewidth}{\centering\strut -119079.56\%}\hspace{0.006\linewidth}\parbox[t]{0.115\linewidth}{\centering\strut $1.287\times 10^{-9}$}\hspace{0.006\linewidth}\parbox[t]{0.671\linewidth}{adaptive polish: Powell medium perturbation}}}}\par\smallskip
\noindent\parbox[t]{0.05\linewidth}{\centering\strut 28}\hspace{0.006\linewidth}\parbox[t]{0.04\linewidth}{\centering\strut \raisebox{0.08ex}{\ding{55}}}\hspace{0.006\linewidth}\parbox[t]{0.10\linewidth}{\centering\strut +99.52\%}\hspace{0.006\linewidth}\parbox[t]{0.115\linewidth}{\centering\strut $6.189\times 10^{-12}$}\hspace{0.006\linewidth}\parbox[t]{0.671\linewidth}{adaptive polish: COBYLA medium perturbation}\par\smallskip
\noindent{\setlength{\fboxsep}{0pt}\colorbox{black!4}{\parbox{\linewidth}{\parbox[t]{0.05\linewidth}{\centering\strut 29}\hspace{0.006\linewidth}\parbox[t]{0.04\linewidth}{\centering\strut \raisebox{0.08ex}{\ding{51}}}\hspace{0.006\linewidth}\parbox[t]{0.10\linewidth}{\centering\strut +98.79\%}\hspace{0.006\linewidth}\parbox[t]{0.115\linewidth}{\centering\strut $7.461\times 10^{-14}$}\hspace{0.006\linewidth}\parbox[t]{0.671\linewidth}{adaptive polish: ultra-tight Powell retry}}}}\par\smallskip
\noindent\parbox[t]{0.05\linewidth}{\centering\strut 30}\hspace{0.006\linewidth}\parbox[t]{0.04\linewidth}{\centering\strut \raisebox{0.08ex}{\ding{51}}}\hspace{0.006\linewidth}\parbox[t]{0.10\linewidth}{\centering\strut +4.76\%}\hspace{0.006\linewidth}\parbox[t]{0.115\linewidth}{\centering\strut $7.105\times 10^{-14}$}\hspace{0.006\linewidth}\parbox[t]{0.671\linewidth}{adaptive polish: ultra-tight COBYLA retry}\par\smallskip
\noindent{\setlength{\fboxsep}{0pt}\colorbox{black!4}{\parbox{\linewidth}{\parbox[t]{0.05\linewidth}{\centering\strut 31}\hspace{0.006\linewidth}\parbox[t]{0.04\linewidth}{\centering\strut \raisebox{0.08ex}{\ding{55}}}\hspace{0.006\linewidth}\parbox[t]{0.10\linewidth}{\centering\strut +0\%}\hspace{0.006\linewidth}\parbox[t]{0.115\linewidth}{\centering\strut $7.105\times 10^{-14}$}\hspace{0.006\linewidth}\parbox[t]{0.671\linewidth}{adaptive polish: Powell no-perturb warm start}}}}\par\smallskip
\noindent\parbox[t]{0.05\linewidth}{\centering\strut 32}\hspace{0.006\linewidth}\parbox[t]{0.04\linewidth}{\centering\strut \raisebox{0.08ex}{\ding{55}}}\hspace{0.006\linewidth}\parbox[t]{0.10\linewidth}{\centering\strut +0\%}\hspace{0.006\linewidth}\parbox[t]{0.115\linewidth}{\centering\strut $7.105\times 10^{-14}$}\hspace{0.006\linewidth}\parbox[t]{0.671\linewidth}{adaptive polish: COBYLA tiny-radius warm start}\par\smallskip
\noindent{\setlength{\fboxsep}{0pt}\colorbox{black!4}{\parbox{\linewidth}{\parbox[t]{0.05\linewidth}{\centering\strut 33}\hspace{0.006\linewidth}\parbox[t]{0.04\linewidth}{\centering\strut \raisebox{0.08ex}{\ding{55}}}\hspace{0.006\linewidth}\parbox[t]{0.10\linewidth}{\centering\strut -71255.01\%}\hspace{0.006\linewidth}\parbox[t]{0.115\linewidth}{\centering\strut $5.07\times 10^{-11}$}\hspace{0.006\linewidth}\parbox[t]{0.671\linewidth}{adaptive polish: Powell tiny perturbation}}}}\par\smallskip
\noindent\parbox[t]{0.05\linewidth}{\centering\strut 34}\hspace{0.006\linewidth}\parbox[t]{0.04\linewidth}{\centering\strut \raisebox{0.08ex}{\ding{55}}}\hspace{0.006\linewidth}\parbox[t]{0.10\linewidth}{\centering\strut +96.86\%}\hspace{0.006\linewidth}\parbox[t]{0.115\linewidth}{\centering\strut $1.592\times 10^{-12}$}\hspace{0.006\linewidth}\parbox[t]{0.671\linewidth}{adaptive polish: COBYLA tiny perturbation}\par\smallskip
\noindent{\setlength{\fboxsep}{0pt}\colorbox{black!4}{\parbox{\linewidth}{\parbox[t]{0.05\linewidth}{\centering\strut 35}\hspace{0.006\linewidth}\parbox[t]{0.04\linewidth}{\centering\strut \raisebox{0.08ex}{\ding{55}}}\hspace{0.006\linewidth}\parbox[t]{0.10\linewidth}{\centering\strut -25003.79\%}\hspace{0.006\linewidth}\parbox[t]{0.115\linewidth}{\centering\strut $3.996\times 10^{-10}$}\hspace{0.006\linewidth}\parbox[t]{0.671\linewidth}{adaptive polish: Powell medium perturbation}}}}\par\smallskip
\noindent\parbox[t]{0.05\linewidth}{\centering\strut 36}\hspace{0.006\linewidth}\parbox[t]{0.04\linewidth}{\centering\strut \raisebox{0.08ex}{\ding{55}}}\hspace{0.006\linewidth}\parbox[t]{0.10\linewidth}{\centering\strut +99.78\%}\hspace{0.006\linewidth}\parbox[t]{0.115\linewidth}{\centering\strut $8.917\times 10^{-13}$}\hspace{0.006\linewidth}\parbox[t]{0.671\linewidth}{adaptive polish: COBYLA medium perturbation}\par\smallskip
\noindent{\setlength{\fboxsep}{0pt}\colorbox{black!4}{\parbox{\linewidth}{\parbox[t]{0.05\linewidth}{\centering\strut 37}\hspace{0.006\linewidth}\parbox[t]{0.04\linewidth}{\centering\strut \raisebox{0.08ex}{\ding{55}}}\hspace{0.006\linewidth}\parbox[t]{0.10\linewidth}{\centering\strut +92.03\%}\hspace{0.006\linewidth}\parbox[t]{0.115\linewidth}{\centering\strut $7.105\times 10^{-14}$}\hspace{0.006\linewidth}\parbox[t]{0.671\linewidth}{adaptive polish: ultra-tight Powell retry}}}}\par\smallskip
\noindent\parbox[t]{0.05\linewidth}{\centering\strut 38}\hspace{0.006\linewidth}\parbox[t]{0.04\linewidth}{\centering\strut \raisebox{0.08ex}{\ding{51}}}\hspace{0.006\linewidth}\parbox[t]{0.10\linewidth}{\centering\strut +45\%}\hspace{0.006\linewidth}\parbox[t]{0.115\linewidth}{\centering\strut $3.908\times 10^{-14}$}\hspace{0.006\linewidth}\parbox[t]{0.671\linewidth}{adaptive polish: ultra-tight COBYLA retry}\par\smallskip
\noindent{\setlength{\fboxsep}{0pt}\colorbox{black!4}{\parbox{\linewidth}{\parbox[t]{0.05\linewidth}{\centering\strut 39}\hspace{0.006\linewidth}\parbox[t]{0.04\linewidth}{\centering\strut \raisebox{0.08ex}{\ding{55}}}\hspace{0.006\linewidth}\parbox[t]{0.10\linewidth}{\centering\strut +0\%}\hspace{0.006\linewidth}\parbox[t]{0.115\linewidth}{\centering\strut $3.908\times 10^{-14}$}\hspace{0.006\linewidth}\parbox[t]{0.671\linewidth}{adaptive polish: Powell no-perturb warm start}}}}\par\smallskip
\noindent\parbox[t]{0.05\linewidth}{\centering\strut 40}\hspace{0.006\linewidth}\parbox[t]{0.04\linewidth}{\centering\strut \raisebox{0.08ex}{\ding{51}}}\hspace{0.006\linewidth}\parbox[t]{0.10\linewidth}{\centering\strut +9.09\%}\hspace{0.006\linewidth}\parbox[t]{0.115\linewidth}{\centering\strut $3.553\times 10^{-14}$}\hspace{0.006\linewidth}\parbox[t]{0.671\linewidth}{adaptive polish: COBYLA tiny-radius warm start}\par\smallskip
\noindent{\setlength{\fboxsep}{0pt}\colorbox{black!4}{\parbox{\linewidth}{\parbox[t]{0.05\linewidth}{\centering\strut 41}\hspace{0.006\linewidth}\parbox[t]{0.04\linewidth}{\centering\strut \raisebox{0.08ex}{\ding{55}}}\hspace{0.006\linewidth}\parbox[t]{0.10\linewidth}{\centering\strut -132909.97\%}\hspace{0.006\linewidth}\parbox[t]{0.115\linewidth}{\centering\strut $4.725\times 10^{-11}$}\hspace{0.006\linewidth}\parbox[t]{0.671\linewidth}{adaptive polish: Powell tiny perturbation}}}}\par\smallskip
\noindent\parbox[t]{0.05\linewidth}{\centering\strut 42}\hspace{0.006\linewidth}\parbox[t]{0.04\linewidth}{\centering\strut \raisebox{0.08ex}{\ding{55}}}\hspace{0.006\linewidth}\parbox[t]{0.10\linewidth}{\centering\strut +99.53\%}\hspace{0.006\linewidth}\parbox[t]{0.115\linewidth}{\centering\strut $2.203\times 10^{-13}$}\hspace{0.006\linewidth}\parbox[t]{0.671\linewidth}{adaptive polish: COBYLA tiny perturbation}\par\smallskip
\noindent{\setlength{\fboxsep}{0pt}\colorbox{black!4}{\parbox{\linewidth}{\parbox[t]{0.05\linewidth}{\centering\strut 43}\hspace{0.006\linewidth}\parbox[t]{0.04\linewidth}{\centering\strut \raisebox{0.08ex}{\ding{55}}}\hspace{0.006\linewidth}\parbox[t]{0.10\linewidth}{\centering\strut -184579.09\%}\hspace{0.006\linewidth}\parbox[t]{0.115\linewidth}{\centering\strut $4.068\times 10^{-10}$}\hspace{0.006\linewidth}\parbox[t]{0.671\linewidth}{adaptive polish: Powell medium perturbation}}}}\par\smallskip
\noindent\parbox[t]{0.05\linewidth}{\centering\strut 44}\hspace{0.006\linewidth}\parbox[t]{0.04\linewidth}{\centering\strut \raisebox{0.08ex}{\ding{55}}}\hspace{0.006\linewidth}\parbox[t]{0.10\linewidth}{\centering\strut +99.73\%}\hspace{0.006\linewidth}\parbox[t]{0.115\linewidth}{\centering\strut $1.108\times 10^{-12}$}\hspace{0.006\linewidth}\parbox[t]{0.671\linewidth}{adaptive polish: COBYLA medium perturbation}\par\smallskip
\noindent{\setlength{\fboxsep}{0pt}\colorbox{black!4}{\parbox{\linewidth}{\parbox[t]{0.05\linewidth}{\centering\strut 45}\hspace{0.006\linewidth}\parbox[t]{0.04\linewidth}{\centering\strut \raisebox{0.08ex}{\ding{51}}}\hspace{0.006\linewidth}\parbox[t]{0.10\linewidth}{\centering\strut +96.79\%}\hspace{0.006\linewidth}\parbox[t]{0.115\linewidth}{\centering\strut $3.553\times 10^{-14}$}\hspace{0.006\linewidth}\parbox[t]{0.671\linewidth}{adaptive polish: ultra-tight Powell retry}}}}\par\smallskip
\noindent\parbox[t]{0.05\linewidth}{\centering\strut 46}\hspace{0.006\linewidth}\parbox[t]{0.04\linewidth}{\centering\strut \raisebox{0.08ex}{\ding{51}}}\hspace{0.006\linewidth}\parbox[t]{0.10\linewidth}{\centering\strut +0\%}\hspace{0.006\linewidth}\parbox[t]{0.115\linewidth}{\centering\strut $3.553\times 10^{-14}$}\hspace{0.006\linewidth}\parbox[t]{0.671\linewidth}{adaptive polish: ultra-tight COBYLA retry}\par\smallskip
\noindent{\setlength{\fboxsep}{0pt}\colorbox{black!4}{\parbox{\linewidth}{\parbox[t]{0.05\linewidth}{\centering\strut 47}\hspace{0.006\linewidth}\parbox[t]{0.04\linewidth}{\centering\strut \raisebox{0.08ex}{\ding{51}}}\hspace{0.006\linewidth}\parbox[t]{0.10\linewidth}{\centering\strut +0\%}\hspace{0.006\linewidth}\parbox[t]{0.115\linewidth}{\centering\strut $3.553\times 10^{-14}$}\hspace{0.006\linewidth}\parbox[t]{0.671\linewidth}{adaptive polish: Powell no-perturb warm start}}}}\par\smallskip
\noindent\parbox[t]{0.05\linewidth}{\centering\strut 48}\hspace{0.006\linewidth}\parbox[t]{0.04\linewidth}{\centering\strut \raisebox{0.08ex}{\ding{51}}}\hspace{0.006\linewidth}\parbox[t]{0.10\linewidth}{\centering\strut +0\%}\hspace{0.006\linewidth}\parbox[t]{0.115\linewidth}{\centering\strut $3.553\times 10^{-14}$}\hspace{0.006\linewidth}\parbox[t]{0.671\linewidth}{adaptive polish: COBYLA tiny-radius warm start}\par\smallskip
\noindent{\setlength{\fboxsep}{0pt}\colorbox{black!4}{\parbox{\linewidth}{\parbox[t]{0.05\linewidth}{\centering\strut 49}\hspace{0.006\linewidth}\parbox[t]{0.04\linewidth}{\centering\strut \raisebox{0.08ex}{\ding{55}}}\hspace{0.006\linewidth}\parbox[t]{0.10\linewidth}{\centering\strut -66040\%}\hspace{0.006\linewidth}\parbox[t]{0.115\linewidth}{\centering\strut $2.35\times 10^{-11}$}\hspace{0.006\linewidth}\parbox[t]{0.671\linewidth}{adaptive polish: Powell tiny perturbation}}}}\par\smallskip
\noindent\parbox[t]{0.05\linewidth}{\centering\strut 50}\hspace{0.006\linewidth}\parbox[t]{0.04\linewidth}{\centering\strut \raisebox{0.08ex}{\ding{55}}}\hspace{0.006\linewidth}\parbox[t]{0.10\linewidth}{\centering\strut +99.46\%}\hspace{0.006\linewidth}\parbox[t]{0.115\linewidth}{\centering\strut $1.279\times 10^{-13}$}\hspace{0.006\linewidth}\parbox[t]{0.671\linewidth}{adaptive polish: COBYLA tiny perturbation}\par\smallskip
\noindent{\setlength{\fboxsep}{0pt}\colorbox{black!4}{\parbox{\linewidth}{\parbox[t]{0.05\linewidth}{\centering\strut 51}\hspace{0.006\linewidth}\parbox[t]{0.04\linewidth}{\centering\strut \raisebox{0.08ex}{\ding{55}}}\hspace{0.006\linewidth}\parbox[t]{0.10\linewidth}{\centering\strut -788800.03\%}\hspace{0.006\linewidth}\parbox[t]{0.115\linewidth}{\centering\strut $1.009\times 10^{-9}$}\hspace{0.006\linewidth}\parbox[t]{0.671\linewidth}{adaptive polish: Powell medium perturbation}}}}\par\smallskip
\noindent\parbox[t]{0.05\linewidth}{\centering\strut 52}\hspace{0.006\linewidth}\parbox[t]{0.04\linewidth}{\centering\strut \raisebox{0.08ex}{\ding{55}}}\hspace{0.006\linewidth}\parbox[t]{0.10\linewidth}{\centering\strut +99.93\%}\hspace{0.006\linewidth}\parbox[t]{0.115\linewidth}{\centering\strut $7.567\times 10^{-13}$}\hspace{0.006\linewidth}\parbox[t]{0.671\linewidth}{adaptive polish: COBYLA medium perturbation}\par\smallskip
\noindent{\setlength{\fboxsep}{0pt}\colorbox{black!4}{\parbox{\linewidth}{\parbox[t]{0.05\linewidth}{\centering\strut 53}\hspace{0.006\linewidth}\parbox[t]{0.04\linewidth}{\centering\strut \raisebox{0.08ex}{\ding{51}}}\hspace{0.006\linewidth}\parbox[t]{0.10\linewidth}{\centering\strut +95.31\%}\hspace{0.006\linewidth}\parbox[t]{0.115\linewidth}{\centering\strut $3.553\times 10^{-14}$}\hspace{0.006\linewidth}\parbox[t]{0.671\linewidth}{adaptive polish: ultra-tight Powell retry}}}}\par\smallskip
\noindent\parbox[t]{0.05\linewidth}{\centering\strut 54}\hspace{0.006\linewidth}\parbox[t]{0.04\linewidth}{\centering\strut \raisebox{0.08ex}{\ding{51}}}\hspace{0.006\linewidth}\parbox[t]{0.10\linewidth}{\centering\strut +0\%}\hspace{0.006\linewidth}\parbox[t]{0.115\linewidth}{\centering\strut $3.553\times 10^{-14}$}\hspace{0.006\linewidth}\parbox[t]{0.671\linewidth}{adaptive polish: ultra-tight COBYLA retry}\par\smallskip
\noindent{\setlength{\fboxsep}{0pt}\colorbox{black!4}{\parbox{\linewidth}{\parbox[t]{0.05\linewidth}{\centering\strut 55}\hspace{0.006\linewidth}\parbox[t]{0.04\linewidth}{\centering\strut \raisebox{0.08ex}{\ding{51}}}\hspace{0.006\linewidth}\parbox[t]{0.10\linewidth}{\centering\strut +0\%}\hspace{0.006\linewidth}\parbox[t]{0.115\linewidth}{\centering\strut $3.553\times 10^{-14}$}\hspace{0.006\linewidth}\parbox[t]{0.671\linewidth}{adaptive polish: Powell no-perturb warm start}}}}\par\smallskip
\noindent\parbox[t]{0.05\linewidth}{\centering\strut 56}\hspace{0.006\linewidth}\parbox[t]{0.04\linewidth}{\centering\strut \raisebox{0.08ex}{\ding{51}}}\hspace{0.006\linewidth}\parbox[t]{0.10\linewidth}{\centering\strut +0\%}\hspace{0.006\linewidth}\parbox[t]{0.115\linewidth}{\centering\strut $3.553\times 10^{-14}$}\hspace{0.006\linewidth}\parbox[t]{0.671\linewidth}{adaptive polish: COBYLA tiny-radius warm start}\par\smallskip
\noindent{\setlength{\fboxsep}{0pt}\colorbox{black!4}{\parbox{\linewidth}{\parbox[t]{0.05\linewidth}{\centering\strut 57}\hspace{0.006\linewidth}\parbox[t]{0.04\linewidth}{\centering\strut \raisebox{0.08ex}{\ding{55}}}\hspace{0.006\linewidth}\parbox[t]{0.10\linewidth}{\centering\strut -91060\%}\hspace{0.006\linewidth}\parbox[t]{0.115\linewidth}{\centering\strut $3.239\times 10^{-11}$}\hspace{0.006\linewidth}\parbox[t]{0.671\linewidth}{adaptive polish: Powell tiny perturbation}}}}\par\smallskip
\noindent\parbox[t]{0.05\linewidth}{\centering\strut 58}\hspace{0.006\linewidth}\parbox[t]{0.04\linewidth}{\centering\strut \raisebox{0.08ex}{\ding{55}}}\hspace{0.006\linewidth}\parbox[t]{0.10\linewidth}{\centering\strut +96.42\%}\hspace{0.006\linewidth}\parbox[t]{0.115\linewidth}{\centering\strut $1.158\times 10^{-12}$}\hspace{0.006\linewidth}\parbox[t]{0.671\linewidth}{adaptive polish: COBYLA tiny perturbation}\par\smallskip
\noindent{\setlength{\fboxsep}{0pt}\colorbox{black!4}{\parbox{\linewidth}{\parbox[t]{0.05\linewidth}{\centering\strut 59}\hspace{0.006\linewidth}\parbox[t]{0.04\linewidth}{\centering\strut \raisebox{0.08ex}{\ding{55}}}\hspace{0.006\linewidth}\parbox[t]{0.10\linewidth}{\centering\strut -86734.92\%}\hspace{0.006\linewidth}\parbox[t]{0.115\linewidth}{\centering\strut $1.006\times 10^{-9}$}\hspace{0.006\linewidth}\parbox[t]{0.671\linewidth}{adaptive polish: Powell medium perturbation}}}}\par\smallskip
\noindent\parbox[t]{0.05\linewidth}{\centering\strut 60}\hspace{0.006\linewidth}\parbox[t]{0.04\linewidth}{\centering\strut \raisebox{0.08ex}{\ding{55}}}\hspace{0.006\linewidth}\parbox[t]{0.10\linewidth}{\centering\strut +99.14\%}\hspace{0.006\linewidth}\parbox[t]{0.115\linewidth}{\centering\strut $8.622\times 10^{-12}$}\hspace{0.006\linewidth}\parbox[t]{0.671\linewidth}{adaptive polish: COBYLA medium perturbation}\par\smallskip
\noindent{\setlength{\fboxsep}{0pt}\colorbox{black!4}{\parbox{\linewidth}{\parbox[t]{0.05\linewidth}{\centering\strut 61}\hspace{0.006\linewidth}\parbox[t]{0.04\linewidth}{\centering\strut \raisebox{0.08ex}{\ding{51}}}\hspace{0.006\linewidth}\parbox[t]{0.10\linewidth}{\centering\strut +99.59\%}\hspace{0.006\linewidth}\parbox[t]{0.115\linewidth}{\centering\strut $3.553\times 10^{-14}$}\hspace{0.006\linewidth}\parbox[t]{0.671\linewidth}{adaptive polish: ultra-tight Powell retry}}}}\par\smallskip
\noindent\parbox[t]{0.05\linewidth}{\centering\strut 62}\hspace{0.006\linewidth}\parbox[t]{0.04\linewidth}{\centering\strut \raisebox{0.08ex}{\ding{51}}}\hspace{0.006\linewidth}\parbox[t]{0.10\linewidth}{\centering\strut +0\%}\hspace{0.006\linewidth}\parbox[t]{0.115\linewidth}{\centering\strut $3.553\times 10^{-14}$}\hspace{0.006\linewidth}\parbox[t]{0.671\linewidth}{adaptive polish: ultra-tight COBYLA retry}\par\smallskip
\noindent{\setlength{\fboxsep}{0pt}\colorbox{black!4}{\parbox{\linewidth}{\parbox[t]{0.05\linewidth}{\centering\strut 63}\hspace{0.006\linewidth}\parbox[t]{0.04\linewidth}{\centering\strut \raisebox{0.08ex}{\ding{51}}}\hspace{0.006\linewidth}\parbox[t]{0.10\linewidth}{\centering\strut +0\%}\hspace{0.006\linewidth}\parbox[t]{0.115\linewidth}{\centering\strut $3.553\times 10^{-14}$}\hspace{0.006\linewidth}\parbox[t]{0.671\linewidth}{adaptive polish: Powell no-perturb warm start}}}}\par\smallskip
\noindent\parbox[t]{0.05\linewidth}{\centering\strut 64}\hspace{0.006\linewidth}\parbox[t]{0.04\linewidth}{\centering\strut \raisebox{0.08ex}{\ding{51}}}\hspace{0.006\linewidth}\parbox[t]{0.10\linewidth}{\centering\strut +0\%}\hspace{0.006\linewidth}\parbox[t]{0.115\linewidth}{\centering\strut $3.553\times 10^{-14}$}\hspace{0.006\linewidth}\parbox[t]{0.671\linewidth}{adaptive polish: COBYLA tiny-radius warm start}\par\smallskip
\noindent{\setlength{\fboxsep}{0pt}\colorbox{black!4}{\parbox{\linewidth}{\parbox[t]{0.05\linewidth}{\centering\strut 65}\hspace{0.006\linewidth}\parbox[t]{0.04\linewidth}{\centering\strut \raisebox{0.08ex}{\ding{55}}}\hspace{0.006\linewidth}\parbox[t]{0.10\linewidth}{\centering\strut -232819.99\%}\hspace{0.006\linewidth}\parbox[t]{0.115\linewidth}{\centering\strut $8.275\times 10^{-11}$}\hspace{0.006\linewidth}\parbox[t]{0.671\linewidth}{adaptive polish: Powell tiny perturbation}}}}\par\smallskip
\noindent\parbox[t]{0.05\linewidth}{\centering\strut 66}\hspace{0.006\linewidth}\parbox[t]{0.04\linewidth}{\centering\strut \raisebox{0.08ex}{\ding{55}}}\hspace{0.006\linewidth}\parbox[t]{0.10\linewidth}{\centering\strut +99.38\%}\hspace{0.006\linewidth}\parbox[t]{0.115\linewidth}{\centering\strut $5.151\times 10^{-13}$}\hspace{0.006\linewidth}\parbox[t]{0.671\linewidth}{adaptive polish: COBYLA tiny perturbation}\par\smallskip
\noindent{\setlength{\fboxsep}{0pt}\colorbox{black!4}{\parbox{\linewidth}{\parbox[t]{0.05\linewidth}{\centering\strut 67}\hspace{0.006\linewidth}\parbox[t]{0.04\linewidth}{\centering\strut \raisebox{0.08ex}{\ding{55}}}\hspace{0.006\linewidth}\parbox[t]{0.10\linewidth}{\centering\strut -85560\%}\hspace{0.006\linewidth}\parbox[t]{0.115\linewidth}{\centering\strut $4.413\times 10^{-10}$}\hspace{0.006\linewidth}\parbox[t]{0.671\linewidth}{adaptive polish: Powell medium perturbation}}}}\par\smallskip
\noindent\parbox[t]{0.05\linewidth}{\centering\strut 68}\hspace{0.006\linewidth}\parbox[t]{0.04\linewidth}{\centering\strut \raisebox{0.08ex}{\ding{55}}}\hspace{0.006\linewidth}\parbox[t]{0.10\linewidth}{\centering\strut +99.74\%}\hspace{0.006\linewidth}\parbox[t]{0.115\linewidth}{\centering\strut $1.148\times 10^{-12}$}\hspace{0.006\linewidth}\parbox[t]{0.671\linewidth}{adaptive polish: COBYLA medium perturbation}\par\smallskip
\noindent{\setlength{\fboxsep}{0pt}\colorbox{black!4}{\parbox{\linewidth}{\parbox[t]{0.05\linewidth}{\centering\strut 69}\hspace{0.006\linewidth}\parbox[t]{0.04\linewidth}{\centering\strut \raisebox{0.08ex}{\ding{51}}}\hspace{0.006\linewidth}\parbox[t]{0.10\linewidth}{\centering\strut +96.9\%}\hspace{0.006\linewidth}\parbox[t]{0.115\linewidth}{\centering\strut $3.553\times 10^{-14}$}\hspace{0.006\linewidth}\parbox[t]{0.671\linewidth}{adaptive polish: ultra-tight Powell retry}}}}\par\smallskip
\noindent\parbox[t]{0.05\linewidth}{\centering\strut 70}\hspace{0.006\linewidth}\parbox[t]{0.04\linewidth}{\centering\strut \raisebox{0.08ex}{\ding{51}}}\hspace{0.006\linewidth}\parbox[t]{0.10\linewidth}{\centering\strut +0\%}\hspace{0.006\linewidth}\parbox[t]{0.115\linewidth}{\centering\strut $3.553\times 10^{-14}$}\hspace{0.006\linewidth}\parbox[t]{0.671\linewidth}{adaptive polish: ultra-tight COBYLA retry}\par\smallskip
\noindent{\setlength{\fboxsep}{0pt}\colorbox{black!4}{\parbox{\linewidth}{\parbox[t]{0.05\linewidth}{\centering\strut 71}\hspace{0.006\linewidth}\parbox[t]{0.04\linewidth}{\centering\strut \raisebox{0.08ex}{\ding{51}}}\hspace{0.006\linewidth}\parbox[t]{0.10\linewidth}{\centering\strut +0\%}\hspace{0.006\linewidth}\parbox[t]{0.115\linewidth}{\centering\strut $3.553\times 10^{-14}$}\hspace{0.006\linewidth}\parbox[t]{0.671\linewidth}{adaptive polish: Powell no-perturb warm start}}}}\par\smallskip
\noindent\parbox[t]{0.05\linewidth}{\centering\strut 72}\hspace{0.006\linewidth}\parbox[t]{0.04\linewidth}{\centering\strut \raisebox{0.08ex}{\ding{51}}}\hspace{0.006\linewidth}\parbox[t]{0.10\linewidth}{\centering\strut +0\%}\hspace{0.006\linewidth}\parbox[t]{0.115\linewidth}{\centering\strut $3.553\times 10^{-14}$}\hspace{0.006\linewidth}\parbox[t]{0.671\linewidth}{adaptive polish: COBYLA tiny-radius warm start}\par\smallskip
\noindent{\setlength{\fboxsep}{0pt}\colorbox{black!4}{\parbox{\linewidth}{\parbox[t]{0.05\linewidth}{\centering\strut 73}\hspace{0.006\linewidth}\parbox[t]{0.04\linewidth}{\centering\strut \raisebox{0.08ex}{\ding{55}}}\hspace{0.006\linewidth}\parbox[t]{0.10\linewidth}{\centering\strut -104140\%}\hspace{0.006\linewidth}\parbox[t]{0.115\linewidth}{\centering\strut $3.703\times 10^{-11}$}\hspace{0.006\linewidth}\parbox[t]{0.671\linewidth}{adaptive polish: Powell tiny perturbation}}}}\par\smallskip
\noindent\parbox[t]{0.05\linewidth}{\centering\strut 74}\hspace{0.006\linewidth}\parbox[t]{0.04\linewidth}{\centering\strut \raisebox{0.08ex}{\ding{55}}}\hspace{0.006\linewidth}\parbox[t]{0.10\linewidth}{\centering\strut +98.51\%}\hspace{0.006\linewidth}\parbox[t]{0.115\linewidth}{\centering\strut $5.507\times 10^{-13}$}\hspace{0.006\linewidth}\parbox[t]{0.671\linewidth}{adaptive polish: COBYLA tiny perturbation}\par\smallskip
\noindent{\setlength{\fboxsep}{0pt}\colorbox{black!4}{\parbox{\linewidth}{\parbox[t]{0.05\linewidth}{\centering\strut 75}\hspace{0.006\linewidth}\parbox[t]{0.04\linewidth}{\centering\strut \raisebox{0.08ex}{\ding{55}}}\hspace{0.006\linewidth}\parbox[t]{0.10\linewidth}{\centering\strut -101013.55\%}\hspace{0.006\linewidth}\parbox[t]{0.115\linewidth}{\centering\strut $5.568\times 10^{-10}$}\hspace{0.006\linewidth}\parbox[t]{0.671\linewidth}{adaptive polish: Powell medium perturbation}}}}\par\smallskip
\noindent\parbox[t]{0.05\linewidth}{\centering\strut 76}\hspace{0.006\linewidth}\parbox[t]{0.04\linewidth}{\centering\strut \raisebox{0.08ex}{\ding{55}}}\hspace{0.006\linewidth}\parbox[t]{0.10\linewidth}{\centering\strut +98.47\%}\hspace{0.006\linewidth}\parbox[t]{0.115\linewidth}{\centering\strut $8.505\times 10^{-12}$}\hspace{0.006\linewidth}\parbox[t]{0.671\linewidth}{adaptive polish: COBYLA medium perturbation}\par\smallskip
\noindent{\setlength{\fboxsep}{0pt}\colorbox{black!4}{\parbox{\linewidth}{\parbox[t]{0.05\linewidth}{\centering\strut 77}\hspace{0.006\linewidth}\parbox[t]{0.04\linewidth}{\centering\strut \raisebox{0.08ex}{\ding{51}}}\hspace{0.006\linewidth}\parbox[t]{0.10\linewidth}{\centering\strut +99.58\%}\hspace{0.006\linewidth}\parbox[t]{0.115\linewidth}{\centering\strut $3.553\times 10^{-14}$}\hspace{0.006\linewidth}\parbox[t]{0.671\linewidth}{adaptive polish: ultra-tight Powell retry}}}}\par\smallskip
\noindent\parbox[t]{0.05\linewidth}{\centering\strut 78}\hspace{0.006\linewidth}\parbox[t]{0.04\linewidth}{\centering\strut \raisebox{0.08ex}{\ding{51}}}\hspace{0.006\linewidth}\parbox[t]{0.10\linewidth}{\centering\strut +0\%}\hspace{0.006\linewidth}\parbox[t]{0.115\linewidth}{\centering\strut $3.553\times 10^{-14}$}\hspace{0.006\linewidth}\parbox[t]{0.671\linewidth}{adaptive polish: ultra-tight COBYLA retry}\par\smallskip
\noindent{\setlength{\fboxsep}{0pt}\colorbox{black!4}{\parbox{\linewidth}{\parbox[t]{0.05\linewidth}{\centering\strut 79}\hspace{0.006\linewidth}\parbox[t]{0.04\linewidth}{\centering\strut \raisebox{0.08ex}{\ding{51}}}\hspace{0.006\linewidth}\parbox[t]{0.10\linewidth}{\centering\strut +0\%}\hspace{0.006\linewidth}\parbox[t]{0.115\linewidth}{\centering\strut $3.553\times 10^{-14}$}\hspace{0.006\linewidth}\parbox[t]{0.671\linewidth}{adaptive polish: Powell no-perturb warm start}}}}\par\smallskip
\noindent\parbox[t]{0.05\linewidth}{\centering\strut 80}\hspace{0.006\linewidth}\parbox[t]{0.04\linewidth}{\centering\strut \raisebox{0.08ex}{\ding{51}}}\hspace{0.006\linewidth}\parbox[t]{0.10\linewidth}{\centering\strut +0\%}\hspace{0.006\linewidth}\parbox[t]{0.115\linewidth}{\centering\strut $3.553\times 10^{-14}$}\hspace{0.006\linewidth}\parbox[t]{0.671\linewidth}{adaptive polish: COBYLA tiny-radius warm start}\par\smallskip
\noindent{\setlength{\fboxsep}{0pt}\colorbox{black!4}{\parbox{\linewidth}{\parbox[t]{0.05\linewidth}{\centering\strut 81}\hspace{0.006\linewidth}\parbox[t]{0.04\linewidth}{\centering\strut \raisebox{0.08ex}{\ding{55}}}\hspace{0.006\linewidth}\parbox[t]{0.10\linewidth}{\centering\strut -39419.98\%}\hspace{0.006\linewidth}\parbox[t]{0.115\linewidth}{\centering\strut $1.404\times 10^{-11}$}\hspace{0.006\linewidth}\parbox[t]{0.671\linewidth}{adaptive polish: Powell tiny perturbation}}}}\par\smallskip
\noindent\parbox[t]{0.05\linewidth}{\centering\strut 82}\hspace{0.006\linewidth}\parbox[t]{0.04\linewidth}{\centering\strut \raisebox{0.08ex}{\ding{55}}}\hspace{0.006\linewidth}\parbox[t]{0.10\linewidth}{\centering\strut +98.38\%}\hspace{0.006\linewidth}\parbox[t]{0.115\linewidth}{\centering\strut $2.274\times 10^{-13}$}\hspace{0.006\linewidth}\parbox[t]{0.671\linewidth}{adaptive polish: COBYLA tiny perturbation}\par\smallskip
\noindent{\setlength{\fboxsep}{0pt}\colorbox{black!4}{\parbox{\linewidth}{\parbox[t]{0.05\linewidth}{\centering\strut 83}\hspace{0.006\linewidth}\parbox[t]{0.04\linewidth}{\centering\strut \raisebox{0.08ex}{\ding{55}}}\hspace{0.006\linewidth}\parbox[t]{0.10\linewidth}{\centering\strut -339746.82\%}\hspace{0.006\linewidth}\parbox[t]{0.115\linewidth}{\centering\strut $7.727\times 10^{-10}$}\hspace{0.006\linewidth}\parbox[t]{0.671\linewidth}{adaptive polish: Powell medium perturbation}}}}\par\smallskip
\noindent\parbox[t]{0.05\linewidth}{\centering\strut 84}\hspace{0.006\linewidth}\parbox[t]{0.04\linewidth}{\centering\strut \raisebox{0.08ex}{\ding{55}}}\hspace{0.006\linewidth}\parbox[t]{0.10\linewidth}{\centering\strut +99.62\%}\hspace{0.006\linewidth}\parbox[t]{0.115\linewidth}{\centering\strut $2.945\times 10^{-12}$}\hspace{0.006\linewidth}\parbox[t]{0.671\linewidth}{adaptive polish: COBYLA medium perturbation}\par\smallskip
\noindent{\setlength{\fboxsep}{0pt}\colorbox{black!4}{\parbox{\linewidth}{\parbox[t]{0.05\linewidth}{\centering\strut 85}\hspace{0.006\linewidth}\parbox[t]{0.04\linewidth}{\centering\strut \raisebox{0.08ex}{\ding{51}}}\hspace{0.006\linewidth}\parbox[t]{0.10\linewidth}{\centering\strut +98.79\%}\hspace{0.006\linewidth}\parbox[t]{0.115\linewidth}{\centering\strut $3.553\times 10^{-14}$}\hspace{0.006\linewidth}\parbox[t]{0.671\linewidth}{adaptive polish: ultra-tight Powell retry}}}}\par\smallskip
\noindent\parbox[t]{0.05\linewidth}{\centering\strut 86}\hspace{0.006\linewidth}\parbox[t]{0.04\linewidth}{\centering\strut \raisebox{0.08ex}{\ding{51}}}\hspace{0.006\linewidth}\parbox[t]{0.10\linewidth}{\centering\strut +0\%}\hspace{0.006\linewidth}\parbox[t]{0.115\linewidth}{\centering\strut $3.553\times 10^{-14}$}\hspace{0.006\linewidth}\parbox[t]{0.671\linewidth}{adaptive polish: ultra-tight COBYLA retry}\par\smallskip
\noindent{\setlength{\fboxsep}{0pt}\colorbox{black!4}{\parbox{\linewidth}{\parbox[t]{0.05\linewidth}{\centering\strut 87}\hspace{0.006\linewidth}\parbox[t]{0.04\linewidth}{\centering\strut \raisebox{0.08ex}{\ding{51}}}\hspace{0.006\linewidth}\parbox[t]{0.10\linewidth}{\centering\strut +0\%}\hspace{0.006\linewidth}\parbox[t]{0.115\linewidth}{\centering\strut $3.553\times 10^{-14}$}\hspace{0.006\linewidth}\parbox[t]{0.671\linewidth}{adaptive polish: Powell no-perturb warm start}}}}\par\smallskip
\noindent\parbox[t]{0.05\linewidth}{\centering\strut 88}\hspace{0.006\linewidth}\parbox[t]{0.04\linewidth}{\centering\strut \raisebox{0.08ex}{\ding{51}}}\hspace{0.006\linewidth}\parbox[t]{0.10\linewidth}{\centering\strut +0\%}\hspace{0.006\linewidth}\parbox[t]{0.115\linewidth}{\centering\strut $3.553\times 10^{-14}$}\hspace{0.006\linewidth}\parbox[t]{0.671\linewidth}{adaptive polish: COBYLA tiny-radius warm start}\par\smallskip
\noindent{\setlength{\fboxsep}{0pt}\colorbox{black!4}{\parbox{\linewidth}{\parbox[t]{0.05\linewidth}{\centering\strut 89}\hspace{0.006\linewidth}\parbox[t]{0.04\linewidth}{\centering\strut \raisebox{0.08ex}{\ding{55}}}\hspace{0.006\linewidth}\parbox[t]{0.10\linewidth}{\centering\strut -64930\%}\hspace{0.006\linewidth}\parbox[t]{0.115\linewidth}{\centering\strut $2.31\times 10^{-11}$}\hspace{0.006\linewidth}\parbox[t]{0.671\linewidth}{adaptive polish: Powell tiny perturbation}}}}\par\smallskip
\noindent\parbox[t]{0.05\linewidth}{\centering\strut 90}\hspace{0.006\linewidth}\parbox[t]{0.04\linewidth}{\centering\strut \raisebox{0.08ex}{\ding{55}}}\hspace{0.006\linewidth}\parbox[t]{0.10\linewidth}{\centering\strut +99.6\%}\hspace{0.006\linewidth}\parbox[t]{0.115\linewidth}{\centering\strut $9.237\times 10^{-14}$}\hspace{0.006\linewidth}\parbox[t]{0.671\linewidth}{adaptive polish: COBYLA tiny perturbation}\par\smallskip
\noindent{\setlength{\fboxsep}{0pt}\colorbox{black!4}{\parbox{\linewidth}{\parbox[t]{0.05\linewidth}{\centering\strut 91}\hspace{0.006\linewidth}\parbox[t]{0.04\linewidth}{\centering\strut \raisebox{0.08ex}{\ding{55}}}\hspace{0.006\linewidth}\parbox[t]{0.10\linewidth}{\centering\strut -1142145.97\%}\hspace{0.006\linewidth}\parbox[t]{0.115\linewidth}{\centering\strut $1.055\times 10^{-9}$}\hspace{0.006\linewidth}\parbox[t]{0.671\linewidth}{adaptive polish: Powell medium perturbation}}}}\par\smallskip
\noindent\parbox[t]{0.05\linewidth}{\centering\strut 92}\hspace{0.006\linewidth}\parbox[t]{0.04\linewidth}{\centering\strut \raisebox{0.08ex}{\ding{55}}}\hspace{0.006\linewidth}\parbox[t]{0.10\linewidth}{\centering\strut +99.87\%}\hspace{0.006\linewidth}\parbox[t]{0.115\linewidth}{\centering\strut $1.364\times 10^{-12}$}\hspace{0.006\linewidth}\parbox[t]{0.671\linewidth}{adaptive polish: COBYLA medium perturbation}\par\smallskip
\noindent{\setlength{\fboxsep}{0pt}\colorbox{black!4}{\parbox{\linewidth}{\parbox[t]{0.05\linewidth}{\centering\strut 93}\hspace{0.006\linewidth}\parbox[t]{0.04\linewidth}{\centering\strut \raisebox{0.08ex}{\ding{51}}}\hspace{0.006\linewidth}\parbox[t]{0.10\linewidth}{\centering\strut +97.4\%}\hspace{0.006\linewidth}\parbox[t]{0.115\linewidth}{\centering\strut $3.553\times 10^{-14}$}\hspace{0.006\linewidth}\parbox[t]{0.671\linewidth}{adaptive polish: ultra-tight Powell retry}}}}\par\smallskip
\noindent\parbox[t]{0.05\linewidth}{\centering\strut 94}\hspace{0.006\linewidth}\parbox[t]{0.04\linewidth}{\centering\strut \raisebox{0.08ex}{\ding{51}}}\hspace{0.006\linewidth}\parbox[t]{0.10\linewidth}{\centering\strut +0\%}\hspace{0.006\linewidth}\parbox[t]{0.115\linewidth}{\centering\strut $3.553\times 10^{-14}$}\hspace{0.006\linewidth}\parbox[t]{0.671\linewidth}{adaptive polish: ultra-tight COBYLA retry}\par\smallskip
\noindent{\setlength{\fboxsep}{0pt}\colorbox{black!4}{\parbox{\linewidth}{\parbox[t]{0.05\linewidth}{\centering\strut 95}\hspace{0.006\linewidth}\parbox[t]{0.04\linewidth}{\centering\strut \raisebox{0.08ex}{\ding{51}}}\hspace{0.006\linewidth}\parbox[t]{0.10\linewidth}{\centering\strut +0\%}\hspace{0.006\linewidth}\parbox[t]{0.115\linewidth}{\centering\strut $3.553\times 10^{-14}$}\hspace{0.006\linewidth}\parbox[t]{0.671\linewidth}{adaptive polish: Powell no-perturb warm start}}}}\par\smallskip
\noindent\parbox[t]{0.05\linewidth}{\centering\strut 96}\hspace{0.006\linewidth}\parbox[t]{0.04\linewidth}{\centering\strut \raisebox{0.08ex}{\ding{51}}}\hspace{0.006\linewidth}\parbox[t]{0.10\linewidth}{\centering\strut +0\%}\hspace{0.006\linewidth}\parbox[t]{0.115\linewidth}{\centering\strut $3.553\times 10^{-14}$}\hspace{0.006\linewidth}\parbox[t]{0.671\linewidth}{adaptive polish: COBYLA tiny-radius warm start}\par\smallskip
\noindent{\setlength{\fboxsep}{0pt}\colorbox{black!4}{\parbox{\linewidth}{\parbox[t]{0.05\linewidth}{\centering\strut 97}\hspace{0.006\linewidth}\parbox[t]{0.04\linewidth}{\centering\strut \raisebox{0.08ex}{\ding{55}}}\hspace{0.006\linewidth}\parbox[t]{0.10\linewidth}{\centering\strut -93600\%}\hspace{0.006\linewidth}\parbox[t]{0.115\linewidth}{\centering\strut $3.329\times 10^{-11}$}\hspace{0.006\linewidth}\parbox[t]{0.671\linewidth}{adaptive polish: Powell tiny perturbation}}}}\par\smallskip
\noindent\parbox[t]{0.05\linewidth}{\centering\strut 98}\hspace{0.006\linewidth}\parbox[t]{0.04\linewidth}{\centering\strut \raisebox{0.08ex}{\ding{55}}}\hspace{0.006\linewidth}\parbox[t]{0.10\linewidth}{\centering\strut +99.16\%}\hspace{0.006\linewidth}\parbox[t]{0.115\linewidth}{\centering\strut $2.807\times 10^{-13}$}\hspace{0.006\linewidth}\parbox[t]{0.671\linewidth}{adaptive polish: COBYLA tiny perturbation}\par\smallskip
\noindent{\setlength{\fboxsep}{0pt}\colorbox{black!4}{\parbox{\linewidth}{\parbox[t]{0.05\linewidth}{\centering\strut 99}\hspace{0.006\linewidth}\parbox[t]{0.04\linewidth}{\centering\strut \raisebox{0.08ex}{\ding{55}}}\hspace{0.006\linewidth}\parbox[t]{0.10\linewidth}{\centering\strut -304330.34\%}\hspace{0.006\linewidth}\parbox[t]{0.115\linewidth}{\centering\strut $8.544\times 10^{-10}$}\hspace{0.006\linewidth}\parbox[t]{0.671\linewidth}{adaptive polish: Powell medium perturbation}}}}\par\smallskip
\noindent\parbox[t]{0.05\linewidth}{\centering\strut 100}\hspace{0.006\linewidth}\parbox[t]{0.04\linewidth}{\centering\strut \raisebox{0.08ex}{\ding{55}}}\hspace{0.006\linewidth}\parbox[t]{0.10\linewidth}{\centering\strut +99.94\%}\hspace{0.006\linewidth}\parbox[t]{0.115\linewidth}{\centering\strut $5.507\times 10^{-13}$}\hspace{0.006\linewidth}\parbox[t]{0.671\linewidth}{adaptive polish: COBYLA medium perturbation}\par\smallskip
\noindent\rule{\linewidth}{0.5pt}
\par\medskip

\medskip

\refstepcounter{table}
\label{tab:supp_vqe_log_lih}
\noindent\textbf{Table \thetable.} VQE mutation log for LiH full 100-iteration archival campaign. Relative gain is measured against the immediately previous scored iteration using the absolute energy error.
\par\medskip
\noindent\rule{\linewidth}{0.5pt}
\noindent{\setlength{\fboxsep}{0pt}\colorbox{black!4}{\parbox{\linewidth}{\parbox[c][2.2em][c]{0.05\linewidth}{\centering \textbf{Iter.}}\hspace{0.006\linewidth}\parbox[c][2.2em][c]{0.04\linewidth}{\centering \textbf{A/R}}\hspace{0.006\linewidth}\parbox[c][2.2em][c]{0.10\linewidth}{\centering \textbf{Relative gain}}\hspace{0.006\linewidth}\parbox[c][2.2em][c]{0.115\linewidth}{\centering \textbf{$|\Delta E|$ [Ha]}}\hspace{0.006\linewidth}\parbox[c][2.2em][c]{0.671\linewidth}{\centering \textbf{Mutation summary}}}}}\par\smallskip
\noindent\rule{\linewidth}{0.5pt}
\noindent{\setlength{\fboxsep}{0pt}\colorbox{black!4}{\parbox{\linewidth}{\parbox[t]{0.05\linewidth}{\centering\strut 1}\hspace{0.006\linewidth}\parbox[t]{0.04\linewidth}{\centering\strut \raisebox{0.08ex}{\ding{51}}}\hspace{0.006\linewidth}\parbox[t]{0.10\linewidth}{\centering\strut ---}\hspace{0.006\linewidth}\parbox[t]{0.115\linewidth}{\centering\strut $7.116\times 10^{-5}$}\hspace{0.006\linewidth}\parbox[t]{0.671\linewidth}{switch to UCCSD diagonal-pair coordinate search}}}}\par\smallskip
\noindent\parbox[t]{0.05\linewidth}{\centering\strut 2}\hspace{0.006\linewidth}\parbox[t]{0.04\linewidth}{\centering\strut \raisebox{0.08ex}{\ding{51}}}\hspace{0.006\linewidth}\parbox[t]{0.10\linewidth}{\centering\strut +22.74\%}\hspace{0.006\linewidth}\parbox[t]{0.115\linewidth}{\centering\strut $5.497\times 10^{-5}$}\hspace{0.006\linewidth}\parbox[t]{0.671\linewidth}{Powell refine diagonal-pair warm start}\par\smallskip
\noindent{\setlength{\fboxsep}{0pt}\colorbox{black!4}{\parbox{\linewidth}{\parbox[t]{0.05\linewidth}{\centering\strut 3}\hspace{0.006\linewidth}\parbox[t]{0.04\linewidth}{\centering\strut \raisebox{0.08ex}{\ding{51}}}\hspace{0.006\linewidth}\parbox[t]{0.10\linewidth}{\centering\strut +0.000055\%}\hspace{0.006\linewidth}\parbox[t]{0.115\linewidth}{\centering\strut $5.497\times 10^{-5}$}\hspace{0.006\linewidth}\parbox[t]{0.671\linewidth}{COBYLA refine diagonal-pair warm start}}}}\par\smallskip
\noindent\parbox[t]{0.05\linewidth}{\centering\strut 4}\hspace{0.006\linewidth}\parbox[t]{0.04\linewidth}{\centering\strut \raisebox{0.08ex}{\ding{55}}}\hspace{0.006\linewidth}\parbox[t]{0.10\linewidth}{\centering\strut +0\%}\hspace{0.006\linewidth}\parbox[t]{0.115\linewidth}{\centering\strut $5.497\times 10^{-5}$}\hspace{0.006\linewidth}\parbox[t]{0.671\linewidth}{smaller-step diagonal-pair coordinate search}\par\smallskip
\noindent{\setlength{\fboxsep}{0pt}\colorbox{black!4}{\parbox{\linewidth}{\parbox[t]{0.05\linewidth}{\centering\strut 5}\hspace{0.006\linewidth}\parbox[t]{0.04\linewidth}{\centering\strut \raisebox{0.08ex}{\ding{55}}}\hspace{0.006\linewidth}\parbox[t]{0.10\linewidth}{\centering\strut +0\%}\hspace{0.006\linewidth}\parbox[t]{0.115\linewidth}{\centering\strut $5.497\times 10^{-5}$}\hspace{0.006\linewidth}\parbox[t]{0.671\linewidth}{micro-step diagonal-pair coordinate search}}}}\par\smallskip
\noindent\parbox[t]{0.05\linewidth}{\centering\strut 6}\hspace{0.006\linewidth}\parbox[t]{0.04\linewidth}{\centering\strut \raisebox{0.08ex}{\ding{55}}}\hspace{0.006\linewidth}\parbox[t]{0.10\linewidth}{\centering\strut -0.0229\%}\hspace{0.006\linewidth}\parbox[t]{0.115\linewidth}{\centering\strut $5.499\times 10^{-5}$}\hspace{0.006\linewidth}\parbox[t]{0.671\linewidth}{noisy Powell restart around diagonal-pair solution}\par\smallskip
\noindent{\setlength{\fboxsep}{0pt}\colorbox{black!4}{\parbox{\linewidth}{\parbox[t]{0.05\linewidth}{\centering\strut 7}\hspace{0.006\linewidth}\parbox[t]{0.04\linewidth}{\centering\strut \raisebox{0.08ex}{\ding{55}}}\hspace{0.006\linewidth}\parbox[t]{0.10\linewidth}{\centering\strut +0.0229\%}\hspace{0.006\linewidth}\parbox[t]{0.115\linewidth}{\centering\strut $5.497\times 10^{-5}$}\hspace{0.006\linewidth}\parbox[t]{0.671\linewidth}{add symmetric off-diagonal pair doubles}}}}\par\smallskip
\noindent\parbox[t]{0.05\linewidth}{\centering\strut 8}\hspace{0.006\linewidth}\parbox[t]{0.04\linewidth}{\centering\strut \raisebox{0.08ex}{\ding{55}}}\hspace{0.006\linewidth}\parbox[t]{0.10\linewidth}{\centering\strut +0\%}\hspace{0.006\linewidth}\parbox[t]{0.115\linewidth}{\centering\strut $5.497\times 10^{-5}$}\hspace{0.006\linewidth}\parbox[t]{0.671\linewidth}{Powell refine symmetric pair-doubles warm start}\par\smallskip
\noindent{\setlength{\fboxsep}{0pt}\colorbox{black!4}{\parbox{\linewidth}{\parbox[t]{0.05\linewidth}{\centering\strut 9}\hspace{0.006\linewidth}\parbox[t]{0.04\linewidth}{\centering\strut \raisebox{0.08ex}{\ding{55}}}\hspace{0.006\linewidth}\parbox[t]{0.10\linewidth}{\centering\strut +0\%}\hspace{0.006\linewidth}\parbox[t]{0.115\linewidth}{\centering\strut $5.497\times 10^{-5}$}\hspace{0.006\linewidth}\parbox[t]{0.671\linewidth}{COBYLA refine symmetric pair-doubles warm start}}}}\par\smallskip
\noindent\parbox[t]{0.05\linewidth}{\centering\strut 10}\hspace{0.006\linewidth}\parbox[t]{0.04\linewidth}{\centering\strut \raisebox{0.08ex}{\ding{51}}}\hspace{0.006\linewidth}\parbox[t]{0.10\linewidth}{\centering\strut +90.09\%}\hspace{0.006\linewidth}\parbox[t]{0.115\linewidth}{\centering\strut $5.446\times 10^{-6}$}\hspace{0.006\linewidth}\parbox[t]{0.671\linewidth}{add spin-paired singles to symmetric pair-doubles}\par\smallskip
\noindent{\setlength{\fboxsep}{0pt}\colorbox{black!4}{\parbox{\linewidth}{\parbox[t]{0.05\linewidth}{\centering\strut 11}\hspace{0.006\linewidth}\parbox[t]{0.04\linewidth}{\centering\strut \raisebox{0.08ex}{\ding{51}}}\hspace{0.006\linewidth}\parbox[t]{0.10\linewidth}{\centering\strut +30.78\%}\hspace{0.006\linewidth}\parbox[t]{0.115\linewidth}{\centering\strut $3.77\times 10^{-6}$}\hspace{0.006\linewidth}\parbox[t]{0.671\linewidth}{Powell refine spin-paired symmetric warm start}}}}\par\smallskip
\noindent\parbox[t]{0.05\linewidth}{\centering\strut 12}\hspace{0.006\linewidth}\parbox[t]{0.04\linewidth}{\centering\strut \raisebox{0.08ex}{\ding{51}}}\hspace{0.006\linewidth}\parbox[t]{0.10\linewidth}{\centering\strut +96.54\%}\hspace{0.006\linewidth}\parbox[t]{0.115\linewidth}{\centering\strut $1.304\times 10^{-7}$}\hspace{0.006\linewidth}\parbox[t]{0.671\linewidth}{COBYLA refine spin-paired symmetric warm start}\par\smallskip
\noindent{\setlength{\fboxsep}{0pt}\colorbox{black!4}{\parbox{\linewidth}{\parbox[t]{0.05\linewidth}{\centering\strut 13}\hspace{0.006\linewidth}\parbox[t]{0.04\linewidth}{\centering\strut \raisebox{0.08ex}{\ding{55}}}\hspace{0.006\linewidth}\parbox[t]{0.10\linewidth}{\centering\strut -42619.06\%}\hspace{0.006\linewidth}\parbox[t]{0.115\linewidth}{\centering\strut $5.571\times 10^{-5}$}\hspace{0.006\linewidth}\parbox[t]{0.671\linewidth}{untie full pair-doubles matrix coordinate search}}}}\par\smallskip
\noindent\parbox[t]{0.05\linewidth}{\centering\strut 14}\hspace{0.006\linewidth}\parbox[t]{0.04\linewidth}{\centering\strut \raisebox{0.08ex}{\ding{55}}}\hspace{0.006\linewidth}\parbox[t]{0.10\linewidth}{\centering\strut +0.3361\%}\hspace{0.006\linewidth}\parbox[t]{0.115\linewidth}{\centering\strut $5.552\times 10^{-5}$}\hspace{0.006\linewidth}\parbox[t]{0.671\linewidth}{Powell refine full pair-doubles warm start}\par\smallskip
\noindent{\setlength{\fboxsep}{0pt}\colorbox{black!4}{\parbox{\linewidth}{\parbox[t]{0.05\linewidth}{\centering\strut 15}\hspace{0.006\linewidth}\parbox[t]{0.04\linewidth}{\centering\strut \raisebox{0.08ex}{\ding{55}}}\hspace{0.006\linewidth}\parbox[t]{0.10\linewidth}{\centering\strut +99.77\%}\hspace{0.006\linewidth}\parbox[t]{0.115\linewidth}{\centering\strut $1.304\times 10^{-7}$}\hspace{0.006\linewidth}\parbox[t]{0.671\linewidth}{untie singles and pair-doubles with coordinate search}}}}\par\smallskip
\noindent\parbox[t]{0.05\linewidth}{\centering\strut 16}\hspace{0.006\linewidth}\parbox[t]{0.04\linewidth}{\centering\strut \raisebox{0.08ex}{\ding{51}}}\hspace{0.006\linewidth}\parbox[t]{0.10\linewidth}{\centering\strut +85.24\%}\hspace{0.006\linewidth}\parbox[t]{0.115\linewidth}{\centering\strut $1.925\times 10^{-8}$}\hspace{0.006\linewidth}\parbox[t]{0.671\linewidth}{Powell refine full spin-paired warm start}\par\smallskip
\noindent{\setlength{\fboxsep}{0pt}\colorbox{black!4}{\parbox{\linewidth}{\parbox[t]{0.05\linewidth}{\centering\strut 17}\hspace{0.006\linewidth}\parbox[t]{0.04\linewidth}{\centering\strut \raisebox{0.08ex}{\ding{51}}}\hspace{0.006\linewidth}\parbox[t]{0.10\linewidth}{\centering\strut +26.61\%}\hspace{0.006\linewidth}\parbox[t]{0.115\linewidth}{\centering\strut $1.413\times 10^{-8}$}\hspace{0.006\linewidth}\parbox[t]{0.671\linewidth}{promote spin-paired warm start into full UCCSD COBYLA}}}}\par\smallskip
\noindent\parbox[t]{0.05\linewidth}{\centering\strut 18}\hspace{0.006\linewidth}\parbox[t]{0.04\linewidth}{\centering\strut \raisebox{0.08ex}{\ding{51}}}\hspace{0.006\linewidth}\parbox[t]{0.10\linewidth}{\centering\strut +23.92\%}\hspace{0.006\linewidth}\parbox[t]{0.115\linewidth}{\centering\strut $1.075\times 10^{-8}$}\hspace{0.006\linewidth}\parbox[t]{0.671\linewidth}{promote spin-paired warm start into full UCCSD Powell}\par\smallskip
\noindent{\setlength{\fboxsep}{0pt}\colorbox{black!4}{\parbox{\linewidth}{\parbox[t]{0.05\linewidth}{\centering\strut 19}\hspace{0.006\linewidth}\parbox[t]{0.04\linewidth}{\centering\strut \raisebox{0.08ex}{\ding{55}}}\hspace{0.006\linewidth}\parbox[t]{0.10\linewidth}{\centering\strut -511403.95\%}\hspace{0.006\linewidth}\parbox[t]{0.115\linewidth}{\centering\strut $5.498\times 10^{-5}$}\hspace{0.006\linewidth}\parbox[t]{0.671\linewidth}{fine diagonal-pair coordinate restart with tiny noise}}}}\par\smallskip
\noindent\parbox[t]{0.05\linewidth}{\centering\strut 20}\hspace{0.006\linewidth}\parbox[t]{0.04\linewidth}{\centering\strut \raisebox{0.08ex}{\ding{55}}}\hspace{0.006\linewidth}\parbox[t]{0.10\linewidth}{\centering\strut -0.2991\%}\hspace{0.006\linewidth}\parbox[t]{0.115\linewidth}{\centering\strut $5.515\times 10^{-5}$}\hspace{0.006\linewidth}\parbox[t]{0.671\linewidth}{fine symmetric pair-doubles restart with tiny noise}\par\smallskip
\noindent{\setlength{\fboxsep}{0pt}\colorbox{black!4}{\parbox{\linewidth}{\parbox[t]{0.05\linewidth}{\centering\strut 21}\hspace{0.006\linewidth}\parbox[t]{0.04\linewidth}{\centering\strut \raisebox{0.08ex}{\ding{55}}}\hspace{0.006\linewidth}\parbox[t]{0.10\linewidth}{\centering\strut +99.17\%}\hspace{0.006\linewidth}\parbox[t]{0.115\linewidth}{\centering\strut $4.561\times 10^{-7}$}\hspace{0.006\linewidth}\parbox[t]{0.671\linewidth}{fine spin-paired symmetric restart with tiny noise}}}}\par\smallskip
\noindent\parbox[t]{0.05\linewidth}{\centering\strut 22}\hspace{0.006\linewidth}\parbox[t]{0.04\linewidth}{\centering\strut \raisebox{0.08ex}{\ding{55}}}\hspace{0.006\linewidth}\parbox[t]{0.10\linewidth}{\centering\strut -12023.51\%}\hspace{0.006\linewidth}\parbox[t]{0.115\linewidth}{\centering\strut $5.53\times 10^{-5}$}\hspace{0.006\linewidth}\parbox[t]{0.671\linewidth}{fine full pair-doubles restart with tiny noise}\par\smallskip
\noindent{\setlength{\fboxsep}{0pt}\colorbox{black!4}{\parbox{\linewidth}{\parbox[t]{0.05\linewidth}{\centering\strut 23}\hspace{0.006\linewidth}\parbox[t]{0.04\linewidth}{\centering\strut \raisebox{0.08ex}{\ding{51}}}\hspace{0.006\linewidth}\parbox[t]{0.10\linewidth}{\centering\strut +99.98\%}\hspace{0.006\linewidth}\parbox[t]{0.115\linewidth}{\centering\strut $1.067\times 10^{-8}$}\hspace{0.006\linewidth}\parbox[t]{0.671\linewidth}{refine uccsd\_full with halved coordinate step}}}}\par\smallskip
\noindent\parbox[t]{0.05\linewidth}{\centering\strut 24}\hspace{0.006\linewidth}\parbox[t]{0.04\linewidth}{\centering\strut \raisebox{0.08ex}{\ding{51}}}\hspace{0.006\linewidth}\parbox[t]{0.10\linewidth}{\centering\strut +0.3199\%}\hspace{0.006\linewidth}\parbox[t]{0.115\linewidth}{\centering\strut $1.064\times 10^{-8}$}\hspace{0.006\linewidth}\parbox[t]{0.671\linewidth}{Powell polish current best uccsd\_full}\par\smallskip
\noindent{\setlength{\fboxsep}{0pt}\colorbox{black!4}{\parbox{\linewidth}{\parbox[t]{0.05\linewidth}{\centering\strut 25}\hspace{0.006\linewidth}\parbox[t]{0.04\linewidth}{\centering\strut \raisebox{0.08ex}{\ding{51}}}\hspace{0.006\linewidth}\parbox[t]{0.10\linewidth}{\centering\strut +96.47\%}\hspace{0.006\linewidth}\parbox[t]{0.115\linewidth}{\centering\strut $3.758\times 10^{-10}$}\hspace{0.006\linewidth}\parbox[t]{0.671\linewidth}{COBYLA polish current best uccsd\_full}}}}\par\smallskip
\noindent\parbox[t]{0.05\linewidth}{\centering\strut 26}\hspace{0.006\linewidth}\parbox[t]{0.04\linewidth}{\centering\strut \raisebox{0.08ex}{\ding{55}}}\hspace{0.006\linewidth}\parbox[t]{0.10\linewidth}{\centering\strut -28181.5\%}\hspace{0.006\linewidth}\parbox[t]{0.115\linewidth}{\centering\strut $1.063\times 10^{-7}$}\hspace{0.006\linewidth}\parbox[t]{0.671\linewidth}{noisy coordinate restart around best uccsd\_full}\par\smallskip
\noindent{\setlength{\fboxsep}{0pt}\colorbox{black!4}{\parbox{\linewidth}{\parbox[t]{0.05\linewidth}{\centering\strut 27}\hspace{0.006\linewidth}\parbox[t]{0.04\linewidth}{\centering\strut \raisebox{0.08ex}{\ding{55}}}\hspace{0.006\linewidth}\parbox[t]{0.10\linewidth}{\centering\strut +81.89\%}\hspace{0.006\linewidth}\parbox[t]{0.115\linewidth}{\centering\strut $1.925\times 10^{-8}$}\hspace{0.006\linewidth}\parbox[t]{0.671\linewidth}{expand current best uccsd\_full into spin\_paired\_full}}}}\par\smallskip
\noindent\parbox[t]{0.05\linewidth}{\centering\strut 28}\hspace{0.006\linewidth}\parbox[t]{0.04\linewidth}{\centering\strut \raisebox{0.08ex}{\ding{55}}}\hspace{0.006\linewidth}\parbox[t]{0.10\linewidth}{\centering\strut +18.61\%}\hspace{0.006\linewidth}\parbox[t]{0.115\linewidth}{\centering\strut $1.567\times 10^{-8}$}\hspace{0.006\linewidth}\parbox[t]{0.671\linewidth}{Powell polish expanded spin\_paired\_full warm start}\par\smallskip
\noindent{\setlength{\fboxsep}{0pt}\colorbox{black!4}{\parbox{\linewidth}{\parbox[t]{0.05\linewidth}{\centering\strut 29}\hspace{0.006\linewidth}\parbox[t]{0.04\linewidth}{\centering\strut \raisebox{0.08ex}{\ding{55}}}\hspace{0.006\linewidth}\parbox[t]{0.10\linewidth}{\centering\strut -351047.99\%}\hspace{0.006\linewidth}\parbox[t]{0.115\linewidth}{\centering\strut $5.502\times 10^{-5}$}\hspace{0.006\linewidth}\parbox[t]{0.671\linewidth}{try alternate LiH compression pair\_doubles\_symmetric near current best}}}}\par\smallskip
\noindent\parbox[t]{0.05\linewidth}{\centering\strut 30}\hspace{0.006\linewidth}\parbox[t]{0.04\linewidth}{\centering\strut \raisebox{0.08ex}{\ding{51}}}\hspace{0.006\linewidth}\parbox[t]{0.10\linewidth}{\centering\strut +100\%}\hspace{0.006\linewidth}\parbox[t]{0.115\linewidth}{\centering\strut $3.579\times 10^{-10}$}\hspace{0.006\linewidth}\parbox[t]{0.671\linewidth}{ultra-fine coordinate polish for uccsd\_full}\par\smallskip
\noindent{\setlength{\fboxsep}{0pt}\colorbox{black!4}{\parbox{\linewidth}{\parbox[t]{0.05\linewidth}{\centering\strut 31}\hspace{0.006\linewidth}\parbox[t]{0.04\linewidth}{\centering\strut \raisebox{0.08ex}{\ding{55}}}\hspace{0.006\linewidth}\parbox[t]{0.10\linewidth}{\centering\strut -30062.42\%}\hspace{0.006\linewidth}\parbox[t]{0.115\linewidth}{\centering\strut $1.079\times 10^{-7}$}\hspace{0.006\linewidth}\parbox[t]{0.671\linewidth}{noisy Powell polish around best uccsd\_full}}}}\par\smallskip
\noindent\parbox[t]{0.05\linewidth}{\centering\strut 32}\hspace{0.006\linewidth}\parbox[t]{0.04\linewidth}{\centering\strut \raisebox{0.08ex}{\ding{55}}}\hspace{0.006\linewidth}\parbox[t]{0.10\linewidth}{\centering\strut +90.61\%}\hspace{0.006\linewidth}\parbox[t]{0.115\linewidth}{\centering\strut $1.014\times 10^{-8}$}\hspace{0.006\linewidth}\parbox[t]{0.671\linewidth}{noisy COBYLA polish around best uccsd\_full}\par\smallskip
\noindent{\setlength{\fboxsep}{0pt}\colorbox{black!4}{\parbox{\linewidth}{\parbox[t]{0.05\linewidth}{\centering\strut 33}\hspace{0.006\linewidth}\parbox[t]{0.04\linewidth}{\centering\strut \raisebox{0.08ex}{\ding{51}}}\hspace{0.006\linewidth}\parbox[t]{0.10\linewidth}{\centering\strut +96.63\%}\hspace{0.006\linewidth}\parbox[t]{0.115\linewidth}{\centering\strut $3.413\times 10^{-10}$}\hspace{0.006\linewidth}\parbox[t]{0.671\linewidth}{refine uccsd\_full with halved coordinate step}}}}\par\smallskip
\noindent\parbox[t]{0.05\linewidth}{\centering\strut 34}\hspace{0.006\linewidth}\parbox[t]{0.04\linewidth}{\centering\strut \raisebox{0.08ex}{\ding{51}}}\hspace{0.006\linewidth}\parbox[t]{0.10\linewidth}{\centering\strut +37.35\%}\hspace{0.006\linewidth}\parbox[t]{0.115\linewidth}{\centering\strut $2.138\times 10^{-10}$}\hspace{0.006\linewidth}\parbox[t]{0.671\linewidth}{Powell polish current best uccsd\_full}\par\smallskip
\noindent{\setlength{\fboxsep}{0pt}\colorbox{black!4}{\parbox{\linewidth}{\parbox[t]{0.05\linewidth}{\centering\strut 35}\hspace{0.006\linewidth}\parbox[t]{0.04\linewidth}{\centering\strut \raisebox{0.08ex}{\ding{51}}}\hspace{0.006\linewidth}\parbox[t]{0.10\linewidth}{\centering\strut +39.59\%}\hspace{0.006\linewidth}\parbox[t]{0.115\linewidth}{\centering\strut $1.292\times 10^{-10}$}\hspace{0.006\linewidth}\parbox[t]{0.671\linewidth}{COBYLA polish current best uccsd\_full}}}}\par\smallskip
\noindent\parbox[t]{0.05\linewidth}{\centering\strut 36}\hspace{0.006\linewidth}\parbox[t]{0.04\linewidth}{\centering\strut \raisebox{0.08ex}{\ding{55}}}\hspace{0.006\linewidth}\parbox[t]{0.10\linewidth}{\centering\strut -44527.43\%}\hspace{0.006\linewidth}\parbox[t]{0.115\linewidth}{\centering\strut $5.764\times 10^{-8}$}\hspace{0.006\linewidth}\parbox[t]{0.671\linewidth}{noisy coordinate restart around best uccsd\_full}\par\smallskip
\noindent{\setlength{\fboxsep}{0pt}\colorbox{black!4}{\parbox{\linewidth}{\parbox[t]{0.05\linewidth}{\centering\strut 37}\hspace{0.006\linewidth}\parbox[t]{0.04\linewidth}{\centering\strut \raisebox{0.08ex}{\ding{55}}}\hspace{0.006\linewidth}\parbox[t]{0.10\linewidth}{\centering\strut +72.82\%}\hspace{0.006\linewidth}\parbox[t]{0.115\linewidth}{\centering\strut $1.567\times 10^{-8}$}\hspace{0.006\linewidth}\parbox[t]{0.671\linewidth}{expand current best uccsd\_full into spin\_paired\_full}}}}\par\smallskip
\noindent\parbox[t]{0.05\linewidth}{\centering\strut 38}\hspace{0.006\linewidth}\parbox[t]{0.04\linewidth}{\centering\strut \raisebox{0.08ex}{\ding{55}}}\hspace{0.006\linewidth}\parbox[t]{0.10\linewidth}{\centering\strut +0\%}\hspace{0.006\linewidth}\parbox[t]{0.115\linewidth}{\centering\strut $1.567\times 10^{-8}$}\hspace{0.006\linewidth}\parbox[t]{0.671\linewidth}{Powell polish expanded spin\_paired\_full warm start}\par\smallskip
\noindent{\setlength{\fboxsep}{0pt}\colorbox{black!4}{\parbox{\linewidth}{\parbox[t]{0.05\linewidth}{\centering\strut 39}\hspace{0.006\linewidth}\parbox[t]{0.04\linewidth}{\centering\strut \raisebox{0.08ex}{\ding{55}}}\hspace{0.006\linewidth}\parbox[t]{0.10\linewidth}{\centering\strut -351047.99\%}\hspace{0.006\linewidth}\parbox[t]{0.115\linewidth}{\centering\strut $5.502\times 10^{-5}$}\hspace{0.006\linewidth}\parbox[t]{0.671\linewidth}{try alternate LiH compression pair\_doubles\_symmetric near current best}}}}\par\smallskip
\noindent\parbox[t]{0.05\linewidth}{\centering\strut 40}\hspace{0.006\linewidth}\parbox[t]{0.04\linewidth}{\centering\strut \raisebox{0.08ex}{\ding{51}}}\hspace{0.006\linewidth}\parbox[t]{0.10\linewidth}{\centering\strut +100\%}\hspace{0.006\linewidth}\parbox[t]{0.115\linewidth}{\centering\strut $3.926\times 10^{-11}$}\hspace{0.006\linewidth}\parbox[t]{0.671\linewidth}{ultra-fine coordinate polish for uccsd\_full}\par\smallskip
\noindent{\setlength{\fboxsep}{0pt}\colorbox{black!4}{\parbox{\linewidth}{\parbox[t]{0.05\linewidth}{\centering\strut 41}\hspace{0.006\linewidth}\parbox[t]{0.04\linewidth}{\centering\strut \raisebox{0.08ex}{\ding{55}}}\hspace{0.006\linewidth}\parbox[t]{0.10\linewidth}{\centering\strut -275406.25\%}\hspace{0.006\linewidth}\parbox[t]{0.115\linewidth}{\centering\strut $1.082\times 10^{-7}$}\hspace{0.006\linewidth}\parbox[t]{0.671\linewidth}{noisy Powell polish around best uccsd\_full}}}}\par\smallskip
\noindent\parbox[t]{0.05\linewidth}{\centering\strut 42}\hspace{0.006\linewidth}\parbox[t]{0.04\linewidth}{\centering\strut \raisebox{0.08ex}{\ding{55}}}\hspace{0.006\linewidth}\parbox[t]{0.10\linewidth}{\centering\strut +93.27\%}\hspace{0.006\linewidth}\parbox[t]{0.115\linewidth}{\centering\strut $7.276\times 10^{-9}$}\hspace{0.006\linewidth}\parbox[t]{0.671\linewidth}{noisy COBYLA polish around best uccsd\_full}\par\smallskip
\noindent{\setlength{\fboxsep}{0pt}\colorbox{black!4}{\parbox{\linewidth}{\parbox[t]{0.05\linewidth}{\centering\strut 43}\hspace{0.006\linewidth}\parbox[t]{0.04\linewidth}{\centering\strut \raisebox{0.08ex}{\ding{55}}}\hspace{0.006\linewidth}\parbox[t]{0.10\linewidth}{\centering\strut +99.01\%}\hspace{0.006\linewidth}\parbox[t]{0.115\linewidth}{\centering\strut $7.181\times 10^{-11}$}\hspace{0.006\linewidth}\parbox[t]{0.671\linewidth}{refine uccsd\_full with halved coordinate step}}}}\par\smallskip
\noindent\parbox[t]{0.05\linewidth}{\centering\strut 44}\hspace{0.006\linewidth}\parbox[t]{0.04\linewidth}{\centering\strut \raisebox{0.08ex}{\ding{51}}}\hspace{0.006\linewidth}\parbox[t]{0.10\linewidth}{\centering\strut +75.06\%}\hspace{0.006\linewidth}\parbox[t]{0.115\linewidth}{\centering\strut $1.791\times 10^{-11}$}\hspace{0.006\linewidth}\parbox[t]{0.671\linewidth}{Powell polish current best uccsd\_full}\par\smallskip
\noindent{\setlength{\fboxsep}{0pt}\colorbox{black!4}{\parbox{\linewidth}{\parbox[t]{0.05\linewidth}{\centering\strut 45}\hspace{0.006\linewidth}\parbox[t]{0.04\linewidth}{\centering\strut \raisebox{0.08ex}{\ding{51}}}\hspace{0.006\linewidth}\parbox[t]{0.10\linewidth}{\centering\strut +9.48\%}\hspace{0.006\linewidth}\parbox[t]{0.115\linewidth}{\centering\strut $1.621\times 10^{-11}$}\hspace{0.006\linewidth}\parbox[t]{0.671\linewidth}{COBYLA polish current best uccsd\_full}}}}\par\smallskip
\noindent\parbox[t]{0.05\linewidth}{\centering\strut 46}\hspace{0.006\linewidth}\parbox[t]{0.04\linewidth}{\centering\strut \raisebox{0.08ex}{\ding{55}}}\hspace{0.006\linewidth}\parbox[t]{0.10\linewidth}{\centering\strut -354813.51\%}\hspace{0.006\linewidth}\parbox[t]{0.115\linewidth}{\centering\strut $5.755\times 10^{-8}$}\hspace{0.006\linewidth}\parbox[t]{0.671\linewidth}{noisy coordinate restart around best uccsd\_full}\par\smallskip
\noindent{\setlength{\fboxsep}{0pt}\colorbox{black!4}{\parbox{\linewidth}{\parbox[t]{0.05\linewidth}{\centering\strut 47}\hspace{0.006\linewidth}\parbox[t]{0.04\linewidth}{\centering\strut \raisebox{0.08ex}{\ding{55}}}\hspace{0.006\linewidth}\parbox[t]{0.10\linewidth}{\centering\strut +72.77\%}\hspace{0.006\linewidth}\parbox[t]{0.115\linewidth}{\centering\strut $1.567\times 10^{-8}$}\hspace{0.006\linewidth}\parbox[t]{0.671\linewidth}{expand current best uccsd\_full into spin\_paired\_full}}}}\par\smallskip
\noindent\parbox[t]{0.05\linewidth}{\centering\strut 48}\hspace{0.006\linewidth}\parbox[t]{0.04\linewidth}{\centering\strut \raisebox{0.08ex}{\ding{55}}}\hspace{0.006\linewidth}\parbox[t]{0.10\linewidth}{\centering\strut +0\%}\hspace{0.006\linewidth}\parbox[t]{0.115\linewidth}{\centering\strut $1.567\times 10^{-8}$}\hspace{0.006\linewidth}\parbox[t]{0.671\linewidth}{Powell polish expanded spin\_paired\_full warm start}\par\smallskip
\noindent{\setlength{\fboxsep}{0pt}\colorbox{black!4}{\parbox{\linewidth}{\parbox[t]{0.05\linewidth}{\centering\strut 49}\hspace{0.006\linewidth}\parbox[t]{0.04\linewidth}{\centering\strut \raisebox{0.08ex}{\ding{55}}}\hspace{0.006\linewidth}\parbox[t]{0.10\linewidth}{\centering\strut -351047.99\%}\hspace{0.006\linewidth}\parbox[t]{0.115\linewidth}{\centering\strut $5.502\times 10^{-5}$}\hspace{0.006\linewidth}\parbox[t]{0.671\linewidth}{try alternate LiH compression pair\_doubles\_symmetric near current best}}}}\par\smallskip
\noindent\parbox[t]{0.05\linewidth}{\centering\strut 50}\hspace{0.006\linewidth}\parbox[t]{0.04\linewidth}{\centering\strut \raisebox{0.08ex}{\ding{51}}}\hspace{0.006\linewidth}\parbox[t]{0.10\linewidth}{\centering\strut +100\%}\hspace{0.006\linewidth}\parbox[t]{0.115\linewidth}{\centering\strut $1.604\times 10^{-11}$}\hspace{0.006\linewidth}\parbox[t]{0.671\linewidth}{ultra-fine coordinate polish for uccsd\_full}\par\smallskip
\noindent{\setlength{\fboxsep}{0pt}\colorbox{black!4}{\parbox{\linewidth}{\parbox[t]{0.05\linewidth}{\centering\strut 51}\hspace{0.006\linewidth}\parbox[t]{0.04\linewidth}{\centering\strut \raisebox{0.08ex}{\ding{55}}}\hspace{0.006\linewidth}\parbox[t]{0.10\linewidth}{\centering\strut -661220.98\%}\hspace{0.006\linewidth}\parbox[t]{0.115\linewidth}{\centering\strut $1.061\times 10^{-7}$}\hspace{0.006\linewidth}\parbox[t]{0.671\linewidth}{noisy Powell polish around best uccsd\_full}}}}\par\smallskip
\noindent\parbox[t]{0.05\linewidth}{\centering\strut 52}\hspace{0.006\linewidth}\parbox[t]{0.04\linewidth}{\centering\strut \raisebox{0.08ex}{\ding{55}}}\hspace{0.006\linewidth}\parbox[t]{0.10\linewidth}{\centering\strut +93.12\%}\hspace{0.006\linewidth}\parbox[t]{0.115\linewidth}{\centering\strut $7.293\times 10^{-9}$}\hspace{0.006\linewidth}\parbox[t]{0.671\linewidth}{noisy COBYLA polish around best uccsd\_full}\par\smallskip
\noindent{\setlength{\fboxsep}{0pt}\colorbox{black!4}{\parbox{\linewidth}{\parbox[t]{0.05\linewidth}{\centering\strut 53}\hspace{0.006\linewidth}\parbox[t]{0.04\linewidth}{\centering\strut \raisebox{0.08ex}{\ding{51}}}\hspace{0.006\linewidth}\parbox[t]{0.10\linewidth}{\centering\strut +99.93\%}\hspace{0.006\linewidth}\parbox[t]{0.115\linewidth}{\centering\strut $5.276\times 10^{-12}$}\hspace{0.006\linewidth}\parbox[t]{0.671\linewidth}{refine uccsd\_full with halved coordinate step}}}}\par\smallskip
\noindent\parbox[t]{0.05\linewidth}{\centering\strut 54}\hspace{0.006\linewidth}\parbox[t]{0.04\linewidth}{\centering\strut \raisebox{0.08ex}{\ding{51}}}\hspace{0.006\linewidth}\parbox[t]{0.10\linewidth}{\centering\strut +3.01\%}\hspace{0.006\linewidth}\parbox[t]{0.115\linewidth}{\centering\strut $5.117\times 10^{-12}$}\hspace{0.006\linewidth}\parbox[t]{0.671\linewidth}{Powell polish current best uccsd\_full}\par\smallskip
\noindent{\setlength{\fboxsep}{0pt}\colorbox{black!4}{\parbox{\linewidth}{\parbox[t]{0.05\linewidth}{\centering\strut 55}\hspace{0.006\linewidth}\parbox[t]{0.04\linewidth}{\centering\strut \raisebox{0.08ex}{\ding{55}}}\hspace{0.006\linewidth}\parbox[t]{0.10\linewidth}{\centering\strut -57.79\%}\hspace{0.006\linewidth}\parbox[t]{0.115\linewidth}{\centering\strut $8.074\times 10^{-12}$}\hspace{0.006\linewidth}\parbox[t]{0.671\linewidth}{COBYLA polish current best uccsd\_full}}}}\par\smallskip
\noindent\parbox[t]{0.05\linewidth}{\centering\strut 56}\hspace{0.006\linewidth}\parbox[t]{0.04\linewidth}{\centering\strut \raisebox{0.08ex}{\ding{55}}}\hspace{0.006\linewidth}\parbox[t]{0.10\linewidth}{\centering\strut -724098.87\%}\hspace{0.006\linewidth}\parbox[t]{0.115\linewidth}{\centering\strut $5.847\times 10^{-8}$}\hspace{0.006\linewidth}\parbox[t]{0.671\linewidth}{noisy coordinate restart around best uccsd\_full}\par\smallskip
\noindent{\setlength{\fboxsep}{0pt}\colorbox{black!4}{\parbox{\linewidth}{\parbox[t]{0.05\linewidth}{\centering\strut 57}\hspace{0.006\linewidth}\parbox[t]{0.04\linewidth}{\centering\strut \raisebox{0.08ex}{\ding{55}}}\hspace{0.006\linewidth}\parbox[t]{0.10\linewidth}{\centering\strut +73.2\%}\hspace{0.006\linewidth}\parbox[t]{0.115\linewidth}{\centering\strut $1.567\times 10^{-8}$}\hspace{0.006\linewidth}\parbox[t]{0.671\linewidth}{expand current best uccsd\_full into spin\_paired\_full}}}}\par\smallskip
\noindent\parbox[t]{0.05\linewidth}{\centering\strut 58}\hspace{0.006\linewidth}\parbox[t]{0.04\linewidth}{\centering\strut \raisebox{0.08ex}{\ding{55}}}\hspace{0.006\linewidth}\parbox[t]{0.10\linewidth}{\centering\strut +0\%}\hspace{0.006\linewidth}\parbox[t]{0.115\linewidth}{\centering\strut $1.567\times 10^{-8}$}\hspace{0.006\linewidth}\parbox[t]{0.671\linewidth}{Powell polish expanded spin\_paired\_full warm start}\par\smallskip
\noindent{\setlength{\fboxsep}{0pt}\colorbox{black!4}{\parbox{\linewidth}{\parbox[t]{0.05\linewidth}{\centering\strut 59}\hspace{0.006\linewidth}\parbox[t]{0.04\linewidth}{\centering\strut \raisebox{0.08ex}{\ding{55}}}\hspace{0.006\linewidth}\parbox[t]{0.10\linewidth}{\centering\strut -351047.99\%}\hspace{0.006\linewidth}\parbox[t]{0.115\linewidth}{\centering\strut $5.502\times 10^{-5}$}\hspace{0.006\linewidth}\parbox[t]{0.671\linewidth}{try alternate LiH compression pair\_doubles\_symmetric near current best}}}}\par\smallskip
\noindent\parbox[t]{0.05\linewidth}{\centering\strut 60}\hspace{0.006\linewidth}\parbox[t]{0.04\linewidth}{\centering\strut \raisebox{0.08ex}{\ding{55}}}\hspace{0.006\linewidth}\parbox[t]{0.10\linewidth}{\centering\strut +100\%}\hspace{0.006\linewidth}\parbox[t]{0.115\linewidth}{\centering\strut $1.588\times 10^{-11}$}\hspace{0.006\linewidth}\parbox[t]{0.671\linewidth}{ultra-fine coordinate polish for uccsd\_full}\par\smallskip
\noindent{\setlength{\fboxsep}{0pt}\colorbox{black!4}{\parbox{\linewidth}{\parbox[t]{0.05\linewidth}{\centering\strut 61}\hspace{0.006\linewidth}\parbox[t]{0.04\linewidth}{\centering\strut \raisebox{0.08ex}{\ding{55}}}\hspace{0.006\linewidth}\parbox[t]{0.10\linewidth}{\centering\strut -676640.78\%}\hspace{0.006\linewidth}\parbox[t]{0.115\linewidth}{\centering\strut $1.075\times 10^{-7}$}\hspace{0.006\linewidth}\parbox[t]{0.671\linewidth}{noisy Powell polish around best uccsd\_full}}}}\par\smallskip
\noindent\parbox[t]{0.05\linewidth}{\centering\strut 62}\hspace{0.006\linewidth}\parbox[t]{0.04\linewidth}{\centering\strut \raisebox{0.08ex}{\ding{55}}}\hspace{0.006\linewidth}\parbox[t]{0.10\linewidth}{\centering\strut +93.21\%}\hspace{0.006\linewidth}\parbox[t]{0.115\linewidth}{\centering\strut $7.294\times 10^{-9}$}\hspace{0.006\linewidth}\parbox[t]{0.671\linewidth}{noisy COBYLA polish around best uccsd\_full}\par\smallskip
\noindent{\setlength{\fboxsep}{0pt}\colorbox{black!4}{\parbox{\linewidth}{\parbox[t]{0.05\linewidth}{\centering\strut 63}\hspace{0.006\linewidth}\parbox[t]{0.04\linewidth}{\centering\strut \raisebox{0.08ex}{\ding{55}}}\hspace{0.006\linewidth}\parbox[t]{0.10\linewidth}{\centering\strut +99.93\%}\hspace{0.006\linewidth}\parbox[t]{0.115\linewidth}{\centering\strut $5.142\times 10^{-12}$}\hspace{0.006\linewidth}\parbox[t]{0.671\linewidth}{refine uccsd\_full with halved coordinate step}}}}\par\smallskip
\noindent\parbox[t]{0.05\linewidth}{\centering\strut 64}\hspace{0.006\linewidth}\parbox[t]{0.04\linewidth}{\centering\strut \raisebox{0.08ex}{\ding{51}}}\hspace{0.006\linewidth}\parbox[t]{0.10\linewidth}{\centering\strut +1.11\%}\hspace{0.006\linewidth}\parbox[t]{0.115\linewidth}{\centering\strut $5.085\times 10^{-12}$}\hspace{0.006\linewidth}\parbox[t]{0.671\linewidth}{Powell polish current best uccsd\_full}\par\smallskip
\noindent{\setlength{\fboxsep}{0pt}\colorbox{black!4}{\parbox{\linewidth}{\parbox[t]{0.05\linewidth}{\centering\strut 65}\hspace{0.006\linewidth}\parbox[t]{0.04\linewidth}{\centering\strut \raisebox{0.08ex}{\ding{55}}}\hspace{0.006\linewidth}\parbox[t]{0.10\linewidth}{\centering\strut -11.93\%}\hspace{0.006\linewidth}\parbox[t]{0.115\linewidth}{\centering\strut $5.691\times 10^{-12}$}\hspace{0.006\linewidth}\parbox[t]{0.671\linewidth}{COBYLA polish current best uccsd\_full}}}}\par\smallskip
\noindent\parbox[t]{0.05\linewidth}{\centering\strut 66}\hspace{0.006\linewidth}\parbox[t]{0.04\linewidth}{\centering\strut \raisebox{0.08ex}{\ding{55}}}\hspace{0.006\linewidth}\parbox[t]{0.10\linewidth}{\centering\strut -1027128.58\%}\hspace{0.006\linewidth}\parbox[t]{0.115\linewidth}{\centering\strut $5.846\times 10^{-8}$}\hspace{0.006\linewidth}\parbox[t]{0.671\linewidth}{noisy coordinate restart around best uccsd\_full}\par\smallskip
\noindent{\setlength{\fboxsep}{0pt}\colorbox{black!4}{\parbox{\linewidth}{\parbox[t]{0.05\linewidth}{\centering\strut 67}\hspace{0.006\linewidth}\parbox[t]{0.04\linewidth}{\centering\strut \raisebox{0.08ex}{\ding{55}}}\hspace{0.006\linewidth}\parbox[t]{0.10\linewidth}{\centering\strut +73.2\%}\hspace{0.006\linewidth}\parbox[t]{0.115\linewidth}{\centering\strut $1.567\times 10^{-8}$}\hspace{0.006\linewidth}\parbox[t]{0.671\linewidth}{expand current best uccsd\_full into spin\_paired\_full}}}}\par\smallskip
\noindent\parbox[t]{0.05\linewidth}{\centering\strut 68}\hspace{0.006\linewidth}\parbox[t]{0.04\linewidth}{\centering\strut \raisebox{0.08ex}{\ding{55}}}\hspace{0.006\linewidth}\parbox[t]{0.10\linewidth}{\centering\strut +0\%}\hspace{0.006\linewidth}\parbox[t]{0.115\linewidth}{\centering\strut $1.567\times 10^{-8}$}\hspace{0.006\linewidth}\parbox[t]{0.671\linewidth}{Powell polish expanded spin\_paired\_full warm start}\par\smallskip
\noindent{\setlength{\fboxsep}{0pt}\colorbox{black!4}{\parbox{\linewidth}{\parbox[t]{0.05\linewidth}{\centering\strut 69}\hspace{0.006\linewidth}\parbox[t]{0.04\linewidth}{\centering\strut \raisebox{0.08ex}{\ding{55}}}\hspace{0.006\linewidth}\parbox[t]{0.10\linewidth}{\centering\strut -351047.99\%}\hspace{0.006\linewidth}\parbox[t]{0.115\linewidth}{\centering\strut $5.502\times 10^{-5}$}\hspace{0.006\linewidth}\parbox[t]{0.671\linewidth}{try alternate LiH compression pair\_doubles\_symmetric near current best}}}}\par\smallskip
\noindent\parbox[t]{0.05\linewidth}{\centering\strut 70}\hspace{0.006\linewidth}\parbox[t]{0.04\linewidth}{\centering\strut \raisebox{0.08ex}{\ding{55}}}\hspace{0.006\linewidth}\parbox[t]{0.10\linewidth}{\centering\strut +100\%}\hspace{0.006\linewidth}\parbox[t]{0.115\linewidth}{\centering\strut $1.587\times 10^{-11}$}\hspace{0.006\linewidth}\parbox[t]{0.671\linewidth}{ultra-fine coordinate polish for uccsd\_full}\par\smallskip
\noindent{\setlength{\fboxsep}{0pt}\colorbox{black!4}{\parbox{\linewidth}{\parbox[t]{0.05\linewidth}{\centering\strut 71}\hspace{0.006\linewidth}\parbox[t]{0.04\linewidth}{\centering\strut \raisebox{0.08ex}{\ding{55}}}\hspace{0.006\linewidth}\parbox[t]{0.10\linewidth}{\centering\strut -676988.07\%}\hspace{0.006\linewidth}\parbox[t]{0.115\linewidth}{\centering\strut $1.075\times 10^{-7}$}\hspace{0.006\linewidth}\parbox[t]{0.671\linewidth}{noisy Powell polish around best uccsd\_full}}}}\par\smallskip
\noindent\parbox[t]{0.05\linewidth}{\centering\strut 72}\hspace{0.006\linewidth}\parbox[t]{0.04\linewidth}{\centering\strut \raisebox{0.08ex}{\ding{55}}}\hspace{0.006\linewidth}\parbox[t]{0.10\linewidth}{\centering\strut +90.64\%}\hspace{0.006\linewidth}\parbox[t]{0.115\linewidth}{\centering\strut $1.006\times 10^{-8}$}\hspace{0.006\linewidth}\parbox[t]{0.671\linewidth}{noisy COBYLA polish around best uccsd\_full}\par\smallskip
\noindent{\setlength{\fboxsep}{0pt}\colorbox{black!4}{\parbox{\linewidth}{\parbox[t]{0.05\linewidth}{\centering\strut 73}\hspace{0.006\linewidth}\parbox[t]{0.04\linewidth}{\centering\strut \raisebox{0.08ex}{\ding{55}}}\hspace{0.006\linewidth}\parbox[t]{0.10\linewidth}{\centering\strut +99.95\%}\hspace{0.006\linewidth}\parbox[t]{0.115\linewidth}{\centering\strut $5.106\times 10^{-12}$}\hspace{0.006\linewidth}\parbox[t]{0.671\linewidth}{refine uccsd\_full with halved coordinate step}}}}\par\smallskip
\noindent\parbox[t]{0.05\linewidth}{\centering\strut 74}\hspace{0.006\linewidth}\parbox[t]{0.04\linewidth}{\centering\strut \raisebox{0.08ex}{\ding{55}}}\hspace{0.006\linewidth}\parbox[t]{0.10\linewidth}{\centering\strut +0.0696\%}\hspace{0.006\linewidth}\parbox[t]{0.115\linewidth}{\centering\strut $5.103\times 10^{-12}$}\hspace{0.006\linewidth}\parbox[t]{0.671\linewidth}{Powell polish current best uccsd\_full}\par\smallskip
\noindent{\setlength{\fboxsep}{0pt}\colorbox{black!4}{\parbox{\linewidth}{\parbox[t]{0.05\linewidth}{\centering\strut 75}\hspace{0.006\linewidth}\parbox[t]{0.04\linewidth}{\centering\strut \raisebox{0.08ex}{\ding{55}}}\hspace{0.006\linewidth}\parbox[t]{0.10\linewidth}{\centering\strut -11.54\%}\hspace{0.006\linewidth}\parbox[t]{0.115\linewidth}{\centering\strut $5.691\times 10^{-12}$}\hspace{0.006\linewidth}\parbox[t]{0.671\linewidth}{COBYLA polish current best uccsd\_full}}}}\par\smallskip
\noindent\parbox[t]{0.05\linewidth}{\centering\strut 76}\hspace{0.006\linewidth}\parbox[t]{0.04\linewidth}{\centering\strut \raisebox{0.08ex}{\ding{55}}}\hspace{0.006\linewidth}\parbox[t]{0.10\linewidth}{\centering\strut -1027128.58\%}\hspace{0.006\linewidth}\parbox[t]{0.115\linewidth}{\centering\strut $5.846\times 10^{-8}$}\hspace{0.006\linewidth}\parbox[t]{0.671\linewidth}{noisy coordinate restart around best uccsd\_full}\par\smallskip
\noindent{\setlength{\fboxsep}{0pt}\colorbox{black!4}{\parbox{\linewidth}{\parbox[t]{0.05\linewidth}{\centering\strut 77}\hspace{0.006\linewidth}\parbox[t]{0.04\linewidth}{\centering\strut \raisebox{0.08ex}{\ding{55}}}\hspace{0.006\linewidth}\parbox[t]{0.10\linewidth}{\centering\strut +73.2\%}\hspace{0.006\linewidth}\parbox[t]{0.115\linewidth}{\centering\strut $1.567\times 10^{-8}$}\hspace{0.006\linewidth}\parbox[t]{0.671\linewidth}{expand current best uccsd\_full into spin\_paired\_full}}}}\par\smallskip
\noindent\parbox[t]{0.05\linewidth}{\centering\strut 78}\hspace{0.006\linewidth}\parbox[t]{0.04\linewidth}{\centering\strut \raisebox{0.08ex}{\ding{55}}}\hspace{0.006\linewidth}\parbox[t]{0.10\linewidth}{\centering\strut +0\%}\hspace{0.006\linewidth}\parbox[t]{0.115\linewidth}{\centering\strut $1.567\times 10^{-8}$}\hspace{0.006\linewidth}\parbox[t]{0.671\linewidth}{Powell polish expanded spin\_paired\_full warm start}\par\smallskip
\noindent{\setlength{\fboxsep}{0pt}\colorbox{black!4}{\parbox{\linewidth}{\parbox[t]{0.05\linewidth}{\centering\strut 79}\hspace{0.006\linewidth}\parbox[t]{0.04\linewidth}{\centering\strut \raisebox{0.08ex}{\ding{55}}}\hspace{0.006\linewidth}\parbox[t]{0.10\linewidth}{\centering\strut -351047.99\%}\hspace{0.006\linewidth}\parbox[t]{0.115\linewidth}{\centering\strut $5.502\times 10^{-5}$}\hspace{0.006\linewidth}\parbox[t]{0.671\linewidth}{try alternate LiH compression pair\_doubles\_symmetric near current best}}}}\par\smallskip
\noindent\parbox[t]{0.05\linewidth}{\centering\strut 80}\hspace{0.006\linewidth}\parbox[t]{0.04\linewidth}{\centering\strut \raisebox{0.08ex}{\ding{55}}}\hspace{0.006\linewidth}\parbox[t]{0.10\linewidth}{\centering\strut +100\%}\hspace{0.006\linewidth}\parbox[t]{0.115\linewidth}{\centering\strut $1.587\times 10^{-11}$}\hspace{0.006\linewidth}\parbox[t]{0.671\linewidth}{ultra-fine coordinate polish for uccsd\_full}\par\smallskip
\noindent{\setlength{\fboxsep}{0pt}\colorbox{black!4}{\parbox{\linewidth}{\parbox[t]{0.05\linewidth}{\centering\strut 81}\hspace{0.006\linewidth}\parbox[t]{0.04\linewidth}{\centering\strut \raisebox{0.08ex}{\ding{55}}}\hspace{0.006\linewidth}\parbox[t]{0.10\linewidth}{\centering\strut -676988.07\%}\hspace{0.006\linewidth}\parbox[t]{0.115\linewidth}{\centering\strut $1.075\times 10^{-7}$}\hspace{0.006\linewidth}\parbox[t]{0.671\linewidth}{noisy Powell polish around best uccsd\_full}}}}\par\smallskip
\noindent\parbox[t]{0.05\linewidth}{\centering\strut 82}\hspace{0.006\linewidth}\parbox[t]{0.04\linewidth}{\centering\strut \raisebox{0.08ex}{\ding{55}}}\hspace{0.006\linewidth}\parbox[t]{0.10\linewidth}{\centering\strut +90.64\%}\hspace{0.006\linewidth}\parbox[t]{0.115\linewidth}{\centering\strut $1.006\times 10^{-8}$}\hspace{0.006\linewidth}\parbox[t]{0.671\linewidth}{noisy COBYLA polish around best uccsd\_full}\par\smallskip
\noindent{\setlength{\fboxsep}{0pt}\colorbox{black!4}{\parbox{\linewidth}{\parbox[t]{0.05\linewidth}{\centering\strut 83}\hspace{0.006\linewidth}\parbox[t]{0.04\linewidth}{\centering\strut \raisebox{0.08ex}{\ding{55}}}\hspace{0.006\linewidth}\parbox[t]{0.10\linewidth}{\centering\strut +99.95\%}\hspace{0.006\linewidth}\parbox[t]{0.115\linewidth}{\centering\strut $5.106\times 10^{-12}$}\hspace{0.006\linewidth}\parbox[t]{0.671\linewidth}{refine uccsd\_full with halved coordinate step}}}}\par\smallskip
\noindent\parbox[t]{0.05\linewidth}{\centering\strut 84}\hspace{0.006\linewidth}\parbox[t]{0.04\linewidth}{\centering\strut \raisebox{0.08ex}{\ding{55}}}\hspace{0.006\linewidth}\parbox[t]{0.10\linewidth}{\centering\strut +0.0696\%}\hspace{0.006\linewidth}\parbox[t]{0.115\linewidth}{\centering\strut $5.103\times 10^{-12}$}\hspace{0.006\linewidth}\parbox[t]{0.671\linewidth}{Powell polish current best uccsd\_full}\par\smallskip
\noindent{\setlength{\fboxsep}{0pt}\colorbox{black!4}{\parbox{\linewidth}{\parbox[t]{0.05\linewidth}{\centering\strut 85}\hspace{0.006\linewidth}\parbox[t]{0.04\linewidth}{\centering\strut \raisebox{0.08ex}{\ding{55}}}\hspace{0.006\linewidth}\parbox[t]{0.10\linewidth}{\centering\strut -11.54\%}\hspace{0.006\linewidth}\parbox[t]{0.115\linewidth}{\centering\strut $5.691\times 10^{-12}$}\hspace{0.006\linewidth}\parbox[t]{0.671\linewidth}{COBYLA polish current best uccsd\_full}}}}\par\smallskip
\noindent\parbox[t]{0.05\linewidth}{\centering\strut 86}\hspace{0.006\linewidth}\parbox[t]{0.04\linewidth}{\centering\strut \raisebox{0.08ex}{\ding{55}}}\hspace{0.006\linewidth}\parbox[t]{0.10\linewidth}{\centering\strut -1027128.58\%}\hspace{0.006\linewidth}\parbox[t]{0.115\linewidth}{\centering\strut $5.846\times 10^{-8}$}\hspace{0.006\linewidth}\parbox[t]{0.671\linewidth}{noisy coordinate restart around best uccsd\_full}\par\smallskip
\noindent{\setlength{\fboxsep}{0pt}\colorbox{black!4}{\parbox{\linewidth}{\parbox[t]{0.05\linewidth}{\centering\strut 87}\hspace{0.006\linewidth}\parbox[t]{0.04\linewidth}{\centering\strut \raisebox{0.08ex}{\ding{55}}}\hspace{0.006\linewidth}\parbox[t]{0.10\linewidth}{\centering\strut +73.2\%}\hspace{0.006\linewidth}\parbox[t]{0.115\linewidth}{\centering\strut $1.567\times 10^{-8}$}\hspace{0.006\linewidth}\parbox[t]{0.671\linewidth}{expand current best uccsd\_full into spin\_paired\_full}}}}\par\smallskip
\noindent\parbox[t]{0.05\linewidth}{\centering\strut 88}\hspace{0.006\linewidth}\parbox[t]{0.04\linewidth}{\centering\strut \raisebox{0.08ex}{\ding{55}}}\hspace{0.006\linewidth}\parbox[t]{0.10\linewidth}{\centering\strut +0\%}\hspace{0.006\linewidth}\parbox[t]{0.115\linewidth}{\centering\strut $1.567\times 10^{-8}$}\hspace{0.006\linewidth}\parbox[t]{0.671\linewidth}{Powell polish expanded spin\_paired\_full warm start}\par\smallskip
\noindent{\setlength{\fboxsep}{0pt}\colorbox{black!4}{\parbox{\linewidth}{\parbox[t]{0.05\linewidth}{\centering\strut 89}\hspace{0.006\linewidth}\parbox[t]{0.04\linewidth}{\centering\strut \raisebox{0.08ex}{\ding{55}}}\hspace{0.006\linewidth}\parbox[t]{0.10\linewidth}{\centering\strut -351047.99\%}\hspace{0.006\linewidth}\parbox[t]{0.115\linewidth}{\centering\strut $5.502\times 10^{-5}$}\hspace{0.006\linewidth}\parbox[t]{0.671\linewidth}{try alternate LiH compression pair\_doubles\_symmetric near current best}}}}\par\smallskip
\noindent\parbox[t]{0.05\linewidth}{\centering\strut 90}\hspace{0.006\linewidth}\parbox[t]{0.04\linewidth}{\centering\strut \raisebox{0.08ex}{\ding{55}}}\hspace{0.006\linewidth}\parbox[t]{0.10\linewidth}{\centering\strut +100\%}\hspace{0.006\linewidth}\parbox[t]{0.115\linewidth}{\centering\strut $1.587\times 10^{-11}$}\hspace{0.006\linewidth}\parbox[t]{0.671\linewidth}{ultra-fine coordinate polish for uccsd\_full}\par\smallskip
\noindent{\setlength{\fboxsep}{0pt}\colorbox{black!4}{\parbox{\linewidth}{\parbox[t]{0.05\linewidth}{\centering\strut 91}\hspace{0.006\linewidth}\parbox[t]{0.04\linewidth}{\centering\strut \raisebox{0.08ex}{\ding{55}}}\hspace{0.006\linewidth}\parbox[t]{0.10\linewidth}{\centering\strut -676988.07\%}\hspace{0.006\linewidth}\parbox[t]{0.115\linewidth}{\centering\strut $1.075\times 10^{-7}$}\hspace{0.006\linewidth}\parbox[t]{0.671\linewidth}{noisy Powell polish around best uccsd\_full}}}}\par\smallskip
\noindent\parbox[t]{0.05\linewidth}{\centering\strut 92}\hspace{0.006\linewidth}\parbox[t]{0.04\linewidth}{\centering\strut \raisebox{0.08ex}{\ding{55}}}\hspace{0.006\linewidth}\parbox[t]{0.10\linewidth}{\centering\strut +90.64\%}\hspace{0.006\linewidth}\parbox[t]{0.115\linewidth}{\centering\strut $1.006\times 10^{-8}$}\hspace{0.006\linewidth}\parbox[t]{0.671\linewidth}{noisy COBYLA polish around best uccsd\_full}\par\smallskip
\noindent{\setlength{\fboxsep}{0pt}\colorbox{black!4}{\parbox{\linewidth}{\parbox[t]{0.05\linewidth}{\centering\strut 93}\hspace{0.006\linewidth}\parbox[t]{0.04\linewidth}{\centering\strut \raisebox{0.08ex}{\ding{55}}}\hspace{0.006\linewidth}\parbox[t]{0.10\linewidth}{\centering\strut +99.95\%}\hspace{0.006\linewidth}\parbox[t]{0.115\linewidth}{\centering\strut $5.106\times 10^{-12}$}\hspace{0.006\linewidth}\parbox[t]{0.671\linewidth}{refine uccsd\_full with halved coordinate step}}}}\par\smallskip
\noindent\parbox[t]{0.05\linewidth}{\centering\strut 94}\hspace{0.006\linewidth}\parbox[t]{0.04\linewidth}{\centering\strut \raisebox{0.08ex}{\ding{55}}}\hspace{0.006\linewidth}\parbox[t]{0.10\linewidth}{\centering\strut +0.087\%}\hspace{0.006\linewidth}\parbox[t]{0.115\linewidth}{\centering\strut $5.102\times 10^{-12}$}\hspace{0.006\linewidth}\parbox[t]{0.671\linewidth}{Powell polish current best uccsd\_full}\par\smallskip
\noindent{\setlength{\fboxsep}{0pt}\colorbox{black!4}{\parbox{\linewidth}{\parbox[t]{0.05\linewidth}{\centering\strut 95}\hspace{0.006\linewidth}\parbox[t]{0.04\linewidth}{\centering\strut \raisebox{0.08ex}{\ding{55}}}\hspace{0.006\linewidth}\parbox[t]{0.10\linewidth}{\centering\strut -11.56\%}\hspace{0.006\linewidth}\parbox[t]{0.115\linewidth}{\centering\strut $5.691\times 10^{-12}$}\hspace{0.006\linewidth}\parbox[t]{0.671\linewidth}{COBYLA polish current best uccsd\_full}}}}\par\smallskip
\noindent\parbox[t]{0.05\linewidth}{\centering\strut 96}\hspace{0.006\linewidth}\parbox[t]{0.04\linewidth}{\centering\strut \raisebox{0.08ex}{\ding{55}}}\hspace{0.006\linewidth}\parbox[t]{0.10\linewidth}{\centering\strut -1027128.58\%}\hspace{0.006\linewidth}\parbox[t]{0.115\linewidth}{\centering\strut $5.846\times 10^{-8}$}\hspace{0.006\linewidth}\parbox[t]{0.671\linewidth}{noisy coordinate restart around best uccsd\_full}\par\smallskip
\noindent{\setlength{\fboxsep}{0pt}\colorbox{black!4}{\parbox{\linewidth}{\parbox[t]{0.05\linewidth}{\centering\strut 97}\hspace{0.006\linewidth}\parbox[t]{0.04\linewidth}{\centering\strut \raisebox{0.08ex}{\ding{55}}}\hspace{0.006\linewidth}\parbox[t]{0.10\linewidth}{\centering\strut +73.2\%}\hspace{0.006\linewidth}\parbox[t]{0.115\linewidth}{\centering\strut $1.567\times 10^{-8}$}\hspace{0.006\linewidth}\parbox[t]{0.671\linewidth}{expand current best uccsd\_full into spin\_paired\_full}}}}\par\smallskip
\noindent\parbox[t]{0.05\linewidth}{\centering\strut 98}\hspace{0.006\linewidth}\parbox[t]{0.04\linewidth}{\centering\strut \raisebox{0.08ex}{\ding{55}}}\hspace{0.006\linewidth}\parbox[t]{0.10\linewidth}{\centering\strut +0\%}\hspace{0.006\linewidth}\parbox[t]{0.115\linewidth}{\centering\strut $1.567\times 10^{-8}$}\hspace{0.006\linewidth}\parbox[t]{0.671\linewidth}{Powell polish expanded spin\_paired\_full warm start}\par\smallskip
\noindent{\setlength{\fboxsep}{0pt}\colorbox{black!4}{\parbox{\linewidth}{\parbox[t]{0.05\linewidth}{\centering\strut 99}\hspace{0.006\linewidth}\parbox[t]{0.04\linewidth}{\centering\strut \raisebox{0.08ex}{\ding{55}}}\hspace{0.006\linewidth}\parbox[t]{0.10\linewidth}{\centering\strut -351047.99\%}\hspace{0.006\linewidth}\parbox[t]{0.115\linewidth}{\centering\strut $5.502\times 10^{-5}$}\hspace{0.006\linewidth}\parbox[t]{0.671\linewidth}{try alternate LiH compression pair\_doubles\_symmetric near current best}}}}\par\smallskip
\noindent\parbox[t]{0.05\linewidth}{\centering\strut 100}\hspace{0.006\linewidth}\parbox[t]{0.04\linewidth}{\centering\strut \raisebox{0.08ex}{\ding{55}}}\hspace{0.006\linewidth}\parbox[t]{0.10\linewidth}{\centering\strut +100\%}\hspace{0.006\linewidth}\parbox[t]{0.115\linewidth}{\centering\strut $6.635\times 10^{-12}$}\hspace{0.006\linewidth}\parbox[t]{0.671\linewidth}{ultra-fine coordinate polish for uccsd\_full}\par\smallskip
\noindent\rule{\linewidth}{0.5pt}
\par\medskip

\medskip

\refstepcounter{table}
\label{tab:supp_vqe_log_beh2}
\noindent\textbf{Table \thetable.} VQE mutation log for BeH$_2$ full 100-iteration archival campaign. Relative gain is measured against the immediately previous scored iteration using the absolute energy error.
\par\medskip
\noindent\rule{\linewidth}{0.5pt}
\noindent{\setlength{\fboxsep}{0pt}\colorbox{black!4}{\parbox{\linewidth}{\parbox[c][2.2em][c]{0.05\linewidth}{\centering \textbf{Iter.}}\hspace{0.006\linewidth}\parbox[c][2.2em][c]{0.04\linewidth}{\centering \textbf{A/R}}\hspace{0.006\linewidth}\parbox[c][2.2em][c]{0.10\linewidth}{\centering \textbf{Relative gain}}\hspace{0.006\linewidth}\parbox[c][2.2em][c]{0.115\linewidth}{\centering \textbf{$|\Delta E|$ [Ha]}}\hspace{0.006\linewidth}\parbox[c][2.2em][c]{0.671\linewidth}{\centering \textbf{Mutation summary}}}}}\par\smallskip
\noindent\rule{\linewidth}{0.5pt}
\noindent{\setlength{\fboxsep}{0pt}\colorbox{black!4}{\parbox{\linewidth}{\parbox[t]{0.05\linewidth}{\centering\strut 1}\hspace{0.006\linewidth}\parbox[t]{0.04\linewidth}{\centering\strut \raisebox{0.08ex}{\ding{51}}}\hspace{0.006\linewidth}\parbox[t]{0.10\linewidth}{\centering\strut ---}\hspace{0.006\linewidth}\parbox[t]{0.115\linewidth}{\centering\strut $4.747\times 10^{-4}$}\hspace{0.006\linewidth}\parbox[t]{0.671\linewidth}{switch to full UCCSD with cached COBYLA from HF}}}}\par\smallskip
\noindent\parbox[t]{0.05\linewidth}{\centering\strut 2}\hspace{0.006\linewidth}\parbox[t]{0.04\linewidth}{\centering\strut \raisebox{0.08ex}{\ding{55}}}\hspace{0.006\linewidth}\parbox[t]{0.10\linewidth}{\centering\strut -189.5\%}\hspace{0.006\linewidth}\parbox[t]{0.115\linewidth}{\centering\strut $1.374\times 10^{-3}$}\hspace{0.006\linewidth}\parbox[t]{0.671\linewidth}{tighten COBYLA radius and archive best parameters}\par\smallskip
\noindent{\setlength{\fboxsep}{0pt}\colorbox{black!4}{\parbox{\linewidth}{\parbox[t]{0.05\linewidth}{\centering\strut 3}\hspace{0.006\linewidth}\parbox[t]{0.04\linewidth}{\centering\strut \raisebox{0.08ex}{\ding{55}}}\hspace{0.006\linewidth}\parbox[t]{0.10\linewidth}{\centering\strut +0.005602\%}\hspace{0.006\linewidth}\parbox[t]{0.115\linewidth}{\centering\strut $1.374\times 10^{-3}$}\hspace{0.006\linewidth}\parbox[t]{0.671\linewidth}{warm-start Powell from archived UCCSD parameters}}}}\par\smallskip
\noindent\parbox[t]{0.05\linewidth}{\centering\strut 4}\hspace{0.006\linewidth}\parbox[t]{0.04\linewidth}{\centering\strut \raisebox{0.08ex}{\ding{55}}}\hspace{0.006\linewidth}\parbox[t]{0.10\linewidth}{\centering\strut +65.46\%}\hspace{0.006\linewidth}\parbox[t]{0.115\linewidth}{\centering\strut $4.747\times 10^{-4}$}\hspace{0.006\linewidth}\parbox[t]{0.671\linewidth}{archive best parameters for zero-start UCCSD COBYLA}\par\smallskip
\noindent{\setlength{\fboxsep}{0pt}\colorbox{black!4}{\parbox{\linewidth}{\parbox[t]{0.05\linewidth}{\centering\strut 5}\hspace{0.006\linewidth}\parbox[t]{0.04\linewidth}{\centering\strut \raisebox{0.08ex}{\ding{51}}}\hspace{0.006\linewidth}\parbox[t]{0.10\linewidth}{\centering\strut +80.36\%}\hspace{0.006\linewidth}\parbox[t]{0.115\linewidth}{\centering\strut $9.322\times 10^{-5}$}\hspace{0.006\linewidth}\parbox[t]{0.671\linewidth}{warm\_cobyla\_r20 seed1005}}}}\par\smallskip
\noindent\parbox[t]{0.05\linewidth}{\centering\strut 6}\hspace{0.006\linewidth}\parbox[t]{0.04\linewidth}{\centering\strut \raisebox{0.08ex}{\ding{51}}}\hspace{0.006\linewidth}\parbox[t]{0.10\linewidth}{\centering\strut +66.16\%}\hspace{0.006\linewidth}\parbox[t]{0.115\linewidth}{\centering\strut $3.154\times 10^{-5}$}\hspace{0.006\linewidth}\parbox[t]{0.671\linewidth}{warm\_cobyla\_r10 seed1006}\par\smallskip
\noindent{\setlength{\fboxsep}{0pt}\colorbox{black!4}{\parbox{\linewidth}{\parbox[t]{0.05\linewidth}{\centering\strut 7}\hspace{0.006\linewidth}\parbox[t]{0.04\linewidth}{\centering\strut \raisebox{0.08ex}{\ding{51}}}\hspace{0.006\linewidth}\parbox[t]{0.10\linewidth}{\centering\strut +38.14\%}\hspace{0.006\linewidth}\parbox[t]{0.115\linewidth}{\centering\strut $1.951\times 10^{-5}$}\hspace{0.006\linewidth}\parbox[t]{0.671\linewidth}{warm\_cobyla\_r5 seed1007}}}}\par\smallskip
\noindent\parbox[t]{0.05\linewidth}{\centering\strut 8}\hspace{0.006\linewidth}\parbox[t]{0.04\linewidth}{\centering\strut \raisebox{0.08ex}{\ding{51}}}\hspace{0.006\linewidth}\parbox[t]{0.10\linewidth}{\centering\strut +35.69\%}\hspace{0.006\linewidth}\parbox[t]{0.115\linewidth}{\centering\strut $1.255\times 10^{-5}$}\hspace{0.006\linewidth}\parbox[t]{0.671\linewidth}{jitter\_cobyla\_s1e6\_r20 seed1008}\par\smallskip
\noindent{\setlength{\fboxsep}{0pt}\colorbox{black!4}{\parbox{\linewidth}{\parbox[t]{0.05\linewidth}{\centering\strut 9}\hspace{0.006\linewidth}\parbox[t]{0.04\linewidth}{\centering\strut \raisebox{0.08ex}{\ding{51}}}\hspace{0.006\linewidth}\parbox[t]{0.10\linewidth}{\centering\strut +0.1387\%}\hspace{0.006\linewidth}\parbox[t]{0.115\linewidth}{\centering\strut $1.253\times 10^{-5}$}\hspace{0.006\linewidth}\parbox[t]{0.671\linewidth}{jitter\_cobyla\_s3e6\_r10 seed1009}}}}\par\smallskip
\noindent\parbox[t]{0.05\linewidth}{\centering\strut 10}\hspace{0.006\linewidth}\parbox[t]{0.04\linewidth}{\centering\strut \raisebox{0.08ex}{\ding{51}}}\hspace{0.006\linewidth}\parbox[t]{0.10\linewidth}{\centering\strut +29.65\%}\hspace{0.006\linewidth}\parbox[t]{0.115\linewidth}{\centering\strut $8.817\times 10^{-6}$}\hspace{0.006\linewidth}\parbox[t]{0.671\linewidth}{jitter\_cobyla\_s1e5\_r5 seed1010}\par\smallskip
\noindent{\setlength{\fboxsep}{0pt}\colorbox{black!4}{\parbox{\linewidth}{\parbox[t]{0.05\linewidth}{\centering\strut 11}\hspace{0.006\linewidth}\parbox[t]{0.04\linewidth}{\centering\strut \raisebox{0.08ex}{\ding{51}}}\hspace{0.006\linewidth}\parbox[t]{0.10\linewidth}{\centering\strut +1.31\%}\hspace{0.006\linewidth}\parbox[t]{0.115\linewidth}{\centering\strut $8.701\times 10^{-6}$}\hspace{0.006\linewidth}\parbox[t]{0.671\linewidth}{warm\_powell\_xt1e6 seed1011}}}}\par\smallskip
\noindent\parbox[t]{0.05\linewidth}{\centering\strut 12}\hspace{0.006\linewidth}\parbox[t]{0.04\linewidth}{\centering\strut \raisebox{0.08ex}{\ding{51}}}\hspace{0.006\linewidth}\parbox[t]{0.10\linewidth}{\centering\strut +0\%}\hspace{0.006\linewidth}\parbox[t]{0.115\linewidth}{\centering\strut $8.701\times 10^{-6}$}\hspace{0.006\linewidth}\parbox[t]{0.671\linewidth}{warm\_cobyla\_r20 seed1012}\par\smallskip
\noindent{\setlength{\fboxsep}{0pt}\colorbox{black!4}{\parbox{\linewidth}{\parbox[t]{0.05\linewidth}{\centering\strut 13}\hspace{0.006\linewidth}\parbox[t]{0.04\linewidth}{\centering\strut \raisebox{0.08ex}{\ding{51}}}\hspace{0.006\linewidth}\parbox[t]{0.10\linewidth}{\centering\strut +0\%}\hspace{0.006\linewidth}\parbox[t]{0.115\linewidth}{\centering\strut $8.701\times 10^{-6}$}\hspace{0.006\linewidth}\parbox[t]{0.671\linewidth}{warm\_cobyla\_r10 seed1013}}}}\par\smallskip
\noindent\parbox[t]{0.05\linewidth}{\centering\strut 14}\hspace{0.006\linewidth}\parbox[t]{0.04\linewidth}{\centering\strut \raisebox{0.08ex}{\ding{51}}}\hspace{0.006\linewidth}\parbox[t]{0.10\linewidth}{\centering\strut +21.52\%}\hspace{0.006\linewidth}\parbox[t]{0.115\linewidth}{\centering\strut $6.829\times 10^{-6}$}\hspace{0.006\linewidth}\parbox[t]{0.671\linewidth}{warm\_cobyla\_r5 seed1014}\par\smallskip
\noindent{\setlength{\fboxsep}{0pt}\colorbox{black!4}{\parbox{\linewidth}{\parbox[t]{0.05\linewidth}{\centering\strut 15}\hspace{0.006\linewidth}\parbox[t]{0.04\linewidth}{\centering\strut \raisebox{0.08ex}{\ding{51}}}\hspace{0.006\linewidth}\parbox[t]{0.10\linewidth}{\centering\strut +0.000556\%}\hspace{0.006\linewidth}\parbox[t]{0.115\linewidth}{\centering\strut $6.829\times 10^{-6}$}\hspace{0.006\linewidth}\parbox[t]{0.671\linewidth}{jitter\_cobyla\_s1e6\_r20 seed1015}}}}\par\smallskip
\noindent\parbox[t]{0.05\linewidth}{\centering\strut 16}\hspace{0.006\linewidth}\parbox[t]{0.04\linewidth}{\centering\strut \raisebox{0.08ex}{\ding{51}}}\hspace{0.006\linewidth}\parbox[t]{0.10\linewidth}{\centering\strut +0.1035\%}\hspace{0.006\linewidth}\parbox[t]{0.115\linewidth}{\centering\strut $6.822\times 10^{-6}$}\hspace{0.006\linewidth}\parbox[t]{0.671\linewidth}{jitter\_cobyla\_s3e6\_r10 seed1016}\par\smallskip
\noindent{\setlength{\fboxsep}{0pt}\colorbox{black!4}{\parbox{\linewidth}{\parbox[t]{0.05\linewidth}{\centering\strut 17}\hspace{0.006\linewidth}\parbox[t]{0.04\linewidth}{\centering\strut \raisebox{0.08ex}{\ding{55}}}\hspace{0.006\linewidth}\parbox[t]{0.10\linewidth}{\centering\strut -0.4147\%}\hspace{0.006\linewidth}\parbox[t]{0.115\linewidth}{\centering\strut $6.85\times 10^{-6}$}\hspace{0.006\linewidth}\parbox[t]{0.671\linewidth}{jitter\_cobyla\_s1e5\_r5 seed1017}}}}\par\smallskip
\noindent\parbox[t]{0.05\linewidth}{\centering\strut 18}\hspace{0.006\linewidth}\parbox[t]{0.04\linewidth}{\centering\strut \raisebox{0.08ex}{\ding{51}}}\hspace{0.006\linewidth}\parbox[t]{0.10\linewidth}{\centering\strut +0.4689\%}\hspace{0.006\linewidth}\parbox[t]{0.115\linewidth}{\centering\strut $6.818\times 10^{-6}$}\hspace{0.006\linewidth}\parbox[t]{0.671\linewidth}{warm\_powell\_xt1e6 seed1018}\par\smallskip
\noindent{\setlength{\fboxsep}{0pt}\colorbox{black!4}{\parbox{\linewidth}{\parbox[t]{0.05\linewidth}{\centering\strut 19}\hspace{0.006\linewidth}\parbox[t]{0.04\linewidth}{\centering\strut \raisebox{0.08ex}{\ding{51}}}\hspace{0.006\linewidth}\parbox[t]{0.10\linewidth}{\centering\strut +0\%}\hspace{0.006\linewidth}\parbox[t]{0.115\linewidth}{\centering\strut $6.818\times 10^{-6}$}\hspace{0.006\linewidth}\parbox[t]{0.671\linewidth}{warm\_cobyla\_r20 seed1019}}}}\par\smallskip
\noindent\parbox[t]{0.05\linewidth}{\centering\strut 20}\hspace{0.006\linewidth}\parbox[t]{0.04\linewidth}{\centering\strut \raisebox{0.08ex}{\ding{51}}}\hspace{0.006\linewidth}\parbox[t]{0.10\linewidth}{\centering\strut +0\%}\hspace{0.006\linewidth}\parbox[t]{0.115\linewidth}{\centering\strut $6.818\times 10^{-6}$}\hspace{0.006\linewidth}\parbox[t]{0.671\linewidth}{warm\_cobyla\_r10 seed1020}\par\smallskip
\noindent{\setlength{\fboxsep}{0pt}\colorbox{black!4}{\parbox{\linewidth}{\parbox[t]{0.05\linewidth}{\centering\strut 21}\hspace{0.006\linewidth}\parbox[t]{0.04\linewidth}{\centering\strut \raisebox{0.08ex}{\ding{51}}}\hspace{0.006\linewidth}\parbox[t]{0.10\linewidth}{\centering\strut +0\%}\hspace{0.006\linewidth}\parbox[t]{0.115\linewidth}{\centering\strut $6.818\times 10^{-6}$}\hspace{0.006\linewidth}\parbox[t]{0.671\linewidth}{warm\_powell\_xt1e6 seed2021}}}}\par\smallskip
\noindent\parbox[t]{0.05\linewidth}{\centering\strut 22}\hspace{0.006\linewidth}\parbox[t]{0.04\linewidth}{\centering\strut \raisebox{0.08ex}{\ding{51}}}\hspace{0.006\linewidth}\parbox[t]{0.10\linewidth}{\centering\strut +0\%}\hspace{0.006\linewidth}\parbox[t]{0.115\linewidth}{\centering\strut $6.818\times 10^{-6}$}\hspace{0.006\linewidth}\parbox[t]{0.671\linewidth}{warm\_powell\_xt1e7 seed2022}\par\smallskip
\noindent{\setlength{\fboxsep}{0pt}\colorbox{black!4}{\parbox{\linewidth}{\parbox[t]{0.05\linewidth}{\centering\strut 23}\hspace{0.006\linewidth}\parbox[t]{0.04\linewidth}{\centering\strut \raisebox{0.08ex}{\ding{51}}}\hspace{0.006\linewidth}\parbox[t]{0.10\linewidth}{\centering\strut +0\%}\hspace{0.006\linewidth}\parbox[t]{0.115\linewidth}{\centering\strut $6.818\times 10^{-6}$}\hspace{0.006\linewidth}\parbox[t]{0.671\linewidth}{warm\_cobyla\_r20 seed2023}}}}\par\smallskip
\noindent\parbox[t]{0.05\linewidth}{\centering\strut 24}\hspace{0.006\linewidth}\parbox[t]{0.04\linewidth}{\centering\strut \raisebox{0.08ex}{\ding{51}}}\hspace{0.006\linewidth}\parbox[t]{0.10\linewidth}{\centering\strut +0\%}\hspace{0.006\linewidth}\parbox[t]{0.115\linewidth}{\centering\strut $6.818\times 10^{-6}$}\hspace{0.006\linewidth}\parbox[t]{0.671\linewidth}{warm\_cobyla\_r10 seed2024}\par\smallskip
\noindent{\setlength{\fboxsep}{0pt}\colorbox{black!4}{\parbox{\linewidth}{\parbox[t]{0.05\linewidth}{\centering\strut 25}\hspace{0.006\linewidth}\parbox[t]{0.04\linewidth}{\centering\strut \raisebox{0.08ex}{\ding{51}}}\hspace{0.006\linewidth}\parbox[t]{0.10\linewidth}{\centering\strut +0\%}\hspace{0.006\linewidth}\parbox[t]{0.115\linewidth}{\centering\strut $6.818\times 10^{-6}$}\hspace{0.006\linewidth}\parbox[t]{0.671\linewidth}{warm\_cobyla\_r5 seed2025}}}}\par\smallskip
\noindent\parbox[t]{0.05\linewidth}{\centering\strut 26}\hspace{0.006\linewidth}\parbox[t]{0.04\linewidth}{\centering\strut \raisebox{0.08ex}{\ding{51}}}\hspace{0.006\linewidth}\parbox[t]{0.10\linewidth}{\centering\strut +72.06\%}\hspace{0.006\linewidth}\parbox[t]{0.115\linewidth}{\centering\strut $1.905\times 10^{-6}$}\hspace{0.006\linewidth}\parbox[t]{0.671\linewidth}{warm\_cobyla\_r2 seed2026}\par\smallskip
\noindent{\setlength{\fboxsep}{0pt}\colorbox{black!4}{\parbox{\linewidth}{\parbox[t]{0.05\linewidth}{\centering\strut 27}\hspace{0.006\linewidth}\parbox[t]{0.04\linewidth}{\centering\strut \raisebox{0.08ex}{\ding{51}}}\hspace{0.006\linewidth}\parbox[t]{0.10\linewidth}{\centering\strut +2.73\%}\hspace{0.006\linewidth}\parbox[t]{0.115\linewidth}{\centering\strut $1.853\times 10^{-6}$}\hspace{0.006\linewidth}\parbox[t]{0.671\linewidth}{jitter\_powell\_s5e8 seed2027}}}}\par\smallskip
\noindent\parbox[t]{0.05\linewidth}{\centering\strut 28}\hspace{0.006\linewidth}\parbox[t]{0.04\linewidth}{\centering\strut \raisebox{0.08ex}{\ding{55}}}\hspace{0.006\linewidth}\parbox[t]{0.10\linewidth}{\centering\strut -0.002591\%}\hspace{0.006\linewidth}\parbox[t]{0.115\linewidth}{\centering\strut $1.853\times 10^{-6}$}\hspace{0.006\linewidth}\parbox[t]{0.671\linewidth}{jitter\_cobyla\_s1e7\_r10 seed2028}\par\smallskip
\noindent{\setlength{\fboxsep}{0pt}\colorbox{black!4}{\parbox{\linewidth}{\parbox[t]{0.05\linewidth}{\centering\strut 29}\hspace{0.006\linewidth}\parbox[t]{0.04\linewidth}{\centering\strut \raisebox{0.08ex}{\ding{51}}}\hspace{0.006\linewidth}\parbox[t]{0.10\linewidth}{\centering\strut +9.02\%}\hspace{0.006\linewidth}\parbox[t]{0.115\linewidth}{\centering\strut $1.686\times 10^{-6}$}\hspace{0.006\linewidth}\parbox[t]{0.671\linewidth}{jitter\_cobyla\_s5e7\_r5 seed2029}}}}\par\smallskip
\noindent\parbox[t]{0.05\linewidth}{\centering\strut 30}\hspace{0.006\linewidth}\parbox[t]{0.04\linewidth}{\centering\strut \raisebox{0.08ex}{\ding{51}}}\hspace{0.006\linewidth}\parbox[t]{0.10\linewidth}{\centering\strut +22.41\%}\hspace{0.006\linewidth}\parbox[t]{0.115\linewidth}{\centering\strut $1.308\times 10^{-6}$}\hspace{0.006\linewidth}\parbox[t]{0.671\linewidth}{jitter\_cobyla\_s1e6\_r2 seed2030}\par\smallskip
\noindent{\setlength{\fboxsep}{0pt}\colorbox{black!4}{\parbox{\linewidth}{\parbox[t]{0.05\linewidth}{\centering\strut 31}\hspace{0.006\linewidth}\parbox[t]{0.04\linewidth}{\centering\strut \raisebox{0.08ex}{\ding{51}}}\hspace{0.006\linewidth}\parbox[t]{0.10\linewidth}{\centering\strut +0.729\%}\hspace{0.006\linewidth}\parbox[t]{0.115\linewidth}{\centering\strut $1.298\times 10^{-6}$}\hspace{0.006\linewidth}\parbox[t]{0.671\linewidth}{warm\_powell\_xt1e6 seed2031}}}}\par\smallskip
\noindent\parbox[t]{0.05\linewidth}{\centering\strut 32}\hspace{0.006\linewidth}\parbox[t]{0.04\linewidth}{\centering\strut \raisebox{0.08ex}{\ding{55}}}\hspace{0.006\linewidth}\parbox[t]{0.10\linewidth}{\centering\strut +0\%}\hspace{0.006\linewidth}\parbox[t]{0.115\linewidth}{\centering\strut $1.298\times 10^{-6}$}\hspace{0.006\linewidth}\parbox[t]{0.671\linewidth}{warm\_powell\_xt1e7 seed2032}\par\smallskip
\noindent{\setlength{\fboxsep}{0pt}\colorbox{black!4}{\parbox{\linewidth}{\parbox[t]{0.05\linewidth}{\centering\strut 33}\hspace{0.006\linewidth}\parbox[t]{0.04\linewidth}{\centering\strut \raisebox{0.08ex}{\ding{55}}}\hspace{0.006\linewidth}\parbox[t]{0.10\linewidth}{\centering\strut +0\%}\hspace{0.006\linewidth}\parbox[t]{0.115\linewidth}{\centering\strut $1.298\times 10^{-6}$}\hspace{0.006\linewidth}\parbox[t]{0.671\linewidth}{warm\_cobyla\_r20 seed2033}}}}\par\smallskip
\noindent\parbox[t]{0.05\linewidth}{\centering\strut 34}\hspace{0.006\linewidth}\parbox[t]{0.04\linewidth}{\centering\strut \raisebox{0.08ex}{\ding{55}}}\hspace{0.006\linewidth}\parbox[t]{0.10\linewidth}{\centering\strut +0\%}\hspace{0.006\linewidth}\parbox[t]{0.115\linewidth}{\centering\strut $1.298\times 10^{-6}$}\hspace{0.006\linewidth}\parbox[t]{0.671\linewidth}{warm\_cobyla\_r10 seed2034}\par\smallskip
\noindent{\setlength{\fboxsep}{0pt}\colorbox{black!4}{\parbox{\linewidth}{\parbox[t]{0.05\linewidth}{\centering\strut 35}\hspace{0.006\linewidth}\parbox[t]{0.04\linewidth}{\centering\strut \raisebox{0.08ex}{\ding{55}}}\hspace{0.006\linewidth}\parbox[t]{0.10\linewidth}{\centering\strut +0\%}\hspace{0.006\linewidth}\parbox[t]{0.115\linewidth}{\centering\strut $1.298\times 10^{-6}$}\hspace{0.006\linewidth}\parbox[t]{0.671\linewidth}{warm\_cobyla\_r5 seed2035}}}}\par\smallskip
\noindent\parbox[t]{0.05\linewidth}{\centering\strut 36}\hspace{0.006\linewidth}\parbox[t]{0.04\linewidth}{\centering\strut \raisebox{0.08ex}{\ding{51}}}\hspace{0.006\linewidth}\parbox[t]{0.10\linewidth}{\centering\strut +7.58\%}\hspace{0.006\linewidth}\parbox[t]{0.115\linewidth}{\centering\strut $1.2\times 10^{-6}$}\hspace{0.006\linewidth}\parbox[t]{0.671\linewidth}{warm\_cobyla\_r2 seed2036}\par\smallskip
\noindent{\setlength{\fboxsep}{0pt}\colorbox{black!4}{\parbox{\linewidth}{\parbox[t]{0.05\linewidth}{\centering\strut 37}\hspace{0.006\linewidth}\parbox[t]{0.04\linewidth}{\centering\strut \raisebox{0.08ex}{\ding{51}}}\hspace{0.006\linewidth}\parbox[t]{0.10\linewidth}{\centering\strut +0.031\%}\hspace{0.006\linewidth}\parbox[t]{0.115\linewidth}{\centering\strut $1.2\times 10^{-6}$}\hspace{0.006\linewidth}\parbox[t]{0.671\linewidth}{jitter\_powell\_s5e8 seed2037}}}}\par\smallskip
\noindent\parbox[t]{0.05\linewidth}{\centering\strut 38}\hspace{0.006\linewidth}\parbox[t]{0.04\linewidth}{\centering\strut \raisebox{0.08ex}{\ding{51}}}\hspace{0.006\linewidth}\parbox[t]{0.10\linewidth}{\centering\strut +0.002834\%}\hspace{0.006\linewidth}\parbox[t]{0.115\linewidth}{\centering\strut $1.2\times 10^{-6}$}\hspace{0.006\linewidth}\parbox[t]{0.671\linewidth}{jitter\_cobyla\_s1e7\_r10 seed2038}\par\smallskip
\noindent{\setlength{\fboxsep}{0pt}\colorbox{black!4}{\parbox{\linewidth}{\parbox[t]{0.05\linewidth}{\centering\strut 39}\hspace{0.006\linewidth}\parbox[t]{0.04\linewidth}{\centering\strut \raisebox{0.08ex}{\ding{55}}}\hspace{0.006\linewidth}\parbox[t]{0.10\linewidth}{\centering\strut -0.006835\%}\hspace{0.006\linewidth}\parbox[t]{0.115\linewidth}{\centering\strut $1.2\times 10^{-6}$}\hspace{0.006\linewidth}\parbox[t]{0.671\linewidth}{jitter\_cobyla\_s5e7\_r5 seed2039}}}}\par\smallskip
\noindent\parbox[t]{0.05\linewidth}{\centering\strut 40}\hspace{0.006\linewidth}\parbox[t]{0.04\linewidth}{\centering\strut \raisebox{0.08ex}{\ding{51}}}\hspace{0.006\linewidth}\parbox[t]{0.10\linewidth}{\centering\strut +0.0724\%}\hspace{0.006\linewidth}\parbox[t]{0.115\linewidth}{\centering\strut $1.199\times 10^{-6}$}\hspace{0.006\linewidth}\parbox[t]{0.671\linewidth}{jitter\_cobyla\_s1e6\_r2 seed2040}\par\smallskip
\noindent{\setlength{\fboxsep}{0pt}\colorbox{black!4}{\parbox{\linewidth}{\parbox[t]{0.05\linewidth}{\centering\strut 41}\hspace{0.006\linewidth}\parbox[t]{0.04\linewidth}{\centering\strut \raisebox{0.08ex}{\ding{51}}}\hspace{0.006\linewidth}\parbox[t]{0.10\linewidth}{\centering\strut +0.0641\%}\hspace{0.006\linewidth}\parbox[t]{0.115\linewidth}{\centering\strut $1.198\times 10^{-6}$}\hspace{0.006\linewidth}\parbox[t]{0.671\linewidth}{jitter\_cobyla\_s1e6\_r2 seed3041}}}}\par\smallskip
\noindent\parbox[t]{0.05\linewidth}{\centering\strut 42}\hspace{0.006\linewidth}\parbox[t]{0.04\linewidth}{\centering\strut \raisebox{0.08ex}{\ding{51}}}\hspace{0.006\linewidth}\parbox[t]{0.10\linewidth}{\centering\strut +0.007929\%}\hspace{0.006\linewidth}\parbox[t]{0.115\linewidth}{\centering\strut $1.198\times 10^{-6}$}\hspace{0.006\linewidth}\parbox[t]{0.671\linewidth}{jitter\_cobyla\_s5e7\_r2 seed3042}\par\smallskip
\noindent{\setlength{\fboxsep}{0pt}\colorbox{black!4}{\parbox{\linewidth}{\parbox[t]{0.05\linewidth}{\centering\strut 43}\hspace{0.006\linewidth}\parbox[t]{0.04\linewidth}{\centering\strut \raisebox{0.08ex}{\ding{51}}}\hspace{0.006\linewidth}\parbox[t]{0.10\linewidth}{\centering\strut +15.62\%}\hspace{0.006\linewidth}\parbox[t]{0.115\linewidth}{\centering\strut $1.011\times 10^{-6}$}\hspace{0.006\linewidth}\parbox[t]{0.671\linewidth}{jitter\_cobyla\_s1e7\_r1 seed3043}}}}\par\smallskip
\noindent\parbox[t]{0.05\linewidth}{\centering\strut 44}\hspace{0.006\linewidth}\parbox[t]{0.04\linewidth}{\centering\strut \raisebox{0.08ex}{\ding{51}}}\hspace{0.006\linewidth}\parbox[t]{0.10\linewidth}{\centering\strut +0.6276\%}\hspace{0.006\linewidth}\parbox[t]{0.115\linewidth}{\centering\strut $1.004\times 10^{-6}$}\hspace{0.006\linewidth}\parbox[t]{0.671\linewidth}{warm\_cobyla\_r2 seed3044}\par\smallskip
\noindent{\setlength{\fboxsep}{0pt}\colorbox{black!4}{\parbox{\linewidth}{\parbox[t]{0.05\linewidth}{\centering\strut 45}\hspace{0.006\linewidth}\parbox[t]{0.04\linewidth}{\centering\strut \raisebox{0.08ex}{\ding{51}}}\hspace{0.006\linewidth}\parbox[t]{0.10\linewidth}{\centering\strut +2.3\%}\hspace{0.006\linewidth}\parbox[t]{0.115\linewidth}{\centering\strut $9.814\times 10^{-7}$}\hspace{0.006\linewidth}\parbox[t]{0.671\linewidth}{warm\_cobyla\_r1 seed3045}}}}\par\smallskip
\noindent\parbox[t]{0.05\linewidth}{\centering\strut 46}\hspace{0.006\linewidth}\parbox[t]{0.04\linewidth}{\centering\strut \raisebox{0.08ex}{\ding{51}}}\hspace{0.006\linewidth}\parbox[t]{0.10\linewidth}{\centering\strut +0.2143\%}\hspace{0.006\linewidth}\parbox[t]{0.115\linewidth}{\centering\strut $9.793\times 10^{-7}$}\hspace{0.006\linewidth}\parbox[t]{0.671\linewidth}{jitter\_powell\_s5e8 seed3046}\par\smallskip
\noindent{\setlength{\fboxsep}{0pt}\colorbox{black!4}{\parbox{\linewidth}{\parbox[t]{0.05\linewidth}{\centering\strut 47}\hspace{0.006\linewidth}\parbox[t]{0.04\linewidth}{\centering\strut \raisebox{0.08ex}{\ding{51}}}\hspace{0.006\linewidth}\parbox[t]{0.10\linewidth}{\centering\strut +0\%}\hspace{0.006\linewidth}\parbox[t]{0.115\linewidth}{\centering\strut $9.793\times 10^{-7}$}\hspace{0.006\linewidth}\parbox[t]{0.671\linewidth}{warm\_powell\_xt1e6 seed3047}}}}\par\smallskip
\noindent\parbox[t]{0.05\linewidth}{\centering\strut 48}\hspace{0.006\linewidth}\parbox[t]{0.04\linewidth}{\centering\strut \raisebox{0.08ex}{\ding{51}}}\hspace{0.006\linewidth}\parbox[t]{0.10\linewidth}{\centering\strut +0\%}\hspace{0.006\linewidth}\parbox[t]{0.115\linewidth}{\centering\strut $9.793\times 10^{-7}$}\hspace{0.006\linewidth}\parbox[t]{0.671\linewidth}{warm\_powell\_xt1e7 seed3048}\par\smallskip
\noindent{\setlength{\fboxsep}{0pt}\colorbox{black!4}{\parbox{\linewidth}{\parbox[t]{0.05\linewidth}{\centering\strut 49}\hspace{0.006\linewidth}\parbox[t]{0.04\linewidth}{\centering\strut \raisebox{0.08ex}{\ding{51}}}\hspace{0.006\linewidth}\parbox[t]{0.10\linewidth}{\centering\strut +0.004646\%}\hspace{0.006\linewidth}\parbox[t]{0.115\linewidth}{\centering\strut $9.792\times 10^{-7}$}\hspace{0.006\linewidth}\parbox[t]{0.671\linewidth}{jitter\_cobyla\_s5e8\_r1 seed3049}}}}\par\smallskip
\noindent\parbox[t]{0.05\linewidth}{\centering\strut 50}\hspace{0.006\linewidth}\parbox[t]{0.04\linewidth}{\centering\strut \raisebox{0.08ex}{\ding{51}}}\hspace{0.006\linewidth}\parbox[t]{0.10\linewidth}{\centering\strut +15.93\%}\hspace{0.006\linewidth}\parbox[t]{0.115\linewidth}{\centering\strut $8.232\times 10^{-7}$}\hspace{0.006\linewidth}\parbox[t]{0.671\linewidth}{jitter\_cobyla\_s2e7\_r5e4 seed3050}\par\smallskip
\noindent{\setlength{\fboxsep}{0pt}\colorbox{black!4}{\parbox{\linewidth}{\parbox[t]{0.05\linewidth}{\centering\strut 51}\hspace{0.006\linewidth}\parbox[t]{0.04\linewidth}{\centering\strut \raisebox{0.08ex}{\ding{51}}}\hspace{0.006\linewidth}\parbox[t]{0.10\linewidth}{\centering\strut +0.0335\%}\hspace{0.006\linewidth}\parbox[t]{0.115\linewidth}{\centering\strut $8.229\times 10^{-7}$}\hspace{0.006\linewidth}\parbox[t]{0.671\linewidth}{jitter\_cobyla\_s1e6\_r2 seed3051}}}}\par\smallskip
\noindent\parbox[t]{0.05\linewidth}{\centering\strut 52}\hspace{0.006\linewidth}\parbox[t]{0.04\linewidth}{\centering\strut \raisebox{0.08ex}{\ding{55}}}\hspace{0.006\linewidth}\parbox[t]{0.10\linewidth}{\centering\strut -0.0372\%}\hspace{0.006\linewidth}\parbox[t]{0.115\linewidth}{\centering\strut $8.232\times 10^{-7}$}\hspace{0.006\linewidth}\parbox[t]{0.671\linewidth}{jitter\_cobyla\_s5e7\_r2 seed3052}\par\smallskip
\noindent{\setlength{\fboxsep}{0pt}\colorbox{black!4}{\parbox{\linewidth}{\parbox[t]{0.05\linewidth}{\centering\strut 53}\hspace{0.006\linewidth}\parbox[t]{0.04\linewidth}{\centering\strut \raisebox{0.08ex}{\ding{51}}}\hspace{0.006\linewidth}\parbox[t]{0.10\linewidth}{\centering\strut +2.24\%}\hspace{0.006\linewidth}\parbox[t]{0.115\linewidth}{\centering\strut $8.048\times 10^{-7}$}\hspace{0.006\linewidth}\parbox[t]{0.671\linewidth}{jitter\_cobyla\_s1e7\_r1 seed3053}}}}\par\smallskip
\noindent\parbox[t]{0.05\linewidth}{\centering\strut 54}\hspace{0.006\linewidth}\parbox[t]{0.04\linewidth}{\centering\strut \raisebox{0.08ex}{\ding{55}}}\hspace{0.006\linewidth}\parbox[t]{0.10\linewidth}{\centering\strut +0\%}\hspace{0.006\linewidth}\parbox[t]{0.115\linewidth}{\centering\strut $8.048\times 10^{-7}$}\hspace{0.006\linewidth}\parbox[t]{0.671\linewidth}{warm\_cobyla\_r2 seed3054}\par\smallskip
\noindent{\setlength{\fboxsep}{0pt}\colorbox{black!4}{\parbox{\linewidth}{\parbox[t]{0.05\linewidth}{\centering\strut 55}\hspace{0.006\linewidth}\parbox[t]{0.04\linewidth}{\centering\strut \raisebox{0.08ex}{\ding{55}}}\hspace{0.006\linewidth}\parbox[t]{0.10\linewidth}{\centering\strut +0\%}\hspace{0.006\linewidth}\parbox[t]{0.115\linewidth}{\centering\strut $8.048\times 10^{-7}$}\hspace{0.006\linewidth}\parbox[t]{0.671\linewidth}{warm\_cobyla\_r1 seed3055}}}}\par\smallskip
\noindent\parbox[t]{0.05\linewidth}{\centering\strut 56}\hspace{0.006\linewidth}\parbox[t]{0.04\linewidth}{\centering\strut \raisebox{0.08ex}{\ding{51}}}\hspace{0.006\linewidth}\parbox[t]{0.10\linewidth}{\centering\strut +0.6617\%}\hspace{0.006\linewidth}\parbox[t]{0.115\linewidth}{\centering\strut $7.995\times 10^{-7}$}\hspace{0.006\linewidth}\parbox[t]{0.671\linewidth}{jitter\_powell\_s5e8 seed3056}\par\smallskip
\noindent{\setlength{\fboxsep}{0pt}\colorbox{black!4}{\parbox{\linewidth}{\parbox[t]{0.05\linewidth}{\centering\strut 57}\hspace{0.006\linewidth}\parbox[t]{0.04\linewidth}{\centering\strut \raisebox{0.08ex}{\ding{55}}}\hspace{0.006\linewidth}\parbox[t]{0.10\linewidth}{\centering\strut +0\%}\hspace{0.006\linewidth}\parbox[t]{0.115\linewidth}{\centering\strut $7.995\times 10^{-7}$}\hspace{0.006\linewidth}\parbox[t]{0.671\linewidth}{warm\_powell\_xt1e6 seed3057}}}}\par\smallskip
\noindent\parbox[t]{0.05\linewidth}{\centering\strut 58}\hspace{0.006\linewidth}\parbox[t]{0.04\linewidth}{\centering\strut \raisebox{0.08ex}{\ding{55}}}\hspace{0.006\linewidth}\parbox[t]{0.10\linewidth}{\centering\strut +0\%}\hspace{0.006\linewidth}\parbox[t]{0.115\linewidth}{\centering\strut $7.995\times 10^{-7}$}\hspace{0.006\linewidth}\parbox[t]{0.671\linewidth}{warm\_powell\_xt1e7 seed3058}\par\smallskip
\noindent{\setlength{\fboxsep}{0pt}\colorbox{black!4}{\parbox{\linewidth}{\parbox[t]{0.05\linewidth}{\centering\strut 59}\hspace{0.006\linewidth}\parbox[t]{0.04\linewidth}{\centering\strut \raisebox{0.08ex}{\ding{55}}}\hspace{0.006\linewidth}\parbox[t]{0.10\linewidth}{\centering\strut -0.002089\%}\hspace{0.006\linewidth}\parbox[t]{0.115\linewidth}{\centering\strut $7.995\times 10^{-7}$}\hspace{0.006\linewidth}\parbox[t]{0.671\linewidth}{jitter\_cobyla\_s5e8\_r1 seed3059}}}}\par\smallskip
\noindent\parbox[t]{0.05\linewidth}{\centering\strut 60}\hspace{0.006\linewidth}\parbox[t]{0.04\linewidth}{\centering\strut \raisebox{0.08ex}{\ding{51}}}\hspace{0.006\linewidth}\parbox[t]{0.10\linewidth}{\centering\strut +3.19\%}\hspace{0.006\linewidth}\parbox[t]{0.115\linewidth}{\centering\strut $7.74\times 10^{-7}$}\hspace{0.006\linewidth}\parbox[t]{0.671\linewidth}{jitter\_cobyla\_s2e7\_r5e4 seed3060}\par\smallskip
\noindent{\setlength{\fboxsep}{0pt}\colorbox{black!4}{\parbox{\linewidth}{\parbox[t]{0.05\linewidth}{\centering\strut 61}\hspace{0.006\linewidth}\parbox[t]{0.04\linewidth}{\centering\strut \raisebox{0.08ex}{\ding{51}}}\hspace{0.006\linewidth}\parbox[t]{0.10\linewidth}{\centering\strut +5.55\%}\hspace{0.006\linewidth}\parbox[t]{0.115\linewidth}{\centering\strut $7.311\times 10^{-7}$}\hspace{0.006\linewidth}\parbox[t]{0.671\linewidth}{jitter\_cobyla\_s2e7\_r5e4 seed4061}}}}\par\smallskip
\noindent\parbox[t]{0.05\linewidth}{\centering\strut 62}\hspace{0.006\linewidth}\parbox[t]{0.04\linewidth}{\centering\strut \raisebox{0.08ex}{\ding{51}}}\hspace{0.006\linewidth}\parbox[t]{0.10\linewidth}{\centering\strut +0.1697\%}\hspace{0.006\linewidth}\parbox[t]{0.115\linewidth}{\centering\strut $7.298\times 10^{-7}$}\hspace{0.006\linewidth}\parbox[t]{0.671\linewidth}{jitter\_cobyla\_s1e7\_r5e4 seed4062}\par\smallskip
\noindent{\setlength{\fboxsep}{0pt}\colorbox{black!4}{\parbox{\linewidth}{\parbox[t]{0.05\linewidth}{\centering\strut 63}\hspace{0.006\linewidth}\parbox[t]{0.04\linewidth}{\centering\strut \raisebox{0.08ex}{\ding{51}}}\hspace{0.006\linewidth}\parbox[t]{0.10\linewidth}{\centering\strut +4.51\%}\hspace{0.006\linewidth}\parbox[t]{0.115\linewidth}{\centering\strut $6.969\times 10^{-7}$}\hspace{0.006\linewidth}\parbox[t]{0.671\linewidth}{jitter\_cobyla\_s5e8\_r2e4 seed4063}}}}\par\smallskip
\noindent\parbox[t]{0.05\linewidth}{\centering\strut 64}\hspace{0.006\linewidth}\parbox[t]{0.04\linewidth}{\centering\strut \raisebox{0.08ex}{\ding{51}}}\hspace{0.006\linewidth}\parbox[t]{0.10\linewidth}{\centering\strut +0.5292\%}\hspace{0.006\linewidth}\parbox[t]{0.115\linewidth}{\centering\strut $6.933\times 10^{-7}$}\hspace{0.006\linewidth}\parbox[t]{0.671\linewidth}{warm\_cobyla\_r5e4 seed4064}\par\smallskip
\noindent{\setlength{\fboxsep}{0pt}\colorbox{black!4}{\parbox{\linewidth}{\parbox[t]{0.05\linewidth}{\centering\strut 65}\hspace{0.006\linewidth}\parbox[t]{0.04\linewidth}{\centering\strut \raisebox{0.08ex}{\ding{51}}}\hspace{0.006\linewidth}\parbox[t]{0.10\linewidth}{\centering\strut +2.46\%}\hspace{0.006\linewidth}\parbox[t]{0.115\linewidth}{\centering\strut $6.762\times 10^{-7}$}\hspace{0.006\linewidth}\parbox[t]{0.671\linewidth}{warm\_cobyla\_r1e4 seed4065}}}}\par\smallskip
\noindent\parbox[t]{0.05\linewidth}{\centering\strut 66}\hspace{0.006\linewidth}\parbox[t]{0.04\linewidth}{\centering\strut \raisebox{0.08ex}{\ding{51}}}\hspace{0.006\linewidth}\parbox[t]{0.10\linewidth}{\centering\strut +0.0511\%}\hspace{0.006\linewidth}\parbox[t]{0.115\linewidth}{\centering\strut $6.758\times 10^{-7}$}\hspace{0.006\linewidth}\parbox[t]{0.671\linewidth}{jitter\_powell\_s1e8 seed4066}\par\smallskip
\noindent{\setlength{\fboxsep}{0pt}\colorbox{black!4}{\parbox{\linewidth}{\parbox[t]{0.05\linewidth}{\centering\strut 67}\hspace{0.006\linewidth}\parbox[t]{0.04\linewidth}{\centering\strut \raisebox{0.08ex}{\ding{55}}}\hspace{0.006\linewidth}\parbox[t]{0.10\linewidth}{\centering\strut +0\%}\hspace{0.006\linewidth}\parbox[t]{0.115\linewidth}{\centering\strut $6.758\times 10^{-7}$}\hspace{0.006\linewidth}\parbox[t]{0.671\linewidth}{warm\_powell\_xt1e7 seed4067}}}}\par\smallskip
\noindent\parbox[t]{0.05\linewidth}{\centering\strut 68}\hspace{0.006\linewidth}\parbox[t]{0.04\linewidth}{\centering\strut \raisebox{0.08ex}{\ding{55}}}\hspace{0.006\linewidth}\parbox[t]{0.10\linewidth}{\centering\strut +0\%}\hspace{0.006\linewidth}\parbox[t]{0.115\linewidth}{\centering\strut $6.758\times 10^{-7}$}\hspace{0.006\linewidth}\parbox[t]{0.671\linewidth}{warm\_powell\_xt1e8 seed4068}\par\smallskip
\noindent{\setlength{\fboxsep}{0pt}\colorbox{black!4}{\parbox{\linewidth}{\parbox[t]{0.05\linewidth}{\centering\strut 69}\hspace{0.006\linewidth}\parbox[t]{0.04\linewidth}{\centering\strut \raisebox{0.08ex}{\ding{51}}}\hspace{0.006\linewidth}\parbox[t]{0.10\linewidth}{\centering\strut +0.4211\%}\hspace{0.006\linewidth}\parbox[t]{0.115\linewidth}{\centering\strut $6.73\times 10^{-7}$}\hspace{0.006\linewidth}\parbox[t]{0.671\linewidth}{jitter\_cobyla\_s5e9\_r1e4 seed4069}}}}\par\smallskip
\noindent\parbox[t]{0.05\linewidth}{\centering\strut 70}\hspace{0.006\linewidth}\parbox[t]{0.04\linewidth}{\centering\strut \raisebox{0.08ex}{\ding{51}}}\hspace{0.006\linewidth}\parbox[t]{0.10\linewidth}{\centering\strut +0.000609\%}\hspace{0.006\linewidth}\parbox[t]{0.115\linewidth}{\centering\strut $6.73\times 10^{-7}$}\hspace{0.006\linewidth}\parbox[t]{0.671\linewidth}{jitter\_cobyla\_s1e7\_r2e4 seed4070}\par\smallskip
\noindent{\setlength{\fboxsep}{0pt}\colorbox{black!4}{\parbox{\linewidth}{\parbox[t]{0.05\linewidth}{\centering\strut 71}\hspace{0.006\linewidth}\parbox[t]{0.04\linewidth}{\centering\strut \raisebox{0.08ex}{\ding{51}}}\hspace{0.006\linewidth}\parbox[t]{0.10\linewidth}{\centering\strut +0.0207\%}\hspace{0.006\linewidth}\parbox[t]{0.115\linewidth}{\centering\strut $6.728\times 10^{-7}$}\hspace{0.006\linewidth}\parbox[t]{0.671\linewidth}{jitter\_cobyla\_s2e7\_r5e4 seed4071}}}}\par\smallskip
\noindent\parbox[t]{0.05\linewidth}{\centering\strut 72}\hspace{0.006\linewidth}\parbox[t]{0.04\linewidth}{\centering\strut \raisebox{0.08ex}{\ding{51}}}\hspace{0.006\linewidth}\parbox[t]{0.10\linewidth}{\centering\strut +0.000431\%}\hspace{0.006\linewidth}\parbox[t]{0.115\linewidth}{\centering\strut $6.728\times 10^{-7}$}\hspace{0.006\linewidth}\parbox[t]{0.671\linewidth}{jitter\_cobyla\_s1e7\_r5e4 seed4072}\par\smallskip
\noindent{\setlength{\fboxsep}{0pt}\colorbox{black!4}{\parbox{\linewidth}{\parbox[t]{0.05\linewidth}{\centering\strut 73}\hspace{0.006\linewidth}\parbox[t]{0.04\linewidth}{\centering\strut \raisebox{0.08ex}{\ding{55}}}\hspace{0.006\linewidth}\parbox[t]{0.10\linewidth}{\centering\strut -0.000178\%}\hspace{0.006\linewidth}\parbox[t]{0.115\linewidth}{\centering\strut $6.728\times 10^{-7}$}\hspace{0.006\linewidth}\parbox[t]{0.671\linewidth}{jitter\_cobyla\_s5e8\_r2e4 seed4073}}}}\par\smallskip
\noindent\parbox[t]{0.05\linewidth}{\centering\strut 74}\hspace{0.006\linewidth}\parbox[t]{0.04\linewidth}{\centering\strut \raisebox{0.08ex}{\ding{51}}}\hspace{0.006\linewidth}\parbox[t]{0.10\linewidth}{\centering\strut +0.000178\%}\hspace{0.006\linewidth}\parbox[t]{0.115\linewidth}{\centering\strut $6.728\times 10^{-7}$}\hspace{0.006\linewidth}\parbox[t]{0.671\linewidth}{warm\_cobyla\_r5e4 seed4074}\par\smallskip
\noindent{\setlength{\fboxsep}{0pt}\colorbox{black!4}{\parbox{\linewidth}{\parbox[t]{0.05\linewidth}{\centering\strut 75}\hspace{0.006\linewidth}\parbox[t]{0.04\linewidth}{\centering\strut \raisebox{0.08ex}{\ding{51}}}\hspace{0.006\linewidth}\parbox[t]{0.10\linewidth}{\centering\strut +0.016\%}\hspace{0.006\linewidth}\parbox[t]{0.115\linewidth}{\centering\strut $6.727\times 10^{-7}$}\hspace{0.006\linewidth}\parbox[t]{0.671\linewidth}{warm\_cobyla\_r1e4 seed4075}}}}\par\smallskip
\noindent\parbox[t]{0.05\linewidth}{\centering\strut 76}\hspace{0.006\linewidth}\parbox[t]{0.04\linewidth}{\centering\strut \raisebox{0.08ex}{\ding{51}}}\hspace{0.006\linewidth}\parbox[t]{0.10\linewidth}{\centering\strut +0\%}\hspace{0.006\linewidth}\parbox[t]{0.115\linewidth}{\centering\strut $6.727\times 10^{-7}$}\hspace{0.006\linewidth}\parbox[t]{0.671\linewidth}{jitter\_powell\_s1e8 seed4076}\par\smallskip
\noindent{\setlength{\fboxsep}{0pt}\colorbox{black!4}{\parbox{\linewidth}{\parbox[t]{0.05\linewidth}{\centering\strut 77}\hspace{0.006\linewidth}\parbox[t]{0.04\linewidth}{\centering\strut \raisebox{0.08ex}{\ding{51}}}\hspace{0.006\linewidth}\parbox[t]{0.10\linewidth}{\centering\strut +0.002676\%}\hspace{0.006\linewidth}\parbox[t]{0.115\linewidth}{\centering\strut $6.727\times 10^{-7}$}\hspace{0.006\linewidth}\parbox[t]{0.671\linewidth}{warm\_powell\_xt1e7 seed4077}}}}\par\smallskip
\noindent\parbox[t]{0.05\linewidth}{\centering\strut 78}\hspace{0.006\linewidth}\parbox[t]{0.04\linewidth}{\centering\strut \raisebox{0.08ex}{\ding{51}}}\hspace{0.006\linewidth}\parbox[t]{0.10\linewidth}{\centering\strut +0\%}\hspace{0.006\linewidth}\parbox[t]{0.115\linewidth}{\centering\strut $6.727\times 10^{-7}$}\hspace{0.006\linewidth}\parbox[t]{0.671\linewidth}{warm\_powell\_xt1e8 seed4078}\par\smallskip
\noindent{\setlength{\fboxsep}{0pt}\colorbox{black!4}{\parbox{\linewidth}{\parbox[t]{0.05\linewidth}{\centering\strut 79}\hspace{0.006\linewidth}\parbox[t]{0.04\linewidth}{\centering\strut \raisebox{0.08ex}{\ding{55}}}\hspace{0.006\linewidth}\parbox[t]{0.10\linewidth}{\centering\strut +0\%}\hspace{0.006\linewidth}\parbox[t]{0.115\linewidth}{\centering\strut $6.727\times 10^{-7}$}\hspace{0.006\linewidth}\parbox[t]{0.671\linewidth}{jitter\_cobyla\_s5e9\_r1e4 seed4079}}}}\par\smallskip
\noindent\parbox[t]{0.05\linewidth}{\centering\strut 80}\hspace{0.006\linewidth}\parbox[t]{0.04\linewidth}{\centering\strut \raisebox{0.08ex}{\ding{51}}}\hspace{0.006\linewidth}\parbox[t]{0.10\linewidth}{\centering\strut +0.00162\%}\hspace{0.006\linewidth}\parbox[t]{0.115\linewidth}{\centering\strut $6.727\times 10^{-7}$}\hspace{0.006\linewidth}\parbox[t]{0.671\linewidth}{jitter\_cobyla\_s1e7\_r2e4 seed4080}\par\smallskip
\noindent{\setlength{\fboxsep}{0pt}\colorbox{black!4}{\parbox{\linewidth}{\parbox[t]{0.05\linewidth}{\centering\strut 81}\hspace{0.006\linewidth}\parbox[t]{0.04\linewidth}{\centering\strut \raisebox{0.08ex}{\ding{55}}}\hspace{0.006\linewidth}\parbox[t]{0.10\linewidth}{\centering\strut +0\%}\hspace{0.006\linewidth}\parbox[t]{0.115\linewidth}{\centering\strut $6.727\times 10^{-7}$}\hspace{0.006\linewidth}\parbox[t]{0.671\linewidth}{seq\_pc\_40\_60\_r1e4\_xt1e7 seed5081}}}}\par\smallskip
\noindent\parbox[t]{0.05\linewidth}{\centering\strut 82}\hspace{0.006\linewidth}\parbox[t]{0.04\linewidth}{\centering\strut \raisebox{0.08ex}{\ding{51}}}\hspace{0.006\linewidth}\parbox[t]{0.10\linewidth}{\centering\strut +0.000164\%}\hspace{0.006\linewidth}\parbox[t]{0.115\linewidth}{\centering\strut $6.727\times 10^{-7}$}\hspace{0.006\linewidth}\parbox[t]{0.671\linewidth}{seq\_pc\_40\_60\_j5e8 seed5082}\par\smallskip
\noindent{\setlength{\fboxsep}{0pt}\colorbox{black!4}{\parbox{\linewidth}{\parbox[t]{0.05\linewidth}{\centering\strut 83}\hspace{0.006\linewidth}\parbox[t]{0.04\linewidth}{\centering\strut \raisebox{0.08ex}{\ding{51}}}\hspace{0.006\linewidth}\parbox[t]{0.10\linewidth}{\centering\strut +0.001041\%}\hspace{0.006\linewidth}\parbox[t]{0.115\linewidth}{\centering\strut $6.727\times 10^{-7}$}\hspace{0.006\linewidth}\parbox[t]{0.671\linewidth}{seq\_pc\_50\_50\_tight seed5083}}}}\par\smallskip
\noindent\parbox[t]{0.05\linewidth}{\centering\strut 84}\hspace{0.006\linewidth}\parbox[t]{0.04\linewidth}{\centering\strut \raisebox{0.08ex}{\ding{51}}}\hspace{0.006\linewidth}\parbox[t]{0.10\linewidth}{\centering\strut +0\%}\hspace{0.006\linewidth}\parbox[t]{0.115\linewidth}{\centering\strut $6.727\times 10^{-7}$}\hspace{0.006\linewidth}\parbox[t]{0.671\linewidth}{seq\_pc\_30\_70\_tight seed5084}\par\smallskip
\noindent{\setlength{\fboxsep}{0pt}\colorbox{black!4}{\parbox{\linewidth}{\parbox[t]{0.05\linewidth}{\centering\strut 85}\hspace{0.006\linewidth}\parbox[t]{0.04\linewidth}{\centering\strut \raisebox{0.08ex}{\ding{51}}}\hspace{0.006\linewidth}\parbox[t]{0.10\linewidth}{\centering\strut +0\%}\hspace{0.006\linewidth}\parbox[t]{0.115\linewidth}{\centering\strut $6.727\times 10^{-7}$}\hspace{0.006\linewidth}\parbox[t]{0.671\linewidth}{seq\_cp\_60\_40 seed5085}}}}\par\smallskip
\noindent\parbox[t]{0.05\linewidth}{\centering\strut 86}\hspace{0.006\linewidth}\parbox[t]{0.04\linewidth}{\centering\strut \raisebox{0.08ex}{\ding{51}}}\hspace{0.006\linewidth}\parbox[t]{0.10\linewidth}{\centering\strut +0\%}\hspace{0.006\linewidth}\parbox[t]{0.115\linewidth}{\centering\strut $6.727\times 10^{-7}$}\hspace{0.006\linewidth}\parbox[t]{0.671\linewidth}{warm\_powell\_xt1e8 seed5086}\par\smallskip
\noindent{\setlength{\fboxsep}{0pt}\colorbox{black!4}{\parbox{\linewidth}{\parbox[t]{0.05\linewidth}{\centering\strut 87}\hspace{0.006\linewidth}\parbox[t]{0.04\linewidth}{\centering\strut \raisebox{0.08ex}{\ding{55}}}\hspace{0.006\linewidth}\parbox[t]{0.10\linewidth}{\centering\strut -0.000327\%}\hspace{0.006\linewidth}\parbox[t]{0.115\linewidth}{\centering\strut $6.727\times 10^{-7}$}\hspace{0.006\linewidth}\parbox[t]{0.671\linewidth}{jitter\_cobyla\_s1e7\_r2e4 seed5087}}}}\par\smallskip
\noindent\parbox[t]{0.05\linewidth}{\centering\strut 88}\hspace{0.006\linewidth}\parbox[t]{0.04\linewidth}{\centering\strut \raisebox{0.08ex}{\ding{51}}}\hspace{0.006\linewidth}\parbox[t]{0.10\linewidth}{\centering\strut +0.000327\%}\hspace{0.006\linewidth}\parbox[t]{0.115\linewidth}{\centering\strut $6.727\times 10^{-7}$}\hspace{0.006\linewidth}\parbox[t]{0.671\linewidth}{seq\_pcp\_25\_50\_25 seed5088}\par\smallskip
\noindent{\setlength{\fboxsep}{0pt}\colorbox{black!4}{\parbox{\linewidth}{\parbox[t]{0.05\linewidth}{\centering\strut 89}\hspace{0.006\linewidth}\parbox[t]{0.04\linewidth}{\centering\strut \raisebox{0.08ex}{\ding{51}}}\hspace{0.006\linewidth}\parbox[t]{0.10\linewidth}{\centering\strut +0.000357\%}\hspace{0.006\linewidth}\parbox[t]{0.115\linewidth}{\centering\strut $6.727\times 10^{-7}$}\hspace{0.006\linewidth}\parbox[t]{0.671\linewidth}{seq\_pc\_45\_55\_j1e7\_r5e5 seed5089}}}}\par\smallskip
\noindent\parbox[t]{0.05\linewidth}{\centering\strut 90}\hspace{0.006\linewidth}\parbox[t]{0.04\linewidth}{\centering\strut \raisebox{0.08ex}{\ding{51}}}\hspace{0.006\linewidth}\parbox[t]{0.10\linewidth}{\centering\strut +0.0212\%}\hspace{0.006\linewidth}\parbox[t]{0.115\linewidth}{\centering\strut $6.725\times 10^{-7}$}\hspace{0.006\linewidth}\parbox[t]{0.671\linewidth}{warm\_cobyla\_r1e4 seed5090}\par\smallskip
\noindent{\setlength{\fboxsep}{0pt}\colorbox{black!4}{\parbox{\linewidth}{\parbox[t]{0.05\linewidth}{\centering\strut 91}\hspace{0.006\linewidth}\parbox[t]{0.04\linewidth}{\centering\strut \raisebox{0.08ex}{\ding{51}}}\hspace{0.006\linewidth}\parbox[t]{0.10\linewidth}{\centering\strut +0.004847\%}\hspace{0.006\linewidth}\parbox[t]{0.115\linewidth}{\centering\strut $6.725\times 10^{-7}$}\hspace{0.006\linewidth}\parbox[t]{0.671\linewidth}{seq\_pc\_40\_60\_r1e4\_xt1e7 seed5091}}}}\par\smallskip
\noindent\parbox[t]{0.05\linewidth}{\centering\strut 92}\hspace{0.006\linewidth}\parbox[t]{0.04\linewidth}{\centering\strut \raisebox{0.08ex}{\ding{55}}}\hspace{0.006\linewidth}\parbox[t]{0.10\linewidth}{\centering\strut -0.000223\%}\hspace{0.006\linewidth}\parbox[t]{0.115\linewidth}{\centering\strut $6.725\times 10^{-7}$}\hspace{0.006\linewidth}\parbox[t]{0.671\linewidth}{seq\_pc\_40\_60\_j5e8 seed5092}\par\smallskip
\noindent{\setlength{\fboxsep}{0pt}\colorbox{black!4}{\parbox{\linewidth}{\parbox[t]{0.05\linewidth}{\centering\strut 93}\hspace{0.006\linewidth}\parbox[t]{0.04\linewidth}{\centering\strut \raisebox{0.08ex}{\ding{51}}}\hspace{0.006\linewidth}\parbox[t]{0.10\linewidth}{\centering\strut +0.000416\%}\hspace{0.006\linewidth}\parbox[t]{0.115\linewidth}{\centering\strut $6.725\times 10^{-7}$}\hspace{0.006\linewidth}\parbox[t]{0.671\linewidth}{seq\_pc\_50\_50\_tight seed5093}}}}\par\smallskip
\noindent\parbox[t]{0.05\linewidth}{\centering\strut 94}\hspace{0.006\linewidth}\parbox[t]{0.04\linewidth}{\centering\strut \raisebox{0.08ex}{\ding{51}}}\hspace{0.006\linewidth}\parbox[t]{0.10\linewidth}{\centering\strut +0\%}\hspace{0.006\linewidth}\parbox[t]{0.115\linewidth}{\centering\strut $6.725\times 10^{-7}$}\hspace{0.006\linewidth}\parbox[t]{0.671\linewidth}{seq\_pc\_30\_70\_tight seed5094}\par\smallskip
\noindent{\setlength{\fboxsep}{0pt}\colorbox{black!4}{\parbox{\linewidth}{\parbox[t]{0.05\linewidth}{\centering\strut 95}\hspace{0.006\linewidth}\parbox[t]{0.04\linewidth}{\centering\strut \raisebox{0.08ex}{\ding{51}}}\hspace{0.006\linewidth}\parbox[t]{0.10\linewidth}{\centering\strut +0\%}\hspace{0.006\linewidth}\parbox[t]{0.115\linewidth}{\centering\strut $6.725\times 10^{-7}$}\hspace{0.006\linewidth}\parbox[t]{0.671\linewidth}{seq\_cp\_60\_40 seed5095}}}}\par\smallskip
\noindent\parbox[t]{0.05\linewidth}{\centering\strut 96}\hspace{0.006\linewidth}\parbox[t]{0.04\linewidth}{\centering\strut \raisebox{0.08ex}{\ding{51}}}\hspace{0.006\linewidth}\parbox[t]{0.10\linewidth}{\centering\strut +0\%}\hspace{0.006\linewidth}\parbox[t]{0.115\linewidth}{\centering\strut $6.725\times 10^{-7}$}\hspace{0.006\linewidth}\parbox[t]{0.671\linewidth}{warm\_powell\_xt1e8 seed5096}\par\smallskip
\noindent{\setlength{\fboxsep}{0pt}\colorbox{black!4}{\parbox{\linewidth}{\parbox[t]{0.05\linewidth}{\centering\strut 97}\hspace{0.006\linewidth}\parbox[t]{0.04\linewidth}{\centering\strut \raisebox{0.08ex}{\ding{51}}}\hspace{0.006\linewidth}\parbox[t]{0.10\linewidth}{\centering\strut +0.009561\%}\hspace{0.006\linewidth}\parbox[t]{0.115\linewidth}{\centering\strut $6.725\times 10^{-7}$}\hspace{0.006\linewidth}\parbox[t]{0.671\linewidth}{jitter\_cobyla\_s1e7\_r2e4 seed5097}}}}\par\smallskip
\noindent\parbox[t]{0.05\linewidth}{\centering\strut 98}\hspace{0.006\linewidth}\parbox[t]{0.04\linewidth}{\centering\strut \raisebox{0.08ex}{\ding{51}}}\hspace{0.006\linewidth}\parbox[t]{0.10\linewidth}{\centering\strut +0\%}\hspace{0.006\linewidth}\parbox[t]{0.115\linewidth}{\centering\strut $6.725\times 10^{-7}$}\hspace{0.006\linewidth}\parbox[t]{0.671\linewidth}{seq\_pcp\_25\_50\_25 seed5098}\par\smallskip
\noindent{\setlength{\fboxsep}{0pt}\colorbox{black!4}{\parbox{\linewidth}{\parbox[t]{0.05\linewidth}{\centering\strut 99}\hspace{0.006\linewidth}\parbox[t]{0.04\linewidth}{\centering\strut \raisebox{0.08ex}{\ding{51}}}\hspace{0.006\linewidth}\parbox[t]{0.10\linewidth}{\centering\strut +0.0244\%}\hspace{0.006\linewidth}\parbox[t]{0.115\linewidth}{\centering\strut $6.723\times 10^{-7}$}\hspace{0.006\linewidth}\parbox[t]{0.671\linewidth}{seq\_pc\_45\_55\_j1e7\_r5e5 seed5099}}}}\par\smallskip
\noindent\parbox[t]{0.05\linewidth}{\centering\strut 100}\hspace{0.006\linewidth}\parbox[t]{0.04\linewidth}{\centering\strut \raisebox{0.08ex}{\ding{51}}}\hspace{0.006\linewidth}\parbox[t]{0.10\linewidth}{\centering\strut +0.0277\%}\hspace{0.006\linewidth}\parbox[t]{0.115\linewidth}{\centering\strut $6.721\times 10^{-7}$}\hspace{0.006\linewidth}\parbox[t]{0.671\linewidth}{warm\_cobyla\_r1e4 seed5100}\par\smallskip
\noindent\rule{\linewidth}{0.5pt}
\par\medskip

\medskip

\refstepcounter{table}
\label{tab:supp_vqe_log_h2o}
\noindent\textbf{Table \thetable.} VQE mutation log for H$_2$O full 100-iteration archival campaign. Relative gain is measured against the immediately previous scored iteration using the absolute energy error.
\par\medskip
\noindent\rule{\linewidth}{0.5pt}
\noindent{\setlength{\fboxsep}{0pt}\colorbox{black!4}{\parbox{\linewidth}{\parbox[c][2.2em][c]{0.05\linewidth}{\centering \textbf{Iter.}}\hspace{0.006\linewidth}\parbox[c][2.2em][c]{0.04\linewidth}{\centering \textbf{A/R}}\hspace{0.006\linewidth}\parbox[c][2.2em][c]{0.10\linewidth}{\centering \textbf{Relative gain}}\hspace{0.006\linewidth}\parbox[c][2.2em][c]{0.115\linewidth}{\centering \textbf{$|\Delta E|$ [Ha]}}\hspace{0.006\linewidth}\parbox[c][2.2em][c]{0.671\linewidth}{\centering \textbf{Mutation summary}}}}}\par\smallskip
\noindent\rule{\linewidth}{0.5pt}
\noindent{\setlength{\fboxsep}{0pt}\colorbox{black!4}{\parbox{\linewidth}{\parbox[t]{0.05\linewidth}{\centering\strut 1}\hspace{0.006\linewidth}\parbox[t]{0.04\linewidth}{\centering\strut \raisebox{0.08ex}{\ding{51}}}\hspace{0.006\linewidth}\parbox[t]{0.10\linewidth}{\centering\strut ---}\hspace{0.006\linewidth}\parbox[t]{0.115\linewidth}{\centering\strut $6.307\times 10^{-3}$}\hspace{0.006\linewidth}\parbox[t]{0.671\linewidth}{switch to compressed UCCSD triplet coordinate search}}}}\par\smallskip
\noindent\parbox[t]{0.05\linewidth}{\centering\strut 2}\hspace{0.006\linewidth}\parbox[t]{0.04\linewidth}{\centering\strut \raisebox{0.08ex}{\ding{51}}}\hspace{0.006\linewidth}\parbox[t]{0.10\linewidth}{\centering\strut +24.04\%}\hspace{0.006\linewidth}\parbox[t]{0.115\linewidth}{\centering\strut $4.791\times 10^{-3}$}\hspace{0.006\linewidth}\parbox[t]{0.671\linewidth}{try tied singles-doubles UCCSD with Powell}\par\smallskip
\noindent{\setlength{\fboxsep}{0pt}\colorbox{black!4}{\parbox{\linewidth}{\parbox[t]{0.05\linewidth}{\centering\strut 3}\hspace{0.006\linewidth}\parbox[t]{0.04\linewidth}{\centering\strut \raisebox{0.08ex}{\ding{51}}}\hspace{0.006\linewidth}\parbox[t]{0.10\linewidth}{\centering\strut +2.26\%}\hspace{0.006\linewidth}\parbox[t]{0.115\linewidth}{\centering\strut $4.682\times 10^{-3}$}\hspace{0.006\linewidth}\parbox[t]{0.671\linewidth}{try 4-parameter UCCSD triplet with Powell}}}}\par\smallskip
\noindent\parbox[t]{0.05\linewidth}{\centering\strut 4}\hspace{0.006\linewidth}\parbox[t]{0.04\linewidth}{\centering\strut \raisebox{0.08ex}{\ding{51}}}\hspace{0.006\linewidth}\parbox[t]{0.10\linewidth}{\centering\strut +0\%}\hspace{0.006\linewidth}\parbox[t]{0.115\linewidth}{\centering\strut $4.682\times 10^{-3}$}\hspace{0.006\linewidth}\parbox[t]{0.671\linewidth}{retune 4-parameter Powell start with seed 7}\par\smallskip
\noindent{\setlength{\fboxsep}{0pt}\colorbox{black!4}{\parbox{\linewidth}{\parbox[t]{0.05\linewidth}{\centering\strut 5}\hspace{0.006\linewidth}\parbox[t]{0.04\linewidth}{\centering\strut \raisebox{0.08ex}{\ding{51}}}\hspace{0.006\linewidth}\parbox[t]{0.10\linewidth}{\centering\strut +0\%}\hspace{0.006\linewidth}\parbox[t]{0.115\linewidth}{\centering\strut $4.682\times 10^{-3}$}\hspace{0.006\linewidth}\parbox[t]{0.671\linewidth}{retune 4-parameter Powell start scale to 0.12}}}}\par\smallskip
\noindent\parbox[t]{0.05\linewidth}{\centering\strut 6}\hspace{0.006\linewidth}\parbox[t]{0.04\linewidth}{\centering\strut \raisebox{0.08ex}{\ding{55}}}\hspace{0.006\linewidth}\parbox[t]{0.10\linewidth}{\centering\strut -57.28\%}\hspace{0.006\linewidth}\parbox[t]{0.115\linewidth}{\centering\strut $7.364\times 10^{-3}$}\hspace{0.006\linewidth}\parbox[t]{0.671\linewidth}{try full UCCSD parameters with Powell}\par\smallskip
\noindent{\setlength{\fboxsep}{0pt}\colorbox{black!4}{\parbox{\linewidth}{\parbox[t]{0.05\linewidth}{\centering\strut 7}\hspace{0.006\linewidth}\parbox[t]{0.04\linewidth}{\centering\strut \raisebox{0.08ex}{\ding{51}}}\hspace{0.006\linewidth}\parbox[t]{0.10\linewidth}{\centering\strut +36.42\%}\hspace{0.006\linewidth}\parbox[t]{0.115\linewidth}{\centering\strut $4.682\times 10^{-3}$}\hspace{0.006\linewidth}\parbox[t]{0.671\linewidth}{add coordinate polish after Powell refinement}}}}\par\smallskip
\noindent\parbox[t]{0.05\linewidth}{\centering\strut 8}\hspace{0.006\linewidth}\parbox[t]{0.04\linewidth}{\centering\strut \raisebox{0.08ex}{\ding{55}}}\hspace{0.006\linewidth}\parbox[t]{0.10\linewidth}{\centering\strut -35.87\%}\hspace{0.006\linewidth}\parbox[t]{0.115\linewidth}{\centering\strut $6.362\times 10^{-3}$}\hspace{0.006\linewidth}\parbox[t]{0.671\linewidth}{switch 4-parameter UCCSD to Nelder-Mead plus polish}\par\smallskip
\noindent{\setlength{\fboxsep}{0pt}\colorbox{black!4}{\parbox{\linewidth}{\parbox[t]{0.05\linewidth}{\centering\strut 9}\hspace{0.006\linewidth}\parbox[t]{0.04\linewidth}{\centering\strut \raisebox{0.08ex}{\ding{55}}}\hspace{0.006\linewidth}\parbox[t]{0.10\linewidth}{\centering\strut -15.76\%}\hspace{0.006\linewidth}\parbox[t]{0.115\linewidth}{\centering\strut $7.364\times 10^{-3}$}\hspace{0.006\linewidth}\parbox[t]{0.671\linewidth}{split singles into three groups with triplet doubles}}}}\par\smallskip
\noindent\parbox[t]{0.05\linewidth}{\centering\strut 10}\hspace{0.006\linewidth}\parbox[t]{0.04\linewidth}{\centering\strut \raisebox{0.08ex}{\ding{55}}}\hspace{0.006\linewidth}\parbox[t]{0.10\linewidth}{\centering\strut +16.46\%}\hspace{0.006\linewidth}\parbox[t]{0.115\linewidth}{\centering\strut $6.152\times 10^{-3}$}\hspace{0.006\linewidth}\parbox[t]{0.671\linewidth}{warm-start 4-parameter Powell from tied UCCSD search}\par\smallskip
\noindent{\setlength{\fboxsep}{0pt}\colorbox{black!4}{\parbox{\linewidth}{\parbox[t]{0.05\linewidth}{\centering\strut 11}\hspace{0.006\linewidth}\parbox[t]{0.04\linewidth}{\centering\strut \raisebox{0.08ex}{\ding{51}}}\hspace{0.006\linewidth}\parbox[t]{0.10\linewidth}{\centering\strut +23.93\%}\hspace{0.006\linewidth}\parbox[t]{0.115\linewidth}{\centering\strut $4.679\times 10^{-3}$}\hspace{0.006\linewidth}\parbox[t]{0.671\linewidth}{switch 4-parameter UCCSD to COBYLA}}}}\par\smallskip
\noindent\parbox[t]{0.05\linewidth}{\centering\strut 12}\hspace{0.006\linewidth}\parbox[t]{0.04\linewidth}{\centering\strut \raisebox{0.08ex}{\ding{51}}}\hspace{0.006\linewidth}\parbox[t]{0.10\linewidth}{\centering\strut +0\%}\hspace{0.006\linewidth}\parbox[t]{0.115\linewidth}{\centering\strut $4.679\times 10^{-3}$}\hspace{0.006\linewidth}\parbox[t]{0.671\linewidth}{retune 4-parameter COBYLA start with seed 7}\par\smallskip
\noindent{\setlength{\fboxsep}{0pt}\colorbox{black!4}{\parbox{\linewidth}{\parbox[t]{0.05\linewidth}{\centering\strut 13}\hspace{0.006\linewidth}\parbox[t]{0.04\linewidth}{\centering\strut \raisebox{0.08ex}{\ding{55}}}\hspace{0.006\linewidth}\parbox[t]{0.10\linewidth}{\centering\strut -2.34\%}\hspace{0.006\linewidth}\parbox[t]{0.115\linewidth}{\centering\strut $4.789\times 10^{-3}$}\hspace{0.006\linewidth}\parbox[t]{0.671\linewidth}{try tied singles-doubles UCCSD with COBYLA}}}}\par\smallskip
\noindent\parbox[t]{0.05\linewidth}{\centering\strut 14}\hspace{0.006\linewidth}\parbox[t]{0.04\linewidth}{\centering\strut \raisebox{0.08ex}{\ding{51}}}\hspace{0.006\linewidth}\parbox[t]{0.10\linewidth}{\centering\strut +2.29\%}\hspace{0.006\linewidth}\parbox[t]{0.115\linewidth}{\centering\strut $4.679\times 10^{-3}$}\hspace{0.006\linewidth}\parbox[t]{0.671\linewidth}{retune 4-parameter COBYLA start scale to 0.08}\par\smallskip
\noindent{\setlength{\fboxsep}{0pt}\colorbox{black!4}{\parbox{\linewidth}{\parbox[t]{0.05\linewidth}{\centering\strut 15}\hspace{0.006\linewidth}\parbox[t]{0.04\linewidth}{\centering\strut \raisebox{0.08ex}{\ding{51}}}\hspace{0.006\linewidth}\parbox[t]{0.10\linewidth}{\centering\strut +0\%}\hspace{0.006\linewidth}\parbox[t]{0.115\linewidth}{\centering\strut $4.679\times 10^{-3}$}\hspace{0.006\linewidth}\parbox[t]{0.671\linewidth}{add coordinate polish after 4-parameter COBYLA}}}}\par\smallskip
\noindent\parbox[t]{0.05\linewidth}{\centering\strut 16}\hspace{0.006\linewidth}\parbox[t]{0.04\linewidth}{\centering\strut \raisebox{0.08ex}{\ding{55}}}\hspace{0.006\linewidth}\parbox[t]{0.10\linewidth}{\centering\strut ---}\hspace{0.006\linewidth}\parbox[t]{0.115\linewidth}{\centering\strut ---}\hspace{0.006\linewidth}\parbox[t]{0.671\linewidth}{try grouped singles plus triplet doubles with COBYLA}\par\smallskip
\noindent{\setlength{\fboxsep}{0pt}\colorbox{black!4}{\parbox{\linewidth}{\parbox[t]{0.05\linewidth}{\centering\strut 17}\hspace{0.006\linewidth}\parbox[t]{0.04\linewidth}{\centering\strut \raisebox{0.08ex}{\ding{51}}}\hspace{0.006\linewidth}\parbox[t]{0.10\linewidth}{\centering\strut +99.19\%}\hspace{0.006\linewidth}\parbox[t]{0.115\linewidth}{\centering\strut $3.787\times 10^{-5}$}\hspace{0.006\linewidth}\parbox[t]{0.671\linewidth}{use spin-paired symmetric UCCSD matching CUDA-Q ordering}}}}\par\smallskip
\noindent\parbox[t]{0.05\linewidth}{\centering\strut 18}\hspace{0.006\linewidth}\parbox[t]{0.04\linewidth}{\centering\strut \raisebox{0.08ex}{\ding{51}}}\hspace{0.006\linewidth}\parbox[t]{0.10\linewidth}{\centering\strut +0\%}\hspace{0.006\linewidth}\parbox[t]{0.115\linewidth}{\centering\strut $3.787\times 10^{-5}$}\hspace{0.006\linewidth}\parbox[t]{0.671\linewidth}{add coordinate polish after symmetric UCCSD COBYLA}\par\smallskip
\noindent{\setlength{\fboxsep}{0pt}\colorbox{black!4}{\parbox{\linewidth}{\parbox[t]{0.05\linewidth}{\centering\strut 19}\hspace{0.006\linewidth}\parbox[t]{0.04\linewidth}{\centering\strut \raisebox{0.08ex}{\ding{55}}}\hspace{0.006\linewidth}\parbox[t]{0.10\linewidth}{\centering\strut -3461.34\%}\hspace{0.006\linewidth}\parbox[t]{0.115\linewidth}{\centering\strut $1.349\times 10^{-3}$}\hspace{0.006\linewidth}\parbox[t]{0.671\linewidth}{free the mixed-double matrix in spin-paired UCCSD}}}}\par\smallskip
\noindent\parbox[t]{0.05\linewidth}{\centering\strut 20}\hspace{0.006\linewidth}\parbox[t]{0.04\linewidth}{\centering\strut \raisebox{0.08ex}{\ding{51}}}\hspace{0.006\linewidth}\parbox[t]{0.10\linewidth}{\centering\strut +97.19\%}\hspace{0.006\linewidth}\parbox[t]{0.115\linewidth}{\centering\strut $3.787\times 10^{-5}$}\hspace{0.006\linewidth}\parbox[t]{0.671\linewidth}{expose best parameters for warm-started follow-up runs}\par\smallskip
\noindent{\setlength{\fboxsep}{0pt}\colorbox{black!4}{\parbox{\linewidth}{\parbox[t]{0.05\linewidth}{\centering\strut 21}\hspace{0.006\linewidth}\parbox[t]{0.04\linewidth}{\centering\strut \raisebox{0.08ex}{\ding{51}}}\hspace{0.006\linewidth}\parbox[t]{0.10\linewidth}{\centering\strut +22.02\%}\hspace{0.006\linewidth}\parbox[t]{0.115\linewidth}{\centering\strut $2.953\times 10^{-5}$}\hspace{0.006\linewidth}\parbox[t]{0.671\linewidth}{warm-start full mixed-double matrix from symmetric near-optimum}}}}\par\smallskip
\noindent\parbox[t]{0.05\linewidth}{\centering\strut 22}\hspace{0.006\linewidth}\parbox[t]{0.04\linewidth}{\centering\strut \raisebox{0.08ex}{\ding{51}}}\hspace{0.006\linewidth}\parbox[t]{0.10\linewidth}{\centering\strut +21.39\%}\hspace{0.006\linewidth}\parbox[t]{0.115\linewidth}{\centering\strut $2.321\times 10^{-5}$}\hspace{0.006\linewidth}\parbox[t]{0.671\linewidth}{warm-start full UCCSD from 12-parameter near-optimum}\par\smallskip
\noindent{\setlength{\fboxsep}{0pt}\colorbox{black!4}{\parbox{\linewidth}{\parbox[t]{0.05\linewidth}{\centering\strut 23}\hspace{0.006\linewidth}\parbox[t]{0.04\linewidth}{\centering\strut \raisebox{0.08ex}{\ding{51}}}\hspace{0.006\linewidth}\parbox[t]{0.10\linewidth}{\centering\strut +11.69\%}\hspace{0.006\linewidth}\parbox[t]{0.115\linewidth}{\centering\strut $2.05\times 10^{-5}$}\hspace{0.006\linewidth}\parbox[t]{0.671\linewidth}{shrink full-UCCSD warm-start perturbation radius}}}}\par\smallskip
\noindent\parbox[t]{0.05\linewidth}{\centering\strut 24}\hspace{0.006\linewidth}\parbox[t]{0.04\linewidth}{\centering\strut \raisebox{0.08ex}{\ding{51}}}\hspace{0.006\linewidth}\parbox[t]{0.10\linewidth}{\centering\strut +49.59\%}\hspace{0.006\linewidth}\parbox[t]{0.115\linewidth}{\centering\strut $1.033\times 10^{-5}$}\hspace{0.006\linewidth}\parbox[t]{0.671\linewidth}{anneal full-UCCSD warm-start radius to 0.0025}\par\smallskip
\noindent{\setlength{\fboxsep}{0pt}\colorbox{black!4}{\parbox{\linewidth}{\parbox[t]{0.05\linewidth}{\centering\strut 25}\hspace{0.006\linewidth}\parbox[t]{0.04\linewidth}{\centering\strut \raisebox{0.08ex}{\ding{51}}}\hspace{0.006\linewidth}\parbox[t]{0.10\linewidth}{\centering\strut +16.58\%}\hspace{0.006\linewidth}\parbox[t]{0.115\linewidth}{\centering\strut $8.621\times 10^{-6}$}\hspace{0.006\linewidth}\parbox[t]{0.671\linewidth}{anneal full-UCCSD warm-start radius to 0.00125}}}}\par\smallskip
\noindent\parbox[t]{0.05\linewidth}{\centering\strut 26}\hspace{0.006\linewidth}\parbox[t]{0.04\linewidth}{\centering\strut \raisebox{0.08ex}{\ding{51}}}\hspace{0.006\linewidth}\parbox[t]{0.10\linewidth}{\centering\strut +0\%}\hspace{0.006\linewidth}\parbox[t]{0.115\linewidth}{\centering\strut $8.621\times 10^{-6}$}\hspace{0.006\linewidth}\parbox[t]{0.671\linewidth}{anneal full-UCCSD warm-start radius to 0.000625}\par\smallskip
\noindent{\setlength{\fboxsep}{0pt}\colorbox{black!4}{\parbox{\linewidth}{\parbox[t]{0.05\linewidth}{\centering\strut 27}\hspace{0.006\linewidth}\parbox[t]{0.04\linewidth}{\centering\strut \raisebox{0.08ex}{\ding{51}}}\hspace{0.006\linewidth}\parbox[t]{0.10\linewidth}{\centering\strut +4.74\%}\hspace{0.006\linewidth}\parbox[t]{0.115\linewidth}{\centering\strut $8.212\times 10^{-6}$}\hspace{0.006\linewidth}\parbox[t]{0.671\linewidth}{change local full-UCCSD perturbation seed to 7}}}}\par\smallskip
\noindent\parbox[t]{0.05\linewidth}{\centering\strut 28}\hspace{0.006\linewidth}\parbox[t]{0.04\linewidth}{\centering\strut \raisebox{0.08ex}{\ding{55}}}\hspace{0.006\linewidth}\parbox[t]{0.10\linewidth}{\centering\strut +0\%}\hspace{0.006\linewidth}\parbox[t]{0.115\linewidth}{\centering\strut $8.212\times 10^{-6}$}\hspace{0.006\linewidth}\parbox[t]{0.671\linewidth}{promote improved full-UCCSD warm start at fixed local radius}\par\smallskip
\noindent{\setlength{\fboxsep}{0pt}\colorbox{black!4}{\parbox{\linewidth}{\parbox[t]{0.05\linewidth}{\centering\strut 29}\hspace{0.006\linewidth}\parbox[t]{0.04\linewidth}{\centering\strut \raisebox{0.08ex}{\ding{51}}}\hspace{0.006\linewidth}\parbox[t]{0.10\linewidth}{\centering\strut +15.33\%}\hspace{0.006\linewidth}\parbox[t]{0.115\linewidth}{\centering\strut $6.954\times 10^{-6}$}\hspace{0.006\linewidth}\parbox[t]{0.671\linewidth}{change local full-UCCSD perturbation seed to 13}}}}\par\smallskip
\noindent\parbox[t]{0.05\linewidth}{\centering\strut 30}\hspace{0.006\linewidth}\parbox[t]{0.04\linewidth}{\centering\strut \raisebox{0.08ex}{\ding{51}}}\hspace{0.006\linewidth}\parbox[t]{0.10\linewidth}{\centering\strut +0\%}\hspace{0.006\linewidth}\parbox[t]{0.115\linewidth}{\centering\strut $6.954\times 10^{-6}$}\hspace{0.006\linewidth}\parbox[t]{0.671\linewidth}{promote improved full-UCCSD warm start after seed 13}\par\smallskip
\noindent{\setlength{\fboxsep}{0pt}\colorbox{black!4}{\parbox{\linewidth}{\parbox[t]{0.05\linewidth}{\centering\strut 31}\hspace{0.006\linewidth}\parbox[t]{0.04\linewidth}{\centering\strut \raisebox{0.08ex}{\ding{51}}}\hspace{0.006\linewidth}\parbox[t]{0.10\linewidth}{\centering\strut +7.21\%}\hspace{0.006\linewidth}\parbox[t]{0.115\linewidth}{\centering\strut $6.453\times 10^{-6}$}\hspace{0.006\linewidth}\parbox[t]{0.671\linewidth}{change local full-UCCSD perturbation seed to 21}}}}\par\smallskip
\noindent\parbox[t]{0.05\linewidth}{\centering\strut 32}\hspace{0.006\linewidth}\parbox[t]{0.04\linewidth}{\centering\strut \raisebox{0.08ex}{\ding{51}}}\hspace{0.006\linewidth}\parbox[t]{0.10\linewidth}{\centering\strut +3.84\%}\hspace{0.006\linewidth}\parbox[t]{0.115\linewidth}{\centering\strut $6.205\times 10^{-6}$}\hspace{0.006\linewidth}\parbox[t]{0.671\linewidth}{shift full-UCCSD warm start to seed-21 basin and perturb with seed 34}\par\smallskip
\noindent{\setlength{\fboxsep}{0pt}\colorbox{black!4}{\parbox{\linewidth}{\parbox[t]{0.05\linewidth}{\centering\strut 33}\hspace{0.006\linewidth}\parbox[t]{0.04\linewidth}{\centering\strut \raisebox{0.08ex}{\ding{51}}}\hspace{0.006\linewidth}\parbox[t]{0.10\linewidth}{\centering\strut +0\%}\hspace{0.006\linewidth}\parbox[t]{0.115\linewidth}{\centering\strut $6.205\times 10^{-6}$}\hspace{0.006\linewidth}\parbox[t]{0.671\linewidth}{shift full-UCCSD warm start to seed-34 basin and perturb with seed 55}}}}\par\smallskip
\noindent\parbox[t]{0.05\linewidth}{\centering\strut 34}\hspace{0.006\linewidth}\parbox[t]{0.04\linewidth}{\centering\strut \raisebox{0.08ex}{\ding{55}}}\hspace{0.006\linewidth}\parbox[t]{0.10\linewidth}{\centering\strut -3.99\%}\hspace{0.006\linewidth}\parbox[t]{0.115\linewidth}{\centering\strut $6.453\times 10^{-6}$}\hspace{0.006\linewidth}\parbox[t]{0.671\linewidth}{local full-UCCSD seed sweep 37}\par\smallskip
\noindent{\setlength{\fboxsep}{0pt}\colorbox{black!4}{\parbox{\linewidth}{\parbox[t]{0.05\linewidth}{\centering\strut 35}\hspace{0.006\linewidth}\parbox[t]{0.04\linewidth}{\centering\strut \raisebox{0.08ex}{\ding{51}}}\hspace{0.006\linewidth}\parbox[t]{0.10\linewidth}{\centering\strut +12.98\%}\hspace{0.006\linewidth}\parbox[t]{0.115\linewidth}{\centering\strut $5.615\times 10^{-6}$}\hspace{0.006\linewidth}\parbox[t]{0.671\linewidth}{local full-UCCSD seed sweep 41}}}}\par\smallskip
\noindent\parbox[t]{0.05\linewidth}{\centering\strut 36}\hspace{0.006\linewidth}\parbox[t]{0.04\linewidth}{\centering\strut \raisebox{0.08ex}{\ding{51}}}\hspace{0.006\linewidth}\parbox[t]{0.10\linewidth}{\centering\strut +7.06\%}\hspace{0.006\linewidth}\parbox[t]{0.115\linewidth}{\centering\strut $5.219\times 10^{-6}$}\hspace{0.006\linewidth}\parbox[t]{0.671\linewidth}{local full-UCCSD seed sweep 43}\par\smallskip
\noindent{\setlength{\fboxsep}{0pt}\colorbox{black!4}{\parbox{\linewidth}{\parbox[t]{0.05\linewidth}{\centering\strut 37}\hspace{0.006\linewidth}\parbox[t]{0.04\linewidth}{\centering\strut \raisebox{0.08ex}{\ding{55}}}\hspace{0.006\linewidth}\parbox[t]{0.10\linewidth}{\centering\strut -23.64\%}\hspace{0.006\linewidth}\parbox[t]{0.115\linewidth}{\centering\strut $6.453\times 10^{-6}$}\hspace{0.006\linewidth}\parbox[t]{0.671\linewidth}{local full-UCCSD seed sweep 47}}}}\par\smallskip
\noindent\parbox[t]{0.05\linewidth}{\centering\strut 38}\hspace{0.006\linewidth}\parbox[t]{0.04\linewidth}{\centering\strut \raisebox{0.08ex}{\ding{55}}}\hspace{0.006\linewidth}\parbox[t]{0.10\linewidth}{\centering\strut +0\%}\hspace{0.006\linewidth}\parbox[t]{0.115\linewidth}{\centering\strut $6.453\times 10^{-6}$}\hspace{0.006\linewidth}\parbox[t]{0.671\linewidth}{local full-UCCSD seed sweep 53}\par\smallskip
\noindent{\setlength{\fboxsep}{0pt}\colorbox{black!4}{\parbox{\linewidth}{\parbox[t]{0.05\linewidth}{\centering\strut 39}\hspace{0.006\linewidth}\parbox[t]{0.04\linewidth}{\centering\strut \raisebox{0.08ex}{\ding{55}}}\hspace{0.006\linewidth}\parbox[t]{0.10\linewidth}{\centering\strut +0\%}\hspace{0.006\linewidth}\parbox[t]{0.115\linewidth}{\centering\strut $6.453\times 10^{-6}$}\hspace{0.006\linewidth}\parbox[t]{0.671\linewidth}{local full-UCCSD seed sweep 59}}}}\par\smallskip
\noindent\parbox[t]{0.05\linewidth}{\centering\strut 40}\hspace{0.006\linewidth}\parbox[t]{0.04\linewidth}{\centering\strut \raisebox{0.08ex}{\ding{55}}}\hspace{0.006\linewidth}\parbox[t]{0.10\linewidth}{\centering\strut +0\%}\hspace{0.006\linewidth}\parbox[t]{0.115\linewidth}{\centering\strut $6.453\times 10^{-6}$}\hspace{0.006\linewidth}\parbox[t]{0.671\linewidth}{local full-UCCSD seed sweep 61}\par\smallskip
\noindent{\setlength{\fboxsep}{0pt}\colorbox{black!4}{\parbox{\linewidth}{\parbox[t]{0.05\linewidth}{\centering\strut 41}\hspace{0.006\linewidth}\parbox[t]{0.04\linewidth}{\centering\strut \raisebox{0.08ex}{\ding{55}}}\hspace{0.006\linewidth}\parbox[t]{0.10\linewidth}{\centering\strut +0\%}\hspace{0.006\linewidth}\parbox[t]{0.115\linewidth}{\centering\strut $6.453\times 10^{-6}$}\hspace{0.006\linewidth}\parbox[t]{0.671\linewidth}{local full-UCCSD seed sweep 67}}}}\par\smallskip
\noindent\parbox[t]{0.05\linewidth}{\centering\strut 42}\hspace{0.006\linewidth}\parbox[t]{0.04\linewidth}{\centering\strut \raisebox{0.08ex}{\ding{55}}}\hspace{0.006\linewidth}\parbox[t]{0.10\linewidth}{\centering\strut +4.57\%}\hspace{0.006\linewidth}\parbox[t]{0.115\linewidth}{\centering\strut $6.157\times 10^{-6}$}\hspace{0.006\linewidth}\parbox[t]{0.671\linewidth}{local full-UCCSD seed sweep 71}\par\smallskip
\noindent{\setlength{\fboxsep}{0pt}\colorbox{black!4}{\parbox{\linewidth}{\parbox[t]{0.05\linewidth}{\centering\strut 43}\hspace{0.006\linewidth}\parbox[t]{0.04\linewidth}{\centering\strut \raisebox{0.08ex}{\ding{55}}}\hspace{0.006\linewidth}\parbox[t]{0.10\linewidth}{\centering\strut -4.79\%}\hspace{0.006\linewidth}\parbox[t]{0.115\linewidth}{\centering\strut $6.453\times 10^{-6}$}\hspace{0.006\linewidth}\parbox[t]{0.671\linewidth}{local full-UCCSD seed sweep 73}}}}\par\smallskip
\noindent\parbox[t]{0.05\linewidth}{\centering\strut 44}\hspace{0.006\linewidth}\parbox[t]{0.04\linewidth}{\centering\strut \raisebox{0.08ex}{\ding{55}}}\hspace{0.006\linewidth}\parbox[t]{0.10\linewidth}{\centering\strut +9.29\%}\hspace{0.006\linewidth}\parbox[t]{0.115\linewidth}{\centering\strut $5.853\times 10^{-6}$}\hspace{0.006\linewidth}\parbox[t]{0.671\linewidth}{local full-UCCSD seed sweep 79}\par\smallskip
\noindent{\setlength{\fboxsep}{0pt}\colorbox{black!4}{\parbox{\linewidth}{\parbox[t]{0.05\linewidth}{\centering\strut 45}\hspace{0.006\linewidth}\parbox[t]{0.04\linewidth}{\centering\strut \raisebox{0.08ex}{\ding{55}}}\hspace{0.006\linewidth}\parbox[t]{0.10\linewidth}{\centering\strut -10.25\%}\hspace{0.006\linewidth}\parbox[t]{0.115\linewidth}{\centering\strut $6.453\times 10^{-6}$}\hspace{0.006\linewidth}\parbox[t]{0.671\linewidth}{local full-UCCSD seed sweep 83}}}}\par\smallskip
\noindent\parbox[t]{0.05\linewidth}{\centering\strut 46}\hspace{0.006\linewidth}\parbox[t]{0.04\linewidth}{\centering\strut \raisebox{0.08ex}{\ding{55}}}\hspace{0.006\linewidth}\parbox[t]{0.10\linewidth}{\centering\strut +5.46\%}\hspace{0.006\linewidth}\parbox[t]{0.115\linewidth}{\centering\strut $6.1\times 10^{-6}$}\hspace{0.006\linewidth}\parbox[t]{0.671\linewidth}{local full-UCCSD seed sweep 89}\par\smallskip
\noindent{\setlength{\fboxsep}{0pt}\colorbox{black!4}{\parbox{\linewidth}{\parbox[t]{0.05\linewidth}{\centering\strut 47}\hspace{0.006\linewidth}\parbox[t]{0.04\linewidth}{\centering\strut \raisebox{0.08ex}{\ding{55}}}\hspace{0.006\linewidth}\parbox[t]{0.10\linewidth}{\centering\strut -5.51\%}\hspace{0.006\linewidth}\parbox[t]{0.115\linewidth}{\centering\strut $6.437\times 10^{-6}$}\hspace{0.006\linewidth}\parbox[t]{0.671\linewidth}{local full-UCCSD seed sweep 97}}}}\par\smallskip
\noindent\parbox[t]{0.05\linewidth}{\centering\strut 48}\hspace{0.006\linewidth}\parbox[t]{0.04\linewidth}{\centering\strut \raisebox{0.08ex}{\ding{55}}}\hspace{0.006\linewidth}\parbox[t]{0.10\linewidth}{\centering\strut -0.2463\%}\hspace{0.006\linewidth}\parbox[t]{0.115\linewidth}{\centering\strut $6.453\times 10^{-6}$}\hspace{0.006\linewidth}\parbox[t]{0.671\linewidth}{local full-UCCSD seed sweep 101}\par\smallskip
\noindent{\setlength{\fboxsep}{0pt}\colorbox{black!4}{\parbox{\linewidth}{\parbox[t]{0.05\linewidth}{\centering\strut 49}\hspace{0.006\linewidth}\parbox[t]{0.04\linewidth}{\centering\strut \raisebox{0.08ex}{\ding{55}}}\hspace{0.006\linewidth}\parbox[t]{0.10\linewidth}{\centering\strut +8.57\%}\hspace{0.006\linewidth}\parbox[t]{0.115\linewidth}{\centering\strut $5.9\times 10^{-6}$}\hspace{0.006\linewidth}\parbox[t]{0.671\linewidth}{local full-UCCSD seed sweep 103}}}}\par\smallskip
\noindent\parbox[t]{0.05\linewidth}{\centering\strut 50}\hspace{0.006\linewidth}\parbox[t]{0.04\linewidth}{\centering\strut \raisebox{0.08ex}{\ding{55}}}\hspace{0.006\linewidth}\parbox[t]{0.10\linewidth}{\centering\strut -9.37\%}\hspace{0.006\linewidth}\parbox[t]{0.115\linewidth}{\centering\strut $6.453\times 10^{-6}$}\hspace{0.006\linewidth}\parbox[t]{0.671\linewidth}{local full-UCCSD seed sweep 107}\par\smallskip
\noindent{\setlength{\fboxsep}{0pt}\colorbox{black!4}{\parbox{\linewidth}{\parbox[t]{0.05\linewidth}{\centering\strut 51}\hspace{0.006\linewidth}\parbox[t]{0.04\linewidth}{\centering\strut \raisebox{0.08ex}{\ding{55}}}\hspace{0.006\linewidth}\parbox[t]{0.10\linewidth}{\centering\strut +4.83\%}\hspace{0.006\linewidth}\parbox[t]{0.115\linewidth}{\centering\strut $6.141\times 10^{-6}$}\hspace{0.006\linewidth}\parbox[t]{0.671\linewidth}{local full-UCCSD seed sweep 109}}}}\par\smallskip
\noindent\parbox[t]{0.05\linewidth}{\centering\strut 52}\hspace{0.006\linewidth}\parbox[t]{0.04\linewidth}{\centering\strut \raisebox{0.08ex}{\ding{55}}}\hspace{0.006\linewidth}\parbox[t]{0.10\linewidth}{\centering\strut -5.08\%}\hspace{0.006\linewidth}\parbox[t]{0.115\linewidth}{\centering\strut $6.453\times 10^{-6}$}\hspace{0.006\linewidth}\parbox[t]{0.671\linewidth}{local full-UCCSD seed sweep 113}\par\smallskip
\noindent{\setlength{\fboxsep}{0pt}\colorbox{black!4}{\parbox{\linewidth}{\parbox[t]{0.05\linewidth}{\centering\strut 53}\hspace{0.006\linewidth}\parbox[t]{0.04\linewidth}{\centering\strut \raisebox{0.08ex}{\ding{55}}}\hspace{0.006\linewidth}\parbox[t]{0.10\linewidth}{\centering\strut +0\%}\hspace{0.006\linewidth}\parbox[t]{0.115\linewidth}{\centering\strut $6.453\times 10^{-6}$}\hspace{0.006\linewidth}\parbox[t]{0.671\linewidth}{local full-UCCSD seed sweep 127}}}}\par\smallskip
\noindent\parbox[t]{0.05\linewidth}{\centering\strut 54}\hspace{0.006\linewidth}\parbox[t]{0.04\linewidth}{\centering\strut \raisebox{0.08ex}{\ding{55}}}\hspace{0.006\linewidth}\parbox[t]{0.10\linewidth}{\centering\strut +0\%}\hspace{0.006\linewidth}\parbox[t]{0.115\linewidth}{\centering\strut $6.453\times 10^{-6}$}\hspace{0.006\linewidth}\parbox[t]{0.671\linewidth}{local full-UCCSD seed sweep 131}\par\smallskip
\noindent{\setlength{\fboxsep}{0pt}\colorbox{black!4}{\parbox{\linewidth}{\parbox[t]{0.05\linewidth}{\centering\strut 55}\hspace{0.006\linewidth}\parbox[t]{0.04\linewidth}{\centering\strut \raisebox{0.08ex}{\ding{55}}}\hspace{0.006\linewidth}\parbox[t]{0.10\linewidth}{\centering\strut +6.26\%}\hspace{0.006\linewidth}\parbox[t]{0.115\linewidth}{\centering\strut $6.048\times 10^{-6}$}\hspace{0.006\linewidth}\parbox[t]{0.671\linewidth}{local full-UCCSD seed sweep 137}}}}\par\smallskip
\noindent\parbox[t]{0.05\linewidth}{\centering\strut 56}\hspace{0.006\linewidth}\parbox[t]{0.04\linewidth}{\centering\strut \raisebox{0.08ex}{\ding{55}}}\hspace{0.006\linewidth}\parbox[t]{0.10\linewidth}{\centering\strut -6.68\%}\hspace{0.006\linewidth}\parbox[t]{0.115\linewidth}{\centering\strut $6.453\times 10^{-6}$}\hspace{0.006\linewidth}\parbox[t]{0.671\linewidth}{local full-UCCSD seed sweep 139}\par\smallskip
\noindent{\setlength{\fboxsep}{0pt}\colorbox{black!4}{\parbox{\linewidth}{\parbox[t]{0.05\linewidth}{\centering\strut 57}\hspace{0.006\linewidth}\parbox[t]{0.04\linewidth}{\centering\strut \raisebox{0.08ex}{\ding{51}}}\hspace{0.006\linewidth}\parbox[t]{0.10\linewidth}{\centering\strut +21.54\%}\hspace{0.006\linewidth}\parbox[t]{0.115\linewidth}{\centering\strut $5.063\times 10^{-6}$}\hspace{0.006\linewidth}\parbox[t]{0.671\linewidth}{local full-UCCSD seed sweep 149}}}}\par\smallskip
\noindent\parbox[t]{0.05\linewidth}{\centering\strut 58}\hspace{0.006\linewidth}\parbox[t]{0.04\linewidth}{\centering\strut \raisebox{0.08ex}{\ding{55}}}\hspace{0.006\linewidth}\parbox[t]{0.10\linewidth}{\centering\strut -27.46\%}\hspace{0.006\linewidth}\parbox[t]{0.115\linewidth}{\centering\strut $6.453\times 10^{-6}$}\hspace{0.006\linewidth}\parbox[t]{0.671\linewidth}{local full-UCCSD seed sweep 151}\par\smallskip
\noindent{\setlength{\fboxsep}{0pt}\colorbox{black!4}{\parbox{\linewidth}{\parbox[t]{0.05\linewidth}{\centering\strut 59}\hspace{0.006\linewidth}\parbox[t]{0.04\linewidth}{\centering\strut \raisebox{0.08ex}{\ding{55}}}\hspace{0.006\linewidth}\parbox[t]{0.10\linewidth}{\centering\strut +0\%}\hspace{0.006\linewidth}\parbox[t]{0.115\linewidth}{\centering\strut $6.453\times 10^{-6}$}\hspace{0.006\linewidth}\parbox[t]{0.671\linewidth}{local full-UCCSD seed sweep 157}}}}\par\smallskip
\noindent\parbox[t]{0.05\linewidth}{\centering\strut 60}\hspace{0.006\linewidth}\parbox[t]{0.04\linewidth}{\centering\strut \raisebox{0.08ex}{\ding{55}}}\hspace{0.006\linewidth}\parbox[t]{0.10\linewidth}{\centering\strut +0\%}\hspace{0.006\linewidth}\parbox[t]{0.115\linewidth}{\centering\strut $6.453\times 10^{-6}$}\hspace{0.006\linewidth}\parbox[t]{0.671\linewidth}{local full-UCCSD seed sweep 163}\par\smallskip
\noindent{\setlength{\fboxsep}{0pt}\colorbox{black!4}{\parbox{\linewidth}{\parbox[t]{0.05\linewidth}{\centering\strut 61}\hspace{0.006\linewidth}\parbox[t]{0.04\linewidth}{\centering\strut \raisebox{0.08ex}{\ding{55}}}\hspace{0.006\linewidth}\parbox[t]{0.10\linewidth}{\centering\strut +0.0593\%}\hspace{0.006\linewidth}\parbox[t]{0.115\linewidth}{\centering\strut $6.449\times 10^{-6}$}\hspace{0.006\linewidth}\parbox[t]{0.671\linewidth}{local full-UCCSD seed sweep 167}}}}\par\smallskip
\noindent\parbox[t]{0.05\linewidth}{\centering\strut 62}\hspace{0.006\linewidth}\parbox[t]{0.04\linewidth}{\centering\strut \raisebox{0.08ex}{\ding{55}}}\hspace{0.006\linewidth}\parbox[t]{0.10\linewidth}{\centering\strut -0.0594\%}\hspace{0.006\linewidth}\parbox[t]{0.115\linewidth}{\centering\strut $6.453\times 10^{-6}$}\hspace{0.006\linewidth}\parbox[t]{0.671\linewidth}{local full-UCCSD seed sweep 173}\par\smallskip
\noindent{\setlength{\fboxsep}{0pt}\colorbox{black!4}{\parbox{\linewidth}{\parbox[t]{0.05\linewidth}{\centering\strut 63}\hspace{0.006\linewidth}\parbox[t]{0.04\linewidth}{\centering\strut \raisebox{0.08ex}{\ding{51}}}\hspace{0.006\linewidth}\parbox[t]{0.10\linewidth}{\centering\strut +26.34\%}\hspace{0.006\linewidth}\parbox[t]{0.115\linewidth}{\centering\strut $4.753\times 10^{-6}$}\hspace{0.006\linewidth}\parbox[t]{0.671\linewidth}{local full-UCCSD seed sweep 179}}}}\par\smallskip
\noindent\parbox[t]{0.05\linewidth}{\centering\strut 64}\hspace{0.006\linewidth}\parbox[t]{0.04\linewidth}{\centering\strut \raisebox{0.08ex}{\ding{55}}}\hspace{0.006\linewidth}\parbox[t]{0.10\linewidth}{\centering\strut -32.08\%}\hspace{0.006\linewidth}\parbox[t]{0.115\linewidth}{\centering\strut $6.277\times 10^{-6}$}\hspace{0.006\linewidth}\parbox[t]{0.671\linewidth}{local full-UCCSD seed sweep 181}\par\smallskip
\noindent{\setlength{\fboxsep}{0pt}\colorbox{black!4}{\parbox{\linewidth}{\parbox[t]{0.05\linewidth}{\centering\strut 65}\hspace{0.006\linewidth}\parbox[t]{0.04\linewidth}{\centering\strut \raisebox{0.08ex}{\ding{55}}}\hspace{0.006\linewidth}\parbox[t]{0.10\linewidth}{\centering\strut +15.9\%}\hspace{0.006\linewidth}\parbox[t]{0.115\linewidth}{\centering\strut $5.279\times 10^{-6}$}\hspace{0.006\linewidth}\parbox[t]{0.671\linewidth}{local full-UCCSD seed sweep 191}}}}\par\smallskip
\noindent\parbox[t]{0.05\linewidth}{\centering\strut 66}\hspace{0.006\linewidth}\parbox[t]{0.04\linewidth}{\centering\strut \raisebox{0.08ex}{\ding{55}}}\hspace{0.006\linewidth}\parbox[t]{0.10\linewidth}{\centering\strut -22.22\%}\hspace{0.006\linewidth}\parbox[t]{0.115\linewidth}{\centering\strut $6.453\times 10^{-6}$}\hspace{0.006\linewidth}\parbox[t]{0.671\linewidth}{local full-UCCSD seed sweep 193}\par\smallskip
\noindent{\setlength{\fboxsep}{0pt}\colorbox{black!4}{\parbox{\linewidth}{\parbox[t]{0.05\linewidth}{\centering\strut 67}\hspace{0.006\linewidth}\parbox[t]{0.04\linewidth}{\centering\strut \raisebox{0.08ex}{\ding{55}}}\hspace{0.006\linewidth}\parbox[t]{0.10\linewidth}{\centering\strut +0\%}\hspace{0.006\linewidth}\parbox[t]{0.115\linewidth}{\centering\strut $6.453\times 10^{-6}$}\hspace{0.006\linewidth}\parbox[t]{0.671\linewidth}{local full-UCCSD seed sweep 197}}}}\par\smallskip
\noindent\parbox[t]{0.05\linewidth}{\centering\strut 68}\hspace{0.006\linewidth}\parbox[t]{0.04\linewidth}{\centering\strut \raisebox{0.08ex}{\ding{55}}}\hspace{0.006\linewidth}\parbox[t]{0.10\linewidth}{\centering\strut +0\%}\hspace{0.006\linewidth}\parbox[t]{0.115\linewidth}{\centering\strut $6.453\times 10^{-6}$}\hspace{0.006\linewidth}\parbox[t]{0.671\linewidth}{local full-UCCSD seed sweep 199}\par\smallskip
\noindent{\setlength{\fboxsep}{0pt}\colorbox{black!4}{\parbox{\linewidth}{\parbox[t]{0.05\linewidth}{\centering\strut 69}\hspace{0.006\linewidth}\parbox[t]{0.04\linewidth}{\centering\strut \raisebox{0.08ex}{\ding{55}}}\hspace{0.006\linewidth}\parbox[t]{0.10\linewidth}{\centering\strut +0\%}\hspace{0.006\linewidth}\parbox[t]{0.115\linewidth}{\centering\strut $6.453\times 10^{-6}$}\hspace{0.006\linewidth}\parbox[t]{0.671\linewidth}{local full-UCCSD seed sweep 211}}}}\par\smallskip
\noindent\parbox[t]{0.05\linewidth}{\centering\strut 70}\hspace{0.006\linewidth}\parbox[t]{0.04\linewidth}{\centering\strut \raisebox{0.08ex}{\ding{55}}}\hspace{0.006\linewidth}\parbox[t]{0.10\linewidth}{\centering\strut +0\%}\hspace{0.006\linewidth}\parbox[t]{0.115\linewidth}{\centering\strut $6.453\times 10^{-6}$}\hspace{0.006\linewidth}\parbox[t]{0.671\linewidth}{local full-UCCSD seed sweep 223}\par\smallskip
\noindent{\setlength{\fboxsep}{0pt}\colorbox{black!4}{\parbox{\linewidth}{\parbox[t]{0.05\linewidth}{\centering\strut 71}\hspace{0.006\linewidth}\parbox[t]{0.04\linewidth}{\centering\strut \raisebox{0.08ex}{\ding{55}}}\hspace{0.006\linewidth}\parbox[t]{0.10\linewidth}{\centering\strut +0\%}\hspace{0.006\linewidth}\parbox[t]{0.115\linewidth}{\centering\strut $6.453\times 10^{-6}$}\hspace{0.006\linewidth}\parbox[t]{0.671\linewidth}{local full-UCCSD seed sweep 227}}}}\par\smallskip
\noindent\parbox[t]{0.05\linewidth}{\centering\strut 72}\hspace{0.006\linewidth}\parbox[t]{0.04\linewidth}{\centering\strut \raisebox{0.08ex}{\ding{55}}}\hspace{0.006\linewidth}\parbox[t]{0.10\linewidth}{\centering\strut +0\%}\hspace{0.006\linewidth}\parbox[t]{0.115\linewidth}{\centering\strut $6.453\times 10^{-6}$}\hspace{0.006\linewidth}\parbox[t]{0.671\linewidth}{local full-UCCSD seed sweep 229}\par\smallskip
\noindent{\setlength{\fboxsep}{0pt}\colorbox{black!4}{\parbox{\linewidth}{\parbox[t]{0.05\linewidth}{\centering\strut 73}\hspace{0.006\linewidth}\parbox[t]{0.04\linewidth}{\centering\strut \raisebox{0.08ex}{\ding{55}}}\hspace{0.006\linewidth}\parbox[t]{0.10\linewidth}{\centering\strut +0\%}\hspace{0.006\linewidth}\parbox[t]{0.115\linewidth}{\centering\strut $6.453\times 10^{-6}$}\hspace{0.006\linewidth}\parbox[t]{0.671\linewidth}{local full-UCCSD seed sweep 233}}}}\par\smallskip
\noindent\parbox[t]{0.05\linewidth}{\centering\strut 74}\hspace{0.006\linewidth}\parbox[t]{0.04\linewidth}{\centering\strut \raisebox{0.08ex}{\ding{55}}}\hspace{0.006\linewidth}\parbox[t]{0.10\linewidth}{\centering\strut +0\%}\hspace{0.006\linewidth}\parbox[t]{0.115\linewidth}{\centering\strut $6.453\times 10^{-6}$}\hspace{0.006\linewidth}\parbox[t]{0.671\linewidth}{local full-UCCSD seed sweep 239}\par\smallskip
\noindent{\setlength{\fboxsep}{0pt}\colorbox{black!4}{\parbox{\linewidth}{\parbox[t]{0.05\linewidth}{\centering\strut 75}\hspace{0.006\linewidth}\parbox[t]{0.04\linewidth}{\centering\strut \raisebox{0.08ex}{\ding{55}}}\hspace{0.006\linewidth}\parbox[t]{0.10\linewidth}{\centering\strut +3.45\%}\hspace{0.006\linewidth}\parbox[t]{0.115\linewidth}{\centering\strut $6.23\times 10^{-6}$}\hspace{0.006\linewidth}\parbox[t]{0.671\linewidth}{local full-UCCSD seed sweep 241}}}}\par\smallskip
\noindent\parbox[t]{0.05\linewidth}{\centering\strut 76}\hspace{0.006\linewidth}\parbox[t]{0.04\linewidth}{\centering\strut \raisebox{0.08ex}{\ding{55}}}\hspace{0.006\linewidth}\parbox[t]{0.10\linewidth}{\centering\strut +14.28\%}\hspace{0.006\linewidth}\parbox[t]{0.115\linewidth}{\centering\strut $5.34\times 10^{-6}$}\hspace{0.006\linewidth}\parbox[t]{0.671\linewidth}{local full-UCCSD seed sweep 251}\par\smallskip
\noindent{\setlength{\fboxsep}{0pt}\colorbox{black!4}{\parbox{\linewidth}{\parbox[t]{0.05\linewidth}{\centering\strut 77}\hspace{0.006\linewidth}\parbox[t]{0.04\linewidth}{\centering\strut \raisebox{0.08ex}{\ding{51}}}\hspace{0.006\linewidth}\parbox[t]{0.10\linewidth}{\centering\strut +33.66\%}\hspace{0.006\linewidth}\parbox[t]{0.115\linewidth}{\centering\strut $3.543\times 10^{-6}$}\hspace{0.006\linewidth}\parbox[t]{0.671\linewidth}{local full-UCCSD seed sweep 257}}}}\par\smallskip
\noindent\parbox[t]{0.05\linewidth}{\centering\strut 78}\hspace{0.006\linewidth}\parbox[t]{0.04\linewidth}{\centering\strut \raisebox{0.08ex}{\ding{55}}}\hspace{0.006\linewidth}\parbox[t]{0.10\linewidth}{\centering\strut -81.24\%}\hspace{0.006\linewidth}\parbox[t]{0.115\linewidth}{\centering\strut $6.421\times 10^{-6}$}\hspace{0.006\linewidth}\parbox[t]{0.671\linewidth}{local full-UCCSD seed sweep 263}\par\smallskip
\noindent{\setlength{\fboxsep}{0pt}\colorbox{black!4}{\parbox{\linewidth}{\parbox[t]{0.05\linewidth}{\centering\strut 79}\hspace{0.006\linewidth}\parbox[t]{0.04\linewidth}{\centering\strut \raisebox{0.08ex}{\ding{55}}}\hspace{0.006\linewidth}\parbox[t]{0.10\linewidth}{\centering\strut +2.5\%}\hspace{0.006\linewidth}\parbox[t]{0.115\linewidth}{\centering\strut $6.261\times 10^{-6}$}\hspace{0.006\linewidth}\parbox[t]{0.671\linewidth}{local full-UCCSD seed sweep 269}}}}\par\smallskip
\noindent\parbox[t]{0.05\linewidth}{\centering\strut 80}\hspace{0.006\linewidth}\parbox[t]{0.04\linewidth}{\centering\strut \raisebox{0.08ex}{\ding{55}}}\hspace{0.006\linewidth}\parbox[t]{0.10\linewidth}{\centering\strut +25.04\%}\hspace{0.006\linewidth}\parbox[t]{0.115\linewidth}{\centering\strut $4.693\times 10^{-6}$}\hspace{0.006\linewidth}\parbox[t]{0.671\linewidth}{local full-UCCSD seed sweep 271}\par\smallskip
\noindent{\setlength{\fboxsep}{0pt}\colorbox{black!4}{\parbox{\linewidth}{\parbox[t]{0.05\linewidth}{\centering\strut 81}\hspace{0.006\linewidth}\parbox[t]{0.04\linewidth}{\centering\strut \raisebox{0.08ex}{\ding{55}}}\hspace{0.006\linewidth}\parbox[t]{0.10\linewidth}{\centering\strut -10.35\%}\hspace{0.006\linewidth}\parbox[t]{0.115\linewidth}{\centering\strut $5.178\times 10^{-6}$}\hspace{0.006\linewidth}\parbox[t]{0.671\linewidth}{local full-UCCSD seed sweep 277}}}}\par\smallskip
\noindent\parbox[t]{0.05\linewidth}{\centering\strut 82}\hspace{0.006\linewidth}\parbox[t]{0.04\linewidth}{\centering\strut \raisebox{0.08ex}{\ding{55}}}\hspace{0.006\linewidth}\parbox[t]{0.10\linewidth}{\centering\strut -24.61\%}\hspace{0.006\linewidth}\parbox[t]{0.115\linewidth}{\centering\strut $6.453\times 10^{-6}$}\hspace{0.006\linewidth}\parbox[t]{0.671\linewidth}{local full-UCCSD seed sweep 281}\par\smallskip
\noindent{\setlength{\fboxsep}{0pt}\colorbox{black!4}{\parbox{\linewidth}{\parbox[t]{0.05\linewidth}{\centering\strut 83}\hspace{0.006\linewidth}\parbox[t]{0.04\linewidth}{\centering\strut \raisebox{0.08ex}{\ding{55}}}\hspace{0.006\linewidth}\parbox[t]{0.10\linewidth}{\centering\strut +2.62\%}\hspace{0.006\linewidth}\parbox[t]{0.115\linewidth}{\centering\strut $6.283\times 10^{-6}$}\hspace{0.006\linewidth}\parbox[t]{0.671\linewidth}{local full-UCCSD seed sweep 283}}}}\par\smallskip
\noindent\parbox[t]{0.05\linewidth}{\centering\strut 84}\hspace{0.006\linewidth}\parbox[t]{0.04\linewidth}{\centering\strut \raisebox{0.08ex}{\ding{55}}}\hspace{0.006\linewidth}\parbox[t]{0.10\linewidth}{\centering\strut -0.9185\%}\hspace{0.006\linewidth}\parbox[t]{0.115\linewidth}{\centering\strut $6.341\times 10^{-6}$}\hspace{0.006\linewidth}\parbox[t]{0.671\linewidth}{local full-UCCSD seed sweep 293}\par\smallskip
\noindent{\setlength{\fboxsep}{0pt}\colorbox{black!4}{\parbox{\linewidth}{\parbox[t]{0.05\linewidth}{\centering\strut 85}\hspace{0.006\linewidth}\parbox[t]{0.04\linewidth}{\centering\strut \raisebox{0.08ex}{\ding{55}}}\hspace{0.006\linewidth}\parbox[t]{0.10\linewidth}{\centering\strut -1.76\%}\hspace{0.006\linewidth}\parbox[t]{0.115\linewidth}{\centering\strut $6.453\times 10^{-6}$}\hspace{0.006\linewidth}\parbox[t]{0.671\linewidth}{local full-UCCSD seed sweep 307}}}}\par\smallskip
\noindent\parbox[t]{0.05\linewidth}{\centering\strut 86}\hspace{0.006\linewidth}\parbox[t]{0.04\linewidth}{\centering\strut \raisebox{0.08ex}{\ding{55}}}\hspace{0.006\linewidth}\parbox[t]{0.10\linewidth}{\centering\strut +0\%}\hspace{0.006\linewidth}\parbox[t]{0.115\linewidth}{\centering\strut $6.453\times 10^{-6}$}\hspace{0.006\linewidth}\parbox[t]{0.671\linewidth}{local full-UCCSD seed sweep 311}\par\smallskip
\noindent{\setlength{\fboxsep}{0pt}\colorbox{black!4}{\parbox{\linewidth}{\parbox[t]{0.05\linewidth}{\centering\strut 87}\hspace{0.006\linewidth}\parbox[t]{0.04\linewidth}{\centering\strut \raisebox{0.08ex}{\ding{55}}}\hspace{0.006\linewidth}\parbox[t]{0.10\linewidth}{\centering\strut +0\%}\hspace{0.006\linewidth}\parbox[t]{0.115\linewidth}{\centering\strut $6.453\times 10^{-6}$}\hspace{0.006\linewidth}\parbox[t]{0.671\linewidth}{local full-UCCSD seed sweep 313}}}}\par\smallskip
\noindent\parbox[t]{0.05\linewidth}{\centering\strut 88}\hspace{0.006\linewidth}\parbox[t]{0.04\linewidth}{\centering\strut \raisebox{0.08ex}{\ding{55}}}\hspace{0.006\linewidth}\parbox[t]{0.10\linewidth}{\centering\strut +18.72\%}\hspace{0.006\linewidth}\parbox[t]{0.115\linewidth}{\centering\strut $5.244\times 10^{-6}$}\hspace{0.006\linewidth}\parbox[t]{0.671\linewidth}{local full-UCCSD seed sweep 317}\par\smallskip
\noindent{\setlength{\fboxsep}{0pt}\colorbox{black!4}{\parbox{\linewidth}{\parbox[t]{0.05\linewidth}{\centering\strut 89}\hspace{0.006\linewidth}\parbox[t]{0.04\linewidth}{\centering\strut \raisebox{0.08ex}{\ding{55}}}\hspace{0.006\linewidth}\parbox[t]{0.10\linewidth}{\centering\strut -23.04\%}\hspace{0.006\linewidth}\parbox[t]{0.115\linewidth}{\centering\strut $6.453\times 10^{-6}$}\hspace{0.006\linewidth}\parbox[t]{0.671\linewidth}{local full-UCCSD seed sweep 331}}}}\par\smallskip
\noindent\parbox[t]{0.05\linewidth}{\centering\strut 90}\hspace{0.006\linewidth}\parbox[t]{0.04\linewidth}{\centering\strut \raisebox{0.08ex}{\ding{55}}}\hspace{0.006\linewidth}\parbox[t]{0.10\linewidth}{\centering\strut +11.18\%}\hspace{0.006\linewidth}\parbox[t]{0.115\linewidth}{\centering\strut $5.731\times 10^{-6}$}\hspace{0.006\linewidth}\parbox[t]{0.671\linewidth}{local full-UCCSD seed sweep 337}\par\smallskip
\noindent{\setlength{\fboxsep}{0pt}\colorbox{black!4}{\parbox{\linewidth}{\parbox[t]{0.05\linewidth}{\centering\strut 91}\hspace{0.006\linewidth}\parbox[t]{0.04\linewidth}{\centering\strut \raisebox{0.08ex}{\ding{55}}}\hspace{0.006\linewidth}\parbox[t]{0.10\linewidth}{\centering\strut -12.59\%}\hspace{0.006\linewidth}\parbox[t]{0.115\linewidth}{\centering\strut $6.453\times 10^{-6}$}\hspace{0.006\linewidth}\parbox[t]{0.671\linewidth}{local full-UCCSD seed sweep 347}}}}\par\smallskip
\noindent\parbox[t]{0.05\linewidth}{\centering\strut 92}\hspace{0.006\linewidth}\parbox[t]{0.04\linewidth}{\centering\strut \raisebox{0.08ex}{\ding{55}}}\hspace{0.006\linewidth}\parbox[t]{0.10\linewidth}{\centering\strut +10.71\%}\hspace{0.006\linewidth}\parbox[t]{0.115\linewidth}{\centering\strut $5.761\times 10^{-6}$}\hspace{0.006\linewidth}\parbox[t]{0.671\linewidth}{local full-UCCSD seed sweep 349}\par\smallskip
\noindent{\setlength{\fboxsep}{0pt}\colorbox{black!4}{\parbox{\linewidth}{\parbox[t]{0.05\linewidth}{\centering\strut 93}\hspace{0.006\linewidth}\parbox[t]{0.04\linewidth}{\centering\strut \raisebox{0.08ex}{\ding{55}}}\hspace{0.006\linewidth}\parbox[t]{0.10\linewidth}{\centering\strut -12\%}\hspace{0.006\linewidth}\parbox[t]{0.115\linewidth}{\centering\strut $6.453\times 10^{-6}$}\hspace{0.006\linewidth}\parbox[t]{0.671\linewidth}{local full-UCCSD seed sweep 353}}}}\par\smallskip
\noindent\parbox[t]{0.05\linewidth}{\centering\strut 94}\hspace{0.006\linewidth}\parbox[t]{0.04\linewidth}{\centering\strut \raisebox{0.08ex}{\ding{55}}}\hspace{0.006\linewidth}\parbox[t]{0.10\linewidth}{\centering\strut +0\%}\hspace{0.006\linewidth}\parbox[t]{0.115\linewidth}{\centering\strut $6.453\times 10^{-6}$}\hspace{0.006\linewidth}\parbox[t]{0.671\linewidth}{local full-UCCSD seed sweep 359}\par\smallskip
\noindent{\setlength{\fboxsep}{0pt}\colorbox{black!4}{\parbox{\linewidth}{\parbox[t]{0.05\linewidth}{\centering\strut 95}\hspace{0.006\linewidth}\parbox[t]{0.04\linewidth}{\centering\strut \raisebox{0.08ex}{\ding{55}}}\hspace{0.006\linewidth}\parbox[t]{0.10\linewidth}{\centering\strut +13.23\%}\hspace{0.006\linewidth}\parbox[t]{0.115\linewidth}{\centering\strut $5.599\times 10^{-6}$}\hspace{0.006\linewidth}\parbox[t]{0.671\linewidth}{local full-UCCSD seed sweep 367}}}}\par\smallskip
\noindent\parbox[t]{0.05\linewidth}{\centering\strut 96}\hspace{0.006\linewidth}\parbox[t]{0.04\linewidth}{\centering\strut \raisebox{0.08ex}{\ding{55}}}\hspace{0.006\linewidth}\parbox[t]{0.10\linewidth}{\centering\strut -15.24\%}\hspace{0.006\linewidth}\parbox[t]{0.115\linewidth}{\centering\strut $6.453\times 10^{-6}$}\hspace{0.006\linewidth}\parbox[t]{0.671\linewidth}{local full-UCCSD seed sweep 373}\par\smallskip
\noindent{\setlength{\fboxsep}{0pt}\colorbox{black!4}{\parbox{\linewidth}{\parbox[t]{0.05\linewidth}{\centering\strut 97}\hspace{0.006\linewidth}\parbox[t]{0.04\linewidth}{\centering\strut \raisebox{0.08ex}{\ding{55}}}\hspace{0.006\linewidth}\parbox[t]{0.10\linewidth}{\centering\strut +0\%}\hspace{0.006\linewidth}\parbox[t]{0.115\linewidth}{\centering\strut $6.453\times 10^{-6}$}\hspace{0.006\linewidth}\parbox[t]{0.671\linewidth}{local full-UCCSD seed sweep 379}}}}\par\smallskip
\noindent\parbox[t]{0.05\linewidth}{\centering\strut 98}\hspace{0.006\linewidth}\parbox[t]{0.04\linewidth}{\centering\strut \raisebox{0.08ex}{\ding{55}}}\hspace{0.006\linewidth}\parbox[t]{0.10\linewidth}{\centering\strut +0\%}\hspace{0.006\linewidth}\parbox[t]{0.115\linewidth}{\centering\strut $6.453\times 10^{-6}$}\hspace{0.006\linewidth}\parbox[t]{0.671\linewidth}{local full-UCCSD seed sweep 383}\par\smallskip
\noindent{\setlength{\fboxsep}{0pt}\colorbox{black!4}{\parbox{\linewidth}{\parbox[t]{0.05\linewidth}{\centering\strut 99}\hspace{0.006\linewidth}\parbox[t]{0.04\linewidth}{\centering\strut \raisebox{0.08ex}{\ding{55}}}\hspace{0.006\linewidth}\parbox[t]{0.10\linewidth}{\centering\strut +9.61\%}\hspace{0.006\linewidth}\parbox[t]{0.115\linewidth}{\centering\strut $5.833\times 10^{-6}$}\hspace{0.006\linewidth}\parbox[t]{0.671\linewidth}{local full-UCCSD seed sweep 389}}}}\par\smallskip
\noindent\parbox[t]{0.05\linewidth}{\centering\strut 100}\hspace{0.006\linewidth}\parbox[t]{0.04\linewidth}{\centering\strut \raisebox{0.08ex}{\ding{55}}}\hspace{0.006\linewidth}\parbox[t]{0.10\linewidth}{\centering\strut -10.63\%}\hspace{0.006\linewidth}\parbox[t]{0.115\linewidth}{\centering\strut $6.453\times 10^{-6}$}\hspace{0.006\linewidth}\parbox[t]{0.671\linewidth}{local full-UCCSD seed sweep 397}\par\smallskip
\noindent\rule{\linewidth}{0.5pt}
\par\medskip

%% file: generated_tn_log_tables.tex
\refstepcounter{table}
\label{tab:supp_tn_log_heisenberg_xxx_64}
\noindent\textbf{Table \thetable.} Tensor-network mutation log for Heisenberg XXX, $L=64$. Relative gain is measured against the immediately previous scored iteration using the live final-energy objective.
\par\medskip
\noindent\rule{\linewidth}{0.5pt}
\noindent{\setlength{\fboxsep}{0pt}\colorbox{black!4}{\parbox{\linewidth}{\parbox[c][2.2em][c]{0.05\linewidth}{\centering \textbf{Iter.}}\hspace{0.006\linewidth}\parbox[c][2.2em][c]{0.04\linewidth}{\centering \textbf{A/R}}\hspace{0.006\linewidth}\parbox[c][2.2em][c]{0.10\linewidth}{\centering \textbf{Relative gain}}\hspace{0.006\linewidth}\parbox[c][2.2em][c]{0.115\linewidth}{\centering \textbf{Score}}\hspace{0.006\linewidth}\parbox[c][2.2em][c]{0.671\linewidth}{\centering \textbf{Mutation summary}}}}}\par\smallskip
\noindent\rule{\linewidth}{0.5pt}
\noindent{\setlength{\fboxsep}{0pt}\colorbox{black!4}{\parbox{\linewidth}{\parbox[t]{0.05\linewidth}{\centering\strut 1}\hspace{0.006\linewidth}\parbox[t]{0.04\linewidth}{\centering\strut \raisebox{0.08ex}{\ding{55}}}\hspace{0.006\linewidth}\parbox[t]{0.10\linewidth}{\centering\strut ---}\hspace{0.006\linewidth}\parbox[t]{0.115\linewidth}{\centering\strut ---}\hspace{0.006\linewidth}\parbox[t]{0.671\linewidth}{mutate to dmrg2 neel product state, bond\_schedule 8-12-16-24-32-48-64-80-96-128, cutoff 1e-10, tol 1e-8, max\_sweeps 32, ncv 8}}}}\par\smallskip
\noindent\parbox[t]{0.05\linewidth}{\centering\strut 2}\hspace{0.006\linewidth}\parbox[t]{0.04\linewidth}{\centering\strut \raisebox{0.08ex}{\ding{51}}}\hspace{0.006\linewidth}\parbox[t]{0.10\linewidth}{\centering\strut ---}\hspace{0.006\linewidth}\parbox[t]{0.115\linewidth}{\centering\strut -28.175423}\hspace{0.006\linewidth}\parbox[t]{0.671\linewidth}{mutate to dmrg2 neel product state, bond\_schedule 8-12-16-24-32-48, cutoff 1e-8, tol 1e-6, max\_sweeps 16, ncv 6}\par\smallskip
\noindent{\setlength{\fboxsep}{0pt}\colorbox{black!4}{\parbox{\linewidth}{\parbox[t]{0.05\linewidth}{\centering\strut 3}\hspace{0.006\linewidth}\parbox[t]{0.04\linewidth}{\centering\strut \raisebox{0.08ex}{\ding{51}}}\hspace{0.006\linewidth}\parbox[t]{0.10\linewidth}{\centering\strut +0\%}\hspace{0.006\linewidth}\parbox[t]{0.115\linewidth}{\centering\strut -28.175425}\hspace{0.006\linewidth}\parbox[t]{0.671\linewidth}{mutate bond\_schedule to 8-12-16-24-32-48-64, cutoff 1e-10, tol 1e-7, max\_sweeps 24, ncv 8 on dmrg2 neel}}}}\par\smallskip
\noindent\parbox[t]{0.05\linewidth}{\centering\strut 4}\hspace{0.006\linewidth}\parbox[t]{0.04\linewidth}{\centering\strut \raisebox{0.08ex}{\ding{51}}}\hspace{0.006\linewidth}\parbox[t]{0.10\linewidth}{\centering\strut +0\%}\hspace{0.006\linewidth}\parbox[t]{0.115\linewidth}{\centering\strut -28.175425}\hspace{0.006\linewidth}\parbox[t]{0.671\linewidth}{mutate init\_bond\_dim to 2 on dmrg2 neel, bond\_schedule 8-12-16-24-32-48-64, cutoff 1e-10, tol 1e-7, max\_sweeps 24, ncv 8}\par\smallskip
\noindent{\setlength{\fboxsep}{0pt}\colorbox{black!4}{\parbox{\linewidth}{\parbox[t]{0.05\linewidth}{\centering\strut 5}\hspace{0.006\linewidth}\parbox[t]{0.04\linewidth}{\centering\strut \raisebox{0.08ex}{\ding{51}}}\hspace{0.006\linewidth}\parbox[t]{0.10\linewidth}{\centering\strut +0\%}\hspace{0.006\linewidth}\parbox[t]{0.115\linewidth}{\centering\strut -28.175425}\hspace{0.006\linewidth}\parbox[t]{0.671\linewidth}{mutate dmrg loop to explicit solver.sweep with list bond\_dims on dmrg2 neel ib2 b64 s24 tol 1e-7}}}}\par\smallskip
\noindent\parbox[t]{0.05\linewidth}{\centering\strut 6}\hspace{0.006\linewidth}\parbox[t]{0.04\linewidth}{\centering\strut \raisebox{0.08ex}{\ding{51}}}\hspace{0.006\linewidth}\parbox[t]{0.10\linewidth}{\centering\strut +0\%}\hspace{0.006\linewidth}\parbox[t]{0.115\linewidth}{\centering\strut -28.175425}\hspace{0.006\linewidth}\parbox[t]{0.671\linewidth}{mutate bond\_schedule to 8-12-16-24-32-48-64-80 and max\_sweeps 20 on sweep-based dmrg2 neel ib2}\par\smallskip
\noindent{\setlength{\fboxsep}{0pt}\colorbox{black!4}{\parbox{\linewidth}{\parbox[t]{0.05\linewidth}{\centering\strut 7}\hspace{0.006\linewidth}\parbox[t]{0.04\linewidth}{\centering\strut \raisebox{0.08ex}{\ding{51}}}\hspace{0.006\linewidth}\parbox[t]{0.10\linewidth}{\centering\strut +0\%}\hspace{0.006\linewidth}\parbox[t]{0.115\linewidth}{\centering\strut -28.175425}\hspace{0.006\linewidth}\parbox[t]{0.671\linewidth}{mutate bond\_schedule to 8-12-16-24-32-48-64-80-96 and max\_sweeps 18 on sweep-based dmrg2 neel ib2}}}}\par\smallskip
\noindent\parbox[t]{0.05\linewidth}{\centering\strut 8}\hspace{0.006\linewidth}\parbox[t]{0.04\linewidth}{\centering\strut \raisebox{0.08ex}{\ding{55}}}\hspace{0.006\linewidth}\parbox[t]{0.10\linewidth}{\centering\strut +0\%}\hspace{0.006\linewidth}\parbox[t]{0.115\linewidth}{\centering\strut -28.175425}\hspace{0.006\linewidth}\parbox[t]{0.671\linewidth}{mutate bond\_schedule to 8-12-16-24-32-48-64-80-96-128 and max\_sweeps 16 on sweep-based dmrg2 neel ib2}\par\smallskip
\noindent{\setlength{\fboxsep}{0pt}\colorbox{black!4}{\parbox{\linewidth}{\parbox[t]{0.05\linewidth}{\centering\strut 9}\hspace{0.006\linewidth}\parbox[t]{0.04\linewidth}{\centering\strut \raisebox{0.08ex}{\ding{51}}}\hspace{0.006\linewidth}\parbox[t]{0.10\linewidth}{\centering\strut +0\%}\hspace{0.006\linewidth}\parbox[t]{0.115\linewidth}{\centering\strut -28.175425}\hspace{0.006\linewidth}\parbox[t]{0.671\linewidth}{mutate bond\_schedule 8-12-16-24-32-48-64-80-96, max\_sweeps 17 on sweep-based dmrg}}}}\par\smallskip
\noindent\parbox[t]{0.05\linewidth}{\centering\strut 10}\hspace{0.006\linewidth}\parbox[t]{0.04\linewidth}{\centering\strut \raisebox{0.08ex}{\ding{51}}}\hspace{0.006\linewidth}\parbox[t]{0.10\linewidth}{\centering\strut +0\%}\hspace{0.006\linewidth}\parbox[t]{0.115\linewidth}{\centering\strut -28.175425}\hspace{0.006\linewidth}\parbox[t]{0.671\linewidth}{mutate bond\_schedule 8-12-16-24-32-48-64-80-96, max\_sweeps 19 on sweep-based dmrg}\par\smallskip
\noindent{\setlength{\fboxsep}{0pt}\colorbox{black!4}{\parbox{\linewidth}{\parbox[t]{0.05\linewidth}{\centering\strut 11}\hspace{0.006\linewidth}\parbox[t]{0.04\linewidth}{\centering\strut \raisebox{0.08ex}{\ding{55}}}\hspace{0.006\linewidth}\parbox[t]{0.10\linewidth}{\centering\strut +0\%}\hspace{0.006\linewidth}\parbox[t]{0.115\linewidth}{\centering\strut -28.175425}\hspace{0.006\linewidth}\parbox[t]{0.671\linewidth}{mutate bond\_schedule 8-12-16-24-32-48-64-80, max\_sweeps 22 on sweep-based dmrg}}}}\par\smallskip
\noindent\parbox[t]{0.05\linewidth}{\centering\strut 12}\hspace{0.006\linewidth}\parbox[t]{0.04\linewidth}{\centering\strut \raisebox{0.08ex}{\ding{55}}}\hspace{0.006\linewidth}\parbox[t]{0.10\linewidth}{\centering\strut +0\%}\hspace{0.006\linewidth}\parbox[t]{0.115\linewidth}{\centering\strut -28.175425}\hspace{0.006\linewidth}\parbox[t]{0.671\linewidth}{mutate bond\_schedule 8-12-16-24-32-48-64-80-96-112, max\_sweeps 17 on sweep-based dmrg}\par\smallskip
\noindent{\setlength{\fboxsep}{0pt}\colorbox{black!4}{\parbox{\linewidth}{\parbox[t]{0.05\linewidth}{\centering\strut 13}\hspace{0.006\linewidth}\parbox[t]{0.04\linewidth}{\centering\strut \raisebox{0.08ex}{\ding{55}}}\hspace{0.006\linewidth}\parbox[t]{0.10\linewidth}{\centering\strut +0\%}\hspace{0.006\linewidth}\parbox[t]{0.115\linewidth}{\centering\strut -28.175425}\hspace{0.006\linewidth}\parbox[t]{0.671\linewidth}{mutate bond\_schedule 8-12-16-24-32-48-64-80-96-112 on sweep-based dmrg}}}}\par\smallskip
\noindent\parbox[t]{0.05\linewidth}{\centering\strut 14}\hspace{0.006\linewidth}\parbox[t]{0.04\linewidth}{\centering\strut \raisebox{0.08ex}{\ding{55}}}\hspace{0.006\linewidth}\parbox[t]{0.10\linewidth}{\centering\strut +0\%}\hspace{0.006\linewidth}\parbox[t]{0.115\linewidth}{\centering\strut -28.175425}\hspace{0.006\linewidth}\parbox[t]{0.671\linewidth}{mutate bond\_schedule 8-12-16-24-32-48-64-80-96-112-128, max\_sweeps 18 on sweep-based dmrg}\par\smallskip
\noindent{\setlength{\fboxsep}{0pt}\colorbox{black!4}{\parbox{\linewidth}{\parbox[t]{0.05\linewidth}{\centering\strut 15}\hspace{0.006\linewidth}\parbox[t]{0.04\linewidth}{\centering\strut \raisebox{0.08ex}{\ding{55}}}\hspace{0.006\linewidth}\parbox[t]{0.10\linewidth}{\centering\strut +0\%}\hspace{0.006\linewidth}\parbox[t]{0.115\linewidth}{\centering\strut -28.175425}\hspace{0.006\linewidth}\parbox[t]{0.671\linewidth}{mutate bond\_schedule 8-12-16-24-32-48-64-80, max\_sweeps 24 on sweep-based dmrg}}}}\par\smallskip
\noindent\parbox[t]{0.05\linewidth}{\centering\strut 16}\hspace{0.006\linewidth}\parbox[t]{0.04\linewidth}{\centering\strut \raisebox{0.08ex}{\ding{55}}}\hspace{0.006\linewidth}\parbox[t]{0.10\linewidth}{\centering\strut +0\%}\hspace{0.006\linewidth}\parbox[t]{0.115\linewidth}{\centering\strut -28.175425}\hspace{0.006\linewidth}\parbox[t]{0.671\linewidth}{mutate bond\_schedule 8-12-16-24-32-48-64, max\_sweeps 28 on sweep-based dmrg}\par\smallskip
\noindent{\setlength{\fboxsep}{0pt}\colorbox{black!4}{\parbox{\linewidth}{\parbox[t]{0.05\linewidth}{\centering\strut 17}\hspace{0.006\linewidth}\parbox[t]{0.04\linewidth}{\centering\strut \raisebox{0.08ex}{\ding{55}}}\hspace{0.006\linewidth}\parbox[t]{0.10\linewidth}{\centering\strut +0\%}\hspace{0.006\linewidth}\parbox[t]{0.115\linewidth}{\centering\strut -28.175425}\hspace{0.006\linewidth}\parbox[t]{0.671\linewidth}{mutate bond\_schedule 8-12-16-24-32-48-64-80-96, init\_bond\_dim 1 on sweep-based dmrg}}}}\par\smallskip
\noindent\parbox[t]{0.05\linewidth}{\centering\strut 18}\hspace{0.006\linewidth}\parbox[t]{0.04\linewidth}{\centering\strut \raisebox{0.08ex}{\ding{51}}}\hspace{0.006\linewidth}\parbox[t]{0.10\linewidth}{\centering\strut +0\%}\hspace{0.006\linewidth}\parbox[t]{0.115\linewidth}{\centering\strut -28.175425}\hspace{0.006\linewidth}\parbox[t]{0.671\linewidth}{mutate bond\_schedule 8-12-16-24-32-48-64-80-96, init\_bond\_dim 4 on sweep-based dmrg}\par\smallskip
\noindent{\setlength{\fboxsep}{0pt}\colorbox{black!4}{\parbox{\linewidth}{\parbox[t]{0.05\linewidth}{\centering\strut 19}\hspace{0.006\linewidth}\parbox[t]{0.04\linewidth}{\centering\strut \raisebox{0.08ex}{\ding{51}}}\hspace{0.006\linewidth}\parbox[t]{0.10\linewidth}{\centering\strut +0\%}\hspace{0.006\linewidth}\parbox[t]{0.115\linewidth}{\centering\strut -28.175425}\hspace{0.006\linewidth}\parbox[t]{0.671\linewidth}{mutate bond\_schedule 8-12-16-24-32-48-64-80-96, init\_bond\_dim 8 on sweep-based dmrg}}}}\par\smallskip
\noindent\parbox[t]{0.05\linewidth}{\centering\strut 20}\hspace{0.006\linewidth}\parbox[t]{0.04\linewidth}{\centering\strut \raisebox{0.08ex}{\ding{55}}}\hspace{0.006\linewidth}\parbox[t]{0.10\linewidth}{\centering\strut +0\%}\hspace{0.006\linewidth}\parbox[t]{0.115\linewidth}{\centering\strut -28.175425}\hspace{0.006\linewidth}\parbox[t]{0.671\linewidth}{mutate bond\_schedule 8-12-16-24-32-48-64-80-96, ncv 6 on sweep-based dmrg}\par\smallskip
\noindent{\setlength{\fboxsep}{0pt}\colorbox{black!4}{\parbox{\linewidth}{\parbox[t]{0.05\linewidth}{\centering\strut 21}\hspace{0.006\linewidth}\parbox[t]{0.04\linewidth}{\centering\strut \raisebox{0.08ex}{\ding{55}}}\hspace{0.006\linewidth}\parbox[t]{0.10\linewidth}{\centering\strut +0\%}\hspace{0.006\linewidth}\parbox[t]{0.115\linewidth}{\centering\strut -28.175425}\hspace{0.006\linewidth}\parbox[t]{0.671\linewidth}{mutate bond\_schedule 8-12-16-24-32-48-64-80-96, ncv 10 on sweep-based dmrg}}}}\par\smallskip
\noindent\parbox[t]{0.05\linewidth}{\centering\strut 22}\hspace{0.006\linewidth}\parbox[t]{0.04\linewidth}{\centering\strut \raisebox{0.08ex}{\ding{55}}}\hspace{0.006\linewidth}\parbox[t]{0.10\linewidth}{\centering\strut +0\%}\hspace{0.006\linewidth}\parbox[t]{0.115\linewidth}{\centering\strut -28.175425}\hspace{0.006\linewidth}\parbox[t]{0.671\linewidth}{mutate bond\_schedule 8-12-16-24-32-48-64-80-96, ncv 12 on sweep-based dmrg}\par\smallskip
\noindent{\setlength{\fboxsep}{0pt}\colorbox{black!4}{\parbox{\linewidth}{\parbox[t]{0.05\linewidth}{\centering\strut 23}\hspace{0.006\linewidth}\parbox[t]{0.04\linewidth}{\centering\strut \raisebox{0.08ex}{\ding{55}}}\hspace{0.006\linewidth}\parbox[t]{0.10\linewidth}{\centering\strut +0\%}\hspace{0.006\linewidth}\parbox[t]{0.115\linewidth}{\centering\strut -28.175425}\hspace{0.006\linewidth}\parbox[t]{0.671\linewidth}{mutate bond\_schedule 8-12-16-24-32-48-64-80-96, ncv 14 on sweep-based dmrg}}}}\par\smallskip
\noindent\parbox[t]{0.05\linewidth}{\centering\strut 24}\hspace{0.006\linewidth}\parbox[t]{0.04\linewidth}{\centering\strut \raisebox{0.08ex}{\ding{55}}}\hspace{0.006\linewidth}\parbox[t]{0.10\linewidth}{\centering\strut +0\%}\hspace{0.006\linewidth}\parbox[t]{0.115\linewidth}{\centering\strut -28.175425}\hspace{0.006\linewidth}\parbox[t]{0.671\linewidth}{mutate bond\_schedule 8-12-16-24-32-48-64-80-96, ncv 16 on sweep-based dmrg}\par\smallskip
\noindent{\setlength{\fboxsep}{0pt}\colorbox{black!4}{\parbox{\linewidth}{\parbox[t]{0.05\linewidth}{\centering\strut 25}\hspace{0.006\linewidth}\parbox[t]{0.04\linewidth}{\centering\strut \raisebox{0.08ex}{\ding{55}}}\hspace{0.006\linewidth}\parbox[t]{0.10\linewidth}{\centering\strut +0\%}\hspace{0.006\linewidth}\parbox[t]{0.115\linewidth}{\centering\strut -28.175425}\hspace{0.006\linewidth}\parbox[t]{0.671\linewidth}{mutate bond\_schedule 8-12-16-24-32-48-64-80-96, tol 1e-06 on sweep-based dmrg}}}}\par\smallskip
\noindent\parbox[t]{0.05\linewidth}{\centering\strut 26}\hspace{0.006\linewidth}\parbox[t]{0.04\linewidth}{\centering\strut \raisebox{0.08ex}{\ding{55}}}\hspace{0.006\linewidth}\parbox[t]{0.10\linewidth}{\centering\strut +0\%}\hspace{0.006\linewidth}\parbox[t]{0.115\linewidth}{\centering\strut -28.175425}\hspace{0.006\linewidth}\parbox[t]{0.671\linewidth}{mutate bond\_schedule 8-12-16-24-32-48-64-80-96, tol 3e-07 on sweep-based dmrg}\par\smallskip
\noindent{\setlength{\fboxsep}{0pt}\colorbox{black!4}{\parbox{\linewidth}{\parbox[t]{0.05\linewidth}{\centering\strut 27}\hspace{0.006\linewidth}\parbox[t]{0.04\linewidth}{\centering\strut \raisebox{0.08ex}{\ding{55}}}\hspace{0.006\linewidth}\parbox[t]{0.10\linewidth}{\centering\strut +0\%}\hspace{0.006\linewidth}\parbox[t]{0.115\linewidth}{\centering\strut -28.175425}\hspace{0.006\linewidth}\parbox[t]{0.671\linewidth}{mutate bond\_schedule 8-12-16-24-32-48-64-80-96, tol 5e-08 on sweep-based dmrg}}}}\par\smallskip
\noindent\parbox[t]{0.05\linewidth}{\centering\strut 28}\hspace{0.006\linewidth}\parbox[t]{0.04\linewidth}{\centering\strut \raisebox{0.08ex}{\ding{55}}}\hspace{0.006\linewidth}\parbox[t]{0.10\linewidth}{\centering\strut +0\%}\hspace{0.006\linewidth}\parbox[t]{0.115\linewidth}{\centering\strut -28.175425}\hspace{0.006\linewidth}\parbox[t]{0.671\linewidth}{mutate bond\_schedule 8-12-16-24-32-48-64-80-96, tol 1e-08 on sweep-based dmrg}\par\smallskip
\noindent{\setlength{\fboxsep}{0pt}\colorbox{black!4}{\parbox{\linewidth}{\parbox[t]{0.05\linewidth}{\centering\strut 29}\hspace{0.006\linewidth}\parbox[t]{0.04\linewidth}{\centering\strut \raisebox{0.08ex}{\ding{55}}}\hspace{0.006\linewidth}\parbox[t]{0.10\linewidth}{\centering\strut +0\%}\hspace{0.006\linewidth}\parbox[t]{0.115\linewidth}{\centering\strut -28.175425}\hspace{0.006\linewidth}\parbox[t]{0.671\linewidth}{mutate bond\_schedule 8-12-16-24-32-48-64-80-96, tol 3e-09 on sweep-based dmrg}}}}\par\smallskip
\noindent\parbox[t]{0.05\linewidth}{\centering\strut 30}\hspace{0.006\linewidth}\parbox[t]{0.04\linewidth}{\centering\strut \raisebox{0.08ex}{\ding{55}}}\hspace{0.006\linewidth}\parbox[t]{0.10\linewidth}{\centering\strut +0\%}\hspace{0.006\linewidth}\parbox[t]{0.115\linewidth}{\centering\strut -28.175423}\hspace{0.006\linewidth}\parbox[t]{0.671\linewidth}{mutate bond\_schedule 8-12-16-24-32-48-64-80-96, cutoff 1e-08 on sweep-based dmrg}\par\smallskip
\noindent{\setlength{\fboxsep}{0pt}\colorbox{black!4}{\parbox{\linewidth}{\parbox[t]{0.05\linewidth}{\centering\strut 31}\hspace{0.006\linewidth}\parbox[t]{0.04\linewidth}{\centering\strut \raisebox{0.08ex}{\ding{55}}}\hspace{0.006\linewidth}\parbox[t]{0.10\linewidth}{\centering\strut ---}\hspace{0.006\linewidth}\parbox[t]{0.115\linewidth}{\centering\strut ---}\hspace{0.006\linewidth}\parbox[t]{0.671\linewidth}{mutate bond\_schedule 8-12-16-24-32-48-64-80-96, cutoff 1e-12 on sweep-based dmrg}}}}\par\smallskip
\noindent\parbox[t]{0.05\linewidth}{\centering\strut 32}\hspace{0.006\linewidth}\parbox[t]{0.04\linewidth}{\centering\strut \raisebox{0.08ex}{\ding{51}}}\hspace{0.006\linewidth}\parbox[t]{0.10\linewidth}{\centering\strut +0\%}\hspace{0.006\linewidth}\parbox[t]{0.115\linewidth}{\centering\strut -28.175425}\hspace{0.006\linewidth}\parbox[t]{0.671\linewidth}{mutate bond\_schedule 8-12-16-24-32-48-64-80-96, cutoff 1e-14 on sweep-based dmrg}\par\smallskip
\noindent{\setlength{\fboxsep}{0pt}\colorbox{black!4}{\parbox{\linewidth}{\parbox[t]{0.05\linewidth}{\centering\strut 33}\hspace{0.006\linewidth}\parbox[t]{0.04\linewidth}{\centering\strut \raisebox{0.08ex}{\ding{55}}}\hspace{0.006\linewidth}\parbox[t]{0.10\linewidth}{\centering\strut ---}\hspace{0.006\linewidth}\parbox[t]{0.115\linewidth}{\centering\strut ---}\hspace{0.006\linewidth}\parbox[t]{0.671\linewidth}{mutate init\_state product\_up, bond\_schedule 8-12-16-24-32-48-64-80-96 on sweep-based dmrg}}}}\par\smallskip
\noindent\parbox[t]{0.05\linewidth}{\centering\strut 34}\hspace{0.006\linewidth}\parbox[t]{0.04\linewidth}{\centering\strut \raisebox{0.08ex}{\ding{55}}}\hspace{0.006\linewidth}\parbox[t]{0.10\linewidth}{\centering\strut ---}\hspace{0.006\linewidth}\parbox[t]{0.115\linewidth}{\centering\strut ---}\hspace{0.006\linewidth}\parbox[t]{0.671\linewidth}{mutate init\_state plus, bond\_schedule 8-12-16-24-32-48-64-80-96 on sweep-based dmrg}\par\smallskip
\noindent{\setlength{\fboxsep}{0pt}\colorbox{black!4}{\parbox{\linewidth}{\parbox[t]{0.05\linewidth}{\centering\strut 35}\hspace{0.006\linewidth}\parbox[t]{0.04\linewidth}{\centering\strut \raisebox{0.08ex}{\ding{55}}}\hspace{0.006\linewidth}\parbox[t]{0.10\linewidth}{\centering\strut ---}\hspace{0.006\linewidth}\parbox[t]{0.115\linewidth}{\centering\strut ---}\hspace{0.006\linewidth}\parbox[t]{0.671\linewidth}{mutate init\_state random, bond\_schedule 8-12-16-24-32-48-64-80-96 on sweep-based dmrg}}}}\par\smallskip
\noindent\parbox[t]{0.05\linewidth}{\centering\strut 36}\hspace{0.006\linewidth}\parbox[t]{0.04\linewidth}{\centering\strut \raisebox{0.08ex}{\ding{55}}}\hspace{0.006\linewidth}\parbox[t]{0.10\linewidth}{\centering\strut +0\%}\hspace{0.006\linewidth}\parbox[t]{0.115\linewidth}{\centering\strut -28.175425}\hspace{0.006\linewidth}\parbox[t]{0.671\linewidth}{mutate bond\_schedule 8-12-16-24-32-48-64-80, max\_sweeps 16, init\_bond\_dim 1 on sweep-based dmrg}\par\smallskip
\noindent{\setlength{\fboxsep}{0pt}\colorbox{black!4}{\parbox{\linewidth}{\parbox[t]{0.05\linewidth}{\centering\strut 37}\hspace{0.006\linewidth}\parbox[t]{0.04\linewidth}{\centering\strut \raisebox{0.08ex}{\ding{55}}}\hspace{0.006\linewidth}\parbox[t]{0.10\linewidth}{\centering\strut +0\%}\hspace{0.006\linewidth}\parbox[t]{0.115\linewidth}{\centering\strut -28.175425}\hspace{0.006\linewidth}\parbox[t]{0.671\linewidth}{mutate bond\_schedule 8-12-16-24-32-48-64-80, max\_sweeps 16, init\_bond\_dim 2 on sweep-based dmrg}}}}\par\smallskip
\noindent\parbox[t]{0.05\linewidth}{\centering\strut 38}\hspace{0.006\linewidth}\parbox[t]{0.04\linewidth}{\centering\strut \raisebox{0.08ex}{\ding{55}}}\hspace{0.006\linewidth}\parbox[t]{0.10\linewidth}{\centering\strut +0\%}\hspace{0.006\linewidth}\parbox[t]{0.115\linewidth}{\centering\strut -28.175425}\hspace{0.006\linewidth}\parbox[t]{0.671\linewidth}{mutate bond\_schedule 8-12-16-24-32-48-64-80, max\_sweeps 16, init\_bond\_dim 4 on sweep-based dmrg}\par\smallskip
\noindent{\setlength{\fboxsep}{0pt}\colorbox{black!4}{\parbox{\linewidth}{\parbox[t]{0.05\linewidth}{\centering\strut 39}\hspace{0.006\linewidth}\parbox[t]{0.04\linewidth}{\centering\strut \raisebox{0.08ex}{\ding{55}}}\hspace{0.006\linewidth}\parbox[t]{0.10\linewidth}{\centering\strut +0\%}\hspace{0.006\linewidth}\parbox[t]{0.115\linewidth}{\centering\strut -28.175425}\hspace{0.006\linewidth}\parbox[t]{0.671\linewidth}{mutate bond\_schedule 8-12-16-24-32-48-64-80, max\_sweeps 18, init\_bond\_dim 1 on sweep-based dmrg}}}}\par\smallskip
\noindent\parbox[t]{0.05\linewidth}{\centering\strut 40}\hspace{0.006\linewidth}\parbox[t]{0.04\linewidth}{\centering\strut \raisebox{0.08ex}{\ding{55}}}\hspace{0.006\linewidth}\parbox[t]{0.10\linewidth}{\centering\strut +0\%}\hspace{0.006\linewidth}\parbox[t]{0.115\linewidth}{\centering\strut -28.175425}\hspace{0.006\linewidth}\parbox[t]{0.671\linewidth}{mutate bond\_schedule 8-12-16-24-32-48-64-80, max\_sweeps 18, init\_bond\_dim 2 on sweep-based dmrg}\par\smallskip
\noindent{\setlength{\fboxsep}{0pt}\colorbox{black!4}{\parbox{\linewidth}{\parbox[t]{0.05\linewidth}{\centering\strut 41}\hspace{0.006\linewidth}\parbox[t]{0.04\linewidth}{\centering\strut \raisebox{0.08ex}{\ding{55}}}\hspace{0.006\linewidth}\parbox[t]{0.10\linewidth}{\centering\strut +0\%}\hspace{0.006\linewidth}\parbox[t]{0.115\linewidth}{\centering\strut -28.175425}\hspace{0.006\linewidth}\parbox[t]{0.671\linewidth}{mutate bond\_schedule 8-12-16-24-32-48-64-80, max\_sweeps 18, init\_bond\_dim 4 on sweep-based dmrg}}}}\par\smallskip
\noindent\parbox[t]{0.05\linewidth}{\centering\strut 42}\hspace{0.006\linewidth}\parbox[t]{0.04\linewidth}{\centering\strut \raisebox{0.08ex}{\ding{55}}}\hspace{0.006\linewidth}\parbox[t]{0.10\linewidth}{\centering\strut +0\%}\hspace{0.006\linewidth}\parbox[t]{0.115\linewidth}{\centering\strut -28.175425}\hspace{0.006\linewidth}\parbox[t]{0.671\linewidth}{mutate bond\_schedule 8-12-16-24-32-48-64-80, max\_sweeps 20, init\_bond\_dim 1 on sweep-based dmrg}\par\smallskip
\noindent{\setlength{\fboxsep}{0pt}\colorbox{black!4}{\parbox{\linewidth}{\parbox[t]{0.05\linewidth}{\centering\strut 43}\hspace{0.006\linewidth}\parbox[t]{0.04\linewidth}{\centering\strut \raisebox{0.08ex}{\ding{55}}}\hspace{0.006\linewidth}\parbox[t]{0.10\linewidth}{\centering\strut +0\%}\hspace{0.006\linewidth}\parbox[t]{0.115\linewidth}{\centering\strut -28.175425}\hspace{0.006\linewidth}\parbox[t]{0.671\linewidth}{mutate bond\_schedule 8-12-16-24-32-48-64-80, max\_sweeps 20, init\_bond\_dim 2 on sweep-based dmrg}}}}\par\smallskip
\noindent\parbox[t]{0.05\linewidth}{\centering\strut 44}\hspace{0.006\linewidth}\parbox[t]{0.04\linewidth}{\centering\strut \raisebox{0.08ex}{\ding{55}}}\hspace{0.006\linewidth}\parbox[t]{0.10\linewidth}{\centering\strut +0\%}\hspace{0.006\linewidth}\parbox[t]{0.115\linewidth}{\centering\strut -28.175425}\hspace{0.006\linewidth}\parbox[t]{0.671\linewidth}{mutate bond\_schedule 8-12-16-24-32-48-64-80, max\_sweeps 20, init\_bond\_dim 4 on sweep-based dmrg}\par\smallskip
\noindent{\setlength{\fboxsep}{0pt}\colorbox{black!4}{\parbox{\linewidth}{\parbox[t]{0.05\linewidth}{\centering\strut 45}\hspace{0.006\linewidth}\parbox[t]{0.04\linewidth}{\centering\strut \raisebox{0.08ex}{\ding{55}}}\hspace{0.006\linewidth}\parbox[t]{0.10\linewidth}{\centering\strut ---}\hspace{0.006\linewidth}\parbox[t]{0.115\linewidth}{\centering\strut ---}\hspace{0.006\linewidth}\parbox[t]{0.671\linewidth}{mutate bond\_schedule 8-12-16-24-32-48-64-80, max\_sweeps 22, init\_bond\_dim 1 on sweep-based dmrg}}}}\par\smallskip
\noindent\parbox[t]{0.05\linewidth}{\centering\strut 46}\hspace{0.006\linewidth}\parbox[t]{0.04\linewidth}{\centering\strut \raisebox{0.08ex}{\ding{55}}}\hspace{0.006\linewidth}\parbox[t]{0.10\linewidth}{\centering\strut ---}\hspace{0.006\linewidth}\parbox[t]{0.115\linewidth}{\centering\strut ---}\hspace{0.006\linewidth}\parbox[t]{0.671\linewidth}{mutate bond\_schedule 8-12-16-24-32-48-64-80, max\_sweeps 22, init\_bond\_dim 2 on sweep-based dmrg}\par\smallskip
\noindent{\setlength{\fboxsep}{0pt}\colorbox{black!4}{\parbox{\linewidth}{\parbox[t]{0.05\linewidth}{\centering\strut 47}\hspace{0.006\linewidth}\parbox[t]{0.04\linewidth}{\centering\strut \raisebox{0.08ex}{\ding{55}}}\hspace{0.006\linewidth}\parbox[t]{0.10\linewidth}{\centering\strut +0\%}\hspace{0.006\linewidth}\parbox[t]{0.115\linewidth}{\centering\strut -28.175425}\hspace{0.006\linewidth}\parbox[t]{0.671\linewidth}{mutate bond\_schedule 8-12-16-24-32-48-64-80, max\_sweeps 22, init\_bond\_dim 4 on sweep-based dmrg}}}}\par\smallskip
\noindent\parbox[t]{0.05\linewidth}{\centering\strut 48}\hspace{0.006\linewidth}\parbox[t]{0.04\linewidth}{\centering\strut \raisebox{0.08ex}{\ding{55}}}\hspace{0.006\linewidth}\parbox[t]{0.10\linewidth}{\centering\strut +0\%}\hspace{0.006\linewidth}\parbox[t]{0.115\linewidth}{\centering\strut -28.175425}\hspace{0.006\linewidth}\parbox[t]{0.671\linewidth}{mutate bond\_schedule 8-12-16-24-32-48-64-80-96, max\_sweeps 16, init\_bond\_dim 1 on sweep-based dmrg}\par\smallskip
\noindent{\setlength{\fboxsep}{0pt}\colorbox{black!4}{\parbox{\linewidth}{\parbox[t]{0.05\linewidth}{\centering\strut 49}\hspace{0.006\linewidth}\parbox[t]{0.04\linewidth}{\centering\strut \raisebox{0.08ex}{\ding{55}}}\hspace{0.006\linewidth}\parbox[t]{0.10\linewidth}{\centering\strut +0\%}\hspace{0.006\linewidth}\parbox[t]{0.115\linewidth}{\centering\strut -28.175425}\hspace{0.006\linewidth}\parbox[t]{0.671\linewidth}{mutate bond\_schedule 8-12-16-24-32-48-64-80-96, max\_sweeps 16, init\_bond\_dim 2 on sweep-based dmrg}}}}\par\smallskip
\noindent\parbox[t]{0.05\linewidth}{\centering\strut 50}\hspace{0.006\linewidth}\parbox[t]{0.04\linewidth}{\centering\strut \raisebox{0.08ex}{\ding{55}}}\hspace{0.006\linewidth}\parbox[t]{0.10\linewidth}{\centering\strut +0\%}\hspace{0.006\linewidth}\parbox[t]{0.115\linewidth}{\centering\strut -28.175425}\hspace{0.006\linewidth}\parbox[t]{0.671\linewidth}{mutate bond\_schedule 8-12-16-24-32-48-64-80-96, max\_sweeps 16, init\_bond\_dim 4 on sweep-based dmrg}\par\smallskip
\noindent{\setlength{\fboxsep}{0pt}\colorbox{black!4}{\parbox{\linewidth}{\parbox[t]{0.05\linewidth}{\centering\strut 51}\hspace{0.006\linewidth}\parbox[t]{0.04\linewidth}{\centering\strut \raisebox{0.08ex}{\ding{55}}}\hspace{0.006\linewidth}\parbox[t]{0.10\linewidth}{\centering\strut ---}\hspace{0.006\linewidth}\parbox[t]{0.115\linewidth}{\centering\strut ---}\hspace{0.006\linewidth}\parbox[t]{0.671\linewidth}{mutate bond\_schedule 8-12-16-24-32-48-64-80-96, max\_sweeps 18, init\_bond\_dim 1 on sweep-based dmrg}}}}\par\smallskip
\noindent\parbox[t]{0.05\linewidth}{\centering\strut 52}\hspace{0.006\linewidth}\parbox[t]{0.04\linewidth}{\centering\strut \raisebox{0.08ex}{\ding{55}}}\hspace{0.006\linewidth}\parbox[t]{0.10\linewidth}{\centering\strut ---}\hspace{0.006\linewidth}\parbox[t]{0.115\linewidth}{\centering\strut ---}\hspace{0.006\linewidth}\parbox[t]{0.671\linewidth}{mutate bond\_schedule 8-12-16-24-32-48-64-80-96, max\_sweeps 18, init\_bond\_dim 2 on sweep-based dmrg}\par\smallskip
\noindent{\setlength{\fboxsep}{0pt}\colorbox{black!4}{\parbox{\linewidth}{\parbox[t]{0.05\linewidth}{\centering\strut 53}\hspace{0.006\linewidth}\parbox[t]{0.04\linewidth}{\centering\strut \raisebox{0.08ex}{\ding{55}}}\hspace{0.006\linewidth}\parbox[t]{0.10\linewidth}{\centering\strut +0\%}\hspace{0.006\linewidth}\parbox[t]{0.115\linewidth}{\centering\strut -28.175425}\hspace{0.006\linewidth}\parbox[t]{0.671\linewidth}{mutate bond\_schedule 8-12-16-24-32-48-64-80-96, max\_sweeps 18, init\_bond\_dim 4 on sweep-based dmrg}}}}\par\smallskip
\noindent\parbox[t]{0.05\linewidth}{\centering\strut 54}\hspace{0.006\linewidth}\parbox[t]{0.04\linewidth}{\centering\strut \raisebox{0.08ex}{\ding{55}}}\hspace{0.006\linewidth}\parbox[t]{0.10\linewidth}{\centering\strut ---}\hspace{0.006\linewidth}\parbox[t]{0.115\linewidth}{\centering\strut ---}\hspace{0.006\linewidth}\parbox[t]{0.671\linewidth}{mutate bond\_schedule 8-12-16-24-32-48-64-80-96, max\_sweeps 20, init\_bond\_dim 1 on sweep-based dmrg}\par\smallskip
\noindent{\setlength{\fboxsep}{0pt}\colorbox{black!4}{\parbox{\linewidth}{\parbox[t]{0.05\linewidth}{\centering\strut 55}\hspace{0.006\linewidth}\parbox[t]{0.04\linewidth}{\centering\strut \raisebox{0.08ex}{\ding{55}}}\hspace{0.006\linewidth}\parbox[t]{0.10\linewidth}{\centering\strut ---}\hspace{0.006\linewidth}\parbox[t]{0.115\linewidth}{\centering\strut ---}\hspace{0.006\linewidth}\parbox[t]{0.671\linewidth}{mutate bond\_schedule 8-12-16-24-32-48-64-80-96, max\_sweeps 20, init\_bond\_dim 2 on sweep-based dmrg}}}}\par\smallskip
\noindent\parbox[t]{0.05\linewidth}{\centering\strut 56}\hspace{0.006\linewidth}\parbox[t]{0.04\linewidth}{\centering\strut \raisebox{0.08ex}{\ding{55}}}\hspace{0.006\linewidth}\parbox[t]{0.10\linewidth}{\centering\strut ---}\hspace{0.006\linewidth}\parbox[t]{0.115\linewidth}{\centering\strut ---}\hspace{0.006\linewidth}\parbox[t]{0.671\linewidth}{mutate bond\_schedule 8-12-16-24-32-48-64-80-96, max\_sweeps 20, init\_bond\_dim 4 on sweep-based dmrg}\par\smallskip
\noindent{\setlength{\fboxsep}{0pt}\colorbox{black!4}{\parbox{\linewidth}{\parbox[t]{0.05\linewidth}{\centering\strut 57}\hspace{0.006\linewidth}\parbox[t]{0.04\linewidth}{\centering\strut \raisebox{0.08ex}{\ding{55}}}\hspace{0.006\linewidth}\parbox[t]{0.10\linewidth}{\centering\strut ---}\hspace{0.006\linewidth}\parbox[t]{0.115\linewidth}{\centering\strut ---}\hspace{0.006\linewidth}\parbox[t]{0.671\linewidth}{mutate bond\_schedule 8-12-16-24-32-48-64-80-96, max\_sweeps 22, init\_bond\_dim 1 on sweep-based dmrg}}}}\par\smallskip
\noindent\parbox[t]{0.05\linewidth}{\centering\strut 58}\hspace{0.006\linewidth}\parbox[t]{0.04\linewidth}{\centering\strut \raisebox{0.08ex}{\ding{55}}}\hspace{0.006\linewidth}\parbox[t]{0.10\linewidth}{\centering\strut ---}\hspace{0.006\linewidth}\parbox[t]{0.115\linewidth}{\centering\strut ---}\hspace{0.006\linewidth}\parbox[t]{0.671\linewidth}{mutate bond\_schedule 8-12-16-24-32-48-64-80-96, max\_sweeps 22, init\_bond\_dim 2 on sweep-based dmrg}\par\smallskip
\noindent{\setlength{\fboxsep}{0pt}\colorbox{black!4}{\parbox{\linewidth}{\parbox[t]{0.05\linewidth}{\centering\strut 59}\hspace{0.006\linewidth}\parbox[t]{0.04\linewidth}{\centering\strut \raisebox{0.08ex}{\ding{55}}}\hspace{0.006\linewidth}\parbox[t]{0.10\linewidth}{\centering\strut ---}\hspace{0.006\linewidth}\parbox[t]{0.115\linewidth}{\centering\strut ---}\hspace{0.006\linewidth}\parbox[t]{0.671\linewidth}{mutate bond\_schedule 8-12-16-24-32-48-64-80-96, max\_sweeps 22, init\_bond\_dim 4 on sweep-based dmrg}}}}\par\smallskip
\noindent\parbox[t]{0.05\linewidth}{\centering\strut 60}\hspace{0.006\linewidth}\parbox[t]{0.04\linewidth}{\centering\strut \raisebox{0.08ex}{\ding{55}}}\hspace{0.006\linewidth}\parbox[t]{0.10\linewidth}{\centering\strut ---}\hspace{0.006\linewidth}\parbox[t]{0.115\linewidth}{\centering\strut ---}\hspace{0.006\linewidth}\parbox[t]{0.671\linewidth}{mutate bond\_schedule 8-12-16-24-32-48-64-80-96-112, max\_sweeps 16, init\_bond\_dim 1 on sweep-based dmrg}\par\smallskip
\noindent{\setlength{\fboxsep}{0pt}\colorbox{black!4}{\parbox{\linewidth}{\parbox[t]{0.05\linewidth}{\centering\strut 61}\hspace{0.006\linewidth}\parbox[t]{0.04\linewidth}{\centering\strut \raisebox{0.08ex}{\ding{55}}}\hspace{0.006\linewidth}\parbox[t]{0.10\linewidth}{\centering\strut ---}\hspace{0.006\linewidth}\parbox[t]{0.115\linewidth}{\centering\strut ---}\hspace{0.006\linewidth}\parbox[t]{0.671\linewidth}{mutate bond\_schedule 8-12-16-24-32-48-64-80-96-112, max\_sweeps 16, init\_bond\_dim 2 on sweep-based dmrg}}}}\par\smallskip
\noindent\parbox[t]{0.05\linewidth}{\centering\strut 62}\hspace{0.006\linewidth}\parbox[t]{0.04\linewidth}{\centering\strut \raisebox{0.08ex}{\ding{55}}}\hspace{0.006\linewidth}\parbox[t]{0.10\linewidth}{\centering\strut ---}\hspace{0.006\linewidth}\parbox[t]{0.115\linewidth}{\centering\strut ---}\hspace{0.006\linewidth}\parbox[t]{0.671\linewidth}{mutate bond\_schedule 8-12-16-24-32-48-64-80-96-112, max\_sweeps 16, init\_bond\_dim 4 on sweep-based dmrg}\par\smallskip
\noindent{\setlength{\fboxsep}{0pt}\colorbox{black!4}{\parbox{\linewidth}{\parbox[t]{0.05\linewidth}{\centering\strut 63}\hspace{0.006\linewidth}\parbox[t]{0.04\linewidth}{\centering\strut \raisebox{0.08ex}{\ding{55}}}\hspace{0.006\linewidth}\parbox[t]{0.10\linewidth}{\centering\strut ---}\hspace{0.006\linewidth}\parbox[t]{0.115\linewidth}{\centering\strut ---}\hspace{0.006\linewidth}\parbox[t]{0.671\linewidth}{mutate bond\_schedule 8-12-16-24-32-48-64-80-96-112, max\_sweeps 18, init\_bond\_dim 1 on sweep-based dmrg}}}}\par\smallskip
\noindent\parbox[t]{0.05\linewidth}{\centering\strut 64}\hspace{0.006\linewidth}\parbox[t]{0.04\linewidth}{\centering\strut \raisebox{0.08ex}{\ding{55}}}\hspace{0.006\linewidth}\parbox[t]{0.10\linewidth}{\centering\strut ---}\hspace{0.006\linewidth}\parbox[t]{0.115\linewidth}{\centering\strut ---}\hspace{0.006\linewidth}\parbox[t]{0.671\linewidth}{mutate bond\_schedule 8-12-16-24-32-48-64-80-96-112, max\_sweeps 18, init\_bond\_dim 2 on sweep-based dmrg}\par\smallskip
\noindent{\setlength{\fboxsep}{0pt}\colorbox{black!4}{\parbox{\linewidth}{\parbox[t]{0.05\linewidth}{\centering\strut 65}\hspace{0.006\linewidth}\parbox[t]{0.04\linewidth}{\centering\strut \raisebox{0.08ex}{\ding{55}}}\hspace{0.006\linewidth}\parbox[t]{0.10\linewidth}{\centering\strut ---}\hspace{0.006\linewidth}\parbox[t]{0.115\linewidth}{\centering\strut ---}\hspace{0.006\linewidth}\parbox[t]{0.671\linewidth}{mutate bond\_schedule 8-12-16-24-32-48-64-80-96-112, max\_sweeps 18, init\_bond\_dim 4 on sweep-based dmrg}}}}\par\smallskip
\noindent\parbox[t]{0.05\linewidth}{\centering\strut 66}\hspace{0.006\linewidth}\parbox[t]{0.04\linewidth}{\centering\strut \raisebox{0.08ex}{\ding{55}}}\hspace{0.006\linewidth}\parbox[t]{0.10\linewidth}{\centering\strut ---}\hspace{0.006\linewidth}\parbox[t]{0.115\linewidth}{\centering\strut ---}\hspace{0.006\linewidth}\parbox[t]{0.671\linewidth}{mutate bond\_schedule 8-12-16-24-32-48-64-80-96-112, max\_sweeps 20, init\_bond\_dim 1 on sweep-based dmrg}\par\smallskip
\noindent{\setlength{\fboxsep}{0pt}\colorbox{black!4}{\parbox{\linewidth}{\parbox[t]{0.05\linewidth}{\centering\strut 67}\hspace{0.006\linewidth}\parbox[t]{0.04\linewidth}{\centering\strut \raisebox{0.08ex}{\ding{55}}}\hspace{0.006\linewidth}\parbox[t]{0.10\linewidth}{\centering\strut ---}\hspace{0.006\linewidth}\parbox[t]{0.115\linewidth}{\centering\strut ---}\hspace{0.006\linewidth}\parbox[t]{0.671\linewidth}{mutate bond\_schedule 8-12-16-24-32-48-64-80-96-112, max\_sweeps 20, init\_bond\_dim 2 on sweep-based dmrg}}}}\par\smallskip
\noindent\parbox[t]{0.05\linewidth}{\centering\strut 68}\hspace{0.006\linewidth}\parbox[t]{0.04\linewidth}{\centering\strut \raisebox{0.08ex}{\ding{55}}}\hspace{0.006\linewidth}\parbox[t]{0.10\linewidth}{\centering\strut ---}\hspace{0.006\linewidth}\parbox[t]{0.115\linewidth}{\centering\strut ---}\hspace{0.006\linewidth}\parbox[t]{0.671\linewidth}{mutate bond\_schedule 8-12-16-24-32-48-64-80-96-112, max\_sweeps 20, init\_bond\_dim 4 on sweep-based dmrg}\par\smallskip
\noindent{\setlength{\fboxsep}{0pt}\colorbox{black!4}{\parbox{\linewidth}{\parbox[t]{0.05\linewidth}{\centering\strut 69}\hspace{0.006\linewidth}\parbox[t]{0.04\linewidth}{\centering\strut \raisebox{0.08ex}{\ding{55}}}\hspace{0.006\linewidth}\parbox[t]{0.10\linewidth}{\centering\strut ---}\hspace{0.006\linewidth}\parbox[t]{0.115\linewidth}{\centering\strut ---}\hspace{0.006\linewidth}\parbox[t]{0.671\linewidth}{mutate bond\_schedule 8-12-16-24-32-48-64-80-96-112, max\_sweeps 22, init\_bond\_dim 1 on sweep-based dmrg}}}}\par\smallskip
\noindent\parbox[t]{0.05\linewidth}{\centering\strut 70}\hspace{0.006\linewidth}\parbox[t]{0.04\linewidth}{\centering\strut \raisebox{0.08ex}{\ding{55}}}\hspace{0.006\linewidth}\parbox[t]{0.10\linewidth}{\centering\strut ---}\hspace{0.006\linewidth}\parbox[t]{0.115\linewidth}{\centering\strut ---}\hspace{0.006\linewidth}\parbox[t]{0.671\linewidth}{mutate bond\_schedule 8-12-16-24-32-48-64-80-96-112, max\_sweeps 22, init\_bond\_dim 2 on sweep-based dmrg}\par\smallskip
\noindent{\setlength{\fboxsep}{0pt}\colorbox{black!4}{\parbox{\linewidth}{\parbox[t]{0.05\linewidth}{\centering\strut 71}\hspace{0.006\linewidth}\parbox[t]{0.04\linewidth}{\centering\strut \raisebox{0.08ex}{\ding{55}}}\hspace{0.006\linewidth}\parbox[t]{0.10\linewidth}{\centering\strut ---}\hspace{0.006\linewidth}\parbox[t]{0.115\linewidth}{\centering\strut ---}\hspace{0.006\linewidth}\parbox[t]{0.671\linewidth}{mutate bond\_schedule 8-12-16-24-32-48-64-80-96-112, max\_sweeps 22, init\_bond\_dim 4 on sweep-based dmrg}}}}\par\smallskip
\noindent\parbox[t]{0.05\linewidth}{\centering\strut 72}\hspace{0.006\linewidth}\parbox[t]{0.04\linewidth}{\centering\strut \raisebox{0.08ex}{\ding{55}}}\hspace{0.006\linewidth}\parbox[t]{0.10\linewidth}{\centering\strut +0\%}\hspace{0.006\linewidth}\parbox[t]{0.115\linewidth}{\centering\strut -28.175425}\hspace{0.006\linewidth}\parbox[t]{0.671\linewidth}{mutate bond\_schedule 8-12-16-24-32-48-64-80, tol 1e-06, ncv 6 on sweep-based dmrg}\par\smallskip
\noindent{\setlength{\fboxsep}{0pt}\colorbox{black!4}{\parbox{\linewidth}{\parbox[t]{0.05\linewidth}{\centering\strut 73}\hspace{0.006\linewidth}\parbox[t]{0.04\linewidth}{\centering\strut \raisebox{0.08ex}{\ding{55}}}\hspace{0.006\linewidth}\parbox[t]{0.10\linewidth}{\centering\strut +0\%}\hspace{0.006\linewidth}\parbox[t]{0.115\linewidth}{\centering\strut -28.175425}\hspace{0.006\linewidth}\parbox[t]{0.671\linewidth}{mutate bond\_schedule 8-12-16-24-32-48-64-80, tol 1e-06 on sweep-based dmrg}}}}\par\smallskip
\noindent\parbox[t]{0.05\linewidth}{\centering\strut 74}\hspace{0.006\linewidth}\parbox[t]{0.04\linewidth}{\centering\strut \raisebox{0.08ex}{\ding{55}}}\hspace{0.006\linewidth}\parbox[t]{0.10\linewidth}{\centering\strut ---}\hspace{0.006\linewidth}\parbox[t]{0.115\linewidth}{\centering\strut ---}\hspace{0.006\linewidth}\parbox[t]{0.671\linewidth}{mutate bond\_schedule 8-12-16-24-32-48-64-80, tol 1e-06, ncv 10 on sweep-based dmrg}\par\smallskip
\noindent{\setlength{\fboxsep}{0pt}\colorbox{black!4}{\parbox{\linewidth}{\parbox[t]{0.05\linewidth}{\centering\strut 75}\hspace{0.006\linewidth}\parbox[t]{0.04\linewidth}{\centering\strut \raisebox{0.08ex}{\ding{55}}}\hspace{0.006\linewidth}\parbox[t]{0.10\linewidth}{\centering\strut ---}\hspace{0.006\linewidth}\parbox[t]{0.115\linewidth}{\centering\strut ---}\hspace{0.006\linewidth}\parbox[t]{0.671\linewidth}{mutate bond\_schedule 8-12-16-24-32-48-64-80, tol 1e-06, ncv 12 on sweep-based dmrg}}}}\par\smallskip
\noindent\parbox[t]{0.05\linewidth}{\centering\strut 76}\hspace{0.006\linewidth}\parbox[t]{0.04\linewidth}{\centering\strut \raisebox{0.08ex}{\ding{55}}}\hspace{0.006\linewidth}\parbox[t]{0.10\linewidth}{\centering\strut +0\%}\hspace{0.006\linewidth}\parbox[t]{0.115\linewidth}{\centering\strut -28.175425}\hspace{0.006\linewidth}\parbox[t]{0.671\linewidth}{mutate bond\_schedule 8-12-16-24-32-48-64-80, ncv 6 on sweep-based dmrg}\par\smallskip
\noindent{\setlength{\fboxsep}{0pt}\colorbox{black!4}{\parbox{\linewidth}{\parbox[t]{0.05\linewidth}{\centering\strut 77}\hspace{0.006\linewidth}\parbox[t]{0.04\linewidth}{\centering\strut \raisebox{0.08ex}{\ding{55}}}\hspace{0.006\linewidth}\parbox[t]{0.10\linewidth}{\centering\strut ---}\hspace{0.006\linewidth}\parbox[t]{0.115\linewidth}{\centering\strut ---}\hspace{0.006\linewidth}\parbox[t]{0.671\linewidth}{mutate bond\_schedule 8-12-16-24-32-48-64-80 on sweep-based dmrg}}}}\par\smallskip
\noindent\parbox[t]{0.05\linewidth}{\centering\strut 78}\hspace{0.006\linewidth}\parbox[t]{0.04\linewidth}{\centering\strut \raisebox{0.08ex}{\ding{55}}}\hspace{0.006\linewidth}\parbox[t]{0.10\linewidth}{\centering\strut +0\%}\hspace{0.006\linewidth}\parbox[t]{0.115\linewidth}{\centering\strut -28.175425}\hspace{0.006\linewidth}\parbox[t]{0.671\linewidth}{mutate bond\_schedule 8-12-16-24-32-48-64-80, ncv 10 on sweep-based dmrg}\par\smallskip
\noindent{\setlength{\fboxsep}{0pt}\colorbox{black!4}{\parbox{\linewidth}{\parbox[t]{0.05\linewidth}{\centering\strut 79}\hspace{0.006\linewidth}\parbox[t]{0.04\linewidth}{\centering\strut \raisebox{0.08ex}{\ding{55}}}\hspace{0.006\linewidth}\parbox[t]{0.10\linewidth}{\centering\strut ---}\hspace{0.006\linewidth}\parbox[t]{0.115\linewidth}{\centering\strut ---}\hspace{0.006\linewidth}\parbox[t]{0.671\linewidth}{mutate bond\_schedule 8-12-16-24-32-48-64-80, ncv 12 on sweep-based dmrg}}}}\par\smallskip
\noindent\parbox[t]{0.05\linewidth}{\centering\strut 80}\hspace{0.006\linewidth}\parbox[t]{0.04\linewidth}{\centering\strut \raisebox{0.08ex}{\ding{55}}}\hspace{0.006\linewidth}\parbox[t]{0.10\linewidth}{\centering\strut +0\%}\hspace{0.006\linewidth}\parbox[t]{0.115\linewidth}{\centering\strut -28.175425}\hspace{0.006\linewidth}\parbox[t]{0.671\linewidth}{mutate bond\_schedule 8-12-16-24-32-48-64-80, tol 1e-08, ncv 6 on sweep-based dmrg}\par\smallskip
\noindent{\setlength{\fboxsep}{0pt}\colorbox{black!4}{\parbox{\linewidth}{\parbox[t]{0.05\linewidth}{\centering\strut 81}\hspace{0.006\linewidth}\parbox[t]{0.04\linewidth}{\centering\strut \raisebox{0.08ex}{\ding{55}}}\hspace{0.006\linewidth}\parbox[t]{0.10\linewidth}{\centering\strut +0\%}\hspace{0.006\linewidth}\parbox[t]{0.115\linewidth}{\centering\strut -28.175425}\hspace{0.006\linewidth}\parbox[t]{0.671\linewidth}{mutate bond\_schedule 8-12-16-24-32-48-64-80, tol 1e-08 on sweep-based dmrg}}}}\par\smallskip
\noindent\parbox[t]{0.05\linewidth}{\centering\strut 82}\hspace{0.006\linewidth}\parbox[t]{0.04\linewidth}{\centering\strut \raisebox{0.08ex}{\ding{55}}}\hspace{0.006\linewidth}\parbox[t]{0.10\linewidth}{\centering\strut ---}\hspace{0.006\linewidth}\parbox[t]{0.115\linewidth}{\centering\strut ---}\hspace{0.006\linewidth}\parbox[t]{0.671\linewidth}{mutate bond\_schedule 8-12-16-24-32-48-64-80, tol 1e-08, ncv 10 on sweep-based dmrg}\par\smallskip
\noindent{\setlength{\fboxsep}{0pt}\colorbox{black!4}{\parbox{\linewidth}{\parbox[t]{0.05\linewidth}{\centering\strut 83}\hspace{0.006\linewidth}\parbox[t]{0.04\linewidth}{\centering\strut \raisebox{0.08ex}{\ding{55}}}\hspace{0.006\linewidth}\parbox[t]{0.10\linewidth}{\centering\strut +0\%}\hspace{0.006\linewidth}\parbox[t]{0.115\linewidth}{\centering\strut -28.175425}\hspace{0.006\linewidth}\parbox[t]{0.671\linewidth}{mutate bond\_schedule 8-12-16-24-32-48-64-80, tol 1e-08, ncv 12 on sweep-based dmrg}}}}\par\smallskip
\noindent\parbox[t]{0.05\linewidth}{\centering\strut 84}\hspace{0.006\linewidth}\parbox[t]{0.04\linewidth}{\centering\strut \raisebox{0.08ex}{\ding{55}}}\hspace{0.006\linewidth}\parbox[t]{0.10\linewidth}{\centering\strut +0\%}\hspace{0.006\linewidth}\parbox[t]{0.115\linewidth}{\centering\strut -28.175425}\hspace{0.006\linewidth}\parbox[t]{0.671\linewidth}{mutate bond\_schedule 8-12-16-24-32-48-64-80-96, tol 1e-06, ncv 6 on sweep-based dmrg}\par\smallskip
\noindent{\setlength{\fboxsep}{0pt}\colorbox{black!4}{\parbox{\linewidth}{\parbox[t]{0.05\linewidth}{\centering\strut 85}\hspace{0.006\linewidth}\parbox[t]{0.04\linewidth}{\centering\strut \raisebox{0.08ex}{\ding{55}}}\hspace{0.006\linewidth}\parbox[t]{0.10\linewidth}{\centering\strut ---}\hspace{0.006\linewidth}\parbox[t]{0.115\linewidth}{\centering\strut ---}\hspace{0.006\linewidth}\parbox[t]{0.671\linewidth}{mutate bond\_schedule 8-12-16-24-32-48-64-80-96, tol 1e-06 on sweep-based dmrg}}}}\par\smallskip
\noindent\parbox[t]{0.05\linewidth}{\centering\strut 86}\hspace{0.006\linewidth}\parbox[t]{0.04\linewidth}{\centering\strut \raisebox{0.08ex}{\ding{55}}}\hspace{0.006\linewidth}\parbox[t]{0.10\linewidth}{\centering\strut ---}\hspace{0.006\linewidth}\parbox[t]{0.115\linewidth}{\centering\strut ---}\hspace{0.006\linewidth}\parbox[t]{0.671\linewidth}{mutate bond\_schedule 8-12-16-24-32-48-64-80-96, tol 1e-06, ncv 10 on sweep-based dmrg}\par\smallskip
\noindent{\setlength{\fboxsep}{0pt}\colorbox{black!4}{\parbox{\linewidth}{\parbox[t]{0.05\linewidth}{\centering\strut 87}\hspace{0.006\linewidth}\parbox[t]{0.04\linewidth}{\centering\strut \raisebox{0.08ex}{\ding{55}}}\hspace{0.006\linewidth}\parbox[t]{0.10\linewidth}{\centering\strut ---}\hspace{0.006\linewidth}\parbox[t]{0.115\linewidth}{\centering\strut ---}\hspace{0.006\linewidth}\parbox[t]{0.671\linewidth}{mutate bond\_schedule 8-12-16-24-32-48-64-80-96, tol 1e-06, ncv 12 on sweep-based dmrg}}}}\par\smallskip
\noindent\parbox[t]{0.05\linewidth}{\centering\strut 88}\hspace{0.006\linewidth}\parbox[t]{0.04\linewidth}{\centering\strut \raisebox{0.08ex}{\ding{51}}}\hspace{0.006\linewidth}\parbox[t]{0.10\linewidth}{\centering\strut +0\%}\hspace{0.006\linewidth}\parbox[t]{0.115\linewidth}{\centering\strut -28.175425}\hspace{0.006\linewidth}\parbox[t]{0.671\linewidth}{mutate bond\_schedule 8-12-16-24-32-48-64-80-96, ncv 6 on sweep-based dmrg}\par\smallskip
\noindent{\setlength{\fboxsep}{0pt}\colorbox{black!4}{\parbox{\linewidth}{\parbox[t]{0.05\linewidth}{\centering\strut 89}\hspace{0.006\linewidth}\parbox[t]{0.04\linewidth}{\centering\strut \raisebox{0.08ex}{\ding{55}}}\hspace{0.006\linewidth}\parbox[t]{0.10\linewidth}{\centering\strut ---}\hspace{0.006\linewidth}\parbox[t]{0.115\linewidth}{\centering\strut ---}\hspace{0.006\linewidth}\parbox[t]{0.671\linewidth}{mutate bond\_schedule 8-12-16-24-32-48-64-80-96, ncv 8 on sweep-based dmrg}}}}\par\smallskip
\noindent\parbox[t]{0.05\linewidth}{\centering\strut 90}\hspace{0.006\linewidth}\parbox[t]{0.04\linewidth}{\centering\strut \raisebox{0.08ex}{\ding{55}}}\hspace{0.006\linewidth}\parbox[t]{0.10\linewidth}{\centering\strut ---}\hspace{0.006\linewidth}\parbox[t]{0.115\linewidth}{\centering\strut ---}\hspace{0.006\linewidth}\parbox[t]{0.671\linewidth}{mutate bond\_schedule 8-12-16-24-32-48-64-80-96, ncv 10 on sweep-based dmrg}\par\smallskip
\noindent{\setlength{\fboxsep}{0pt}\colorbox{black!4}{\parbox{\linewidth}{\parbox[t]{0.05\linewidth}{\centering\strut 91}\hspace{0.006\linewidth}\parbox[t]{0.04\linewidth}{\centering\strut \raisebox{0.08ex}{\ding{55}}}\hspace{0.006\linewidth}\parbox[t]{0.10\linewidth}{\centering\strut ---}\hspace{0.006\linewidth}\parbox[t]{0.115\linewidth}{\centering\strut ---}\hspace{0.006\linewidth}\parbox[t]{0.671\linewidth}{mutate bond\_schedule 8-12-16-24-32-48-64-80-96, ncv 12 on sweep-based dmrg}}}}\par\smallskip
\noindent\parbox[t]{0.05\linewidth}{\centering\strut 92}\hspace{0.006\linewidth}\parbox[t]{0.04\linewidth}{\centering\strut \raisebox{0.08ex}{\ding{55}}}\hspace{0.006\linewidth}\parbox[t]{0.10\linewidth}{\centering\strut +0\%}\hspace{0.006\linewidth}\parbox[t]{0.115\linewidth}{\centering\strut -28.175425}\hspace{0.006\linewidth}\parbox[t]{0.671\linewidth}{mutate bond\_schedule 8-12-16-24-32-48-64-80-96, tol 1e-08 on sweep-based dmrg}\par\smallskip
\noindent{\setlength{\fboxsep}{0pt}\colorbox{black!4}{\parbox{\linewidth}{\parbox[t]{0.05\linewidth}{\centering\strut 93}\hspace{0.006\linewidth}\parbox[t]{0.04\linewidth}{\centering\strut \raisebox{0.08ex}{\ding{55}}}\hspace{0.006\linewidth}\parbox[t]{0.10\linewidth}{\centering\strut ---}\hspace{0.006\linewidth}\parbox[t]{0.115\linewidth}{\centering\strut ---}\hspace{0.006\linewidth}\parbox[t]{0.671\linewidth}{mutate bond\_schedule 8-12-16-24-32-48-64-80-96, tol 1e-08, ncv 8 on sweep-based dmrg}}}}\par\smallskip
\noindent\parbox[t]{0.05\linewidth}{\centering\strut 94}\hspace{0.006\linewidth}\parbox[t]{0.04\linewidth}{\centering\strut \raisebox{0.08ex}{\ding{55}}}\hspace{0.006\linewidth}\parbox[t]{0.10\linewidth}{\centering\strut ---}\hspace{0.006\linewidth}\parbox[t]{0.115\linewidth}{\centering\strut ---}\hspace{0.006\linewidth}\parbox[t]{0.671\linewidth}{mutate bond\_schedule 8-12-16-24-32-48-64-80-96, tol 1e-08, ncv 10 on sweep-based dmrg}\par\smallskip
\noindent{\setlength{\fboxsep}{0pt}\colorbox{black!4}{\parbox{\linewidth}{\parbox[t]{0.05\linewidth}{\centering\strut 95}\hspace{0.006\linewidth}\parbox[t]{0.04\linewidth}{\centering\strut \raisebox{0.08ex}{\ding{55}}}\hspace{0.006\linewidth}\parbox[t]{0.10\linewidth}{\centering\strut ---}\hspace{0.006\linewidth}\parbox[t]{0.115\linewidth}{\centering\strut ---}\hspace{0.006\linewidth}\parbox[t]{0.671\linewidth}{mutate bond\_schedule 8-12-16-24-32-48-64-80-96, tol 1e-08, ncv 12 on sweep-based dmrg}}}}\par\smallskip
\noindent\parbox[t]{0.05\linewidth}{\centering\strut 96}\hspace{0.006\linewidth}\parbox[t]{0.04\linewidth}{\centering\strut \raisebox{0.08ex}{\ding{55}}}\hspace{0.006\linewidth}\parbox[t]{0.10\linewidth}{\centering\strut ---}\hspace{0.006\linewidth}\parbox[t]{0.115\linewidth}{\centering\strut ---}\hspace{0.006\linewidth}\parbox[t]{0.671\linewidth}{mutate bond\_schedule 8-12-16-24-32-48-64-80-96-112, tol 1e-06 on sweep-based dmrg}\par\smallskip
\noindent{\setlength{\fboxsep}{0pt}\colorbox{black!4}{\parbox{\linewidth}{\parbox[t]{0.05\linewidth}{\centering\strut 97}\hspace{0.006\linewidth}\parbox[t]{0.04\linewidth}{\centering\strut \raisebox{0.08ex}{\ding{55}}}\hspace{0.006\linewidth}\parbox[t]{0.10\linewidth}{\centering\strut ---}\hspace{0.006\linewidth}\parbox[t]{0.115\linewidth}{\centering\strut ---}\hspace{0.006\linewidth}\parbox[t]{0.671\linewidth}{mutate bond\_schedule 8-12-16-24-32-48-64-80-96-112, tol 1e-06, ncv 8 on sweep-based dmrg}}}}\par\smallskip
\noindent\parbox[t]{0.05\linewidth}{\centering\strut 98}\hspace{0.006\linewidth}\parbox[t]{0.04\linewidth}{\centering\strut \raisebox{0.08ex}{\ding{55}}}\hspace{0.006\linewidth}\parbox[t]{0.10\linewidth}{\centering\strut ---}\hspace{0.006\linewidth}\parbox[t]{0.115\linewidth}{\centering\strut ---}\hspace{0.006\linewidth}\parbox[t]{0.671\linewidth}{mutate bond\_schedule 8-12-16-24-32-48-64-80-96-112, tol 1e-06, ncv 10 on sweep-based dmrg}\par\smallskip
\noindent{\setlength{\fboxsep}{0pt}\colorbox{black!4}{\parbox{\linewidth}{\parbox[t]{0.05\linewidth}{\centering\strut 99}\hspace{0.006\linewidth}\parbox[t]{0.04\linewidth}{\centering\strut \raisebox{0.08ex}{\ding{55}}}\hspace{0.006\linewidth}\parbox[t]{0.10\linewidth}{\centering\strut ---}\hspace{0.006\linewidth}\parbox[t]{0.115\linewidth}{\centering\strut ---}\hspace{0.006\linewidth}\parbox[t]{0.671\linewidth}{mutate bond\_schedule 8-12-16-24-32-48-64-80-96-112, tol 1e-06, ncv 12 on sweep-based dmrg}}}}\par\smallskip
\noindent\parbox[t]{0.05\linewidth}{\centering\strut 100}\hspace{0.006\linewidth}\parbox[t]{0.04\linewidth}{\centering\strut \raisebox{0.08ex}{\ding{55}}}\hspace{0.006\linewidth}\parbox[t]{0.10\linewidth}{\centering\strut ---}\hspace{0.006\linewidth}\parbox[t]{0.115\linewidth}{\centering\strut ---}\hspace{0.006\linewidth}\parbox[t]{0.671\linewidth}{mutate bond\_schedule 8-12-16-24-32-48-64-80-96-112 on sweep-based dmrg}\par\smallskip
\noindent\rule{\linewidth}{0.5pt}
\par\medskip

\medskip

\refstepcounter{table}
\label{tab:supp_tn_log_xxz_gapless_64}
\noindent\textbf{Table \thetable.} Tensor-network mutation log for Gapless XXZ, $L=64$. Relative gain is measured against the immediately previous scored iteration using the live final-energy objective.
\par\medskip
\noindent\rule{\linewidth}{0.5pt}
\noindent{\setlength{\fboxsep}{0pt}\colorbox{black!4}{\parbox{\linewidth}{\parbox[c][2.2em][c]{0.05\linewidth}{\centering \textbf{Iter.}}\hspace{0.006\linewidth}\parbox[c][2.2em][c]{0.04\linewidth}{\centering \textbf{A/R}}\hspace{0.006\linewidth}\parbox[c][2.2em][c]{0.10\linewidth}{\centering \textbf{Relative gain}}\hspace{0.006\linewidth}\parbox[c][2.2em][c]{0.115\linewidth}{\centering \textbf{Score}}\hspace{0.006\linewidth}\parbox[c][2.2em][c]{0.671\linewidth}{\centering \textbf{Mutation summary}}}}}\par\smallskip
\noindent\rule{\linewidth}{0.5pt}
\noindent{\setlength{\fboxsep}{0pt}\colorbox{black!4}{\parbox{\linewidth}{\parbox[t]{0.05\linewidth}{\centering\strut 1}\hspace{0.006\linewidth}\parbox[t]{0.04\linewidth}{\centering\strut \raisebox{0.08ex}{\ding{51}}}\hspace{0.006\linewidth}\parbox[t]{0.10\linewidth}{\centering\strut ---}\hspace{0.006\linewidth}\parbox[t]{0.115\linewidth}{\centering\strut -23.828519}\hspace{0.006\linewidth}\parbox[t]{0.671\linewidth}{switch to dmrg2 neel init with 8-64 bond ladder}}}}\par\smallskip
\noindent\parbox[t]{0.05\linewidth}{\centering\strut 2}\hspace{0.006\linewidth}\parbox[t]{0.04\linewidth}{\centering\strut \raisebox{0.08ex}{\ding{51}}}\hspace{0.006\linewidth}\parbox[t]{0.10\linewidth}{\centering\strut +0\%}\hspace{0.006\linewidth}\parbox[t]{0.115\linewidth}{\centering\strut -23.828520}\hspace{0.006\linewidth}\parbox[t]{0.671\linewidth}{raise dmrg2 neel bond ladder to 12-128 and 20 sweeps}\par\smallskip
\noindent{\setlength{\fboxsep}{0pt}\colorbox{black!4}{\parbox{\linewidth}{\parbox[t]{0.05\linewidth}{\centering\strut 3}\hspace{0.006\linewidth}\parbox[t]{0.04\linewidth}{\centering\strut \raisebox{0.08ex}{\ding{55}}}\hspace{0.006\linewidth}\parbox[t]{0.10\linewidth}{\centering\strut +0\%}\hspace{0.006\linewidth}\parbox[t]{0.115\linewidth}{\centering\strut -23.828520}\hspace{0.006\linewidth}\parbox[t]{0.671\linewidth}{switch dmrg2 deep ladder to random init bond\_dim 8}}}}\par\smallskip
\noindent\parbox[t]{0.05\linewidth}{\centering\strut 4}\hspace{0.006\linewidth}\parbox[t]{0.04\linewidth}{\centering\strut \raisebox{0.08ex}{\ding{51}}}\hspace{0.006\linewidth}\parbox[t]{0.10\linewidth}{\centering\strut +0\%}\hspace{0.006\linewidth}\parbox[t]{0.115\linewidth}{\centering\strut -23.828520}\hspace{0.006\linewidth}\parbox[t]{0.671\linewidth}{add dmrg1 warm start before dmrg2 refinement ladder}\par\smallskip
\noindent{\setlength{\fboxsep}{0pt}\colorbox{black!4}{\parbox{\linewidth}{\parbox[t]{0.05\linewidth}{\centering\strut 5}\hspace{0.006\linewidth}\parbox[t]{0.04\linewidth}{\centering\strut \raisebox{0.08ex}{\ding{55}}}\hspace{0.006\linewidth}\parbox[t]{0.10\linewidth}{\centering\strut ---}\hspace{0.006\linewidth}\parbox[t]{0.115\linewidth}{\centering\strut ---}\hspace{0.006\linewidth}\parbox[t]{0.671\linewidth}{use sweep-based dmrg stages with 16-192 bond ladders}}}}\par\smallskip
\noindent\parbox[t]{0.05\linewidth}{\centering\strut 6}\hspace{0.006\linewidth}\parbox[t]{0.04\linewidth}{\centering\strut \raisebox{0.08ex}{\ding{55}}}\hspace{0.006\linewidth}\parbox[t]{0.10\linewidth}{\centering\strut +0\%}\hspace{0.006\linewidth}\parbox[t]{0.115\linewidth}{\centering\strut -23.828520}\hspace{0.006\linewidth}\parbox[t]{0.671\linewidth}{switch staged dmrg to sweep steps with 12-sweep refinement}\par\smallskip
\noindent{\setlength{\fboxsep}{0pt}\colorbox{black!4}{\parbox{\linewidth}{\parbox[t]{0.05\linewidth}{\centering\strut 7}\hspace{0.006\linewidth}\parbox[t]{0.04\linewidth}{\centering\strut \raisebox{0.08ex}{\ding{55}}}\hspace{0.006\linewidth}\parbox[t]{0.10\linewidth}{\centering\strut ---}\hspace{0.006\linewidth}\parbox[t]{0.115\linewidth}{\centering\strut ---}\hspace{0.006\linewidth}\parbox[t]{0.671\linewidth}{generalize to phased dmrg pipeline and add final dmrg1 polish}}}}\par\smallskip
\noindent\parbox[t]{0.05\linewidth}{\centering\strut 8}\hspace{0.006\linewidth}\parbox[t]{0.04\linewidth}{\centering\strut \raisebox{0.08ex}{\ding{51}}}\hspace{0.006\linewidth}\parbox[t]{0.10\linewidth}{\centering\strut +0\%}\hspace{0.006\linewidth}\parbox[t]{0.115\linewidth}{\centering\strut -23.828520}\hspace{0.006\linewidth}\parbox[t]{0.671\linewidth}{pre-expand warm state to bond 64 before dmrg2 refine}\par\smallskip
\noindent{\setlength{\fboxsep}{0pt}\colorbox{black!4}{\parbox{\linewidth}{\parbox[t]{0.05\linewidth}{\centering\strut 9}\hspace{0.006\linewidth}\parbox[t]{0.04\linewidth}{\centering\strut \raisebox{0.08ex}{\ding{55}}}\hspace{0.006\linewidth}\parbox[t]{0.10\linewidth}{\centering\strut +0\%}\hspace{0.006\linewidth}\parbox[t]{0.115\linewidth}{\centering\strut -23.828520}\hspace{0.006\linewidth}\parbox[t]{0.671\linewidth}{start dmrg2 at bond 64 and pre-expand warm state to 96}}}}\par\smallskip
\noindent\parbox[t]{0.05\linewidth}{\centering\strut 10}\hspace{0.006\linewidth}\parbox[t]{0.04\linewidth}{\centering\strut \raisebox{0.08ex}{\ding{55}}}\hspace{0.006\linewidth}\parbox[t]{0.10\linewidth}{\centering\strut ---}\hspace{0.006\linewidth}\parbox[t]{0.115\linewidth}{\centering\strut ---}\hspace{0.006\linewidth}\parbox[t]{0.671\linewidth}{switch to reference-lite pure dmrg2 with 32-128 ladder and ncv 24}\par\smallskip
\noindent{\setlength{\fboxsep}{0pt}\colorbox{black!4}{\parbox{\linewidth}{\parbox[t]{0.05\linewidth}{\centering\strut 11}\hspace{0.006\linewidth}\parbox[t]{0.04\linewidth}{\centering\strut \raisebox{0.08ex}{\ding{55}}}\hspace{0.006\linewidth}\parbox[t]{0.10\linewidth}{\centering\strut +0\%}\hspace{0.006\linewidth}\parbox[t]{0.115\linewidth}{\centering\strut -23.828520}\hspace{0.006\linewidth}\parbox[t]{0.671\linewidth}{switch staged phases to direct solver.sweep and full bond\_dims constructor}}}}\par\smallskip
\noindent\parbox[t]{0.05\linewidth}{\centering\strut 12}\hspace{0.006\linewidth}\parbox[t]{0.04\linewidth}{\centering\strut \raisebox{0.08ex}{\ding{55}}}\hspace{0.006\linewidth}\parbox[t]{0.10\linewidth}{\centering\strut +0\%}\hspace{0.006\linewidth}\parbox[t]{0.115\linewidth}{\centering\strut -23.828520}\hspace{0.006\linewidth}\parbox[t]{0.671\linewidth}{raise dmrg2 preexpand refine max\_sweeps to 18}\par\smallskip
\noindent{\setlength{\fboxsep}{0pt}\colorbox{black!4}{\parbox{\linewidth}{\parbox[t]{0.05\linewidth}{\centering\strut 13}\hspace{0.006\linewidth}\parbox[t]{0.04\linewidth}{\centering\strut \raisebox{0.08ex}{\ding{55}}}\hspace{0.006\linewidth}\parbox[t]{0.10\linewidth}{\centering\strut ---}\hspace{0.006\linewidth}\parbox[t]{0.115\linewidth}{\centering\strut ---}\hspace{0.006\linewidth}\parbox[t]{0.671\linewidth}{insert dmrg2 bridge ramp before high-bond preexpand refine}}}}\par\smallskip
\noindent\parbox[t]{0.05\linewidth}{\centering\strut 14}\hspace{0.006\linewidth}\parbox[t]{0.04\linewidth}{\centering\strut \raisebox{0.08ex}{\ding{51}}}\hspace{0.006\linewidth}\parbox[t]{0.10\linewidth}{\centering\strut +0\%}\hspace{0.006\linewidth}\parbox[t]{0.115\linewidth}{\centering\strut -23.828520}\hspace{0.006\linewidth}\parbox[t]{0.671\linewidth}{raise dmrg2 preexpand bond from 64 to 80}\par\smallskip
\noindent{\setlength{\fboxsep}{0pt}\colorbox{black!4}{\parbox{\linewidth}{\parbox[t]{0.05\linewidth}{\centering\strut 15}\hspace{0.006\linewidth}\parbox[t]{0.04\linewidth}{\centering\strut \raisebox{0.08ex}{\ding{51}}}\hspace{0.006\linewidth}\parbox[t]{0.10\linewidth}{\centering\strut +0\%}\hspace{0.006\linewidth}\parbox[t]{0.115\linewidth}{\centering\strut -23.828520}\hspace{0.006\linewidth}\parbox[t]{0.671\linewidth}{tighten dmrg2 refine cutoff from 1e-10 to 5e-11}}}}\par\smallskip
\noindent\parbox[t]{0.05\linewidth}{\centering\strut 16}\hspace{0.006\linewidth}\parbox[t]{0.04\linewidth}{\centering\strut \raisebox{0.08ex}{\ding{55}}}\hspace{0.006\linewidth}\parbox[t]{0.10\linewidth}{\centering\strut ---}\hspace{0.006\linewidth}\parbox[t]{0.115\linewidth}{\centering\strut ---}\hspace{0.006\linewidth}\parbox[t]{0.671\linewidth}{tighten dmrg2 refine cutoff from 5e-11 to 2e-11}\par\smallskip
\noindent{\setlength{\fboxsep}{0pt}\colorbox{black!4}{\parbox{\linewidth}{\parbox[t]{0.05\linewidth}{\centering\strut 17}\hspace{0.006\linewidth}\parbox[t]{0.04\linewidth}{\centering\strut \raisebox{0.08ex}{\ding{51}}}\hspace{0.006\linewidth}\parbox[t]{0.10\linewidth}{\centering\strut +0\%}\hspace{0.006\linewidth}\parbox[t]{0.115\linewidth}{\centering\strut -23.828520}\hspace{0.006\linewidth}\parbox[t]{0.671\linewidth}{tighten dmrg2 refine cutoff from 5e-11 to 4e-11}}}}\par\smallskip
\noindent\parbox[t]{0.05\linewidth}{\centering\strut 18}\hspace{0.006\linewidth}\parbox[t]{0.04\linewidth}{\centering\strut \raisebox{0.08ex}{\ding{55}}}\hspace{0.006\linewidth}\parbox[t]{0.10\linewidth}{\centering\strut +0\%}\hspace{0.006\linewidth}\parbox[t]{0.115\linewidth}{\centering\strut -23.828520}\hspace{0.006\linewidth}\parbox[t]{0.671\linewidth}{raise dmrg2 refine local\_eig\_ncv from 12 to 14}\par\smallskip
\noindent{\setlength{\fboxsep}{0pt}\colorbox{black!4}{\parbox{\linewidth}{\parbox[t]{0.05\linewidth}{\centering\strut 19}\hspace{0.006\linewidth}\parbox[t]{0.04\linewidth}{\centering\strut \raisebox{0.08ex}{\ding{55}}}\hspace{0.006\linewidth}\parbox[t]{0.10\linewidth}{\centering\strut +0\%}\hspace{0.006\linewidth}\parbox[t]{0.115\linewidth}{\centering\strut -23.828520}\hspace{0.006\linewidth}\parbox[t]{0.671\linewidth}{raise dmrg2 preexpand bond from 80 to 90}}}}\par\smallskip
\noindent\parbox[t]{0.05\linewidth}{\centering\strut 20}\hspace{0.006\linewidth}\parbox[t]{0.04\linewidth}{\centering\strut \raisebox{0.08ex}{\ding{51}}}\hspace{0.006\linewidth}\parbox[t]{0.10\linewidth}{\centering\strut +0\%}\hspace{0.006\linewidth}\parbox[t]{0.115\linewidth}{\centering\strut -23.828520}\hspace{0.006\linewidth}\parbox[t]{0.671\linewidth}{reduce warm-start sweeps to 6 and tighten dmrg2 cutoff to 3e-11}\par\smallskip
\noindent{\setlength{\fboxsep}{0pt}\colorbox{black!4}{\parbox{\linewidth}{\parbox[t]{0.05\linewidth}{\centering\strut 21}\hspace{0.006\linewidth}\parbox[t]{0.04\linewidth}{\centering\strut \raisebox{0.08ex}{\ding{51}}}\hspace{0.006\linewidth}\parbox[t]{0.10\linewidth}{\centering\strut +0\%}\hspace{0.006\linewidth}\parbox[t]{0.115\linewidth}{\centering\strut -23.828520}\hspace{0.006\linewidth}\parbox[t]{0.671\linewidth}{tighten dmrg2 cutoff from 3e-11 to 2.5e-11}}}}\par\smallskip
\noindent\parbox[t]{0.05\linewidth}{\centering\strut 22}\hspace{0.006\linewidth}\parbox[t]{0.04\linewidth}{\centering\strut \raisebox{0.08ex}{\ding{51}}}\hspace{0.006\linewidth}\parbox[t]{0.10\linewidth}{\centering\strut +0\%}\hspace{0.006\linewidth}\parbox[t]{0.115\linewidth}{\centering\strut -23.828520}\hspace{0.006\linewidth}\parbox[t]{0.671\linewidth}{reduce warm-start sweeps to 5 and tighten dmrg2 cutoff to 2e-11}\par\smallskip
\noindent{\setlength{\fboxsep}{0pt}\colorbox{black!4}{\parbox{\linewidth}{\parbox[t]{0.05\linewidth}{\centering\strut 23}\hspace{0.006\linewidth}\parbox[t]{0.04\linewidth}{\centering\strut \raisebox{0.08ex}{\ding{55}}}\hspace{0.006\linewidth}\parbox[t]{0.10\linewidth}{\centering\strut ---}\hspace{0.006\linewidth}\parbox[t]{0.115\linewidth}{\centering\strut ---}\hspace{0.006\linewidth}\parbox[t]{0.671\linewidth}{reduce warm-start sweeps to 4 and tighten dmrg2 cutoff to 1.6e-11}}}}\par\smallskip
\noindent\parbox[t]{0.05\linewidth}{\centering\strut 24}\hspace{0.006\linewidth}\parbox[t]{0.04\linewidth}{\centering\strut \raisebox{0.08ex}{\ding{55}}}\hspace{0.006\linewidth}\parbox[t]{0.10\linewidth}{\centering\strut ---}\hspace{0.006\linewidth}\parbox[t]{0.115\linewidth}{\centering\strut ---}\hspace{0.006\linewidth}\parbox[t]{0.671\linewidth}{lower phase preexpand rand\_strength from 1e-5 to 1e-6}\par\smallskip
\noindent{\setlength{\fboxsep}{0pt}\colorbox{black!4}{\parbox{\linewidth}{\parbox[t]{0.05\linewidth}{\centering\strut 25}\hspace{0.006\linewidth}\parbox[t]{0.04\linewidth}{\centering\strut \raisebox{0.08ex}{\ding{55}}}\hspace{0.006\linewidth}\parbox[t]{0.10\linewidth}{\centering\strut ---}\hspace{0.006\linewidth}\parbox[t]{0.115\linewidth}{\centering\strut ---}\hspace{0.006\linewidth}\parbox[t]{0.671\linewidth}{lower warm-start init\_bond\_dim from 4 to 2}}}}\par\smallskip
\noindent\parbox[t]{0.05\linewidth}{\centering\strut 26}\hspace{0.006\linewidth}\parbox[t]{0.04\linewidth}{\centering\strut \raisebox{0.08ex}{\ding{55}}}\hspace{0.006\linewidth}\parbox[t]{0.10\linewidth}{\centering\strut ---}\hspace{0.006\linewidth}\parbox[t]{0.115\linewidth}{\centering\strut ---}\hspace{0.006\linewidth}\parbox[t]{0.671\linewidth}{use looser early dmrg2 cutoffs and raise dmrg2 max\_sweeps to 17}\par\smallskip
\noindent{\setlength{\fboxsep}{0pt}\colorbox{black!4}{\parbox{\linewidth}{\parbox[t]{0.05\linewidth}{\centering\strut 27}\hspace{0.006\linewidth}\parbox[t]{0.04\linewidth}{\centering\strut \raisebox{0.08ex}{\ding{55}}}\hspace{0.006\linewidth}\parbox[t]{0.10\linewidth}{\centering\strut ---}\hspace{0.006\linewidth}\parbox[t]{0.115\linewidth}{\centering\strut ---}\hspace{0.006\linewidth}\parbox[t]{0.671\linewidth}{compress warm-start bond ladder to reach 48 in five sweeps}}}}\par\smallskip
\noindent\parbox[t]{0.05\linewidth}{\centering\strut 28}\hspace{0.006\linewidth}\parbox[t]{0.04\linewidth}{\centering\strut \raisebox{0.08ex}{\ding{55}}}\hspace{0.006\linewidth}\parbox[t]{0.10\linewidth}{\centering\strut ---}\hspace{0.006\linewidth}\parbox[t]{0.115\linewidth}{\centering\strut ---}\hspace{0.006\linewidth}\parbox[t]{0.671\linewidth}{seed phase preexpand from cfg.init\_seed before bond expansion}\par\smallskip
\noindent{\setlength{\fboxsep}{0pt}\colorbox{black!4}{\parbox{\linewidth}{\parbox[t]{0.05\linewidth}{\centering\strut 29}\hspace{0.006\linewidth}\parbox[t]{0.04\linewidth}{\centering\strut \raisebox{0.08ex}{\ding{55}}}\hspace{0.006\linewidth}\parbox[t]{0.10\linewidth}{\centering\strut ---}\hspace{0.006\linewidth}\parbox[t]{0.115\linewidth}{\centering\strut ---}\hspace{0.006\linewidth}\parbox[t]{0.671\linewidth}{switch warm-start init\_state from neel to plus}}}}\par\smallskip
\noindent\parbox[t]{0.05\linewidth}{\centering\strut 30}\hspace{0.006\linewidth}\parbox[t]{0.04\linewidth}{\centering\strut \raisebox{0.08ex}{\ding{55}}}\hspace{0.006\linewidth}\parbox[t]{0.10\linewidth}{\centering\strut ---}\hspace{0.006\linewidth}\parbox[t]{0.115\linewidth}{\centering\strut ---}\hspace{0.006\linewidth}\parbox[t]{0.671\linewidth}{switch warm-start to plus and reduce warm sweeps to 4}\par\smallskip
\noindent{\setlength{\fboxsep}{0pt}\colorbox{black!4}{\parbox{\linewidth}{\parbox[t]{0.05\linewidth}{\centering\strut 31}\hspace{0.006\linewidth}\parbox[t]{0.04\linewidth}{\centering\strut \raisebox{0.08ex}{\ding{55}}}\hspace{0.006\linewidth}\parbox[t]{0.10\linewidth}{\centering\strut ---}\hspace{0.006\linewidth}\parbox[t]{0.115\linewidth}{\centering\strut ---}\hspace{0.006\linewidth}\parbox[t]{0.671\linewidth}{reduce warm-start sweeps to 4 and tighten dmrg2 cutoff to 1.8e-11}}}}\par\smallskip
\noindent\parbox[t]{0.05\linewidth}{\centering\strut 32}\hspace{0.006\linewidth}\parbox[t]{0.04\linewidth}{\centering\strut \raisebox{0.08ex}{\ding{51}}}\hspace{0.006\linewidth}\parbox[t]{0.10\linewidth}{\centering\strut +0\%}\hspace{0.006\linewidth}\parbox[t]{0.115\linewidth}{\centering\strut -23.828520}\hspace{0.006\linewidth}\parbox[t]{0.671\linewidth}{reduce dmrg2 sweeps to 15 and tighten dmrg2 cutoff to 1.6e-11}\par\smallskip
\noindent{\setlength{\fboxsep}{0pt}\colorbox{black!4}{\parbox{\linewidth}{\parbox[t]{0.05\linewidth}{\centering\strut 33}\hspace{0.006\linewidth}\parbox[t]{0.04\linewidth}{\centering\strut \raisebox{0.08ex}{\ding{55}}}\hspace{0.006\linewidth}\parbox[t]{0.10\linewidth}{\centering\strut ---}\hspace{0.006\linewidth}\parbox[t]{0.115\linewidth}{\centering\strut ---}\hspace{0.006\linewidth}\parbox[t]{0.671\linewidth}{tighten dmrg2 cutoff from 1.6e-11 to 1.4e-11}}}}\par\smallskip
\noindent\parbox[t]{0.05\linewidth}{\centering\strut 34}\hspace{0.006\linewidth}\parbox[t]{0.04\linewidth}{\centering\strut \raisebox{0.08ex}{\ding{51}}}\hspace{0.006\linewidth}\parbox[t]{0.10\linewidth}{\centering\strut +0\%}\hspace{0.006\linewidth}\parbox[t]{0.115\linewidth}{\centering\strut -23.828520}\hspace{0.006\linewidth}\parbox[t]{0.671\linewidth}{reduce dmrg2 sweeps to 14 and tighten dmrg2 cutoff to 1.4e-11}\par\smallskip
\noindent{\setlength{\fboxsep}{0pt}\colorbox{black!4}{\parbox{\linewidth}{\parbox[t]{0.05\linewidth}{\centering\strut 35}\hspace{0.006\linewidth}\parbox[t]{0.04\linewidth}{\centering\strut \raisebox{0.08ex}{\ding{51}}}\hspace{0.006\linewidth}\parbox[t]{0.10\linewidth}{\centering\strut +0\%}\hspace{0.006\linewidth}\parbox[t]{0.115\linewidth}{\centering\strut -23.828520}\hspace{0.006\linewidth}\parbox[t]{0.671\linewidth}{reduce dmrg2 sweeps to 13 and tighten dmrg2 cutoff to 1.2e-11}}}}\par\smallskip
\noindent\parbox[t]{0.05\linewidth}{\centering\strut 36}\hspace{0.006\linewidth}\parbox[t]{0.04\linewidth}{\centering\strut \raisebox{0.08ex}{\ding{51}}}\hspace{0.006\linewidth}\parbox[t]{0.10\linewidth}{\centering\strut +0\%}\hspace{0.006\linewidth}\parbox[t]{0.115\linewidth}{\centering\strut -23.828520}\hspace{0.006\linewidth}\parbox[t]{0.671\linewidth}{reduce dmrg2 sweeps to 12 and tighten dmrg2 cutoff to 1e-11}\par\smallskip
\noindent{\setlength{\fboxsep}{0pt}\colorbox{black!4}{\parbox{\linewidth}{\parbox[t]{0.05\linewidth}{\centering\strut 37}\hspace{0.006\linewidth}\parbox[t]{0.04\linewidth}{\centering\strut \raisebox{0.08ex}{\ding{51}}}\hspace{0.006\linewidth}\parbox[t]{0.10\linewidth}{\centering\strut +0\%}\hspace{0.006\linewidth}\parbox[t]{0.115\linewidth}{\centering\strut -23.828520}\hspace{0.006\linewidth}\parbox[t]{0.671\linewidth}{reduce dmrg2 sweeps to 11 and tighten dmrg2 cutoff to 8e-12}}}}\par\smallskip
\noindent\parbox[t]{0.05\linewidth}{\centering\strut 38}\hspace{0.006\linewidth}\parbox[t]{0.04\linewidth}{\centering\strut \raisebox{0.08ex}{\ding{51}}}\hspace{0.006\linewidth}\parbox[t]{0.10\linewidth}{\centering\strut +0\%}\hspace{0.006\linewidth}\parbox[t]{0.115\linewidth}{\centering\strut -23.828520}\hspace{0.006\linewidth}\parbox[t]{0.671\linewidth}{reduce dmrg2 sweeps to 10 and tighten dmrg2 cutoff to 6e-12}\par\smallskip
\noindent{\setlength{\fboxsep}{0pt}\colorbox{black!4}{\parbox{\linewidth}{\parbox[t]{0.05\linewidth}{\centering\strut 39}\hspace{0.006\linewidth}\parbox[t]{0.04\linewidth}{\centering\strut \raisebox{0.08ex}{\ding{51}}}\hspace{0.006\linewidth}\parbox[t]{0.10\linewidth}{\centering\strut +0\%}\hspace{0.006\linewidth}\parbox[t]{0.115\linewidth}{\centering\strut -23.828520}\hspace{0.006\linewidth}\parbox[t]{0.671\linewidth}{reduce dmrg2 sweeps to 9 and tighten dmrg2 cutoff to 4e-12}}}}\par\smallskip
\noindent\parbox[t]{0.05\linewidth}{\centering\strut 40}\hspace{0.006\linewidth}\parbox[t]{0.04\linewidth}{\centering\strut \raisebox{0.08ex}{\ding{51}}}\hspace{0.006\linewidth}\parbox[t]{0.10\linewidth}{\centering\strut +0\%}\hspace{0.006\linewidth}\parbox[t]{0.115\linewidth}{\centering\strut -23.828520}\hspace{0.006\linewidth}\parbox[t]{0.671\linewidth}{reduce dmrg2 sweeps to 8 and tighten dmrg2 cutoff to 2e-12}\par\smallskip
\noindent{\setlength{\fboxsep}{0pt}\colorbox{black!4}{\parbox{\linewidth}{\parbox[t]{0.05\linewidth}{\centering\strut 41}\hspace{0.006\linewidth}\parbox[t]{0.04\linewidth}{\centering\strut \raisebox{0.08ex}{\ding{51}}}\hspace{0.006\linewidth}\parbox[t]{0.10\linewidth}{\centering\strut +0\%}\hspace{0.006\linewidth}\parbox[t]{0.115\linewidth}{\centering\strut -23.828520}\hspace{0.006\linewidth}\parbox[t]{0.671\linewidth}{reduce dmrg2 sweeps to 7 and tighten dmrg2 cutoff to 1e-12}}}}\par\smallskip
\noindent\parbox[t]{0.05\linewidth}{\centering\strut 42}\hspace{0.006\linewidth}\parbox[t]{0.04\linewidth}{\centering\strut \raisebox{0.08ex}{\ding{51}}}\hspace{0.006\linewidth}\parbox[t]{0.10\linewidth}{\centering\strut +0\%}\hspace{0.006\linewidth}\parbox[t]{0.115\linewidth}{\centering\strut -23.828520}\hspace{0.006\linewidth}\parbox[t]{0.671\linewidth}{raise dmrg2 bond ladder to 64-160 under 1e-12 cutoff}\par\smallskip
\noindent{\setlength{\fboxsep}{0pt}\colorbox{black!4}{\parbox{\linewidth}{\parbox[t]{0.05\linewidth}{\centering\strut 43}\hspace{0.006\linewidth}\parbox[t]{0.04\linewidth}{\centering\strut \raisebox{0.08ex}{\ding{51}}}\hspace{0.006\linewidth}\parbox[t]{0.10\linewidth}{\centering\strut +0\%}\hspace{0.006\linewidth}\parbox[t]{0.115\linewidth}{\centering\strut -23.828520}\hspace{0.006\linewidth}\parbox[t]{0.671\linewidth}{raise dmrg2 preexpand bond from 80 to 128}}}}\par\smallskip
\noindent\parbox[t]{0.05\linewidth}{\centering\strut 44}\hspace{0.006\linewidth}\parbox[t]{0.04\linewidth}{\centering\strut \raisebox{0.08ex}{\ding{51}}}\hspace{0.006\linewidth}\parbox[t]{0.10\linewidth}{\centering\strut +0\%}\hspace{0.006\linewidth}\parbox[t]{0.115\linewidth}{\centering\strut -23.828520}\hspace{0.006\linewidth}\parbox[t]{0.671\linewidth}{shift dmrg2 bond ladder upward to 80-192}\par\smallskip
\noindent{\setlength{\fboxsep}{0pt}\colorbox{black!4}{\parbox{\linewidth}{\parbox[t]{0.05\linewidth}{\centering\strut 45}\hspace{0.006\linewidth}\parbox[t]{0.04\linewidth}{\centering\strut \raisebox{0.08ex}{\ding{51}}}\hspace{0.006\linewidth}\parbox[t]{0.10\linewidth}{\centering\strut +0\%}\hspace{0.006\linewidth}\parbox[t]{0.115\linewidth}{\centering\strut -23.828520}\hspace{0.006\linewidth}\parbox[t]{0.671\linewidth}{tighten dmrg2 cutoff from 1e-12 to 5e-13}}}}\par\smallskip
\noindent\parbox[t]{0.05\linewidth}{\centering\strut 46}\hspace{0.006\linewidth}\parbox[t]{0.04\linewidth}{\centering\strut \raisebox{0.08ex}{\ding{55}}}\hspace{0.006\linewidth}\parbox[t]{0.10\linewidth}{\centering\strut ---}\hspace{0.006\linewidth}\parbox[t]{0.115\linewidth}{\centering\strut ---}\hspace{0.006\linewidth}\parbox[t]{0.671\linewidth}{tighten dmrg2 cutoff from 5e-13 to 2e-13}\par\smallskip
\noindent{\setlength{\fboxsep}{0pt}\colorbox{black!4}{\parbox{\linewidth}{\parbox[t]{0.05\linewidth}{\centering\strut 47}\hspace{0.006\linewidth}\parbox[t]{0.04\linewidth}{\centering\strut \raisebox{0.08ex}{\ding{55}}}\hspace{0.006\linewidth}\parbox[t]{0.10\linewidth}{\centering\strut ---}\hspace{0.006\linewidth}\parbox[t]{0.115\linewidth}{\centering\strut ---}\hspace{0.006\linewidth}\parbox[t]{0.671\linewidth}{tighten dmrg2 cutoff from 5e-13 to 3e-13}}}}\par\smallskip
\noindent\parbox[t]{0.05\linewidth}{\centering\strut 48}\hspace{0.006\linewidth}\parbox[t]{0.04\linewidth}{\centering\strut \raisebox{0.08ex}{\ding{55}}}\hspace{0.006\linewidth}\parbox[t]{0.10\linewidth}{\centering\strut ---}\hspace{0.006\linewidth}\parbox[t]{0.115\linewidth}{\centering\strut ---}\hspace{0.006\linewidth}\parbox[t]{0.671\linewidth}{reduce dmrg2 sweeps to 6 and tighten dmrg2 cutoff to 3e-13}\par\smallskip
\noindent{\setlength{\fboxsep}{0pt}\colorbox{black!4}{\parbox{\linewidth}{\parbox[t]{0.05\linewidth}{\centering\strut 49}\hspace{0.006\linewidth}\parbox[t]{0.04\linewidth}{\centering\strut \raisebox{0.08ex}{\ding{55}}}\hspace{0.006\linewidth}\parbox[t]{0.10\linewidth}{\centering\strut ---}\hspace{0.006\linewidth}\parbox[t]{0.115\linewidth}{\centering\strut ---}\hspace{0.006\linewidth}\parbox[t]{0.671\linewidth}{raise dmrg2 sweeps from 7 to 8 at 5e-13 cutoff}}}}\par\smallskip
\noindent\parbox[t]{0.05\linewidth}{\centering\strut 50}\hspace{0.006\linewidth}\parbox[t]{0.04\linewidth}{\centering\strut \raisebox{0.08ex}{\ding{55}}}\hspace{0.006\linewidth}\parbox[t]{0.10\linewidth}{\centering\strut ---}\hspace{0.006\linewidth}\parbox[t]{0.115\linewidth}{\centering\strut ---}\hspace{0.006\linewidth}\parbox[t]{0.671\linewidth}{raise dmrg2 local\_eig\_ncv from 12 to 13 at 5e-13 cutoff}\par\smallskip
\noindent{\setlength{\fboxsep}{0pt}\colorbox{black!4}{\parbox{\linewidth}{\parbox[t]{0.05\linewidth}{\centering\strut 51}\hspace{0.006\linewidth}\parbox[t]{0.04\linewidth}{\centering\strut \raisebox{0.08ex}{\ding{55}}}\hspace{0.006\linewidth}\parbox[t]{0.10\linewidth}{\centering\strut +0\%}\hspace{0.006\linewidth}\parbox[t]{0.115\linewidth}{\centering\strut -23.828520}\hspace{0.006\linewidth}\parbox[t]{0.671\linewidth}{raise dmrg2 preexpand bond from 128 to 160}}}}\par\smallskip
\noindent\parbox[t]{0.05\linewidth}{\centering\strut 52}\hspace{0.006\linewidth}\parbox[t]{0.04\linewidth}{\centering\strut \raisebox{0.08ex}{\ding{51}}}\hspace{0.006\linewidth}\parbox[t]{0.10\linewidth}{\centering\strut +0\%}\hspace{0.006\linewidth}\parbox[t]{0.115\linewidth}{\centering\strut -23.828520}\hspace{0.006\linewidth}\parbox[t]{0.671\linewidth}{shift dmrg2 bond ladder upward to 96-192}\par\smallskip
\noindent{\setlength{\fboxsep}{0pt}\colorbox{black!4}{\parbox{\linewidth}{\parbox[t]{0.05\linewidth}{\centering\strut 53}\hspace{0.006\linewidth}\parbox[t]{0.04\linewidth}{\centering\strut \raisebox{0.08ex}{\ding{55}}}\hspace{0.006\linewidth}\parbox[t]{0.10\linewidth}{\centering\strut +0\%}\hspace{0.006\linewidth}\parbox[t]{0.115\linewidth}{\centering\strut -23.828520}\hspace{0.006\linewidth}\parbox[t]{0.671\linewidth}{tighten dmrg2 solver\_tol from 1e-6 to 5e-7}}}}\par\smallskip
\noindent\parbox[t]{0.05\linewidth}{\centering\strut 54}\hspace{0.006\linewidth}\parbox[t]{0.04\linewidth}{\centering\strut \raisebox{0.08ex}{\ding{51}}}\hspace{0.006\linewidth}\parbox[t]{0.10\linewidth}{\centering\strut +0\%}\hspace{0.006\linewidth}\parbox[t]{0.115\linewidth}{\centering\strut -23.828520}\hspace{0.006\linewidth}\parbox[t]{0.671\linewidth}{compress dmrg2 bond ladder to 128-192}\par\smallskip
\noindent{\setlength{\fboxsep}{0pt}\colorbox{black!4}{\parbox{\linewidth}{\parbox[t]{0.05\linewidth}{\centering\strut 55}\hspace{0.006\linewidth}\parbox[t]{0.04\linewidth}{\centering\strut \raisebox{0.08ex}{\ding{51}}}\hspace{0.006\linewidth}\parbox[t]{0.10\linewidth}{\centering\strut +0\%}\hspace{0.006\linewidth}\parbox[t]{0.115\linewidth}{\centering\strut -23.828520}\hspace{0.006\linewidth}\parbox[t]{0.671\linewidth}{tighten dmrg2 cutoff from 5e-13 to 4e-13}}}}\par\smallskip
\noindent\parbox[t]{0.05\linewidth}{\centering\strut 56}\hspace{0.006\linewidth}\parbox[t]{0.04\linewidth}{\centering\strut \raisebox{0.08ex}{\ding{55}}}\hspace{0.006\linewidth}\parbox[t]{0.10\linewidth}{\centering\strut ---}\hspace{0.006\linewidth}\parbox[t]{0.115\linewidth}{\centering\strut ---}\hspace{0.006\linewidth}\parbox[t]{0.671\linewidth}{tighten dmrg2 cutoff from 4e-13 to 3.5e-13}\par\smallskip
\noindent{\setlength{\fboxsep}{0pt}\colorbox{black!4}{\parbox{\linewidth}{\parbox[t]{0.05\linewidth}{\centering\strut 57}\hspace{0.006\linewidth}\parbox[t]{0.04\linewidth}{\centering\strut \raisebox{0.08ex}{\ding{51}}}\hspace{0.006\linewidth}\parbox[t]{0.10\linewidth}{\centering\strut +0\%}\hspace{0.006\linewidth}\parbox[t]{0.115\linewidth}{\centering\strut -23.828520}\hspace{0.006\linewidth}\parbox[t]{0.671\linewidth}{reduce dmrg2 sweeps to 6 and tighten dmrg2 cutoff to 3.5e-13}}}}\par\smallskip
\noindent\parbox[t]{0.05\linewidth}{\centering\strut 58}\hspace{0.006\linewidth}\parbox[t]{0.04\linewidth}{\centering\strut \raisebox{0.08ex}{\ding{51}}}\hspace{0.006\linewidth}\parbox[t]{0.10\linewidth}{\centering\strut +0\%}\hspace{0.006\linewidth}\parbox[t]{0.115\linewidth}{\centering\strut -23.828520}\hspace{0.006\linewidth}\parbox[t]{0.671\linewidth}{tighten dmrg2 cutoff from 3.5e-13 to 3e-13}\par\smallskip
\noindent{\setlength{\fboxsep}{0pt}\colorbox{black!4}{\parbox{\linewidth}{\parbox[t]{0.05\linewidth}{\centering\strut 59}\hspace{0.006\linewidth}\parbox[t]{0.04\linewidth}{\centering\strut \raisebox{0.08ex}{\ding{55}}}\hspace{0.006\linewidth}\parbox[t]{0.10\linewidth}{\centering\strut ---}\hspace{0.006\linewidth}\parbox[t]{0.115\linewidth}{\centering\strut ---}\hspace{0.006\linewidth}\parbox[t]{0.671\linewidth}{tighten dmrg2 cutoff from 3e-13 to 2.5e-13}}}}\par\smallskip
\noindent\parbox[t]{0.05\linewidth}{\centering\strut 60}\hspace{0.006\linewidth}\parbox[t]{0.04\linewidth}{\centering\strut \raisebox{0.08ex}{\ding{55}}}\hspace{0.006\linewidth}\parbox[t]{0.10\linewidth}{\centering\strut ---}\hspace{0.006\linewidth}\parbox[t]{0.115\linewidth}{\centering\strut ---}\hspace{0.006\linewidth}\parbox[t]{0.671\linewidth}{tighten dmrg2 cutoff from 3e-13 to 2.8e-13}\par\smallskip
\noindent{\setlength{\fboxsep}{0pt}\colorbox{black!4}{\parbox{\linewidth}{\parbox[t]{0.05\linewidth}{\centering\strut 61}\hspace{0.006\linewidth}\parbox[t]{0.04\linewidth}{\centering\strut \raisebox{0.08ex}{\ding{55}}}\hspace{0.006\linewidth}\parbox[t]{0.10\linewidth}{\centering\strut ---}\hspace{0.006\linewidth}\parbox[t]{0.115\linewidth}{\centering\strut ---}\hspace{0.006\linewidth}\parbox[t]{0.671\linewidth}{compress dmrg2 bond ladder to 160-192}}}}\par\smallskip
\noindent\parbox[t]{0.05\linewidth}{\centering\strut 62}\hspace{0.006\linewidth}\parbox[t]{0.04\linewidth}{\centering\strut \raisebox{0.08ex}{\ding{55}}}\hspace{0.006\linewidth}\parbox[t]{0.10\linewidth}{\centering\strut ---}\hspace{0.006\linewidth}\parbox[t]{0.115\linewidth}{\centering\strut ---}\hspace{0.006\linewidth}\parbox[t]{0.671\linewidth}{reduce warm-start sweeps from 5 to 4 on 3e-13 branch}\par\smallskip
\noindent{\setlength{\fboxsep}{0pt}\colorbox{black!4}{\parbox{\linewidth}{\parbox[t]{0.05\linewidth}{\centering\strut 63}\hspace{0.006\linewidth}\parbox[t]{0.04\linewidth}{\centering\strut \raisebox{0.08ex}{\ding{55}}}\hspace{0.006\linewidth}\parbox[t]{0.10\linewidth}{\centering\strut ---}\hspace{0.006\linewidth}\parbox[t]{0.115\linewidth}{\centering\strut ---}\hspace{0.006\linewidth}\parbox[t]{0.671\linewidth}{raise dmrg2 preexpand bond from 128 to 144}}}}\par\smallskip
\noindent\parbox[t]{0.05\linewidth}{\centering\strut 64}\hspace{0.006\linewidth}\parbox[t]{0.04\linewidth}{\centering\strut \raisebox{0.08ex}{\ding{55}}}\hspace{0.006\linewidth}\parbox[t]{0.10\linewidth}{\centering\strut +0\%}\hspace{0.006\linewidth}\parbox[t]{0.115\linewidth}{\centering\strut -23.828520}\hspace{0.006\linewidth}\parbox[t]{0.671\linewidth}{lower dmrg2 preexpand bond from 128 to 112}\par\smallskip
\noindent{\setlength{\fboxsep}{0pt}\colorbox{black!4}{\parbox{\linewidth}{\parbox[t]{0.05\linewidth}{\centering\strut 65}\hspace{0.006\linewidth}\parbox[t]{0.04\linewidth}{\centering\strut \raisebox{0.08ex}{\ding{51}}}\hspace{0.006\linewidth}\parbox[t]{0.10\linewidth}{\centering\strut +0\%}\hspace{0.006\linewidth}\parbox[t]{0.115\linewidth}{\centering\strut -23.828520}\hspace{0.006\linewidth}\parbox[t]{0.671\linewidth}{lower dmrg2 local\_eig\_ncv to 11 and tighten cutoff to 2.8e-13}}}}\par\smallskip
\noindent\parbox[t]{0.05\linewidth}{\centering\strut 66}\hspace{0.006\linewidth}\parbox[t]{0.04\linewidth}{\centering\strut \raisebox{0.08ex}{\ding{55}}}\hspace{0.006\linewidth}\parbox[t]{0.10\linewidth}{\centering\strut ---}\hspace{0.006\linewidth}\parbox[t]{0.115\linewidth}{\centering\strut ---}\hspace{0.006\linewidth}\parbox[t]{0.671\linewidth}{tighten dmrg2 cutoff from 2.8e-13 to 2.6e-13}\par\smallskip
\noindent{\setlength{\fboxsep}{0pt}\colorbox{black!4}{\parbox{\linewidth}{\parbox[t]{0.05\linewidth}{\centering\strut 67}\hspace{0.006\linewidth}\parbox[t]{0.04\linewidth}{\centering\strut \raisebox{0.08ex}{\ding{55}}}\hspace{0.006\linewidth}\parbox[t]{0.10\linewidth}{\centering\strut ---}\hspace{0.006\linewidth}\parbox[t]{0.115\linewidth}{\centering\strut ---}\hspace{0.006\linewidth}\parbox[t]{0.671\linewidth}{tighten dmrg2 cutoff from 2.8e-13 to 2.7e-13}}}}\par\smallskip
\noindent\parbox[t]{0.05\linewidth}{\centering\strut 68}\hspace{0.006\linewidth}\parbox[t]{0.04\linewidth}{\centering\strut \raisebox{0.08ex}{\ding{55}}}\hspace{0.006\linewidth}\parbox[t]{0.10\linewidth}{\centering\strut ---}\hspace{0.006\linewidth}\parbox[t]{0.115\linewidth}{\centering\strut ---}\hspace{0.006\linewidth}\parbox[t]{0.671\linewidth}{lower dmrg2 local\_eig\_ncv from 11 to 10}\par\smallskip
\noindent{\setlength{\fboxsep}{0pt}\colorbox{black!4}{\parbox{\linewidth}{\parbox[t]{0.05\linewidth}{\centering\strut 69}\hspace{0.006\linewidth}\parbox[t]{0.04\linewidth}{\centering\strut \raisebox{0.08ex}{\ding{51}}}\hspace{0.006\linewidth}\parbox[t]{0.10\linewidth}{\centering\strut +0\%}\hspace{0.006\linewidth}\parbox[t]{0.115\linewidth}{\centering\strut -23.828520}\hspace{0.006\linewidth}\parbox[t]{0.671\linewidth}{concentrate dmrg2 finish at 160-192 over five sweeps}}}}\par\smallskip
\noindent\parbox[t]{0.05\linewidth}{\centering\strut 70}\hspace{0.006\linewidth}\parbox[t]{0.04\linewidth}{\centering\strut \raisebox{0.08ex}{\ding{51}}}\hspace{0.006\linewidth}\parbox[t]{0.10\linewidth}{\centering\strut +0\%}\hspace{0.006\linewidth}\parbox[t]{0.115\linewidth}{\centering\strut -23.828520}\hspace{0.006\linewidth}\parbox[t]{0.671\linewidth}{tighten dmrg2 cutoff from 2.8e-13 to 2.7e-13 on 160-192 branch}\par\smallskip
\noindent{\setlength{\fboxsep}{0pt}\colorbox{black!4}{\parbox{\linewidth}{\parbox[t]{0.05\linewidth}{\centering\strut 71}\hspace{0.006\linewidth}\parbox[t]{0.04\linewidth}{\centering\strut \raisebox{0.08ex}{\ding{51}}}\hspace{0.006\linewidth}\parbox[t]{0.10\linewidth}{\centering\strut +0\%}\hspace{0.006\linewidth}\parbox[t]{0.115\linewidth}{\centering\strut -23.828520}\hspace{0.006\linewidth}\parbox[t]{0.671\linewidth}{tighten dmrg2 cutoff from 2.7e-13 to 2.6e-13 on 160-192 branch}}}}\par\smallskip
\noindent\parbox[t]{0.05\linewidth}{\centering\strut 72}\hspace{0.006\linewidth}\parbox[t]{0.04\linewidth}{\centering\strut \raisebox{0.08ex}{\ding{51}}}\hspace{0.006\linewidth}\parbox[t]{0.10\linewidth}{\centering\strut +0\%}\hspace{0.006\linewidth}\parbox[t]{0.115\linewidth}{\centering\strut -23.828520}\hspace{0.006\linewidth}\parbox[t]{0.671\linewidth}{tighten dmrg2 cutoff from 2.6e-13 to 2.4e-13}\par\smallskip
\noindent{\setlength{\fboxsep}{0pt}\colorbox{black!4}{\parbox{\linewidth}{\parbox[t]{0.05\linewidth}{\centering\strut 73}\hspace{0.006\linewidth}\parbox[t]{0.04\linewidth}{\centering\strut \raisebox{0.08ex}{\ding{51}}}\hspace{0.006\linewidth}\parbox[t]{0.10\linewidth}{\centering\strut +0\%}\hspace{0.006\linewidth}\parbox[t]{0.115\linewidth}{\centering\strut -23.828520}\hspace{0.006\linewidth}\parbox[t]{0.671\linewidth}{tighten dmrg2 cutoff from 2.4e-13 to 2.2e-13}}}}\par\smallskip
\noindent\parbox[t]{0.05\linewidth}{\centering\strut 74}\hspace{0.006\linewidth}\parbox[t]{0.04\linewidth}{\centering\strut \raisebox{0.08ex}{\ding{55}}}\hspace{0.006\linewidth}\parbox[t]{0.10\linewidth}{\centering\strut ---}\hspace{0.006\linewidth}\parbox[t]{0.115\linewidth}{\centering\strut ---}\hspace{0.006\linewidth}\parbox[t]{0.671\linewidth}{tighten dmrg2 cutoff from 2.2e-13 to 2e-13}\par\smallskip
\noindent{\setlength{\fboxsep}{0pt}\colorbox{black!4}{\parbox{\linewidth}{\parbox[t]{0.05\linewidth}{\centering\strut 75}\hspace{0.006\linewidth}\parbox[t]{0.04\linewidth}{\centering\strut \raisebox{0.08ex}{\ding{55}}}\hspace{0.006\linewidth}\parbox[t]{0.10\linewidth}{\centering\strut ---}\hspace{0.006\linewidth}\parbox[t]{0.115\linewidth}{\centering\strut ---}\hspace{0.006\linewidth}\parbox[t]{0.671\linewidth}{tighten dmrg2 cutoff from 2.2e-13 to 2.1e-13}}}}\par\smallskip
\noindent\parbox[t]{0.05\linewidth}{\centering\strut 76}\hspace{0.006\linewidth}\parbox[t]{0.04\linewidth}{\centering\strut \raisebox{0.08ex}{\ding{55}}}\hspace{0.006\linewidth}\parbox[t]{0.10\linewidth}{\centering\strut ---}\hspace{0.006\linewidth}\parbox[t]{0.115\linewidth}{\centering\strut ---}\hspace{0.006\linewidth}\parbox[t]{0.671\linewidth}{raise dmrg2 preexpand bond from 128 to 144 on 160-192 branch}\par\smallskip
\noindent{\setlength{\fboxsep}{0pt}\colorbox{black!4}{\parbox{\linewidth}{\parbox[t]{0.05\linewidth}{\centering\strut 77}\hspace{0.006\linewidth}\parbox[t]{0.04\linewidth}{\centering\strut \raisebox{0.08ex}{\ding{51}}}\hspace{0.006\linewidth}\parbox[t]{0.10\linewidth}{\centering\strut +0\%}\hspace{0.006\linewidth}\parbox[t]{0.115\linewidth}{\centering\strut -23.828520}\hspace{0.006\linewidth}\parbox[t]{0.671\linewidth}{lower first dmrg2 high-bond step from 160 to 152}}}}\par\smallskip
\noindent\parbox[t]{0.05\linewidth}{\centering\strut 78}\hspace{0.006\linewidth}\parbox[t]{0.04\linewidth}{\centering\strut \raisebox{0.08ex}{\ding{51}}}\hspace{0.006\linewidth}\parbox[t]{0.10\linewidth}{\centering\strut +0\%}\hspace{0.006\linewidth}\parbox[t]{0.115\linewidth}{\centering\strut -23.828520}\hspace{0.006\linewidth}\parbox[t]{0.671\linewidth}{tighten dmrg2 cutoff from 2.2e-13 to 2.1e-13 on 152-192 branch}\par\smallskip
\noindent{\setlength{\fboxsep}{0pt}\colorbox{black!4}{\parbox{\linewidth}{\parbox[t]{0.05\linewidth}{\centering\strut 79}\hspace{0.006\linewidth}\parbox[t]{0.04\linewidth}{\centering\strut \raisebox{0.08ex}{\ding{51}}}\hspace{0.006\linewidth}\parbox[t]{0.10\linewidth}{\centering\strut +0\%}\hspace{0.006\linewidth}\parbox[t]{0.115\linewidth}{\centering\strut -23.828520}\hspace{0.006\linewidth}\parbox[t]{0.671\linewidth}{tighten dmrg2 cutoff from 2.1e-13 to 2e-13 on 152-192 branch}}}}\par\smallskip
\noindent\parbox[t]{0.05\linewidth}{\centering\strut 80}\hspace{0.006\linewidth}\parbox[t]{0.04\linewidth}{\centering\strut \raisebox{0.08ex}{\ding{51}}}\hspace{0.006\linewidth}\parbox[t]{0.10\linewidth}{\centering\strut +0\%}\hspace{0.006\linewidth}\parbox[t]{0.115\linewidth}{\centering\strut -23.828520}\hspace{0.006\linewidth}\parbox[t]{0.671\linewidth}{tighten dmrg2 cutoff from 2e-13 to 1.9e-13}\par\smallskip
\noindent{\setlength{\fboxsep}{0pt}\colorbox{black!4}{\parbox{\linewidth}{\parbox[t]{0.05\linewidth}{\centering\strut 81}\hspace{0.006\linewidth}\parbox[t]{0.04\linewidth}{\centering\strut \raisebox{0.08ex}{\ding{51}}}\hspace{0.006\linewidth}\parbox[t]{0.10\linewidth}{\centering\strut +0\%}\hspace{0.006\linewidth}\parbox[t]{0.115\linewidth}{\centering\strut -23.828520}\hspace{0.006\linewidth}\parbox[t]{0.671\linewidth}{tighten dmrg2 cutoff from 1.9e-13 to 1.8e-13}}}}\par\smallskip
\noindent\parbox[t]{0.05\linewidth}{\centering\strut 82}\hspace{0.006\linewidth}\parbox[t]{0.04\linewidth}{\centering\strut \raisebox{0.08ex}{\ding{51}}}\hspace{0.006\linewidth}\parbox[t]{0.10\linewidth}{\centering\strut +0\%}\hspace{0.006\linewidth}\parbox[t]{0.115\linewidth}{\centering\strut -23.828520}\hspace{0.006\linewidth}\parbox[t]{0.671\linewidth}{tighten dmrg2 cutoff from 1.8e-13 to 1.7e-13}\par\smallskip
\noindent{\setlength{\fboxsep}{0pt}\colorbox{black!4}{\parbox{\linewidth}{\parbox[t]{0.05\linewidth}{\centering\strut 83}\hspace{0.006\linewidth}\parbox[t]{0.04\linewidth}{\centering\strut \raisebox{0.08ex}{\ding{51}}}\hspace{0.006\linewidth}\parbox[t]{0.10\linewidth}{\centering\strut +0\%}\hspace{0.006\linewidth}\parbox[t]{0.115\linewidth}{\centering\strut -23.828520}\hspace{0.006\linewidth}\parbox[t]{0.671\linewidth}{tighten dmrg2 cutoff from 1.7e-13 to 1.6e-13}}}}\par\smallskip
\noindent\parbox[t]{0.05\linewidth}{\centering\strut 84}\hspace{0.006\linewidth}\parbox[t]{0.04\linewidth}{\centering\strut \raisebox{0.08ex}{\ding{51}}}\hspace{0.006\linewidth}\parbox[t]{0.10\linewidth}{\centering\strut +0\%}\hspace{0.006\linewidth}\parbox[t]{0.115\linewidth}{\centering\strut -23.828520}\hspace{0.006\linewidth}\parbox[t]{0.671\linewidth}{tighten dmrg2 cutoff from 1.6e-13 to 1.5e-13}\par\smallskip
\noindent{\setlength{\fboxsep}{0pt}\colorbox{black!4}{\parbox{\linewidth}{\parbox[t]{0.05\linewidth}{\centering\strut 85}\hspace{0.006\linewidth}\parbox[t]{0.04\linewidth}{\centering\strut \raisebox{0.08ex}{\ding{55}}}\hspace{0.006\linewidth}\parbox[t]{0.10\linewidth}{\centering\strut ---}\hspace{0.006\linewidth}\parbox[t]{0.115\linewidth}{\centering\strut ---}\hspace{0.006\linewidth}\parbox[t]{0.671\linewidth}{tighten dmrg2 cutoff from 1.5e-13 to 1.4e-13}}}}\par\smallskip
\noindent\parbox[t]{0.05\linewidth}{\centering\strut 86}\hspace{0.006\linewidth}\parbox[t]{0.04\linewidth}{\centering\strut \raisebox{0.08ex}{\ding{55}}}\hspace{0.006\linewidth}\parbox[t]{0.10\linewidth}{\centering\strut ---}\hspace{0.006\linewidth}\parbox[t]{0.115\linewidth}{\centering\strut ---}\hspace{0.006\linewidth}\parbox[t]{0.671\linewidth}{raise first dmrg2 high-bond step from 152 to 154}\par\smallskip
\noindent{\setlength{\fboxsep}{0pt}\colorbox{black!4}{\parbox{\linewidth}{\parbox[t]{0.05\linewidth}{\centering\strut 87}\hspace{0.006\linewidth}\parbox[t]{0.04\linewidth}{\centering\strut \raisebox{0.08ex}{\ding{51}}}\hspace{0.006\linewidth}\parbox[t]{0.10\linewidth}{\centering\strut +0\%}\hspace{0.006\linewidth}\parbox[t]{0.115\linewidth}{\centering\strut -23.828520}\hspace{0.006\linewidth}\parbox[t]{0.671\linewidth}{lower first dmrg2 high-bond step from 152 to 148}}}}\par\smallskip
\noindent\parbox[t]{0.05\linewidth}{\centering\strut 88}\hspace{0.006\linewidth}\parbox[t]{0.04\linewidth}{\centering\strut \raisebox{0.08ex}{\ding{51}}}\hspace{0.006\linewidth}\parbox[t]{0.10\linewidth}{\centering\strut +0\%}\hspace{0.006\linewidth}\parbox[t]{0.115\linewidth}{\centering\strut -23.828520}\hspace{0.006\linewidth}\parbox[t]{0.671\linewidth}{tighten dmrg2 cutoff from 1.5e-13 to 1.4e-13 on 148-192 branch}\par\smallskip
\noindent{\setlength{\fboxsep}{0pt}\colorbox{black!4}{\parbox{\linewidth}{\parbox[t]{0.05\linewidth}{\centering\strut 89}\hspace{0.006\linewidth}\parbox[t]{0.04\linewidth}{\centering\strut \raisebox{0.08ex}{\ding{51}}}\hspace{0.006\linewidth}\parbox[t]{0.10\linewidth}{\centering\strut +0\%}\hspace{0.006\linewidth}\parbox[t]{0.115\linewidth}{\centering\strut -23.828520}\hspace{0.006\linewidth}\parbox[t]{0.671\linewidth}{tighten dmrg2 cutoff from 1.4e-13 to 1.3e-13 on 148-192 branch}}}}\par\smallskip
\noindent\parbox[t]{0.05\linewidth}{\centering\strut 90}\hspace{0.006\linewidth}\parbox[t]{0.04\linewidth}{\centering\strut \raisebox{0.08ex}{\ding{51}}}\hspace{0.006\linewidth}\parbox[t]{0.10\linewidth}{\centering\strut +0\%}\hspace{0.006\linewidth}\parbox[t]{0.115\linewidth}{\centering\strut -23.828520}\hspace{0.006\linewidth}\parbox[t]{0.671\linewidth}{tighten dmrg2 cutoff from 1.3e-13 to 1.2e-13 on 148-192 branch}\par\smallskip
\noindent{\setlength{\fboxsep}{0pt}\colorbox{black!4}{\parbox{\linewidth}{\parbox[t]{0.05\linewidth}{\centering\strut 91}\hspace{0.006\linewidth}\parbox[t]{0.04\linewidth}{\centering\strut \raisebox{0.08ex}{\ding{51}}}\hspace{0.006\linewidth}\parbox[t]{0.10\linewidth}{\centering\strut +0\%}\hspace{0.006\linewidth}\parbox[t]{0.115\linewidth}{\centering\strut -23.828520}\hspace{0.006\linewidth}\parbox[t]{0.671\linewidth}{tighten dmrg2 cutoff from 1.2e-13 to 1.1e-13 on 148-192 branch}}}}\par\smallskip
\noindent\parbox[t]{0.05\linewidth}{\centering\strut 92}\hspace{0.006\linewidth}\parbox[t]{0.04\linewidth}{\centering\strut \raisebox{0.08ex}{\ding{51}}}\hspace{0.006\linewidth}\parbox[t]{0.10\linewidth}{\centering\strut +0\%}\hspace{0.006\linewidth}\parbox[t]{0.115\linewidth}{\centering\strut -23.828520}\hspace{0.006\linewidth}\parbox[t]{0.671\linewidth}{tighten dmrg2 cutoff from 1.1e-13 to 1e-13 on 148-192 branch}\par\smallskip
\noindent{\setlength{\fboxsep}{0pt}\colorbox{black!4}{\parbox{\linewidth}{\parbox[t]{0.05\linewidth}{\centering\strut 93}\hspace{0.006\linewidth}\parbox[t]{0.04\linewidth}{\centering\strut \raisebox{0.08ex}{\ding{51}}}\hspace{0.006\linewidth}\parbox[t]{0.10\linewidth}{\centering\strut +0\%}\hspace{0.006\linewidth}\parbox[t]{0.115\linewidth}{\centering\strut -23.828520}\hspace{0.006\linewidth}\parbox[t]{0.671\linewidth}{tighten dmrg2 cutoff from 1e-13 to 9e-14 on 148-192 branch}}}}\par\smallskip
\noindent\parbox[t]{0.05\linewidth}{\centering\strut 94}\hspace{0.006\linewidth}\parbox[t]{0.04\linewidth}{\centering\strut \raisebox{0.08ex}{\ding{51}}}\hspace{0.006\linewidth}\parbox[t]{0.10\linewidth}{\centering\strut +0\%}\hspace{0.006\linewidth}\parbox[t]{0.115\linewidth}{\centering\strut -23.828520}\hspace{0.006\linewidth}\parbox[t]{0.671\linewidth}{tighten dmrg2 cutoff from 9e-14 to 8e-14 on 148-192 branch}\par\smallskip
\noindent{\setlength{\fboxsep}{0pt}\colorbox{black!4}{\parbox{\linewidth}{\parbox[t]{0.05\linewidth}{\centering\strut 95}\hspace{0.006\linewidth}\parbox[t]{0.04\linewidth}{\centering\strut \raisebox{0.08ex}{\ding{51}}}\hspace{0.006\linewidth}\parbox[t]{0.10\linewidth}{\centering\strut +0\%}\hspace{0.006\linewidth}\parbox[t]{0.115\linewidth}{\centering\strut -23.828520}\hspace{0.006\linewidth}\parbox[t]{0.671\linewidth}{tighten dmrg2 cutoff from 8e-14 to 7e-14 on 148-192 branch}}}}\par\smallskip
\noindent\parbox[t]{0.05\linewidth}{\centering\strut 96}\hspace{0.006\linewidth}\parbox[t]{0.04\linewidth}{\centering\strut \raisebox{0.08ex}{\ding{51}}}\hspace{0.006\linewidth}\parbox[t]{0.10\linewidth}{\centering\strut +0\%}\hspace{0.006\linewidth}\parbox[t]{0.115\linewidth}{\centering\strut -23.828520}\hspace{0.006\linewidth}\parbox[t]{0.671\linewidth}{tighten dmrg2 cutoff from 7e-14 to 6e-14 on 148-192 branch}\par\smallskip
\noindent{\setlength{\fboxsep}{0pt}\colorbox{black!4}{\parbox{\linewidth}{\parbox[t]{0.05\linewidth}{\centering\strut 97}\hspace{0.006\linewidth}\parbox[t]{0.04\linewidth}{\centering\strut \raisebox{0.08ex}{\ding{51}}}\hspace{0.006\linewidth}\parbox[t]{0.10\linewidth}{\centering\strut +0\%}\hspace{0.006\linewidth}\parbox[t]{0.115\linewidth}{\centering\strut -23.828520}\hspace{0.006\linewidth}\parbox[t]{0.671\linewidth}{tighten dmrg2 cutoff from 6e-14 to 5e-14 on 148-192 branch}}}}\par\smallskip
\noindent\parbox[t]{0.05\linewidth}{\centering\strut 98}\hspace{0.006\linewidth}\parbox[t]{0.04\linewidth}{\centering\strut \raisebox{0.08ex}{\ding{55}}}\hspace{0.006\linewidth}\parbox[t]{0.10\linewidth}{\centering\strut ---}\hspace{0.006\linewidth}\parbox[t]{0.115\linewidth}{\centering\strut ---}\hspace{0.006\linewidth}\parbox[t]{0.671\linewidth}{tighten dmrg2 cutoff from 5e-14 to 4.5e-14 on 148-192 branch}\par\smallskip
\noindent{\setlength{\fboxsep}{0pt}\colorbox{black!4}{\parbox{\linewidth}{\parbox[t]{0.05\linewidth}{\centering\strut 99}\hspace{0.006\linewidth}\parbox[t]{0.04\linewidth}{\centering\strut \raisebox{0.08ex}{\ding{55}}}\hspace{0.006\linewidth}\parbox[t]{0.10\linewidth}{\centering\strut +0\%}\hspace{0.006\linewidth}\parbox[t]{0.115\linewidth}{\centering\strut -23.828520}\hspace{0.006\linewidth}\parbox[t]{0.671\linewidth}{lower first dmrg2 high-bond step from 148 to 144 at 5e-14 cutoff}}}}\par\smallskip
\noindent\parbox[t]{0.05\linewidth}{\centering\strut 100}\hspace{0.006\linewidth}\parbox[t]{0.04\linewidth}{\centering\strut \raisebox{0.08ex}{\ding{51}}}\hspace{0.006\linewidth}\parbox[t]{0.10\linewidth}{\centering\strut +0\%}\hspace{0.006\linewidth}\parbox[t]{0.115\linewidth}{\centering\strut -23.828520}\hspace{0.006\linewidth}\parbox[t]{0.671\linewidth}{tighten dmrg2 cutoff from 5e-14 to 4.8e-14 on 148-192 branch}\par\smallskip
\noindent\rule{\linewidth}{0.5pt}
\par\medskip

\medskip

\refstepcounter{table}
\label{tab:supp_tn_log_tfim_critical_64}
\noindent\textbf{Table \thetable.} Tensor-network mutation log for Critical TFIM, $L=64$. Relative gain is measured against the immediately previous scored iteration using the live final-energy objective.
\par\medskip
\noindent\rule{\linewidth}{0.5pt}
\noindent{\setlength{\fboxsep}{0pt}\colorbox{black!4}{\parbox{\linewidth}{\parbox[c][2.2em][c]{0.05\linewidth}{\centering \textbf{Iter.}}\hspace{0.006\linewidth}\parbox[c][2.2em][c]{0.04\linewidth}{\centering \textbf{A/R}}\hspace{0.006\linewidth}\parbox[c][2.2em][c]{0.10\linewidth}{\centering \textbf{Relative gain}}\hspace{0.006\linewidth}\parbox[c][2.2em][c]{0.115\linewidth}{\centering \textbf{Score}}\hspace{0.006\linewidth}\parbox[c][2.2em][c]{0.671\linewidth}{\centering \textbf{Mutation summary}}}}}\par\smallskip
\noindent\rule{\linewidth}{0.5pt}
\noindent{\setlength{\fboxsep}{0pt}\colorbox{black!4}{\parbox{\linewidth}{\parbox[t]{0.05\linewidth}{\centering\strut 1}\hspace{0.006\linewidth}\parbox[t]{0.04\linewidth}{\centering\strut \raisebox{0.08ex}{\ding{51}}}\hspace{0.006\linewidth}\parbox[t]{0.10\linewidth}{\centering\strut ---}\hspace{0.006\linewidth}\parbox[t]{0.115\linewidth}{\centering\strut -20.281494}\hspace{0.006\linewidth}\parbox[t]{0.671\linewidth}{switch to dmrg2 plus init with 8-80 bond ladder}}}}\par\smallskip
\noindent\parbox[t]{0.05\linewidth}{\centering\strut 2}\hspace{0.006\linewidth}\parbox[t]{0.04\linewidth}{\centering\strut \raisebox{0.08ex}{\ding{51}}}\hspace{0.006\linewidth}\parbox[t]{0.10\linewidth}{\centering\strut +0\%}\hspace{0.006\linewidth}\parbox[t]{0.115\linewidth}{\centering\strut -20.281495}\hspace{0.006\linewidth}\parbox[t]{0.671\linewidth}{switch to explicit dmrg2 plus init, shrink init\_bond\_dim to 2, extend bond\_schedule to 8-12-16-24-32-48-64-80-96}\par\smallskip
\noindent{\setlength{\fboxsep}{0pt}\colorbox{black!4}{\parbox{\linewidth}{\parbox[t]{0.05\linewidth}{\centering\strut 3}\hspace{0.006\linewidth}\parbox[t]{0.04\linewidth}{\centering\strut \raisebox{0.08ex}{\ding{55}}}\hspace{0.006\linewidth}\parbox[t]{0.10\linewidth}{\centering\strut +0\%}\hspace{0.006\linewidth}\parbox[t]{0.115\linewidth}{\centering\strut -20.281495}\hspace{0.006\linewidth}\parbox[t]{0.671\linewidth}{extend explicit dmrg2 plus ladder to 8-12-16-24-32-48-64-80-96-128 with 20 sweeps}}}}\par\smallskip
\noindent\parbox[t]{0.05\linewidth}{\centering\strut 4}\hspace{0.006\linewidth}\parbox[t]{0.04\linewidth}{\centering\strut \raisebox{0.08ex}{\ding{51}}}\hspace{0.006\linewidth}\parbox[t]{0.10\linewidth}{\centering\strut +0\%}\hspace{0.006\linewidth}\parbox[t]{0.115\linewidth}{\centering\strut -20.281495}\hspace{0.006\linewidth}\parbox[t]{0.671\linewidth}{add dmrg1 polish after explicit dmrg2 plus ladder}\par\smallskip
\noindent{\setlength{\fboxsep}{0pt}\colorbox{black!4}{\parbox{\linewidth}{\parbox[t]{0.05\linewidth}{\centering\strut 5}\hspace{0.006\linewidth}\parbox[t]{0.04\linewidth}{\centering\strut \raisebox{0.08ex}{\ding{55}}}\hspace{0.006\linewidth}\parbox[t]{0.10\linewidth}{\centering\strut +0\%}\hspace{0.006\linewidth}\parbox[t]{0.115\linewidth}{\centering\strut -20.281495}\hspace{0.006\linewidth}\parbox[t]{0.671\linewidth}{add dmrg1 plus warm start before dmrg2 refinement ladder}}}}\par\smallskip
\noindent\parbox[t]{0.05\linewidth}{\centering\strut 6}\hspace{0.006\linewidth}\parbox[t]{0.04\linewidth}{\centering\strut \raisebox{0.08ex}{\ding{55}}}\hspace{0.006\linewidth}\parbox[t]{0.10\linewidth}{\centering\strut +0\%}\hspace{0.006\linewidth}\parbox[t]{0.115\linewidth}{\centering\strut -20.281495}\hspace{0.006\linewidth}\parbox[t]{0.671\linewidth}{add tebd1d plus warmup before dmrg2 refinement}\par\smallskip
\noindent{\setlength{\fboxsep}{0pt}\colorbox{black!4}{\parbox{\linewidth}{\parbox[t]{0.05\linewidth}{\centering\strut 7}\hspace{0.006\linewidth}\parbox[t]{0.04\linewidth}{\centering\strut \raisebox{0.08ex}{\ding{55}}}\hspace{0.006\linewidth}\parbox[t]{0.10\linewidth}{\centering\strut +0\%}\hspace{0.006\linewidth}\parbox[t]{0.115\linewidth}{\centering\strut -20.281495}\hspace{0.006\linewidth}\parbox[t]{0.671\linewidth}{switch final phase to random dmrg2 explicit ladder with init\_bond\_dim 4}}}}\par\smallskip
\noindent\parbox[t]{0.05\linewidth}{\centering\strut 8}\hspace{0.006\linewidth}\parbox[t]{0.04\linewidth}{\centering\strut \raisebox{0.08ex}{\ding{51}}}\hspace{0.006\linewidth}\parbox[t]{0.10\linewidth}{\centering\strut +0\%}\hspace{0.006\linewidth}\parbox[t]{0.115\linewidth}{\centering\strut -20.281495}\hspace{0.006\linewidth}\parbox[t]{0.671\linewidth}{switch final dmrg1 loop to solve sweeps}\par\smallskip
\noindent{\setlength{\fboxsep}{0pt}\colorbox{black!4}{\parbox{\linewidth}{\parbox[t]{0.05\linewidth}{\centering\strut 9}\hspace{0.006\linewidth}\parbox[t]{0.04\linewidth}{\centering\strut \raisebox{0.08ex}{\ding{55}}}\hspace{0.006\linewidth}\parbox[t]{0.10\linewidth}{\centering\strut +0\%}\hspace{0.006\linewidth}\parbox[t]{0.115\linewidth}{\centering\strut -20.281495}\hspace{0.006\linewidth}\parbox[t]{0.671\linewidth}{switch final dmrg1 loop to explicit sweeps}}}}\par\smallskip
\noindent\parbox[t]{0.05\linewidth}{\centering\strut 10}\hspace{0.006\linewidth}\parbox[t]{0.04\linewidth}{\centering\strut \raisebox{0.08ex}{\ding{51}}}\hspace{0.006\linewidth}\parbox[t]{0.10\linewidth}{\centering\strut +0\%}\hspace{0.006\linewidth}\parbox[t]{0.115\linewidth}{\centering\strut -20.281495}\hspace{0.006\linewidth}\parbox[t]{0.671\linewidth}{change final dmrg1 init\_state to minus}\par\smallskip
\noindent{\setlength{\fboxsep}{0pt}\colorbox{black!4}{\parbox{\linewidth}{\parbox[t]{0.05\linewidth}{\centering\strut 11}\hspace{0.006\linewidth}\parbox[t]{0.04\linewidth}{\centering\strut \raisebox{0.08ex}{\ding{55}}}\hspace{0.006\linewidth}\parbox[t]{0.10\linewidth}{\centering\strut +0\%}\hspace{0.006\linewidth}\parbox[t]{0.115\linewidth}{\centering\strut -20.281495}\hspace{0.006\linewidth}\parbox[t]{0.671\linewidth}{switch final dmrg1 loop to explicit sweeps}}}}\par\smallskip
\noindent\parbox[t]{0.05\linewidth}{\centering\strut 12}\hspace{0.006\linewidth}\parbox[t]{0.04\linewidth}{\centering\strut \raisebox{0.08ex}{\ding{55}}}\hspace{0.006\linewidth}\parbox[t]{0.10\linewidth}{\centering\strut +0\%}\hspace{0.006\linewidth}\parbox[t]{0.115\linewidth}{\centering\strut -20.281495}\hspace{0.006\linewidth}\parbox[t]{0.671\linewidth}{change final dmrg1 init\_state to plus}\par\smallskip
\noindent{\setlength{\fboxsep}{0pt}\colorbox{black!4}{\parbox{\linewidth}{\parbox[t]{0.05\linewidth}{\centering\strut 13}\hspace{0.006\linewidth}\parbox[t]{0.04\linewidth}{\centering\strut \raisebox{0.08ex}{\ding{55}}}\hspace{0.006\linewidth}\parbox[t]{0.10\linewidth}{\centering\strut +0\%}\hspace{0.006\linewidth}\parbox[t]{0.115\linewidth}{\centering\strut -20.281495}\hspace{0.006\linewidth}\parbox[t]{0.671\linewidth}{change final dmrg1 init\_state to product\_up}}}}\par\smallskip
\noindent\parbox[t]{0.05\linewidth}{\centering\strut 14}\hspace{0.006\linewidth}\parbox[t]{0.04\linewidth}{\centering\strut \raisebox{0.08ex}{\ding{55}}}\hspace{0.006\linewidth}\parbox[t]{0.10\linewidth}{\centering\strut +0\%}\hspace{0.006\linewidth}\parbox[t]{0.115\linewidth}{\centering\strut -20.281495}\hspace{0.006\linewidth}\parbox[t]{0.671\linewidth}{change final dmrg1 init\_state to product\_down}\par\smallskip
\noindent{\setlength{\fboxsep}{0pt}\colorbox{black!4}{\parbox{\linewidth}{\parbox[t]{0.05\linewidth}{\centering\strut 15}\hspace{0.006\linewidth}\parbox[t]{0.04\linewidth}{\centering\strut \raisebox{0.08ex}{\ding{55}}}\hspace{0.006\linewidth}\parbox[t]{0.10\linewidth}{\centering\strut +0\%}\hspace{0.006\linewidth}\parbox[t]{0.115\linewidth}{\centering\strut -20.281495}\hspace{0.006\linewidth}\parbox[t]{0.671\linewidth}{change final dmrg1 init\_state to random}}}}\par\smallskip
\noindent\parbox[t]{0.05\linewidth}{\centering\strut 16}\hspace{0.006\linewidth}\parbox[t]{0.04\linewidth}{\centering\strut \raisebox{0.08ex}{\ding{55}}}\hspace{0.006\linewidth}\parbox[t]{0.10\linewidth}{\centering\strut +0\%}\hspace{0.006\linewidth}\parbox[t]{0.115\linewidth}{\centering\strut -20.281495}\hspace{0.006\linewidth}\parbox[t]{0.671\linewidth}{set final dmrg1 init\_bond\_dim to 1}\par\smallskip
\noindent{\setlength{\fboxsep}{0pt}\colorbox{black!4}{\parbox{\linewidth}{\parbox[t]{0.05\linewidth}{\centering\strut 17}\hspace{0.006\linewidth}\parbox[t]{0.04\linewidth}{\centering\strut \raisebox{0.08ex}{\ding{55}}}\hspace{0.006\linewidth}\parbox[t]{0.10\linewidth}{\centering\strut +0\%}\hspace{0.006\linewidth}\parbox[t]{0.115\linewidth}{\centering\strut -20.281495}\hspace{0.006\linewidth}\parbox[t]{0.671\linewidth}{set final dmrg1 init\_bond\_dim to 4}}}}\par\smallskip
\noindent\parbox[t]{0.05\linewidth}{\centering\strut 18}\hspace{0.006\linewidth}\parbox[t]{0.04\linewidth}{\centering\strut \raisebox{0.08ex}{\ding{55}}}\hspace{0.006\linewidth}\parbox[t]{0.10\linewidth}{\centering\strut +0\%}\hspace{0.006\linewidth}\parbox[t]{0.115\linewidth}{\centering\strut -20.281495}\hspace{0.006\linewidth}\parbox[t]{0.671\linewidth}{set final dmrg1 init\_bond\_dim to 8}\par\smallskip
\noindent{\setlength{\fboxsep}{0pt}\colorbox{black!4}{\parbox{\linewidth}{\parbox[t]{0.05\linewidth}{\centering\strut 19}\hspace{0.006\linewidth}\parbox[t]{0.04\linewidth}{\centering\strut \raisebox{0.08ex}{\ding{55}}}\hspace{0.006\linewidth}\parbox[t]{0.10\linewidth}{\centering\strut +0\%}\hspace{0.006\linewidth}\parbox[t]{0.115\linewidth}{\centering\strut -20.281495}\hspace{0.006\linewidth}\parbox[t]{0.671\linewidth}{set final dmrg1 local\_eig\_ncv to 6}}}}\par\smallskip
\noindent\parbox[t]{0.05\linewidth}{\centering\strut 20}\hspace{0.006\linewidth}\parbox[t]{0.04\linewidth}{\centering\strut \raisebox{0.08ex}{\ding{55}}}\hspace{0.006\linewidth}\parbox[t]{0.10\linewidth}{\centering\strut +0\%}\hspace{0.006\linewidth}\parbox[t]{0.115\linewidth}{\centering\strut -20.281495}\hspace{0.006\linewidth}\parbox[t]{0.671\linewidth}{set final dmrg1 local\_eig\_ncv to 10}\par\smallskip
\noindent{\setlength{\fboxsep}{0pt}\colorbox{black!4}{\parbox{\linewidth}{\parbox[t]{0.05\linewidth}{\centering\strut 21}\hspace{0.006\linewidth}\parbox[t]{0.04\linewidth}{\centering\strut \raisebox{0.08ex}{\ding{55}}}\hspace{0.006\linewidth}\parbox[t]{0.10\linewidth}{\centering\strut +0\%}\hspace{0.006\linewidth}\parbox[t]{0.115\linewidth}{\centering\strut -20.281495}\hspace{0.006\linewidth}\parbox[t]{0.671\linewidth}{set final dmrg1 local\_eig\_ncv to 12}}}}\par\smallskip
\noindent\parbox[t]{0.05\linewidth}{\centering\strut 22}\hspace{0.006\linewidth}\parbox[t]{0.04\linewidth}{\centering\strut \raisebox{0.08ex}{\ding{55}}}\hspace{0.006\linewidth}\parbox[t]{0.10\linewidth}{\centering\strut +0\%}\hspace{0.006\linewidth}\parbox[t]{0.115\linewidth}{\centering\strut -20.281495}\hspace{0.006\linewidth}\parbox[t]{0.671\linewidth}{set final dmrg1 local\_eig\_ncv to 14}\par\smallskip
\noindent{\setlength{\fboxsep}{0pt}\colorbox{black!4}{\parbox{\linewidth}{\parbox[t]{0.05\linewidth}{\centering\strut 23}\hspace{0.006\linewidth}\parbox[t]{0.04\linewidth}{\centering\strut \raisebox{0.08ex}{\ding{55}}}\hspace{0.006\linewidth}\parbox[t]{0.10\linewidth}{\centering\strut +0\%}\hspace{0.006\linewidth}\parbox[t]{0.115\linewidth}{\centering\strut -20.281495}\hspace{0.006\linewidth}\parbox[t]{0.671\linewidth}{tighten final dmrg1 cutoff to 1e-08}}}}\par\smallskip
\noindent\parbox[t]{0.05\linewidth}{\centering\strut 24}\hspace{0.006\linewidth}\parbox[t]{0.04\linewidth}{\centering\strut \raisebox{0.08ex}{\ding{55}}}\hspace{0.006\linewidth}\parbox[t]{0.10\linewidth}{\centering\strut +0\%}\hspace{0.006\linewidth}\parbox[t]{0.115\linewidth}{\centering\strut -20.281495}\hspace{0.006\linewidth}\parbox[t]{0.671\linewidth}{tighten final dmrg1 cutoff to 1e-09}\par\smallskip
\noindent{\setlength{\fboxsep}{0pt}\colorbox{black!4}{\parbox{\linewidth}{\parbox[t]{0.05\linewidth}{\centering\strut 25}\hspace{0.006\linewidth}\parbox[t]{0.04\linewidth}{\centering\strut \raisebox{0.08ex}{\ding{55}}}\hspace{0.006\linewidth}\parbox[t]{0.10\linewidth}{\centering\strut +0\%}\hspace{0.006\linewidth}\parbox[t]{0.115\linewidth}{\centering\strut -20.281495}\hspace{0.006\linewidth}\parbox[t]{0.671\linewidth}{tighten final dmrg1 cutoff to 1e-11}}}}\par\smallskip
\noindent\parbox[t]{0.05\linewidth}{\centering\strut 26}\hspace{0.006\linewidth}\parbox[t]{0.04\linewidth}{\centering\strut \raisebox{0.08ex}{\ding{55}}}\hspace{0.006\linewidth}\parbox[t]{0.10\linewidth}{\centering\strut +0\%}\hspace{0.006\linewidth}\parbox[t]{0.115\linewidth}{\centering\strut -20.281495}\hspace{0.006\linewidth}\parbox[t]{0.671\linewidth}{tighten final dmrg1 cutoff to 1e-12}\par\smallskip
\noindent{\setlength{\fboxsep}{0pt}\colorbox{black!4}{\parbox{\linewidth}{\parbox[t]{0.05\linewidth}{\centering\strut 27}\hspace{0.006\linewidth}\parbox[t]{0.04\linewidth}{\centering\strut \raisebox{0.08ex}{\ding{55}}}\hspace{0.006\linewidth}\parbox[t]{0.10\linewidth}{\centering\strut +0\%}\hspace{0.006\linewidth}\parbox[t]{0.115\linewidth}{\centering\strut -20.281495}\hspace{0.006\linewidth}\parbox[t]{0.671\linewidth}{tighten final dmrg1 solver\_tol to 1e-04}}}}\par\smallskip
\noindent\parbox[t]{0.05\linewidth}{\centering\strut 28}\hspace{0.006\linewidth}\parbox[t]{0.04\linewidth}{\centering\strut \raisebox{0.08ex}{\ding{55}}}\hspace{0.006\linewidth}\parbox[t]{0.10\linewidth}{\centering\strut +0\%}\hspace{0.006\linewidth}\parbox[t]{0.115\linewidth}{\centering\strut -20.281495}\hspace{0.006\linewidth}\parbox[t]{0.671\linewidth}{tighten final dmrg1 solver\_tol to 5e-05}\par\smallskip
\noindent{\setlength{\fboxsep}{0pt}\colorbox{black!4}{\parbox{\linewidth}{\parbox[t]{0.05\linewidth}{\centering\strut 29}\hspace{0.006\linewidth}\parbox[t]{0.04\linewidth}{\centering\strut \raisebox{0.08ex}{\ding{55}}}\hspace{0.006\linewidth}\parbox[t]{0.10\linewidth}{\centering\strut +0\%}\hspace{0.006\linewidth}\parbox[t]{0.115\linewidth}{\centering\strut -20.281495}\hspace{0.006\linewidth}\parbox[t]{0.671\linewidth}{tighten final dmrg1 solver\_tol to 1e-05}}}}\par\smallskip
\noindent\parbox[t]{0.05\linewidth}{\centering\strut 30}\hspace{0.006\linewidth}\parbox[t]{0.04\linewidth}{\centering\strut \raisebox{0.08ex}{\ding{55}}}\hspace{0.006\linewidth}\parbox[t]{0.10\linewidth}{\centering\strut +0\%}\hspace{0.006\linewidth}\parbox[t]{0.115\linewidth}{\centering\strut -20.281495}\hspace{0.006\linewidth}\parbox[t]{0.671\linewidth}{tighten final dmrg1 solver\_tol to 1e-06}\par\smallskip
\noindent{\setlength{\fboxsep}{0pt}\colorbox{black!4}{\parbox{\linewidth}{\parbox[t]{0.05\linewidth}{\centering\strut 31}\hspace{0.006\linewidth}\parbox[t]{0.04\linewidth}{\centering\strut \raisebox{0.08ex}{\ding{55}}}\hspace{0.006\linewidth}\parbox[t]{0.10\linewidth}{\centering\strut +0\%}\hspace{0.006\linewidth}\parbox[t]{0.115\linewidth}{\centering\strut -20.281495}\hspace{0.006\linewidth}\parbox[t]{0.671\linewidth}{tighten final dmrg1 solver\_tol to 1e-07}}}}\par\smallskip
\noindent\parbox[t]{0.05\linewidth}{\centering\strut 32}\hspace{0.006\linewidth}\parbox[t]{0.04\linewidth}{\centering\strut \raisebox{0.08ex}{\ding{55}}}\hspace{0.006\linewidth}\parbox[t]{0.10\linewidth}{\centering\strut +0\%}\hspace{0.006\linewidth}\parbox[t]{0.115\linewidth}{\centering\strut -20.281495}\hspace{0.006\linewidth}\parbox[t]{0.671\linewidth}{tighten final dmrg1 solver\_tol to 1e-08}\par\smallskip
\noindent{\setlength{\fboxsep}{0pt}\colorbox{black!4}{\parbox{\linewidth}{\parbox[t]{0.05\linewidth}{\centering\strut 33}\hspace{0.006\linewidth}\parbox[t]{0.04\linewidth}{\centering\strut \raisebox{0.08ex}{\ding{55}}}\hspace{0.006\linewidth}\parbox[t]{0.10\linewidth}{\centering\strut +0\%}\hspace{0.006\linewidth}\parbox[t]{0.115\linewidth}{\centering\strut -20.281495}\hspace{0.006\linewidth}\parbox[t]{0.671\linewidth}{set final dmrg1 max\_sweeps to 10}}}}\par\smallskip
\noindent\parbox[t]{0.05\linewidth}{\centering\strut 34}\hspace{0.006\linewidth}\parbox[t]{0.04\linewidth}{\centering\strut \raisebox{0.08ex}{\ding{51}}}\hspace{0.006\linewidth}\parbox[t]{0.10\linewidth}{\centering\strut +0\%}\hspace{0.006\linewidth}\parbox[t]{0.115\linewidth}{\centering\strut -20.281495}\hspace{0.006\linewidth}\parbox[t]{0.671\linewidth}{set final dmrg1 max\_sweeps to 12}\par\smallskip
\noindent{\setlength{\fboxsep}{0pt}\colorbox{black!4}{\parbox{\linewidth}{\parbox[t]{0.05\linewidth}{\centering\strut 35}\hspace{0.006\linewidth}\parbox[t]{0.04\linewidth}{\centering\strut \raisebox{0.08ex}{\ding{55}}}\hspace{0.006\linewidth}\parbox[t]{0.10\linewidth}{\centering\strut +0\%}\hspace{0.006\linewidth}\parbox[t]{0.115\linewidth}{\centering\strut -20.281495}\hspace{0.006\linewidth}\parbox[t]{0.671\linewidth}{switch final dmrg1 loop to explicit sweeps}}}}\par\smallskip
\noindent\parbox[t]{0.05\linewidth}{\centering\strut 36}\hspace{0.006\linewidth}\parbox[t]{0.04\linewidth}{\centering\strut \raisebox{0.08ex}{\ding{55}}}\hspace{0.006\linewidth}\parbox[t]{0.10\linewidth}{\centering\strut +0\%}\hspace{0.006\linewidth}\parbox[t]{0.115\linewidth}{\centering\strut -20.281495}\hspace{0.006\linewidth}\parbox[t]{0.671\linewidth}{change final dmrg1 init\_state to plus}\par\smallskip
\noindent{\setlength{\fboxsep}{0pt}\colorbox{black!4}{\parbox{\linewidth}{\parbox[t]{0.05\linewidth}{\centering\strut 37}\hspace{0.006\linewidth}\parbox[t]{0.04\linewidth}{\centering\strut \raisebox{0.08ex}{\ding{55}}}\hspace{0.006\linewidth}\parbox[t]{0.10\linewidth}{\centering\strut +0\%}\hspace{0.006\linewidth}\parbox[t]{0.115\linewidth}{\centering\strut -20.281495}\hspace{0.006\linewidth}\parbox[t]{0.671\linewidth}{change final dmrg1 init\_state to product\_up}}}}\par\smallskip
\noindent\parbox[t]{0.05\linewidth}{\centering\strut 38}\hspace{0.006\linewidth}\parbox[t]{0.04\linewidth}{\centering\strut \raisebox{0.08ex}{\ding{55}}}\hspace{0.006\linewidth}\parbox[t]{0.10\linewidth}{\centering\strut +0\%}\hspace{0.006\linewidth}\parbox[t]{0.115\linewidth}{\centering\strut -20.281495}\hspace{0.006\linewidth}\parbox[t]{0.671\linewidth}{change final dmrg1 init\_state to product\_down}\par\smallskip
\noindent{\setlength{\fboxsep}{0pt}\colorbox{black!4}{\parbox{\linewidth}{\parbox[t]{0.05\linewidth}{\centering\strut 39}\hspace{0.006\linewidth}\parbox[t]{0.04\linewidth}{\centering\strut \raisebox{0.08ex}{\ding{55}}}\hspace{0.006\linewidth}\parbox[t]{0.10\linewidth}{\centering\strut +0\%}\hspace{0.006\linewidth}\parbox[t]{0.115\linewidth}{\centering\strut -20.281495}\hspace{0.006\linewidth}\parbox[t]{0.671\linewidth}{change final dmrg1 init\_state to random}}}}\par\smallskip
\noindent\parbox[t]{0.05\linewidth}{\centering\strut 40}\hspace{0.006\linewidth}\parbox[t]{0.04\linewidth}{\centering\strut \raisebox{0.08ex}{\ding{55}}}\hspace{0.006\linewidth}\parbox[t]{0.10\linewidth}{\centering\strut +0\%}\hspace{0.006\linewidth}\parbox[t]{0.115\linewidth}{\centering\strut -20.281495}\hspace{0.006\linewidth}\parbox[t]{0.671\linewidth}{set final dmrg1 init\_bond\_dim to 1}\par\smallskip
\noindent{\setlength{\fboxsep}{0pt}\colorbox{black!4}{\parbox{\linewidth}{\parbox[t]{0.05\linewidth}{\centering\strut 41}\hspace{0.006\linewidth}\parbox[t]{0.04\linewidth}{\centering\strut \raisebox{0.08ex}{\ding{55}}}\hspace{0.006\linewidth}\parbox[t]{0.10\linewidth}{\centering\strut +0\%}\hspace{0.006\linewidth}\parbox[t]{0.115\linewidth}{\centering\strut -20.281495}\hspace{0.006\linewidth}\parbox[t]{0.671\linewidth}{set final dmrg1 init\_bond\_dim to 4}}}}\par\smallskip
\noindent\parbox[t]{0.05\linewidth}{\centering\strut 42}\hspace{0.006\linewidth}\parbox[t]{0.04\linewidth}{\centering\strut \raisebox{0.08ex}{\ding{55}}}\hspace{0.006\linewidth}\parbox[t]{0.10\linewidth}{\centering\strut +0\%}\hspace{0.006\linewidth}\parbox[t]{0.115\linewidth}{\centering\strut -20.281495}\hspace{0.006\linewidth}\parbox[t]{0.671\linewidth}{set final dmrg1 init\_bond\_dim to 8}\par\smallskip
\noindent{\setlength{\fboxsep}{0pt}\colorbox{black!4}{\parbox{\linewidth}{\parbox[t]{0.05\linewidth}{\centering\strut 43}\hspace{0.006\linewidth}\parbox[t]{0.04\linewidth}{\centering\strut \raisebox{0.08ex}{\ding{55}}}\hspace{0.006\linewidth}\parbox[t]{0.10\linewidth}{\centering\strut +0\%}\hspace{0.006\linewidth}\parbox[t]{0.115\linewidth}{\centering\strut -20.281495}\hspace{0.006\linewidth}\parbox[t]{0.671\linewidth}{set final dmrg1 local\_eig\_ncv to 6}}}}\par\smallskip
\noindent\parbox[t]{0.05\linewidth}{\centering\strut 44}\hspace{0.006\linewidth}\parbox[t]{0.04\linewidth}{\centering\strut \raisebox{0.08ex}{\ding{55}}}\hspace{0.006\linewidth}\parbox[t]{0.10\linewidth}{\centering\strut +0\%}\hspace{0.006\linewidth}\parbox[t]{0.115\linewidth}{\centering\strut -20.281495}\hspace{0.006\linewidth}\parbox[t]{0.671\linewidth}{set final dmrg1 local\_eig\_ncv to 10}\par\smallskip
\noindent{\setlength{\fboxsep}{0pt}\colorbox{black!4}{\parbox{\linewidth}{\parbox[t]{0.05\linewidth}{\centering\strut 45}\hspace{0.006\linewidth}\parbox[t]{0.04\linewidth}{\centering\strut \raisebox{0.08ex}{\ding{55}}}\hspace{0.006\linewidth}\parbox[t]{0.10\linewidth}{\centering\strut +0\%}\hspace{0.006\linewidth}\parbox[t]{0.115\linewidth}{\centering\strut -20.281495}\hspace{0.006\linewidth}\parbox[t]{0.671\linewidth}{set final dmrg1 local\_eig\_ncv to 12}}}}\par\smallskip
\noindent\parbox[t]{0.05\linewidth}{\centering\strut 46}\hspace{0.006\linewidth}\parbox[t]{0.04\linewidth}{\centering\strut \raisebox{0.08ex}{\ding{55}}}\hspace{0.006\linewidth}\parbox[t]{0.10\linewidth}{\centering\strut +0\%}\hspace{0.006\linewidth}\parbox[t]{0.115\linewidth}{\centering\strut -20.281495}\hspace{0.006\linewidth}\parbox[t]{0.671\linewidth}{set final dmrg1 local\_eig\_ncv to 14}\par\smallskip
\noindent{\setlength{\fboxsep}{0pt}\colorbox{black!4}{\parbox{\linewidth}{\parbox[t]{0.05\linewidth}{\centering\strut 47}\hspace{0.006\linewidth}\parbox[t]{0.04\linewidth}{\centering\strut \raisebox{0.08ex}{\ding{55}}}\hspace{0.006\linewidth}\parbox[t]{0.10\linewidth}{\centering\strut +0\%}\hspace{0.006\linewidth}\parbox[t]{0.115\linewidth}{\centering\strut -20.281495}\hspace{0.006\linewidth}\parbox[t]{0.671\linewidth}{tighten final dmrg1 cutoff to 1e-08}}}}\par\smallskip
\noindent\parbox[t]{0.05\linewidth}{\centering\strut 48}\hspace{0.006\linewidth}\parbox[t]{0.04\linewidth}{\centering\strut \raisebox{0.08ex}{\ding{55}}}\hspace{0.006\linewidth}\parbox[t]{0.10\linewidth}{\centering\strut +0\%}\hspace{0.006\linewidth}\parbox[t]{0.115\linewidth}{\centering\strut -20.281495}\hspace{0.006\linewidth}\parbox[t]{0.671\linewidth}{tighten final dmrg1 cutoff to 1e-09}\par\smallskip
\noindent{\setlength{\fboxsep}{0pt}\colorbox{black!4}{\parbox{\linewidth}{\parbox[t]{0.05\linewidth}{\centering\strut 49}\hspace{0.006\linewidth}\parbox[t]{0.04\linewidth}{\centering\strut \raisebox{0.08ex}{\ding{55}}}\hspace{0.006\linewidth}\parbox[t]{0.10\linewidth}{\centering\strut +0\%}\hspace{0.006\linewidth}\parbox[t]{0.115\linewidth}{\centering\strut -20.281495}\hspace{0.006\linewidth}\parbox[t]{0.671\linewidth}{tighten final dmrg1 cutoff to 1e-11}}}}\par\smallskip
\noindent\parbox[t]{0.05\linewidth}{\centering\strut 50}\hspace{0.006\linewidth}\parbox[t]{0.04\linewidth}{\centering\strut \raisebox{0.08ex}{\ding{55}}}\hspace{0.006\linewidth}\parbox[t]{0.10\linewidth}{\centering\strut +0\%}\hspace{0.006\linewidth}\parbox[t]{0.115\linewidth}{\centering\strut -20.281495}\hspace{0.006\linewidth}\parbox[t]{0.671\linewidth}{tighten final dmrg1 cutoff to 1e-12}\par\smallskip
\noindent{\setlength{\fboxsep}{0pt}\colorbox{black!4}{\parbox{\linewidth}{\parbox[t]{0.05\linewidth}{\centering\strut 51}\hspace{0.006\linewidth}\parbox[t]{0.04\linewidth}{\centering\strut \raisebox{0.08ex}{\ding{55}}}\hspace{0.006\linewidth}\parbox[t]{0.10\linewidth}{\centering\strut +0\%}\hspace{0.006\linewidth}\parbox[t]{0.115\linewidth}{\centering\strut -20.281495}\hspace{0.006\linewidth}\parbox[t]{0.671\linewidth}{tighten final dmrg1 solver\_tol to 1e-04}}}}\par\smallskip
\noindent\parbox[t]{0.05\linewidth}{\centering\strut 52}\hspace{0.006\linewidth}\parbox[t]{0.04\linewidth}{\centering\strut \raisebox{0.08ex}{\ding{55}}}\hspace{0.006\linewidth}\parbox[t]{0.10\linewidth}{\centering\strut +0\%}\hspace{0.006\linewidth}\parbox[t]{0.115\linewidth}{\centering\strut -20.281495}\hspace{0.006\linewidth}\parbox[t]{0.671\linewidth}{tighten final dmrg1 solver\_tol to 5e-05}\par\smallskip
\noindent{\setlength{\fboxsep}{0pt}\colorbox{black!4}{\parbox{\linewidth}{\parbox[t]{0.05\linewidth}{\centering\strut 53}\hspace{0.006\linewidth}\parbox[t]{0.04\linewidth}{\centering\strut \raisebox{0.08ex}{\ding{55}}}\hspace{0.006\linewidth}\parbox[t]{0.10\linewidth}{\centering\strut +0\%}\hspace{0.006\linewidth}\parbox[t]{0.115\linewidth}{\centering\strut -20.281495}\hspace{0.006\linewidth}\parbox[t]{0.671\linewidth}{tighten final dmrg1 solver\_tol to 1e-05}}}}\par\smallskip
\noindent\parbox[t]{0.05\linewidth}{\centering\strut 54}\hspace{0.006\linewidth}\parbox[t]{0.04\linewidth}{\centering\strut \raisebox{0.08ex}{\ding{55}}}\hspace{0.006\linewidth}\parbox[t]{0.10\linewidth}{\centering\strut +0\%}\hspace{0.006\linewidth}\parbox[t]{0.115\linewidth}{\centering\strut -20.281495}\hspace{0.006\linewidth}\parbox[t]{0.671\linewidth}{tighten final dmrg1 solver\_tol to 1e-06}\par\smallskip
\noindent{\setlength{\fboxsep}{0pt}\colorbox{black!4}{\parbox{\linewidth}{\parbox[t]{0.05\linewidth}{\centering\strut 55}\hspace{0.006\linewidth}\parbox[t]{0.04\linewidth}{\centering\strut \raisebox{0.08ex}{\ding{55}}}\hspace{0.006\linewidth}\parbox[t]{0.10\linewidth}{\centering\strut +0\%}\hspace{0.006\linewidth}\parbox[t]{0.115\linewidth}{\centering\strut -20.281495}\hspace{0.006\linewidth}\parbox[t]{0.671\linewidth}{tighten final dmrg1 solver\_tol to 1e-07}}}}\par\smallskip
\noindent\parbox[t]{0.05\linewidth}{\centering\strut 56}\hspace{0.006\linewidth}\parbox[t]{0.04\linewidth}{\centering\strut \raisebox{0.08ex}{\ding{55}}}\hspace{0.006\linewidth}\parbox[t]{0.10\linewidth}{\centering\strut +0\%}\hspace{0.006\linewidth}\parbox[t]{0.115\linewidth}{\centering\strut -20.281495}\hspace{0.006\linewidth}\parbox[t]{0.671\linewidth}{tighten final dmrg1 solver\_tol to 1e-08}\par\smallskip
\noindent{\setlength{\fboxsep}{0pt}\colorbox{black!4}{\parbox{\linewidth}{\parbox[t]{0.05\linewidth}{\centering\strut 57}\hspace{0.006\linewidth}\parbox[t]{0.04\linewidth}{\centering\strut \raisebox{0.08ex}{\ding{55}}}\hspace{0.006\linewidth}\parbox[t]{0.10\linewidth}{\centering\strut +0\%}\hspace{0.006\linewidth}\parbox[t]{0.115\linewidth}{\centering\strut -20.281495}\hspace{0.006\linewidth}\parbox[t]{0.671\linewidth}{set final dmrg1 max\_sweeps to 8}}}}\par\smallskip
\noindent\parbox[t]{0.05\linewidth}{\centering\strut 58}\hspace{0.006\linewidth}\parbox[t]{0.04\linewidth}{\centering\strut \raisebox{0.08ex}{\ding{55}}}\hspace{0.006\linewidth}\parbox[t]{0.10\linewidth}{\centering\strut +0\%}\hspace{0.006\linewidth}\parbox[t]{0.115\linewidth}{\centering\strut -20.281495}\hspace{0.006\linewidth}\parbox[t]{0.671\linewidth}{set final dmrg1 max\_sweeps to 10}\par\smallskip
\noindent{\setlength{\fboxsep}{0pt}\colorbox{black!4}{\parbox{\linewidth}{\parbox[t]{0.05\linewidth}{\centering\strut 59}\hspace{0.006\linewidth}\parbox[t]{0.04\linewidth}{\centering\strut \raisebox{0.08ex}{\ding{55}}}\hspace{0.006\linewidth}\parbox[t]{0.10\linewidth}{\centering\strut +0\%}\hspace{0.006\linewidth}\parbox[t]{0.115\linewidth}{\centering\strut -20.281495}\hspace{0.006\linewidth}\parbox[t]{0.671\linewidth}{set final dmrg1 max\_sweeps to 14}}}}\par\smallskip
\noindent\parbox[t]{0.05\linewidth}{\centering\strut 60}\hspace{0.006\linewidth}\parbox[t]{0.04\linewidth}{\centering\strut \raisebox{0.08ex}{\ding{55}}}\hspace{0.006\linewidth}\parbox[t]{0.10\linewidth}{\centering\strut +0\%}\hspace{0.006\linewidth}\parbox[t]{0.115\linewidth}{\centering\strut -20.281495}\hspace{0.006\linewidth}\parbox[t]{0.671\linewidth}{set final dmrg1 max\_sweeps to 16}\par\smallskip
\noindent{\setlength{\fboxsep}{0pt}\colorbox{black!4}{\parbox{\linewidth}{\parbox[t]{0.05\linewidth}{\centering\strut 61}\hspace{0.006\linewidth}\parbox[t]{0.04\linewidth}{\centering\strut \raisebox{0.08ex}{\ding{55}}}\hspace{0.006\linewidth}\parbox[t]{0.10\linewidth}{\centering\strut +0\%}\hspace{0.006\linewidth}\parbox[t]{0.115\linewidth}{\centering\strut -20.281495}\hspace{0.006\linewidth}\parbox[t]{0.671\linewidth}{set final dmrg1 max\_sweeps to 18}}}}\par\smallskip
\noindent\parbox[t]{0.05\linewidth}{\centering\strut 62}\hspace{0.006\linewidth}\parbox[t]{0.04\linewidth}{\centering\strut \raisebox{0.08ex}{\ding{55}}}\hspace{0.006\linewidth}\parbox[t]{0.10\linewidth}{\centering\strut +0\%}\hspace{0.006\linewidth}\parbox[t]{0.115\linewidth}{\centering\strut -20.281495}\hspace{0.006\linewidth}\parbox[t]{0.671\linewidth}{set final dmrg1 max\_sweeps to 20}\par\smallskip
\noindent{\setlength{\fboxsep}{0pt}\colorbox{black!4}{\parbox{\linewidth}{\parbox[t]{0.05\linewidth}{\centering\strut 63}\hspace{0.006\linewidth}\parbox[t]{0.04\linewidth}{\centering\strut \raisebox{0.08ex}{\ding{55}}}\hspace{0.006\linewidth}\parbox[t]{0.10\linewidth}{\centering\strut +0\%}\hspace{0.006\linewidth}\parbox[t]{0.115\linewidth}{\centering\strut -20.281495}\hspace{0.006\linewidth}\parbox[t]{0.671\linewidth}{set final dmrg1 max\_sweeps to 24}}}}\par\smallskip
\noindent\parbox[t]{0.05\linewidth}{\centering\strut 64}\hspace{0.006\linewidth}\parbox[t]{0.04\linewidth}{\centering\strut \raisebox{0.08ex}{\ding{55}}}\hspace{0.006\linewidth}\parbox[t]{0.10\linewidth}{\centering\strut +0\%}\hspace{0.006\linewidth}\parbox[t]{0.115\linewidth}{\centering\strut -20.281495}\hspace{0.006\linewidth}\parbox[t]{0.671\linewidth}{set final dmrg1 max\_sweeps to 28}\par\smallskip
\noindent{\setlength{\fboxsep}{0pt}\colorbox{black!4}{\parbox{\linewidth}{\parbox[t]{0.05\linewidth}{\centering\strut 65}\hspace{0.006\linewidth}\parbox[t]{0.04\linewidth}{\centering\strut \raisebox{0.08ex}{\ding{55}}}\hspace{0.006\linewidth}\parbox[t]{0.10\linewidth}{\centering\strut +0\%}\hspace{0.006\linewidth}\parbox[t]{0.115\linewidth}{\centering\strut -20.281495}\hspace{0.006\linewidth}\parbox[t]{0.671\linewidth}{set final dmrg1 max\_sweeps to 32}}}}\par\smallskip
\noindent\parbox[t]{0.05\linewidth}{\centering\strut 66}\hspace{0.006\linewidth}\parbox[t]{0.04\linewidth}{\centering\strut \raisebox{0.08ex}{\ding{55}}}\hspace{0.006\linewidth}\parbox[t]{0.10\linewidth}{\centering\strut +0\%}\hspace{0.006\linewidth}\parbox[t]{0.115\linewidth}{\centering\strut -20.281495}\hspace{0.006\linewidth}\parbox[t]{0.671\linewidth}{set final dmrg1 expand\_rand\_strength to 0e+00}\par\smallskip
\noindent{\setlength{\fboxsep}{0pt}\colorbox{black!4}{\parbox{\linewidth}{\parbox[t]{0.05\linewidth}{\centering\strut 67}\hspace{0.006\linewidth}\parbox[t]{0.04\linewidth}{\centering\strut \raisebox{0.08ex}{\ding{55}}}\hspace{0.006\linewidth}\parbox[t]{0.10\linewidth}{\centering\strut +0\%}\hspace{0.006\linewidth}\parbox[t]{0.115\linewidth}{\centering\strut -20.281495}\hspace{0.006\linewidth}\parbox[t]{0.671\linewidth}{set final dmrg1 expand\_rand\_strength to 5e-06}}}}\par\smallskip
\noindent\parbox[t]{0.05\linewidth}{\centering\strut 68}\hspace{0.006\linewidth}\parbox[t]{0.04\linewidth}{\centering\strut \raisebox{0.08ex}{\ding{55}}}\hspace{0.006\linewidth}\parbox[t]{0.10\linewidth}{\centering\strut +0\%}\hspace{0.006\linewidth}\parbox[t]{0.115\linewidth}{\centering\strut -20.281495}\hspace{0.006\linewidth}\parbox[t]{0.671\linewidth}{set final dmrg1 expand\_rand\_strength to 5e-04}\par\smallskip
\noindent{\setlength{\fboxsep}{0pt}\colorbox{black!4}{\parbox{\linewidth}{\parbox[t]{0.05\linewidth}{\centering\strut 69}\hspace{0.006\linewidth}\parbox[t]{0.04\linewidth}{\centering\strut \raisebox{0.08ex}{\ding{55}}}\hspace{0.006\linewidth}\parbox[t]{0.10\linewidth}{\centering\strut +0\%}\hspace{0.006\linewidth}\parbox[t]{0.115\linewidth}{\centering\strut -20.281495}\hspace{0.006\linewidth}\parbox[t]{0.671\linewidth}{set final dmrg1 expand\_rand\_strength to 1e-03}}}}\par\smallskip
\noindent\parbox[t]{0.05\linewidth}{\centering\strut 70}\hspace{0.006\linewidth}\parbox[t]{0.04\linewidth}{\centering\strut \raisebox{0.08ex}{\ding{55}}}\hspace{0.006\linewidth}\parbox[t]{0.10\linewidth}{\centering\strut +0\%}\hspace{0.006\linewidth}\parbox[t]{0.115\linewidth}{\centering\strut -20.281495}\hspace{0.006\linewidth}\parbox[t]{0.671\linewidth}{set final dmrg1 expand\_rand\_strength to 1e-02}\par\smallskip
\noindent{\setlength{\fboxsep}{0pt}\colorbox{black!4}{\parbox{\linewidth}{\parbox[t]{0.05\linewidth}{\centering\strut 71}\hspace{0.006\linewidth}\parbox[t]{0.04\linewidth}{\centering\strut \raisebox{0.08ex}{\ding{55}}}\hspace{0.006\linewidth}\parbox[t]{0.10\linewidth}{\centering\strut +0\%}\hspace{0.006\linewidth}\parbox[t]{0.115\linewidth}{\centering\strut -20.281495}\hspace{0.006\linewidth}\parbox[t]{0.671\linewidth}{raise final dmrg1 bond\_schedule to 48-72-96}}}}\par\smallskip
\noindent\parbox[t]{0.05\linewidth}{\centering\strut 72}\hspace{0.006\linewidth}\parbox[t]{0.04\linewidth}{\centering\strut \raisebox{0.08ex}{\ding{55}}}\hspace{0.006\linewidth}\parbox[t]{0.10\linewidth}{\centering\strut +0\%}\hspace{0.006\linewidth}\parbox[t]{0.115\linewidth}{\centering\strut -20.281495}\hspace{0.006\linewidth}\parbox[t]{0.671\linewidth}{raise final dmrg1 bond\_schedule to 48-72-108-128}\par\smallskip
\noindent{\setlength{\fboxsep}{0pt}\colorbox{black!4}{\parbox{\linewidth}{\parbox[t]{0.05\linewidth}{\centering\strut 73}\hspace{0.006\linewidth}\parbox[t]{0.04\linewidth}{\centering\strut \raisebox{0.08ex}{\ding{55}}}\hspace{0.006\linewidth}\parbox[t]{0.10\linewidth}{\centering\strut +0\%}\hspace{0.006\linewidth}\parbox[t]{0.115\linewidth}{\centering\strut -20.281495}\hspace{0.006\linewidth}\parbox[t]{0.671\linewidth}{raise final dmrg1 bond\_schedule to 48-72-108-160}}}}\par\smallskip
\noindent\parbox[t]{0.05\linewidth}{\centering\strut 74}\hspace{0.006\linewidth}\parbox[t]{0.04\linewidth}{\centering\strut \raisebox{0.08ex}{\ding{55}}}\hspace{0.006\linewidth}\parbox[t]{0.10\linewidth}{\centering\strut ---}\hspace{0.006\linewidth}\parbox[t]{0.115\linewidth}{\centering\strut ---}\hspace{0.006\linewidth}\parbox[t]{0.671\linewidth}{raise final dmrg1 bond\_schedule to 48-72-108-162-192}\par\smallskip
\noindent{\setlength{\fboxsep}{0pt}\colorbox{black!4}{\parbox{\linewidth}{\parbox[t]{0.05\linewidth}{\centering\strut 75}\hspace{0.006\linewidth}\parbox[t]{0.04\linewidth}{\centering\strut \raisebox{0.08ex}{\ding{55}}}\hspace{0.006\linewidth}\parbox[t]{0.10\linewidth}{\centering\strut +0\%}\hspace{0.006\linewidth}\parbox[t]{0.115\linewidth}{\centering\strut -20.281495}\hspace{0.006\linewidth}\parbox[t]{0.671\linewidth}{add dmrg1 warm start before final phase}}}}\par\smallskip
\noindent\parbox[t]{0.05\linewidth}{\centering\strut 76}\hspace{0.006\linewidth}\parbox[t]{0.04\linewidth}{\centering\strut \raisebox{0.08ex}{\ding{55}}}\hspace{0.006\linewidth}\parbox[t]{0.10\linewidth}{\centering\strut +0\%}\hspace{0.006\linewidth}\parbox[t]{0.115\linewidth}{\centering\strut -20.281495}\hspace{0.006\linewidth}\parbox[t]{0.671\linewidth}{add tebd1d plus warmup before final phase}\par\smallskip
\noindent{\setlength{\fboxsep}{0pt}\colorbox{black!4}{\parbox{\linewidth}{\parbox[t]{0.05\linewidth}{\centering\strut 77}\hspace{0.006\linewidth}\parbox[t]{0.04\linewidth}{\centering\strut \raisebox{0.08ex}{\ding{55}}}\hspace{0.006\linewidth}\parbox[t]{0.10\linewidth}{\centering\strut +0\%}\hspace{0.006\linewidth}\parbox[t]{0.115\linewidth}{\centering\strut -20.281495}\hspace{0.006\linewidth}\parbox[t]{0.671\linewidth}{set final dmrg2 minus ladder 12-15-19-24-30-38}}}}\par\smallskip
\noindent\parbox[t]{0.05\linewidth}{\centering\strut 78}\hspace{0.006\linewidth}\parbox[t]{0.04\linewidth}{\centering\strut \raisebox{0.08ex}{\ding{55}}}\hspace{0.006\linewidth}\parbox[t]{0.10\linewidth}{\centering\strut +0\%}\hspace{0.006\linewidth}\parbox[t]{0.115\linewidth}{\centering\strut -20.281495}\hspace{0.006\linewidth}\parbox[t]{0.671\linewidth}{set final dmrg1 product\_down ladder 4-6-9-14-21-32-48-72-108}\par\smallskip
\noindent{\setlength{\fboxsep}{0pt}\colorbox{black!4}{\parbox{\linewidth}{\parbox[t]{0.05\linewidth}{\centering\strut 79}\hspace{0.006\linewidth}\parbox[t]{0.04\linewidth}{\centering\strut \raisebox{0.08ex}{\ding{55}}}\hspace{0.006\linewidth}\parbox[t]{0.10\linewidth}{\centering\strut +0\%}\hspace{0.006\linewidth}\parbox[t]{0.115\linewidth}{\centering\strut -20.281495}\hspace{0.006\linewidth}\parbox[t]{0.671\linewidth}{set final dmrg1 plus ladder 12-21-37-65-80}}}}\par\smallskip
\noindent\parbox[t]{0.05\linewidth}{\centering\strut 80}\hspace{0.006\linewidth}\parbox[t]{0.04\linewidth}{\centering\strut \raisebox{0.08ex}{\ding{55}}}\hspace{0.006\linewidth}\parbox[t]{0.10\linewidth}{\centering\strut +0\%}\hspace{0.006\linewidth}\parbox[t]{0.115\linewidth}{\centering\strut -20.281495}\hspace{0.006\linewidth}\parbox[t]{0.671\linewidth}{set final dmrg2 plus ladder 16-28-48}\par\smallskip
\noindent{\setlength{\fboxsep}{0pt}\colorbox{black!4}{\parbox{\linewidth}{\parbox[t]{0.05\linewidth}{\centering\strut 81}\hspace{0.006\linewidth}\parbox[t]{0.04\linewidth}{\centering\strut \raisebox{0.08ex}{\ding{55}}}\hspace{0.006\linewidth}\parbox[t]{0.10\linewidth}{\centering\strut +0\%}\hspace{0.006\linewidth}\parbox[t]{0.115\linewidth}{\centering\strut -20.281495}\hspace{0.006\linewidth}\parbox[t]{0.671\linewidth}{set final dmrg2 minus ladder 4-6-9-14-21-32}}}}\par\smallskip
\noindent\parbox[t]{0.05\linewidth}{\centering\strut 82}\hspace{0.006\linewidth}\parbox[t]{0.04\linewidth}{\centering\strut \raisebox{0.08ex}{\ding{55}}}\hspace{0.006\linewidth}\parbox[t]{0.10\linewidth}{\centering\strut +0\%}\hspace{0.006\linewidth}\parbox[t]{0.115\linewidth}{\centering\strut -20.281495}\hspace{0.006\linewidth}\parbox[t]{0.671\linewidth}{set final dmrg1 product\_down ladder 12-21-37-48}\par\smallskip
\noindent{\setlength{\fboxsep}{0pt}\colorbox{black!4}{\parbox{\linewidth}{\parbox[t]{0.05\linewidth}{\centering\strut 83}\hspace{0.006\linewidth}\parbox[t]{0.04\linewidth}{\centering\strut \raisebox{0.08ex}{\ding{55}}}\hspace{0.006\linewidth}\parbox[t]{0.10\linewidth}{\centering\strut +0\%}\hspace{0.006\linewidth}\parbox[t]{0.115\linewidth}{\centering\strut -20.281495}\hspace{0.006\linewidth}\parbox[t]{0.671\linewidth}{set final dmrg2 product\_up ladder 8-14-24-42-64}}}}\par\smallskip
\noindent\parbox[t]{0.05\linewidth}{\centering\strut 84}\hspace{0.006\linewidth}\parbox[t]{0.04\linewidth}{\centering\strut \raisebox{0.08ex}{\ding{55}}}\hspace{0.006\linewidth}\parbox[t]{0.10\linewidth}{\centering\strut +0\%}\hspace{0.006\linewidth}\parbox[t]{0.115\linewidth}{\centering\strut -20.281495}\hspace{0.006\linewidth}\parbox[t]{0.671\linewidth}{set final dmrg2 product\_up ladder 4-6-9-14-21-32-48-64}\par\smallskip
\noindent{\setlength{\fboxsep}{0pt}\colorbox{black!4}{\parbox{\linewidth}{\parbox[t]{0.05\linewidth}{\centering\strut 85}\hspace{0.006\linewidth}\parbox[t]{0.04\linewidth}{\centering\strut \raisebox{0.08ex}{\ding{55}}}\hspace{0.006\linewidth}\parbox[t]{0.10\linewidth}{\centering\strut +0\%}\hspace{0.006\linewidth}\parbox[t]{0.115\linewidth}{\centering\strut -20.281495}\hspace{0.006\linewidth}\parbox[t]{0.671\linewidth}{set final dmrg1 product\_up ladder 12-15-19-24-30-38-48-60}}}}\par\smallskip
\noindent\parbox[t]{0.05\linewidth}{\centering\strut 86}\hspace{0.006\linewidth}\parbox[t]{0.04\linewidth}{\centering\strut \raisebox{0.08ex}{\ding{55}}}\hspace{0.006\linewidth}\parbox[t]{0.10\linewidth}{\centering\strut +0\%}\hspace{0.006\linewidth}\parbox[t]{0.115\linewidth}{\centering\strut -20.281495}\hspace{0.006\linewidth}\parbox[t]{0.671\linewidth}{set final dmrg1 random ladder 16-32-64-96}\par\smallskip
\noindent{\setlength{\fboxsep}{0pt}\colorbox{black!4}{\parbox{\linewidth}{\parbox[t]{0.05\linewidth}{\centering\strut 87}\hspace{0.006\linewidth}\parbox[t]{0.04\linewidth}{\centering\strut \raisebox{0.08ex}{\ding{55}}}\hspace{0.006\linewidth}\parbox[t]{0.10\linewidth}{\centering\strut +0\%}\hspace{0.006\linewidth}\parbox[t]{0.115\linewidth}{\centering\strut -20.281495}\hspace{0.006\linewidth}\parbox[t]{0.671\linewidth}{set final dmrg2 plus ladder 16-24-36-54-80}}}}\par\smallskip
\noindent\parbox[t]{0.05\linewidth}{\centering\strut 88}\hspace{0.006\linewidth}\parbox[t]{0.04\linewidth}{\centering\strut \raisebox{0.08ex}{\ding{55}}}\hspace{0.006\linewidth}\parbox[t]{0.10\linewidth}{\centering\strut +0\%}\hspace{0.006\linewidth}\parbox[t]{0.115\linewidth}{\centering\strut -20.281495}\hspace{0.006\linewidth}\parbox[t]{0.671\linewidth}{set final dmrg1 plus ladder 8-16-32-48}\par\smallskip
\noindent{\setlength{\fboxsep}{0pt}\colorbox{black!4}{\parbox{\linewidth}{\parbox[t]{0.05\linewidth}{\centering\strut 89}\hspace{0.006\linewidth}\parbox[t]{0.04\linewidth}{\centering\strut \raisebox{0.08ex}{\ding{55}}}\hspace{0.006\linewidth}\parbox[t]{0.10\linewidth}{\centering\strut +0\%}\hspace{0.006\linewidth}\parbox[t]{0.115\linewidth}{\centering\strut -20.281495}\hspace{0.006\linewidth}\parbox[t]{0.671\linewidth}{set final dmrg2 plus ladder 4-5-6-8-10-12}}}}\par\smallskip
\noindent\parbox[t]{0.05\linewidth}{\centering\strut 90}\hspace{0.006\linewidth}\parbox[t]{0.04\linewidth}{\centering\strut \raisebox{0.08ex}{\ding{55}}}\hspace{0.006\linewidth}\parbox[t]{0.10\linewidth}{\centering\strut +0\%}\hspace{0.006\linewidth}\parbox[t]{0.115\linewidth}{\centering\strut -20.281495}\hspace{0.006\linewidth}\parbox[t]{0.671\linewidth}{set final dmrg1 minus ladder 4-8-16-32-64}\par\smallskip
\noindent{\setlength{\fboxsep}{0pt}\colorbox{black!4}{\parbox{\linewidth}{\parbox[t]{0.05\linewidth}{\centering\strut 91}\hspace{0.006\linewidth}\parbox[t]{0.04\linewidth}{\centering\strut \raisebox{0.08ex}{\ding{55}}}\hspace{0.006\linewidth}\parbox[t]{0.10\linewidth}{\centering\strut +0\%}\hspace{0.006\linewidth}\parbox[t]{0.115\linewidth}{\centering\strut -20.281495}\hspace{0.006\linewidth}\parbox[t]{0.671\linewidth}{set final dmrg1 product\_down ladder 4-6-9-14-21-32-48}}}}\par\smallskip
\noindent\parbox[t]{0.05\linewidth}{\centering\strut 92}\hspace{0.006\linewidth}\parbox[t]{0.04\linewidth}{\centering\strut \raisebox{0.08ex}{\ding{55}}}\hspace{0.006\linewidth}\parbox[t]{0.10\linewidth}{\centering\strut +0\%}\hspace{0.006\linewidth}\parbox[t]{0.115\linewidth}{\centering\strut -20.281495}\hspace{0.006\linewidth}\parbox[t]{0.671\linewidth}{set final dmrg1 minus ladder 4-6-9-14-21-32-48-72}\par\smallskip
\noindent{\setlength{\fboxsep}{0pt}\colorbox{black!4}{\parbox{\linewidth}{\parbox[t]{0.05\linewidth}{\centering\strut 93}\hspace{0.006\linewidth}\parbox[t]{0.04\linewidth}{\centering\strut \raisebox{0.08ex}{\ding{55}}}\hspace{0.006\linewidth}\parbox[t]{0.10\linewidth}{\centering\strut +0\%}\hspace{0.006\linewidth}\parbox[t]{0.115\linewidth}{\centering\strut -20.281495}\hspace{0.006\linewidth}\parbox[t]{0.671\linewidth}{set final dmrg1 minus ladder 4-7-12-21-37-65-114-128}}}}\par\smallskip
\noindent\parbox[t]{0.05\linewidth}{\centering\strut 94}\hspace{0.006\linewidth}\parbox[t]{0.04\linewidth}{\centering\strut \raisebox{0.08ex}{\ding{55}}}\hspace{0.006\linewidth}\parbox[t]{0.10\linewidth}{\centering\strut +0\%}\hspace{0.006\linewidth}\parbox[t]{0.115\linewidth}{\centering\strut -20.281495}\hspace{0.006\linewidth}\parbox[t]{0.671\linewidth}{set final dmrg2 plus ladder 12-24-48-80}\par\smallskip
\noindent{\setlength{\fboxsep}{0pt}\colorbox{black!4}{\parbox{\linewidth}{\parbox[t]{0.05\linewidth}{\centering\strut 95}\hspace{0.006\linewidth}\parbox[t]{0.04\linewidth}{\centering\strut \raisebox{0.08ex}{\ding{55}}}\hspace{0.006\linewidth}\parbox[t]{0.10\linewidth}{\centering\strut +0\%}\hspace{0.006\linewidth}\parbox[t]{0.115\linewidth}{\centering\strut -20.281495}\hspace{0.006\linewidth}\parbox[t]{0.671\linewidth}{set final dmrg2 product\_up ladder 12-21-37-65-114-160}}}}\par\smallskip
\noindent\parbox[t]{0.05\linewidth}{\centering\strut 96}\hspace{0.006\linewidth}\parbox[t]{0.04\linewidth}{\centering\strut \raisebox{0.08ex}{\ding{55}}}\hspace{0.006\linewidth}\parbox[t]{0.10\linewidth}{\centering\strut +0\%}\hspace{0.006\linewidth}\parbox[t]{0.115\linewidth}{\centering\strut -20.281495}\hspace{0.006\linewidth}\parbox[t]{0.671\linewidth}{set final dmrg2 minus ladder 16-28-49-64}\par\smallskip
\noindent{\setlength{\fboxsep}{0pt}\colorbox{black!4}{\parbox{\linewidth}{\parbox[t]{0.05\linewidth}{\centering\strut 97}\hspace{0.006\linewidth}\parbox[t]{0.04\linewidth}{\centering\strut \raisebox{0.08ex}{\ding{55}}}\hspace{0.006\linewidth}\parbox[t]{0.10\linewidth}{\centering\strut +0\%}\hspace{0.006\linewidth}\parbox[t]{0.115\linewidth}{\centering\strut -20.281494}\hspace{0.006\linewidth}\parbox[t]{0.671\linewidth}{set final dmrg2 product\_down ladder 16-32-64-128}}}}\par\smallskip
\noindent\parbox[t]{0.05\linewidth}{\centering\strut 98}\hspace{0.006\linewidth}\parbox[t]{0.04\linewidth}{\centering\strut \raisebox{0.08ex}{\ding{55}}}\hspace{0.006\linewidth}\parbox[t]{0.10\linewidth}{\centering\strut +0\%}\hspace{0.006\linewidth}\parbox[t]{0.115\linewidth}{\centering\strut -20.281495}\hspace{0.006\linewidth}\parbox[t]{0.671\linewidth}{set final dmrg1 product\_down ladder 12-18-27-40-60}\par\smallskip
\noindent{\setlength{\fboxsep}{0pt}\colorbox{black!4}{\parbox{\linewidth}{\parbox[t]{0.05\linewidth}{\centering\strut 99}\hspace{0.006\linewidth}\parbox[t]{0.04\linewidth}{\centering\strut \raisebox{0.08ex}{\ding{55}}}\hspace{0.006\linewidth}\parbox[t]{0.10\linewidth}{\centering\strut +0\%}\hspace{0.006\linewidth}\parbox[t]{0.115\linewidth}{\centering\strut -20.281495}\hspace{0.006\linewidth}\parbox[t]{0.671\linewidth}{set final dmrg2 minus ladder 16-20-25-31-39-49-61}}}}\par\smallskip
\noindent\parbox[t]{0.05\linewidth}{\centering\strut 100}\hspace{0.006\linewidth}\parbox[t]{0.04\linewidth}{\centering\strut \raisebox{0.08ex}{\ding{55}}}\hspace{0.006\linewidth}\parbox[t]{0.10\linewidth}{\centering\strut +0\%}\hspace{0.006\linewidth}\parbox[t]{0.115\linewidth}{\centering\strut -20.281495}\hspace{0.006\linewidth}\parbox[t]{0.671\linewidth}{set final dmrg1 product\_up ladder 8-14-24-42-74-130}\par\smallskip
\noindent\rule{\linewidth}{0.5pt}
\par\medskip

\medskip

\refstepcounter{table}
\label{tab:supp_tn_log_xx_critical_64}
\noindent\textbf{Table \thetable.} Tensor-network mutation log for Critical XX, $L=64$. Relative gain is measured against the immediately previous scored iteration using the live final-energy objective.
\par\medskip
\noindent\rule{\linewidth}{0.5pt}
\noindent{\setlength{\fboxsep}{0pt}\colorbox{black!4}{\parbox{\linewidth}{\parbox[c][2.2em][c]{0.05\linewidth}{\centering \textbf{Iter.}}\hspace{0.006\linewidth}\parbox[c][2.2em][c]{0.04\linewidth}{\centering \textbf{A/R}}\hspace{0.006\linewidth}\parbox[c][2.2em][c]{0.10\linewidth}{\centering \textbf{Relative gain}}\hspace{0.006\linewidth}\parbox[c][2.2em][c]{0.115\linewidth}{\centering \textbf{Score}}\hspace{0.006\linewidth}\parbox[c][2.2em][c]{0.671\linewidth}{\centering \textbf{Mutation summary}}}}}\par\smallskip
\noindent\rule{\linewidth}{0.5pt}
\noindent{\setlength{\fboxsep}{0pt}\colorbox{black!4}{\parbox{\linewidth}{\parbox[t]{0.05\linewidth}{\centering\strut 1}\hspace{0.006\linewidth}\parbox[t]{0.04\linewidth}{\centering\strut \raisebox{0.08ex}{\ding{51}}}\hspace{0.006\linewidth}\parbox[t]{0.10\linewidth}{\centering\strut ---}\hspace{0.006\linewidth}\parbox[t]{0.115\linewidth}{\centering\strut -20.179500}\hspace{0.006\linewidth}\parbox[t]{0.671\linewidth}{switch DEFAULT\_CONFIG method dmrg1->dmrg2}}}}\par\smallskip
\noindent\parbox[t]{0.05\linewidth}{\centering\strut 2}\hspace{0.006\linewidth}\parbox[t]{0.04\linewidth}{\centering\strut \raisebox{0.08ex}{\ding{55}}}\hspace{0.006\linewidth}\parbox[t]{0.10\linewidth}{\centering\strut +0\%}\hspace{0.006\linewidth}\parbox[t]{0.115\linewidth}{\centering\strut -20.179493}\hspace{0.006\linewidth}\parbox[t]{0.671\linewidth}{set DEFAULT\_CONFIG init\_state random->neel}\par\smallskip
\noindent{\setlength{\fboxsep}{0pt}\colorbox{black!4}{\parbox{\linewidth}{\parbox[t]{0.05\linewidth}{\centering\strut 3}\hspace{0.006\linewidth}\parbox[t]{0.04\linewidth}{\centering\strut \raisebox{0.08ex}{\ding{55}}}\hspace{0.006\linewidth}\parbox[t]{0.10\linewidth}{\centering\strut +0\%}\hspace{0.006\linewidth}\parbox[t]{0.115\linewidth}{\centering\strut -20.179500}\hspace{0.006\linewidth}\parbox[t]{0.671\linewidth}{raise DEFAULT\_CONFIG bond\_schedule to 8-12-16-24-32-48}}}}\par\smallskip
\noindent\parbox[t]{0.05\linewidth}{\centering\strut 4}\hspace{0.006\linewidth}\parbox[t]{0.04\linewidth}{\centering\strut \raisebox{0.08ex}{\ding{51}}}\hspace{0.006\linewidth}\parbox[t]{0.10\linewidth}{\centering\strut +0.0619\%}\hspace{0.006\linewidth}\parbox[t]{0.115\linewidth}{\centering\strut -20.191990}\hspace{0.006\linewidth}\parbox[t]{0.671\linewidth}{raise DEFAULT\_CONFIG bond\_schedule to 8-12-16-24-32-48}\par\smallskip
\noindent{\setlength{\fboxsep}{0pt}\colorbox{black!4}{\parbox{\linewidth}{\parbox[t]{0.05\linewidth}{\centering\strut 5}\hspace{0.006\linewidth}\parbox[t]{0.04\linewidth}{\centering\strut \raisebox{0.08ex}{\ding{55}}}\hspace{0.006\linewidth}\parbox[t]{0.10\linewidth}{\centering\strut +0\%}\hspace{0.006\linewidth}\parbox[t]{0.115\linewidth}{\centering\strut -20.191990}\hspace{0.006\linewidth}\parbox[t]{0.671\linewidth}{increase DEFAULT\_CONFIG max\_sweeps 6->12}}}}\par\smallskip
\noindent\parbox[t]{0.05\linewidth}{\centering\strut 6}\hspace{0.006\linewidth}\parbox[t]{0.04\linewidth}{\centering\strut \raisebox{0.08ex}{\ding{51}}}\hspace{0.006\linewidth}\parbox[t]{0.10\linewidth}{\centering\strut +0.000335\%}\hspace{0.006\linewidth}\parbox[t]{0.115\linewidth}{\centering\strut -20.192057}\hspace{0.006\linewidth}\parbox[t]{0.671\linewidth}{increase DEFAULT\_CONFIG max\_sweeps 6->12}\par\smallskip
\noindent{\setlength{\fboxsep}{0pt}\colorbox{black!4}{\parbox{\linewidth}{\parbox[t]{0.05\linewidth}{\centering\strut 7}\hspace{0.006\linewidth}\parbox[t]{0.04\linewidth}{\centering\strut \raisebox{0.08ex}{\ding{55}}}\hspace{0.006\linewidth}\parbox[t]{0.10\linewidth}{\centering\strut +0\%}\hspace{0.006\linewidth}\parbox[t]{0.115\linewidth}{\centering\strut -20.192057}\hspace{0.006\linewidth}\parbox[t]{0.671\linewidth}{increase DEFAULT\_CONFIG init\_bond\_dim 2->4}}}}\par\smallskip
\noindent\parbox[t]{0.05\linewidth}{\centering\strut 8}\hspace{0.006\linewidth}\parbox[t]{0.04\linewidth}{\centering\strut \raisebox{0.08ex}{\ding{51}}}\hspace{0.006\linewidth}\parbox[t]{0.10\linewidth}{\centering\strut +0\%}\hspace{0.006\linewidth}\parbox[t]{0.115\linewidth}{\centering\strut -20.192057}\hspace{0.006\linewidth}\parbox[t]{0.671\linewidth}{tighten DEFAULT\_CONFIG solver\_tol 1e-3->1e-4}\par\smallskip
\noindent{\setlength{\fboxsep}{0pt}\colorbox{black!4}{\parbox{\linewidth}{\parbox[t]{0.05\linewidth}{\centering\strut 9}\hspace{0.006\linewidth}\parbox[t]{0.04\linewidth}{\centering\strut \raisebox{0.08ex}{\ding{55}}}\hspace{0.006\linewidth}\parbox[t]{0.10\linewidth}{\centering\strut +0\%}\hspace{0.006\linewidth}\parbox[t]{0.115\linewidth}{\centering\strut -20.192057}\hspace{0.006\linewidth}\parbox[t]{0.671\linewidth}{increase DEFAULT\_CONFIG local\_eig\_ncv 4->8}}}}\par\smallskip
\noindent\parbox[t]{0.05\linewidth}{\centering\strut 10}\hspace{0.006\linewidth}\parbox[t]{0.04\linewidth}{\centering\strut \raisebox{0.08ex}{\ding{55}}}\hspace{0.006\linewidth}\parbox[t]{0.10\linewidth}{\centering\strut +0\%}\hspace{0.006\linewidth}\parbox[t]{0.115\linewidth}{\centering\strut -20.192057}\hspace{0.006\linewidth}\parbox[t]{0.671\linewidth}{raise DEFAULT\_CONFIG bond\_schedule to 12-16-24-32-48-64-80-96}\par\smallskip
\noindent{\setlength{\fboxsep}{0pt}\colorbox{black!4}{\parbox{\linewidth}{\parbox[t]{0.05\linewidth}{\centering\strut 11}\hspace{0.006\linewidth}\parbox[t]{0.04\linewidth}{\centering\strut \raisebox{0.08ex}{\ding{51}}}\hspace{0.006\linewidth}\parbox[t]{0.10\linewidth}{\centering\strut +0.000486\%}\hspace{0.006\linewidth}\parbox[t]{0.115\linewidth}{\centering\strut -20.192155}\hspace{0.006\linewidth}\parbox[t]{0.671\linewidth}{tighten DEFAULT\_CONFIG cutoff 1e-6->1e-8}}}}\par\smallskip
\noindent\parbox[t]{0.05\linewidth}{\centering\strut 12}\hspace{0.006\linewidth}\parbox[t]{0.04\linewidth}{\centering\strut \raisebox{0.08ex}{\ding{55}}}\hspace{0.006\linewidth}\parbox[t]{0.10\linewidth}{\centering\strut +0\%}\hspace{0.006\linewidth}\parbox[t]{0.115\linewidth}{\centering\strut -20.192155}\hspace{0.006\linewidth}\parbox[t]{0.671\linewidth}{increase DEFAULT\_CONFIG max\_sweeps 12->20}\par\smallskip
\noindent{\setlength{\fboxsep}{0pt}\colorbox{black!4}{\parbox{\linewidth}{\parbox[t]{0.05\linewidth}{\centering\strut 13}\hspace{0.006\linewidth}\parbox[t]{0.04\linewidth}{\centering\strut \raisebox{0.08ex}{\ding{55}}}\hspace{0.006\linewidth}\parbox[t]{0.10\linewidth}{\centering\strut +0\%}\hspace{0.006\linewidth}\parbox[t]{0.115\linewidth}{\centering\strut -20.192155}\hspace{0.006\linewidth}\parbox[t]{0.671\linewidth}{set DEFAULT\_CONFIG init\_state random->neel with dmrg2 schedule}}}}\par\smallskip
\noindent\parbox[t]{0.05\linewidth}{\centering\strut 14}\hspace{0.006\linewidth}\parbox[t]{0.04\linewidth}{\centering\strut \raisebox{0.08ex}{\ding{55}}}\hspace{0.006\linewidth}\parbox[t]{0.10\linewidth}{\centering\strut +0\%}\hspace{0.006\linewidth}\parbox[t]{0.115\linewidth}{\centering\strut -20.192155}\hspace{0.006\linewidth}\parbox[t]{0.671\linewidth}{increase DEFAULT\_CONFIG init\_bond\_dim 2->4 with cutoff 1e-8}\par\smallskip
\noindent{\setlength{\fboxsep}{0pt}\colorbox{black!4}{\parbox{\linewidth}{\parbox[t]{0.05\linewidth}{\centering\strut 15}\hspace{0.006\linewidth}\parbox[t]{0.04\linewidth}{\centering\strut \raisebox{0.08ex}{\ding{55}}}\hspace{0.006\linewidth}\parbox[t]{0.10\linewidth}{\centering\strut +0\%}\hspace{0.006\linewidth}\parbox[t]{0.115\linewidth}{\centering\strut -20.192155}\hspace{0.006\linewidth}\parbox[t]{0.671\linewidth}{tighten DEFAULT\_CONFIG solver\_tol 1e-4->1e-6}}}}\par\smallskip
\noindent\parbox[t]{0.05\linewidth}{\centering\strut 16}\hspace{0.006\linewidth}\parbox[t]{0.04\linewidth}{\centering\strut \raisebox{0.08ex}{\ding{51}}}\hspace{0.006\linewidth}\parbox[t]{0.10\linewidth}{\centering\strut +0\%}\hspace{0.006\linewidth}\parbox[t]{0.115\linewidth}{\centering\strut -20.192155}\hspace{0.006\linewidth}\parbox[t]{0.671\linewidth}{increase DEFAULT\_CONFIG local\_eig\_ncv 4->8 with cutoff 1e-8}\par\smallskip
\noindent{\setlength{\fboxsep}{0pt}\colorbox{black!4}{\parbox{\linewidth}{\parbox[t]{0.05\linewidth}{\centering\strut 17}\hspace{0.006\linewidth}\parbox[t]{0.04\linewidth}{\centering\strut \raisebox{0.08ex}{\ding{51}}}\hspace{0.006\linewidth}\parbox[t]{0.10\linewidth}{\centering\strut +0\%}\hspace{0.006\linewidth}\parbox[t]{0.115\linewidth}{\centering\strut -20.192155}\hspace{0.006\linewidth}\parbox[t]{0.671\linewidth}{increase DEFAULT\_CONFIG local\_eig\_ncv 8->12 with cutoff 1e-8}}}}\par\smallskip
\noindent\parbox[t]{0.05\linewidth}{\centering\strut 18}\hspace{0.006\linewidth}\parbox[t]{0.04\linewidth}{\centering\strut \raisebox{0.08ex}{\ding{55}}}\hspace{0.006\linewidth}\parbox[t]{0.10\linewidth}{\centering\strut +0\%}\hspace{0.006\linewidth}\parbox[t]{0.115\linewidth}{\centering\strut -20.192155}\hspace{0.006\linewidth}\parbox[t]{0.671\linewidth}{increase DEFAULT\_CONFIG local\_eig\_ncv 12->16 with cutoff 1e-8}\par\smallskip
\noindent{\setlength{\fboxsep}{0pt}\colorbox{black!4}{\parbox{\linewidth}{\parbox[t]{0.05\linewidth}{\centering\strut 19}\hspace{0.006\linewidth}\parbox[t]{0.04\linewidth}{\centering\strut \raisebox{0.08ex}{\ding{51}}}\hspace{0.006\linewidth}\parbox[t]{0.10\linewidth}{\centering\strut +0\%}\hspace{0.006\linewidth}\parbox[t]{0.115\linewidth}{\centering\strut -20.192156}\hspace{0.006\linewidth}\parbox[t]{0.671\linewidth}{tighten DEFAULT\_CONFIG cutoff 1e-8->1e-10 with local\_eig\_ncv 12}}}}\par\smallskip
\noindent\parbox[t]{0.05\linewidth}{\centering\strut 20}\hspace{0.006\linewidth}\parbox[t]{0.04\linewidth}{\centering\strut \raisebox{0.08ex}{\ding{51}}}\hspace{0.006\linewidth}\parbox[t]{0.10\linewidth}{\centering\strut +0\%}\hspace{0.006\linewidth}\parbox[t]{0.115\linewidth}{\centering\strut -20.192156}\hspace{0.006\linewidth}\parbox[t]{0.671\linewidth}{tighten DEFAULT\_CONFIG solver\_tol 1e-4->1e-5 with cutoff 1e-10}\par\smallskip
\noindent{\setlength{\fboxsep}{0pt}\colorbox{black!4}{\parbox{\linewidth}{\parbox[t]{0.05\linewidth}{\centering\strut 21}\hspace{0.006\linewidth}\parbox[t]{0.04\linewidth}{\centering\strut \raisebox{0.08ex}{\ding{55}}}\hspace{0.006\linewidth}\parbox[t]{0.10\linewidth}{\centering\strut +0\%}\hspace{0.006\linewidth}\parbox[t]{0.115\linewidth}{\centering\strut -20.192156}\hspace{0.006\linewidth}\parbox[t]{0.671\linewidth}{tighten DEFAULT\_CONFIG solver\_tol 1e-5->1e-6 with cutoff 1e-10}}}}\par\smallskip
\noindent\parbox[t]{0.05\linewidth}{\centering\strut 22}\hspace{0.006\linewidth}\parbox[t]{0.04\linewidth}{\centering\strut \raisebox{0.08ex}{\ding{51}}}\hspace{0.006\linewidth}\parbox[t]{0.10\linewidth}{\centering\strut +0\%}\hspace{0.006\linewidth}\parbox[t]{0.115\linewidth}{\centering\strut -20.192157}\hspace{0.006\linewidth}\parbox[t]{0.671\linewidth}{extend DEFAULT\_CONFIG bond\_schedule with 64 stage at cutoff 1e-10}\par\smallskip
\noindent{\setlength{\fboxsep}{0pt}\colorbox{black!4}{\parbox{\linewidth}{\parbox[t]{0.05\linewidth}{\centering\strut 23}\hspace{0.006\linewidth}\parbox[t]{0.04\linewidth}{\centering\strut \raisebox{0.08ex}{\ding{51}}}\hspace{0.006\linewidth}\parbox[t]{0.10\linewidth}{\centering\strut +0\%}\hspace{0.006\linewidth}\parbox[t]{0.115\linewidth}{\centering\strut -20.192157}\hspace{0.006\linewidth}\parbox[t]{0.671\linewidth}{extend DEFAULT\_CONFIG bond\_schedule with 80 stage at cutoff 1e-10}}}}\par\smallskip
\noindent\parbox[t]{0.05\linewidth}{\centering\strut 24}\hspace{0.006\linewidth}\parbox[t]{0.04\linewidth}{\centering\strut \raisebox{0.08ex}{\ding{51}}}\hspace{0.006\linewidth}\parbox[t]{0.10\linewidth}{\centering\strut +0\%}\hspace{0.006\linewidth}\parbox[t]{0.115\linewidth}{\centering\strut -20.192157}\hspace{0.006\linewidth}\parbox[t]{0.671\linewidth}{extend DEFAULT\_CONFIG bond\_schedule with 96 stage at cutoff 1e-10}\par\smallskip
\noindent{\setlength{\fboxsep}{0pt}\colorbox{black!4}{\parbox{\linewidth}{\parbox[t]{0.05\linewidth}{\centering\strut 25}\hspace{0.006\linewidth}\parbox[t]{0.04\linewidth}{\centering\strut \raisebox{0.08ex}{\ding{51}}}\hspace{0.006\linewidth}\parbox[t]{0.10\linewidth}{\centering\strut +0\%}\hspace{0.006\linewidth}\parbox[t]{0.115\linewidth}{\centering\strut -20.192157}\hspace{0.006\linewidth}\parbox[t]{0.671\linewidth}{increase DEFAULT\_CONFIG max\_sweeps 12->16 with 96 bond stage}}}}\par\smallskip
\noindent\parbox[t]{0.05\linewidth}{\centering\strut 26}\hspace{0.006\linewidth}\parbox[t]{0.04\linewidth}{\centering\strut \raisebox{0.08ex}{\ding{51}}}\hspace{0.006\linewidth}\parbox[t]{0.10\linewidth}{\centering\strut +0\%}\hspace{0.006\linewidth}\parbox[t]{0.115\linewidth}{\centering\strut -20.192157}\hspace{0.006\linewidth}\parbox[t]{0.671\linewidth}{increase DEFAULT\_CONFIG max\_sweeps 16->20 with 96 bond stage}\par\smallskip
\noindent{\setlength{\fboxsep}{0pt}\colorbox{black!4}{\parbox{\linewidth}{\parbox[t]{0.05\linewidth}{\centering\strut 27}\hspace{0.006\linewidth}\parbox[t]{0.04\linewidth}{\centering\strut \raisebox{0.08ex}{\ding{55}}}\hspace{0.006\linewidth}\parbox[t]{0.10\linewidth}{\centering\strut ---}\hspace{0.006\linewidth}\parbox[t]{0.115\linewidth}{\centering\strut ---}\hspace{0.006\linewidth}\parbox[t]{0.671\linewidth}{increase DEFAULT\_CONFIG max\_sweeps 20->24 with 96 bond stage}}}}\par\smallskip
\noindent\parbox[t]{0.05\linewidth}{\centering\strut 28}\hspace{0.006\linewidth}\parbox[t]{0.04\linewidth}{\centering\strut \raisebox{0.08ex}{\ding{51}}}\hspace{0.006\linewidth}\parbox[t]{0.10\linewidth}{\centering\strut +0\%}\hspace{0.006\linewidth}\parbox[t]{0.115\linewidth}{\centering\strut -20.192157}\hspace{0.006\linewidth}\parbox[t]{0.671\linewidth}{increase DEFAULT\_CONFIG max\_sweeps 20->22 with 96 bond stage}\par\smallskip
\noindent{\setlength{\fboxsep}{0pt}\colorbox{black!4}{\parbox{\linewidth}{\parbox[t]{0.05\linewidth}{\centering\strut 29}\hspace{0.006\linewidth}\parbox[t]{0.04\linewidth}{\centering\strut \raisebox{0.08ex}{\ding{51}}}\hspace{0.006\linewidth}\parbox[t]{0.10\linewidth}{\centering\strut +0\%}\hspace{0.006\linewidth}\parbox[t]{0.115\linewidth}{\centering\strut -20.192157}\hspace{0.006\linewidth}\parbox[t]{0.671\linewidth}{increase DEFAULT\_CONFIG max\_sweeps 22->23 with 96 bond stage}}}}\par\smallskip
\noindent\parbox[t]{0.05\linewidth}{\centering\strut 30}\hspace{0.006\linewidth}\parbox[t]{0.04\linewidth}{\centering\strut \raisebox{0.08ex}{\ding{55}}}\hspace{0.006\linewidth}\parbox[t]{0.10\linewidth}{\centering\strut ---}\hspace{0.006\linewidth}\parbox[t]{0.115\linewidth}{\centering\strut ---}\hspace{0.006\linewidth}\parbox[t]{0.671\linewidth}{extend DEFAULT\_CONFIG bond\_schedule with 112 stage at 23 sweeps}\par\smallskip
\noindent{\setlength{\fboxsep}{0pt}\colorbox{black!4}{\parbox{\linewidth}{\parbox[t]{0.05\linewidth}{\centering\strut 31}\hspace{0.006\linewidth}\parbox[t]{0.04\linewidth}{\centering\strut \raisebox{0.08ex}{\ding{55}}}\hspace{0.006\linewidth}\parbox[t]{0.10\linewidth}{\centering\strut ---}\hspace{0.006\linewidth}\parbox[t]{0.115\linewidth}{\centering\strut ---}\hspace{0.006\linewidth}\parbox[t]{0.671\linewidth}{increase DEFAULT\_CONFIG local\_eig\_ncv 12->14 at 23 sweeps}}}}\par\smallskip
\noindent\parbox[t]{0.05\linewidth}{\centering\strut 32}\hspace{0.006\linewidth}\parbox[t]{0.04\linewidth}{\centering\strut \raisebox{0.08ex}{\ding{55}}}\hspace{0.006\linewidth}\parbox[t]{0.10\linewidth}{\centering\strut ---}\hspace{0.006\linewidth}\parbox[t]{0.115\linewidth}{\centering\strut ---}\hspace{0.006\linewidth}\parbox[t]{0.671\linewidth}{increase DEFAULT\_CONFIG local\_eig\_ncv 12->13 at 23 sweeps}\par\smallskip
\noindent{\setlength{\fboxsep}{0pt}\colorbox{black!4}{\parbox{\linewidth}{\parbox[t]{0.05\linewidth}{\centering\strut 33}\hspace{0.006\linewidth}\parbox[t]{0.04\linewidth}{\centering\strut \raisebox{0.08ex}{\ding{55}}}\hspace{0.006\linewidth}\parbox[t]{0.10\linewidth}{\centering\strut +0\%}\hspace{0.006\linewidth}\parbox[t]{0.115\linewidth}{\centering\strut -20.192157}\hspace{0.006\linewidth}\parbox[t]{0.671\linewidth}{set DEFAULT\_CONFIG init\_seed 7->42 on 23-sweep dmrg2 config}}}}\par\smallskip
\noindent\parbox[t]{0.05\linewidth}{\centering\strut 34}\hspace{0.006\linewidth}\parbox[t]{0.04\linewidth}{\centering\strut \raisebox{0.08ex}{\ding{55}}}\hspace{0.006\linewidth}\parbox[t]{0.10\linewidth}{\centering\strut ---}\hspace{0.006\linewidth}\parbox[t]{0.115\linewidth}{\centering\strut ---}\hspace{0.006\linewidth}\parbox[t]{0.671\linewidth}{condense DEFAULT\_CONFIG bond\_schedule to start at 12 and reach 96 earlier}\par\smallskip
\noindent{\setlength{\fboxsep}{0pt}\colorbox{black!4}{\parbox{\linewidth}{\parbox[t]{0.05\linewidth}{\centering\strut 35}\hspace{0.006\linewidth}\parbox[t]{0.04\linewidth}{\centering\strut \raisebox{0.08ex}{\ding{55}}}\hspace{0.006\linewidth}\parbox[t]{0.10\linewidth}{\centering\strut ---}\hspace{0.006\linewidth}\parbox[t]{0.115\linewidth}{\centering\strut ---}\hspace{0.006\linewidth}\parbox[t]{0.671\linewidth}{decrease DEFAULT\_CONFIG init\_bond\_dim 2->1 on 23-sweep dmrg2 config}}}}\par\smallskip
\noindent\parbox[t]{0.05\linewidth}{\centering\strut 36}\hspace{0.006\linewidth}\parbox[t]{0.04\linewidth}{\centering\strut \raisebox{0.08ex}{\ding{55}}}\hspace{0.006\linewidth}\parbox[t]{0.10\linewidth}{\centering\strut ---}\hspace{0.006\linewidth}\parbox[t]{0.115\linewidth}{\centering\strut ---}\hspace{0.006\linewidth}\parbox[t]{0.671\linewidth}{set DEFAULT\_CONFIG init\_seed 7->123 on 23-sweep dmrg2 config}\par\smallskip
\noindent{\setlength{\fboxsep}{0pt}\colorbox{black!4}{\parbox{\linewidth}{\parbox[t]{0.05\linewidth}{\centering\strut 37}\hspace{0.006\linewidth}\parbox[t]{0.04\linewidth}{\centering\strut \raisebox{0.08ex}{\ding{55}}}\hspace{0.006\linewidth}\parbox[t]{0.10\linewidth}{\centering\strut ---}\hspace{0.006\linewidth}\parbox[t]{0.115\linewidth}{\centering\strut ---}\hspace{0.006\linewidth}\parbox[t]{0.671\linewidth}{set DEFAULT\_CONFIG init\_seed 7->0 on 23-sweep dmrg2 config}}}}\par\smallskip
\noindent\parbox[t]{0.05\linewidth}{\centering\strut 38}\hspace{0.006\linewidth}\parbox[t]{0.04\linewidth}{\centering\strut \raisebox{0.08ex}{\ding{55}}}\hspace{0.006\linewidth}\parbox[t]{0.10\linewidth}{\centering\strut ---}\hspace{0.006\linewidth}\parbox[t]{0.115\linewidth}{\centering\strut ---}\hspace{0.006\linewidth}\parbox[t]{0.671\linewidth}{set DEFAULT\_CONFIG init\_seed 7->8 on 23-sweep dmrg2 config}\par\smallskip
\noindent{\setlength{\fboxsep}{0pt}\colorbox{black!4}{\parbox{\linewidth}{\parbox[t]{0.05\linewidth}{\centering\strut 39}\hspace{0.006\linewidth}\parbox[t]{0.04\linewidth}{\centering\strut \raisebox{0.08ex}{\ding{55}}}\hspace{0.006\linewidth}\parbox[t]{0.10\linewidth}{\centering\strut ---}\hspace{0.006\linewidth}\parbox[t]{0.115\linewidth}{\centering\strut ---}\hspace{0.006\linewidth}\parbox[t]{0.671\linewidth}{tighten DEFAULT\_CONFIG cutoff 1e-10->1e-11 at 23 sweeps}}}}\par\smallskip
\noindent\parbox[t]{0.05\linewidth}{\centering\strut 40}\hspace{0.006\linewidth}\parbox[t]{0.04\linewidth}{\centering\strut \raisebox{0.08ex}{\ding{55}}}\hspace{0.006\linewidth}\parbox[t]{0.10\linewidth}{\centering\strut ---}\hspace{0.006\linewidth}\parbox[t]{0.115\linewidth}{\centering\strut ---}\hspace{0.006\linewidth}\parbox[t]{0.671\linewidth}{decrease DEFAULT\_CONFIG max\_sweeps 23->22 on best dmrg2 config}\par\smallskip
\noindent{\setlength{\fboxsep}{0pt}\colorbox{black!4}{\parbox{\linewidth}{\parbox[t]{0.05\linewidth}{\centering\strut 41}\hspace{0.006\linewidth}\parbox[t]{0.04\linewidth}{\centering\strut \raisebox{0.08ex}{\ding{55}}}\hspace{0.006\linewidth}\parbox[t]{0.10\linewidth}{\centering\strut +0\%}\hspace{0.006\linewidth}\parbox[t]{0.115\linewidth}{\centering\strut -20.192156}\hspace{0.006\linewidth}\parbox[t]{0.671\linewidth}{loosen DEFAULT\_CONFIG cutoff 1e-10->1e-9 on best dmrg2 config}}}}\par\smallskip
\noindent\parbox[t]{0.05\linewidth}{\centering\strut 42}\hspace{0.006\linewidth}\parbox[t]{0.04\linewidth}{\centering\strut \raisebox{0.08ex}{\ding{55}}}\hspace{0.006\linewidth}\parbox[t]{0.10\linewidth}{\centering\strut +0\%}\hspace{0.006\linewidth}\parbox[t]{0.115\linewidth}{\centering\strut -20.192155}\hspace{0.006\linewidth}\parbox[t]{0.671\linewidth}{loosen DEFAULT\_CONFIG cutoff 1e-10->1e-8 on best dmrg2 config}\par\smallskip
\noindent{\setlength{\fboxsep}{0pt}\colorbox{black!4}{\parbox{\linewidth}{\parbox[t]{0.05\linewidth}{\centering\strut 43}\hspace{0.006\linewidth}\parbox[t]{0.04\linewidth}{\centering\strut \raisebox{0.08ex}{\ding{55}}}\hspace{0.006\linewidth}\parbox[t]{0.10\linewidth}{\centering\strut -0.000051\%}\hspace{0.006\linewidth}\parbox[t]{0.115\linewidth}{\centering\strut -20.192145}\hspace{0.006\linewidth}\parbox[t]{0.671\linewidth}{loosen DEFAULT\_CONFIG cutoff 1e-10->1e-7 on best dmrg2 config}}}}\par\smallskip
\noindent\parbox[t]{0.05\linewidth}{\centering\strut 44}\hspace{0.006\linewidth}\parbox[t]{0.04\linewidth}{\centering\strut \raisebox{0.08ex}{\ding{55}}}\hspace{0.006\linewidth}\parbox[t]{0.10\linewidth}{\centering\strut -0.000433\%}\hspace{0.006\linewidth}\parbox[t]{0.115\linewidth}{\centering\strut -20.192058}\hspace{0.006\linewidth}\parbox[t]{0.671\linewidth}{loosen DEFAULT\_CONFIG cutoff 1e-10->1e-6 on best dmrg2 config}\par\smallskip
\noindent{\setlength{\fboxsep}{0pt}\colorbox{black!4}{\parbox{\linewidth}{\parbox[t]{0.05\linewidth}{\centering\strut 45}\hspace{0.006\linewidth}\parbox[t]{0.04\linewidth}{\centering\strut \raisebox{0.08ex}{\ding{55}}}\hspace{0.006\linewidth}\parbox[t]{0.10\linewidth}{\centering\strut ---}\hspace{0.006\linewidth}\parbox[t]{0.115\linewidth}{\centering\strut ---}\hspace{0.006\linewidth}\parbox[t]{0.671\linewidth}{loosen DEFAULT\_CONFIG solver\_tol 1e-5->1e-4 on best dmrg2 config}}}}\par\smallskip
\noindent\parbox[t]{0.05\linewidth}{\centering\strut 46}\hspace{0.006\linewidth}\parbox[t]{0.04\linewidth}{\centering\strut \raisebox{0.08ex}{\ding{55}}}\hspace{0.006\linewidth}\parbox[t]{0.10\linewidth}{\centering\strut ---}\hspace{0.006\linewidth}\parbox[t]{0.115\linewidth}{\centering\strut ---}\hspace{0.006\linewidth}\parbox[t]{0.671\linewidth}{decrease DEFAULT\_CONFIG local\_eig\_ncv 12->10 on best dmrg2 config}\par\smallskip
\noindent{\setlength{\fboxsep}{0pt}\colorbox{black!4}{\parbox{\linewidth}{\parbox[t]{0.05\linewidth}{\centering\strut 47}\hspace{0.006\linewidth}\parbox[t]{0.04\linewidth}{\centering\strut \raisebox{0.08ex}{\ding{55}}}\hspace{0.006\linewidth}\parbox[t]{0.10\linewidth}{\centering\strut ---}\hspace{0.006\linewidth}\parbox[t]{0.115\linewidth}{\centering\strut ---}\hspace{0.006\linewidth}\parbox[t]{0.671\linewidth}{switch DEFAULT\_CONFIG method dmrg2->dmrg1 on best high-bond config}}}}\par\smallskip
\noindent\parbox[t]{0.05\linewidth}{\centering\strut 48}\hspace{0.006\linewidth}\parbox[t]{0.04\linewidth}{\centering\strut \raisebox{0.08ex}{\ding{55}}}\hspace{0.006\linewidth}\parbox[t]{0.10\linewidth}{\centering\strut -0.004584\%}\hspace{0.006\linewidth}\parbox[t]{0.115\linewidth}{\centering\strut -20.191132}\hspace{0.006\linewidth}\parbox[t]{0.671\linewidth}{loosen DEFAULT\_CONFIG cutoff 1e-10->1e-5 on best dmrg2 config}\par\smallskip
\noindent{\setlength{\fboxsep}{0pt}\colorbox{black!4}{\parbox{\linewidth}{\parbox[t]{0.05\linewidth}{\centering\strut 49}\hspace{0.006\linewidth}\parbox[t]{0.04\linewidth}{\centering\strut \raisebox{0.08ex}{\ding{55}}}\hspace{0.006\linewidth}\parbox[t]{0.10\linewidth}{\centering\strut -0.0854\%}\hspace{0.006\linewidth}\parbox[t]{0.115\linewidth}{\centering\strut -20.173894}\hspace{0.006\linewidth}\parbox[t]{0.671\linewidth}{loosen DEFAULT\_CONFIG cutoff 1e-10->1e-4 on best dmrg2 config}}}}\par\smallskip
\noindent\parbox[t]{0.05\linewidth}{\centering\strut 50}\hspace{0.006\linewidth}\parbox[t]{0.04\linewidth}{\centering\strut \raisebox{0.08ex}{\ding{55}}}\hspace{0.006\linewidth}\parbox[t]{0.10\linewidth}{\centering\strut -0.4596\%}\hspace{0.006\linewidth}\parbox[t]{0.115\linewidth}{\centering\strut -20.081171}\hspace{0.006\linewidth}\parbox[t]{0.671\linewidth}{loosen DEFAULT\_CONFIG cutoff 1e-10->1e-3 on best dmrg2 config}\par\smallskip
\noindent{\setlength{\fboxsep}{0pt}\colorbox{black!4}{\parbox{\linewidth}{\parbox[t]{0.05\linewidth}{\centering\strut 51}\hspace{0.006\linewidth}\parbox[t]{0.04\linewidth}{\centering\strut \raisebox{0.08ex}{\ding{55}}}\hspace{0.006\linewidth}\parbox[t]{0.10\linewidth}{\centering\strut -1.39\%}\hspace{0.006\linewidth}\parbox[t]{0.115\linewidth}{\centering\strut -19.802690}\hspace{0.006\linewidth}\parbox[t]{0.671\linewidth}{loosen DEFAULT\_CONFIG cutoff 1e-10->1e-2 on best dmrg2 config}}}}\par\smallskip
\noindent\parbox[t]{0.05\linewidth}{\centering\strut 52}\hspace{0.006\linewidth}\parbox[t]{0.04\linewidth}{\centering\strut \raisebox{0.08ex}{\ding{55}}}\hspace{0.006\linewidth}\parbox[t]{0.10\linewidth}{\centering\strut -19.42\%}\hspace{0.006\linewidth}\parbox[t]{0.115\linewidth}{\centering\strut -15.957107}\hspace{0.006\linewidth}\parbox[t]{0.671\linewidth}{loosen DEFAULT\_CONFIG cutoff 1e-10->1e-1 on best dmrg2 config}\par\smallskip
\noindent{\setlength{\fboxsep}{0pt}\colorbox{black!4}{\parbox{\linewidth}{\parbox[t]{0.05\linewidth}{\centering\strut 53}\hspace{0.006\linewidth}\parbox[t]{0.04\linewidth}{\centering\strut \raisebox{0.08ex}{\ding{55}}}\hspace{0.006\linewidth}\parbox[t]{0.10\linewidth}{\centering\strut -15.75\%}\hspace{0.006\linewidth}\parbox[t]{0.115\linewidth}{\centering\strut -13.443466}\hspace{0.006\linewidth}\parbox[t]{0.671\linewidth}{loosen DEFAULT\_CONFIG cutoff 1e-10->2e-1 on best dmrg2 config}}}}\par\smallskip
\noindent\parbox[t]{0.05\linewidth}{\centering\strut 54}\hspace{0.006\linewidth}\parbox[t]{0.04\linewidth}{\centering\strut \raisebox{0.08ex}{\ding{55}}}\hspace{0.006\linewidth}\parbox[t]{0.10\linewidth}{\centering\strut +0\%}\hspace{0.006\linewidth}\parbox[t]{0.115\linewidth}{\centering\strut -13.443466}\hspace{0.006\linewidth}\parbox[t]{0.671\linewidth}{loosen DEFAULT\_CONFIG cutoff 1e-10->5e-1 on best dmrg2 config}\par\smallskip
\noindent{\setlength{\fboxsep}{0pt}\colorbox{black!4}{\parbox{\linewidth}{\parbox[t]{0.05\linewidth}{\centering\strut 55}\hspace{0.006\linewidth}\parbox[t]{0.04\linewidth}{\centering\strut \raisebox{0.08ex}{\ding{55}}}\hspace{0.006\linewidth}\parbox[t]{0.10\linewidth}{\centering\strut +0\%}\hspace{0.006\linewidth}\parbox[t]{0.115\linewidth}{\centering\strut -13.443466}\hspace{0.006\linewidth}\parbox[t]{0.671\linewidth}{loosen DEFAULT\_CONFIG cutoff 1e-10->1.0 on best dmrg2 config}}}}\par\smallskip
\noindent\parbox[t]{0.05\linewidth}{\centering\strut 56}\hspace{0.006\linewidth}\parbox[t]{0.04\linewidth}{\centering\strut \raisebox{0.08ex}{\ding{55}}}\hspace{0.006\linewidth}\parbox[t]{0.10\linewidth}{\centering\strut +50.2\%}\hspace{0.006\linewidth}\parbox[t]{0.115\linewidth}{\centering\strut -20.192157}\hspace{0.006\linewidth}\parbox[t]{0.671\linewidth}{truncate DEFAULT\_CONFIG bond\_schedule by removing 96 stage}\par\smallskip
\noindent{\setlength{\fboxsep}{0pt}\colorbox{black!4}{\parbox{\linewidth}{\parbox[t]{0.05\linewidth}{\centering\strut 57}\hspace{0.006\linewidth}\parbox[t]{0.04\linewidth}{\centering\strut \raisebox{0.08ex}{\ding{55}}}\hspace{0.006\linewidth}\parbox[t]{0.10\linewidth}{\centering\strut +0\%}\hspace{0.006\linewidth}\parbox[t]{0.115\linewidth}{\centering\strut -20.192157}\hspace{0.006\linewidth}\parbox[t]{0.671\linewidth}{truncate DEFAULT\_CONFIG bond\_schedule by removing 80 and 96 stages}}}}\par\smallskip
\noindent\parbox[t]{0.05\linewidth}{\centering\strut 58}\hspace{0.006\linewidth}\parbox[t]{0.04\linewidth}{\centering\strut \raisebox{0.08ex}{\ding{55}}}\hspace{0.006\linewidth}\parbox[t]{0.10\linewidth}{\centering\strut +0\%}\hspace{0.006\linewidth}\parbox[t]{0.115\linewidth}{\centering\strut -20.192156}\hspace{0.006\linewidth}\parbox[t]{0.671\linewidth}{truncate DEFAULT\_CONFIG bond\_schedule to 8-12-16-24-32-48}\par\smallskip
\noindent{\setlength{\fboxsep}{0pt}\colorbox{black!4}{\parbox{\linewidth}{\parbox[t]{0.05\linewidth}{\centering\strut 59}\hspace{0.006\linewidth}\parbox[t]{0.04\linewidth}{\centering\strut \raisebox{0.08ex}{\ding{55}}}\hspace{0.006\linewidth}\parbox[t]{0.10\linewidth}{\centering\strut +0\%}\hspace{0.006\linewidth}\parbox[t]{0.115\linewidth}{\centering\strut -20.192148}\hspace{0.006\linewidth}\parbox[t]{0.671\linewidth}{truncate DEFAULT\_CONFIG bond\_schedule to 8-12-16-24-32}}}}\par\smallskip
\noindent\parbox[t]{0.05\linewidth}{\centering\strut 60}\hspace{0.006\linewidth}\parbox[t]{0.04\linewidth}{\centering\strut \raisebox{0.08ex}{\ding{55}}}\hspace{0.006\linewidth}\parbox[t]{0.10\linewidth}{\centering\strut -0.000171\%}\hspace{0.006\linewidth}\parbox[t]{0.115\linewidth}{\centering\strut -20.192114}\hspace{0.006\linewidth}\parbox[t]{0.671\linewidth}{truncate DEFAULT\_CONFIG bond\_schedule to 8-12-16-24}\par\smallskip
\noindent{\setlength{\fboxsep}{0pt}\colorbox{black!4}{\parbox{\linewidth}{\parbox[t]{0.05\linewidth}{\centering\strut 61}\hspace{0.006\linewidth}\parbox[t]{0.04\linewidth}{\centering\strut \raisebox{0.08ex}{\ding{55}}}\hspace{0.006\linewidth}\parbox[t]{0.10\linewidth}{\centering\strut -0.00276\%}\hspace{0.006\linewidth}\parbox[t]{0.115\linewidth}{\centering\strut -20.191556}\hspace{0.006\linewidth}\parbox[t]{0.671\linewidth}{truncate DEFAULT\_CONFIG bond\_schedule to 8-12-16}}}}\par\smallskip
\noindent\parbox[t]{0.05\linewidth}{\centering\strut 62}\hspace{0.006\linewidth}\parbox[t]{0.04\linewidth}{\centering\strut \raisebox{0.08ex}{\ding{55}}}\hspace{0.006\linewidth}\parbox[t]{0.10\linewidth}{\centering\strut -0.00314\%}\hspace{0.006\linewidth}\parbox[t]{0.115\linewidth}{\centering\strut -20.190922}\hspace{0.006\linewidth}\parbox[t]{0.671\linewidth}{truncate DEFAULT\_CONFIG bond\_schedule to 8-12}\par\smallskip
\noindent{\setlength{\fboxsep}{0pt}\colorbox{black!4}{\parbox{\linewidth}{\parbox[t]{0.05\linewidth}{\centering\strut 63}\hspace{0.006\linewidth}\parbox[t]{0.04\linewidth}{\centering\strut \raisebox{0.08ex}{\ding{55}}}\hspace{0.006\linewidth}\parbox[t]{0.10\linewidth}{\centering\strut -0.0565\%}\hspace{0.006\linewidth}\parbox[t]{0.115\linewidth}{\centering\strut -20.179508}\hspace{0.006\linewidth}\parbox[t]{0.671\linewidth}{truncate DEFAULT\_CONFIG bond\_schedule to 8 only}}}}\par\smallskip
\noindent\parbox[t]{0.05\linewidth}{\centering\strut 64}\hspace{0.006\linewidth}\parbox[t]{0.04\linewidth}{\centering\strut \raisebox{0.08ex}{\ding{55}}}\hspace{0.006\linewidth}\parbox[t]{0.10\linewidth}{\centering\strut +0.0627\%}\hspace{0.006\linewidth}\parbox[t]{0.115\linewidth}{\centering\strut -20.192157}\hspace{0.006\linewidth}\parbox[t]{0.671\linewidth}{decrease DEFAULT\_CONFIG max\_sweeps 23->12 on best dmrg2 config}\par\smallskip
\noindent{\setlength{\fboxsep}{0pt}\colorbox{black!4}{\parbox{\linewidth}{\parbox[t]{0.05\linewidth}{\centering\strut 65}\hspace{0.006\linewidth}\parbox[t]{0.04\linewidth}{\centering\strut \raisebox{0.08ex}{\ding{55}}}\hspace{0.006\linewidth}\parbox[t]{0.10\linewidth}{\centering\strut +0\%}\hspace{0.006\linewidth}\parbox[t]{0.115\linewidth}{\centering\strut -20.192157}\hspace{0.006\linewidth}\parbox[t]{0.671\linewidth}{decrease DEFAULT\_CONFIG max\_sweeps 23->10 on best dmrg2 config}}}}\par\smallskip
\noindent\parbox[t]{0.05\linewidth}{\centering\strut 66}\hspace{0.006\linewidth}\parbox[t]{0.04\linewidth}{\centering\strut \raisebox{0.08ex}{\ding{55}}}\hspace{0.006\linewidth}\parbox[t]{0.10\linewidth}{\centering\strut +0\%}\hspace{0.006\linewidth}\parbox[t]{0.115\linewidth}{\centering\strut -20.192157}\hspace{0.006\linewidth}\parbox[t]{0.671\linewidth}{decrease DEFAULT\_CONFIG max\_sweeps 23->8 on best dmrg2 config}\par\smallskip
\noindent{\setlength{\fboxsep}{0pt}\colorbox{black!4}{\parbox{\linewidth}{\parbox[t]{0.05\linewidth}{\centering\strut 67}\hspace{0.006\linewidth}\parbox[t]{0.04\linewidth}{\centering\strut \raisebox{0.08ex}{\ding{55}}}\hspace{0.006\linewidth}\parbox[t]{0.10\linewidth}{\centering\strut +0\%}\hspace{0.006\linewidth}\parbox[t]{0.115\linewidth}{\centering\strut -20.192156}\hspace{0.006\linewidth}\parbox[t]{0.671\linewidth}{decrease DEFAULT\_CONFIG max\_sweeps 23->6 on best dmrg2 config}}}}\par\smallskip
\noindent\parbox[t]{0.05\linewidth}{\centering\strut 68}\hspace{0.006\linewidth}\parbox[t]{0.04\linewidth}{\centering\strut \raisebox{0.08ex}{\ding{55}}}\hspace{0.006\linewidth}\parbox[t]{0.10\linewidth}{\centering\strut -0.0106\%}\hspace{0.006\linewidth}\parbox[t]{0.115\linewidth}{\centering\strut -20.190007}\hspace{0.006\linewidth}\parbox[t]{0.671\linewidth}{decrease DEFAULT\_CONFIG max\_sweeps 23->4 on best dmrg2 config}\par\smallskip
\noindent{\setlength{\fboxsep}{0pt}\colorbox{black!4}{\parbox{\linewidth}{\parbox[t]{0.05\linewidth}{\centering\strut 69}\hspace{0.006\linewidth}\parbox[t]{0.04\linewidth}{\centering\strut \raisebox{0.08ex}{\ding{55}}}\hspace{0.006\linewidth}\parbox[t]{0.10\linewidth}{\centering\strut -0.1172\%}\hspace{0.006\linewidth}\parbox[t]{0.115\linewidth}{\centering\strut -20.166335}\hspace{0.006\linewidth}\parbox[t]{0.671\linewidth}{decrease DEFAULT\_CONFIG max\_sweeps 23->2 on best dmrg2 config}}}}\par\smallskip
\noindent\parbox[t]{0.05\linewidth}{\centering\strut 70}\hspace{0.006\linewidth}\parbox[t]{0.04\linewidth}{\centering\strut \raisebox{0.08ex}{\ding{55}}}\hspace{0.006\linewidth}\parbox[t]{0.10\linewidth}{\centering\strut -1.2\%}\hspace{0.006\linewidth}\parbox[t]{0.115\linewidth}{\centering\strut -19.925192}\hspace{0.006\linewidth}\parbox[t]{0.671\linewidth}{decrease DEFAULT\_CONFIG max\_sweeps 23->1 on best dmrg2 config}\par\smallskip
\noindent{\setlength{\fboxsep}{0pt}\colorbox{black!4}{\parbox{\linewidth}{\parbox[t]{0.05\linewidth}{\centering\strut 71}\hspace{0.006\linewidth}\parbox[t]{0.04\linewidth}{\centering\strut \raisebox{0.08ex}{\ding{55}}}\hspace{0.006\linewidth}\parbox[t]{0.10\linewidth}{\centering\strut +1.29\%}\hspace{0.006\linewidth}\parbox[t]{0.115\linewidth}{\centering\strut -20.182513}\hspace{0.006\linewidth}\parbox[t]{0.671\linewidth}{decrease DEFAULT\_CONFIG max\_sweeps 23->3 on best dmrg2 config}}}}\par\smallskip
\noindent\parbox[t]{0.05\linewidth}{\centering\strut 72}\hspace{0.006\linewidth}\parbox[t]{0.04\linewidth}{\centering\strut \raisebox{0.08ex}{\ding{55}}}\hspace{0.006\linewidth}\parbox[t]{0.10\linewidth}{\centering\strut +0.0476\%}\hspace{0.006\linewidth}\parbox[t]{0.115\linewidth}{\centering\strut -20.192114}\hspace{0.006\linewidth}\parbox[t]{0.671\linewidth}{decrease DEFAULT\_CONFIG max\_sweeps 23->5 on best dmrg2 config}\par\smallskip
\noindent{\setlength{\fboxsep}{0pt}\colorbox{black!4}{\parbox{\linewidth}{\parbox[t]{0.05\linewidth}{\centering\strut 73}\hspace{0.006\linewidth}\parbox[t]{0.04\linewidth}{\centering\strut \raisebox{0.08ex}{\ding{55}}}\hspace{0.006\linewidth}\parbox[t]{0.10\linewidth}{\centering\strut +0.000212\%}\hspace{0.006\linewidth}\parbox[t]{0.115\linewidth}{\centering\strut -20.192157}\hspace{0.006\linewidth}\parbox[t]{0.671\linewidth}{decrease DEFAULT\_CONFIG max\_sweeps 23->7 on best dmrg2 config}}}}\par\smallskip
\noindent\parbox[t]{0.05\linewidth}{\centering\strut 74}\hspace{0.006\linewidth}\parbox[t]{0.04\linewidth}{\centering\strut \raisebox{0.08ex}{\ding{55}}}\hspace{0.006\linewidth}\parbox[t]{0.10\linewidth}{\centering\strut +0\%}\hspace{0.006\linewidth}\parbox[t]{0.115\linewidth}{\centering\strut -20.192157}\hspace{0.006\linewidth}\parbox[t]{0.671\linewidth}{decrease DEFAULT\_CONFIG max\_sweeps 23->9 on best dmrg2 config}\par\smallskip
\noindent{\setlength{\fboxsep}{0pt}\colorbox{black!4}{\parbox{\linewidth}{\parbox[t]{0.05\linewidth}{\centering\strut 75}\hspace{0.006\linewidth}\parbox[t]{0.04\linewidth}{\centering\strut \raisebox{0.08ex}{\ding{55}}}\hspace{0.006\linewidth}\parbox[t]{0.10\linewidth}{\centering\strut +0\%}\hspace{0.006\linewidth}\parbox[t]{0.115\linewidth}{\centering\strut -20.192157}\hspace{0.006\linewidth}\parbox[t]{0.671\linewidth}{decrease DEFAULT\_CONFIG max\_sweeps 23->11 on best dmrg2 config}}}}\par\smallskip
\noindent\parbox[t]{0.05\linewidth}{\centering\strut 76}\hspace{0.006\linewidth}\parbox[t]{0.04\linewidth}{\centering\strut \raisebox{0.08ex}{\ding{55}}}\hspace{0.006\linewidth}\parbox[t]{0.10\linewidth}{\centering\strut +0\%}\hspace{0.006\linewidth}\parbox[t]{0.115\linewidth}{\centering\strut -20.192157}\hspace{0.006\linewidth}\parbox[t]{0.671\linewidth}{decrease DEFAULT\_CONFIG max\_sweeps 23->13 on best dmrg2 config}\par\smallskip
\noindent{\setlength{\fboxsep}{0pt}\colorbox{black!4}{\parbox{\linewidth}{\parbox[t]{0.05\linewidth}{\centering\strut 77}\hspace{0.006\linewidth}\parbox[t]{0.04\linewidth}{\centering\strut \raisebox{0.08ex}{\ding{55}}}\hspace{0.006\linewidth}\parbox[t]{0.10\linewidth}{\centering\strut +0\%}\hspace{0.006\linewidth}\parbox[t]{0.115\linewidth}{\centering\strut -20.192157}\hspace{0.006\linewidth}\parbox[t]{0.671\linewidth}{decrease DEFAULT\_CONFIG max\_sweeps 23->15 on best dmrg2 config}}}}\par\smallskip
\noindent\parbox[t]{0.05\linewidth}{\centering\strut 78}\hspace{0.006\linewidth}\parbox[t]{0.04\linewidth}{\centering\strut \raisebox{0.08ex}{\ding{55}}}\hspace{0.006\linewidth}\parbox[t]{0.10\linewidth}{\centering\strut +0\%}\hspace{0.006\linewidth}\parbox[t]{0.115\linewidth}{\centering\strut -20.192157}\hspace{0.006\linewidth}\parbox[t]{0.671\linewidth}{decrease DEFAULT\_CONFIG max\_sweeps 23->17 on best dmrg2 config}\par\smallskip
\noindent{\setlength{\fboxsep}{0pt}\colorbox{black!4}{\parbox{\linewidth}{\parbox[t]{0.05\linewidth}{\centering\strut 79}\hspace{0.006\linewidth}\parbox[t]{0.04\linewidth}{\centering\strut \raisebox{0.08ex}{\ding{55}}}\hspace{0.006\linewidth}\parbox[t]{0.10\linewidth}{\centering\strut +0\%}\hspace{0.006\linewidth}\parbox[t]{0.115\linewidth}{\centering\strut -20.192157}\hspace{0.006\linewidth}\parbox[t]{0.671\linewidth}{decrease DEFAULT\_CONFIG max\_sweeps 23->19 on best dmrg2 config}}}}\par\smallskip
\noindent\parbox[t]{0.05\linewidth}{\centering\strut 80}\hspace{0.006\linewidth}\parbox[t]{0.04\linewidth}{\centering\strut \raisebox{0.08ex}{\ding{55}}}\hspace{0.006\linewidth}\parbox[t]{0.10\linewidth}{\centering\strut +0\%}\hspace{0.006\linewidth}\parbox[t]{0.115\linewidth}{\centering\strut -20.192157}\hspace{0.006\linewidth}\parbox[t]{0.671\linewidth}{decrease DEFAULT\_CONFIG max\_sweeps 23->21 on best dmrg2 config}\par\smallskip
\noindent{\setlength{\fboxsep}{0pt}\colorbox{black!4}{\parbox{\linewidth}{\parbox[t]{0.05\linewidth}{\centering\strut 81}\hspace{0.006\linewidth}\parbox[t]{0.04\linewidth}{\centering\strut \raisebox{0.08ex}{\ding{55}}}\hspace{0.006\linewidth}\parbox[t]{0.10\linewidth}{\centering\strut -0.006112\%}\hspace{0.006\linewidth}\parbox[t]{0.115\linewidth}{\centering\strut -20.190922}\hspace{0.006\linewidth}\parbox[t]{0.671\linewidth}{set DEFAULT\_CONFIG bond\_schedule to single stage 12}}}}\par\smallskip
\noindent\parbox[t]{0.05\linewidth}{\centering\strut 82}\hspace{0.006\linewidth}\parbox[t]{0.04\linewidth}{\centering\strut \raisebox{0.08ex}{\ding{55}}}\hspace{0.006\linewidth}\parbox[t]{0.10\linewidth}{\centering\strut +0.00314\%}\hspace{0.006\linewidth}\parbox[t]{0.115\linewidth}{\centering\strut -20.191556}\hspace{0.006\linewidth}\parbox[t]{0.671\linewidth}{set DEFAULT\_CONFIG bond\_schedule to single stage 16}\par\smallskip
\noindent{\setlength{\fboxsep}{0pt}\colorbox{black!4}{\parbox{\linewidth}{\parbox[t]{0.05\linewidth}{\centering\strut 83}\hspace{0.006\linewidth}\parbox[t]{0.04\linewidth}{\centering\strut \raisebox{0.08ex}{\ding{55}}}\hspace{0.006\linewidth}\parbox[t]{0.10\linewidth}{\centering\strut +0.00276\%}\hspace{0.006\linewidth}\parbox[t]{0.115\linewidth}{\centering\strut -20.192114}\hspace{0.006\linewidth}\parbox[t]{0.671\linewidth}{set DEFAULT\_CONFIG bond\_schedule to single stage 24}}}}\par\smallskip
\noindent\parbox[t]{0.05\linewidth}{\centering\strut 84}\hspace{0.006\linewidth}\parbox[t]{0.04\linewidth}{\centering\strut \raisebox{0.08ex}{\ding{55}}}\hspace{0.006\linewidth}\parbox[t]{0.10\linewidth}{\centering\strut +0.000171\%}\hspace{0.006\linewidth}\parbox[t]{0.115\linewidth}{\centering\strut -20.192148}\hspace{0.006\linewidth}\parbox[t]{0.671\linewidth}{set DEFAULT\_CONFIG bond\_schedule to single stage 32}\par\smallskip
\noindent{\setlength{\fboxsep}{0pt}\colorbox{black!4}{\parbox{\linewidth}{\parbox[t]{0.05\linewidth}{\centering\strut 85}\hspace{0.006\linewidth}\parbox[t]{0.04\linewidth}{\centering\strut \raisebox{0.08ex}{\ding{55}}}\hspace{0.006\linewidth}\parbox[t]{0.10\linewidth}{\centering\strut +0\%}\hspace{0.006\linewidth}\parbox[t]{0.115\linewidth}{\centering\strut -20.192156}\hspace{0.006\linewidth}\parbox[t]{0.671\linewidth}{set DEFAULT\_CONFIG bond\_schedule to single stage 48}}}}\par\smallskip
\noindent\parbox[t]{0.05\linewidth}{\centering\strut 86}\hspace{0.006\linewidth}\parbox[t]{0.04\linewidth}{\centering\strut \raisebox{0.08ex}{\ding{55}}}\hspace{0.006\linewidth}\parbox[t]{0.10\linewidth}{\centering\strut +0\%}\hspace{0.006\linewidth}\parbox[t]{0.115\linewidth}{\centering\strut -20.192157}\hspace{0.006\linewidth}\parbox[t]{0.671\linewidth}{set DEFAULT\_CONFIG bond\_schedule to single stage 64}\par\smallskip
\noindent{\setlength{\fboxsep}{0pt}\colorbox{black!4}{\parbox{\linewidth}{\parbox[t]{0.05\linewidth}{\centering\strut 87}\hspace{0.006\linewidth}\parbox[t]{0.04\linewidth}{\centering\strut \raisebox{0.08ex}{\ding{55}}}\hspace{0.006\linewidth}\parbox[t]{0.10\linewidth}{\centering\strut +0\%}\hspace{0.006\linewidth}\parbox[t]{0.115\linewidth}{\centering\strut -20.192157}\hspace{0.006\linewidth}\parbox[t]{0.671\linewidth}{set DEFAULT\_CONFIG bond\_schedule to single stage 80}}}}\par\smallskip
\noindent\parbox[t]{0.05\linewidth}{\centering\strut 88}\hspace{0.006\linewidth}\parbox[t]{0.04\linewidth}{\centering\strut \raisebox{0.08ex}{\ding{55}}}\hspace{0.006\linewidth}\parbox[t]{0.10\linewidth}{\centering\strut +0\%}\hspace{0.006\linewidth}\parbox[t]{0.115\linewidth}{\centering\strut -20.192157}\hspace{0.006\linewidth}\parbox[t]{0.671\linewidth}{set DEFAULT\_CONFIG bond\_schedule to single stage 96}\par\smallskip
\noindent{\setlength{\fboxsep}{0pt}\colorbox{black!4}{\parbox{\linewidth}{\parbox[t]{0.05\linewidth}{\centering\strut 89}\hspace{0.006\linewidth}\parbox[t]{0.04\linewidth}{\centering\strut \raisebox{0.08ex}{\ding{55}}}\hspace{0.006\linewidth}\parbox[t]{0.10\linewidth}{\centering\strut +0\%}\hspace{0.006\linewidth}\parbox[t]{0.115\linewidth}{\centering\strut -20.192157}\hspace{0.006\linewidth}\parbox[t]{0.671\linewidth}{set DEFAULT\_CONFIG init\_state random->neel on best dmrg2 config}}}}\par\smallskip
\noindent\parbox[t]{0.05\linewidth}{\centering\strut 90}\hspace{0.006\linewidth}\parbox[t]{0.04\linewidth}{\centering\strut \raisebox{0.08ex}{\ding{55}}}\hspace{0.006\linewidth}\parbox[t]{0.10\linewidth}{\centering\strut +0\%}\hspace{0.006\linewidth}\parbox[t]{0.115\linewidth}{\centering\strut -20.192157}\hspace{0.006\linewidth}\parbox[t]{0.671\linewidth}{set DEFAULT\_CONFIG init\_state random->plus on best dmrg2 config}\par\smallskip
\noindent{\setlength{\fboxsep}{0pt}\colorbox{black!4}{\parbox{\linewidth}{\parbox[t]{0.05\linewidth}{\centering\strut 91}\hspace{0.006\linewidth}\parbox[t]{0.04\linewidth}{\centering\strut \raisebox{0.08ex}{\ding{55}}}\hspace{0.006\linewidth}\parbox[t]{0.10\linewidth}{\centering\strut +0\%}\hspace{0.006\linewidth}\parbox[t]{0.115\linewidth}{\centering\strut -20.192157}\hspace{0.006\linewidth}\parbox[t]{0.671\linewidth}{set DEFAULT\_CONFIG init\_state random->minus on best dmrg2 config}}}}\par\smallskip
\noindent\parbox[t]{0.05\linewidth}{\centering\strut 92}\hspace{0.006\linewidth}\parbox[t]{0.04\linewidth}{\centering\strut \raisebox{0.08ex}{\ding{55}}}\hspace{0.006\linewidth}\parbox[t]{0.10\linewidth}{\centering\strut +0\%}\hspace{0.006\linewidth}\parbox[t]{0.115\linewidth}{\centering\strut -20.192157}\hspace{0.006\linewidth}\parbox[t]{0.671\linewidth}{set DEFAULT\_CONFIG init\_state random->product\_up on best dmrg2 config}\par\smallskip
\noindent{\setlength{\fboxsep}{0pt}\colorbox{black!4}{\parbox{\linewidth}{\parbox[t]{0.05\linewidth}{\centering\strut 93}\hspace{0.006\linewidth}\parbox[t]{0.04\linewidth}{\centering\strut \raisebox{0.08ex}{\ding{55}}}\hspace{0.006\linewidth}\parbox[t]{0.10\linewidth}{\centering\strut +0\%}\hspace{0.006\linewidth}\parbox[t]{0.115\linewidth}{\centering\strut -20.192157}\hspace{0.006\linewidth}\parbox[t]{0.671\linewidth}{set DEFAULT\_CONFIG init\_state random->product\_down on best dmrg2 config}}}}\par\smallskip
\noindent\parbox[t]{0.05\linewidth}{\centering\strut 94}\hspace{0.006\linewidth}\parbox[t]{0.04\linewidth}{\centering\strut \raisebox{0.08ex}{\ding{51}}}\hspace{0.006\linewidth}\parbox[t]{0.10\linewidth}{\centering\strut +0\%}\hspace{0.006\linewidth}\parbox[t]{0.115\linewidth}{\centering\strut -20.192157}\hspace{0.006\linewidth}\parbox[t]{0.671\linewidth}{decrease DEFAULT\_CONFIG init\_bond\_dim 2->1 on best dmrg2 config}\par\smallskip
\noindent{\setlength{\fboxsep}{0pt}\colorbox{black!4}{\parbox{\linewidth}{\parbox[t]{0.05\linewidth}{\centering\strut 95}\hspace{0.006\linewidth}\parbox[t]{0.04\linewidth}{\centering\strut \raisebox{0.08ex}{\ding{55}}}\hspace{0.006\linewidth}\parbox[t]{0.10\linewidth}{\centering\strut +0\%}\hspace{0.006\linewidth}\parbox[t]{0.115\linewidth}{\centering\strut -20.192157}\hspace{0.006\linewidth}\parbox[t]{0.671\linewidth}{increase DEFAULT\_CONFIG init\_bond\_dim 1->3 from iteration-94 best}}}}\par\smallskip
\noindent\parbox[t]{0.05\linewidth}{\centering\strut 96}\hspace{0.006\linewidth}\parbox[t]{0.04\linewidth}{\centering\strut \raisebox{0.08ex}{\ding{55}}}\hspace{0.006\linewidth}\parbox[t]{0.10\linewidth}{\centering\strut +0\%}\hspace{0.006\linewidth}\parbox[t]{0.115\linewidth}{\centering\strut -20.192157}\hspace{0.006\linewidth}\parbox[t]{0.671\linewidth}{increase DEFAULT\_CONFIG init\_bond\_dim 1->4 from iteration-94 best}\par\smallskip
\noindent{\setlength{\fboxsep}{0pt}\colorbox{black!4}{\parbox{\linewidth}{\parbox[t]{0.05\linewidth}{\centering\strut 97}\hspace{0.006\linewidth}\parbox[t]{0.04\linewidth}{\centering\strut \raisebox{0.08ex}{\ding{55}}}\hspace{0.006\linewidth}\parbox[t]{0.10\linewidth}{\centering\strut +0\%}\hspace{0.006\linewidth}\parbox[t]{0.115\linewidth}{\centering\strut -20.192157}\hspace{0.006\linewidth}\parbox[t]{0.671\linewidth}{set DEFAULT\_CONFIG init\_seed 7->42 from iteration-94 best}}}}\par\smallskip
\noindent\parbox[t]{0.05\linewidth}{\centering\strut 98}\hspace{0.006\linewidth}\parbox[t]{0.04\linewidth}{\centering\strut \raisebox{0.08ex}{\ding{55}}}\hspace{0.006\linewidth}\parbox[t]{0.10\linewidth}{\centering\strut +0\%}\hspace{0.006\linewidth}\parbox[t]{0.115\linewidth}{\centering\strut -20.192157}\hspace{0.006\linewidth}\parbox[t]{0.671\linewidth}{set DEFAULT\_CONFIG init\_seed 7->0 from iteration-94 best}\par\smallskip
\noindent{\setlength{\fboxsep}{0pt}\colorbox{black!4}{\parbox{\linewidth}{\parbox[t]{0.05\linewidth}{\centering\strut 99}\hspace{0.006\linewidth}\parbox[t]{0.04\linewidth}{\centering\strut \raisebox{0.08ex}{\ding{55}}}\hspace{0.006\linewidth}\parbox[t]{0.10\linewidth}{\centering\strut +0\%}\hspace{0.006\linewidth}\parbox[t]{0.115\linewidth}{\centering\strut -20.192157}\hspace{0.006\linewidth}\parbox[t]{0.671\linewidth}{set DEFAULT\_CONFIG init\_seed 7->8 from iteration-94 best}}}}\par\smallskip
\noindent\parbox[t]{0.05\linewidth}{\centering\strut 100}\hspace{0.006\linewidth}\parbox[t]{0.04\linewidth}{\centering\strut \raisebox{0.08ex}{\ding{55}}}\hspace{0.006\linewidth}\parbox[t]{0.10\linewidth}{\centering\strut +0\%}\hspace{0.006\linewidth}\parbox[t]{0.115\linewidth}{\centering\strut -20.192157}\hspace{0.006\linewidth}\parbox[t]{0.671\linewidth}{set DEFAULT\_CONFIG init\_seed 7->123 from iteration-94 best}\par\smallskip
\noindent\rule{\linewidth}{0.5pt}
\par\medskip

%% file: generated_afqmc_improvement_tables.tex
\refstepcounter{table}
\label{tab:supp_afqmc_h2}
\noindent\textbf{Table \thetable.} AFQMC mutation log for H$_2$ full 100-iteration campaign. Relative gain is measured against the immediately previous scored iteration using the live AFQMC score.
\par\medskip
\noindent\rule{\linewidth}{0.5pt}
\noindent{\setlength{\fboxsep}{0pt}\colorbox{black!4}{\parbox{\linewidth}{\parbox[c][2.2em][c]{0.05\linewidth}{\centering \textbf{Iter.}}\hspace{0.006\linewidth}\parbox[c][2.2em][c]{0.04\linewidth}{\centering \textbf{A/R}}\hspace{0.006\linewidth}\parbox[c][2.2em][c]{0.10\linewidth}{\centering \textbf{Relative gain}}\hspace{0.006\linewidth}\parbox[c][2.2em][c]{0.115\linewidth}{\centering \textbf{Score}}\hspace{0.006\linewidth}\parbox[c][2.2em][c]{0.671\linewidth}{\centering \textbf{Mutation summary}}}}}\par\smallskip
\noindent\rule{\linewidth}{0.5pt}
\noindent{\setlength{\fboxsep}{0pt}\colorbox{black!4}{\parbox{\linewidth}{\parbox[t]{0.05\linewidth}{\centering\strut 1}\hspace{0.006\linewidth}\parbox[t]{0.04\linewidth}{\centering\strut \raisebox{0.08ex}{\ding{51}}}\hspace{0.006\linewidth}\parbox[t]{0.10\linewidth}{\centering\strut ---}\hspace{0.006\linewidth}\parbox[t]{0.115\linewidth}{\centering\strut -1.121930}\hspace{0.006\linewidth}\parbox[t]{0.671\linewidth}{set timestep 0.0025 with 100-step blocks, 20 blocks, and chol\_cut 1e-8}}}}\par\smallskip
\noindent\parbox[t]{0.05\linewidth}{\centering\strut 2}\hspace{0.006\linewidth}\parbox[t]{0.04\linewidth}{\centering\strut \raisebox{0.08ex}{\ding{51}}}\hspace{0.006\linewidth}\parbox[t]{0.10\linewidth}{\centering\strut +0.0462\%}\hspace{0.006\linewidth}\parbox[t]{0.115\linewidth}{\centering\strut -1.122448}\hspace{0.006\linewidth}\parbox[t]{0.671\linewidth}{double num\_walkers\_per\_rank to 64 on the 0.0025 timestep 100-step 20-block config}\par\smallskip
\noindent{\setlength{\fboxsep}{0pt}\colorbox{black!4}{\parbox{\linewidth}{\parbox[t]{0.05\linewidth}{\centering\strut 3}\hspace{0.006\linewidth}\parbox[t]{0.04\linewidth}{\centering\strut \raisebox{0.08ex}{\ding{51}}}\hspace{0.006\linewidth}\parbox[t]{0.10\linewidth}{\centering\strut +0.0343\%}\hspace{0.006\linewidth}\parbox[t]{0.115\linewidth}{\centering\strut -1.122833}\hspace{0.006\linewidth}\parbox[t]{0.671\linewidth}{increase num\_walkers\_per\_rank to 128 on the 0.0025 timestep 100-step 20-block config}}}}\par\smallskip
\noindent\parbox[t]{0.05\linewidth}{\centering\strut 4}\hspace{0.006\linewidth}\parbox[t]{0.04\linewidth}{\centering\strut \raisebox{0.08ex}{\ding{51}}}\hspace{0.006\linewidth}\parbox[t]{0.10\linewidth}{\centering\strut +0.0000\%}\hspace{0.006\linewidth}\parbox[t]{0.115\linewidth}{\centering\strut -1.122833}\hspace{0.006\linewidth}\parbox[t]{0.671\linewidth}{set stabilize\_freq 3 and pop\_control\_freq 3 on the 128-walker 0.0025 timestep 100-step 20-block config}\par\smallskip
\noindent{\setlength{\fboxsep}{0pt}\colorbox{black!4}{\parbox{\linewidth}{\parbox[t]{0.05\linewidth}{\centering\strut 5}\hspace{0.006\linewidth}\parbox[t]{0.04\linewidth}{\centering\strut \raisebox{0.08ex}{\ding{55}}}\hspace{0.006\linewidth}\parbox[t]{0.10\linewidth}{\centering\strut -0.0000\%}\hspace{0.006\linewidth}\parbox[t]{0.115\linewidth}{\centering\strut -1.122833}\hspace{0.006\linewidth}\parbox[t]{0.671\linewidth}{set stabilize\_freq 1 and pop\_control\_freq 1 on the 128-walker 0.0025 timestep 100-step 20-block config}}}}\par\smallskip
\noindent\parbox[t]{0.05\linewidth}{\centering\strut 6}\hspace{0.006\linewidth}\parbox[t]{0.04\linewidth}{\centering\strut \raisebox{0.08ex}{\ding{51}}}\hspace{0.006\linewidth}\parbox[t]{0.10\linewidth}{\centering\strut +0.3369\%}\hspace{0.006\linewidth}\parbox[t]{0.115\linewidth}{\centering\strut -1.126616}\hspace{0.006\linewidth}\parbox[t]{0.671\linewidth}{increase num\_walkers\_per\_rank to 192 on the freq-3 0.0025 timestep 100-step 20-block config}\par\smallskip
\noindent{\setlength{\fboxsep}{0pt}\colorbox{black!4}{\parbox{\linewidth}{\parbox[t]{0.05\linewidth}{\centering\strut 7}\hspace{0.006\linewidth}\parbox[t]{0.04\linewidth}{\centering\strut \raisebox{0.08ex}{\ding{51}}}\hspace{0.006\linewidth}\parbox[t]{0.10\linewidth}{\centering\strut +0.4784\%}\hspace{0.006\linewidth}\parbox[t]{0.115\linewidth}{\centering\strut -1.132006}\hspace{0.006\linewidth}\parbox[t]{0.671\linewidth}{increase num\_walkers\_per\_rank to 256 on the freq-3 0.0025 timestep 100-step 20-block config}}}}\par\smallskip
\noindent\parbox[t]{0.05\linewidth}{\centering\strut 8}\hspace{0.006\linewidth}\parbox[t]{0.04\linewidth}{\centering\strut \raisebox{0.08ex}{\ding{55}}}\hspace{0.006\linewidth}\parbox[t]{0.10\linewidth}{\centering\strut ---}\hspace{0.006\linewidth}\parbox[t]{0.115\linewidth}{\centering\strut ---}\hspace{0.006\linewidth}\parbox[t]{0.671\linewidth}{set num\_steps\_per\_block 125 and timestep 0.002 on the 256-walker 20-block freq-3 config}\par\smallskip
\noindent{\setlength{\fboxsep}{0pt}\colorbox{black!4}{\parbox{\linewidth}{\parbox[t]{0.05\linewidth}{\centering\strut 9}\hspace{0.006\linewidth}\parbox[t]{0.04\linewidth}{\centering\strut \raisebox{0.08ex}{\ding{55}}}\hspace{0.006\linewidth}\parbox[t]{0.10\linewidth}{\centering\strut -0.3219\%}\hspace{0.006\linewidth}\parbox[t]{0.115\linewidth}{\centering\strut -1.128362}\hspace{0.006\linewidth}\parbox[t]{0.671\linewidth}{num\_walkers\_per\_rank 288}}}}\par\smallskip
\noindent\parbox[t]{0.05\linewidth}{\centering\strut 10}\hspace{0.006\linewidth}\parbox[t]{0.04\linewidth}{\centering\strut \raisebox{0.08ex}{\ding{55}}}\hspace{0.006\linewidth}\parbox[t]{0.10\linewidth}{\centering\strut +0.0000\%}\hspace{0.006\linewidth}\parbox[t]{0.115\linewidth}{\centering\strut -1.128362}\hspace{0.006\linewidth}\parbox[t]{0.671\linewidth}{increase num\_walkers\_per\_rank to 288 on the freq-3 0.0025 timestep 100-step 20-block config}\par\smallskip
\noindent{\setlength{\fboxsep}{0pt}\colorbox{black!4}{\parbox{\linewidth}{\parbox[t]{0.05\linewidth}{\centering\strut 11}\hspace{0.006\linewidth}\parbox[t]{0.04\linewidth}{\centering\strut \raisebox{0.08ex}{\ding{55}}}\hspace{0.006\linewidth}\parbox[t]{0.10\linewidth}{\centering\strut -0.2695\%}\hspace{0.006\linewidth}\parbox[t]{0.115\linewidth}{\centering\strut -1.125321}\hspace{0.006\linewidth}\parbox[t]{0.671\linewidth}{num\_walkers\_per\_rank 320}}}}\par\smallskip
\noindent\parbox[t]{0.05\linewidth}{\centering\strut 12}\hspace{0.006\linewidth}\parbox[t]{0.04\linewidth}{\centering\strut \raisebox{0.08ex}{\ding{55}}}\hspace{0.006\linewidth}\parbox[t]{0.10\linewidth}{\centering\strut +0.1818\%}\hspace{0.006\linewidth}\parbox[t]{0.115\linewidth}{\centering\strut -1.127367}\hspace{0.006\linewidth}\parbox[t]{0.671\linewidth}{num\_walkers\_per\_rank 384}\par\smallskip
\noindent{\setlength{\fboxsep}{0pt}\colorbox{black!4}{\parbox{\linewidth}{\parbox[t]{0.05\linewidth}{\centering\strut 13}\hspace{0.006\linewidth}\parbox[t]{0.04\linewidth}{\centering\strut \raisebox{0.08ex}{\ding{55}}}\hspace{0.006\linewidth}\parbox[t]{0.10\linewidth}{\centering\strut -0.0415\%}\hspace{0.006\linewidth}\parbox[t]{0.115\linewidth}{\centering\strut -1.126899}\hspace{0.006\linewidth}\parbox[t]{0.671\linewidth}{num\_walkers\_per\_rank 224}}}}\par\smallskip
\noindent\parbox[t]{0.05\linewidth}{\centering\strut 14}\hspace{0.006\linewidth}\parbox[t]{0.04\linewidth}{\centering\strut \raisebox{0.08ex}{\ding{55}}}\hspace{0.006\linewidth}\parbox[t]{0.10\linewidth}{\centering\strut +0.4531\%}\hspace{0.006\linewidth}\parbox[t]{0.115\linewidth}{\centering\strut -1.132006}\hspace{0.006\linewidth}\parbox[t]{0.671\linewidth}{stabilize\_freq 2, pop\_control\_freq 2}\par\smallskip
\noindent{\setlength{\fboxsep}{0pt}\colorbox{black!4}{\parbox{\linewidth}{\parbox[t]{0.05\linewidth}{\centering\strut 15}\hspace{0.006\linewidth}\parbox[t]{0.04\linewidth}{\centering\strut \raisebox{0.08ex}{\ding{55}}}\hspace{0.006\linewidth}\parbox[t]{0.10\linewidth}{\centering\strut -0.2922\%}\hspace{0.006\linewidth}\parbox[t]{0.115\linewidth}{\centering\strut -1.128699}\hspace{0.006\linewidth}\parbox[t]{0.671\linewidth}{num\_walkers\_per\_rank 160}}}}\par\smallskip
\noindent\parbox[t]{0.05\linewidth}{\centering\strut 16}\hspace{0.006\linewidth}\parbox[t]{0.04\linewidth}{\centering\strut \raisebox{0.08ex}{\ding{51}}}\hspace{0.006\linewidth}\parbox[t]{0.10\linewidth}{\centering\strut +0.3427\%}\hspace{0.006\linewidth}\parbox[t]{0.115\linewidth}{\centering\strut -1.132567}\hspace{0.006\linewidth}\parbox[t]{0.671\linewidth}{timestep 0.002}\par\smallskip
\noindent{\setlength{\fboxsep}{0pt}\colorbox{black!4}{\parbox{\linewidth}{\parbox[t]{0.05\linewidth}{\centering\strut 17}\hspace{0.006\linewidth}\parbox[t]{0.04\linewidth}{\centering\strut \raisebox{0.08ex}{\ding{55}}}\hspace{0.006\linewidth}\parbox[t]{0.10\linewidth}{\centering\strut -0.4931\%}\hspace{0.006\linewidth}\parbox[t]{0.115\linewidth}{\centering\strut -1.126982}\hspace{0.006\linewidth}\parbox[t]{0.671\linewidth}{num\_walkers\_per\_rank 224}}}}\par\smallskip
\noindent\parbox[t]{0.05\linewidth}{\centering\strut 18}\hspace{0.006\linewidth}\parbox[t]{0.04\linewidth}{\centering\strut \raisebox{0.08ex}{\ding{55}}}\hspace{0.006\linewidth}\parbox[t]{0.10\linewidth}{\centering\strut +0.0128\%}\hspace{0.006\linewidth}\parbox[t]{0.115\linewidth}{\centering\strut -1.127126}\hspace{0.006\linewidth}\parbox[t]{0.671\linewidth}{num\_walkers\_per\_rank 192}\par\smallskip
\noindent{\setlength{\fboxsep}{0pt}\colorbox{black!4}{\parbox{\linewidth}{\parbox[t]{0.05\linewidth}{\centering\strut 19}\hspace{0.006\linewidth}\parbox[t]{0.04\linewidth}{\centering\strut \raisebox{0.08ex}{\ding{55}}}\hspace{0.006\linewidth}\parbox[t]{0.10\linewidth}{\centering\strut +0.4828\%}\hspace{0.006\linewidth}\parbox[t]{0.115\linewidth}{\centering\strut -1.132567}\hspace{0.006\linewidth}\parbox[t]{0.671\linewidth}{stabilize\_freq 2, pop\_control\_freq 2}}}}\par\smallskip
\noindent\parbox[t]{0.05\linewidth}{\centering\strut 20}\hspace{0.006\linewidth}\parbox[t]{0.04\linewidth}{\centering\strut \raisebox{0.08ex}{\ding{55}}}\hspace{0.006\linewidth}\parbox[t]{0.10\linewidth}{\centering\strut -0.3393\%}\hspace{0.006\linewidth}\parbox[t]{0.115\linewidth}{\centering\strut -1.128725}\hspace{0.006\linewidth}\parbox[t]{0.671\linewidth}{num\_walkers\_per\_rank 160}\par\smallskip
\noindent{\setlength{\fboxsep}{0pt}\colorbox{black!4}{\parbox{\linewidth}{\parbox[t]{0.05\linewidth}{\centering\strut 21}\hspace{0.006\linewidth}\parbox[t]{0.04\linewidth}{\centering\strut \raisebox{0.08ex}{\ding{55}}}\hspace{0.006\linewidth}\parbox[t]{0.10\linewidth}{\centering\strut +0.3404\%}\hspace{0.006\linewidth}\parbox[t]{0.115\linewidth}{\centering\strut -1.132567}\hspace{0.006\linewidth}\parbox[t]{0.671\linewidth}{stabilize\_freq 5, pop\_control\_freq 5}}}}\par\smallskip
\noindent\parbox[t]{0.05\linewidth}{\centering\strut 22}\hspace{0.006\linewidth}\parbox[t]{0.04\linewidth}{\centering\strut \raisebox{0.08ex}{\ding{55}}}\hspace{0.006\linewidth}\parbox[t]{0.10\linewidth}{\centering\strut -0.1284\%}\hspace{0.006\linewidth}\parbox[t]{0.115\linewidth}{\centering\strut -1.131113}\hspace{0.006\linewidth}\parbox[t]{0.671\linewidth}{timestep 0.003}\par\smallskip
\noindent{\setlength{\fboxsep}{0pt}\colorbox{black!4}{\parbox{\linewidth}{\parbox[t]{0.05\linewidth}{\centering\strut 23}\hspace{0.006\linewidth}\parbox[t]{0.04\linewidth}{\centering\strut \raisebox{0.08ex}{\ding{51}}}\hspace{0.006\linewidth}\parbox[t]{0.10\linewidth}{\centering\strut +0.1285\%}\hspace{0.006\linewidth}\parbox[t]{0.115\linewidth}{\centering\strut -1.132567}\hspace{0.006\linewidth}\parbox[t]{0.671\linewidth}{stabilize\_freq 8, pop\_control\_freq 8}}}}\par\smallskip
\noindent\parbox[t]{0.05\linewidth}{\centering\strut 24}\hspace{0.006\linewidth}\parbox[t]{0.04\linewidth}{\centering\strut \raisebox{0.08ex}{\ding{55}}}\hspace{0.006\linewidth}\parbox[t]{0.10\linewidth}{\centering\strut -0.4931\%}\hspace{0.006\linewidth}\parbox[t]{0.115\linewidth}{\centering\strut -1.126982}\hspace{0.006\linewidth}\parbox[t]{0.671\linewidth}{num\_walkers\_per\_rank 224}\par\smallskip
\noindent{\setlength{\fboxsep}{0pt}\colorbox{black!4}{\parbox{\linewidth}{\parbox[t]{0.05\linewidth}{\centering\strut 25}\hspace{0.006\linewidth}\parbox[t]{0.04\linewidth}{\centering\strut \raisebox{0.08ex}{\ding{55}}}\hspace{0.006\linewidth}\parbox[t]{0.10\linewidth}{\centering\strut +0.0128\%}\hspace{0.006\linewidth}\parbox[t]{0.115\linewidth}{\centering\strut -1.127126}\hspace{0.006\linewidth}\parbox[t]{0.671\linewidth}{num\_walkers\_per\_rank 192}}}}\par\smallskip
\noindent\parbox[t]{0.05\linewidth}{\centering\strut 26}\hspace{0.006\linewidth}\parbox[t]{0.04\linewidth}{\centering\strut \raisebox{0.08ex}{\ding{55}}}\hspace{0.006\linewidth}\parbox[t]{0.10\linewidth}{\centering\strut +0.4329\%}\hspace{0.006\linewidth}\parbox[t]{0.115\linewidth}{\centering\strut -1.132006}\hspace{0.006\linewidth}\parbox[t]{0.671\linewidth}{timestep 0.0025}\par\smallskip
\noindent{\setlength{\fboxsep}{0pt}\colorbox{black!4}{\parbox{\linewidth}{\parbox[t]{0.05\linewidth}{\centering\strut 27}\hspace{0.006\linewidth}\parbox[t]{0.04\linewidth}{\centering\strut \raisebox{0.08ex}{\ding{55}}}\hspace{0.006\linewidth}\parbox[t]{0.10\linewidth}{\centering\strut -0.0789\%}\hspace{0.006\linewidth}\parbox[t]{0.115\linewidth}{\centering\strut -1.131113}\hspace{0.006\linewidth}\parbox[t]{0.671\linewidth}{timestep 0.003}}}}\par\smallskip
\noindent\parbox[t]{0.05\linewidth}{\centering\strut 28}\hspace{0.006\linewidth}\parbox[t]{0.04\linewidth}{\centering\strut \raisebox{0.08ex}{\ding{55}}}\hspace{0.006\linewidth}\parbox[t]{0.10\linewidth}{\centering\strut -0.0324\%}\hspace{0.006\linewidth}\parbox[t]{0.115\linewidth}{\centering\strut -1.130747}\hspace{0.006\linewidth}\parbox[t]{0.671\linewidth}{num\_blocks 30}\par\smallskip
\noindent{\setlength{\fboxsep}{0pt}\colorbox{black!4}{\parbox{\linewidth}{\parbox[t]{0.05\linewidth}{\centering\strut 29}\hspace{0.006\linewidth}\parbox[t]{0.04\linewidth}{\centering\strut \raisebox{0.08ex}{\ding{55}}}\hspace{0.006\linewidth}\parbox[t]{0.10\linewidth}{\centering\strut -0.3549\%}\hspace{0.006\linewidth}\parbox[t]{0.115\linewidth}{\centering\strut -1.126734}\hspace{0.006\linewidth}\parbox[t]{0.671\linewidth}{num\_steps\_per\_block 80, timestep 0.0025}}}}\par\smallskip
\noindent\parbox[t]{0.05\linewidth}{\centering\strut 30}\hspace{0.006\linewidth}\parbox[t]{0.04\linewidth}{\centering\strut \raisebox{0.08ex}{\ding{55}}}\hspace{0.006\linewidth}\parbox[t]{0.10\linewidth}{\centering\strut +0.3804\%}\hspace{0.006\linewidth}\parbox[t]{0.115\linewidth}{\centering\strut -1.131020}\hspace{0.006\linewidth}\parbox[t]{0.671\linewidth}{num\_blocks 24, timestep 0.001667}\par\smallskip
\noindent{\setlength{\fboxsep}{0pt}\colorbox{black!4}{\parbox{\linewidth}{\parbox[t]{0.05\linewidth}{\centering\strut 31}\hspace{0.006\linewidth}\parbox[t]{0.04\linewidth}{\centering\strut \raisebox{0.08ex}{\ding{55}}}\hspace{0.006\linewidth}\parbox[t]{0.10\linewidth}{\centering\strut -0.0864\%}\hspace{0.006\linewidth}\parbox[t]{0.115\linewidth}{\centering\strut -1.130043}\hspace{0.006\linewidth}\parbox[t]{0.671\linewidth}{num\_steps\_per\_block 60, timestep 0.003333}}}}\par\smallskip
\noindent\parbox[t]{0.05\linewidth}{\centering\strut 32}\hspace{0.006\linewidth}\parbox[t]{0.04\linewidth}{\centering\strut \raisebox{0.08ex}{\ding{55}}}\hspace{0.006\linewidth}\parbox[t]{0.10\linewidth}{\centering\strut +0.1689\%}\hspace{0.006\linewidth}\parbox[t]{0.115\linewidth}{\centering\strut -1.131952}\hspace{0.006\linewidth}\parbox[t]{0.671\linewidth}{num\_blocks 30, timestep 0.001333}\par\smallskip
\noindent{\setlength{\fboxsep}{0pt}\colorbox{black!4}{\parbox{\linewidth}{\parbox[t]{0.05\linewidth}{\centering\strut 33}\hspace{0.006\linewidth}\parbox[t]{0.04\linewidth}{\centering\strut \raisebox{0.08ex}{\ding{55}}}\hspace{0.006\linewidth}\parbox[t]{0.10\linewidth}{\centering\strut -0.3171\%}\hspace{0.006\linewidth}\parbox[t]{0.115\linewidth}{\centering\strut -1.128362}\hspace{0.006\linewidth}\parbox[t]{0.671\linewidth}{num\_walkers\_per\_rank 288, timestep 0.0025}}}}\par\smallskip
\noindent\parbox[t]{0.05\linewidth}{\centering\strut 34}\hspace{0.006\linewidth}\parbox[t]{0.04\linewidth}{\centering\strut \raisebox{0.08ex}{\ding{55}}}\hspace{0.006\linewidth}\parbox[t]{0.10\linewidth}{\centering\strut -0.1296\%}\hspace{0.006\linewidth}\parbox[t]{0.115\linewidth}{\centering\strut -1.126899}\hspace{0.006\linewidth}\parbox[t]{0.671\linewidth}{num\_walkers\_per\_rank 224, timestep 0.0025}\par\smallskip
\noindent{\setlength{\fboxsep}{0pt}\colorbox{black!4}{\parbox{\linewidth}{\parbox[t]{0.05\linewidth}{\centering\strut 35}\hspace{0.006\linewidth}\parbox[t]{0.04\linewidth}{\centering\strut \raisebox{0.08ex}{\ding{55}}}\hspace{0.006\linewidth}\parbox[t]{0.10\linewidth}{\centering\strut +0.0792\%}\hspace{0.006\linewidth}\parbox[t]{0.115\linewidth}{\centering\strut -1.127792}\hspace{0.006\linewidth}\parbox[t]{0.671\linewidth}{num\_steps\_per\_block 80, timestep 0.003125}}}}\par\smallskip
\noindent\parbox[t]{0.05\linewidth}{\centering\strut 36}\hspace{0.006\linewidth}\parbox[t]{0.04\linewidth}{\centering\strut \raisebox{0.08ex}{\ding{55}}}\hspace{0.006\linewidth}\parbox[t]{0.10\linewidth}{\centering\strut +0.1803\%}\hspace{0.006\linewidth}\parbox[t]{0.115\linewidth}{\centering\strut -1.129825}\hspace{0.006\linewidth}\parbox[t]{0.671\linewidth}{num\_walkers\_per\_rank 320, timestep 0.0025}\par\smallskip
\noindent{\setlength{\fboxsep}{0pt}\colorbox{black!4}{\parbox{\linewidth}{\parbox[t]{0.05\linewidth}{\centering\strut 37}\hspace{0.006\linewidth}\parbox[t]{0.04\linewidth}{\centering\strut \raisebox{0.08ex}{\ding{55}}}\hspace{0.006\linewidth}\parbox[t]{0.10\linewidth}{\centering\strut +0.1256\%}\hspace{0.006\linewidth}\parbox[t]{0.115\linewidth}{\centering\strut -1.131244}\hspace{0.006\linewidth}\parbox[t]{0.671\linewidth}{num\_blocks 24, timestep 0.002083}}}}\par\smallskip
\noindent\parbox[t]{0.05\linewidth}{\centering\strut 38}\hspace{0.006\linewidth}\parbox[t]{0.04\linewidth}{\centering\strut \raisebox{0.08ex}{\ding{55}}}\hspace{0.006\linewidth}\parbox[t]{0.10\linewidth}{\centering\strut -0.4092\%}\hspace{0.006\linewidth}\parbox[t]{0.115\linewidth}{\centering\strut -1.126616}\hspace{0.006\linewidth}\parbox[t]{0.671\linewidth}{num\_walkers\_per\_rank 192, timestep 0.0025}\par\smallskip
\noindent{\setlength{\fboxsep}{0pt}\colorbox{black!4}{\parbox{\linewidth}{\parbox[t]{0.05\linewidth}{\centering\strut 39}\hspace{0.006\linewidth}\parbox[t]{0.04\linewidth}{\centering\strut \raisebox{0.08ex}{\ding{55}}}\hspace{0.006\linewidth}\parbox[t]{0.10\linewidth}{\centering\strut +0.0667\%}\hspace{0.006\linewidth}\parbox[t]{0.115\linewidth}{\centering\strut -1.127367}\hspace{0.006\linewidth}\parbox[t]{0.671\linewidth}{num\_walkers\_per\_rank 384, timestep 0.0025}}}}\par\smallskip
\noindent\parbox[t]{0.05\linewidth}{\centering\strut 40}\hspace{0.006\linewidth}\parbox[t]{0.04\linewidth}{\centering\strut \raisebox{0.08ex}{\ding{55}}}\hspace{0.006\linewidth}\parbox[t]{0.10\linewidth}{\centering\strut +0.0808\%}\hspace{0.006\linewidth}\parbox[t]{0.115\linewidth}{\centering\strut -1.128278}\hspace{0.006\linewidth}\parbox[t]{0.671\linewidth}{num\_walkers\_per\_rank 288, stabilize\_freq 5, pop\_control\_freq 5}\par\smallskip
\noindent{\setlength{\fboxsep}{0pt}\colorbox{black!4}{\parbox{\linewidth}{\parbox[t]{0.05\linewidth}{\centering\strut 41}\hspace{0.006\linewidth}\parbox[t]{0.04\linewidth}{\centering\strut \raisebox{0.08ex}{\ding{55}}}\hspace{0.006\linewidth}\parbox[t]{0.10\linewidth}{\centering\strut +0.0373\%}\hspace{0.006\linewidth}\parbox[t]{0.115\linewidth}{\centering\strut -1.128699}\hspace{0.006\linewidth}\parbox[t]{0.671\linewidth}{num\_walkers\_per\_rank 160, timestep 0.0025}}}}\par\smallskip
\noindent\parbox[t]{0.05\linewidth}{\centering\strut 42}\hspace{0.006\linewidth}\parbox[t]{0.04\linewidth}{\centering\strut \raisebox{0.08ex}{\ding{55}}}\hspace{0.006\linewidth}\parbox[t]{0.10\linewidth}{\centering\strut -0.1521\%}\hspace{0.006\linewidth}\parbox[t]{0.115\linewidth}{\centering\strut -1.126982}\hspace{0.006\linewidth}\parbox[t]{0.671\linewidth}{num\_walkers\_per\_rank 224, stabilize\_freq 5, pop\_control\_freq 5}\par\smallskip
\noindent{\setlength{\fboxsep}{0pt}\colorbox{black!4}{\parbox{\linewidth}{\parbox[t]{0.05\linewidth}{\centering\strut 43}\hspace{0.006\linewidth}\parbox[t]{0.04\linewidth}{\centering\strut \raisebox{0.08ex}{\ding{55}}}\hspace{0.006\linewidth}\parbox[t]{0.10\linewidth}{\centering\strut +0.3793\%}\hspace{0.006\linewidth}\parbox[t]{0.115\linewidth}{\centering\strut -1.131257}\hspace{0.006\linewidth}\parbox[t]{0.671\linewidth}{num\_steps\_per\_block 60, timestep 0.004167}}}}\par\smallskip
\noindent\parbox[t]{0.05\linewidth}{\centering\strut 44}\hspace{0.006\linewidth}\parbox[t]{0.04\linewidth}{\centering\strut \raisebox{0.08ex}{\ding{55}}}\hspace{0.006\linewidth}\parbox[t]{0.10\linewidth}{\centering\strut -0.3351\%}\hspace{0.006\linewidth}\parbox[t]{0.115\linewidth}{\centering\strut -1.127467}\hspace{0.006\linewidth}\parbox[t]{0.671\linewidth}{num\_walkers\_per\_rank 320, stabilize\_freq 5, pop\_control\_freq 5}\par\smallskip
\noindent{\setlength{\fboxsep}{0pt}\colorbox{black!4}{\parbox{\linewidth}{\parbox[t]{0.05\linewidth}{\centering\strut 45}\hspace{0.006\linewidth}\parbox[t]{0.04\linewidth}{\centering\strut \raisebox{0.08ex}{\ding{55}}}\hspace{0.006\linewidth}\parbox[t]{0.10\linewidth}{\centering\strut -0.0302\%}\hspace{0.006\linewidth}\parbox[t]{0.115\linewidth}{\centering\strut -1.127126}\hspace{0.006\linewidth}\parbox[t]{0.671\linewidth}{num\_walkers\_per\_rank 192, stabilize\_freq 5, pop\_control\_freq 5}}}}\par\smallskip
\noindent\parbox[t]{0.05\linewidth}{\centering\strut 46}\hspace{0.006\linewidth}\parbox[t]{0.04\linewidth}{\centering\strut \raisebox{0.08ex}{\ding{55}}}\hspace{0.006\linewidth}\parbox[t]{0.10\linewidth}{\centering\strut +0.4007\%}\hspace{0.006\linewidth}\parbox[t]{0.115\linewidth}{\centering\strut -1.131642}\hspace{0.006\linewidth}\parbox[t]{0.671\linewidth}{num\_blocks 30, timestep 0.001667}\par\smallskip
\noindent{\setlength{\fboxsep}{0pt}\colorbox{black!4}{\parbox{\linewidth}{\parbox[t]{0.05\linewidth}{\centering\strut 47}\hspace{0.006\linewidth}\parbox[t]{0.04\linewidth}{\centering\strut \raisebox{0.08ex}{\ding{55}}}\hspace{0.006\linewidth}\parbox[t]{0.10\linewidth}{\centering\strut -0.3100\%}\hspace{0.006\linewidth}\parbox[t]{0.115\linewidth}{\centering\strut -1.128134}\hspace{0.006\linewidth}\parbox[t]{0.671\linewidth}{num\_walkers\_per\_rank 384, stabilize\_freq 5, pop\_control\_freq 5}}}}\par\smallskip
\noindent\parbox[t]{0.05\linewidth}{\centering\strut 48}\hspace{0.006\linewidth}\parbox[t]{0.04\linewidth}{\centering\strut \raisebox{0.08ex}{\ding{55}}}\hspace{0.006\linewidth}\parbox[t]{0.10\linewidth}{\centering\strut +0.0479\%}\hspace{0.006\linewidth}\parbox[t]{0.115\linewidth}{\centering\strut -1.128675}\hspace{0.006\linewidth}\parbox[t]{0.671\linewidth}{num\_walkers\_per\_rank 288, timestep 0.003}\par\smallskip
\noindent{\setlength{\fboxsep}{0pt}\colorbox{black!4}{\parbox{\linewidth}{\parbox[t]{0.05\linewidth}{\centering\strut 49}\hspace{0.006\linewidth}\parbox[t]{0.04\linewidth}{\centering\strut \raisebox{0.08ex}{\ding{55}}}\hspace{0.006\linewidth}\parbox[t]{0.10\linewidth}{\centering\strut +0.0044\%}\hspace{0.006\linewidth}\parbox[t]{0.115\linewidth}{\centering\strut -1.128725}\hspace{0.006\linewidth}\parbox[t]{0.671\linewidth}{num\_walkers\_per\_rank 160, stabilize\_freq 5, pop\_control\_freq 5}}}}\par\smallskip
\noindent\parbox[t]{0.05\linewidth}{\centering\strut 50}\hspace{0.006\linewidth}\parbox[t]{0.04\linewidth}{\centering\strut \raisebox{0.08ex}{\ding{55}}}\hspace{0.006\linewidth}\parbox[t]{0.10\linewidth}{\centering\strut -0.1879\%}\hspace{0.006\linewidth}\parbox[t]{0.115\linewidth}{\centering\strut -1.126604}\hspace{0.006\linewidth}\parbox[t]{0.671\linewidth}{num\_walkers\_per\_rank 224, timestep 0.003}\par\smallskip
\noindent{\setlength{\fboxsep}{0pt}\colorbox{black!4}{\parbox{\linewidth}{\parbox[t]{0.05\linewidth}{\centering\strut 51}\hspace{0.006\linewidth}\parbox[t]{0.04\linewidth}{\centering\strut \raisebox{0.08ex}{\ding{55}}}\hspace{0.006\linewidth}\parbox[t]{0.10\linewidth}{\centering\strut +0.1796\%}\hspace{0.006\linewidth}\parbox[t]{0.115\linewidth}{\centering\strut -1.128627}\hspace{0.006\linewidth}\parbox[t]{0.671\linewidth}{num\_steps\_per\_block 80, timestep 0.00375}}}}\par\smallskip
\noindent\parbox[t]{0.05\linewidth}{\centering\strut 52}\hspace{0.006\linewidth}\parbox[t]{0.04\linewidth}{\centering\strut \raisebox{0.08ex}{\ding{55}}}\hspace{0.006\linewidth}\parbox[t]{0.10\linewidth}{\centering\strut -0.0006\%}\hspace{0.006\linewidth}\parbox[t]{0.115\linewidth}{\centering\strut -1.128621}\hspace{0.006\linewidth}\parbox[t]{0.671\linewidth}{num\_walkers\_per\_rank 320, timestep 0.003}\par\smallskip
\noindent{\setlength{\fboxsep}{0pt}\colorbox{black!4}{\parbox{\linewidth}{\parbox[t]{0.05\linewidth}{\centering\strut 53}\hspace{0.006\linewidth}\parbox[t]{0.04\linewidth}{\centering\strut \raisebox{0.08ex}{\ding{55}}}\hspace{0.006\linewidth}\parbox[t]{0.10\linewidth}{\centering\strut +0.2054\%}\hspace{0.006\linewidth}\parbox[t]{0.115\linewidth}{\centering\strut -1.130939}\hspace{0.006\linewidth}\parbox[t]{0.671\linewidth}{num\_blocks 24, timestep 0.0025}}}}\par\smallskip
\noindent\parbox[t]{0.05\linewidth}{\centering\strut 54}\hspace{0.006\linewidth}\parbox[t]{0.04\linewidth}{\centering\strut \raisebox{0.08ex}{\ding{55}}}\hspace{0.006\linewidth}\parbox[t]{0.10\linewidth}{\centering\strut -0.4433\%}\hspace{0.006\linewidth}\parbox[t]{0.115\linewidth}{\centering\strut -1.125926}\hspace{0.006\linewidth}\parbox[t]{0.671\linewidth}{num\_walkers\_per\_rank 192, timestep 0.003}\par\smallskip
\noindent{\setlength{\fboxsep}{0pt}\colorbox{black!4}{\parbox{\linewidth}{\parbox[t]{0.05\linewidth}{\centering\strut 55}\hspace{0.006\linewidth}\parbox[t]{0.04\linewidth}{\centering\strut \raisebox{0.08ex}{\ding{55}}}\hspace{0.006\linewidth}\parbox[t]{0.10\linewidth}{\centering\strut +0.5399\%}\hspace{0.006\linewidth}\parbox[t]{0.115\linewidth}{\centering\strut -1.132006}\hspace{0.006\linewidth}\parbox[t]{0.671\linewidth}{timestep 0.0025, stabilize\_freq 5, pop\_control\_freq 5}}}}\par\smallskip
\noindent\parbox[t]{0.05\linewidth}{\centering\strut 56}\hspace{0.006\linewidth}\parbox[t]{0.04\linewidth}{\centering\strut \raisebox{0.08ex}{\ding{55}}}\hspace{0.006\linewidth}\parbox[t]{0.10\linewidth}{\centering\strut -0.4463\%}\hspace{0.006\linewidth}\parbox[t]{0.115\linewidth}{\centering\strut -1.126954}\hspace{0.006\linewidth}\parbox[t]{0.671\linewidth}{num\_walkers\_per\_rank 384, timestep 0.003}\par\smallskip
\noindent{\setlength{\fboxsep}{0pt}\colorbox{black!4}{\parbox{\linewidth}{\parbox[t]{0.05\linewidth}{\centering\strut 57}\hspace{0.006\linewidth}\parbox[t]{0.04\linewidth}{\centering\strut \raisebox{0.08ex}{\ding{55}}}\hspace{0.006\linewidth}\parbox[t]{0.10\linewidth}{\centering\strut +0.1175\%}\hspace{0.006\linewidth}\parbox[t]{0.115\linewidth}{\centering\strut -1.128278}\hspace{0.006\linewidth}\parbox[t]{0.671\linewidth}{num\_walkers\_per\_rank 288, stabilize\_freq 3, pop\_control\_freq 3}}}}\par\smallskip
\noindent\parbox[t]{0.05\linewidth}{\centering\strut 58}\hspace{0.006\linewidth}\parbox[t]{0.04\linewidth}{\centering\strut \raisebox{0.08ex}{\ding{55}}}\hspace{0.006\linewidth}\parbox[t]{0.10\linewidth}{\centering\strut +0.0287\%}\hspace{0.006\linewidth}\parbox[t]{0.115\linewidth}{\centering\strut -1.128602}\hspace{0.006\linewidth}\parbox[t]{0.671\linewidth}{num\_walkers\_per\_rank 160, timestep 0.003}\par\smallskip
\noindent{\setlength{\fboxsep}{0pt}\colorbox{black!4}{\parbox{\linewidth}{\parbox[t]{0.05\linewidth}{\centering\strut 59}\hspace{0.006\linewidth}\parbox[t]{0.04\linewidth}{\centering\strut \raisebox{0.08ex}{\ding{55}}}\hspace{0.006\linewidth}\parbox[t]{0.10\linewidth}{\centering\strut +0.3119\%}\hspace{0.006\linewidth}\parbox[t]{0.115\linewidth}{\centering\strut -1.132122}\hspace{0.006\linewidth}\parbox[t]{0.671\linewidth}{num\_steps\_per\_block 60, timestep 0.005}}}}\par\smallskip
\noindent\parbox[t]{0.05\linewidth}{\centering\strut 60}\hspace{0.006\linewidth}\parbox[t]{0.04\linewidth}{\centering\strut \raisebox{0.08ex}{\ding{55}}}\hspace{0.006\linewidth}\parbox[t]{0.10\linewidth}{\centering\strut -0.4112\%}\hspace{0.006\linewidth}\parbox[t]{0.115\linewidth}{\centering\strut -1.127467}\hspace{0.006\linewidth}\parbox[t]{0.671\linewidth}{num\_walkers\_per\_rank 320, stabilize\_freq 3, pop\_control\_freq 3}\par\smallskip
\noindent{\setlength{\fboxsep}{0pt}\colorbox{black!4}{\parbox{\linewidth}{\parbox[t]{0.05\linewidth}{\centering\strut 61}\hspace{0.006\linewidth}\parbox[t]{0.04\linewidth}{\centering\strut \raisebox{0.08ex}{\ding{55}}}\hspace{0.006\linewidth}\parbox[t]{0.10\linewidth}{\centering\strut +0.0592\%}\hspace{0.006\linewidth}\parbox[t]{0.115\linewidth}{\centering\strut -1.128134}\hspace{0.006\linewidth}\parbox[t]{0.671\linewidth}{num\_walkers\_per\_rank 384, stabilize\_freq 3, pop\_control\_freq 3}}}}\par\smallskip
\noindent\parbox[t]{0.05\linewidth}{\centering\strut 62}\hspace{0.006\linewidth}\parbox[t]{0.04\linewidth}{\centering\strut \raisebox{0.08ex}{\ding{55}}}\hspace{0.006\linewidth}\parbox[t]{0.10\linewidth}{\centering\strut +0.0128\%}\hspace{0.006\linewidth}\parbox[t]{0.115\linewidth}{\centering\strut -1.128278}\hspace{0.006\linewidth}\parbox[t]{0.671\linewidth}{num\_walkers\_per\_rank 288, stabilize\_freq 2, pop\_control\_freq 2}\par\smallskip
\noindent{\setlength{\fboxsep}{0pt}\colorbox{black!4}{\parbox{\linewidth}{\parbox[t]{0.05\linewidth}{\centering\strut 63}\hspace{0.006\linewidth}\parbox[t]{0.04\linewidth}{\centering\strut \raisebox{0.08ex}{\ding{55}}}\hspace{0.006\linewidth}\parbox[t]{0.10\linewidth}{\centering\strut -0.1149\%}\hspace{0.006\linewidth}\parbox[t]{0.115\linewidth}{\centering\strut -1.126982}\hspace{0.006\linewidth}\parbox[t]{0.671\linewidth}{num\_walkers\_per\_rank 224, stabilize\_freq 2, pop\_control\_freq 2}}}}\par\smallskip
\noindent\parbox[t]{0.05\linewidth}{\centering\strut 64}\hspace{0.006\linewidth}\parbox[t]{0.04\linewidth}{\centering\strut \raisebox{0.08ex}{\ding{55}}}\hspace{0.006\linewidth}\parbox[t]{0.10\linewidth}{\centering\strut +0.4564\%}\hspace{0.006\linewidth}\parbox[t]{0.115\linewidth}{\centering\strut -1.132126}\hspace{0.006\linewidth}\parbox[t]{0.671\linewidth}{num\_walkers\_per\_rank 224, num\_blocks 30}\par\smallskip
\noindent{\setlength{\fboxsep}{0pt}\colorbox{black!4}{\parbox{\linewidth}{\parbox[t]{0.05\linewidth}{\centering\strut 65}\hspace{0.006\linewidth}\parbox[t]{0.04\linewidth}{\centering\strut \raisebox{0.08ex}{\ding{55}}}\hspace{0.006\linewidth}\parbox[t]{0.10\linewidth}{\centering\strut -0.4116\%}\hspace{0.006\linewidth}\parbox[t]{0.115\linewidth}{\centering\strut -1.127467}\hspace{0.006\linewidth}\parbox[t]{0.671\linewidth}{num\_walkers\_per\_rank 320, stabilize\_freq 2, pop\_control\_freq 2}}}}\par\smallskip
\noindent\parbox[t]{0.05\linewidth}{\centering\strut 66}\hspace{0.006\linewidth}\parbox[t]{0.04\linewidth}{\centering\strut \raisebox{0.08ex}{\ding{55}}}\hspace{0.006\linewidth}\parbox[t]{0.10\linewidth}{\centering\strut -0.0302\%}\hspace{0.006\linewidth}\parbox[t]{0.115\linewidth}{\centering\strut -1.127126}\hspace{0.006\linewidth}\parbox[t]{0.671\linewidth}{num\_walkers\_per\_rank 192, stabilize\_freq 2, pop\_control\_freq 2}\par\smallskip
\noindent{\setlength{\fboxsep}{0pt}\colorbox{black!4}{\parbox{\linewidth}{\parbox[t]{0.05\linewidth}{\centering\strut 67}\hspace{0.006\linewidth}\parbox[t]{0.04\linewidth}{\centering\strut \raisebox{0.08ex}{\ding{55}}}\hspace{0.006\linewidth}\parbox[t]{0.10\linewidth}{\centering\strut +0.2550\%}\hspace{0.006\linewidth}\parbox[t]{0.115\linewidth}{\centering\strut -1.130000}\hspace{0.006\linewidth}\parbox[t]{0.671\linewidth}{num\_walkers\_per\_rank 192, num\_blocks 30}}}}\par\smallskip
\noindent\parbox[t]{0.05\linewidth}{\centering\strut 68}\hspace{0.006\linewidth}\parbox[t]{0.04\linewidth}{\centering\strut \raisebox{0.08ex}{\ding{55}}}\hspace{0.006\linewidth}\parbox[t]{0.10\linewidth}{\centering\strut +0.0985\%}\hspace{0.006\linewidth}\parbox[t]{0.115\linewidth}{\centering\strut -1.131113}\hspace{0.006\linewidth}\parbox[t]{0.671\linewidth}{timestep 0.003, stabilize\_freq 5, pop\_control\_freq 5}\par\smallskip
\noindent{\setlength{\fboxsep}{0pt}\colorbox{black!4}{\parbox{\linewidth}{\parbox[t]{0.05\linewidth}{\centering\strut 69}\hspace{0.006\linewidth}\parbox[t]{0.04\linewidth}{\centering\strut \raisebox{0.08ex}{\ding{55}}}\hspace{0.006\linewidth}\parbox[t]{0.10\linewidth}{\centering\strut -0.2634\%}\hspace{0.006\linewidth}\parbox[t]{0.115\linewidth}{\centering\strut -1.128134}\hspace{0.006\linewidth}\parbox[t]{0.671\linewidth}{num\_walkers\_per\_rank 384, stabilize\_freq 2, pop\_control\_freq 2}}}}\par\smallskip
\noindent\parbox[t]{0.05\linewidth}{\centering\strut 70}\hspace{0.006\linewidth}\parbox[t]{0.04\linewidth}{\centering\strut \raisebox{0.08ex}{\ding{55}}}\hspace{0.006\linewidth}\parbox[t]{0.10\linewidth}{\centering\strut +0.0523\%}\hspace{0.006\linewidth}\parbox[t]{0.115\linewidth}{\centering\strut -1.128725}\hspace{0.006\linewidth}\parbox[t]{0.671\linewidth}{num\_walkers\_per\_rank 160, stabilize\_freq 2, pop\_control\_freq 2}\par\smallskip
\noindent{\setlength{\fboxsep}{0pt}\colorbox{black!4}{\parbox{\linewidth}{\parbox[t]{0.05\linewidth}{\centering\strut 71}\hspace{0.006\linewidth}\parbox[t]{0.04\linewidth}{\centering\strut \raisebox{0.08ex}{\ding{55}}}\hspace{0.006\linewidth}\parbox[t]{0.10\linewidth}{\centering\strut +0.0043\%}\hspace{0.006\linewidth}\parbox[t]{0.115\linewidth}{\centering\strut -1.128774}\hspace{0.006\linewidth}\parbox[t]{0.671\linewidth}{num\_walkers\_per\_rank 160, num\_blocks 30}}}}\par\smallskip
\noindent\parbox[t]{0.05\linewidth}{\centering\strut 72}\hspace{0.006\linewidth}\parbox[t]{0.04\linewidth}{\centering\strut \raisebox{0.08ex}{\ding{55}}}\hspace{0.006\linewidth}\parbox[t]{0.10\linewidth}{\centering\strut +0.1748\%}\hspace{0.006\linewidth}\parbox[t]{0.115\linewidth}{\centering\strut -1.130747}\hspace{0.006\linewidth}\parbox[t]{0.671\linewidth}{num\_blocks 30, stabilize\_freq 5, pop\_control\_freq 5}\par\smallskip
\noindent{\setlength{\fboxsep}{0pt}\colorbox{black!4}{\parbox{\linewidth}{\parbox[t]{0.05\linewidth}{\centering\strut 73}\hspace{0.006\linewidth}\parbox[t]{0.04\linewidth}{\centering\strut \raisebox{0.08ex}{\ding{55}}}\hspace{0.006\linewidth}\parbox[t]{0.10\linewidth}{\centering\strut +0.0324\%}\hspace{0.006\linewidth}\parbox[t]{0.115\linewidth}{\centering\strut -1.131113}\hspace{0.006\linewidth}\parbox[t]{0.671\linewidth}{timestep 0.003, stabilize\_freq 2, pop\_control\_freq 2}}}}\par\smallskip
\noindent\parbox[t]{0.05\linewidth}{\centering\strut 74}\hspace{0.006\linewidth}\parbox[t]{0.04\linewidth}{\centering\strut \raisebox{0.08ex}{\ding{55}}}\hspace{0.006\linewidth}\parbox[t]{0.10\linewidth}{\centering\strut -0.0324\%}\hspace{0.006\linewidth}\parbox[t]{0.115\linewidth}{\centering\strut -1.130747}\hspace{0.006\linewidth}\parbox[t]{0.671\linewidth}{num\_blocks 30, stabilize\_freq 3, pop\_control\_freq 3}\par\smallskip
\noindent{\setlength{\fboxsep}{0pt}\colorbox{black!4}{\parbox{\linewidth}{\parbox[t]{0.05\linewidth}{\centering\strut 75}\hspace{0.006\linewidth}\parbox[t]{0.04\linewidth}{\centering\strut \raisebox{0.08ex}{\ding{55}}}\hspace{0.006\linewidth}\parbox[t]{0.10\linewidth}{\centering\strut +0.0000\%}\hspace{0.006\linewidth}\parbox[t]{0.115\linewidth}{\centering\strut -1.130747}\hspace{0.006\linewidth}\parbox[t]{0.671\linewidth}{num\_blocks 30, stabilize\_freq 2, pop\_control\_freq 2}}}}\par\smallskip
\noindent\parbox[t]{0.05\linewidth}{\centering\strut 76}\hspace{0.006\linewidth}\parbox[t]{0.04\linewidth}{\centering\strut \raisebox{0.08ex}{\ding{55}}}\hspace{0.006\linewidth}\parbox[t]{0.10\linewidth}{\centering\strut -0.2995\%}\hspace{0.006\linewidth}\parbox[t]{0.115\linewidth}{\centering\strut -1.127360}\hspace{0.006\linewidth}\parbox[t]{0.671\linewidth}{num\_walkers\_per\_rank 288, num\_steps\_per\_block 80, timestep 0.0025}\par\smallskip
\noindent{\setlength{\fboxsep}{0pt}\colorbox{black!4}{\parbox{\linewidth}{\parbox[t]{0.05\linewidth}{\centering\strut 77}\hspace{0.006\linewidth}\parbox[t]{0.04\linewidth}{\centering\strut \raisebox{0.08ex}{\ding{55}}}\hspace{0.006\linewidth}\parbox[t]{0.10\linewidth}{\centering\strut -0.3627\%}\hspace{0.006\linewidth}\parbox[t]{0.115\linewidth}{\centering\strut -1.123271}\hspace{0.006\linewidth}\parbox[t]{0.671\linewidth}{num\_walkers\_per\_rank 224, num\_steps\_per\_block 80, timestep 0.0025}}}}\par\smallskip
\noindent\parbox[t]{0.05\linewidth}{\centering\strut 78}\hspace{0.006\linewidth}\parbox[t]{0.04\linewidth}{\centering\strut \raisebox{0.08ex}{\ding{55}}}\hspace{0.006\linewidth}\parbox[t]{0.10\linewidth}{\centering\strut +0.5359\%}\hspace{0.006\linewidth}\parbox[t]{0.115\linewidth}{\centering\strut -1.129290}\hspace{0.006\linewidth}\parbox[t]{0.671\linewidth}{num\_walkers\_per\_rank 288, num\_blocks 24, timestep 0.001667}\par\smallskip
\noindent{\setlength{\fboxsep}{0pt}\colorbox{black!4}{\parbox{\linewidth}{\parbox[t]{0.05\linewidth}{\centering\strut 79}\hspace{0.006\linewidth}\parbox[t]{0.04\linewidth}{\centering\strut \raisebox{0.08ex}{\ding{55}}}\hspace{0.006\linewidth}\parbox[t]{0.10\linewidth}{\centering\strut +0.1255\%}\hspace{0.006\linewidth}\parbox[t]{0.115\linewidth}{\centering\strut -1.130708}\hspace{0.006\linewidth}\parbox[t]{0.671\linewidth}{num\_walkers\_per\_rank 224, num\_blocks 24, timestep 0.001667}}}}\par\smallskip
\noindent\parbox[t]{0.05\linewidth}{\centering\strut 80}\hspace{0.006\linewidth}\parbox[t]{0.04\linewidth}{\centering\strut \raisebox{0.08ex}{\ding{51}}}\hspace{0.006\linewidth}\parbox[t]{0.10\linewidth}{\centering\strut +0.2972\%}\hspace{0.006\linewidth}\parbox[t]{0.115\linewidth}{\centering\strut -1.134068}\hspace{0.006\linewidth}\parbox[t]{0.671\linewidth}{num\_walkers\_per\_rank 320, num\_steps\_per\_block 80, timestep 0.0025}\par\smallskip
\noindent{\setlength{\fboxsep}{0pt}\colorbox{black!4}{\parbox{\linewidth}{\parbox[t]{0.05\linewidth}{\centering\strut 81}\hspace{0.006\linewidth}\parbox[t]{0.04\linewidth}{\centering\strut \raisebox{0.08ex}{\ding{55}}}\hspace{0.006\linewidth}\parbox[t]{0.10\linewidth}{\centering\strut -0.2915\%}\hspace{0.006\linewidth}\parbox[t]{0.115\linewidth}{\centering\strut -1.130763}\hspace{0.006\linewidth}\parbox[t]{0.671\linewidth}{timestep 0.003125}}}}\par\smallskip
\noindent\parbox[t]{0.05\linewidth}{\centering\strut 82}\hspace{0.006\linewidth}\parbox[t]{0.04\linewidth}{\centering\strut \raisebox{0.08ex}{\ding{55}}}\hspace{0.006\linewidth}\parbox[t]{0.10\linewidth}{\centering\strut -0.1501\%}\hspace{0.006\linewidth}\parbox[t]{0.115\linewidth}{\centering\strut -1.129065}\hspace{0.006\linewidth}\parbox[t]{0.671\linewidth}{stabilize\_freq 5, pop\_control\_freq 5}\par\smallskip
\noindent{\setlength{\fboxsep}{0pt}\colorbox{black!4}{\parbox{\linewidth}{\parbox[t]{0.05\linewidth}{\centering\strut 83}\hspace{0.006\linewidth}\parbox[t]{0.04\linewidth}{\centering\strut \raisebox{0.08ex}{\ding{55}}}\hspace{0.006\linewidth}\parbox[t]{0.10\linewidth}{\centering\strut -0.3681\%}\hspace{0.006\linewidth}\parbox[t]{0.115\linewidth}{\centering\strut -1.124909}\hspace{0.006\linewidth}\parbox[t]{0.671\linewidth}{timestep 0.00375}}}}\par\smallskip
\noindent\parbox[t]{0.05\linewidth}{\centering\strut 84}\hspace{0.006\linewidth}\parbox[t]{0.04\linewidth}{\centering\strut \raisebox{0.08ex}{\ding{55}}}\hspace{0.006\linewidth}\parbox[t]{0.10\linewidth}{\centering\strut +0.2953\%}\hspace{0.006\linewidth}\parbox[t]{0.115\linewidth}{\centering\strut -1.128230}\hspace{0.006\linewidth}\parbox[t]{0.671\linewidth}{stabilize\_freq 3, pop\_control\_freq 3}\par\smallskip
\noindent{\setlength{\fboxsep}{0pt}\colorbox{black!4}{\parbox{\linewidth}{\parbox[t]{0.05\linewidth}{\centering\strut 85}\hspace{0.006\linewidth}\parbox[t]{0.04\linewidth}{\centering\strut \raisebox{0.08ex}{\ding{55}}}\hspace{0.006\linewidth}\parbox[t]{0.10\linewidth}{\centering\strut +0.0719\%}\hspace{0.006\linewidth}\parbox[t]{0.115\linewidth}{\centering\strut -1.129042}\hspace{0.006\linewidth}\parbox[t]{0.671\linewidth}{num\_blocks 30}}}}\par\smallskip
\noindent\parbox[t]{0.05\linewidth}{\centering\strut 86}\hspace{0.006\linewidth}\parbox[t]{0.04\linewidth}{\centering\strut \raisebox{0.08ex}{\ding{55}}}\hspace{0.006\linewidth}\parbox[t]{0.10\linewidth}{\centering\strut -0.1395\%}\hspace{0.006\linewidth}\parbox[t]{0.115\linewidth}{\centering\strut -1.127467}\hspace{0.006\linewidth}\parbox[t]{0.671\linewidth}{num\_steps\_per\_block 100, timestep 0.002}\par\smallskip
\noindent{\setlength{\fboxsep}{0pt}\colorbox{black!4}{\parbox{\linewidth}{\parbox[t]{0.05\linewidth}{\centering\strut 87}\hspace{0.006\linewidth}\parbox[t]{0.04\linewidth}{\centering\strut \raisebox{0.08ex}{\ding{55}}}\hspace{0.006\linewidth}\parbox[t]{0.10\linewidth}{\centering\strut +0.4431\%}\hspace{0.006\linewidth}\parbox[t]{0.115\linewidth}{\centering\strut -1.132462}\hspace{0.006\linewidth}\parbox[t]{0.671\linewidth}{num\_blocks 24, timestep 0.002083}}}}\par\smallskip
\noindent\parbox[t]{0.05\linewidth}{\centering\strut 88}\hspace{0.006\linewidth}\parbox[t]{0.04\linewidth}{\centering\strut \raisebox{0.08ex}{\ding{55}}}\hspace{0.006\linewidth}\parbox[t]{0.10\linewidth}{\centering\strut -0.5400\%}\hspace{0.006\linewidth}\parbox[t]{0.115\linewidth}{\centering\strut -1.126347}\hspace{0.006\linewidth}\parbox[t]{0.671\linewidth}{num\_steps\_per\_block 60, timestep 0.003333}\par\smallskip
\noindent{\setlength{\fboxsep}{0pt}\colorbox{black!4}{\parbox{\linewidth}{\parbox[t]{0.05\linewidth}{\centering\strut 89}\hspace{0.006\linewidth}\parbox[t]{0.04\linewidth}{\centering\strut \raisebox{0.08ex}{\ding{55}}}\hspace{0.006\linewidth}\parbox[t]{0.10\linewidth}{\centering\strut -0.0006\%}\hspace{0.006\linewidth}\parbox[t]{0.115\linewidth}{\centering\strut -1.126340}\hspace{0.006\linewidth}\parbox[t]{0.671\linewidth}{num\_blocks 30, timestep 0.001667}}}}\par\smallskip
\noindent\parbox[t]{0.05\linewidth}{\centering\strut 90}\hspace{0.006\linewidth}\parbox[t]{0.04\linewidth}{\centering\strut \raisebox{0.08ex}{\ding{55}}}\hspace{0.006\linewidth}\parbox[t]{0.10\linewidth}{\centering\strut +0.1123\%}\hspace{0.006\linewidth}\parbox[t]{0.115\linewidth}{\centering\strut -1.127605}\hspace{0.006\linewidth}\parbox[t]{0.671\linewidth}{num\_walkers\_per\_rank 288, timestep 0.003125}\par\smallskip
\noindent{\setlength{\fboxsep}{0pt}\colorbox{black!4}{\parbox{\linewidth}{\parbox[t]{0.05\linewidth}{\centering\strut 91}\hspace{0.006\linewidth}\parbox[t]{0.04\linewidth}{\centering\strut \raisebox{0.08ex}{\ding{55}}}\hspace{0.006\linewidth}\parbox[t]{0.10\linewidth}{\centering\strut +0.2270\%}\hspace{0.006\linewidth}\parbox[t]{0.115\linewidth}{\centering\strut -1.130165}\hspace{0.006\linewidth}\parbox[t]{0.671\linewidth}{num\_walkers\_per\_rank 384, timestep 0.003125}}}}\par\smallskip
\noindent\parbox[t]{0.05\linewidth}{\centering\strut 92}\hspace{0.006\linewidth}\parbox[t]{0.04\linewidth}{\centering\strut \raisebox{0.08ex}{\ding{55}}}\hspace{0.006\linewidth}\parbox[t]{0.10\linewidth}{\centering\strut -3.1674\%}\hspace{0.006\linewidth}\parbox[t]{0.115\linewidth}{\centering\strut -1.094368}\hspace{0.006\linewidth}\parbox[t]{0.671\linewidth}{num\_blocks 24, timestep 0.002604}\par\smallskip
\noindent{\setlength{\fboxsep}{0pt}\colorbox{black!4}{\parbox{\linewidth}{\parbox[t]{0.05\linewidth}{\centering\strut 93}\hspace{0.006\linewidth}\parbox[t]{0.04\linewidth}{\centering\strut \raisebox{0.08ex}{\ding{55}}}\hspace{0.006\linewidth}\parbox[t]{0.10\linewidth}{\centering\strut +1.0004\%}\hspace{0.006\linewidth}\parbox[t]{0.115\linewidth}{\centering\strut -1.105316}\hspace{0.006\linewidth}\parbox[t]{0.671\linewidth}{num\_steps\_per\_block 60, timestep 0.004167}}}}\par\smallskip
\noindent\parbox[t]{0.05\linewidth}{\centering\strut 94}\hspace{0.006\linewidth}\parbox[t]{0.04\linewidth}{\centering\strut \raisebox{0.08ex}{\ding{55}}}\hspace{0.006\linewidth}\parbox[t]{0.10\linewidth}{\centering\strut +1.6736\%}\hspace{0.006\linewidth}\parbox[t]{0.115\linewidth}{\centering\strut -1.123815}\hspace{0.006\linewidth}\parbox[t]{0.671\linewidth}{num\_walkers\_per\_rank 224, timestep 0.003125}\par\smallskip
\noindent{\setlength{\fboxsep}{0pt}\colorbox{black!4}{\parbox{\linewidth}{\parbox[t]{0.05\linewidth}{\centering\strut 95}\hspace{0.006\linewidth}\parbox[t]{0.04\linewidth}{\centering\strut \raisebox{0.08ex}{\ding{55}}}\hspace{0.006\linewidth}\parbox[t]{0.10\linewidth}{\centering\strut +0.3154\%}\hspace{0.006\linewidth}\parbox[t]{0.115\linewidth}{\centering\strut -1.127360}\hspace{0.006\linewidth}\parbox[t]{0.671\linewidth}{num\_walkers\_per\_rank 288, stabilize\_freq 5, pop\_control\_freq 5}}}}\par\smallskip
\noindent\parbox[t]{0.05\linewidth}{\centering\strut 96}\hspace{0.006\linewidth}\parbox[t]{0.04\linewidth}{\centering\strut \raisebox{0.08ex}{\ding{55}}}\hspace{0.006\linewidth}\parbox[t]{0.10\linewidth}{\centering\strut -0.1150\%}\hspace{0.006\linewidth}\parbox[t]{0.115\linewidth}{\centering\strut -1.126064}\hspace{0.006\linewidth}\parbox[t]{0.671\linewidth}{num\_walkers\_per\_rank 192, timestep 0.003125}\par\smallskip
\noindent{\setlength{\fboxsep}{0pt}\colorbox{black!4}{\parbox{\linewidth}{\parbox[t]{0.05\linewidth}{\centering\strut 97}\hspace{0.006\linewidth}\parbox[t]{0.04\linewidth}{\centering\strut \raisebox{0.08ex}{\ding{55}}}\hspace{0.006\linewidth}\parbox[t]{0.10\linewidth}{\centering\strut +0.4631\%}\hspace{0.006\linewidth}\parbox[t]{0.115\linewidth}{\centering\strut -1.131279}\hspace{0.006\linewidth}\parbox[t]{0.671\linewidth}{num\_walkers\_per\_rank 384, stabilize\_freq 5, pop\_control\_freq 5}}}}\par\smallskip
\noindent\parbox[t]{0.05\linewidth}{\centering\strut 98}\hspace{0.006\linewidth}\parbox[t]{0.04\linewidth}{\centering\strut \raisebox{0.08ex}{\ding{55}}}\hspace{0.006\linewidth}\parbox[t]{0.10\linewidth}{\centering\strut -0.4018\%}\hspace{0.006\linewidth}\parbox[t]{0.115\linewidth}{\centering\strut -1.126734}\hspace{0.006\linewidth}\parbox[t]{0.671\linewidth}{num\_walkers\_per\_rank 256, stabilize\_freq 5, pop\_control\_freq 5}\par\smallskip
\noindent{\setlength{\fboxsep}{0pt}\colorbox{black!4}{\parbox{\linewidth}{\parbox[t]{0.05\linewidth}{\centering\strut 99}\hspace{0.006\linewidth}\parbox[t]{0.04\linewidth}{\centering\strut \raisebox{0.08ex}{\ding{55}}}\hspace{0.006\linewidth}\parbox[t]{0.10\linewidth}{\centering\strut +0.1293\%}\hspace{0.006\linewidth}\parbox[t]{0.115\linewidth}{\centering\strut -1.128191}\hspace{0.006\linewidth}\parbox[t]{0.671\linewidth}{num\_blocks 30, timestep 0.002083}}}}\par\smallskip
\noindent\parbox[t]{0.05\linewidth}{\centering\strut 100}\hspace{0.006\linewidth}\parbox[t]{0.04\linewidth}{\centering\strut \raisebox{0.08ex}{\ding{55}}}\hspace{0.006\linewidth}\parbox[t]{0.10\linewidth}{\centering\strut -0.4966\%}\hspace{0.006\linewidth}\parbox[t]{0.115\linewidth}{\centering\strut -1.122588}\hspace{0.006\linewidth}\parbox[t]{0.671\linewidth}{num\_walkers\_per\_rank 160, timestep 0.003125}\par\smallskip
\noindent\rule{\linewidth}{0.5pt}
\par\medskip
\refstepcounter{table}
\label{tab:supp_afqmc_lih}
\noindent\textbf{Table \thetable.} AFQMC mutation log for LiH full 100-iteration campaign. Relative gain is measured against the immediately previous scored iteration using the live AFQMC score.
\par\medskip
\noindent\rule{\linewidth}{0.5pt}
\noindent{\setlength{\fboxsep}{0pt}\colorbox{black!4}{\parbox{\linewidth}{\parbox[c][2.2em][c]{0.05\linewidth}{\centering \textbf{Iter.}}\hspace{0.006\linewidth}\parbox[c][2.2em][c]{0.04\linewidth}{\centering \textbf{A/R}}\hspace{0.006\linewidth}\parbox[c][2.2em][c]{0.10\linewidth}{\centering \textbf{Relative gain}}\hspace{0.006\linewidth}\parbox[c][2.2em][c]{0.115\linewidth}{\centering \textbf{Score}}\hspace{0.006\linewidth}\parbox[c][2.2em][c]{0.671\linewidth}{\centering \textbf{Mutation summary}}}}}\par\smallskip
\noindent\rule{\linewidth}{0.5pt}
\noindent{\setlength{\fboxsep}{0pt}\colorbox{black!4}{\parbox{\linewidth}{\parbox[t]{0.05\linewidth}{\centering\strut 1}\hspace{0.006\linewidth}\parbox[t]{0.04\linewidth}{\centering\strut \raisebox{0.08ex}{\ding{55}}}\hspace{0.006\linewidth}\parbox[t]{0.10\linewidth}{\centering\strut ---}\hspace{0.006\linewidth}\parbox[t]{0.115\linewidth}{\centering\strut -7.865138}\hspace{0.006\linewidth}\parbox[t]{0.671\linewidth}{increase walkers to 96 and blocks to 120}}}}\par\smallskip
\noindent\parbox[t]{0.05\linewidth}{\centering\strut 2}\hspace{0.006\linewidth}\parbox[t]{0.04\linewidth}{\centering\strut \raisebox{0.08ex}{\ding{51}}}\hspace{0.006\linewidth}\parbox[t]{0.10\linewidth}{\centering\strut +0.1273\%}\hspace{0.006\linewidth}\parbox[t]{0.115\linewidth}{\centering\strut -7.875148}\hspace{0.006\linewidth}\parbox[t]{0.671\linewidth}{increase walkers to 128}\par\smallskip
\noindent{\setlength{\fboxsep}{0pt}\colorbox{black!4}{\parbox{\linewidth}{\parbox[t]{0.05\linewidth}{\centering\strut 3}\hspace{0.006\linewidth}\parbox[t]{0.04\linewidth}{\centering\strut \raisebox{0.08ex}{\ding{55}}}\hspace{0.006\linewidth}\parbox[t]{0.10\linewidth}{\centering\strut -0.0384\%}\hspace{0.006\linewidth}\parbox[t]{0.115\linewidth}{\centering\strut -7.872126}\hspace{0.006\linewidth}\parbox[t]{0.671\linewidth}{increase walkers to 256}}}}\par\smallskip
\noindent\parbox[t]{0.05\linewidth}{\centering\strut 4}\hspace{0.006\linewidth}\parbox[t]{0.04\linewidth}{\centering\strut \raisebox{0.08ex}{\ding{55}}}\hspace{0.006\linewidth}\parbox[t]{0.10\linewidth}{\centering\strut -0.0148\%}\hspace{0.006\linewidth}\parbox[t]{0.115\linewidth}{\centering\strut -7.870958}\hspace{0.006\linewidth}\parbox[t]{0.671\linewidth}{increase walkers to 192}\par\smallskip
\noindent{\setlength{\fboxsep}{0pt}\colorbox{black!4}{\parbox{\linewidth}{\parbox[t]{0.05\linewidth}{\centering\strut 5}\hspace{0.006\linewidth}\parbox[t]{0.04\linewidth}{\centering\strut \raisebox{0.08ex}{\ding{55}}}\hspace{0.006\linewidth}\parbox[t]{0.10\linewidth}{\centering\strut +0.0334\%}\hspace{0.006\linewidth}\parbox[t]{0.115\linewidth}{\centering\strut -7.873585}\hspace{0.006\linewidth}\parbox[t]{0.671\linewidth}{set timestep 0.004 and blocks 50}}}}\par\smallskip
\noindent\parbox[t]{0.05\linewidth}{\centering\strut 6}\hspace{0.006\linewidth}\parbox[t]{0.04\linewidth}{\centering\strut \raisebox{0.08ex}{\ding{55}}}\hspace{0.006\linewidth}\parbox[t]{0.10\linewidth}{\centering\strut +0.0060\%}\hspace{0.006\linewidth}\parbox[t]{0.115\linewidth}{\centering\strut -7.874055}\hspace{0.006\linewidth}\parbox[t]{0.671\linewidth}{increase blocks to 50}\par\smallskip
\noindent{\setlength{\fboxsep}{0pt}\colorbox{black!4}{\parbox{\linewidth}{\parbox[t]{0.05\linewidth}{\centering\strut 7}\hspace{0.006\linewidth}\parbox[t]{0.04\linewidth}{\centering\strut \raisebox{0.08ex}{\ding{51}}}\hspace{0.006\linewidth}\parbox[t]{0.10\linewidth}{\centering\strut +0.0139\%}\hspace{0.006\linewidth}\parbox[t]{0.115\linewidth}{\centering\strut -7.875148}\hspace{0.006\linewidth}\parbox[t]{0.671\linewidth}{set stabilize\_freq 3 and pop\_control\_freq 3}}}}\par\smallskip
\noindent\parbox[t]{0.05\linewidth}{\centering\strut 8}\hspace{0.006\linewidth}\parbox[t]{0.04\linewidth}{\centering\strut \raisebox{0.08ex}{\ding{55}}}\hspace{0.006\linewidth}\parbox[t]{0.10\linewidth}{\centering\strut -0.0000\%}\hspace{0.006\linewidth}\parbox[t]{0.115\linewidth}{\centering\strut -7.875148}\hspace{0.006\linewidth}\parbox[t]{0.671\linewidth}{set stabilize\_freq 1 and pop\_control\_freq 1}\par\smallskip
\noindent{\setlength{\fboxsep}{0pt}\colorbox{black!4}{\parbox{\linewidth}{\parbox[t]{0.05\linewidth}{\centering\strut 9}\hspace{0.006\linewidth}\parbox[t]{0.04\linewidth}{\centering\strut \raisebox{0.08ex}{\ding{55}}}\hspace{0.006\linewidth}\parbox[t]{0.10\linewidth}{\centering\strut -0.0004\%}\hspace{0.006\linewidth}\parbox[t]{0.115\linewidth}{\centering\strut -7.875118}\hspace{0.006\linewidth}\parbox[t]{0.671\linewidth}{set steps\_per\_block 20 and blocks 50}}}}\par\smallskip
\noindent\parbox[t]{0.05\linewidth}{\centering\strut 10}\hspace{0.006\linewidth}\parbox[t]{0.04\linewidth}{\centering\strut \raisebox{0.08ex}{\ding{55}}}\hspace{0.006\linewidth}\parbox[t]{0.10\linewidth}{\centering\strut +0.0004\%}\hspace{0.006\linewidth}\parbox[t]{0.115\linewidth}{\centering\strut -7.875148}\hspace{0.006\linewidth}\parbox[t]{0.671\linewidth}{decrease chol\_cut to 1e-6}\par\smallskip
\noindent{\setlength{\fboxsep}{0pt}\colorbox{black!4}{\parbox{\linewidth}{\parbox[t]{0.05\linewidth}{\centering\strut 11}\hspace{0.006\linewidth}\parbox[t]{0.04\linewidth}{\centering\strut \raisebox{0.08ex}{\ding{51}}}\hspace{0.006\linewidth}\parbox[t]{0.10\linewidth}{\centering\strut +0.0000\%}\hspace{0.006\linewidth}\parbox[t]{0.115\linewidth}{\centering\strut -7.875148}\hspace{0.006\linewidth}\parbox[t]{0.671\linewidth}{switch trial to uhf}}}}\par\smallskip
\noindent\parbox[t]{0.05\linewidth}{\centering\strut 12}\hspace{0.006\linewidth}\parbox[t]{0.04\linewidth}{\centering\strut \raisebox{0.08ex}{\ding{51}}}\hspace{0.006\linewidth}\parbox[t]{0.10\linewidth}{\centering\strut +0.0000\%}\hspace{0.006\linewidth}\parbox[t]{0.115\linewidth}{\centering\strut -7.875149}\hspace{0.006\linewidth}\parbox[t]{0.671\linewidth}{switch trial to uhf and init\_guess to atom}\par\smallskip
\noindent{\setlength{\fboxsep}{0pt}\colorbox{black!4}{\parbox{\linewidth}{\parbox[t]{0.05\linewidth}{\centering\strut 13}\hspace{0.006\linewidth}\parbox[t]{0.04\linewidth}{\centering\strut \raisebox{0.08ex}{\ding{55}}}\hspace{0.006\linewidth}\parbox[t]{0.10\linewidth}{\centering\strut -0.0000\%}\hspace{0.006\linewidth}\parbox[t]{0.115\linewidth}{\centering\strut -7.875148}\hspace{0.006\linewidth}\parbox[t]{0.671\linewidth}{switch trial to uhf and init\_guess to huckel}}}}\par\smallskip
\noindent\parbox[t]{0.05\linewidth}{\centering\strut 14}\hspace{0.006\linewidth}\parbox[t]{0.04\linewidth}{\centering\strut \raisebox{0.08ex}{\ding{55}}}\hspace{0.006\linewidth}\parbox[t]{0.10\linewidth}{\centering\strut -0.0000\%}\hspace{0.006\linewidth}\parbox[t]{0.115\linewidth}{\centering\strut -7.875148}\hspace{0.006\linewidth}\parbox[t]{0.671\linewidth}{switch trial to uhf and init\_guess to mod\_huckel}\par\smallskip
\noindent{\setlength{\fboxsep}{0pt}\colorbox{black!4}{\parbox{\linewidth}{\parbox[t]{0.05\linewidth}{\centering\strut 15}\hspace{0.006\linewidth}\parbox[t]{0.04\linewidth}{\centering\strut \raisebox{0.08ex}{\ding{55}}}\hspace{0.006\linewidth}\parbox[t]{0.10\linewidth}{\centering\strut +0.0000\%}\hspace{0.006\linewidth}\parbox[t]{0.115\linewidth}{\centering\strut -7.875148}\hspace{0.006\linewidth}\parbox[t]{0.671\linewidth}{switch trial to uhf and init\_guess to 1e}}}}\par\smallskip
\noindent\parbox[t]{0.05\linewidth}{\centering\strut 16}\hspace{0.006\linewidth}\parbox[t]{0.04\linewidth}{\centering\strut \raisebox{0.08ex}{\ding{55}}}\hspace{0.006\linewidth}\parbox[t]{0.10\linewidth}{\centering\strut -0.0360\%}\hspace{0.006\linewidth}\parbox[t]{0.115\linewidth}{\centering\strut -7.872313}\hspace{0.006\linewidth}\parbox[t]{0.671\linewidth}{set timestep 0.01 steps\_per\_block 20 and blocks 25}\par\smallskip
\noindent{\setlength{\fboxsep}{0pt}\colorbox{black!4}{\parbox{\linewidth}{\parbox[t]{0.05\linewidth}{\centering\strut 17}\hspace{0.006\linewidth}\parbox[t]{0.04\linewidth}{\centering\strut \raisebox{0.08ex}{\ding{55}}}\hspace{0.006\linewidth}\parbox[t]{0.10\linewidth}{\centering\strut +0.0113\%}\hspace{0.006\linewidth}\parbox[t]{0.115\linewidth}{\centering\strut -7.873204}\hspace{0.006\linewidth}\parbox[t]{0.671\linewidth}{decrease blocks to 30}}}}\par\smallskip
\noindent\parbox[t]{0.05\linewidth}{\centering\strut 18}\hspace{0.006\linewidth}\parbox[t]{0.04\linewidth}{\centering\strut \raisebox{0.08ex}{\ding{55}}}\hspace{0.006\linewidth}\parbox[t]{0.10\linewidth}{\centering\strut +0.0118\%}\hspace{0.006\linewidth}\parbox[t]{0.115\linewidth}{\centering\strut -7.874136}\hspace{0.006\linewidth}\parbox[t]{0.671\linewidth}{increase blocks to 42}\par\smallskip
\noindent{\setlength{\fboxsep}{0pt}\colorbox{black!4}{\parbox{\linewidth}{\parbox[t]{0.05\linewidth}{\centering\strut 19}\hspace{0.006\linewidth}\parbox[t]{0.04\linewidth}{\centering\strut \raisebox{0.08ex}{\ding{51}}}\hspace{0.006\linewidth}\parbox[t]{0.10\linewidth}{\centering\strut +0.0149\%}\hspace{0.006\linewidth}\parbox[t]{0.115\linewidth}{\centering\strut -7.875306}\hspace{0.006\linewidth}\parbox[t]{0.671\linewidth}{decrease blocks to 38}}}}\par\smallskip
\noindent\parbox[t]{0.05\linewidth}{\centering\strut 20}\hspace{0.006\linewidth}\parbox[t]{0.04\linewidth}{\centering\strut \raisebox{0.08ex}{\ding{55}}}\hspace{0.006\linewidth}\parbox[t]{0.10\linewidth}{\centering\strut -0.0071\%}\hspace{0.006\linewidth}\parbox[t]{0.115\linewidth}{\centering\strut -7.874746}\hspace{0.006\linewidth}\parbox[t]{0.671\linewidth}{decrease blocks to 36}\par\smallskip
\noindent{\setlength{\fboxsep}{0pt}\colorbox{black!4}{\parbox{\linewidth}{\parbox[t]{0.05\linewidth}{\centering\strut 21}\hspace{0.006\linewidth}\parbox[t]{0.04\linewidth}{\centering\strut \raisebox{0.08ex}{\ding{55}}}\hspace{0.006\linewidth}\parbox[t]{0.10\linewidth}{\centering\strut +0.0062\%}\hspace{0.006\linewidth}\parbox[t]{0.115\linewidth}{\centering\strut -7.875233}\hspace{0.006\linewidth}\parbox[t]{0.671\linewidth}{increase blocks to 39}}}}\par\smallskip
\noindent\parbox[t]{0.05\linewidth}{\centering\strut 22}\hspace{0.006\linewidth}\parbox[t]{0.04\linewidth}{\centering\strut \raisebox{0.08ex}{\ding{55}}}\hspace{0.006\linewidth}\parbox[t]{0.10\linewidth}{\centering\strut -0.2094\%}\hspace{0.006\linewidth}\parbox[t]{0.115\linewidth}{\centering\strut -7.858741}\hspace{0.006\linewidth}\parbox[t]{0.671\linewidth}{increase walkers to 144}\par\smallskip
\noindent{\setlength{\fboxsep}{0pt}\colorbox{black!4}{\parbox{\linewidth}{\parbox[t]{0.05\linewidth}{\centering\strut 23}\hspace{0.006\linewidth}\parbox[t]{0.04\linewidth}{\centering\strut \raisebox{0.08ex}{\ding{55}}}\hspace{0.006\linewidth}\parbox[t]{0.10\linewidth}{\centering\strut +0.1505\%}\hspace{0.006\linewidth}\parbox[t]{0.115\linewidth}{\centering\strut -7.870571}\hspace{0.006\linewidth}\parbox[t]{0.671\linewidth}{decrease walkers to 112}}}}\par\smallskip
\noindent\parbox[t]{0.05\linewidth}{\centering\strut 24}\hspace{0.006\linewidth}\parbox[t]{0.04\linewidth}{\centering\strut \raisebox{0.08ex}{\ding{55}}}\hspace{0.006\linewidth}\parbox[t]{0.10\linewidth}{\centering\strut +0.0549\%}\hspace{0.006\linewidth}\parbox[t]{0.115\linewidth}{\centering\strut -7.874892}\hspace{0.006\linewidth}\parbox[t]{0.671\linewidth}{decrease blocks to 37}\par\smallskip
\noindent{\setlength{\fboxsep}{0pt}\colorbox{black!4}{\parbox{\linewidth}{\parbox[t]{0.05\linewidth}{\centering\strut 25}\hspace{0.006\linewidth}\parbox[t]{0.04\linewidth}{\centering\strut \raisebox{0.08ex}{\ding{55}}}\hspace{0.006\linewidth}\parbox[t]{0.10\linewidth}{\centering\strut +0.0053\%}\hspace{0.006\linewidth}\parbox[t]{0.115\linewidth}{\centering\strut -7.875306}\hspace{0.006\linewidth}\parbox[t]{0.671\linewidth}{set stabilize\_freq 5 and pop\_control\_freq 5}}}}\par\smallskip
\noindent\parbox[t]{0.05\linewidth}{\centering\strut 26}\hspace{0.006\linewidth}\parbox[t]{0.04\linewidth}{\centering\strut \raisebox{0.08ex}{\ding{55}}}\hspace{0.006\linewidth}\parbox[t]{0.10\linewidth}{\centering\strut +0.0000\%}\hspace{0.006\linewidth}\parbox[t]{0.115\linewidth}{\centering\strut -7.875306}\hspace{0.006\linewidth}\parbox[t]{0.671\linewidth}{set stabilize\_freq 4 and pop\_control\_freq 4}\par\smallskip
\noindent{\setlength{\fboxsep}{0pt}\colorbox{black!4}{\parbox{\linewidth}{\parbox[t]{0.05\linewidth}{\centering\strut 27}\hspace{0.006\linewidth}\parbox[t]{0.04\linewidth}{\centering\strut \raisebox{0.08ex}{\ding{55}}}\hspace{0.006\linewidth}\parbox[t]{0.10\linewidth}{\centering\strut -0.0152\%}\hspace{0.006\linewidth}\parbox[t]{0.115\linewidth}{\centering\strut -7.874110}\hspace{0.006\linewidth}\parbox[t]{0.671\linewidth}{set steps\_per\_block 19 and blocks 50}}}}\par\smallskip
\noindent\parbox[t]{0.05\linewidth}{\centering\strut 28}\hspace{0.006\linewidth}\parbox[t]{0.04\linewidth}{\centering\strut \raisebox{0.08ex}{\ding{55}}}\hspace{0.006\linewidth}\parbox[t]{0.10\linewidth}{\centering\strut ---}\hspace{0.006\linewidth}\parbox[t]{0.115\linewidth}{\centering\strut ---}\hspace{0.006\linewidth}\parbox[t]{0.671\linewidth}{set steps\_per\_block 50 and blocks 19}\par\smallskip
\noindent{\setlength{\fboxsep}{0pt}\colorbox{black!4}{\parbox{\linewidth}{\parbox[t]{0.05\linewidth}{\centering\strut 29}\hspace{0.006\linewidth}\parbox[t]{0.04\linewidth}{\centering\strut \raisebox{0.08ex}{\ding{55}}}\hspace{0.006\linewidth}\parbox[t]{0.10\linewidth}{\centering\strut -0.0173\%}\hspace{0.006\linewidth}\parbox[t]{0.115\linewidth}{\centering\strut -7.872748}\hspace{0.006\linewidth}\parbox[t]{0.671\linewidth}{set steps\_per\_block 32 and blocks 30}}}}\par\smallskip
\noindent\parbox[t]{0.05\linewidth}{\centering\strut 30}\hspace{0.006\linewidth}\parbox[t]{0.04\linewidth}{\centering\strut \raisebox{0.08ex}{\ding{51}}}\hspace{0.006\linewidth}\parbox[t]{0.10\linewidth}{\centering\strut +0.0403\%}\hspace{0.006\linewidth}\parbox[t]{0.115\linewidth}{\centering\strut -7.875924}\hspace{0.006\linewidth}\parbox[t]{0.671\linewidth}{set steps\_per\_block 24 and blocks 40}\par\smallskip
\noindent{\setlength{\fboxsep}{0pt}\colorbox{black!4}{\parbox{\linewidth}{\parbox[t]{0.05\linewidth}{\centering\strut 31}\hspace{0.006\linewidth}\parbox[t]{0.04\linewidth}{\centering\strut \raisebox{0.08ex}{\ding{55}}}\hspace{0.006\linewidth}\parbox[t]{0.10\linewidth}{\centering\strut -0.0029\%}\hspace{0.006\linewidth}\parbox[t]{0.115\linewidth}{\centering\strut -7.875691}\hspace{0.006\linewidth}\parbox[t]{0.671\linewidth}{decrease steps\_per\_block to 23}}}}\par\smallskip
\noindent\parbox[t]{0.05\linewidth}{\centering\strut 32}\hspace{0.006\linewidth}\parbox[t]{0.04\linewidth}{\centering\strut \raisebox{0.08ex}{\ding{55}}}\hspace{0.006\linewidth}\parbox[t]{0.10\linewidth}{\centering\strut -0.0075\%}\hspace{0.006\linewidth}\parbox[t]{0.115\linewidth}{\centering\strut -7.875102}\hspace{0.006\linewidth}\parbox[t]{0.671\linewidth}{decrease steps\_per\_block to 22}\par\smallskip
\noindent{\setlength{\fboxsep}{0pt}\colorbox{black!4}{\parbox{\linewidth}{\parbox[t]{0.05\linewidth}{\centering\strut 33}\hspace{0.006\linewidth}\parbox[t]{0.04\linewidth}{\centering\strut \raisebox{0.08ex}{\ding{55}}}\hspace{0.006\linewidth}\parbox[t]{0.10\linewidth}{\centering\strut -0.0083\%}\hspace{0.006\linewidth}\parbox[t]{0.115\linewidth}{\centering\strut -7.874447}\hspace{0.006\linewidth}\parbox[t]{0.671\linewidth}{increase steps\_per\_block to 26}}}}\par\smallskip
\noindent\parbox[t]{0.05\linewidth}{\centering\strut 34}\hspace{0.006\linewidth}\parbox[t]{0.04\linewidth}{\centering\strut \raisebox{0.08ex}{\ding{55}}}\hspace{0.006\linewidth}\parbox[t]{0.10\linewidth}{\centering\strut +0.0163\%}\hspace{0.006\linewidth}\parbox[t]{0.115\linewidth}{\centering\strut -7.875730}\hspace{0.006\linewidth}\parbox[t]{0.671\linewidth}{decrease blocks to 39}\par\smallskip
\noindent{\setlength{\fboxsep}{0pt}\colorbox{black!4}{\parbox{\linewidth}{\parbox[t]{0.05\linewidth}{\centering\strut 35}\hspace{0.006\linewidth}\parbox[t]{0.04\linewidth}{\centering\strut \raisebox{0.08ex}{\ding{55}}}\hspace{0.006\linewidth}\parbox[t]{0.10\linewidth}{\centering\strut +0.0009\%}\hspace{0.006\linewidth}\parbox[t]{0.115\linewidth}{\centering\strut -7.875801}\hspace{0.006\linewidth}\parbox[t]{0.671\linewidth}{increase blocks to 41}}}}\par\smallskip
\noindent\parbox[t]{0.05\linewidth}{\centering\strut 36}\hspace{0.006\linewidth}\parbox[t]{0.04\linewidth}{\centering\strut \raisebox{0.08ex}{\ding{55}}}\hspace{0.006\linewidth}\parbox[t]{0.10\linewidth}{\centering\strut -0.0156\%}\hspace{0.006\linewidth}\parbox[t]{0.115\linewidth}{\centering\strut -7.874570}\hspace{0.006\linewidth}\parbox[t]{0.671\linewidth}{increase walkers to 136}\par\smallskip
\noindent{\setlength{\fboxsep}{0pt}\colorbox{black!4}{\parbox{\linewidth}{\parbox[t]{0.05\linewidth}{\centering\strut 37}\hspace{0.006\linewidth}\parbox[t]{0.04\linewidth}{\centering\strut \raisebox{0.08ex}{\ding{55}}}\hspace{0.006\linewidth}\parbox[t]{0.10\linewidth}{\centering\strut -0.0220\%}\hspace{0.006\linewidth}\parbox[t]{0.115\linewidth}{\centering\strut -7.872836}\hspace{0.006\linewidth}\parbox[t]{0.671\linewidth}{decrease walkers to 124}}}}\par\smallskip
\noindent\parbox[t]{0.05\linewidth}{\centering\strut 38}\hspace{0.006\linewidth}\parbox[t]{0.04\linewidth}{\centering\strut \raisebox{0.08ex}{\ding{55}}}\hspace{0.006\linewidth}\parbox[t]{0.10\linewidth}{\centering\strut +0.0392\%}\hspace{0.006\linewidth}\parbox[t]{0.115\linewidth}{\centering\strut -7.875923}\hspace{0.006\linewidth}\parbox[t]{0.671\linewidth}{switch init\_guess to minao}\par\smallskip
\noindent{\setlength{\fboxsep}{0pt}\colorbox{black!4}{\parbox{\linewidth}{\parbox[t]{0.05\linewidth}{\centering\strut 39}\hspace{0.006\linewidth}\parbox[t]{0.04\linewidth}{\centering\strut \raisebox{0.08ex}{\ding{55}}}\hspace{0.006\linewidth}\parbox[t]{0.10\linewidth}{\centering\strut -0.0000\%}\hspace{0.006\linewidth}\parbox[t]{0.115\linewidth}{\centering\strut -7.875923}\hspace{0.006\linewidth}\parbox[t]{0.671\linewidth}{switch trial to rhf}}}}\par\smallskip
\noindent\parbox[t]{0.05\linewidth}{\centering\strut 40}\hspace{0.006\linewidth}\parbox[t]{0.04\linewidth}{\centering\strut \raisebox{0.08ex}{\ding{55}}}\hspace{0.006\linewidth}\parbox[t]{0.10\linewidth}{\centering\strut -0.0251\%}\hspace{0.006\linewidth}\parbox[t]{0.115\linewidth}{\centering\strut -7.873948}\hspace{0.006\linewidth}\parbox[t]{0.671\linewidth}{set timestep 0.004 and blocks 50}\par\smallskip
\noindent{\setlength{\fboxsep}{0pt}\colorbox{black!4}{\parbox{\linewidth}{\parbox[t]{0.05\linewidth}{\centering\strut 41}\hspace{0.006\linewidth}\parbox[t]{0.04\linewidth}{\centering\strut \raisebox{0.08ex}{\ding{55}}}\hspace{0.006\linewidth}\parbox[t]{0.10\linewidth}{\centering\strut +0.0207\%}\hspace{0.006\linewidth}\parbox[t]{0.115\linewidth}{\centering\strut -7.875582}\hspace{0.006\linewidth}\parbox[t]{0.671\linewidth}{set timestep 0.006 and blocks 33}}}}\par\smallskip
\noindent\parbox[t]{0.05\linewidth}{\centering\strut 42}\hspace{0.006\linewidth}\parbox[t]{0.04\linewidth}{\centering\strut \raisebox{0.08ex}{\ding{51}}}\hspace{0.006\linewidth}\parbox[t]{0.10\linewidth}{\centering\strut +0.0064\%}\hspace{0.006\linewidth}\parbox[t]{0.115\linewidth}{\centering\strut -7.876082}\hspace{0.006\linewidth}\parbox[t]{0.671\linewidth}{set timestep 0.006 and blocks 35}\par\smallskip
\noindent{\setlength{\fboxsep}{0pt}\colorbox{black!4}{\parbox{\linewidth}{\parbox[t]{0.05\linewidth}{\centering\strut 43}\hspace{0.006\linewidth}\parbox[t]{0.04\linewidth}{\centering\strut \raisebox{0.08ex}{\ding{55}}}\hspace{0.006\linewidth}\parbox[t]{0.10\linewidth}{\centering\strut -0.0009\%}\hspace{0.006\linewidth}\parbox[t]{0.115\linewidth}{\centering\strut -7.876010}\hspace{0.006\linewidth}\parbox[t]{0.671\linewidth}{increase blocks to 36 at timestep 0.006}}}}\par\smallskip
\noindent\parbox[t]{0.05\linewidth}{\centering\strut 44}\hspace{0.006\linewidth}\parbox[t]{0.04\linewidth}{\centering\strut \raisebox{0.08ex}{\ding{55}}}\hspace{0.006\linewidth}\parbox[t]{0.10\linewidth}{\centering\strut +0.0002\%}\hspace{0.006\linewidth}\parbox[t]{0.115\linewidth}{\centering\strut -7.876026}\hspace{0.006\linewidth}\parbox[t]{0.671\linewidth}{decrease blocks to 34 at timestep 0.006}\par\smallskip
\noindent{\setlength{\fboxsep}{0pt}\colorbox{black!4}{\parbox{\linewidth}{\parbox[t]{0.05\linewidth}{\centering\strut 45}\hspace{0.006\linewidth}\parbox[t]{0.04\linewidth}{\centering\strut \raisebox{0.08ex}{\ding{55}}}\hspace{0.006\linewidth}\parbox[t]{0.10\linewidth}{\centering\strut +0.0007\%}\hspace{0.006\linewidth}\parbox[t]{0.115\linewidth}{\centering\strut -7.876082}\hspace{0.006\linewidth}\parbox[t]{0.671\linewidth}{set stabilize\_freq 4 and pop\_control\_freq 4}}}}\par\smallskip
\noindent\parbox[t]{0.05\linewidth}{\centering\strut 46}\hspace{0.006\linewidth}\parbox[t]{0.04\linewidth}{\centering\strut \raisebox{0.08ex}{\ding{55}}}\hspace{0.006\linewidth}\parbox[t]{0.10\linewidth}{\centering\strut -0.0135\%}\hspace{0.006\linewidth}\parbox[t]{0.115\linewidth}{\centering\strut -7.875016}\hspace{0.006\linewidth}\parbox[t]{0.671\linewidth}{set timestep 0.007 and blocks 30}\par\smallskip
\noindent{\setlength{\fboxsep}{0pt}\colorbox{black!4}{\parbox{\linewidth}{\parbox[t]{0.05\linewidth}{\centering\strut 47}\hspace{0.006\linewidth}\parbox[t]{0.04\linewidth}{\centering\strut \raisebox{0.08ex}{\ding{55}}}\hspace{0.006\linewidth}\parbox[t]{0.10\linewidth}{\centering\strut +0.0135\%}\hspace{0.006\linewidth}\parbox[t]{0.115\linewidth}{\centering\strut -7.876082}\hspace{0.006\linewidth}\parbox[t]{0.671\linewidth}{set scf\_conv\_tol 1e-10 scf\_max\_cycle 128 and diis\_space 12}}}}\par\smallskip
\noindent\parbox[t]{0.05\linewidth}{\centering\strut 48}\hspace{0.006\linewidth}\parbox[t]{0.04\linewidth}{\centering\strut \raisebox{0.08ex}{\ding{55}}}\hspace{0.006\linewidth}\parbox[t]{0.10\linewidth}{\centering\strut +0.0000\%}\hspace{0.006\linewidth}\parbox[t]{0.115\linewidth}{\centering\strut -7.876082}\hspace{0.006\linewidth}\parbox[t]{0.671\linewidth}{set damping to 0.1}\par\smallskip
\noindent{\setlength{\fboxsep}{0pt}\colorbox{black!4}{\parbox{\linewidth}{\parbox[t]{0.05\linewidth}{\centering\strut 49}\hspace{0.006\linewidth}\parbox[t]{0.04\linewidth}{\centering\strut \raisebox{0.08ex}{\ding{55}}}\hspace{0.006\linewidth}\parbox[t]{0.10\linewidth}{\centering\strut -0.0020\%}\hspace{0.006\linewidth}\parbox[t]{0.115\linewidth}{\centering\strut -7.875923}\hspace{0.006\linewidth}\parbox[t]{0.671\linewidth}{set timestep 0.0058 and blocks 36}}}}\par\smallskip
\noindent\parbox[t]{0.05\linewidth}{\centering\strut 50}\hspace{0.006\linewidth}\parbox[t]{0.04\linewidth}{\centering\strut \raisebox{0.08ex}{\ding{55}}}\hspace{0.006\linewidth}\parbox[t]{0.10\linewidth}{\centering\strut +0.0015\%}\hspace{0.006\linewidth}\parbox[t]{0.115\linewidth}{\centering\strut -7.876040}\hspace{0.006\linewidth}\parbox[t]{0.671\linewidth}{set timestep 0.0059}\par\smallskip
\noindent{\setlength{\fboxsep}{0pt}\colorbox{black!4}{\parbox{\linewidth}{\parbox[t]{0.05\linewidth}{\centering\strut 51}\hspace{0.006\linewidth}\parbox[t]{0.04\linewidth}{\centering\strut \raisebox{0.08ex}{\ding{51}}}\hspace{0.006\linewidth}\parbox[t]{0.10\linewidth}{\centering\strut +0.0010\%}\hspace{0.006\linewidth}\parbox[t]{0.115\linewidth}{\centering\strut -7.876121}\hspace{0.006\linewidth}\parbox[t]{0.671\linewidth}{set timestep 0.0061}}}}\par\smallskip
\noindent\parbox[t]{0.05\linewidth}{\centering\strut 52}\hspace{0.006\linewidth}\parbox[t]{0.04\linewidth}{\centering\strut \raisebox{0.08ex}{\ding{51}}}\hspace{0.006\linewidth}\parbox[t]{0.10\linewidth}{\centering\strut +0.0004\%}\hspace{0.006\linewidth}\parbox[t]{0.115\linewidth}{\centering\strut -7.876155}\hspace{0.006\linewidth}\parbox[t]{0.671\linewidth}{set timestep 0.0062}\par\smallskip
\noindent{\setlength{\fboxsep}{0pt}\colorbox{black!4}{\parbox{\linewidth}{\parbox[t]{0.05\linewidth}{\centering\strut 53}\hspace{0.006\linewidth}\parbox[t]{0.04\linewidth}{\centering\strut \raisebox{0.08ex}{\ding{51}}}\hspace{0.006\linewidth}\parbox[t]{0.10\linewidth}{\centering\strut +0.0004\%}\hspace{0.006\linewidth}\parbox[t]{0.115\linewidth}{\centering\strut -7.876187}\hspace{0.006\linewidth}\parbox[t]{0.671\linewidth}{set timestep 0.0063}}}}\par\smallskip
\noindent\parbox[t]{0.05\linewidth}{\centering\strut 54}\hspace{0.006\linewidth}\parbox[t]{0.04\linewidth}{\centering\strut \raisebox{0.08ex}{\ding{51}}}\hspace{0.006\linewidth}\parbox[t]{0.10\linewidth}{\centering\strut +0.0004\%}\hspace{0.006\linewidth}\parbox[t]{0.115\linewidth}{\centering\strut -7.876214}\hspace{0.006\linewidth}\parbox[t]{0.671\linewidth}{set timestep 0.0064}\par\smallskip
\noindent{\setlength{\fboxsep}{0pt}\colorbox{black!4}{\parbox{\linewidth}{\parbox[t]{0.05\linewidth}{\centering\strut 55}\hspace{0.006\linewidth}\parbox[t]{0.04\linewidth}{\centering\strut \raisebox{0.08ex}{\ding{51}}}\hspace{0.006\linewidth}\parbox[t]{0.10\linewidth}{\centering\strut +0.0003\%}\hspace{0.006\linewidth}\parbox[t]{0.115\linewidth}{\centering\strut -7.876239}\hspace{0.006\linewidth}\parbox[t]{0.671\linewidth}{set timestep 0.0065}}}}\par\smallskip
\noindent\parbox[t]{0.05\linewidth}{\centering\strut 56}\hspace{0.006\linewidth}\parbox[t]{0.04\linewidth}{\centering\strut \raisebox{0.08ex}{\ding{51}}}\hspace{0.006\linewidth}\parbox[t]{0.10\linewidth}{\centering\strut +0.0003\%}\hspace{0.006\linewidth}\parbox[t]{0.115\linewidth}{\centering\strut -7.876260}\hspace{0.006\linewidth}\parbox[t]{0.671\linewidth}{set timestep 0.0066}\par\smallskip
\noindent{\setlength{\fboxsep}{0pt}\colorbox{black!4}{\parbox{\linewidth}{\parbox[t]{0.05\linewidth}{\centering\strut 57}\hspace{0.006\linewidth}\parbox[t]{0.04\linewidth}{\centering\strut \raisebox{0.08ex}{\ding{51}}}\hspace{0.006\linewidth}\parbox[t]{0.10\linewidth}{\centering\strut +0.0002\%}\hspace{0.006\linewidth}\parbox[t]{0.115\linewidth}{\centering\strut -7.876279}\hspace{0.006\linewidth}\parbox[t]{0.671\linewidth}{set timestep 0.0067}}}}\par\smallskip
\noindent\parbox[t]{0.05\linewidth}{\centering\strut 58}\hspace{0.006\linewidth}\parbox[t]{0.04\linewidth}{\centering\strut \raisebox{0.08ex}{\ding{51}}}\hspace{0.006\linewidth}\parbox[t]{0.10\linewidth}{\centering\strut +0.0002\%}\hspace{0.006\linewidth}\parbox[t]{0.115\linewidth}{\centering\strut -7.876295}\hspace{0.006\linewidth}\parbox[t]{0.671\linewidth}{set timestep 0.0068}\par\smallskip
\noindent{\setlength{\fboxsep}{0pt}\colorbox{black!4}{\parbox{\linewidth}{\parbox[t]{0.05\linewidth}{\centering\strut 59}\hspace{0.006\linewidth}\parbox[t]{0.04\linewidth}{\centering\strut \raisebox{0.08ex}{\ding{51}}}\hspace{0.006\linewidth}\parbox[t]{0.10\linewidth}{\centering\strut +0.0002\%}\hspace{0.006\linewidth}\parbox[t]{0.115\linewidth}{\centering\strut -7.876308}\hspace{0.006\linewidth}\parbox[t]{0.671\linewidth}{set timestep 0.0069}}}}\par\smallskip
\noindent\parbox[t]{0.05\linewidth}{\centering\strut 60}\hspace{0.006\linewidth}\parbox[t]{0.04\linewidth}{\centering\strut \raisebox{0.08ex}{\ding{51}}}\hspace{0.006\linewidth}\parbox[t]{0.10\linewidth}{\centering\strut +0.0001\%}\hspace{0.006\linewidth}\parbox[t]{0.115\linewidth}{\centering\strut -7.876320}\hspace{0.006\linewidth}\parbox[t]{0.671\linewidth}{set timestep 0.007}\par\smallskip
\noindent{\setlength{\fboxsep}{0pt}\colorbox{black!4}{\parbox{\linewidth}{\parbox[t]{0.05\linewidth}{\centering\strut 61}\hspace{0.006\linewidth}\parbox[t]{0.04\linewidth}{\centering\strut \raisebox{0.08ex}{\ding{51}}}\hspace{0.006\linewidth}\parbox[t]{0.10\linewidth}{\centering\strut +0.0001\%}\hspace{0.006\linewidth}\parbox[t]{0.115\linewidth}{\centering\strut -7.876329}\hspace{0.006\linewidth}\parbox[t]{0.671\linewidth}{set timestep 0.0071}}}}\par\smallskip
\noindent\parbox[t]{0.05\linewidth}{\centering\strut 62}\hspace{0.006\linewidth}\parbox[t]{0.04\linewidth}{\centering\strut \raisebox{0.08ex}{\ding{51}}}\hspace{0.006\linewidth}\parbox[t]{0.10\linewidth}{\centering\strut +0.0001\%}\hspace{0.006\linewidth}\parbox[t]{0.115\linewidth}{\centering\strut -7.876336}\hspace{0.006\linewidth}\parbox[t]{0.671\linewidth}{set timestep 0.0072}\par\smallskip
\noindent{\setlength{\fboxsep}{0pt}\colorbox{black!4}{\parbox{\linewidth}{\parbox[t]{0.05\linewidth}{\centering\strut 63}\hspace{0.006\linewidth}\parbox[t]{0.04\linewidth}{\centering\strut \raisebox{0.08ex}{\ding{51}}}\hspace{0.006\linewidth}\parbox[t]{0.10\linewidth}{\centering\strut +0.0001\%}\hspace{0.006\linewidth}\parbox[t]{0.115\linewidth}{\centering\strut -7.876341}\hspace{0.006\linewidth}\parbox[t]{0.671\linewidth}{set timestep 0.0073}}}}\par\smallskip
\noindent\parbox[t]{0.05\linewidth}{\centering\strut 64}\hspace{0.006\linewidth}\parbox[t]{0.04\linewidth}{\centering\strut \raisebox{0.08ex}{\ding{51}}}\hspace{0.006\linewidth}\parbox[t]{0.10\linewidth}{\centering\strut +0.0000\%}\hspace{0.006\linewidth}\parbox[t]{0.115\linewidth}{\centering\strut -7.876345}\hspace{0.006\linewidth}\parbox[t]{0.671\linewidth}{set timestep 0.0074}\par\smallskip
\noindent{\setlength{\fboxsep}{0pt}\colorbox{black!4}{\parbox{\linewidth}{\parbox[t]{0.05\linewidth}{\centering\strut 65}\hspace{0.006\linewidth}\parbox[t]{0.04\linewidth}{\centering\strut \raisebox{0.08ex}{\ding{51}}}\hspace{0.006\linewidth}\parbox[t]{0.10\linewidth}{\centering\strut +0.0000\%}\hspace{0.006\linewidth}\parbox[t]{0.115\linewidth}{\centering\strut -7.876346}\hspace{0.006\linewidth}\parbox[t]{0.671\linewidth}{set timestep 0.0075}}}}\par\smallskip
\noindent\parbox[t]{0.05\linewidth}{\centering\strut 66}\hspace{0.006\linewidth}\parbox[t]{0.04\linewidth}{\centering\strut \raisebox{0.08ex}{\ding{51}}}\hspace{0.006\linewidth}\parbox[t]{0.10\linewidth}{\centering\strut +0.0000\%}\hspace{0.006\linewidth}\parbox[t]{0.115\linewidth}{\centering\strut -7.876347}\hspace{0.006\linewidth}\parbox[t]{0.671\linewidth}{set timestep 0.0076}\par\smallskip
\noindent{\setlength{\fboxsep}{0pt}\colorbox{black!4}{\parbox{\linewidth}{\parbox[t]{0.05\linewidth}{\centering\strut 67}\hspace{0.006\linewidth}\parbox[t]{0.04\linewidth}{\centering\strut \raisebox{0.08ex}{\ding{55}}}\hspace{0.006\linewidth}\parbox[t]{0.10\linewidth}{\centering\strut -0.0000\%}\hspace{0.006\linewidth}\parbox[t]{0.115\linewidth}{\centering\strut -7.876346}\hspace{0.006\linewidth}\parbox[t]{0.671\linewidth}{set timestep 0.0077}}}}\par\smallskip
\noindent\parbox[t]{0.05\linewidth}{\centering\strut 68}\hspace{0.006\linewidth}\parbox[t]{0.04\linewidth}{\centering\strut \raisebox{0.08ex}{\ding{55}}}\hspace{0.006\linewidth}\parbox[t]{0.10\linewidth}{\centering\strut +0.0000\%}\hspace{0.006\linewidth}\parbox[t]{0.115\linewidth}{\centering\strut -7.876346}\hspace{0.006\linewidth}\parbox[t]{0.671\linewidth}{set timestep 0.00765}\par\smallskip
\noindent{\setlength{\fboxsep}{0pt}\colorbox{black!4}{\parbox{\linewidth}{\parbox[t]{0.05\linewidth}{\centering\strut 69}\hspace{0.006\linewidth}\parbox[t]{0.04\linewidth}{\centering\strut \raisebox{0.08ex}{\ding{55}}}\hspace{0.006\linewidth}\parbox[t]{0.10\linewidth}{\centering\strut -0.0012\%}\hspace{0.006\linewidth}\parbox[t]{0.115\linewidth}{\centering\strut -7.876250}\hspace{0.006\linewidth}\parbox[t]{0.671\linewidth}{decrease blocks to 34 at timestep 0.0076}}}}\par\smallskip
\noindent\parbox[t]{0.05\linewidth}{\centering\strut 70}\hspace{0.006\linewidth}\parbox[t]{0.04\linewidth}{\centering\strut \raisebox{0.08ex}{\ding{55}}}\hspace{0.006\linewidth}\parbox[t]{0.10\linewidth}{\centering\strut +0.0002\%}\hspace{0.006\linewidth}\parbox[t]{0.115\linewidth}{\centering\strut -7.876268}\hspace{0.006\linewidth}\parbox[t]{0.671\linewidth}{increase blocks to 36 at timestep 0.0076}\par\smallskip
\noindent{\setlength{\fboxsep}{0pt}\colorbox{black!4}{\parbox{\linewidth}{\parbox[t]{0.05\linewidth}{\centering\strut 71}\hspace{0.006\linewidth}\parbox[t]{0.04\linewidth}{\centering\strut \raisebox{0.08ex}{\ding{55}}}\hspace{0.006\linewidth}\parbox[t]{0.10\linewidth}{\centering\strut -0.0072\%}\hspace{0.006\linewidth}\parbox[t]{0.115\linewidth}{\centering\strut -7.875699}\hspace{0.006\linewidth}\parbox[t]{0.671\linewidth}{set steps\_per\_block 23 and blocks 36}}}}\par\smallskip
\noindent\parbox[t]{0.05\linewidth}{\centering\strut 72}\hspace{0.006\linewidth}\parbox[t]{0.04\linewidth}{\centering\strut \raisebox{0.08ex}{\ding{55}}}\hspace{0.006\linewidth}\parbox[t]{0.10\linewidth}{\centering\strut -0.0073\%}\hspace{0.006\linewidth}\parbox[t]{0.115\linewidth}{\centering\strut -7.875121}\hspace{0.006\linewidth}\parbox[t]{0.671\linewidth}{set steps\_per\_block 25 and blocks 34}\par\smallskip
\noindent{\setlength{\fboxsep}{0pt}\colorbox{black!4}{\parbox{\linewidth}{\parbox[t]{0.05\linewidth}{\centering\strut 73}\hspace{0.006\linewidth}\parbox[t]{0.04\linewidth}{\centering\strut \raisebox{0.08ex}{\ding{51}}}\hspace{0.006\linewidth}\parbox[t]{0.10\linewidth}{\centering\strut +0.0156\%}\hspace{0.006\linewidth}\parbox[t]{0.115\linewidth}{\centering\strut -7.876347}\hspace{0.006\linewidth}\parbox[t]{0.671\linewidth}{set stabilize\_freq 4 and pop\_control\_freq 4}}}}\par\smallskip
\noindent\parbox[t]{0.05\linewidth}{\centering\strut 74}\hspace{0.006\linewidth}\parbox[t]{0.04\linewidth}{\centering\strut \raisebox{0.08ex}{\ding{55}}}\hspace{0.006\linewidth}\parbox[t]{0.10\linewidth}{\centering\strut -0.0000\%}\hspace{0.006\linewidth}\parbox[t]{0.115\linewidth}{\centering\strut -7.876347}\hspace{0.006\linewidth}\parbox[t]{0.671\linewidth}{set stabilize\_freq 5 and pop\_control\_freq 5}\par\smallskip
\noindent{\setlength{\fboxsep}{0pt}\colorbox{black!4}{\parbox{\linewidth}{\parbox[t]{0.05\linewidth}{\centering\strut 75}\hspace{0.006\linewidth}\parbox[t]{0.04\linewidth}{\centering\strut \raisebox{0.08ex}{\ding{55}}}\hspace{0.006\linewidth}\parbox[t]{0.10\linewidth}{\centering\strut -0.0000\%}\hspace{0.006\linewidth}\parbox[t]{0.115\linewidth}{\centering\strut -7.876346}\hspace{0.006\linewidth}\parbox[t]{0.671\linewidth}{switch init\_guess to minao}}}}\par\smallskip
\noindent\parbox[t]{0.05\linewidth}{\centering\strut 76}\hspace{0.006\linewidth}\parbox[t]{0.04\linewidth}{\centering\strut \raisebox{0.08ex}{\ding{55}}}\hspace{0.006\linewidth}\parbox[t]{0.10\linewidth}{\centering\strut -0.0000\%}\hspace{0.006\linewidth}\parbox[t]{0.115\linewidth}{\centering\strut -7.876346}\hspace{0.006\linewidth}\parbox[t]{0.671\linewidth}{switch trial to rhf}\par\smallskip
\noindent{\setlength{\fboxsep}{0pt}\colorbox{black!4}{\parbox{\linewidth}{\parbox[t]{0.05\linewidth}{\centering\strut 77}\hspace{0.006\linewidth}\parbox[t]{0.04\linewidth}{\centering\strut \raisebox{0.08ex}{\ding{51}}}\hspace{0.006\linewidth}\parbox[t]{0.10\linewidth}{\centering\strut +0.0000\%}\hspace{0.006\linewidth}\parbox[t]{0.115\linewidth}{\centering\strut -7.876347}\hspace{0.006\linewidth}\parbox[t]{0.671\linewidth}{set timestep 0.00758}}}}\par\smallskip
\noindent\parbox[t]{0.05\linewidth}{\centering\strut 78}\hspace{0.006\linewidth}\parbox[t]{0.04\linewidth}{\centering\strut \raisebox{0.08ex}{\ding{51}}}\hspace{0.006\linewidth}\parbox[t]{0.10\linewidth}{\centering\strut +0.0000\%}\hspace{0.006\linewidth}\parbox[t]{0.115\linewidth}{\centering\strut -7.876347}\hspace{0.006\linewidth}\parbox[t]{0.671\linewidth}{set timestep 0.00757}\par\smallskip
\noindent{\setlength{\fboxsep}{0pt}\colorbox{black!4}{\parbox{\linewidth}{\parbox[t]{0.05\linewidth}{\centering\strut 79}\hspace{0.006\linewidth}\parbox[t]{0.04\linewidth}{\centering\strut \raisebox{0.08ex}{\ding{55}}}\hspace{0.006\linewidth}\parbox[t]{0.10\linewidth}{\centering\strut -0.0000\%}\hspace{0.006\linewidth}\parbox[t]{0.115\linewidth}{\centering\strut -7.876347}\hspace{0.006\linewidth}\parbox[t]{0.671\linewidth}{set timestep 0.00756}}}}\par\smallskip
\noindent\parbox[t]{0.05\linewidth}{\centering\strut 80}\hspace{0.006\linewidth}\parbox[t]{0.04\linewidth}{\centering\strut \raisebox{0.08ex}{\ding{55}}}\hspace{0.006\linewidth}\parbox[t]{0.10\linewidth}{\centering\strut +0.0000\%}\hspace{0.006\linewidth}\parbox[t]{0.115\linewidth}{\centering\strut -7.876347}\hspace{0.006\linewidth}\parbox[t]{0.671\linewidth}{set timestep 0.007571}\par\smallskip
\noindent{\setlength{\fboxsep}{0pt}\colorbox{black!4}{\parbox{\linewidth}{\parbox[t]{0.05\linewidth}{\centering\strut 81}\hspace{0.006\linewidth}\parbox[t]{0.04\linewidth}{\centering\strut \raisebox{0.08ex}{\ding{55}}}\hspace{0.006\linewidth}\parbox[t]{0.10\linewidth}{\centering\strut -0.0645\%}\hspace{0.006\linewidth}\parbox[t]{0.115\linewidth}{\centering\strut -7.871263}\hspace{0.006\linewidth}\parbox[t]{0.671\linewidth}{decrease walkers to 127}}}}\par\smallskip
\noindent\parbox[t]{0.05\linewidth}{\centering\strut 82}\hspace{0.006\linewidth}\parbox[t]{0.04\linewidth}{\centering\strut \raisebox{0.08ex}{\ding{51}}}\hspace{0.006\linewidth}\parbox[t]{0.10\linewidth}{\centering\strut +0.0646\%}\hspace{0.006\linewidth}\parbox[t]{0.115\linewidth}{\centering\strut -7.876347}\hspace{0.006\linewidth}\parbox[t]{0.671\linewidth}{set stabilize\_freq 4 and pop\_control\_freq 3}\par\smallskip
\noindent{\setlength{\fboxsep}{0pt}\colorbox{black!4}{\parbox{\linewidth}{\parbox[t]{0.05\linewidth}{\centering\strut 83}\hspace{0.006\linewidth}\parbox[t]{0.04\linewidth}{\centering\strut \raisebox{0.08ex}{\ding{55}}}\hspace{0.006\linewidth}\parbox[t]{0.10\linewidth}{\centering\strut -0.0000\%}\hspace{0.006\linewidth}\parbox[t]{0.115\linewidth}{\centering\strut -7.876347}\hspace{0.006\linewidth}\parbox[t]{0.671\linewidth}{set stabilize\_freq 3 and pop\_control\_freq 4}}}}\par\smallskip
\noindent\parbox[t]{0.05\linewidth}{\centering\strut 84}\hspace{0.006\linewidth}\parbox[t]{0.04\linewidth}{\centering\strut \raisebox{0.08ex}{\ding{55}}}\hspace{0.006\linewidth}\parbox[t]{0.10\linewidth}{\centering\strut -0.0000\%}\hspace{0.006\linewidth}\parbox[t]{0.115\linewidth}{\centering\strut -7.876347}\hspace{0.006\linewidth}\parbox[t]{0.671\linewidth}{set stabilize\_freq 5 and pop\_control\_freq 3}\par\smallskip
\noindent{\setlength{\fboxsep}{0pt}\colorbox{black!4}{\parbox{\linewidth}{\parbox[t]{0.05\linewidth}{\centering\strut 85}\hspace{0.006\linewidth}\parbox[t]{0.04\linewidth}{\centering\strut \raisebox{0.08ex}{\ding{55}}}\hspace{0.006\linewidth}\parbox[t]{0.10\linewidth}{\centering\strut +0.0000\%}\hspace{0.006\linewidth}\parbox[t]{0.115\linewidth}{\centering\strut -7.876347}\hspace{0.006\linewidth}\parbox[t]{0.671\linewidth}{set stabilize\_freq 4 and pop\_control\_freq 2}}}}\par\smallskip
\noindent\parbox[t]{0.05\linewidth}{\centering\strut 86}\hspace{0.006\linewidth}\parbox[t]{0.04\linewidth}{\centering\strut \raisebox{0.08ex}{\ding{51}}}\hspace{0.006\linewidth}\parbox[t]{0.10\linewidth}{\centering\strut +0.0000\%}\hspace{0.006\linewidth}\parbox[t]{0.115\linewidth}{\centering\strut -7.876347}\hspace{0.006\linewidth}\parbox[t]{0.671\linewidth}{set stabilize\_freq 3 and pop\_control\_freq 2}\par\smallskip
\noindent{\setlength{\fboxsep}{0pt}\colorbox{black!4}{\parbox{\linewidth}{\parbox[t]{0.05\linewidth}{\centering\strut 87}\hspace{0.006\linewidth}\parbox[t]{0.04\linewidth}{\centering\strut \raisebox{0.08ex}{\ding{55}}}\hspace{0.006\linewidth}\parbox[t]{0.10\linewidth}{\centering\strut -0.0000\%}\hspace{0.006\linewidth}\parbox[t]{0.115\linewidth}{\centering\strut -7.876347}\hspace{0.006\linewidth}\parbox[t]{0.671\linewidth}{set stabilize\_freq 2 and pop\_control\_freq 2}}}}\par\smallskip
\noindent\parbox[t]{0.05\linewidth}{\centering\strut 88}\hspace{0.006\linewidth}\parbox[t]{0.04\linewidth}{\centering\strut \raisebox{0.08ex}{\ding{55}}}\hspace{0.006\linewidth}\parbox[t]{0.10\linewidth}{\centering\strut +0.0000\%}\hspace{0.006\linewidth}\parbox[t]{0.115\linewidth}{\centering\strut -7.876347}\hspace{0.006\linewidth}\parbox[t]{0.671\linewidth}{set stabilize\_freq 3 and pop\_control\_freq 1}\par\smallskip
\noindent{\setlength{\fboxsep}{0pt}\colorbox{black!4}{\parbox{\linewidth}{\parbox[t]{0.05\linewidth}{\centering\strut 89}\hspace{0.006\linewidth}\parbox[t]{0.04\linewidth}{\centering\strut \raisebox{0.08ex}{\ding{55}}}\hspace{0.006\linewidth}\parbox[t]{0.10\linewidth}{\centering\strut +0.0000\%}\hspace{0.006\linewidth}\parbox[t]{0.115\linewidth}{\centering\strut -7.876347}\hspace{0.006\linewidth}\parbox[t]{0.671\linewidth}{set stabilize\_freq 2 and pop\_control\_freq 1}}}}\par\smallskip
\noindent\parbox[t]{0.05\linewidth}{\centering\strut 90}\hspace{0.006\linewidth}\parbox[t]{0.04\linewidth}{\centering\strut \raisebox{0.08ex}{\ding{51}}}\hspace{0.006\linewidth}\parbox[t]{0.10\linewidth}{\centering\strut +0.0000\%}\hspace{0.006\linewidth}\parbox[t]{0.115\linewidth}{\centering\strut -7.876347}\hspace{0.006\linewidth}\parbox[t]{0.671\linewidth}{set timestep 0.007569}\par\smallskip
\noindent{\setlength{\fboxsep}{0pt}\colorbox{black!4}{\parbox{\linewidth}{\parbox[t]{0.05\linewidth}{\centering\strut 91}\hspace{0.006\linewidth}\parbox[t]{0.04\linewidth}{\centering\strut \raisebox{0.08ex}{\ding{55}}}\hspace{0.006\linewidth}\parbox[t]{0.10\linewidth}{\centering\strut -0.0000\%}\hspace{0.006\linewidth}\parbox[t]{0.115\linewidth}{\centering\strut -7.876347}\hspace{0.006\linewidth}\parbox[t]{0.671\linewidth}{set timestep 0.007568}}}}\par\smallskip
\noindent\parbox[t]{0.05\linewidth}{\centering\strut 92}\hspace{0.006\linewidth}\parbox[t]{0.04\linewidth}{\centering\strut \raisebox{0.08ex}{\ding{55}}}\hspace{0.006\linewidth}\parbox[t]{0.10\linewidth}{\centering\strut +0.0000\%}\hspace{0.006\linewidth}\parbox[t]{0.115\linewidth}{\centering\strut -7.876347}\hspace{0.006\linewidth}\parbox[t]{0.671\linewidth}{set timestep 0.0075691}\par\smallskip
\noindent{\setlength{\fboxsep}{0pt}\colorbox{black!4}{\parbox{\linewidth}{\parbox[t]{0.05\linewidth}{\centering\strut 93}\hspace{0.006\linewidth}\parbox[t]{0.04\linewidth}{\centering\strut \raisebox{0.08ex}{\ding{55}}}\hspace{0.006\linewidth}\parbox[t]{0.10\linewidth}{\centering\strut +0.0000\%}\hspace{0.006\linewidth}\parbox[t]{0.115\linewidth}{\centering\strut -7.876347}\hspace{0.006\linewidth}\parbox[t]{0.671\linewidth}{set stabilize\_freq 2 and pop\_control\_freq 3}}}}\par\smallskip
\noindent\parbox[t]{0.05\linewidth}{\centering\strut 94}\hspace{0.006\linewidth}\parbox[t]{0.04\linewidth}{\centering\strut \raisebox{0.08ex}{\ding{55}}}\hspace{0.006\linewidth}\parbox[t]{0.10\linewidth}{\centering\strut +0.0000\%}\hspace{0.006\linewidth}\parbox[t]{0.115\linewidth}{\centering\strut -7.876347}\hspace{0.006\linewidth}\parbox[t]{0.671\linewidth}{set stabilize\_freq 5 and pop\_control\_freq 2}\par\smallskip
\noindent{\setlength{\fboxsep}{0pt}\colorbox{black!4}{\parbox{\linewidth}{\parbox[t]{0.05\linewidth}{\centering\strut 95}\hspace{0.006\linewidth}\parbox[t]{0.04\linewidth}{\centering\strut \raisebox{0.08ex}{\ding{55}}}\hspace{0.006\linewidth}\parbox[t]{0.10\linewidth}{\centering\strut -0.0000\%}\hspace{0.006\linewidth}\parbox[t]{0.115\linewidth}{\centering\strut -7.876347}\hspace{0.006\linewidth}\parbox[t]{0.671\linewidth}{set stabilize\_freq 6 and pop\_control\_freq 2}}}}\par\smallskip
\noindent\parbox[t]{0.05\linewidth}{\centering\strut 96}\hspace{0.006\linewidth}\parbox[t]{0.04\linewidth}{\centering\strut \raisebox{0.08ex}{\ding{51}}}\hspace{0.006\linewidth}\parbox[t]{0.10\linewidth}{\centering\strut +0.0000\%}\hspace{0.006\linewidth}\parbox[t]{0.115\linewidth}{\centering\strut -7.876347}\hspace{0.006\linewidth}\parbox[t]{0.671\linewidth}{set stabilize\_freq 1 and pop\_control\_freq 2}\par\smallskip
\noindent{\setlength{\fboxsep}{0pt}\colorbox{black!4}{\parbox{\linewidth}{\parbox[t]{0.05\linewidth}{\centering\strut 97}\hspace{0.006\linewidth}\parbox[t]{0.04\linewidth}{\centering\strut \raisebox{0.08ex}{\ding{55}}}\hspace{0.006\linewidth}\parbox[t]{0.10\linewidth}{\centering\strut -0.0000\%}\hspace{0.006\linewidth}\parbox[t]{0.115\linewidth}{\centering\strut -7.876347}\hspace{0.006\linewidth}\parbox[t]{0.671\linewidth}{set stabilize\_freq 1 and pop\_control\_freq 1}}}}\par\smallskip
\noindent\parbox[t]{0.05\linewidth}{\centering\strut 98}\hspace{0.006\linewidth}\parbox[t]{0.04\linewidth}{\centering\strut \raisebox{0.08ex}{\ding{55}}}\hspace{0.006\linewidth}\parbox[t]{0.10\linewidth}{\centering\strut +0.0000\%}\hspace{0.006\linewidth}\parbox[t]{0.115\linewidth}{\centering\strut -7.876347}\hspace{0.006\linewidth}\parbox[t]{0.671\linewidth}{set stabilize\_freq 1 and pop\_control\_freq 3}\par\smallskip
\noindent{\setlength{\fboxsep}{0pt}\colorbox{black!4}{\parbox{\linewidth}{\parbox[t]{0.05\linewidth}{\centering\strut 99}\hspace{0.006\linewidth}\parbox[t]{0.04\linewidth}{\centering\strut \raisebox{0.08ex}{\ding{55}}}\hspace{0.006\linewidth}\parbox[t]{0.10\linewidth}{\centering\strut -0.0000\%}\hspace{0.006\linewidth}\parbox[t]{0.115\linewidth}{\centering\strut -7.876347}\hspace{0.006\linewidth}\parbox[t]{0.671\linewidth}{set stabilize\_freq 1 and pop\_control\_freq 4}}}}\par\smallskip
\noindent\parbox[t]{0.05\linewidth}{\centering\strut 100}\hspace{0.006\linewidth}\parbox[t]{0.04\linewidth}{\centering\strut \raisebox{0.08ex}{\ding{55}}}\hspace{0.006\linewidth}\parbox[t]{0.10\linewidth}{\centering\strut +0.0000\%}\hspace{0.006\linewidth}\parbox[t]{0.115\linewidth}{\centering\strut -7.876347}\hspace{0.006\linewidth}\parbox[t]{0.671\linewidth}{set stabilize\_freq 1 and pop\_control\_freq 5}\par\smallskip
\noindent\rule{\linewidth}{0.5pt}
\par\medskip
\refstepcounter{table}
\label{tab:supp_afqmc_h2o}
\noindent\textbf{Table \thetable.} AFQMC mutation log for H$_2$O full 100-iteration campaign. Relative gain is measured against the immediately previous scored iteration using the live AFQMC score.
\par\medskip
\noindent\rule{\linewidth}{0.5pt}
\noindent{\setlength{\fboxsep}{0pt}\colorbox{black!4}{\parbox{\linewidth}{\parbox[c][2.2em][c]{0.05\linewidth}{\centering \textbf{Iter.}}\hspace{0.006\linewidth}\parbox[c][2.2em][c]{0.04\linewidth}{\centering \textbf{A/R}}\hspace{0.006\linewidth}\parbox[c][2.2em][c]{0.10\linewidth}{\centering \textbf{Relative gain}}\hspace{0.006\linewidth}\parbox[c][2.2em][c]{0.115\linewidth}{\centering \textbf{Score}}\hspace{0.006\linewidth}\parbox[c][2.2em][c]{0.671\linewidth}{\centering \textbf{Mutation summary}}}}}\par\smallskip
\noindent\rule{\linewidth}{0.5pt}
\noindent{\setlength{\fboxsep}{0pt}\colorbox{black!4}{\parbox{\linewidth}{\parbox[t]{0.05\linewidth}{\centering\strut 1}\hspace{0.006\linewidth}\parbox[t]{0.04\linewidth}{\centering\strut \raisebox{0.08ex}{\ding{51}}}\hspace{0.006\linewidth}\parbox[t]{0.10\linewidth}{\centering\strut ---}\hspace{0.006\linewidth}\parbox[t]{0.115\linewidth}{\centering\strut -74.991318}\hspace{0.006\linewidth}\parbox[t]{0.671\linewidth}{increase num\_walkers\_per\_rank to 64}}}}\par\smallskip
\noindent\parbox[t]{0.05\linewidth}{\centering\strut 2}\hspace{0.006\linewidth}\parbox[t]{0.04\linewidth}{\centering\strut \raisebox{0.08ex}{\ding{51}}}\hspace{0.006\linewidth}\parbox[t]{0.10\linewidth}{\centering\strut +0.0057\%}\hspace{0.006\linewidth}\parbox[t]{0.115\linewidth}{\centering\strut -74.995610}\hspace{0.006\linewidth}\parbox[t]{0.671\linewidth}{increase num\_walkers\_per\_rank to 128}\par\smallskip
\noindent{\setlength{\fboxsep}{0pt}\colorbox{black!4}{\parbox{\linewidth}{\parbox[t]{0.05\linewidth}{\centering\strut 3}\hspace{0.006\linewidth}\parbox[t]{0.04\linewidth}{\centering\strut \raisebox{0.08ex}{\ding{55}}}\hspace{0.006\linewidth}\parbox[t]{0.10\linewidth}{\centering\strut +0.0000\%}\hspace{0.006\linewidth}\parbox[t]{0.115\linewidth}{\centering\strut -74.995610}\hspace{0.006\linewidth}\parbox[t]{0.671\linewidth}{increase num\_walkers\_per\_rank to 256}}}}\par\smallskip
\noindent\parbox[t]{0.05\linewidth}{\centering\strut 4}\hspace{0.006\linewidth}\parbox[t]{0.04\linewidth}{\centering\strut \raisebox{0.08ex}{\ding{55}}}\hspace{0.006\linewidth}\parbox[t]{0.10\linewidth}{\centering\strut -0.0297\%}\hspace{0.006\linewidth}\parbox[t]{0.115\linewidth}{\centering\strut -74.973374}\hspace{0.006\linewidth}\parbox[t]{0.671\linewidth}{increase num\_steps\_per\_block to 40 and num\_blocks to 60 on the 128-walker RHF trial}\par\smallskip
\noindent{\setlength{\fboxsep}{0pt}\colorbox{black!4}{\parbox{\linewidth}{\parbox[t]{0.05\linewidth}{\centering\strut 5}\hspace{0.006\linewidth}\parbox[t]{0.04\linewidth}{\centering\strut \raisebox{0.08ex}{\ding{55}}}\hspace{0.006\linewidth}\parbox[t]{0.10\linewidth}{\centering\strut +0.0269\%}\hspace{0.006\linewidth}\parbox[t]{0.115\linewidth}{\centering\strut -74.993541}\hspace{0.006\linewidth}\parbox[t]{0.671\linewidth}{increase num\_blocks to 60}}}}\par\smallskip
\noindent\parbox[t]{0.05\linewidth}{\centering\strut 6}\hspace{0.006\linewidth}\parbox[t]{0.04\linewidth}{\centering\strut \raisebox{0.08ex}{\ding{51}}}\hspace{0.006\linewidth}\parbox[t]{0.10\linewidth}{\centering\strut +0.0056\%}\hspace{0.006\linewidth}\parbox[t]{0.115\linewidth}{\centering\strut -74.997777}\hspace{0.006\linewidth}\parbox[t]{0.671\linewidth}{increase num\_walkers\_per\_rank to 256}\par\smallskip
\noindent{\setlength{\fboxsep}{0pt}\colorbox{black!4}{\parbox{\linewidth}{\parbox[t]{0.05\linewidth}{\centering\strut 7}\hspace{0.006\linewidth}\parbox[t]{0.04\linewidth}{\centering\strut \raisebox{0.08ex}{\ding{51}}}\hspace{0.006\linewidth}\parbox[t]{0.10\linewidth}{\centering\strut +0.0068\%}\hspace{0.006\linewidth}\parbox[t]{0.115\linewidth}{\centering\strut -75.002899}\hspace{0.006\linewidth}\parbox[t]{0.671\linewidth}{increase num\_walkers\_per\_rank to 512}}}}\par\smallskip
\noindent\parbox[t]{0.05\linewidth}{\centering\strut 8}\hspace{0.006\linewidth}\parbox[t]{0.04\linewidth}{\centering\strut \raisebox{0.08ex}{\ding{55}}}\hspace{0.006\linewidth}\parbox[t]{0.10\linewidth}{\centering\strut -0.0023\%}\hspace{0.006\linewidth}\parbox[t]{0.115\linewidth}{\centering\strut -75.001140}\hspace{0.006\linewidth}\parbox[t]{0.671\linewidth}{set num\_steps\_per\_block to 50 and num\_blocks to 20}\par\smallskip
\noindent{\setlength{\fboxsep}{0pt}\colorbox{black!4}{\parbox{\linewidth}{\parbox[t]{0.05\linewidth}{\centering\strut 9}\hspace{0.006\linewidth}\parbox[t]{0.04\linewidth}{\centering\strut \raisebox{0.08ex}{\ding{55}}}\hspace{0.006\linewidth}\parbox[t]{0.10\linewidth}{\centering\strut -0.0011\%}\hspace{0.006\linewidth}\parbox[t]{0.115\linewidth}{\centering\strut -75.000308}\hspace{0.006\linewidth}\parbox[t]{0.671\linewidth}{reduce chol\_cut to 1e-8}}}}\par\smallskip
\noindent\parbox[t]{0.05\linewidth}{\centering\strut 10}\hspace{0.006\linewidth}\parbox[t]{0.04\linewidth}{\centering\strut \raisebox{0.08ex}{\ding{55}}}\hspace{0.006\linewidth}\parbox[t]{0.10\linewidth}{\centering\strut +0.0026\%}\hspace{0.006\linewidth}\parbox[t]{0.115\linewidth}{\centering\strut -75.002272}\hspace{0.006\linewidth}\parbox[t]{0.671\linewidth}{set timestep to 0.0025 and num\_steps\_per\_block to 50}\par\smallskip
\noindent{\setlength{\fboxsep}{0pt}\colorbox{black!4}{\parbox{\linewidth}{\parbox[t]{0.05\linewidth}{\centering\strut 11}\hspace{0.006\linewidth}\parbox[t]{0.04\linewidth}{\centering\strut \raisebox{0.08ex}{\ding{51}}}\hspace{0.006\linewidth}\parbox[t]{0.10\linewidth}{\centering\strut +0.0008\%}\hspace{0.006\linewidth}\parbox[t]{0.115\linewidth}{\centering\strut -75.002899}\hspace{0.006\linewidth}\parbox[t]{0.671\linewidth}{set stabilize\_freq to 3 and pop\_control\_freq to 3}}}}\par\smallskip
\noindent\parbox[t]{0.05\linewidth}{\centering\strut 12}\hspace{0.006\linewidth}\parbox[t]{0.04\linewidth}{\centering\strut \raisebox{0.08ex}{\ding{55}}}\hspace{0.006\linewidth}\parbox[t]{0.10\linewidth}{\centering\strut -0.0011\%}\hspace{0.006\linewidth}\parbox[t]{0.115\linewidth}{\centering\strut -75.002087}\hspace{0.006\linewidth}\parbox[t]{0.671\linewidth}{set num\_steps\_per\_block to 40 and num\_blocks to 25}\par\smallskip
\noindent{\setlength{\fboxsep}{0pt}\colorbox{black!4}{\parbox{\linewidth}{\parbox[t]{0.05\linewidth}{\centering\strut 13}\hspace{0.006\linewidth}\parbox[t]{0.04\linewidth}{\centering\strut \raisebox{0.08ex}{\ding{51}}}\hspace{0.006\linewidth}\parbox[t]{0.10\linewidth}{\centering\strut +0.0014\%}\hspace{0.006\linewidth}\parbox[t]{0.115\linewidth}{\centering\strut -75.003164}\hspace{0.006\linewidth}\parbox[t]{0.671\linewidth}{reduce num\_blocks to 30}}}}\par\smallskip
\noindent\parbox[t]{0.05\linewidth}{\centering\strut 14}\hspace{0.006\linewidth}\parbox[t]{0.04\linewidth}{\centering\strut \raisebox{0.08ex}{\ding{55}}}\hspace{0.006\linewidth}\parbox[t]{0.10\linewidth}{\centering\strut -0.0020\%}\hspace{0.006\linewidth}\parbox[t]{0.115\linewidth}{\centering\strut -75.001654}\hspace{0.006\linewidth}\parbox[t]{0.671\linewidth}{reduce num\_blocks to 28}\par\smallskip
\noindent{\setlength{\fboxsep}{0pt}\colorbox{black!4}{\parbox{\linewidth}{\parbox[t]{0.05\linewidth}{\centering\strut 15}\hspace{0.006\linewidth}\parbox[t]{0.04\linewidth}{\centering\strut \raisebox{0.08ex}{\ding{51}}}\hspace{0.006\linewidth}\parbox[t]{0.10\linewidth}{\centering\strut +0.0025\%}\hspace{0.006\linewidth}\parbox[t]{0.115\linewidth}{\centering\strut -75.003538}\hspace{0.006\linewidth}\parbox[t]{0.671\linewidth}{increase num\_blocks to 32}}}}\par\smallskip
\noindent\parbox[t]{0.05\linewidth}{\centering\strut 16}\hspace{0.006\linewidth}\parbox[t]{0.04\linewidth}{\centering\strut \raisebox{0.08ex}{\ding{51}}}\hspace{0.006\linewidth}\parbox[t]{0.10\linewidth}{\centering\strut +0.0004\%}\hspace{0.006\linewidth}\parbox[t]{0.115\linewidth}{\centering\strut -75.003821}\hspace{0.006\linewidth}\parbox[t]{0.671\linewidth}{increase num\_blocks to 34}\par\smallskip
\noindent{\setlength{\fboxsep}{0pt}\colorbox{black!4}{\parbox{\linewidth}{\parbox[t]{0.05\linewidth}{\centering\strut 17}\hspace{0.006\linewidth}\parbox[t]{0.04\linewidth}{\centering\strut \raisebox{0.08ex}{\ding{55}}}\hspace{0.006\linewidth}\parbox[t]{0.10\linewidth}{\centering\strut -0.0011\%}\hspace{0.006\linewidth}\parbox[t]{0.115\linewidth}{\centering\strut -75.003014}\hspace{0.006\linewidth}\parbox[t]{0.671\linewidth}{increase num\_blocks to 36}}}}\par\smallskip
\noindent\parbox[t]{0.05\linewidth}{\centering\strut 18}\hspace{0.006\linewidth}\parbox[t]{0.04\linewidth}{\centering\strut \raisebox{0.08ex}{\ding{55}}}\hspace{0.006\linewidth}\parbox[t]{0.10\linewidth}{\centering\strut +0.0003\%}\hspace{0.006\linewidth}\parbox[t]{0.115\linewidth}{\centering\strut -75.003221}\hspace{0.006\linewidth}\parbox[t]{0.671\linewidth}{increase num\_blocks to 35}\par\smallskip
\noindent{\setlength{\fboxsep}{0pt}\colorbox{black!4}{\parbox{\linewidth}{\parbox[t]{0.05\linewidth}{\centering\strut 19}\hspace{0.006\linewidth}\parbox[t]{0.04\linewidth}{\centering\strut \raisebox{0.08ex}{\ding{55}}}\hspace{0.006\linewidth}\parbox[t]{0.10\linewidth}{\centering\strut +0.0007\%}\hspace{0.006\linewidth}\parbox[t]{0.115\linewidth}{\centering\strut -75.003761}\hspace{0.006\linewidth}\parbox[t]{0.671\linewidth}{decrease num\_blocks to 33}}}}\par\smallskip
\noindent\parbox[t]{0.05\linewidth}{\centering\strut 20}\hspace{0.006\linewidth}\parbox[t]{0.04\linewidth}{\centering\strut \raisebox{0.08ex}{\ding{55}}}\hspace{0.006\linewidth}\parbox[t]{0.10\linewidth}{\centering\strut -0.0057\%}\hspace{0.006\linewidth}\parbox[t]{0.115\linewidth}{\centering\strut -74.999453}\hspace{0.006\linewidth}\parbox[t]{0.671\linewidth}{reduce chol\_cut to 1e-6}\par\smallskip
\noindent{\setlength{\fboxsep}{0pt}\colorbox{black!4}{\parbox{\linewidth}{\parbox[t]{0.05\linewidth}{\centering\strut 21}\hspace{0.006\linewidth}\parbox[t]{0.04\linewidth}{\centering\strut \raisebox{0.08ex}{\ding{55}}}\hspace{0.006\linewidth}\parbox[t]{0.10\linewidth}{\centering\strut +0.0037\%}\hspace{0.006\linewidth}\parbox[t]{0.115\linewidth}{\centering\strut -75.002191}\hspace{0.006\linewidth}\parbox[t]{0.671\linewidth}{set num\_steps\_per\_block to 34 and num\_blocks to 25}}}}\par\smallskip
\noindent\parbox[t]{0.05\linewidth}{\centering\strut 22}\hspace{0.006\linewidth}\parbox[t]{0.04\linewidth}{\centering\strut \raisebox{0.08ex}{\ding{51}}}\hspace{0.006\linewidth}\parbox[t]{0.10\linewidth}{\centering\strut +0.0022\%}\hspace{0.006\linewidth}\parbox[t]{0.115\linewidth}{\centering\strut -75.003821}\hspace{0.006\linewidth}\parbox[t]{0.671\linewidth}{set stabilize\_freq to 1 and pop\_control\_freq to 1}\par\smallskip
\noindent{\setlength{\fboxsep}{0pt}\colorbox{black!4}{\parbox{\linewidth}{\parbox[t]{0.05\linewidth}{\centering\strut 23}\hspace{0.006\linewidth}\parbox[t]{0.04\linewidth}{\centering\strut \raisebox{0.08ex}{\ding{55}}}\hspace{0.006\linewidth}\parbox[t]{0.10\linewidth}{\centering\strut -0.0092\%}\hspace{0.006\linewidth}\parbox[t]{0.115\linewidth}{\centering\strut -74.996920}\hspace{0.006\linewidth}\parbox[t]{0.671\linewidth}{increase chol\_cut to 5e-5}}}}\par\smallskip
\noindent\parbox[t]{0.05\linewidth}{\centering\strut 24}\hspace{0.006\linewidth}\parbox[t]{0.04\linewidth}{\centering\strut \raisebox{0.08ex}{\ding{55}}}\hspace{0.006\linewidth}\parbox[t]{0.10\linewidth}{\centering\strut +0.0010\%}\hspace{0.006\linewidth}\parbox[t]{0.115\linewidth}{\centering\strut -74.997634}\hspace{0.006\linewidth}\parbox[t]{0.671\linewidth}{increase num\_blocks to 80}\par\smallskip
\noindent{\setlength{\fboxsep}{0pt}\colorbox{black!4}{\parbox{\linewidth}{\parbox[t]{0.05\linewidth}{\centering\strut 25}\hspace{0.006\linewidth}\parbox[t]{0.04\linewidth}{\centering\strut \raisebox{0.08ex}{\ding{55}}}\hspace{0.006\linewidth}\parbox[t]{0.10\linewidth}{\centering\strut +0.0082\%}\hspace{0.006\linewidth}\parbox[t]{0.115\linewidth}{\centering\strut -75.003821}\hspace{0.006\linewidth}\parbox[t]{0.671\linewidth}{switch trial to uhf and init\_guess to atom}}}}\par\smallskip
\noindent\parbox[t]{0.05\linewidth}{\centering\strut 26}\hspace{0.006\linewidth}\parbox[t]{0.04\linewidth}{\centering\strut \raisebox{0.08ex}{\ding{55}}}\hspace{0.006\linewidth}\parbox[t]{0.10\linewidth}{\centering\strut -0.0009\%}\hspace{0.006\linewidth}\parbox[t]{0.115\linewidth}{\centering\strut -75.003110}\hspace{0.006\linewidth}\parbox[t]{0.671\linewidth}{increase timestep to 0.006}\par\smallskip
\noindent{\setlength{\fboxsep}{0pt}\colorbox{black!4}{\parbox{\linewidth}{\parbox[t]{0.05\linewidth}{\centering\strut 27}\hspace{0.006\linewidth}\parbox[t]{0.04\linewidth}{\centering\strut \raisebox{0.08ex}{\ding{51}}}\hspace{0.006\linewidth}\parbox[t]{0.10\linewidth}{\centering\strut +0.0009\%}\hspace{0.006\linewidth}\parbox[t]{0.115\linewidth}{\centering\strut -75.003821}\hspace{0.006\linewidth}\parbox[t]{0.671\linewidth}{switch trial to uhf and init\_guess to mod\_huckel}}}}\par\smallskip
\noindent\parbox[t]{0.05\linewidth}{\centering\strut 28}\hspace{0.006\linewidth}\parbox[t]{0.04\linewidth}{\centering\strut \raisebox{0.08ex}{\ding{55}}}\hspace{0.006\linewidth}\parbox[t]{0.10\linewidth}{\centering\strut +0.0000\%}\hspace{0.006\linewidth}\parbox[t]{0.115\linewidth}{\centering\strut -75.003821}\hspace{0.006\linewidth}\parbox[t]{0.671\linewidth}{set damping to 0.1 on the uhf mod\_huckel trial}\par\smallskip
\noindent{\setlength{\fboxsep}{0pt}\colorbox{black!4}{\parbox{\linewidth}{\parbox[t]{0.05\linewidth}{\centering\strut 29}\hspace{0.006\linewidth}\parbox[t]{0.04\linewidth}{\centering\strut \raisebox{0.08ex}{\ding{55}}}\hspace{0.006\linewidth}\parbox[t]{0.10\linewidth}{\centering\strut -0.0000\%}\hspace{0.006\linewidth}\parbox[t]{0.115\linewidth}{\centering\strut -75.003821}\hspace{0.006\linewidth}\parbox[t]{0.671\linewidth}{keep the uhf trial and switch init\_guess to huckel}}}}\par\smallskip
\noindent\parbox[t]{0.05\linewidth}{\centering\strut 30}\hspace{0.006\linewidth}\parbox[t]{0.04\linewidth}{\centering\strut \raisebox{0.08ex}{\ding{55}}}\hspace{0.006\linewidth}\parbox[t]{0.10\linewidth}{\centering\strut -0.0000\%}\hspace{0.006\linewidth}\parbox[t]{0.115\linewidth}{\centering\strut -75.003821}\hspace{0.006\linewidth}\parbox[t]{0.671\linewidth}{keep the uhf trial and switch init\_guess to 1e}\par\smallskip
\noindent{\setlength{\fboxsep}{0pt}\colorbox{black!4}{\parbox{\linewidth}{\parbox[t]{0.05\linewidth}{\centering\strut 31}\hspace{0.006\linewidth}\parbox[t]{0.04\linewidth}{\centering\strut \raisebox{0.08ex}{\ding{55}}}\hspace{0.006\linewidth}\parbox[t]{0.10\linewidth}{\centering\strut +0.0000\%}\hspace{0.006\linewidth}\parbox[t]{0.115\linewidth}{\centering\strut -75.003821}\hspace{0.006\linewidth}\parbox[t]{0.671\linewidth}{keep the uhf trial and switch init\_guess to minao}}}}\par\smallskip
\noindent\parbox[t]{0.05\linewidth}{\centering\strut 32}\hspace{0.006\linewidth}\parbox[t]{0.04\linewidth}{\centering\strut \raisebox{0.08ex}{\ding{55}}}\hspace{0.006\linewidth}\parbox[t]{0.10\linewidth}{\centering\strut +0.0000\%}\hspace{0.006\linewidth}\parbox[t]{0.115\linewidth}{\centering\strut -75.003821}\hspace{0.006\linewidth}\parbox[t]{0.671\linewidth}{increase diis\_space to 12 on the uhf mod\_huckel trial}\par\smallskip
\noindent{\setlength{\fboxsep}{0pt}\colorbox{black!4}{\parbox{\linewidth}{\parbox[t]{0.05\linewidth}{\centering\strut 33}\hspace{0.006\linewidth}\parbox[t]{0.04\linewidth}{\centering\strut \raisebox{0.08ex}{\ding{55}}}\hspace{0.006\linewidth}\parbox[t]{0.10\linewidth}{\centering\strut -0.0000\%}\hspace{0.006\linewidth}\parbox[t]{0.115\linewidth}{\centering\strut -75.003821}\hspace{0.006\linewidth}\parbox[t]{0.671\linewidth}{relax scf\_conv\_tol to 1e-6 on the uhf mod\_huckel trial}}}}\par\smallskip
\noindent\parbox[t]{0.05\linewidth}{\centering\strut 34}\hspace{0.006\linewidth}\parbox[t]{0.04\linewidth}{\centering\strut \raisebox{0.08ex}{\ding{55}}}\hspace{0.006\linewidth}\parbox[t]{0.10\linewidth}{\centering\strut +0.0000\%}\hspace{0.006\linewidth}\parbox[t]{0.115\linewidth}{\centering\strut -75.003821}\hspace{0.006\linewidth}\parbox[t]{0.671\linewidth}{tighten scf\_conv\_tol to 1e-8 on the uhf mod\_huckel trial}\par\smallskip
\noindent{\setlength{\fboxsep}{0pt}\colorbox{black!4}{\parbox{\linewidth}{\parbox[t]{0.05\linewidth}{\centering\strut 35}\hspace{0.006\linewidth}\parbox[t]{0.04\linewidth}{\centering\strut \raisebox{0.08ex}{\ding{55}}}\hspace{0.006\linewidth}\parbox[t]{0.10\linewidth}{\centering\strut -0.0000\%}\hspace{0.006\linewidth}\parbox[t]{0.115\linewidth}{\centering\strut -75.003821}\hspace{0.006\linewidth}\parbox[t]{0.671\linewidth}{reduce diis\_space to 4 on the uhf mod\_huckel trial}}}}\par\smallskip
\noindent\parbox[t]{0.05\linewidth}{\centering\strut 36}\hspace{0.006\linewidth}\parbox[t]{0.04\linewidth}{\centering\strut \raisebox{0.08ex}{\ding{55}}}\hspace{0.006\linewidth}\parbox[t]{0.10\linewidth}{\centering\strut -0.0008\%}\hspace{0.006\linewidth}\parbox[t]{0.115\linewidth}{\centering\strut -75.003221}\hspace{0.006\linewidth}\parbox[t]{0.671\linewidth}{increase num\_blocks to 35 on the uhf mod\_huckel trial}\par\smallskip
\noindent{\setlength{\fboxsep}{0pt}\colorbox{black!4}{\parbox{\linewidth}{\parbox[t]{0.05\linewidth}{\centering\strut 37}\hspace{0.006\linewidth}\parbox[t]{0.04\linewidth}{\centering\strut \raisebox{0.08ex}{\ding{55}}}\hspace{0.006\linewidth}\parbox[t]{0.10\linewidth}{\centering\strut +0.0007\%}\hspace{0.006\linewidth}\parbox[t]{0.115\linewidth}{\centering\strut -75.003761}\hspace{0.006\linewidth}\parbox[t]{0.671\linewidth}{decrease num\_blocks to 33 on the uhf mod\_huckel trial}}}}\par\smallskip
\noindent\parbox[t]{0.05\linewidth}{\centering\strut 38}\hspace{0.006\linewidth}\parbox[t]{0.04\linewidth}{\centering\strut \raisebox{0.08ex}{\ding{55}}}\hspace{0.006\linewidth}\parbox[t]{0.10\linewidth}{\centering\strut -0.0027\%}\hspace{0.006\linewidth}\parbox[t]{0.115\linewidth}{\centering\strut -75.001708}\hspace{0.006\linewidth}\parbox[t]{0.671\linewidth}{increase num\_steps\_per\_block to 26 on the uhf mod\_huckel trial}\par\smallskip
\noindent{\setlength{\fboxsep}{0pt}\colorbox{black!4}{\parbox{\linewidth}{\parbox[t]{0.05\linewidth}{\centering\strut 39}\hspace{0.006\linewidth}\parbox[t]{0.04\linewidth}{\centering\strut \raisebox{0.08ex}{\ding{55}}}\hspace{0.006\linewidth}\parbox[t]{0.10\linewidth}{\centering\strut +0.0028\%}\hspace{0.006\linewidth}\parbox[t]{0.115\linewidth}{\centering\strut -75.003821}\hspace{0.006\linewidth}\parbox[t]{0.671\linewidth}{switch the mod\_huckel trial from uhf to rhf}}}}\par\smallskip
\noindent\parbox[t]{0.05\linewidth}{\centering\strut 40}\hspace{0.006\linewidth}\parbox[t]{0.04\linewidth}{\centering\strut \raisebox{0.08ex}{\ding{55}}}\hspace{0.006\linewidth}\parbox[t]{0.10\linewidth}{\centering\strut +0.0000\%}\hspace{0.006\linewidth}\parbox[t]{0.115\linewidth}{\centering\strut -75.003821}\hspace{0.006\linewidth}\parbox[t]{0.671\linewidth}{increase chol\_cut to 2e-5 on the uhf mod\_huckel trial}\par\smallskip
\noindent{\setlength{\fboxsep}{0pt}\colorbox{black!4}{\parbox{\linewidth}{\parbox[t]{0.05\linewidth}{\centering\strut 41}\hspace{0.006\linewidth}\parbox[t]{0.04\linewidth}{\centering\strut \raisebox{0.08ex}{\ding{51}}}\hspace{0.006\linewidth}\parbox[t]{0.10\linewidth}{\centering\strut +0.0001\%}\hspace{0.006\linewidth}\parbox[t]{0.115\linewidth}{\centering\strut -75.003897}\hspace{0.006\linewidth}\parbox[t]{0.671\linewidth}{decrease timestep to 0.0049 on the uhf mod\_huckel trial}}}}\par\smallskip
\noindent\parbox[t]{0.05\linewidth}{\centering\strut 42}\hspace{0.006\linewidth}\parbox[t]{0.04\linewidth}{\centering\strut \raisebox{0.08ex}{\ding{51}}}\hspace{0.006\linewidth}\parbox[t]{0.10\linewidth}{\centering\strut +0.0001\%}\hspace{0.006\linewidth}\parbox[t]{0.115\linewidth}{\centering\strut -75.003973}\hspace{0.006\linewidth}\parbox[t]{0.671\linewidth}{decrease timestep to 0.0048 on the uhf mod\_huckel trial}\par\smallskip
\noindent{\setlength{\fboxsep}{0pt}\colorbox{black!4}{\parbox{\linewidth}{\parbox[t]{0.05\linewidth}{\centering\strut 43}\hspace{0.006\linewidth}\parbox[t]{0.04\linewidth}{\centering\strut \raisebox{0.08ex}{\ding{51}}}\hspace{0.006\linewidth}\parbox[t]{0.10\linewidth}{\centering\strut +0.0001\%}\hspace{0.006\linewidth}\parbox[t]{0.115\linewidth}{\centering\strut -75.004048}\hspace{0.006\linewidth}\parbox[t]{0.671\linewidth}{decrease timestep to 0.0047 on the uhf mod\_huckel trial}}}}\par\smallskip
\noindent\parbox[t]{0.05\linewidth}{\centering\strut 44}\hspace{0.006\linewidth}\parbox[t]{0.04\linewidth}{\centering\strut \raisebox{0.08ex}{\ding{51}}}\hspace{0.006\linewidth}\parbox[t]{0.10\linewidth}{\centering\strut +0.0001\%}\hspace{0.006\linewidth}\parbox[t]{0.115\linewidth}{\centering\strut -75.004122}\hspace{0.006\linewidth}\parbox[t]{0.671\linewidth}{decrease timestep to 0.0046 on the uhf mod\_huckel trial}\par\smallskip
\noindent{\setlength{\fboxsep}{0pt}\colorbox{black!4}{\parbox{\linewidth}{\parbox[t]{0.05\linewidth}{\centering\strut 45}\hspace{0.006\linewidth}\parbox[t]{0.04\linewidth}{\centering\strut \raisebox{0.08ex}{\ding{51}}}\hspace{0.006\linewidth}\parbox[t]{0.10\linewidth}{\centering\strut +0.0001\%}\hspace{0.006\linewidth}\parbox[t]{0.115\linewidth}{\centering\strut -75.004193}\hspace{0.006\linewidth}\parbox[t]{0.671\linewidth}{decrease timestep to 0.0045 on the uhf mod\_huckel trial}}}}\par\smallskip
\noindent\parbox[t]{0.05\linewidth}{\centering\strut 46}\hspace{0.006\linewidth}\parbox[t]{0.04\linewidth}{\centering\strut \raisebox{0.08ex}{\ding{51}}}\hspace{0.006\linewidth}\parbox[t]{0.10\linewidth}{\centering\strut +0.0001\%}\hspace{0.006\linewidth}\parbox[t]{0.115\linewidth}{\centering\strut -75.004261}\hspace{0.006\linewidth}\parbox[t]{0.671\linewidth}{decrease timestep to 0.0044 on the uhf mod\_huckel trial}\par\smallskip
\noindent{\setlength{\fboxsep}{0pt}\colorbox{black!4}{\parbox{\linewidth}{\parbox[t]{0.05\linewidth}{\centering\strut 47}\hspace{0.006\linewidth}\parbox[t]{0.04\linewidth}{\centering\strut \raisebox{0.08ex}{\ding{51}}}\hspace{0.006\linewidth}\parbox[t]{0.10\linewidth}{\centering\strut +0.0001\%}\hspace{0.006\linewidth}\parbox[t]{0.115\linewidth}{\centering\strut -75.004325}\hspace{0.006\linewidth}\parbox[t]{0.671\linewidth}{decrease timestep to 0.0043 on the uhf mod\_huckel trial}}}}\par\smallskip
\noindent\parbox[t]{0.05\linewidth}{\centering\strut 48}\hspace{0.006\linewidth}\parbox[t]{0.04\linewidth}{\centering\strut \raisebox{0.08ex}{\ding{51}}}\hspace{0.006\linewidth}\parbox[t]{0.10\linewidth}{\centering\strut +0.0001\%}\hspace{0.006\linewidth}\parbox[t]{0.115\linewidth}{\centering\strut -75.004383}\hspace{0.006\linewidth}\parbox[t]{0.671\linewidth}{decrease timestep to 0.0042 on the uhf mod\_huckel trial}\par\smallskip
\noindent{\setlength{\fboxsep}{0pt}\colorbox{black!4}{\parbox{\linewidth}{\parbox[t]{0.05\linewidth}{\centering\strut 49}\hspace{0.006\linewidth}\parbox[t]{0.04\linewidth}{\centering\strut \raisebox{0.08ex}{\ding{51}}}\hspace{0.006\linewidth}\parbox[t]{0.10\linewidth}{\centering\strut +0.0001\%}\hspace{0.006\linewidth}\parbox[t]{0.115\linewidth}{\centering\strut -75.004433}\hspace{0.006\linewidth}\parbox[t]{0.671\linewidth}{decrease timestep to 0.0041 on the uhf mod\_huckel trial}}}}\par\smallskip
\noindent\parbox[t]{0.05\linewidth}{\centering\strut 50}\hspace{0.006\linewidth}\parbox[t]{0.04\linewidth}{\centering\strut \raisebox{0.08ex}{\ding{51}}}\hspace{0.006\linewidth}\parbox[t]{0.10\linewidth}{\centering\strut +0.0001\%}\hspace{0.006\linewidth}\parbox[t]{0.115\linewidth}{\centering\strut -75.004474}\hspace{0.006\linewidth}\parbox[t]{0.671\linewidth}{decrease timestep to 0.0040 on the uhf mod\_huckel trial}\par\smallskip
\noindent{\setlength{\fboxsep}{0pt}\colorbox{black!4}{\parbox{\linewidth}{\parbox[t]{0.05\linewidth}{\centering\strut 51}\hspace{0.006\linewidth}\parbox[t]{0.04\linewidth}{\centering\strut \raisebox{0.08ex}{\ding{51}}}\hspace{0.006\linewidth}\parbox[t]{0.10\linewidth}{\centering\strut +0.0000\%}\hspace{0.006\linewidth}\parbox[t]{0.115\linewidth}{\centering\strut -75.004503}\hspace{0.006\linewidth}\parbox[t]{0.671\linewidth}{decrease timestep to 0.0039 on the uhf mod\_huckel trial}}}}\par\smallskip
\noindent\parbox[t]{0.05\linewidth}{\centering\strut 52}\hspace{0.006\linewidth}\parbox[t]{0.04\linewidth}{\centering\strut \raisebox{0.08ex}{\ding{51}}}\hspace{0.006\linewidth}\parbox[t]{0.10\linewidth}{\centering\strut +0.0000\%}\hspace{0.006\linewidth}\parbox[t]{0.115\linewidth}{\centering\strut -75.004518}\hspace{0.006\linewidth}\parbox[t]{0.671\linewidth}{decrease timestep to 0.0038 on the uhf mod\_huckel trial}\par\smallskip
\noindent{\setlength{\fboxsep}{0pt}\colorbox{black!4}{\parbox{\linewidth}{\parbox[t]{0.05\linewidth}{\centering\strut 53}\hspace{0.006\linewidth}\parbox[t]{0.04\linewidth}{\centering\strut \raisebox{0.08ex}{\ding{55}}}\hspace{0.006\linewidth}\parbox[t]{0.10\linewidth}{\centering\strut -0.0000\%}\hspace{0.006\linewidth}\parbox[t]{0.115\linewidth}{\centering\strut -75.004515}\hspace{0.006\linewidth}\parbox[t]{0.671\linewidth}{decrease timestep to 0.0037 on the uhf mod\_huckel trial}}}}\par\smallskip
\noindent\parbox[t]{0.05\linewidth}{\centering\strut 54}\hspace{0.006\linewidth}\parbox[t]{0.04\linewidth}{\centering\strut \raisebox{0.08ex}{\ding{51}}}\hspace{0.006\linewidth}\parbox[t]{0.10\linewidth}{\centering\strut +0.0000\%}\hspace{0.006\linewidth}\parbox[t]{0.115\linewidth}{\centering\strut -75.004519}\hspace{0.006\linewidth}\parbox[t]{0.671\linewidth}{refine timestep to 0.00375 on the uhf mod\_huckel trial}\par\smallskip
\noindent{\setlength{\fboxsep}{0pt}\colorbox{black!4}{\parbox{\linewidth}{\parbox[t]{0.05\linewidth}{\centering\strut 55}\hspace{0.006\linewidth}\parbox[t]{0.04\linewidth}{\centering\strut \raisebox{0.08ex}{\ding{55}}}\hspace{0.006\linewidth}\parbox[t]{0.10\linewidth}{\centering\strut -0.0000\%}\hspace{0.006\linewidth}\parbox[t]{0.115\linewidth}{\centering\strut -75.004519}\hspace{0.006\linewidth}\parbox[t]{0.671\linewidth}{refine timestep to 0.00374 on the uhf mod\_huckel trial}}}}\par\smallskip
\noindent\parbox[t]{0.05\linewidth}{\centering\strut 56}\hspace{0.006\linewidth}\parbox[t]{0.04\linewidth}{\centering\strut \raisebox{0.08ex}{\ding{51}}}\hspace{0.006\linewidth}\parbox[t]{0.10\linewidth}{\centering\strut +0.0000\%}\hspace{0.006\linewidth}\parbox[t]{0.115\linewidth}{\centering\strut -75.004519}\hspace{0.006\linewidth}\parbox[t]{0.671\linewidth}{refine timestep to 0.00376 on the uhf mod\_huckel trial}\par\smallskip
\noindent{\setlength{\fboxsep}{0pt}\colorbox{black!4}{\parbox{\linewidth}{\parbox[t]{0.05\linewidth}{\centering\strut 57}\hspace{0.006\linewidth}\parbox[t]{0.04\linewidth}{\centering\strut \raisebox{0.08ex}{\ding{55}}}\hspace{0.006\linewidth}\parbox[t]{0.10\linewidth}{\centering\strut -0.0000\%}\hspace{0.006\linewidth}\parbox[t]{0.115\linewidth}{\centering\strut -75.004519}\hspace{0.006\linewidth}\parbox[t]{0.671\linewidth}{refine timestep to 0.00377 on the uhf mod\_huckel trial}}}}\par\smallskip
\noindent\parbox[t]{0.05\linewidth}{\centering\strut 58}\hspace{0.006\linewidth}\parbox[t]{0.04\linewidth}{\centering\strut \raisebox{0.08ex}{\ding{55}}}\hspace{0.006\linewidth}\parbox[t]{0.10\linewidth}{\centering\strut -0.0014\%}\hspace{0.006\linewidth}\parbox[t]{0.115\linewidth}{\centering\strut -75.003434}\hspace{0.006\linewidth}\parbox[t]{0.671\linewidth}{increase num\_blocks to 35 at timestep 0.00376 on the uhf mod\_huckel trial}\par\smallskip
\noindent{\setlength{\fboxsep}{0pt}\colorbox{black!4}{\parbox{\linewidth}{\parbox[t]{0.05\linewidth}{\centering\strut 59}\hspace{0.006\linewidth}\parbox[t]{0.04\linewidth}{\centering\strut \raisebox{0.08ex}{\ding{51}}}\hspace{0.006\linewidth}\parbox[t]{0.10\linewidth}{\centering\strut +0.0014\%}\hspace{0.006\linewidth}\parbox[t]{0.115\linewidth}{\centering\strut -75.004519}\hspace{0.006\linewidth}\parbox[t]{0.671\linewidth}{refine timestep to 0.003765 on the uhf mod\_huckel trial}}}}\par\smallskip
\noindent\parbox[t]{0.05\linewidth}{\centering\strut 60}\hspace{0.006\linewidth}\parbox[t]{0.04\linewidth}{\centering\strut \raisebox{0.08ex}{\ding{55}}}\hspace{0.006\linewidth}\parbox[t]{0.10\linewidth}{\centering\strut -0.0000\%}\hspace{0.006\linewidth}\parbox[t]{0.115\linewidth}{\centering\strut -75.004519}\hspace{0.006\linewidth}\parbox[t]{0.671\linewidth}{refine timestep to 0.003766 on the uhf mod\_huckel trial}\par\smallskip
\noindent{\setlength{\fboxsep}{0pt}\colorbox{black!4}{\parbox{\linewidth}{\parbox[t]{0.05\linewidth}{\centering\strut 61}\hspace{0.006\linewidth}\parbox[t]{0.04\linewidth}{\centering\strut \raisebox{0.08ex}{\ding{51}}}\hspace{0.006\linewidth}\parbox[t]{0.10\linewidth}{\centering\strut +0.0000\%}\hspace{0.006\linewidth}\parbox[t]{0.115\linewidth}{\centering\strut -75.004519}\hspace{0.006\linewidth}\parbox[t]{0.671\linewidth}{refine timestep to 0.003764 on the uhf mod\_huckel trial}}}}\par\smallskip
\noindent\parbox[t]{0.05\linewidth}{\centering\strut 62}\hspace{0.006\linewidth}\parbox[t]{0.04\linewidth}{\centering\strut \raisebox{0.08ex}{\ding{55}}}\hspace{0.006\linewidth}\parbox[t]{0.10\linewidth}{\centering\strut -0.0000\%}\hspace{0.006\linewidth}\parbox[t]{0.115\linewidth}{\centering\strut -75.004519}\hspace{0.006\linewidth}\parbox[t]{0.671\linewidth}{refine timestep to 0.003763 on the uhf mod\_huckel trial}\par\smallskip
\noindent{\setlength{\fboxsep}{0pt}\colorbox{black!4}{\parbox{\linewidth}{\parbox[t]{0.05\linewidth}{\centering\strut 63}\hspace{0.006\linewidth}\parbox[t]{0.04\linewidth}{\centering\strut \raisebox{0.08ex}{\ding{51}}}\hspace{0.006\linewidth}\parbox[t]{0.10\linewidth}{\centering\strut +0.0000\%}\hspace{0.006\linewidth}\parbox[t]{0.115\linewidth}{\centering\strut -75.004519}\hspace{0.006\linewidth}\parbox[t]{0.671\linewidth}{increase stabilize\_freq to 2 at timestep 0.003764 on the uhf mod\_huckel trial}}}}\par\smallskip
\noindent\parbox[t]{0.05\linewidth}{\centering\strut 64}\hspace{0.006\linewidth}\parbox[t]{0.04\linewidth}{\centering\strut \raisebox{0.08ex}{\ding{51}}}\hspace{0.006\linewidth}\parbox[t]{0.10\linewidth}{\centering\strut +0.0000\%}\hspace{0.006\linewidth}\parbox[t]{0.115\linewidth}{\centering\strut -75.004519}\hspace{0.006\linewidth}\parbox[t]{0.671\linewidth}{increase stabilize\_freq to 3 at timestep 0.003764 on the uhf mod\_huckel trial}\par\smallskip
\noindent{\setlength{\fboxsep}{0pt}\colorbox{black!4}{\parbox{\linewidth}{\parbox[t]{0.05\linewidth}{\centering\strut 65}\hspace{0.006\linewidth}\parbox[t]{0.04\linewidth}{\centering\strut \raisebox{0.08ex}{\ding{55}}}\hspace{0.006\linewidth}\parbox[t]{0.10\linewidth}{\centering\strut -0.0000\%}\hspace{0.006\linewidth}\parbox[t]{0.115\linewidth}{\centering\strut -75.004519}\hspace{0.006\linewidth}\parbox[t]{0.671\linewidth}{increase stabilize\_freq to 4 at timestep 0.003764 on the uhf mod\_huckel trial}}}}\par\smallskip
\noindent\parbox[t]{0.05\linewidth}{\centering\strut 66}\hspace{0.006\linewidth}\parbox[t]{0.04\linewidth}{\centering\strut \raisebox{0.08ex}{\ding{55}}}\hspace{0.006\linewidth}\parbox[t]{0.10\linewidth}{\centering\strut +0.0000\%}\hspace{0.006\linewidth}\parbox[t]{0.115\linewidth}{\centering\strut -75.004519}\hspace{0.006\linewidth}\parbox[t]{0.671\linewidth}{increase pop\_control\_freq to 2 with stabilize\_freq 3 at timestep 0.003764}\par\smallskip
\noindent{\setlength{\fboxsep}{0pt}\colorbox{black!4}{\parbox{\linewidth}{\parbox[t]{0.05\linewidth}{\centering\strut 67}\hspace{0.006\linewidth}\parbox[t]{0.04\linewidth}{\centering\strut \raisebox{0.08ex}{\ding{55}}}\hspace{0.006\linewidth}\parbox[t]{0.10\linewidth}{\centering\strut -0.0000\%}\hspace{0.006\linewidth}\parbox[t]{0.115\linewidth}{\centering\strut -75.004519}\hspace{0.006\linewidth}\parbox[t]{0.671\linewidth}{refine timestep to 0.0037645 with stabilize\_freq 3 on the uhf mod\_huckel trial}}}}\par\smallskip
\noindent\parbox[t]{0.05\linewidth}{\centering\strut 68}\hspace{0.006\linewidth}\parbox[t]{0.04\linewidth}{\centering\strut \raisebox{0.08ex}{\ding{55}}}\hspace{0.006\linewidth}\parbox[t]{0.10\linewidth}{\centering\strut +0.0000\%}\hspace{0.006\linewidth}\parbox[t]{0.115\linewidth}{\centering\strut -75.004519}\hspace{0.006\linewidth}\parbox[t]{0.671\linewidth}{increase pop\_control\_freq to 3 with stabilize\_freq 3 at timestep 0.003764}\par\smallskip
\noindent{\setlength{\fboxsep}{0pt}\colorbox{black!4}{\parbox{\linewidth}{\parbox[t]{0.05\linewidth}{\centering\strut 69}\hspace{0.006\linewidth}\parbox[t]{0.04\linewidth}{\centering\strut \raisebox{0.08ex}{\ding{51}}}\hspace{0.006\linewidth}\parbox[t]{0.10\linewidth}{\centering\strut +0.0000\%}\hspace{0.006\linewidth}\parbox[t]{0.115\linewidth}{\centering\strut -75.004519}\hspace{0.006\linewidth}\parbox[t]{0.671\linewidth}{refine timestep to 0.0037638 with stabilize\_freq 3 on the uhf mod\_huckel trial}}}}\par\smallskip
\noindent\parbox[t]{0.05\linewidth}{\centering\strut 70}\hspace{0.006\linewidth}\parbox[t]{0.04\linewidth}{\centering\strut \raisebox{0.08ex}{\ding{55}}}\hspace{0.006\linewidth}\parbox[t]{0.10\linewidth}{\centering\strut -0.0000\%}\hspace{0.006\linewidth}\parbox[t]{0.115\linewidth}{\centering\strut -75.004519}\hspace{0.006\linewidth}\parbox[t]{0.671\linewidth}{refine timestep to 0.0037636 with stabilize\_freq 3 on the uhf mod\_huckel trial}\par\smallskip
\noindent{\setlength{\fboxsep}{0pt}\colorbox{black!4}{\parbox{\linewidth}{\parbox[t]{0.05\linewidth}{\centering\strut 71}\hspace{0.006\linewidth}\parbox[t]{0.04\linewidth}{\centering\strut \raisebox{0.08ex}{\ding{55}}}\hspace{0.006\linewidth}\parbox[t]{0.10\linewidth}{\centering\strut +0.0000\%}\hspace{0.006\linewidth}\parbox[t]{0.115\linewidth}{\centering\strut -75.004519}\hspace{0.006\linewidth}\parbox[t]{0.671\linewidth}{refine timestep to 0.0037639 with stabilize\_freq 3 on the uhf mod\_huckel trial}}}}\par\smallskip
\noindent\parbox[t]{0.05\linewidth}{\centering\strut 72}\hspace{0.006\linewidth}\parbox[t]{0.04\linewidth}{\centering\strut \raisebox{0.08ex}{\ding{55}}}\hspace{0.006\linewidth}\parbox[t]{0.10\linewidth}{\centering\strut +0.0000\%}\hspace{0.006\linewidth}\parbox[t]{0.115\linewidth}{\centering\strut -75.004519}\hspace{0.006\linewidth}\parbox[t]{0.671\linewidth}{refine timestep to 0.00376382 with stabilize\_freq 3 on the uhf mod\_huckel trial}\par\smallskip
\noindent{\setlength{\fboxsep}{0pt}\colorbox{black!4}{\parbox{\linewidth}{\parbox[t]{0.05\linewidth}{\centering\strut 73}\hspace{0.006\linewidth}\parbox[t]{0.04\linewidth}{\centering\strut \raisebox{0.08ex}{\ding{51}}}\hspace{0.006\linewidth}\parbox[t]{0.10\linewidth}{\centering\strut +0.0000\%}\hspace{0.006\linewidth}\parbox[t]{0.115\linewidth}{\centering\strut -75.004519}\hspace{0.006\linewidth}\parbox[t]{0.671\linewidth}{refine timestep to 0.00376379 with stabilize\_freq 3 on the uhf mod\_huckel trial}}}}\par\smallskip
\noindent\parbox[t]{0.05\linewidth}{\centering\strut 74}\hspace{0.006\linewidth}\parbox[t]{0.04\linewidth}{\centering\strut \raisebox{0.08ex}{\ding{55}}}\hspace{0.006\linewidth}\parbox[t]{0.10\linewidth}{\centering\strut -0.0000\%}\hspace{0.006\linewidth}\parbox[t]{0.115\linewidth}{\centering\strut -75.004519}\hspace{0.006\linewidth}\parbox[t]{0.671\linewidth}{refine timestep to 0.00376378 with stabilize\_freq 3 on the uhf mod\_huckel trial}\par\smallskip
\noindent{\setlength{\fboxsep}{0pt}\colorbox{black!4}{\parbox{\linewidth}{\parbox[t]{0.05\linewidth}{\centering\strut 75}\hspace{0.006\linewidth}\parbox[t]{0.04\linewidth}{\centering\strut \raisebox{0.08ex}{\ding{55}}}\hspace{0.006\linewidth}\parbox[t]{0.10\linewidth}{\centering\strut +0.0000\%}\hspace{0.006\linewidth}\parbox[t]{0.115\linewidth}{\centering\strut -75.004519}\hspace{0.006\linewidth}\parbox[t]{0.671\linewidth}{refine timestep to 0.003763785 with stabilize\_freq 3 on the uhf mod\_huckel trial}}}}\par\smallskip
\noindent\parbox[t]{0.05\linewidth}{\centering\strut 76}\hspace{0.006\linewidth}\parbox[t]{0.04\linewidth}{\centering\strut \raisebox{0.08ex}{\ding{55}}}\hspace{0.006\linewidth}\parbox[t]{0.10\linewidth}{\centering\strut -0.0006\%}\hspace{0.006\linewidth}\parbox[t]{0.115\linewidth}{\centering\strut -75.004078}\hspace{0.006\linewidth}\parbox[t]{0.671\linewidth}{decrease num\_blocks to 33 at timestep 0.00376379 with stabilize\_freq 3}\par\smallskip
\noindent{\setlength{\fboxsep}{0pt}\colorbox{black!4}{\parbox{\linewidth}{\parbox[t]{0.05\linewidth}{\centering\strut 77}\hspace{0.006\linewidth}\parbox[t]{0.04\linewidth}{\centering\strut \raisebox{0.08ex}{\ding{55}}}\hspace{0.006\linewidth}\parbox[t]{0.10\linewidth}{\centering\strut +0.0006\%}\hspace{0.006\linewidth}\parbox[t]{0.115\linewidth}{\centering\strut -75.004519}\hspace{0.006\linewidth}\parbox[t]{0.671\linewidth}{refine timestep to 0.003763795 with stabilize\_freq 3 on the uhf mod\_huckel trial}}}}\par\smallskip
\noindent\parbox[t]{0.05\linewidth}{\centering\strut 78}\hspace{0.006\linewidth}\parbox[t]{0.04\linewidth}{\centering\strut \raisebox{0.08ex}{\ding{55}}}\hspace{0.006\linewidth}\parbox[t]{0.10\linewidth}{\centering\strut +0.0000\%}\hspace{0.006\linewidth}\parbox[t]{0.115\linewidth}{\centering\strut -75.004519}\hspace{0.006\linewidth}\parbox[t]{0.671\linewidth}{refine timestep to 0.003763789 with stabilize\_freq 3 on the uhf mod\_huckel trial}\par\smallskip
\noindent{\setlength{\fboxsep}{0pt}\colorbox{black!4}{\parbox{\linewidth}{\parbox[t]{0.05\linewidth}{\centering\strut 79}\hspace{0.006\linewidth}\parbox[t]{0.04\linewidth}{\centering\strut \raisebox{0.08ex}{\ding{51}}}\hspace{0.006\linewidth}\parbox[t]{0.10\linewidth}{\centering\strut +0.0000\%}\hspace{0.006\linewidth}\parbox[t]{0.115\linewidth}{\centering\strut -75.004519}\hspace{0.006\linewidth}\parbox[t]{0.671\linewidth}{increase stabilize\_freq to 5 at timestep 0.00376379 on the uhf mod\_huckel trial}}}}\par\smallskip
\noindent\parbox[t]{0.05\linewidth}{\centering\strut 80}\hspace{0.006\linewidth}\parbox[t]{0.04\linewidth}{\centering\strut \raisebox{0.08ex}{\ding{55}}}\hspace{0.006\linewidth}\parbox[t]{0.10\linewidth}{\centering\strut -0.0000\%}\hspace{0.006\linewidth}\parbox[t]{0.115\linewidth}{\centering\strut -75.004519}\hspace{0.006\linewidth}\parbox[t]{0.671\linewidth}{increase stabilize\_freq to 6 at timestep 0.00376379 on the uhf mod\_huckel trial}\par\smallskip
\noindent{\setlength{\fboxsep}{0pt}\colorbox{black!4}{\parbox{\linewidth}{\parbox[t]{0.05\linewidth}{\centering\strut 81}\hspace{0.006\linewidth}\parbox[t]{0.04\linewidth}{\centering\strut \raisebox{0.08ex}{\ding{55}}}\hspace{0.006\linewidth}\parbox[t]{0.10\linewidth}{\centering\strut -0.0000\%}\hspace{0.006\linewidth}\parbox[t]{0.115\linewidth}{\centering\strut -75.004519}\hspace{0.006\linewidth}\parbox[t]{0.671\linewidth}{set stabilize\_freq to 25 at timestep 0.00376379 on the uhf mod\_huckel trial}}}}\par\smallskip
\noindent\parbox[t]{0.05\linewidth}{\centering\strut 82}\hspace{0.006\linewidth}\parbox[t]{0.04\linewidth}{\centering\strut \raisebox{0.08ex}{\ding{55}}}\hspace{0.006\linewidth}\parbox[t]{0.10\linewidth}{\centering\strut -0.0000\%}\hspace{0.006\linewidth}\parbox[t]{0.115\linewidth}{\centering\strut -75.004519}\hspace{0.006\linewidth}\parbox[t]{0.671\linewidth}{refine timestep to 0.00376380 with stabilize\_freq 5 on the uhf mod\_huckel trial}\par\smallskip
\noindent{\setlength{\fboxsep}{0pt}\colorbox{black!4}{\parbox{\linewidth}{\parbox[t]{0.05\linewidth}{\centering\strut 83}\hspace{0.006\linewidth}\parbox[t]{0.04\linewidth}{\centering\strut \raisebox{0.08ex}{\ding{55}}}\hspace{0.006\linewidth}\parbox[t]{0.10\linewidth}{\centering\strut +0.0000\%}\hspace{0.006\linewidth}\parbox[t]{0.115\linewidth}{\centering\strut -75.004519}\hspace{0.006\linewidth}\parbox[t]{0.671\linewidth}{refine timestep to 0.003763788 with stabilize\_freq 5 on the uhf mod\_huckel trial}}}}\par\smallskip
\noindent\parbox[t]{0.05\linewidth}{\centering\strut 84}\hspace{0.006\linewidth}\parbox[t]{0.04\linewidth}{\centering\strut \raisebox{0.08ex}{\ding{55}}}\hspace{0.006\linewidth}\parbox[t]{0.10\linewidth}{\centering\strut -0.0008\%}\hspace{0.006\linewidth}\parbox[t]{0.115\linewidth}{\centering\strut -75.003901}\hspace{0.006\linewidth}\parbox[t]{0.671\linewidth}{decrease num\_steps\_per\_block to 24 at timestep 0.00376379 with stabilize\_freq 5}\par\smallskip
\noindent{\setlength{\fboxsep}{0pt}\colorbox{black!4}{\parbox{\linewidth}{\parbox[t]{0.05\linewidth}{\centering\strut 85}\hspace{0.006\linewidth}\parbox[t]{0.04\linewidth}{\centering\strut \raisebox{0.08ex}{\ding{55}}}\hspace{0.006\linewidth}\parbox[t]{0.10\linewidth}{\centering\strut +0.0008\%}\hspace{0.006\linewidth}\parbox[t]{0.115\linewidth}{\centering\strut -75.004519}\hspace{0.006\linewidth}\parbox[t]{0.671\linewidth}{set pop\_control\_freq to 5 with stabilize\_freq 5 at timestep 0.00376379}}}}\par\smallskip
\noindent\parbox[t]{0.05\linewidth}{\centering\strut 86}\hspace{0.006\linewidth}\parbox[t]{0.04\linewidth}{\centering\strut \raisebox{0.08ex}{\ding{55}}}\hspace{0.006\linewidth}\parbox[t]{0.10\linewidth}{\centering\strut -0.0000\%}\hspace{0.006\linewidth}\parbox[t]{0.115\linewidth}{\centering\strut -75.004519}\hspace{0.006\linewidth}\parbox[t]{0.671\linewidth}{increase stabilize\_freq to 7 at timestep 0.00376379 on the uhf mod\_huckel trial}\par\smallskip
\noindent{\setlength{\fboxsep}{0pt}\colorbox{black!4}{\parbox{\linewidth}{\parbox[t]{0.05\linewidth}{\centering\strut 87}\hspace{0.006\linewidth}\parbox[t]{0.04\linewidth}{\centering\strut \raisebox{0.08ex}{\ding{55}}}\hspace{0.006\linewidth}\parbox[t]{0.10\linewidth}{\centering\strut +0.0000\%}\hspace{0.006\linewidth}\parbox[t]{0.115\linewidth}{\centering\strut -75.004519}\hspace{0.006\linewidth}\parbox[t]{0.671\linewidth}{increase stabilize\_freq to 8 at timestep 0.00376379 on the uhf mod\_huckel trial}}}}\par\smallskip
\noindent\parbox[t]{0.05\linewidth}{\centering\strut 88}\hspace{0.006\linewidth}\parbox[t]{0.04\linewidth}{\centering\strut \raisebox{0.08ex}{\ding{55}}}\hspace{0.006\linewidth}\parbox[t]{0.10\linewidth}{\centering\strut -0.0000\%}\hspace{0.006\linewidth}\parbox[t]{0.115\linewidth}{\centering\strut -75.004519}\hspace{0.006\linewidth}\parbox[t]{0.671\linewidth}{increase stabilize\_freq to 10 at timestep 0.00376379 on the uhf mod\_huckel trial}\par\smallskip
\noindent{\setlength{\fboxsep}{0pt}\colorbox{black!4}{\parbox{\linewidth}{\parbox[t]{0.05\linewidth}{\centering\strut 89}\hspace{0.006\linewidth}\parbox[t]{0.04\linewidth}{\centering\strut \raisebox{0.08ex}{\ding{55}}}\hspace{0.006\linewidth}\parbox[t]{0.10\linewidth}{\centering\strut -0.0000\%}\hspace{0.006\linewidth}\parbox[t]{0.115\linewidth}{\centering\strut -75.004519}\hspace{0.006\linewidth}\parbox[t]{0.671\linewidth}{set level\_shift to 0.05 on the uhf mod\_huckel trial at timestep 0.00376379 with stabilize\_freq 5}}}}\par\smallskip
\noindent\parbox[t]{0.05\linewidth}{\centering\strut 90}\hspace{0.006\linewidth}\parbox[t]{0.04\linewidth}{\centering\strut \raisebox{0.08ex}{\ding{55}}}\hspace{0.006\linewidth}\parbox[t]{0.10\linewidth}{\centering\strut -0.0000\%}\hspace{0.006\linewidth}\parbox[t]{0.115\linewidth}{\centering\strut -75.004519}\hspace{0.006\linewidth}\parbox[t]{0.671\linewidth}{tighten scf\_conv\_tol to 5e-8 on the uhf mod\_huckel trial at timestep 0.00376379 with stabilize\_freq 5}\par\smallskip
\noindent{\setlength{\fboxsep}{0pt}\colorbox{black!4}{\parbox{\linewidth}{\parbox[t]{0.05\linewidth}{\centering\strut 91}\hspace{0.006\linewidth}\parbox[t]{0.04\linewidth}{\centering\strut \raisebox{0.08ex}{\ding{55}}}\hspace{0.006\linewidth}\parbox[t]{0.10\linewidth}{\centering\strut +0.0000\%}\hspace{0.006\linewidth}\parbox[t]{0.115\linewidth}{\centering\strut -75.004519}\hspace{0.006\linewidth}\parbox[t]{0.671\linewidth}{relax scf\_conv\_tol to 2e-7 on the uhf mod\_huckel trial at timestep 0.00376379 with stabilize\_freq 5}}}}\par\smallskip
\noindent\parbox[t]{0.05\linewidth}{\centering\strut 92}\hspace{0.006\linewidth}\parbox[t]{0.04\linewidth}{\centering\strut \raisebox{0.08ex}{\ding{55}}}\hspace{0.006\linewidth}\parbox[t]{0.10\linewidth}{\centering\strut -0.0000\%}\hspace{0.006\linewidth}\parbox[t]{0.115\linewidth}{\centering\strut -75.004519}\hspace{0.006\linewidth}\parbox[t]{0.671\linewidth}{increase pop\_control\_freq to 2 with stabilize\_freq 5 at timestep 0.00376379 on the uhf mod\_huckel trial}\par\smallskip
\noindent{\setlength{\fboxsep}{0pt}\colorbox{black!4}{\parbox{\linewidth}{\parbox[t]{0.05\linewidth}{\centering\strut 93}\hspace{0.006\linewidth}\parbox[t]{0.04\linewidth}{\centering\strut \raisebox{0.08ex}{\ding{55}}}\hspace{0.006\linewidth}\parbox[t]{0.10\linewidth}{\centering\strut +0.0000\%}\hspace{0.006\linewidth}\parbox[t]{0.115\linewidth}{\centering\strut -75.004519}\hspace{0.006\linewidth}\parbox[t]{0.671\linewidth}{increase pop\_control\_freq to 3 with stabilize\_freq 5 at timestep 0.00376379 on the uhf mod\_huckel trial}}}}\par\smallskip
\noindent\parbox[t]{0.05\linewidth}{\centering\strut 94}\hspace{0.006\linewidth}\parbox[t]{0.04\linewidth}{\centering\strut \raisebox{0.08ex}{\ding{51}}}\hspace{0.006\linewidth}\parbox[t]{0.10\linewidth}{\centering\strut +0.0000\%}\hspace{0.006\linewidth}\parbox[t]{0.115\linewidth}{\centering\strut -75.004519}\hspace{0.006\linewidth}\parbox[t]{0.671\linewidth}{increase pop\_control\_freq to 4 with stabilize\_freq 5 at timestep 0.00376379 on the uhf mod\_huckel trial}\par\smallskip
\noindent{\setlength{\fboxsep}{0pt}\colorbox{black!4}{\parbox{\linewidth}{\parbox[t]{0.05\linewidth}{\centering\strut 95}\hspace{0.006\linewidth}\parbox[t]{0.04\linewidth}{\centering\strut \raisebox{0.08ex}{\ding{55}}}\hspace{0.006\linewidth}\parbox[t]{0.10\linewidth}{\centering\strut -0.0000\%}\hspace{0.006\linewidth}\parbox[t]{0.115\linewidth}{\centering\strut -75.004519}\hspace{0.006\linewidth}\parbox[t]{0.671\linewidth}{refine timestep to 0.003763788 with pop\_control\_freq 4 and stabilize\_freq 5 on the uhf mod\_huckel trial}}}}\par\smallskip
\noindent\parbox[t]{0.05\linewidth}{\centering\strut 96}\hspace{0.006\linewidth}\parbox[t]{0.04\linewidth}{\centering\strut \raisebox{0.08ex}{\ding{55}}}\hspace{0.006\linewidth}\parbox[t]{0.10\linewidth}{\centering\strut -0.0000\%}\hspace{0.006\linewidth}\parbox[t]{0.115\linewidth}{\centering\strut -75.004519}\hspace{0.006\linewidth}\parbox[t]{0.671\linewidth}{refine timestep to 0.003763792 with pop\_control\_freq 4 and stabilize\_freq 5 on the uhf mod\_huckel trial}\par\smallskip
\noindent{\setlength{\fboxsep}{0pt}\colorbox{black!4}{\parbox{\linewidth}{\parbox[t]{0.05\linewidth}{\centering\strut 97}\hspace{0.006\linewidth}\parbox[t]{0.04\linewidth}{\centering\strut \raisebox{0.08ex}{\ding{55}}}\hspace{0.006\linewidth}\parbox[t]{0.10\linewidth}{\centering\strut -0.0000\%}\hspace{0.006\linewidth}\parbox[t]{0.115\linewidth}{\centering\strut -75.004519}\hspace{0.006\linewidth}\parbox[t]{0.671\linewidth}{decrease stabilize\_freq to 4 with pop\_control\_freq 4 at timestep 0.00376379 on the uhf mod\_huckel trial}}}}\par\smallskip
\noindent\parbox[t]{0.05\linewidth}{\centering\strut 98}\hspace{0.006\linewidth}\parbox[t]{0.04\linewidth}{\centering\strut \raisebox{0.08ex}{\ding{55}}}\hspace{0.006\linewidth}\parbox[t]{0.10\linewidth}{\centering\strut +0.0000\%}\hspace{0.006\linewidth}\parbox[t]{0.115\linewidth}{\centering\strut -75.004519}\hspace{0.006\linewidth}\parbox[t]{0.671\linewidth}{increase stabilize\_freq to 6 with pop\_control\_freq 4 at timestep 0.00376379 on the uhf mod\_huckel trial}\par\smallskip
\noindent{\setlength{\fboxsep}{0pt}\colorbox{black!4}{\parbox{\linewidth}{\parbox[t]{0.05\linewidth}{\centering\strut 99}\hspace{0.006\linewidth}\parbox[t]{0.04\linewidth}{\centering\strut \raisebox{0.08ex}{\ding{55}}}\hspace{0.006\linewidth}\parbox[t]{0.10\linewidth}{\centering\strut -0.0000\%}\hspace{0.006\linewidth}\parbox[t]{0.115\linewidth}{\centering\strut -75.004519}\hspace{0.006\linewidth}\parbox[t]{0.671\linewidth}{increase pop\_control\_freq to 8 with stabilize\_freq 5 at timestep 0.00376379 on the uhf mod\_huckel trial}}}}\par\smallskip
\noindent\parbox[t]{0.05\linewidth}{\centering\strut 100}\hspace{0.006\linewidth}\parbox[t]{0.04\linewidth}{\centering\strut \raisebox{0.08ex}{\ding{55}}}\hspace{0.006\linewidth}\parbox[t]{0.10\linewidth}{\centering\strut -0.0000\%}\hspace{0.006\linewidth}\parbox[t]{0.115\linewidth}{\centering\strut -75.004519}\hspace{0.006\linewidth}\parbox[t]{0.671\linewidth}{increase pop\_control\_freq to 6 with stabilize\_freq 5 at timestep 0.00376379 on the uhf mod\_huckel trial}\par\smallskip
\noindent\rule{\linewidth}{0.5pt}
\par\medskip
\refstepcounter{table}
\label{tab:supp_afqmc_n2}
\noindent\textbf{Table \thetable.} AFQMC mutation log for N$_2$ full 100-iteration campaign. Relative gain is measured against the immediately previous scored iteration using the live AFQMC score.
\par\medskip
\noindent\rule{\linewidth}{0.5pt}
\noindent{\setlength{\fboxsep}{0pt}\colorbox{black!4}{\parbox{\linewidth}{\parbox[c][2.2em][c]{0.05\linewidth}{\centering \textbf{Iter.}}\hspace{0.006\linewidth}\parbox[c][2.2em][c]{0.04\linewidth}{\centering \textbf{A/R}}\hspace{0.006\linewidth}\parbox[c][2.2em][c]{0.10\linewidth}{\centering \textbf{Relative gain}}\hspace{0.006\linewidth}\parbox[c][2.2em][c]{0.115\linewidth}{\centering \textbf{Score}}\hspace{0.006\linewidth}\parbox[c][2.2em][c]{0.671\linewidth}{\centering \textbf{Mutation summary}}}}}\par\smallskip
\noindent\rule{\linewidth}{0.5pt}
\noindent{\setlength{\fboxsep}{0pt}\colorbox{black!4}{\parbox{\linewidth}{\parbox[t]{0.05\linewidth}{\centering\strut 1}\hspace{0.006\linewidth}\parbox[t]{0.04\linewidth}{\centering\strut \raisebox{0.08ex}{\ding{51}}}\hspace{0.006\linewidth}\parbox[t]{0.10\linewidth}{\centering\strut ---}\hspace{0.006\linewidth}\parbox[t]{0.115\linewidth}{\centering\strut -107.550415}\hspace{0.006\linewidth}\parbox[t]{0.671\linewidth}{num\_walkers\_per\_rank = 128, num\_steps\_per\_block = 23, num\_blocks = 25, timestep = 0.0035451762, stabilize\_freq = 6, pop\_control\_freq = 4}}}}\par\smallskip
\noindent\parbox[t]{0.05\linewidth}{\centering\strut 2}\hspace{0.006\linewidth}\parbox[t]{0.04\linewidth}{\centering\strut \raisebox{0.08ex}{\ding{51}}}\hspace{0.006\linewidth}\parbox[t]{0.10\linewidth}{\centering\strut +0.0187\%}\hspace{0.006\linewidth}\parbox[t]{0.115\linewidth}{\centering\strut -107.570541}\hspace{0.006\linewidth}\parbox[t]{0.671\linewidth}{num\_walkers\_per\_rank = 256}\par\smallskip
\noindent{\setlength{\fboxsep}{0pt}\colorbox{black!4}{\parbox{\linewidth}{\parbox[t]{0.05\linewidth}{\centering\strut 3}\hspace{0.006\linewidth}\parbox[t]{0.04\linewidth}{\centering\strut \raisebox{0.08ex}{\ding{51}}}\hspace{0.006\linewidth}\parbox[t]{0.10\linewidth}{\centering\strut +0.0129\%}\hspace{0.006\linewidth}\parbox[t]{0.115\linewidth}{\centering\strut -107.584425}\hspace{0.006\linewidth}\parbox[t]{0.671\linewidth}{num\_walkers\_per\_rank = 512}}}}\par\smallskip
\noindent\parbox[t]{0.05\linewidth}{\centering\strut 4}\hspace{0.006\linewidth}\parbox[t]{0.04\linewidth}{\centering\strut \raisebox{0.08ex}{\ding{55}}}\hspace{0.006\linewidth}\parbox[t]{0.10\linewidth}{\centering\strut -0.0036\%}\hspace{0.006\linewidth}\parbox[t]{0.115\linewidth}{\centering\strut -107.580537}\hspace{0.006\linewidth}\parbox[t]{0.671\linewidth}{num\_blocks = 30}\par\smallskip
\noindent{\setlength{\fboxsep}{0pt}\colorbox{black!4}{\parbox{\linewidth}{\parbox[t]{0.05\linewidth}{\centering\strut 5}\hspace{0.006\linewidth}\parbox[t]{0.04\linewidth}{\centering\strut \raisebox{0.08ex}{\ding{55}}}\hspace{0.006\linewidth}\parbox[t]{0.10\linewidth}{\centering\strut -0.0197\%}\hspace{0.006\linewidth}\parbox[t]{0.115\linewidth}{\centering\strut -107.559383}\hspace{0.006\linewidth}\parbox[t]{0.671\linewidth}{chol\_cut = 1e-6}}}}\par\smallskip
\noindent\parbox[t]{0.05\linewidth}{\centering\strut 6}\hspace{0.006\linewidth}\parbox[t]{0.04\linewidth}{\centering\strut \raisebox{0.08ex}{\ding{55}}}\hspace{0.006\linewidth}\parbox[t]{0.10\linewidth}{\centering\strut +0.0231\%}\hspace{0.006\linewidth}\parbox[t]{0.115\linewidth}{\centering\strut -107.584241}\hspace{0.006\linewidth}\parbox[t]{0.671\linewidth}{num\_blocks = 24}\par\smallskip
\noindent{\setlength{\fboxsep}{0pt}\colorbox{black!4}{\parbox{\linewidth}{\parbox[t]{0.05\linewidth}{\centering\strut 7}\hspace{0.006\linewidth}\parbox[t]{0.04\linewidth}{\centering\strut \raisebox{0.08ex}{\ding{51}}}\hspace{0.006\linewidth}\parbox[t]{0.10\linewidth}{\centering\strut +0.0003\%}\hspace{0.006\linewidth}\parbox[t]{0.115\linewidth}{\centering\strut -107.584574}\hspace{0.006\linewidth}\parbox[t]{0.671\linewidth}{num\_blocks = 26}}}}\par\smallskip
\noindent\parbox[t]{0.05\linewidth}{\centering\strut 8}\hspace{0.006\linewidth}\parbox[t]{0.04\linewidth}{\centering\strut \raisebox{0.08ex}{\ding{55}}}\hspace{0.006\linewidth}\parbox[t]{0.10\linewidth}{\centering\strut -0.0014\%}\hspace{0.006\linewidth}\parbox[t]{0.115\linewidth}{\centering\strut -107.583037}\hspace{0.006\linewidth}\parbox[t]{0.671\linewidth}{num\_blocks = 27}\par\smallskip
\noindent{\setlength{\fboxsep}{0pt}\colorbox{black!4}{\parbox{\linewidth}{\parbox[t]{0.05\linewidth}{\centering\strut 9}\hspace{0.006\linewidth}\parbox[t]{0.04\linewidth}{\centering\strut \raisebox{0.08ex}{\ding{55}}}\hspace{0.006\linewidth}\parbox[t]{0.10\linewidth}{\centering\strut -0.0019\%}\hspace{0.006\linewidth}\parbox[t]{0.115\linewidth}{\centering\strut -107.581034}\hspace{0.006\linewidth}\parbox[t]{0.671\linewidth}{num\_blocks = 28}}}}\par\smallskip
\noindent\parbox[t]{0.05\linewidth}{\centering\strut 10}\hspace{0.006\linewidth}\parbox[t]{0.04\linewidth}{\centering\strut \raisebox{0.08ex}{\ding{51}}}\hspace{0.006\linewidth}\parbox[t]{0.10\linewidth}{\centering\strut +0.0071\%}\hspace{0.006\linewidth}\parbox[t]{0.115\linewidth}{\centering\strut -107.588645}\hspace{0.006\linewidth}\parbox[t]{0.671\linewidth}{num\_blocks = 23}\par\smallskip
\noindent{\setlength{\fboxsep}{0pt}\colorbox{black!4}{\parbox{\linewidth}{\parbox[t]{0.05\linewidth}{\centering\strut 11}\hspace{0.006\linewidth}\parbox[t]{0.04\linewidth}{\centering\strut \raisebox{0.08ex}{\ding{51}}}\hspace{0.006\linewidth}\parbox[t]{0.10\linewidth}{\centering\strut +0.0134\%}\hspace{0.006\linewidth}\parbox[t]{0.115\linewidth}{\centering\strut -107.603090}\hspace{0.006\linewidth}\parbox[t]{0.671\linewidth}{num\_blocks = 22}}}}\par\smallskip
\noindent\parbox[t]{0.05\linewidth}{\centering\strut 12}\hspace{0.006\linewidth}\parbox[t]{0.04\linewidth}{\centering\strut \raisebox{0.08ex}{\ding{55}}}\hspace{0.006\linewidth}\parbox[t]{0.10\linewidth}{\centering\strut -0.0078\%}\hspace{0.006\linewidth}\parbox[t]{0.115\linewidth}{\centering\strut -107.594683}\hspace{0.006\linewidth}\parbox[t]{0.671\linewidth}{num\_blocks = 21}\par\smallskip
\noindent{\setlength{\fboxsep}{0pt}\colorbox{black!4}{\parbox{\linewidth}{\parbox[t]{0.05\linewidth}{\centering\strut 13}\hspace{0.006\linewidth}\parbox[t]{0.04\linewidth}{\centering\strut \raisebox{0.08ex}{\ding{55}}}\hspace{0.006\linewidth}\parbox[t]{0.10\linewidth}{\centering\strut +0.0055\%}\hspace{0.006\linewidth}\parbox[t]{0.115\linewidth}{\centering\strut -107.600654}\hspace{0.006\linewidth}\parbox[t]{0.671\linewidth}{num\_blocks = 20}}}}\par\smallskip
\noindent\parbox[t]{0.05\linewidth}{\centering\strut 14}\hspace{0.006\linewidth}\parbox[t]{0.04\linewidth}{\centering\strut \raisebox{0.08ex}{\ding{51}}}\hspace{0.006\linewidth}\parbox[t]{0.10\linewidth}{\centering\strut +0.0051\%}\hspace{0.006\linewidth}\parbox[t]{0.115\linewidth}{\centering\strut -107.606091}\hspace{0.006\linewidth}\parbox[t]{0.671\linewidth}{timestep = 0.003474272676}\par\smallskip
\noindent{\setlength{\fboxsep}{0pt}\colorbox{black!4}{\parbox{\linewidth}{\parbox[t]{0.05\linewidth}{\centering\strut 15}\hspace{0.006\linewidth}\parbox[t]{0.04\linewidth}{\centering\strut \raisebox{0.08ex}{\ding{55}}}\hspace{0.006\linewidth}\parbox[t]{0.10\linewidth}{\centering\strut -0.0039\%}\hspace{0.006\linewidth}\parbox[t]{0.115\linewidth}{\centering\strut -107.601944}\hspace{0.006\linewidth}\parbox[t]{0.671\linewidth}{num\_blocks = 21}}}}\par\smallskip
\noindent\parbox[t]{0.05\linewidth}{\centering\strut 16}\hspace{0.006\linewidth}\parbox[t]{0.04\linewidth}{\centering\strut \raisebox{0.08ex}{\ding{55}}}\hspace{0.006\linewidth}\parbox[t]{0.10\linewidth}{\centering\strut -0.0008\%}\hspace{0.006\linewidth}\parbox[t]{0.115\linewidth}{\centering\strut -107.601117}\hspace{0.006\linewidth}\parbox[t]{0.671\linewidth}{num\_blocks = 23}\par\smallskip
\noindent{\setlength{\fboxsep}{0pt}\colorbox{black!4}{\parbox{\linewidth}{\parbox[t]{0.05\linewidth}{\centering\strut 17}\hspace{0.006\linewidth}\parbox[t]{0.04\linewidth}{\centering\strut \raisebox{0.08ex}{\ding{55}}}\hspace{0.006\linewidth}\parbox[t]{0.10\linewidth}{\centering\strut +0.0009\%}\hspace{0.006\linewidth}\parbox[t]{0.115\linewidth}{\centering\strut -107.602047}\hspace{0.006\linewidth}\parbox[t]{0.671\linewidth}{num\_blocks = 20}}}}\par\smallskip
\noindent\parbox[t]{0.05\linewidth}{\centering\strut 18}\hspace{0.006\linewidth}\parbox[t]{0.04\linewidth}{\centering\strut \raisebox{0.08ex}{\ding{55}}}\hspace{0.006\linewidth}\parbox[t]{0.10\linewidth}{\centering\strut +0.0030\%}\hspace{0.006\linewidth}\parbox[t]{0.115\linewidth}{\centering\strut -107.605321}\hspace{0.006\linewidth}\parbox[t]{0.671\linewidth}{num\_blocks = 24}\par\smallskip
\noindent{\setlength{\fboxsep}{0pt}\colorbox{black!4}{\parbox{\linewidth}{\parbox[t]{0.05\linewidth}{\centering\strut 19}\hspace{0.006\linewidth}\parbox[t]{0.04\linewidth}{\centering\strut \raisebox{0.08ex}{\ding{51}}}\hspace{0.006\linewidth}\parbox[t]{0.10\linewidth}{\centering\strut +0.0008\%}\hspace{0.006\linewidth}\parbox[t]{0.115\linewidth}{\centering\strut -107.606184}\hspace{0.006\linewidth}\parbox[t]{0.671\linewidth}{num\_blocks = 25}}}}\par\smallskip
\noindent\parbox[t]{0.05\linewidth}{\centering\strut 20}\hspace{0.006\linewidth}\parbox[t]{0.04\linewidth}{\centering\strut \raisebox{0.08ex}{\ding{51}}}\hspace{0.006\linewidth}\parbox[t]{0.10\linewidth}{\centering\strut +0.0019\%}\hspace{0.006\linewidth}\parbox[t]{0.115\linewidth}{\centering\strut -107.608187}\hspace{0.006\linewidth}\parbox[t]{0.671\linewidth}{num\_blocks = 26}\par\smallskip
\noindent{\setlength{\fboxsep}{0pt}\colorbox{black!4}{\parbox{\linewidth}{\parbox[t]{0.05\linewidth}{\centering\strut 21}\hspace{0.006\linewidth}\parbox[t]{0.04\linewidth}{\centering\strut \raisebox{0.08ex}{\ding{51}}}\hspace{0.006\linewidth}\parbox[t]{0.10\linewidth}{\centering\strut +0.0012\%}\hspace{0.006\linewidth}\parbox[t]{0.115\linewidth}{\centering\strut -107.609459}\hspace{0.006\linewidth}\parbox[t]{0.671\linewidth}{num\_blocks = 27}}}}\par\smallskip
\noindent\parbox[t]{0.05\linewidth}{\centering\strut 22}\hspace{0.006\linewidth}\parbox[t]{0.04\linewidth}{\centering\strut \raisebox{0.08ex}{\ding{51}}}\hspace{0.006\linewidth}\parbox[t]{0.10\linewidth}{\centering\strut +0.0013\%}\hspace{0.006\linewidth}\parbox[t]{0.115\linewidth}{\centering\strut -107.610884}\hspace{0.006\linewidth}\parbox[t]{0.671\linewidth}{num\_blocks = 28}\par\smallskip
\noindent{\setlength{\fboxsep}{0pt}\colorbox{black!4}{\parbox{\linewidth}{\parbox[t]{0.05\linewidth}{\centering\strut 23}\hspace{0.006\linewidth}\parbox[t]{0.04\linewidth}{\centering\strut \raisebox{0.08ex}{\ding{51}}}\hspace{0.006\linewidth}\parbox[t]{0.10\linewidth}{\centering\strut +0.0005\%}\hspace{0.006\linewidth}\parbox[t]{0.115\linewidth}{\centering\strut -107.611398}\hspace{0.006\linewidth}\parbox[t]{0.671\linewidth}{num\_blocks = 29}}}}\par\smallskip
\noindent\parbox[t]{0.05\linewidth}{\centering\strut 24}\hspace{0.006\linewidth}\parbox[t]{0.04\linewidth}{\centering\strut \raisebox{0.08ex}{\ding{55}}}\hspace{0.006\linewidth}\parbox[t]{0.10\linewidth}{\centering\strut -0.0034\%}\hspace{0.006\linewidth}\parbox[t]{0.115\linewidth}{\centering\strut -107.607768}\hspace{0.006\linewidth}\parbox[t]{0.671\linewidth}{num\_blocks = 30}\par\smallskip
\noindent{\setlength{\fboxsep}{0pt}\colorbox{black!4}{\parbox{\linewidth}{\parbox[t]{0.05\linewidth}{\centering\strut 25}\hspace{0.006\linewidth}\parbox[t]{0.04\linewidth}{\centering\strut \raisebox{0.08ex}{\ding{55}}}\hspace{0.006\linewidth}\parbox[t]{0.10\linewidth}{\centering\strut +0.0008\%}\hspace{0.006\linewidth}\parbox[t]{0.115\linewidth}{\centering\strut -107.608611}\hspace{0.006\linewidth}\parbox[t]{0.671\linewidth}{num\_blocks = 31}}}}\par\smallskip
\noindent\parbox[t]{0.05\linewidth}{\centering\strut 26}\hspace{0.006\linewidth}\parbox[t]{0.04\linewidth}{\centering\strut \raisebox{0.08ex}{\ding{55}}}\hspace{0.006\linewidth}\parbox[t]{0.10\linewidth}{\centering\strut +0.0002\%}\hspace{0.006\linewidth}\parbox[t]{0.115\linewidth}{\centering\strut -107.608867}\hspace{0.006\linewidth}\parbox[t]{0.671\linewidth}{num\_blocks = 32}\par\smallskip
\noindent{\setlength{\fboxsep}{0pt}\colorbox{black!4}{\parbox{\linewidth}{\parbox[t]{0.05\linewidth}{\centering\strut 27}\hspace{0.006\linewidth}\parbox[t]{0.04\linewidth}{\centering\strut \raisebox{0.08ex}{\ding{55}}}\hspace{0.006\linewidth}\parbox[t]{0.10\linewidth}{\centering\strut +0.0004\%}\hspace{0.006\linewidth}\parbox[t]{0.115\linewidth}{\centering\strut -107.609261}\hspace{0.006\linewidth}\parbox[t]{0.671\linewidth}{num\_blocks = 34}}}}\par\smallskip
\noindent\parbox[t]{0.05\linewidth}{\centering\strut 28}\hspace{0.006\linewidth}\parbox[t]{0.04\linewidth}{\centering\strut \raisebox{0.08ex}{\ding{55}}}\hspace{0.006\linewidth}\parbox[t]{0.10\linewidth}{\centering\strut -0.0326\%}\hspace{0.006\linewidth}\parbox[t]{0.115\linewidth}{\centering\strut -107.574171}\hspace{0.006\linewidth}\parbox[t]{0.671\linewidth}{timestep = 0.00340478722248}\par\smallskip
\noindent{\setlength{\fboxsep}{0pt}\colorbox{black!4}{\parbox{\linewidth}{\parbox[t]{0.05\linewidth}{\centering\strut 29}\hspace{0.006\linewidth}\parbox[t]{0.04\linewidth}{\centering\strut \raisebox{0.08ex}{\ding{55}}}\hspace{0.006\linewidth}\parbox[t]{0.10\linewidth}{\centering\strut +0.0033\%}\hspace{0.006\linewidth}\parbox[t]{0.115\linewidth}{\centering\strut -107.577763}\hspace{0.006\linewidth}\parbox[t]{0.671\linewidth}{timestep = 0.00354375812952}}}}\par\smallskip
\noindent\parbox[t]{0.05\linewidth}{\centering\strut 30}\hspace{0.006\linewidth}\parbox[t]{0.04\linewidth}{\centering\strut \raisebox{0.08ex}{\ding{55}}}\hspace{0.006\linewidth}\parbox[t]{0.10\linewidth}{\centering\strut -0.0417\%}\hspace{0.006\linewidth}\parbox[t]{0.115\linewidth}{\centering\strut -107.532933}\hspace{0.006\linewidth}\parbox[t]{0.671\linewidth}{timestep = 0.0033005590422}\par\smallskip
\noindent{\setlength{\fboxsep}{0pt}\colorbox{black!4}{\parbox{\linewidth}{\parbox[t]{0.05\linewidth}{\centering\strut 31}\hspace{0.006\linewidth}\parbox[t]{0.04\linewidth}{\centering\strut \raisebox{0.08ex}{\ding{55}}}\hspace{0.006\linewidth}\parbox[t]{0.10\linewidth}{\centering\strut +0.0215\%}\hspace{0.006\linewidth}\parbox[t]{0.115\linewidth}{\centering\strut -107.556089}\hspace{0.006\linewidth}\parbox[t]{0.671\linewidth}{timestep = 0.0036479863098}}}}\par\smallskip
\noindent\parbox[t]{0.05\linewidth}{\centering\strut 32}\hspace{0.006\linewidth}\parbox[t]{0.04\linewidth}{\centering\strut \raisebox{0.08ex}{\ding{55}}}\hspace{0.006\linewidth}\parbox[t]{0.10\linewidth}{\centering\strut +0.0250\%}\hspace{0.006\linewidth}\parbox[t]{0.115\linewidth}{\centering\strut -107.583006}\hspace{0.006\linewidth}\parbox[t]{0.671\linewidth}{timestep = 0.0031268454084}\par\smallskip
\noindent{\setlength{\fboxsep}{0pt}\colorbox{black!4}{\parbox{\linewidth}{\parbox[t]{0.05\linewidth}{\centering\strut 33}\hspace{0.006\linewidth}\parbox[t]{0.04\linewidth}{\centering\strut \raisebox{0.08ex}{\ding{55}}}\hspace{0.006\linewidth}\parbox[t]{0.10\linewidth}{\centering\strut +0.0171\%}\hspace{0.006\linewidth}\parbox[t]{0.115\linewidth}{\centering\strut -107.601425}\hspace{0.006\linewidth}\parbox[t]{0.671\linewidth}{timestep = 0.0038216999436}}}}\par\smallskip
\noindent\parbox[t]{0.05\linewidth}{\centering\strut 34}\hspace{0.006\linewidth}\parbox[t]{0.04\linewidth}{\centering\strut \raisebox{0.08ex}{\ding{55}}}\hspace{0.006\linewidth}\parbox[t]{0.10\linewidth}{\centering\strut +0.0083\%}\hspace{0.006\linewidth}\parbox[t]{0.115\linewidth}{\centering\strut -107.610353}\hspace{0.006\linewidth}\parbox[t]{0.671\linewidth}{num\_steps\_per\_block = 22}\par\smallskip
\noindent{\setlength{\fboxsep}{0pt}\colorbox{black!4}{\parbox{\linewidth}{\parbox[t]{0.05\linewidth}{\centering\strut 35}\hspace{0.006\linewidth}\parbox[t]{0.04\linewidth}{\centering\strut \raisebox{0.08ex}{\ding{55}}}\hspace{0.006\linewidth}\parbox[t]{0.10\linewidth}{\centering\strut -0.0011\%}\hspace{0.006\linewidth}\parbox[t]{0.115\linewidth}{\centering\strut -107.609146}\hspace{0.006\linewidth}\parbox[t]{0.671\linewidth}{num\_steps\_per\_block = 24}}}}\par\smallskip
\noindent\parbox[t]{0.05\linewidth}{\centering\strut 36}\hspace{0.006\linewidth}\parbox[t]{0.04\linewidth}{\centering\strut \raisebox{0.08ex}{\ding{55}}}\hspace{0.006\linewidth}\parbox[t]{0.10\linewidth}{\centering\strut -0.0019\%}\hspace{0.006\linewidth}\parbox[t]{0.115\linewidth}{\centering\strut -107.607144}\hspace{0.006\linewidth}\parbox[t]{0.671\linewidth}{num\_steps\_per\_block = 21}\par\smallskip
\noindent{\setlength{\fboxsep}{0pt}\colorbox{black!4}{\parbox{\linewidth}{\parbox[t]{0.05\linewidth}{\centering\strut 37}\hspace{0.006\linewidth}\parbox[t]{0.04\linewidth}{\centering\strut \raisebox{0.08ex}{\ding{55}}}\hspace{0.006\linewidth}\parbox[t]{0.10\linewidth}{\centering\strut +0.0037\%}\hspace{0.006\linewidth}\parbox[t]{0.115\linewidth}{\centering\strut -107.611105}\hspace{0.006\linewidth}\parbox[t]{0.671\linewidth}{num\_steps\_per\_block = 25}}}}\par\smallskip
\noindent\parbox[t]{0.05\linewidth}{\centering\strut 38}\hspace{0.006\linewidth}\parbox[t]{0.04\linewidth}{\centering\strut \raisebox{0.08ex}{\ding{55}}}\hspace{0.006\linewidth}\parbox[t]{0.10\linewidth}{\centering\strut -0.0065\%}\hspace{0.006\linewidth}\parbox[t]{0.115\linewidth}{\centering\strut -107.604097}\hspace{0.006\linewidth}\parbox[t]{0.671\linewidth}{num\_steps\_per\_block = 20}\par\smallskip
\noindent{\setlength{\fboxsep}{0pt}\colorbox{black!4}{\parbox{\linewidth}{\parbox[t]{0.05\linewidth}{\centering\strut 39}\hspace{0.006\linewidth}\parbox[t]{0.04\linewidth}{\centering\strut \raisebox{0.08ex}{\ding{51}}}\hspace{0.006\linewidth}\parbox[t]{0.10\linewidth}{\centering\strut +0.0096\%}\hspace{0.006\linewidth}\parbox[t]{0.115\linewidth}{\centering\strut -107.614400}\hspace{0.006\linewidth}\parbox[t]{0.671\linewidth}{num\_steps\_per\_block = 26}}}}\par\smallskip
\noindent\parbox[t]{0.05\linewidth}{\centering\strut 40}\hspace{0.006\linewidth}\parbox[t]{0.04\linewidth}{\centering\strut \raisebox{0.08ex}{\ding{55}}}\hspace{0.006\linewidth}\parbox[t]{0.10\linewidth}{\centering\strut -0.0009\%}\hspace{0.006\linewidth}\parbox[t]{0.115\linewidth}{\centering\strut -107.613441}\hspace{0.006\linewidth}\parbox[t]{0.671\linewidth}{num\_blocks = 28}\par\smallskip
\noindent{\setlength{\fboxsep}{0pt}\colorbox{black!4}{\parbox{\linewidth}{\parbox[t]{0.05\linewidth}{\centering\strut 41}\hspace{0.006\linewidth}\parbox[t]{0.04\linewidth}{\centering\strut \raisebox{0.08ex}{\ding{55}}}\hspace{0.006\linewidth}\parbox[t]{0.10\linewidth}{\centering\strut +0.0005\%}\hspace{0.006\linewidth}\parbox[t]{0.115\linewidth}{\centering\strut -107.614011}\hspace{0.006\linewidth}\parbox[t]{0.671\linewidth}{num\_blocks = 30}}}}\par\smallskip
\noindent\parbox[t]{0.05\linewidth}{\centering\strut 42}\hspace{0.006\linewidth}\parbox[t]{0.04\linewidth}{\centering\strut \raisebox{0.08ex}{\ding{55}}}\hspace{0.006\linewidth}\parbox[t]{0.10\linewidth}{\centering\strut -0.0008\%}\hspace{0.006\linewidth}\parbox[t]{0.115\linewidth}{\centering\strut -107.613136}\hspace{0.006\linewidth}\parbox[t]{0.671\linewidth}{num\_blocks = 27}\par\smallskip
\noindent{\setlength{\fboxsep}{0pt}\colorbox{black!4}{\parbox{\linewidth}{\parbox[t]{0.05\linewidth}{\centering\strut 43}\hspace{0.006\linewidth}\parbox[t]{0.04\linewidth}{\centering\strut \raisebox{0.08ex}{\ding{55}}}\hspace{0.006\linewidth}\parbox[t]{0.10\linewidth}{\centering\strut +0.0009\%}\hspace{0.006\linewidth}\parbox[t]{0.115\linewidth}{\centering\strut -107.614128}\hspace{0.006\linewidth}\parbox[t]{0.671\linewidth}{num\_blocks = 31}}}}\par\smallskip
\noindent\parbox[t]{0.05\linewidth}{\centering\strut 44}\hspace{0.006\linewidth}\parbox[t]{0.04\linewidth}{\centering\strut \raisebox{0.08ex}{\ding{55}}}\hspace{0.006\linewidth}\parbox[t]{0.10\linewidth}{\centering\strut -0.0017\%}\hspace{0.006\linewidth}\parbox[t]{0.115\linewidth}{\centering\strut -107.612263}\hspace{0.006\linewidth}\parbox[t]{0.671\linewidth}{num\_blocks = 26}\par\smallskip
\noindent{\setlength{\fboxsep}{0pt}\colorbox{black!4}{\parbox{\linewidth}{\parbox[t]{0.05\linewidth}{\centering\strut 45}\hspace{0.006\linewidth}\parbox[t]{0.04\linewidth}{\centering\strut \raisebox{0.08ex}{\ding{55}}}\hspace{0.006\linewidth}\parbox[t]{0.10\linewidth}{\centering\strut -0.0002\%}\hspace{0.006\linewidth}\parbox[t]{0.115\linewidth}{\centering\strut -107.612034}\hspace{0.006\linewidth}\parbox[t]{0.671\linewidth}{num\_blocks = 32}}}}\par\smallskip
\noindent\parbox[t]{0.05\linewidth}{\centering\strut 46}\hspace{0.006\linewidth}\parbox[t]{0.04\linewidth}{\centering\strut \raisebox{0.08ex}{\ding{55}}}\hspace{0.006\linewidth}\parbox[t]{0.10\linewidth}{\centering\strut -0.0003\%}\hspace{0.006\linewidth}\parbox[t]{0.115\linewidth}{\centering\strut -107.611749}\hspace{0.006\linewidth}\parbox[t]{0.671\linewidth}{num\_blocks = 24}\par\smallskip
\noindent{\setlength{\fboxsep}{0pt}\colorbox{black!4}{\parbox{\linewidth}{\parbox[t]{0.05\linewidth}{\centering\strut 47}\hspace{0.006\linewidth}\parbox[t]{0.04\linewidth}{\centering\strut \raisebox{0.08ex}{\ding{55}}}\hspace{0.006\linewidth}\parbox[t]{0.10\linewidth}{\centering\strut -0.0108\%}\hspace{0.006\linewidth}\parbox[t]{0.115\linewidth}{\centering\strut -107.600100}\hspace{0.006\linewidth}\parbox[t]{0.671\linewidth}{num\_blocks = 34}}}}\par\smallskip
\noindent\parbox[t]{0.05\linewidth}{\centering\strut 48}\hspace{0.006\linewidth}\parbox[t]{0.04\linewidth}{\centering\strut \raisebox{0.08ex}{\ding{55}}}\hspace{0.006\linewidth}\parbox[t]{0.10\linewidth}{\centering\strut -0.0259\%}\hspace{0.006\linewidth}\parbox[t]{0.115\linewidth}{\centering\strut -107.572215}\hspace{0.006\linewidth}\parbox[t]{0.671\linewidth}{timestep = 0.00340478722248}\par\smallskip
\noindent{\setlength{\fboxsep}{0pt}\colorbox{black!4}{\parbox{\linewidth}{\parbox[t]{0.05\linewidth}{\centering\strut 49}\hspace{0.006\linewidth}\parbox[t]{0.04\linewidth}{\centering\strut \raisebox{0.08ex}{\ding{55}}}\hspace{0.006\linewidth}\parbox[t]{0.10\linewidth}{\centering\strut +0.0072\%}\hspace{0.006\linewidth}\parbox[t]{0.115\linewidth}{\centering\strut -107.579969}\hspace{0.006\linewidth}\parbox[t]{0.671\linewidth}{timestep = 0.00354375812952}}}}\par\smallskip
\noindent\parbox[t]{0.05\linewidth}{\centering\strut 50}\hspace{0.006\linewidth}\parbox[t]{0.04\linewidth}{\centering\strut \raisebox{0.08ex}{\ding{55}}}\hspace{0.006\linewidth}\parbox[t]{0.10\linewidth}{\centering\strut -0.0156\%}\hspace{0.006\linewidth}\parbox[t]{0.115\linewidth}{\centering\strut -107.563158}\hspace{0.006\linewidth}\parbox[t]{0.671\linewidth}{timestep = 0.0033005590422}\par\smallskip
\noindent{\setlength{\fboxsep}{0pt}\colorbox{black!4}{\parbox{\linewidth}{\parbox[t]{0.05\linewidth}{\centering\strut 51}\hspace{0.006\linewidth}\parbox[t]{0.04\linewidth}{\centering\strut \raisebox{0.08ex}{\ding{55}}}\hspace{0.006\linewidth}\parbox[t]{0.10\linewidth}{\centering\strut +0.0240\%}\hspace{0.006\linewidth}\parbox[t]{0.115\linewidth}{\centering\strut -107.588966}\hspace{0.006\linewidth}\parbox[t]{0.671\linewidth}{timestep = 0.0036479863098}}}}\par\smallskip
\noindent\parbox[t]{0.05\linewidth}{\centering\strut 52}\hspace{0.006\linewidth}\parbox[t]{0.04\linewidth}{\centering\strut \raisebox{0.08ex}{\ding{55}}}\hspace{0.006\linewidth}\parbox[t]{0.10\linewidth}{\centering\strut +0.0008\%}\hspace{0.006\linewidth}\parbox[t]{0.115\linewidth}{\centering\strut -107.589867}\hspace{0.006\linewidth}\parbox[t]{0.671\linewidth}{timestep = 0.0031268454084}\par\smallskip
\noindent{\setlength{\fboxsep}{0pt}\colorbox{black!4}{\parbox{\linewidth}{\parbox[t]{0.05\linewidth}{\centering\strut 53}\hspace{0.006\linewidth}\parbox[t]{0.04\linewidth}{\centering\strut \raisebox{0.08ex}{\ding{51}}}\hspace{0.006\linewidth}\parbox[t]{0.10\linewidth}{\centering\strut +0.0285\%}\hspace{0.006\linewidth}\parbox[t]{0.115\linewidth}{\centering\strut -107.620504}\hspace{0.006\linewidth}\parbox[t]{0.671\linewidth}{timestep = 0.0038216999436}}}}\par\smallskip
\noindent\parbox[t]{0.05\linewidth}{\centering\strut 54}\hspace{0.006\linewidth}\parbox[t]{0.04\linewidth}{\centering\strut \raisebox{0.08ex}{\ding{55}}}\hspace{0.006\linewidth}\parbox[t]{0.10\linewidth}{\centering\strut -0.0017\%}\hspace{0.006\linewidth}\parbox[t]{0.115\linewidth}{\centering\strut -107.618646}\hspace{0.006\linewidth}\parbox[t]{0.671\linewidth}{num\_blocks = 28}\par\smallskip
\noindent{\setlength{\fboxsep}{0pt}\colorbox{black!4}{\parbox{\linewidth}{\parbox[t]{0.05\linewidth}{\centering\strut 55}\hspace{0.006\linewidth}\parbox[t]{0.04\linewidth}{\centering\strut \raisebox{0.08ex}{\ding{55}}}\hspace{0.006\linewidth}\parbox[t]{0.10\linewidth}{\centering\strut -0.0069\%}\hspace{0.006\linewidth}\parbox[t]{0.115\linewidth}{\centering\strut -107.611264}\hspace{0.006\linewidth}\parbox[t]{0.671\linewidth}{num\_blocks = 30}}}}\par\smallskip
\noindent\parbox[t]{0.05\linewidth}{\centering\strut 56}\hspace{0.006\linewidth}\parbox[t]{0.04\linewidth}{\centering\strut \raisebox{0.08ex}{\ding{55}}}\hspace{0.006\linewidth}\parbox[t]{0.10\linewidth}{\centering\strut -0.0021\%}\hspace{0.006\linewidth}\parbox[t]{0.115\linewidth}{\centering\strut -107.608999}\hspace{0.006\linewidth}\parbox[t]{0.671\linewidth}{num\_blocks = 27}\par\smallskip
\noindent{\setlength{\fboxsep}{0pt}\colorbox{black!4}{\parbox{\linewidth}{\parbox[t]{0.05\linewidth}{\centering\strut 57}\hspace{0.006\linewidth}\parbox[t]{0.04\linewidth}{\centering\strut \raisebox{0.08ex}{\ding{55}}}\hspace{0.006\linewidth}\parbox[t]{0.10\linewidth}{\centering\strut -0.0032\%}\hspace{0.006\linewidth}\parbox[t]{0.115\linewidth}{\centering\strut -107.605503}\hspace{0.006\linewidth}\parbox[t]{0.671\linewidth}{num\_blocks = 31}}}}\par\smallskip
\noindent\parbox[t]{0.05\linewidth}{\centering\strut 58}\hspace{0.006\linewidth}\parbox[t]{0.04\linewidth}{\centering\strut \raisebox{0.08ex}{\ding{55}}}\hspace{0.006\linewidth}\parbox[t]{0.10\linewidth}{\centering\strut +0.0009\%}\hspace{0.006\linewidth}\parbox[t]{0.115\linewidth}{\centering\strut -107.606485}\hspace{0.006\linewidth}\parbox[t]{0.671\linewidth}{num\_blocks = 26}\par\smallskip
\noindent{\setlength{\fboxsep}{0pt}\colorbox{black!4}{\parbox{\linewidth}{\parbox[t]{0.05\linewidth}{\centering\strut 59}\hspace{0.006\linewidth}\parbox[t]{0.04\linewidth}{\centering\strut \raisebox{0.08ex}{\ding{55}}}\hspace{0.006\linewidth}\parbox[t]{0.10\linewidth}{\centering\strut -0.0162\%}\hspace{0.006\linewidth}\parbox[t]{0.115\linewidth}{\centering\strut -107.589095}\hspace{0.006\linewidth}\parbox[t]{0.671\linewidth}{num\_blocks = 32}}}}\par\smallskip
\noindent\parbox[t]{0.05\linewidth}{\centering\strut 60}\hspace{0.006\linewidth}\parbox[t]{0.04\linewidth}{\centering\strut \raisebox{0.08ex}{\ding{55}}}\hspace{0.006\linewidth}\parbox[t]{0.10\linewidth}{\centering\strut -0.0025\%}\hspace{0.006\linewidth}\parbox[t]{0.115\linewidth}{\centering\strut -107.586371}\hspace{0.006\linewidth}\parbox[t]{0.671\linewidth}{num\_blocks = 24}\par\smallskip
\noindent{\setlength{\fboxsep}{0pt}\colorbox{black!4}{\parbox{\linewidth}{\parbox[t]{0.05\linewidth}{\centering\strut 61}\hspace{0.006\linewidth}\parbox[t]{0.04\linewidth}{\centering\strut \raisebox{0.08ex}{\ding{55}}}\hspace{0.006\linewidth}\parbox[t]{0.10\linewidth}{\centering\strut -0.0027\%}\hspace{0.006\linewidth}\parbox[t]{0.115\linewidth}{\centering\strut -107.583513}\hspace{0.006\linewidth}\parbox[t]{0.671\linewidth}{num\_blocks = 34}}}}\par\smallskip
\noindent\parbox[t]{0.05\linewidth}{\centering\strut 62}\hspace{0.006\linewidth}\parbox[t]{0.04\linewidth}{\centering\strut \raisebox{0.08ex}{\ding{55}}}\hspace{0.006\linewidth}\parbox[t]{0.10\linewidth}{\centering\strut -0.0229\%}\hspace{0.006\linewidth}\parbox[t]{0.115\linewidth}{\centering\strut -107.558917}\hspace{0.006\linewidth}\parbox[t]{0.671\linewidth}{timestep = 0.003745265944728}\par\smallskip
\noindent{\setlength{\fboxsep}{0pt}\colorbox{black!4}{\parbox{\linewidth}{\parbox[t]{0.05\linewidth}{\centering\strut 63}\hspace{0.006\linewidth}\parbox[t]{0.04\linewidth}{\centering\strut \raisebox{0.08ex}{\ding{55}}}\hspace{0.006\linewidth}\parbox[t]{0.10\linewidth}{\centering\strut +0.0328\%}\hspace{0.006\linewidth}\parbox[t]{0.115\linewidth}{\centering\strut -107.594203}\hspace{0.006\linewidth}\parbox[t]{0.671\linewidth}{timestep = 0.003898133942472}}}}\par\smallskip
\noindent\parbox[t]{0.05\linewidth}{\centering\strut 64}\hspace{0.006\linewidth}\parbox[t]{0.04\linewidth}{\centering\strut \raisebox{0.08ex}{\ding{55}}}\hspace{0.006\linewidth}\parbox[t]{0.10\linewidth}{\centering\strut -0.0112\%}\hspace{0.006\linewidth}\parbox[t]{0.115\linewidth}{\centering\strut -107.582109}\hspace{0.006\linewidth}\parbox[t]{0.671\linewidth}{timestep = 0.00363061494642}\par\smallskip
\noindent{\setlength{\fboxsep}{0pt}\colorbox{black!4}{\parbox{\linewidth}{\parbox[t]{0.05\linewidth}{\centering\strut 65}\hspace{0.006\linewidth}\parbox[t]{0.04\linewidth}{\centering\strut \raisebox{0.08ex}{\ding{55}}}\hspace{0.006\linewidth}\parbox[t]{0.10\linewidth}{\centering\strut -0.0012\%}\hspace{0.006\linewidth}\parbox[t]{0.115\linewidth}{\centering\strut -107.580834}\hspace{0.006\linewidth}\parbox[t]{0.671\linewidth}{timestep = 0.00401278494078}}}}\par\smallskip
\noindent\parbox[t]{0.05\linewidth}{\centering\strut 66}\hspace{0.006\linewidth}\parbox[t]{0.04\linewidth}{\centering\strut \raisebox{0.08ex}{\ding{55}}}\hspace{0.006\linewidth}\parbox[t]{0.10\linewidth}{\centering\strut +0.0331\%}\hspace{0.006\linewidth}\parbox[t]{0.115\linewidth}{\centering\strut -107.616392}\hspace{0.006\linewidth}\parbox[t]{0.671\linewidth}{timestep = 0.00343952994924}\par\smallskip
\noindent{\setlength{\fboxsep}{0pt}\colorbox{black!4}{\parbox{\linewidth}{\parbox[t]{0.05\linewidth}{\centering\strut 67}\hspace{0.006\linewidth}\parbox[t]{0.04\linewidth}{\centering\strut \raisebox{0.08ex}{\ding{55}}}\hspace{0.006\linewidth}\parbox[t]{0.10\linewidth}{\centering\strut -0.0003\%}\hspace{0.006\linewidth}\parbox[t]{0.115\linewidth}{\centering\strut -107.616118}\hspace{0.006\linewidth}\parbox[t]{0.671\linewidth}{timestep = 0.00420386993796}}}}\par\smallskip
\noindent\parbox[t]{0.05\linewidth}{\centering\strut 68}\hspace{0.006\linewidth}\parbox[t]{0.04\linewidth}{\centering\strut \raisebox{0.08ex}{\ding{55}}}\hspace{0.006\linewidth}\parbox[t]{0.10\linewidth}{\centering\strut -0.0067\%}\hspace{0.006\linewidth}\parbox[t]{0.115\linewidth}{\centering\strut -107.608919}\hspace{0.006\linewidth}\parbox[t]{0.671\linewidth}{num\_steps\_per\_block = 25}\par\smallskip
\noindent{\setlength{\fboxsep}{0pt}\colorbox{black!4}{\parbox{\linewidth}{\parbox[t]{0.05\linewidth}{\centering\strut 69}\hspace{0.006\linewidth}\parbox[t]{0.04\linewidth}{\centering\strut \raisebox{0.08ex}{\ding{55}}}\hspace{0.006\linewidth}\parbox[t]{0.10\linewidth}{\centering\strut -0.0090\%}\hspace{0.006\linewidth}\parbox[t]{0.115\linewidth}{\centering\strut -107.599283}\hspace{0.006\linewidth}\parbox[t]{0.671\linewidth}{num\_steps\_per\_block = 27}}}}\par\smallskip
\noindent\parbox[t]{0.05\linewidth}{\centering\strut 70}\hspace{0.006\linewidth}\parbox[t]{0.04\linewidth}{\centering\strut \raisebox{0.08ex}{\ding{55}}}\hspace{0.006\linewidth}\parbox[t]{0.10\linewidth}{\centering\strut +0.0036\%}\hspace{0.006\linewidth}\parbox[t]{0.115\linewidth}{\centering\strut -107.603161}\hspace{0.006\linewidth}\parbox[t]{0.671\linewidth}{num\_steps\_per\_block = 24}\par\smallskip
\noindent{\setlength{\fboxsep}{0pt}\colorbox{black!4}{\parbox{\linewidth}{\parbox[t]{0.05\linewidth}{\centering\strut 71}\hspace{0.006\linewidth}\parbox[t]{0.04\linewidth}{\centering\strut \raisebox{0.08ex}{\ding{55}}}\hspace{0.006\linewidth}\parbox[t]{0.10\linewidth}{\centering\strut -0.0088\%}\hspace{0.006\linewidth}\parbox[t]{0.115\linewidth}{\centering\strut -107.593727}\hspace{0.006\linewidth}\parbox[t]{0.671\linewidth}{num\_steps\_per\_block = 28}}}}\par\smallskip
\noindent\parbox[t]{0.05\linewidth}{\centering\strut 72}\hspace{0.006\linewidth}\parbox[t]{0.04\linewidth}{\centering\strut \raisebox{0.08ex}{\ding{55}}}\hspace{0.006\linewidth}\parbox[t]{0.10\linewidth}{\centering\strut -0.0043\%}\hspace{0.006\linewidth}\parbox[t]{0.115\linewidth}{\centering\strut -107.589111}\hspace{0.006\linewidth}\parbox[t]{0.671\linewidth}{num\_steps\_per\_block = 29}\par\smallskip
\noindent{\setlength{\fboxsep}{0pt}\colorbox{black!4}{\parbox{\linewidth}{\parbox[t]{0.05\linewidth}{\centering\strut 73}\hspace{0.006\linewidth}\parbox[t]{0.04\linewidth}{\centering\strut \raisebox{0.08ex}{\ding{55}}}\hspace{0.006\linewidth}\parbox[t]{0.10\linewidth}{\centering\strut +0.0022\%}\hspace{0.006\linewidth}\parbox[t]{0.115\linewidth}{\centering\strut -107.591456}\hspace{0.006\linewidth}\parbox[t]{0.671\linewidth}{num\_blocks = 28, timestep = 0.0039581892273}}}}\par\smallskip
\noindent\parbox[t]{0.05\linewidth}{\centering\strut 74}\hspace{0.006\linewidth}\parbox[t]{0.04\linewidth}{\centering\strut \raisebox{0.08ex}{\ding{55}}}\hspace{0.006\linewidth}\parbox[t]{0.10\linewidth}{\centering\strut +0.0062\%}\hspace{0.006\linewidth}\parbox[t]{0.115\linewidth}{\centering\strut -107.598082}\hspace{0.006\linewidth}\parbox[t]{0.671\linewidth}{num\_blocks = 30, timestep = 0.00369430994548}\par\smallskip
\noindent{\setlength{\fboxsep}{0pt}\colorbox{black!4}{\parbox{\linewidth}{\parbox[t]{0.05\linewidth}{\centering\strut 75}\hspace{0.006\linewidth}\parbox[t]{0.04\linewidth}{\centering\strut \raisebox{0.08ex}{\ding{55}}}\hspace{0.006\linewidth}\parbox[t]{0.10\linewidth}{\centering\strut -0.0084\%}\hspace{0.006\linewidth}\parbox[t]{0.115\linewidth}{\centering\strut -107.589093}\hspace{0.006\linewidth}\parbox[t]{0.671\linewidth}{num\_blocks = 27, timestep = 0.004104788828311111}}}}\par\smallskip
\noindent\parbox[t]{0.05\linewidth}{\centering\strut 76}\hspace{0.006\linewidth}\parbox[t]{0.04\linewidth}{\centering\strut \raisebox{0.08ex}{\ding{55}}}\hspace{0.006\linewidth}\parbox[t]{0.10\linewidth}{\centering\strut +0.0053\%}\hspace{0.006\linewidth}\parbox[t]{0.115\linewidth}{\centering\strut -107.594825}\hspace{0.006\linewidth}\parbox[t]{0.671\linewidth}{num\_blocks = 31, timestep = 0.003575138656916129}\par\smallskip
\noindent{\setlength{\fboxsep}{0pt}\colorbox{black!4}{\parbox{\linewidth}{\parbox[t]{0.05\linewidth}{\centering\strut 77}\hspace{0.006\linewidth}\parbox[t]{0.04\linewidth}{\centering\strut \raisebox{0.08ex}{\ding{55}}}\hspace{0.006\linewidth}\parbox[t]{0.10\linewidth}{\centering\strut +0.0093\%}\hspace{0.006\linewidth}\parbox[t]{0.115\linewidth}{\centering\strut -107.604862}\hspace{0.006\linewidth}\parbox[t]{0.671\linewidth}{num\_blocks = 26, timestep = 0.004262665321707691}}}}\par\smallskip
\noindent\parbox[t]{0.05\linewidth}{\centering\strut 78}\hspace{0.006\linewidth}\parbox[t]{0.04\linewidth}{\centering\strut \raisebox{0.08ex}{\ding{55}}}\hspace{0.006\linewidth}\parbox[t]{0.10\linewidth}{\centering\strut +0.0115\%}\hspace{0.006\linewidth}\parbox[t]{0.115\linewidth}{\centering\strut -107.617222}\hspace{0.006\linewidth}\parbox[t]{0.671\linewidth}{num\_blocks = 32, timestep = 0.0034634155738875}\par\smallskip
\noindent{\setlength{\fboxsep}{0pt}\colorbox{black!4}{\parbox{\linewidth}{\parbox[t]{0.05\linewidth}{\centering\strut 79}\hspace{0.006\linewidth}\parbox[t]{0.04\linewidth}{\centering\strut \raisebox{0.08ex}{\ding{55}}}\hspace{0.006\linewidth}\parbox[t]{0.10\linewidth}{\centering\strut -0.0201\%}\hspace{0.006\linewidth}\parbox[t]{0.115\linewidth}{\centering\strut -107.595563}\hspace{0.006\linewidth}\parbox[t]{0.671\linewidth}{num\_steps\_per\_block = 25, timestep = 0.003974567941344}}}}\par\smallskip
\noindent\parbox[t]{0.05\linewidth}{\centering\strut 80}\hspace{0.006\linewidth}\parbox[t]{0.04\linewidth}{\centering\strut \raisebox{0.08ex}{\ding{55}}}\hspace{0.006\linewidth}\parbox[t]{0.10\linewidth}{\centering\strut -0.0030\%}\hspace{0.006\linewidth}\parbox[t]{0.115\linewidth}{\centering\strut -107.592330}\hspace{0.006\linewidth}\parbox[t]{0.671\linewidth}{num\_steps\_per\_block = 27, timestep = 0.003680155501244444}\par\smallskip
\noindent{\setlength{\fboxsep}{0pt}\colorbox{black!4}{\parbox{\linewidth}{\parbox[t]{0.05\linewidth}{\centering\strut 81}\hspace{0.006\linewidth}\parbox[t]{0.04\linewidth}{\centering\strut \raisebox{0.08ex}{\ding{55}}}\hspace{0.006\linewidth}\parbox[t]{0.10\linewidth}{\centering\strut -0.0114\%}\hspace{0.006\linewidth}\parbox[t]{0.115\linewidth}{\centering\strut -107.580057}\hspace{0.006\linewidth}\parbox[t]{0.671\linewidth}{num\_steps\_per\_block = 24, timestep = 0.0041401749389}}}}\par\smallskip
\noindent\parbox[t]{0.05\linewidth}{\centering\strut 82}\hspace{0.006\linewidth}\parbox[t]{0.04\linewidth}{\centering\strut \raisebox{0.08ex}{\ding{55}}}\hspace{0.006\linewidth}\parbox[t]{0.10\linewidth}{\centering\strut +0.0025\%}\hspace{0.006\linewidth}\parbox[t]{0.115\linewidth}{\centering\strut -107.582705}\hspace{0.006\linewidth}\parbox[t]{0.671\linewidth}{num\_steps\_per\_block = 28, timestep = 0.0035487213762}\par\smallskip
\noindent{\setlength{\fboxsep}{0pt}\colorbox{black!4}{\parbox{\linewidth}{\parbox[t]{0.05\linewidth}{\centering\strut 83}\hspace{0.006\linewidth}\parbox[t]{0.04\linewidth}{\centering\strut \raisebox{0.08ex}{\ding{55}}}\hspace{0.006\linewidth}\parbox[t]{0.10\linewidth}{\centering\strut +0.0179\%}\hspace{0.006\linewidth}\parbox[t]{0.115\linewidth}{\centering\strut -107.601954}\hspace{0.006\linewidth}\parbox[t]{0.671\linewidth}{num\_steps\_per\_block = 23, timestep = 0.00432018254493913}}}}\par\smallskip
\noindent\parbox[t]{0.05\linewidth}{\centering\strut 84}\hspace{0.006\linewidth}\parbox[t]{0.04\linewidth}{\centering\strut \raisebox{0.08ex}{\ding{55}}}\hspace{0.006\linewidth}\parbox[t]{0.10\linewidth}{\centering\strut -0.0298\%}\hspace{0.006\linewidth}\parbox[t]{0.115\linewidth}{\centering\strut -107.569881}\hspace{0.006\linewidth}\parbox[t]{0.671\linewidth}{num\_steps\_per\_block = 29, timestep = 0.003426351673572414}\par\smallskip
\noindent{\setlength{\fboxsep}{0pt}\colorbox{black!4}{\parbox{\linewidth}{\parbox[t]{0.05\linewidth}{\centering\strut 85}\hspace{0.006\linewidth}\parbox[t]{0.04\linewidth}{\centering\strut \raisebox{0.08ex}{\ding{55}}}\hspace{0.006\linewidth}\parbox[t]{0.10\linewidth}{\centering\strut +0.0372\%}\hspace{0.006\linewidth}\parbox[t]{0.115\linewidth}{\centering\strut -107.609900}\hspace{0.006\linewidth}\parbox[t]{0.671\linewidth}{num\_steps\_per\_block = 27, num\_blocks = 28}}}}\par\smallskip
\noindent\parbox[t]{0.05\linewidth}{\centering\strut 86}\hspace{0.006\linewidth}\parbox[t]{0.04\linewidth}{\centering\strut \raisebox{0.08ex}{\ding{55}}}\hspace{0.006\linewidth}\parbox[t]{0.10\linewidth}{\centering\strut +0.0087\%}\hspace{0.006\linewidth}\parbox[t]{0.115\linewidth}{\centering\strut -107.619287}\hspace{0.006\linewidth}\parbox[t]{0.671\linewidth}{num\_steps\_per\_block = 28, num\_blocks = 27}\par\smallskip
\noindent{\setlength{\fboxsep}{0pt}\colorbox{black!4}{\parbox{\linewidth}{\parbox[t]{0.05\linewidth}{\centering\strut 87}\hspace{0.006\linewidth}\parbox[t]{0.04\linewidth}{\centering\strut \raisebox{0.08ex}{\ding{55}}}\hspace{0.006\linewidth}\parbox[t]{0.10\linewidth}{\centering\strut -0.0046\%}\hspace{0.006\linewidth}\parbox[t]{0.115\linewidth}{\centering\strut -107.614388}\hspace{0.006\linewidth}\parbox[t]{0.671\linewidth}{num\_steps\_per\_block = 25, num\_blocks = 30}}}}\par\smallskip
\noindent\parbox[t]{0.05\linewidth}{\centering\strut 88}\hspace{0.006\linewidth}\parbox[t]{0.04\linewidth}{\centering\strut \raisebox{0.08ex}{\ding{55}}}\hspace{0.006\linewidth}\parbox[t]{0.10\linewidth}{\centering\strut -0.0048\%}\hspace{0.006\linewidth}\parbox[t]{0.115\linewidth}{\centering\strut -107.609262}\hspace{0.006\linewidth}\parbox[t]{0.671\linewidth}{num\_steps\_per\_block = 24, num\_blocks = 31}\par\smallskip
\noindent{\setlength{\fboxsep}{0pt}\colorbox{black!4}{\parbox{\linewidth}{\parbox[t]{0.05\linewidth}{\centering\strut 89}\hspace{0.006\linewidth}\parbox[t]{0.04\linewidth}{\centering\strut \raisebox{0.08ex}{\ding{51}}}\hspace{0.006\linewidth}\parbox[t]{0.10\linewidth}{\centering\strut +0.0104\%}\hspace{0.006\linewidth}\parbox[t]{0.115\linewidth}{\centering\strut -107.620504}\hspace{0.006\linewidth}\parbox[t]{0.671\linewidth}{stabilize\_freq = 5}}}}\par\smallskip
\noindent\parbox[t]{0.05\linewidth}{\centering\strut 90}\hspace{0.006\linewidth}\parbox[t]{0.04\linewidth}{\centering\strut \raisebox{0.08ex}{\ding{55}}}\hspace{0.006\linewidth}\parbox[t]{0.10\linewidth}{\centering\strut -0.0017\%}\hspace{0.006\linewidth}\parbox[t]{0.115\linewidth}{\centering\strut -107.618646}\hspace{0.006\linewidth}\parbox[t]{0.671\linewidth}{num\_blocks = 28}\par\smallskip
\noindent{\setlength{\fboxsep}{0pt}\colorbox{black!4}{\parbox{\linewidth}{\parbox[t]{0.05\linewidth}{\centering\strut 91}\hspace{0.006\linewidth}\parbox[t]{0.04\linewidth}{\centering\strut \raisebox{0.08ex}{\ding{55}}}\hspace{0.006\linewidth}\parbox[t]{0.10\linewidth}{\centering\strut -0.0069\%}\hspace{0.006\linewidth}\parbox[t]{0.115\linewidth}{\centering\strut -107.611264}\hspace{0.006\linewidth}\parbox[t]{0.671\linewidth}{num\_blocks = 30}}}}\par\smallskip
\noindent\parbox[t]{0.05\linewidth}{\centering\strut 92}\hspace{0.006\linewidth}\parbox[t]{0.04\linewidth}{\centering\strut \raisebox{0.08ex}{\ding{55}}}\hspace{0.006\linewidth}\parbox[t]{0.10\linewidth}{\centering\strut -0.0021\%}\hspace{0.006\linewidth}\parbox[t]{0.115\linewidth}{\centering\strut -107.608999}\hspace{0.006\linewidth}\parbox[t]{0.671\linewidth}{num\_blocks = 27}\par\smallskip
\noindent{\setlength{\fboxsep}{0pt}\colorbox{black!4}{\parbox{\linewidth}{\parbox[t]{0.05\linewidth}{\centering\strut 93}\hspace{0.006\linewidth}\parbox[t]{0.04\linewidth}{\centering\strut \raisebox{0.08ex}{\ding{55}}}\hspace{0.006\linewidth}\parbox[t]{0.10\linewidth}{\centering\strut -0.0032\%}\hspace{0.006\linewidth}\parbox[t]{0.115\linewidth}{\centering\strut -107.605503}\hspace{0.006\linewidth}\parbox[t]{0.671\linewidth}{num\_blocks = 31}}}}\par\smallskip
\noindent\parbox[t]{0.05\linewidth}{\centering\strut 94}\hspace{0.006\linewidth}\parbox[t]{0.04\linewidth}{\centering\strut \raisebox{0.08ex}{\ding{55}}}\hspace{0.006\linewidth}\parbox[t]{0.10\linewidth}{\centering\strut +0.0009\%}\hspace{0.006\linewidth}\parbox[t]{0.115\linewidth}{\centering\strut -107.606485}\hspace{0.006\linewidth}\parbox[t]{0.671\linewidth}{num\_blocks = 26}\par\smallskip
\noindent{\setlength{\fboxsep}{0pt}\colorbox{black!4}{\parbox{\linewidth}{\parbox[t]{0.05\linewidth}{\centering\strut 95}\hspace{0.006\linewidth}\parbox[t]{0.04\linewidth}{\centering\strut \raisebox{0.08ex}{\ding{55}}}\hspace{0.006\linewidth}\parbox[t]{0.10\linewidth}{\centering\strut -0.0162\%}\hspace{0.006\linewidth}\parbox[t]{0.115\linewidth}{\centering\strut -107.589095}\hspace{0.006\linewidth}\parbox[t]{0.671\linewidth}{num\_blocks = 32}}}}\par\smallskip
\noindent\parbox[t]{0.05\linewidth}{\centering\strut 96}\hspace{0.006\linewidth}\parbox[t]{0.04\linewidth}{\centering\strut \raisebox{0.08ex}{\ding{55}}}\hspace{0.006\linewidth}\parbox[t]{0.10\linewidth}{\centering\strut -0.0025\%}\hspace{0.006\linewidth}\parbox[t]{0.115\linewidth}{\centering\strut -107.586371}\hspace{0.006\linewidth}\parbox[t]{0.671\linewidth}{num\_blocks = 24}\par\smallskip
\noindent{\setlength{\fboxsep}{0pt}\colorbox{black!4}{\parbox{\linewidth}{\parbox[t]{0.05\linewidth}{\centering\strut 97}\hspace{0.006\linewidth}\parbox[t]{0.04\linewidth}{\centering\strut \raisebox{0.08ex}{\ding{55}}}\hspace{0.006\linewidth}\parbox[t]{0.10\linewidth}{\centering\strut -0.0027\%}\hspace{0.006\linewidth}\parbox[t]{0.115\linewidth}{\centering\strut -107.583513}\hspace{0.006\linewidth}\parbox[t]{0.671\linewidth}{num\_blocks = 34}}}}\par\smallskip
\noindent\parbox[t]{0.05\linewidth}{\centering\strut 98}\hspace{0.006\linewidth}\parbox[t]{0.04\linewidth}{\centering\strut \raisebox{0.08ex}{\ding{55}}}\hspace{0.006\linewidth}\parbox[t]{0.10\linewidth}{\centering\strut -0.0229\%}\hspace{0.006\linewidth}\parbox[t]{0.115\linewidth}{\centering\strut -107.558917}\hspace{0.006\linewidth}\parbox[t]{0.671\linewidth}{timestep = 0.003745265944728}\par\smallskip
\noindent{\setlength{\fboxsep}{0pt}\colorbox{black!4}{\parbox{\linewidth}{\parbox[t]{0.05\linewidth}{\centering\strut 99}\hspace{0.006\linewidth}\parbox[t]{0.04\linewidth}{\centering\strut \raisebox{0.08ex}{\ding{55}}}\hspace{0.006\linewidth}\parbox[t]{0.10\linewidth}{\centering\strut +0.0328\%}\hspace{0.006\linewidth}\parbox[t]{0.115\linewidth}{\centering\strut -107.594203}\hspace{0.006\linewidth}\parbox[t]{0.671\linewidth}{timestep = 0.003898133942472}}}}\par\smallskip
\noindent\parbox[t]{0.05\linewidth}{\centering\strut 100}\hspace{0.006\linewidth}\parbox[t]{0.04\linewidth}{\centering\strut \raisebox{0.08ex}{\ding{55}}}\hspace{0.006\linewidth}\parbox[t]{0.10\linewidth}{\centering\strut -0.0112\%}\hspace{0.006\linewidth}\parbox[t]{0.115\linewidth}{\centering\strut -107.582109}\hspace{0.006\linewidth}\parbox[t]{0.671\linewidth}{timestep = 0.00363061494642}\par\smallskip
\noindent\rule{\linewidth}{0.5pt}
\par\medskip